\RequirePackage{lineno}
\documentclass[aps,prd,longbibliography,notitlepage,superscriptaddress,nofootinbib]{revtex4-1}[11pt]
\usepackage{graphicx}
\usepackage{amssymb,mathrsfs,empheq}
\usepackage{dsfont}
\usepackage{bm}
\usepackage{slashed}
\usepackage{dcolumn}
\usepackage[linktocpage=true]{hyperref}
\usepackage{amsmath}
\usepackage{float}
\usepackage[templates]{genealogytree}
\usepackage[utf8]{inputenc}
\usetikzlibrary{arrows,shapes,positioning,shadows,trees}
\usepackage{xcolor}
\usepackage{verbatim}

\begin{document}

\title{NLO impact factor for inclusive photon$+$dijet production in $e+A$ DIS at small $x$}

\author{Kaushik Roy}
\email{kaushik.roy.1@stonybrook.edu}
\affiliation{Department of Physics and Astronomy, Stony Brook University, Stony Brook, NY 11794, USA}
\affiliation{Physics Department, Brookhaven National Laboratory, Bldg. 510A, Upton, NY 11973, USA}

\author{Raju Venugopalan}
\email{raju@bnl.gov}
\affiliation{Physics Department, Brookhaven National Laboratory, Bldg. 510A, Upton, NY 11973, USA}

\date{\today}

\begin{abstract}
We compute the next-to-leading order (NLO) impact factor for inclusive photon $+$dijet production in electron-nucleus (e+A) deeply inelastic scattering (DIS) at small $x$. An important ingredient in our computation is the simple structure of ``shock wave" fermion and gluon propagators. This allows one to employ standard momentum space Feynman diagram techniques for higher order computations in the Regge limit of fixed $Q^2\gg \Lambda_{\rm QCD}^2$ and $x\rightarrow 0$. Our computations in the Color Glass Condensate (CGC) effective field theory include the resummation of all-twist power corrections $Q_s^2/Q^2$, where $Q_s$ is the saturation scale in the nucleus. We discuss the structure of ultraviolet, collinear and soft divergences in the CGC, and extract the leading logs in $x$; the structure of the corresponding rapidity divergences gives a nontrivial first principles derivation of the JIMWLK renormalization group evolution equation for multiparton lightlike Wilson line correlators. Explicit expressions are given for the $x$-independent  $O(\alpha_s)$ contributions that constitute the NLO impact factor. These results, combined with extant results on NLO JIMWLK evolution, provide the ingredients to compute the inclusive photon $+$ dijet cross-section at small $x$ to $O(\alpha_s^3 \ln(x))$.  First results for the NLO impact factor in inclusive dijet production are recovered in the soft photon limit. A byproduct of our computation is the LO photon+ 3 jet (quark-antiquark-gluon) cross-section. 

\end{abstract}

\maketitle

\tableofcontents 

\makeatletter
\let\toc@pre\relax
\let\toc@post\relax
\makeatother 

\section{Introduction} \label{sec:introduction}
An important discovery of the electron-proton ($e+p$) deep inelastic scattering (DIS) experiments at HERA was the rapid growth of the gluon distribution with decreasing Bjorken $x$, for fixed large momentum transfer squared $Q^2$. This demonstrated that the proton wavefunction in the corresponding high energy Regge limit is dominated by Fock state configurations containing large numbers of gluons. Their number grows via bremsstrahlung with increasing energy or decreasing $x$. It was conjectured that in this Regge limit repulsive gluon recombination and screening effects~\cite{Gribov:1984tu,Mueller:1985wy} conspire to slow down the growth of cross-sections.  A remarkable consequence is that a semi-hard saturation scale $Q_{s}(x)$ is generated dynamically by these competing many-body effects. 

If the saturation scale is large compared to the intrinsic QCD scale, asymptotic freedom suggests that the coupling $\alpha_s(Q_s)\ll 1$; this allows weak coupling effective field theory techniques to be employed in systematically computing cross-sections in a regime of QCD where field strengths are nonperturbatively large. The physics of this nonlinear regime of QCD can be quantified in a classical effective field theory (EFT) called the Color Glass Condensate (CGC)~\cite{McLerran:1993ni,McLerran:1993ka,McLerran:1994vd,Iancu:2003xm,Gelis:2010nm,Kovchegov:2012mbw,Blaizot:2016qgz} which implements a Born-Oppenheimer separation of the relevant degrees of freedom into static color sources at large $x$ and gauge fields at small $x$. 

A further important element in the EFT is that the separation in $x$ between color sources and fields satisfies a renormalization group (RG) equation as the scale separation is evolved towards smaller $x$. This can be understood in a Wilsonian picture wherein, with each small change $\delta x$ in the scale separation, the dynamical gauge degrees of freedom within  $\delta x$ are absorbed into the static light cone color sources in the classical EFT at the lower $x-\delta x$ scale. The RG equation correspondingly describes the change in a nonperturbative weight functional describing the distribution of color sources, from large $x$ towards smaller $x$, and efficiently resums simultaneously logarithms in $\alpha_s \ln(1/x)$ and power corrections $Q_s^2(x)/Q^2$ when they become large. To leading logarithmic accuracy in $x$, this RG equation is the JIMWLK equation~\cite{JalilianMarian:1997gr,JalilianMarian:1997dw,Iancu:2000hn,Ferreiro:2001qy}  with a corresponding JIMWLK Hamiltonian~\cite{Weigert:2000gi} that contains all the relevant information regarding the rapidity evolution of n-point Wilson line correlators. 

A number of well-known results are obtained as limits of the JIMWLK Hamiltonian. In the limit of large number of colors $N_c$, and for large atomic nuclei with $A \gg 1$, the JIMWLK equation for the simplest ``dipole" correlator of light-like Wilson lines describing fully inclusive DIS is the Balitsky-Kovchegov (BK) equation~\cite{Balitsky:1995ub,Kovchegov:1999yj}. In the leading twist limit, where $Q_s^2 (x) /Q^2 \ll 1$, this reduces to the BFKL equation~\cite{Kuraev:1977fs,Balitsky:1978ic}. The latter was first derived by explicit computation of  perturbative QCD Feynman diagrams in Regge asymptotics.

The CGC EFT has been applied to compute a large number of final states in both DIS and in hadron-hadron collisions. An excellent introductory review of the formalism and applications can be found in \cite{Blaizot:2016qgz}. An important test of the framework will be its ability to make predictions of high accuracy that can be compared to experiment. The ideal experimental conditions for such tests require access to large $Q^2$ and small $x$. Further, since the saturation scale is enhanced in nuclei as $Q_{s,A}^2\sim A^{1/3} \, Q_{s,{\rm proton}}^2$ \cite{Gribov:1984tu,McLerran:1993ni,Kovchegov:2012mbw} , heavy nuclear targets are preferred. In principle, these conditions are achieved in proton-nucleus ($p+A$) collisions at the LHC. However large final state interactions may be present in these experiments that will complicate interpretations of the data~\cite{Dusling:2018hsg}. These considerations provide a major motivation for future Electron-Ion Collider (EIC) experiments~\cite{Accardi:2012qut,AbelleiraFernandez:2012cc,Aschenauer:2017jsk}. 

Such experiments at an EIC are envisaged to have much higher luminosities than were available at HERA; this, in combination with the nuclear beams, greatly enhances the possibility that DIS $e+A$ collider measurements may uncover definitive evidence for gluon saturation. These precision DIS experiments demand higher order computations in Regge asymptotics in close analogy to how higher order computations in the Bjorken limit provided powerful tests of perturbative QCD. To further the analogy, just as one computes universal splitting functions and process dependent coefficient functions in the Bjorken limit of DIS, one needs to compute both universal multiparton lightlike correlators (or equivalently, as we shall discuss, their generating weight functional describing color charge distributions) and process dependent impact factors to higher order accuracy in the Regge limit. 

The next-to-leading order (NLO) computation of the BFKL kernel has now been available for over twenty years~\cite{Fadin:1998py,Ciafaloni:1998gs}. Subtleties in the treatment of singularities in the NLO BFKL kernel were noticed shortly after~\cite{Salam:1998tj,Ciafaloni:1999yw}--for a recent comprehensive discussion, see \cite{Ducloue:2019ezk}. Computations of higher order corrections to the BK equations followed in short order~\cite{Balitsky:2008zza,Kovchegov:2006vj,Braun:2007vi} and specific prescriptions for the running coupling following from these studies were implemented~\cite{Albacete:2007yr} in phenomenological studies. More recently, NLO computations of multi-point Wilson line correlators, or equivalently, the NLO JIMWLK Hamiltonian have become available~\cite{Balitsky:2013fea,Kovner:2013ona,Balitsky:2014mca,Lublinsky:2016meo,Caron-Huot:2013fea} and next-to-next-to-leading order (NNLO) computations of BK and JIMWLK are underway~\cite{Caron-Huot:2016tzz,Caron-Huot:2017fxr}.

Because of the absence of final state interactions, isolated photons are clean probes of this strongly correlated gluonic matter. Several computations~\cite{Gelis:2002ki,Benic:2016yqt,Benic:2016uku,Benic:2018hvb,Altinoluk:2018uax} of inclusive photon production have been performed in the CGC framework in the context of proton-nucleus ($p+A$) collisions. In a recent paper~\cite{Roy:2018jxq}, henceforth referred to as Paper I, we reported on a first CGC computation of the leading order differential cross-section for inclusive prompt photon production in conjunction with two jets in electron-nucleus ($e+A$) DIS at small $x$. This process has clean initial and final states and is the simplest non-trivial process besides fully inclusive DIS to study the physics of gluon saturation in $e+A$ collisions. This computation provides more differential phase space distributions, thereby going beyond  existing small $x$ computations~\cite{Bartels:2000gt,Bartels:2002uz,Bartels:2001mv,Balitsky:2010ze,Balitsky:2012bs,Beuf:2011xd,Beuf:2016wdz,Beuf:2017bpd,Hanninen:2017ddy} on the total cross-section for fully inclusive DIS; exceptions are the NLO differential cross-section computations by Boussarie \textit{et al.}~\cite{Boussarie:2014lxa,Boussarie:2016ogo,Boussarie:2016bkq}, albeit for diffractive DIS. 

In Paper I, we showed explicitly how in the limit of the final state photon momentum $k_{\gamma} \rightarrow 0$, the Low-Burnett-Kroll soft photon theorem~\cite{Low:1958sn,Burnett:1967km,Bell:1969yw} allows one to recover existing results for inclusive dijet production in DIS~\cite{Dominguez:2011wm}. In the leading twist limit, we also obtained the $k_{\perp}$ and collinear factorized expressions which match the dominant NLO small $x$ perturbative QCD (pQCD) contributions. In particular, our result in the collinear limit is directly proportional to the nuclear gluon distribution at small $x$.

For precision physics in pQCD, it is essential to go beyond leading order descriptions for quantitative studies of data. NLO computations will be especially important for the discovery and characterization of gluon saturation in $e+A$ DIS where its effects are anticipated to be larger than in $e+p$ DIS\footnote{Moreover, as our discussion of Paper I suggests, the leading twist limits of these NLO computations can also be matched to results for the same in the collinear factorization framework.}. As we observed in our LO photon+dijet computation, there are novel quadrupole gauge invariant correlators of lightlike Wilson lines that appear, whose energy evolution, in addition to the dipole correlators measured in fully inclusive DIS, will be a sensitive test of JIMWLK evolution. We will show how such correlators, in combination with the dipole correlators, violate the soft gluon theorem. A quantitative understanding of this violation can provide deeper insight into the infrared structure of QCD in the 
Regge limit~\cite{Strominger:2017zoo}.

Byproducts of our computation of the differential photon+dijet cross-section are the first NLO results for inclusive dijet, inclusive photon and photon+jet measurements at an EIC. Further, the NLO graphs for real gluon emission provide the complete LO results for the photon+3-jet ($\gamma+q\bar{q} g$) and 3-jet $q\bar{q}g$ final states~\cite{Ayala:2016lhd}. We will point to the steps necessary for extracting ``numbers" from our computation; though much more computationally challenging than comparable computations in the highly developed collinear factorization pQCD framework, 
such a program is feasible and essential for precision physics at an EIC.  

As we will discuss further shortly, all computations in the CGC EFT rely on a separation of scales between static color sources and dynamical gauge fields. Thus perturbative computations at small $x$ in this framework are performed in a background of such static color sources and physical quantities are obtained by subsequent gauge invariant averaging over these sources. The first principles formalism in quantum field theory underlying such computations in strong background fields has been discussed previously~\cite{Gelis:2006yv,Gelis:2006cr,Gelis:2007kn,Gelis:2008rw,Jeon:2013zga}; in particular, 
\cite{Gelis:2006yv} and \cite{Jeon:2013zga} provide pedagogical discussions in complementary approaches. 

We begin our discussion of the NLO DIS photon+dijet computation with the starting point of all CGC computations, the classical Yang-Mills equations,  \begin{equation}
[D_{\mu},F^{\mu \nu}](x)=g \,  \delta^{\nu +} \delta(x^{-}) \rho_{A}(\bm{x}_{\perp}) \, .
\label{eq:Yang-Mills}
\end{equation}
Here the covariant derivative $D_\mu = \partial_\mu - ig A_\mu$, $g$ represents the QCD gauge coupling, and $\rho_{A}(x^{-},\bm{x}_{\perp})\approx \rho_{A}(\bm{x}_{\perp})\delta(x^-)$ represents the color charge density of large $x$ static sources for the small $x$ dynamical fields $A^\mu$. The delta-function in the color charge density denotes that we are working in a frame where the nucleus is moving in the positive $z$-direction at nearly the speed of light with large light cone longitudinal momentum $P^{+}_{N}\rightarrow\infty$. (See Appendix \ref{sec:conventions} for details of the conventions adopted in this work.) In addition we will choose the frame in which the virtual photon has a large longitudinal momentum $q^{-}$ and transverse momentum $\bm{q}_{\perp}=0$.

The solution of the classical equations in Lorenz gauge $\partial_\mu A^\mu=0$ is given by 
\begin{equation}
A_{\rm cl}^+ = \int \frac{\mathrm{d}^2 \bm{z}_\perp}{ 4\pi}\, \ln\frac{1}{(\bm{x}_\perp-\bm{z}_\perp)^2\Lambda^2}\, \rho_A(x^-,\bm{z}_\perp)\,\,;\,\, A_{\rm cl}^{-}=0\,\,;\,\, A_{{\rm cl},\perp} =0\,,
\label{eq:A+}
\end{equation}
where $\Lambda$ is an infrared cutoff that is necessary to invert the Laplace equation $-\nabla_\perp^2 A_{\rm cl}^+ = g\rho_A$ in two dimensions. This solution to the Yang-Mills equations in Lorenz gauge is simply related to the solution in the light cone gauge ${\tilde A}^+=0$, where one obtains likewise that ${\tilde A}_{\rm cl}^-=0$ and 
${\tilde A}_{\rm cl}^i = \frac{i}{g} U\partial^i U^\dagger$, and  
%The corresponding Feynman diagrams for the dressed propagators and the vertex factors are depicted in Fig.~\ref{fig:effective-vertices}. 
\begin{equation}
U(\bm{x}_{\perp})=P_{-} \Bigg( \text{exp} \Bigg\{ -ig \int_{-\infty}^{+\infty} \mathrm{d}z^{-} A_{\rm cl}^{+,a} (z^{-},\bm{x}_{\perp}) \,T^a\Bigg\} \Bigg) \, ,  
\label{eq:Wilson-line}
\end{equation}
denotes the adjoint Wilson line expressed in terms of the the large $x$ static color source densities via Eq.~(\ref{eq:A+}). Note that $T^a$, $a=1,\cdots,8$  are the generators of color $SU(3)$ in the adjoint representation. 

These Wilson lines, and their counterparts ${\tilde U}$ in the fundamental representation represent respectively, the path ordered phase acquired by a gluon and a quark in their eikonal multiple scattering off the classical background field of a nucleus. The Wilson line ${\tilde U}$ is obtained by replacing the adjoint generators in Eq.~(\ref{eq:Wilson-line}) with the fundamental generators: $T^a\rightarrow t^a$. 
In the LO photon+dijet cross-section of Paper I, the virtual photon fluctuates into a quark-antiquark pair that multiple scatters off the classical background field of the nucleus. In the Feynman diagram computations of this LO process, the Wilson lines are incorporated in the momentum space structure of the dressed quark propagators. At NLO, there are real and virtual gluon contributions to the leading order process. The corresponding gluon propagators are also dressed by multiple scattering off the classical background field of the nucleus. 

\begin{figure}[!htbp]
\begin{minipage}[b]{0.8\textwidth}
\includegraphics[width=\textwidth]{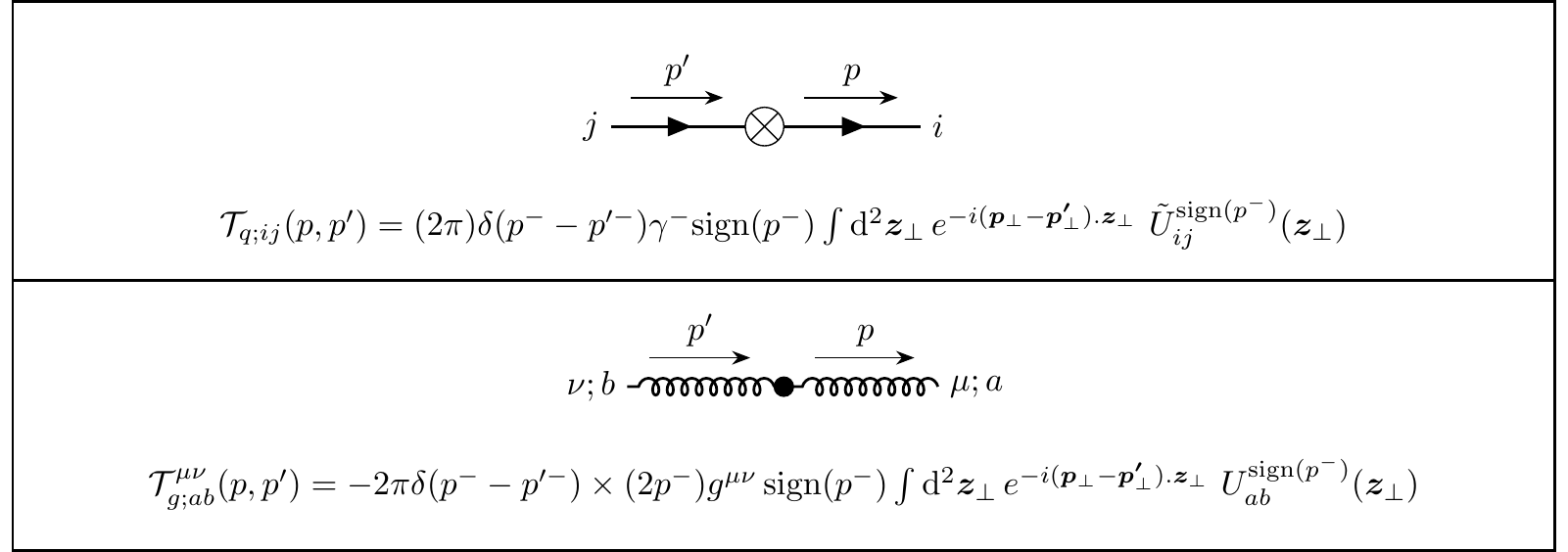}
\end{minipage}
\caption{Feynman diagrams for the dressed quark and gluon propagators with effective vertices denoted respectively by crossed and filled blobs. The respective effective vertices are also shown. $i$ and $j$ represent color indices in the fundamental representation of $SU(N_{c})$ whereas $a$ and $b$ stand for the same in the adjoint representation. In the saturation regime these vertices are effectively of order unity. \label{fig:effective-vertices}}
\end{figure}

The structure of the dressed quark and gluon propagators in the classical background field, is particularly simple in the ``wrong'' light cone (LC) gauge $A^{-}=0$. As indicated by Eq.~(\ref{eq:A+}), this gauge shares the same classical field solution with Lorenz gauge. However, unlike the LC gauge $A^+=0$ for $P^{+}_{N}\rightarrow\infty$, it does not provide a simple physical interpretation of parton distribution functions. Any concern in this regard is however far outweighed by the advantages provided by the simple forms of the dressed propagators that were computed previously in Refs.~\cite{McLerran:1994vd,Ayala:1995hx,Ayala:1995kg,Balitsky:1995ub,McLerran:1998nk,Balitsky:2001mr}. 
The effective vertices for these dressed quark and gluon  ``shockwave" propagators, incorporating the fundamental and adjoint Wilson lines respectively, are shown in Fig.~\ref{fig:effective-vertices}. 

As discussed in Paper I, these effective vertices are identical to the quark-quark-reggeon and gluon-gluon-reggeon effective vertices~\cite{Caron-Huot:2013fea,Bondarenko:2017vfc,Ayala:2017rmh,Hentschinski:2018rrf,Bondarenko:2018kqs,Bondarenko:2018eid} in Lipatov's reggeon field theory~\cite{Lipatov:1996ts}. Another salient feature in the expressions below is that we do not subtract the unit matrix in the expansion of Wilson lines. This allows for the possibility that the quarks and gluons do not scatter in addition to all their possible multiple scatterings encoded in the higher terms in the Wilson line expansion. Consequently, we draw Feynman diagrams with all dressed fermion and gluon propagators and for each such kinematically allowed process, we only need to subtract the ``no scattering'' contribution (obtained by putting $\tilde{U}$ and $U$'s to unity) to get the physical amplitude. As we will see, this significantly aids in the NLO computation where the number of contributing processes is large.

Note further that since our computations do not employ light-front perturbation theory like many of the NLO computations in the literature, and are carried out entirely in momentum space (also unlike many computations), they also provide a useful cross-check on extant NLO results on fully inclusive DIS. The techniques employed here may also provide a pathway to carrying out higher order computations to NNLO in the Regge limit. 

%They incorporate the all-twist resummation in $g \rho_{A} \sim 1$ for the eikonal interactions of the quark/gluon with the background classical field which is a reflection of the fact that the physics in the saturation regime is intrinsically non-perturbative, $\rho_{A} \sim O(1/g) \gg 1$ although the coupling constant is small. 

To proceed with our NLO computation, it is important to elaborate further on the RG procedure for resummation of large logarithms in $x$, specifically with regard to how it applies to the inclusive photon+dijet computation of interest. As noted, the Wilsonian RG ideology on which the CGC EFT is based naturally involves a cutoff scale in rapidity or longitudinal momentum separating the soft and hard partons in a hadron/nucleus. 
%The argument for this separation is well founded based on the experimental data on proton structure function, $F_{2}(x,Q^{2})$ measured at HERA~\cite{Abt:1993cb,Ahmed:1995fd,Derrick:1993fta,Derrick:1994sz,Martin:1994kn,Lai:1994bb}. 
At LO, this scale $\Lambda_{0}^{+}$ (or rapidity $Y_{0}=\ln (\Lambda_{beam}^+ /\Lambda_{0}^{+})$) is arbitrary and the fast or valence modes with longitudinal momenta $k^{+} \gg \Lambda_{0}^{+}$ are represented by the stochastic color charge density, $\rho_{A}(x^{-},\bm{x}_{\perp})$. A gauge invariant weight functional $W_{\Lambda_{0}^{+} (Y_{0}) } [\rho_{A}]$ describes the probability density corresponding to this charge density. As also noted, the soft modes are represented by classical color fields that are solutions of the classical Yang-Mills equations with appropriate gauge fixing conditions.

\begin{figure}[!htbp]
\begin{minipage}[b]{0.4\textwidth}
\includegraphics[width=\textwidth]{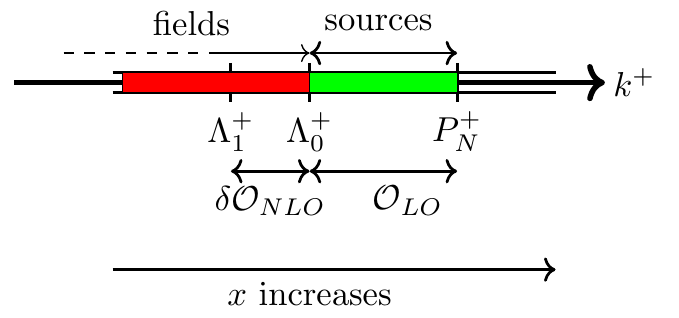}
\end{minipage}
\caption{Schematic illustration of sources and fields in the CGC effective theory in terms of the cutoff scale $\Lambda_{0}^{+}$, or equivalently the rapidity $Y_{0}$.  At NLO, contributions from field modes in the range $\Lambda_1^+ < k^+ <\Lambda_{0}^{+}$, such that $\alpha_S\ln(\Lambda_0^+/\Lambda_1^+) \equiv \alpha_S\Delta Y < 1$, are integrated out and absorbed into the source densities at the scale $\Lambda_1^+$. This self-similar renormalization group (RG) pattern is repeated successively generating the JIMWLK RG equation for the source densities.  \label{fig:NLO-evolution}} 
\end{figure}

As we boost the nucleus towards the small $x$ scale of interest (or towards higher energies), modes that were below the cutoff $\Lambda_{0}^{+}$ start contributing to the scattering process. We therefore have to consider quantum effects induced by these ``semi-fast'' gluons~\cite{Iancu:2000hn} (see Fig.~\ref{fig:NLO-evolution}) which can be defined as the nearly on-shell fluctuations with momenta deeply inside the strip $\Lambda_{1}^{+}(=b\Lambda^{+}_{0}) \ll \vert l^{+} \vert \ll  \Lambda_{0}^{+}$ or conversely, energies in the range $\Lambda_{0}^{-} \ll \vert l^{-} \vert \ll \Lambda^{-}_{0}/b$ where
\begin{equation}
\Lambda^{-}_{0}=\frac{Q^{2}_{0}}{2x_{0}P^{+}_{N}}= \frac{Q^{2}_{0}}{x_{0}}.\frac{x}{Q^{2}} q^{-} \, .
\label{eq:initial-energy-scale}
\end{equation} 
Here $Q^{2}_{0}$ and $x_{0}$ are respectively the virtuality of the nucleus  and the Bjorken-$x$ at the initial scale. Further, $Q^{2}=-2q^{+}q^{-}$ is the fixed virtuality of the exchanged virtual photon in DIS (in Regge kinematics) and $x$ is the Bjorken-$x$ of interest determined by the kinematics of the process. The effect of integrating out these fluctuations manifests itself in the appearance of large logarithms $\ln(1/b)$ which for $\alpha_{S} \ln(1/b) \sim 1$ must be resummed to all orders in $\alpha_{S}$. 

\begin{figure}[!htbp]
\begin{center}
\includegraphics[scale=1]{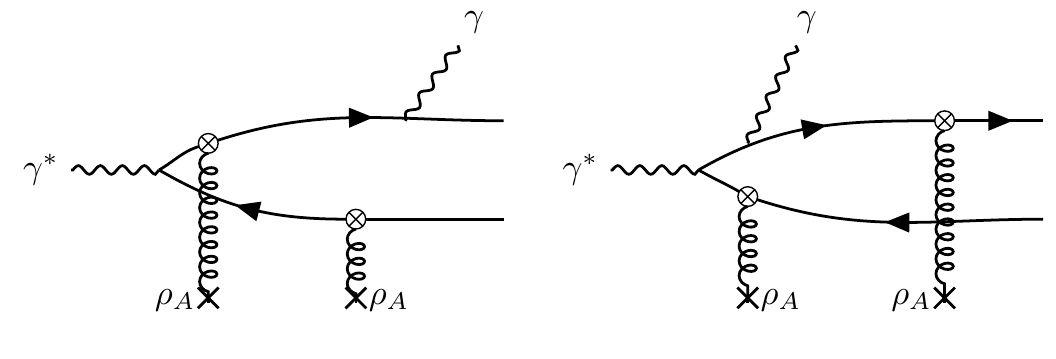}
\caption{Processes contributing to the leading order amplitude and hence the cross-section $\mathrm{d} \hat{\sigma}_{\rm LO}$ in Eq.~\ref{eq:leading-order-dsigma}. The other two diagrams are obtained simply by interchanging the quark and antiquark lines. \label{fig:LO-intro}}
\end{center}
\end{figure}

Denoting the differential cross-section for inclusive photon plus dijet production by $\mathrm{d} \sigma$ for simplicity, we  can write down its 
expectation value in the CGC EFT at LO as~\cite{Roy:2018jxq}
\begin{equation}
\langle \mathrm{d} \sigma_{\text{LO}} \rangle = \int [\mathcal{D} \rho_{A} ] \, W_{\Lambda_{0}^{-}} [\rho_{A}]\, \mathrm{d} \hat{\sigma}_{\text{LO}} [\rho_{A}]
%= \int [\mathcal{D} \rho_{A} ] \, W_{\Lambda_{0}^{-}} [\rho_{A}]\,  C_{0} \otimes   \, \mathcal{R}_{0}[\rho_{A}] 
\, .
\label{eq:leading-order-dsigma}
\end{equation}
The r.h.s represents the fact that the LO cross-section is first computed for a fixed distribution of color sources $\rho_{A}$ with $\Lambda^- < \Lambda_{0}^{-}$. This object $\mathrm{d} \hat{\sigma}_{\text{LO}} [\rho_{A}]$ shown in Fig.~\ref{fig:LO-intro}, is computed using standard techniques in perturbative QCD albeit, as noted, with the modified propagators listed in Fig.~\ref{fig:effective-vertices}, wherein the dependence on $\rho_{A}$ enters (at LO) via the fundamental Wilson line ${\tilde U}$. 

The process independent weight functional 
$W_{\Lambda_{0}^{-}} [\rho_{A}]$ is a nonperturbative object that contains fundamental information about $n$-body correlations amongst gluons at the initial scale $\Lambda_{0}^{-}$. It can be understood as representing large $x$ diagonal elements of the density matrix of QCD in the Regge limit; a recent discussion of $W[\rho_A]$, and generalizations thereof, can be found in \cite{Armesto:2019mna}.

%\begin{equation}
%\Xi(\bm{x}_{\perp},\bm{y}_{\perp};\bm{y'}_{\perp},\bm{x'}_{\perp})=1-D_{xy}-D_{y'x'}+Q_{y'x';xy} \, ,
%\label{eq:color-structure-LO}
%\end{equation}

\begin{figure}[!htbp]
\begin{minipage}[b]{0.7\textwidth}
\includegraphics[width=\textwidth]{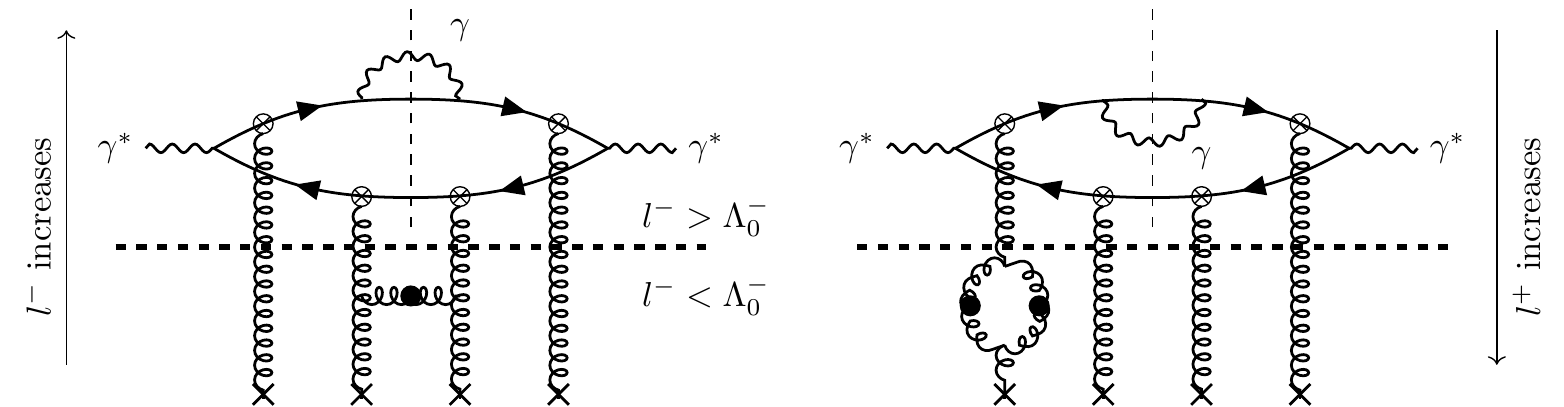}
\end{minipage}
\caption{NLO contributions from gluon modes with $l^{-}< \Lambda^{-}_{0}$ fluctuating within the target. The dashed horizontal line represents the EFT scale $\Lambda_{0}^{-}$ separating the gluon modes in the target and the projectile dipole. The diagrams shown are representative of those containing the large logarithms $\ln(\Lambda_{1}^{-}/\Lambda_{0}^{-})$ that contribute towards the LL$x$ evolution of $W[\rho_A]$.
\label{fig:LLx-evolution-NLO-1}}
\end{figure}

At NLO ($\equiv O(\alpha_{S})$) in the CGC power counting, we have to account for quantum fluctuations of both the quark-antiquark dipole as well as the wavefunction of the nuclear target. In the EFT language, these are distinguished by the magnitude of the LC momentum of the gluon modes relative to the initial scale $\Lambda_{0}^{-}$. As shown in Fig.~\ref{fig:LLx-evolution-NLO-1}, the modes with $l^{-} < \Lambda_{0}^{-}$ (which we shall denote as NLO:1) can be interpreted as contributions from Fock states dressing the target wavefunction. The contribution to the cross-section for processes of this kind can in general be written as 
\begin{equation}
\langle \mathrm{d} \sigma_{\text{NLO}:1} \rangle=  \int [\mathcal{D} \rho_{A} ] \, W_{\Lambda_{0}^{-}} [\rho_{A}]\, \mathrm{d} \hat{\sigma}_{\text{NLO}:1} [\rho_{A}]
%\Big( C_{1} \otimes \mathcal{R}_{1}[\rho_{A}] \Big) 
 \, ,
\label{eq:NLO-1-dsigma}
\end{equation}
where $\mathrm{d} \hat{\sigma}_{\text{NLO}:1} [\rho_{A}]$, for a fixed configuration of $\rho_A$, is comprised of nontrivial  combinations of Dirac traces and Fourier transforms of color traces over products of the Wilson lines ${\tilde U}$ and $U$. From these contributions, we are interested in collecting only  pieces that contain large logarithms in $\Lambda_{0}^{-}$. These can be written as
\begin{equation}
\langle \delta \sigma_{\text{NLO:1}} \rangle=  \int [\mathcal{D} \rho_{A}]  \, W_{\Lambda^{-}_{0} } [\rho_{A}]  \ln(\Lambda^{-}_{1} /\Lambda^{-}_{0}) \,\, \mathcal{H}_{\text{LO}}  \,  \mathrm{d} \hat{\sigma}_{\text{LO}}[\rho_{A}] \, ,
\label{eq:NLO-1-dsigma-LLx}
\end{equation}
where $\Lambda_{1}^{-}$ is the scale to which the target gluon modes are evolved. Here $\mathcal{H}_{\text{LO}}$ represents the JIMWLK Hamiltonian~\cite{JalilianMarian:1997gr,JalilianMarian:1997dw,Iancu:2000hn,Ferreiro:2001qy,Weigert:2000gi}; its explicit form will be discussed later in the paper.  For our purposes here, it suffices to note that  $\mathcal{H}_{\text{LO}} \, \mathrm{d} \hat{\sigma}_{\text{LO}} $ is of order $\alpha_{S} \, \mathrm{d}\hat{\sigma}_{LO} [\rho_{A}]$. Combining the above contribution with the LO result in Eq.~(\ref{eq:leading-order-dsigma}), and using the Hermiticity of $W$ with respect to the functional integration over $\rho_{A}$, we can write the result as
\begin{equation}
\langle \mathrm{d} \sigma_{\text{LO}} + \delta \sigma_{\text{NLO:1}} \rangle = \int      [\mathcal{D}\rho_{A}] \Big\{\Big(  1+ \ln(\Lambda_{1}^{-}/ \Lambda_{0}^{-}) \mathcal{H}_{\text{LO}} \Big) W_{\Lambda^{-}_{0}} [\rho_{A}] \Big\}\, \mathrm{d} \hat{\sigma}_{\text{LO}} [\rho_{A}] \, .
\label{eq:LO-JIMWLK}
\end{equation}
Further redefining
\begin{equation}
 \Big( 1+ \ln(\Lambda_{1}^{-}/ \Lambda_{0}^{-}) \mathcal{H}_{\text{LO}} \Big) W_{\Lambda_{0}^{-}}[\rho_{A}] = W_{\Lambda_{1}^{-}}[\rho_{A}] \, ,
 \label{eq:W-LLx}
\end{equation}
and thereby absorbing the effects of the semi-fast gluons in terms of a modification of the probability distribution of the fast color sources, one obtains the leading log in $x$ JIMWLK equation
\begin{equation}
\frac{\partial}{\partial(\ln \Lambda^{-})} W_{\Lambda^{-}} [\rho_{A}] = \mathcal{H}_{\text{LO}}\, W_{\Lambda^{-}} [\rho_{A}] \, .
\label{eq:LO-JIMWLK_final}
\end{equation}
To derive this result, we employed the essential RG philosophy that the observable on the l.h.s of Eq.~\ref{eq:LO-JIMWLK} must be independent of the  arbitrary scale $\Lambda_{0}^{-}$ separating the static color sources $\rho_A$ from the dynamical gauge fields. Replacing the expression in curly brackets in Eq.~(\ref{eq:LO-JIMWLK}) by the r.h.s of Eq.~(\ref{eq:W-LLx}) 
%satisfying Eq.~(\ref{eq:LO-JIMWLK_final}) 
is equivalent to summing the leading logarithmic terms $\alpha_S\ln(1/x)$ to all orders in perturbation theory. 
We will henceforth label the weight functional that satisfies Eq.~(\ref{eq:LO-JIMWLK_final}) as $W^{LLx}[\rho_{A}]$.

\begin{figure}[!htbp]
\begin{minipage}[b]{0.8\textwidth}
\includegraphics[width=\textwidth]{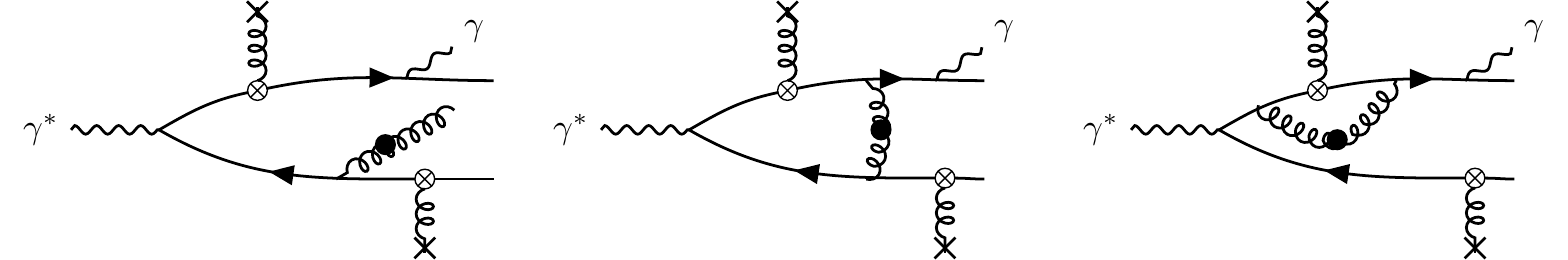}
\end{minipage}
\caption{Representative NLO contributions from gluon modes in the projectile with $l^{-} > \Lambda_{0}^{-}$. Both real and virtual emission diagrams are shown. For the latter case, we have to consider interference diagrams with LO processes described in \cite{Roy:2018jxq}. \label{fig:NLO-representative-projectile}}
\end{figure}

One also has NLO contributions from gluon modes with $l^{-} >\Lambda_{0}^{-}$ corresponding to quantum fluctuations of the dipole projectile.  As shown in Fig.~\ref{fig:NLO-representative-projectile}, these include real gluon emission and virtual gluon exchange processes; the filled blobs represent the dressed gluon propagators allowing for the possibility that the gluon can scatter off the background classical field of the nucleus. We will refer to contributions from the quantum fluctuations of the projectile as NLO:2 contributions to distinguish these from the NLO:1 quantum fluctuations of the target below the scale $\Lambda_0^-$. 
 
 As in any loop computation, the intermediate steps of our calculation will contain soft, collinear and ultraviolet (UV) singularities depending on the region of phase space of the gluon that we are integrating over. In the virtual graphs, UV divergences appear from integrals over the transverse momentum of the gluon in the loop;  these are isolated using dimensional regularization in $d=2-\epsilon$ dimensions. At this NLO order of quantum fluctuations of the projectile quark-antiquark pair, all UV divergences must vanish or cancel without the necessity of renormalizing the parameters of the EFT. This is 
 because we will work the limit of massless quarks and there is no running of the QCD coupling constant in the projectile wavefunction at this order in the CGC power counting. 
 
 The small $x$ divergences arise from integrating over the quantum fluctuations induced by ``slow" or ``semi-fast" gluons with `--' longitudinal momenta that are  small relative to the large $q^-$ momentum of the virtual photon. These divergences are regulated by imposing a cutoff at the initial scale, $\Lambda_{0}^{-}$ of the evolution which is defined in Eq.~\ref{eq:initial-energy-scale}. The resulting logarithms in $\Lambda_{0}^{-}$ (or equivalently $x$) are absorbed into small $x$ renormalization group evolution of the weight functional $W[\rho_{A}]$ as shown in Eq.~\ref{eq:LO-JIMWLK_final}. At higher orders, it may be necessary to employ more sophisticated regularization schemes for these rapidity divergences as well~\cite{Liu:2019iml}. 

For gluon emission diagrams, in addition to small $x$ divergences, there are also singularities that arise from the region of phase space where the unscattered gluon is soft or collinear to the quark or antiquark. 
%When we allow the phase space to be differential in the (anti) quark momenta there are 
In particular, there are residual collinear divergences that survive after real and virtual contributions are combined. These divergences are absorbed into the evolution of fragmentation functions. Conversely, we can regulate the phase space integration over final states by promoting partons to jets where the latter are defined using a cone algorithm~\cite{Sterman:1977wj,Furman:1981kf}. We 
will show explicitly in the limit of small jet cone size~\cite{Ivanov:2012ms,Aversa:1988vb,Aversa:1989xw,Jager:2004jh} that collinear divergences between real and virtual graphs cancel completely enabling the extraction of the dominant contributions towards the jet cross-section.

With all the divergences in the ($\rm NLO:2$) quantum fluctuations of the virtual photon projectile accounted for, 
%and the small $x$ logarithms absorbed in the RG evolution of the weight functional $W_{\Lambda_{0}^{-}}[\rho_{A}]$ 
one can write the infrared (IR) safe jet cross-section as
\begin{align}
\langle \mathrm{d} \sigma_{\text{NLO}:2}^{\rm jet} \rangle = \langle \delta \sigma_{\rm NLO:2}^{\rm jet} \rangle+    \int [\mathcal{D} \rho_{A} ] \, W_{\Lambda_{0}^{-}} [\rho_{A}] \, \mathrm{d} \hat{\sigma}_{\text{NLO}}^{\rm jet;finite} [\rho_{A}] 
 \, .
\label{eq:NLO-2-dsigma}
\end{align} 
These NLO contributions (shown in Fig.~\ref{fig:NLO-representative-projectile}) 
%which we call $\mathrm{d} \hat{\sigma}_{\text{NLO}:2} [\rho_{A}]$ 
can be broken up into two pieces. The first piece is obtained by taking the ``slow" gluon limit, $l^{-} \rightarrow 0$ and is identical ($ \delta \sigma_{\rm NLO:2}^{\rm jet} =\delta \sigma_{\rm NLO:1}^{\rm jet}
   $) to the expression in Eq.~\ref{eq:NLO-1-dsigma-LLx} at the momentum scale $\Lambda_{0}^{-}$. We will show this explicitly later in the paper. Specifically, this matching corresponds to a first principles derivation of leading log JIMWLK evolution for a nontrivial final state of the projectile. While the NLO:2 derivation represents the slow IR limit of the projectile, the RG corresponds to 
 matching it to quantum fluctuations at their fast scale $\Lambda_0^-$ in the target.

The second term on the r.h.s of  Eq.~\ref{eq:NLO-2-dsigma} ($\mathrm{d} \hat{\sigma}_{\text{NLO}}^{\text{jet;finite}} [\rho_{A}]$) contains genuine $\alpha_{S}$ suppressed (without logs in $x$) contributions  to the differential cross-section from real and virtual graphs. For the latter, it is possible to deduce analytical expressions because the divergent structures can be isolated at the level of the amplitude. In contrast, divergent structures in the real NLO contributions are manifest only at the level of the squared amplitude and obtaining analytical expressions for the finite terms is challenging. However they can be evaluated numerically using the fact that rapidity divergences can be isolated in the slow gluon limit; these can then be subtracted from the squared amplitudes (using a numerical 
cutoff procedure) to obtain the desired finite pieces. 

Further, by replacing parton momenta in these contributions  with those of jets (using a jet algorithm) gets rid of the remaining collinear divergences. The finite contributions $\mathrm{d} \hat{\sigma}_{\text{NLO}}^{\text{jet;finite}} [\rho_{A}]$ that we will compute explicitly in this paper are, in the language of Regge theory, the NLO ``impact factor" corrections to the LO impact factor $\mathrm{d} \hat{\sigma}_{\rm LO}^{\rm jet} [\rho_{A}]$.
 
\begin{figure}[!htbp]
\begin{minipage}[b]{0.8\textwidth}
\includegraphics[width=\textwidth]{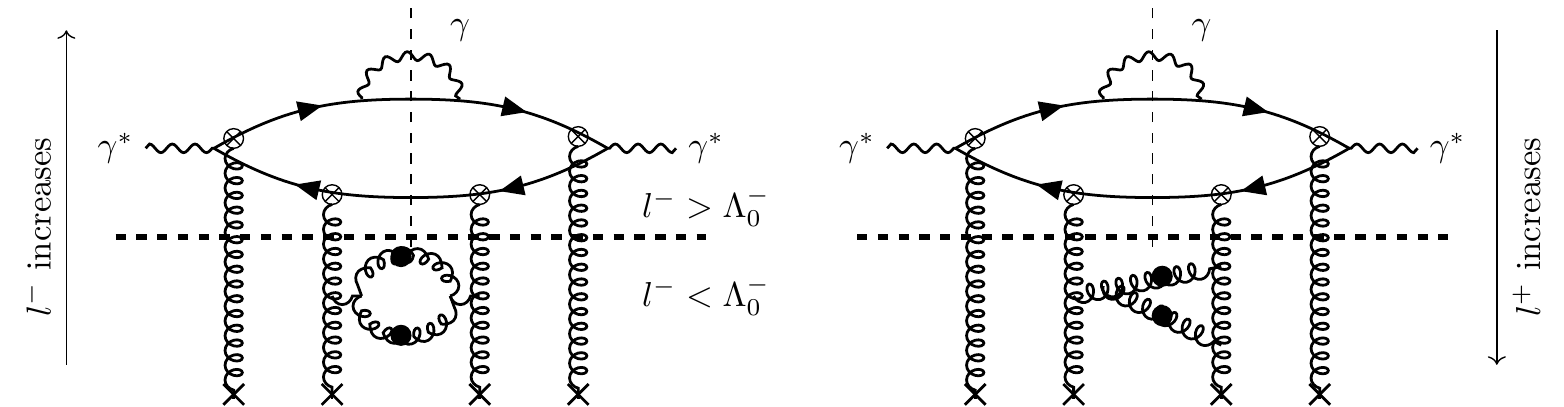}
\end{minipage}
\caption{NNLO diagrams that contain $\alpha_{S}$ magnitude contributions for $\alpha_{S}\text{ln}(\Lambda^{-}_{1}/\Lambda^{-}_{0}) \sim 1$. These diagrams contribute to the NLO JIMWLK kernel. \label{fig:NLLx-evolution-1}}
\end{figure}

Computing this process-dependent NLO impact factor $\mathrm{d} \hat{\sigma}_{\text{NLO}}^{\text{jet;finite}} [\rho_{A}]$ for photon+dijet production is important because it allows us to  go one step further in precision and consider relevant (two loop) NNLO contributions to the cross-section that have terms proportional to $\alpha_{S}^{2} \ln(\Lambda_{1}^{-}/\Lambda_{0}^{-})$. These contributions are effectively of NLO magnitude if $\alpha_{S} \ln(\Lambda_{1}^{-}/\Lambda_{0}^{-}) \sim 1$. 
Diagrams corresponding to a two loop fluctuation of the target are shown in Fig.~\ref{fig:NLLx-evolution-1}. In the class of such two loop diagrams, there are contributions of order $\alpha_{S}^{2} \ln^2(\Lambda_{1}^{-}/\Lambda_{0}^{-})\sim O(1)$ which are included in the leading log JIMWLK resummation, as represented by $W^{LLx}[\rho_{A}]$. There are also contributions from  two-loop QCD diagrams proportional to $\alpha_{S}^{2}$ alone (without leading logs in $x$) but these are suppressed at the desired accuracy of our problem. We will consider here only those two loop contributions in Fig.~\ref{fig:NLLx-evolution-1} that contain next-to-leading logarithms in $x$ (NLL$x$) contributions to the result in Eq.~\ref{eq:LO-JIMWLK}. This in turn gives us the LO+NLL$x$ result which can be expressed in terms of a modified weight-functional as
\begin{equation}
\langle \mathrm{d} \sigma^{\rm jet}   \rangle_{\text{LO}+\text{NLL$x$}}  = \int      [\mathcal{D}\rho_{A}] \,  W_{\Lambda_1^-}^{NLLx}  [\rho_{A}]  \, \mathrm{d} \hat{\sigma}_{\rm LO}^{\rm jet} [\rho_{A}] \, ,
\label{eq:NLL-JIMWLK}
\end{equation}
where
\begin{equation}
W_{\Lambda_1^-}^{NLLx}  [\rho_{A}] =\Big\{  1+ \ln(\Lambda_{1}^{-}/ \Lambda_{0}^{-}) (\mathcal{H}_{\text{LO}}+\mathcal{H}_{\text{NLO}})  \Big\}\,W_{\Lambda_{0}^{-}}[\rho_{A}]\, ,
\label{eq:NLL-WF}
\end{equation}
and the NLO JIMWLK Hamiltonian $\mathcal{H}_{\text{NLO}}$  \cite{Balitsky:2013fea,Kovner:2013ona,Balitsky:2014mca,Lublinsky:2016meo,Caron-Huot:2013fea} is of order $\alpha_{S}^{2}$.

\begin{figure}[!htbp]
\begin{minipage}[b]{0.8\textwidth}
\includegraphics[width=\textwidth]{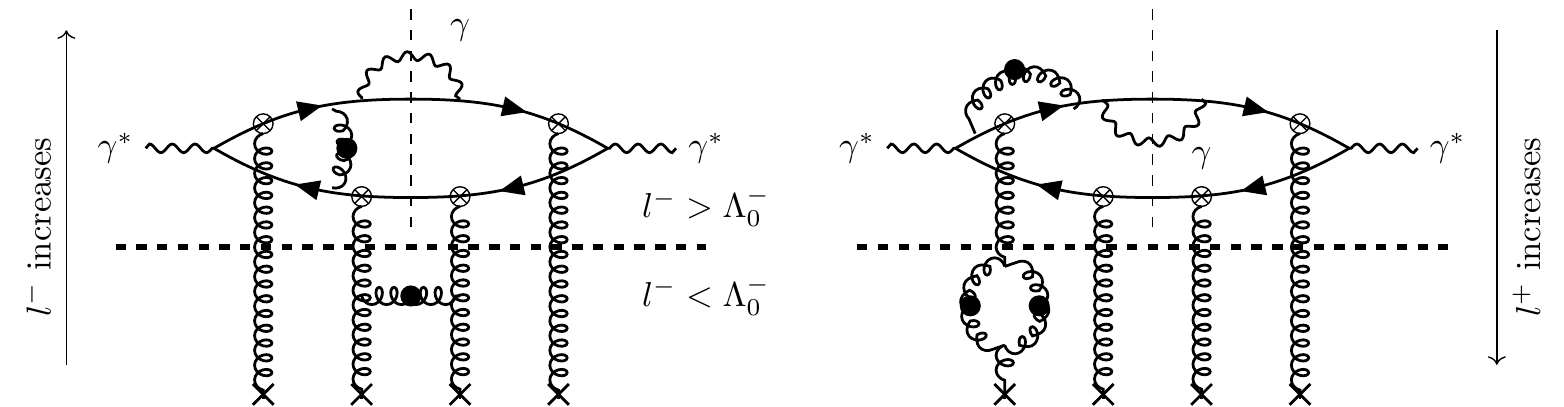}
\end{minipage}
\caption{Representative NNLO diagrams whose leading logarithmic pieces are combined with the $\alpha_{S}$ suppressed contributions to show the LL$x$ evolution of the weight functional for the large-$x$ color sources at NLO. \label{fig:NLLx-evolution-2}}
\end{figure}

There are however a second class of two loop NNLO contributions of the sort shown in Fig.~\ref{fig:NLLx-evolution-2} which contain contributions that are parametrically of order $\alpha_{S}^{2} \ln(\Lambda^{-}_{1}/\Lambda_{0}^{-})$. These correspond to one loop fluctuations of both the projectile and the target. From these processes, we have to extract the LL$x$ pieces from the gluon fluctuations below the cut $\Lambda_{0}^{-}$ and match them with the O($\alpha_S$)  NLO ``impact factor'' expression in Eq.~\ref{eq:NLO-2-dsigma} to obtain
\begin{equation}
\langle \delta \sigma_{\text{NNLO}}^{\rm jet}  \rangle=  \int [\mathcal{D} \rho_{A}] \,  W^{LLx}[\rho_{A}] \, \mathrm{d} \hat{\sigma}_{\text{NLO}}^{\text{jet;finite}} [\rho_{A}]  \,.
\label{eq:LOJIMWLK-NLO}
\end{equation}

As a result of our power counting in powers of $\alpha_S$ and $\alpha_S  \ln(\Lambda^{-}_{1}/\Lambda_{0}^{-}$), one can finally write the complete NLO result for the differential cross-section at NLL$x$ accuracy by combining the results in Eqs.~\ref{eq:NLL-JIMWLK} and \ref{eq:LOJIMWLK-NLO} respectively as
\begin{align}
\langle \mathrm{d} \sigma^{\rm jet}  \rangle_{\text{NLO}+\text{NLL$x$}}  &= \int [\mathcal{D} \rho_{A}] \, \Big\{ W^{NLLx}[\rho_{A}]  \,  \mathrm{d} \hat{\sigma}_{\text{LO}}^{\rm jet} [\rho_{A}]  + W^{LLx}[\rho_{A}] \, \mathrm{d} \hat{\sigma}_{\text{NLO}}^{\rm jet;finite} [\rho_{A}] \Big\}  \nonumber \\
&\simeq  \int [\mathcal{D} \rho_{A}] \,  \Big( W^{NLLx} [\rho_{A}] \, \Big\{  \mathrm{d} \hat{\sigma}_{\text{LO}}^{\rm jet} [\rho_{A}] +\mathrm{d} \hat{\sigma}_{\text{NLO}}^{\text{jet;finite}} [\rho_{A}] \Big\} + O(\alpha_{S}^{3} \ln(\Lambda^{-}_{1}/\Lambda^{-}_{0}) ) \Big) \, .
\label{eq:dsigma-NLO-NLLx}
\end{align}
As the $\simeq$ in the above equation indicates, we can go one step further by promoting the second term on the r.h.s of the first equality from $W^{LLx}\rightarrow W^{NLLx}$. This  extends the scope of computation to $O(\alpha_S^3 \ln(1/x))$ with the understanding that we will miss terms at that order of accuracy.  

The weight functional $W^{NLLx} [\rho_{A}]$ in Eq.~(\ref{eq:NLL-WF}) can be obtained by adapting extant results for LO and NLO($\equiv$O($\alpha_{S}^{2}$)) JIMWLK~\cite{Lublinsky:2016meo,Kovner:2013ona,Kovner:2014lca,Grabovsky:2013mba,Caron-Huot:2013fea} and BK~\cite{Balitsky:2008zza,Balitsky:2013fea,Kovchegov:2006vj} evolution into our approach. Thus to obtain results for photon+dijet production up to $O(\alpha_{S}^{3} \ln(\Lambda^{-}_{1}/\Lambda^{-}_{0}) ) $ accuracy, it is sufficient to compute the NLO impact factor $\mathrm{d} \hat{\sigma}_{\text{NLO}}^{\text{jet;finite}} [\rho_{A}]$. To go beyond this level of accuracy, we would have to compute fluctuations in the projectile involving the emission of two gluons, two loop virtual gluon processes, as well as interference diagrams of the genuine NLO processes shown in Fig.~\ref{fig:NLO-representative-projectile}. This will be reserved for a future project. 

The rest of the paper is organized as follows. In Sec.~\ref{sec:LO-review}, we shall briefly review the essential elements of the leading order (LO) computation of Paper I~\cite{Roy:2018jxq} and revisit some of the key results obtained there. In Sec.~\ref{sec:contributions-NLO}, we will outline the structure of the NLO computation, structured for convenience in two subsections. In Sec.~\ref{sec:structure-NLO-contributions}  we categorize contributions to the NLO amplitudes from real gluon emission and virtual gluon exchange diagrams in terms of their color structure. We also provide an interactive flowchart (see Fig.~\ref{fig:computational-tree}) in this subsection furnished with hyperlinks that direct the reader to the final expressions later in the text for the various components constituting the real and virtual contributions to the NLO amplitude. In Sec.~\ref{sec:assembly}, we organize these contributions at the level of the squared amplitude in Table~\ref{tab:NLO-assembly} in terms of common Wilson line structures. Doing so enables one to see cancellations of divergences in a transparent manner; this in turn facilitates the computation of the finite NLO impact factor in the photon+dijet inclusive cross-section. 
 
Section~\ref{sec:real-emission-details} contains a detailed computation of the amplitude for real gluon emission processes. These are discussed separately for the case when the real gluon either  crosses or does not cross the nuclear ``shock wave'' using a  representative diagram from each category. The final expression for the amplitude is given by Eq.~\ref{eq:master-NLOamp-real-emission}. In section~\ref{sec:virtual-emission-details}, we describe in detail the computation of the amplitude for virtual gluon exchange processes; these are broadly classified into the topologies of self-energy and vertex corrections. Specifically, sections~\ref{sec:virtual-self-energy-dressed} and \ref{sec:virtual-self-energy} deal with the amplitudes for self-energy graphs with dressed and free gluon propagators respectively. There are ultraviolet and rapidity singularities associated with these processes which are carefully isolated from the finite parts. For each class of diagrams, we use a representative graph to show the explicit computation. The results for the amplitudes from these processes are given respectively by Eqs.~\ref{eq:amplitude-SE1}, \ref{eq:amplitude-SE2} and \ref{eq:amplitude-self-energy-SE3}. A similar exercise is performed in sections~\ref{sec:virtual-corrections-vertex} and \ref{sec:vertex-corrections-free-gluon} respectively for the vertex correction processes with dressed and free gluon propagators. The final expressions for the amplitudes are given by Eqs.~\ref{eq:amplitude-V1-generic} and \ref{eq:amplitude-V2-generic} for the case of dressed gluon propagators and by Eqs.~\ref{eq:amplitude-V3-generic} and \ref{eq:amplitude-V4-generic} for the case in which the gluon does not cross the shock wave.
 
Section~\ref{sec:jet-cross-section} combines the results obtained in the earlier sections to obtain the final result for the principal goal of our study, the NLO impact factor for photon $+$ dijet production in $e+A$ DIS. We demonstrate here the cancellation of collinear divergences between real and virtual processes resulting in an infrared safe differential cross-section. To facilitate this, we introduce jet definitions and work in the approximation of a jet with small cone radius~\cite{Ivanov:2012ms} to explicitly extract the collinearly divergent contributions from the squared amplitudes of real gluon emission graphs which contain the possibility of a gluon being collinear to the (anti) quark. We note that there is no Sudakov suppression of the cross-section  because we have not imposed any kinematic constraints. Interestingly, we observe that (unlike the case of diffractive DIS~\cite{Boussarie:2016ogo}) the NLO cross-section does not factorize into the LO result and kinematic factors in the soft gluon limit. The implications of this result will be addressed in future work. 

In Section~\ref{sec:JIMWLK-evolution}, we take the slow gluon limit of our general expressions for the cross-sections and show that these provide a first principles derivation of the JIMWLK RG equation. While there exist several derivations of the JIMWLK equation in the literature going back to the original papers, many of these begin at the outset in the slow gluon limit.  It is therefore interesting to see how the JIMWLK equation arises in the explicit computation of the nontrivial photon+dijet  cross-section.  This exercise also helps lay the groundwork for an independent derivation of the NLL$x$ JIMWLK equation. 

We will end this paper with a brief summary and outlook. With regard to the latter, an important next step is to provide quantitative predictions for measurements at a future EIC. These are 
significantly more challenging even though our use of the wrong light cone ($A^-=0$) gauge allows us to present our computations in a manner analogous to comparable NLO computations in collinear factorization computations. This is firstly because going away from the collinear limit introduces additional nontrivial integrals in the computations. Further, a quantitative 
computation of the dipole and quadrupole correlations is much more complex than their parton distribution (pdf) counterparts. This is unsurprising because the former contain a tremendous amount of information on the physics of many-body correlations in QCD that are not contained in the pdfs. Nevertheless, the technology to achieve the desired goal has advanced considerably to bring it within reach. 

The  principal results and conclusions of this paper are spelled out in an accompanying letter~\cite{Roy:2019cux}. 

Appendices~\ref{sec:conventions} through \ref{sec:non-collinear-contributions} supplement the material in the body of the paper. The notations and conventions used throughout the paper are summarized  in Appendix~\ref{sec:conventions}. In the computation of the amplitude for the various processes, we will encounter tensor integrals over transverse components of the gluon loop momenta. General expressions for these constituent integrals along with details for special cases are provided in Appendix~\ref{sec:constituent-integrals-real-emission} for both processes with gluon emissions and gluon loops. Appendix~\ref{sec:R-factors-real-emission} contains detailed expressions for the amplitudes for gluon emission processes that are too cumbersome to include  in the main text. Likewise, in Appendix~\ref{sec:quark-self-energy}, we provide a detailed computation of the quark self-energy, which provides the template to compute the amplitudes of self-energy graphs where the gluon propagator is not dressed. The expression obtained in Eq.~\ref{eq:quark-selfenergy-loop-contribution} for the gluon loop contribution is very general and can be straightforwardly used in any pQCD computation performed using light cone coordinates and in the light cone gauge. A similar computation for the virtual gluon corrections to the $\gamma q q$ and $\gamma^{*}q \bar{q}$ vertices is provided in Appendices~\ref{sec:quark-real photon-quark-vertex-gluon-correction} and \ref{sec:vertex-correction-computation}.

Appendix~\ref{sec:T-V1-div-parts} contains the rapidity divergent pieces, discussed in Sec.~\ref{sec:virtual-corrections-vertex}, for the vertex corrections with the dressed gluon propagator that are not provided in the main text. Similar expressions for the amplitudes with final state interactions (discussed in Sec.~\ref{sec:vertex-corrections-free-gluon}) are provided in Appendix~\ref{sec:T-V4-div-parts}. The expressions for the finite pieces of the amplitudes are distributed over seven subsections in Appendix~\ref{sec:finite-pieces-virtual-graphs}. Finally, Appendix~\ref{sec:non-collinear-contributions} provides a short proof of the sub-dominance of non-collinearly divergent contributions to the cross-section for real gluon emissions when we work in the limit of small jet cone radius.

\section{General definitions and brief review of LO computation} \label{sec:LO-review}

We will work in the light cone (LC) gauge $A^{-}=0$ throughout this computation. The highly energetic nucleus is considered to be right moving so that it has a large `$+$' component of LC momentum $P_{N}^{+}$. The virtual photon exchanged between the electron and nucleus is considered to be left moving and consequently has a large `$-$' component of LC momentum $q^-$. The mass of the electron is neglected throughout the calculation. 

Following the LO computation in \cite{Roy:2018jxq}, we can write the amplitude for inclusive photon$+$dijet production in DIS as
\begin{equation}
\mathcal{M}(\bm{\tilde{l}}, \bm{\tilde{l'}},\bm{q},\bm{k},\bm{p},\bm{k}_{\gamma})=\frac{e}{Q^{2}} \overline{u}(\tilde{l'}) \,  \mathcal{G}^{\mu}_{q} \,  u(\tilde{l}) \mathcal{M}_{\mu}(\bm{q},\bm{k},\bm{p},\bm{k}_{\gamma};\lambda) \enskip,
\label{eq:amplitude-master}
\end{equation}
where
\begin{equation}
\mathcal{G}^{\mu}_{q}= \gamma^{\mu} - \frac{\slashed{q} n^{\mu}+ q^{\mu} \gamma^{-}}{q^{-}} , \qquad n^{\mu}=\delta^{\mu +} \, ,
\label{eq:gamma-contraction-with-virtual-photon-propagator}
\end{equation}
is obtained by index contraction with the propagator for the exchanged photon with momentum\footnote{See Appendix~\ref{sec:conventions} for the conventions used in this paper. }  $q=(-Q^{2}/2q^{-},q^{-},\bm{0}_{\perp})$. The amplitude for the hadronic subprocess is given by
\begin{equation}
\mathcal{M}_{\mu}(\bm{q},\bm{k},\bm{p},\bm{k}_{\gamma};\lambda)= \epsilon^{* \alpha}(\bm{k}_{\gamma},\lambda) \mathcal{M}_{\mu \alpha}(\bm{q},\bm{k},\bm{p},\bm{k}_{\gamma}) \enskip, \label{eq:hadronic-amp-master}
\end{equation}
and is the quantity of interest. Here $\epsilon(\bm{k}_{\gamma},\lambda) $ is the polarization vector for the outgoing  photon. The 4-momentum\footnote{For the outward directed external momenta, we have $\{k^{-}, p^{-}, k_{\gamma}^{-}, k^{-}_{g} \} > 0$.} assignments are given in Table~\ref{tab:example} and boldface letters denote 3-momentum vectors.

\begin{table*}[!htbp]
\caption{\label{tab:example}4-momentum assignments used in the calculation}
\begin{ruledtabular}
\begin{tabular}{lll}
$q$: Exchanged virtual photon & $\tilde{l}$: Incoming electron &\, ${\tilde l}^\prime$: Outgoing electron   \\
$k$: Quark, directed outward  & $p$: Antiquark, directed outward &\, $k_{\gamma}$:  Outgoing photon \\
$l_{i}$: Quark or gluon internal (loop) momentum to be integrated over  &\, $k_{g}$: Outgoing real gluon & \\
$P_{\text{tot}}$: Total momentum of final state in real emission= $p+k+k_{\gamma}+k_{g}$ \\
 $P$: Total momentum of final state in virtual emission and LO= $p+k+k_{\gamma}$
\end{tabular}
\end{ruledtabular}
\end{table*}

We define the following ratios of the outgoing momenta to the dominant component $q^{-}$ of the  incoming virtual photon momentum 
\begin{equation}
z_{q}=\frac{k^{-}}{q^{-}}, \quad z_{\bar{q}}=\frac{p^{-}}{q^{-}}, \quad z_{\gamma}=\frac{k^{-}_{\gamma}}{q^{-}}, \quad z_{g}=\frac{k_{g}^{-}}{q^{-}} \, .
\end{equation}
 We will work in the limit of light quarks and neglect their masses in the present computation. Since we are dealing with prompt photon production in DIS at small $x$, the dominant contribution is indeed expected to involve light quarks. 
 
Squaring the expression for the amplitude in Eq.~\eqref{eq:amplitude-master}, and performing the necessary averaging and sum over electron spins and photon polarizations\footnote{We use here the identity
\begin{equation*}
\sum_{\lambda}  \epsilon^{ \beta}(\bm{k}_{\gamma},\lambda) \epsilon^{* \alpha}(\bm{k}_{\gamma},\lambda)=-g^{\alpha \beta} + \frac{k^{\alpha}_{\gamma}n^{\beta}+k^{\beta}_{\gamma}n^{\alpha}}{k^{-}_{\gamma}} \enskip ,
\end{equation*} 
as the sum over outgoing photon polarizations. By virtue of the Ward identity, we can easily show that only terms proportional to $g^{\alpha \beta}$ contribute, thereby leading to Eq.~\ref{eq:H-tensor}.}, we can write 
\begin{equation}
\frac{1}{2} \sum_{\text{spins}, \lambda} \vert \mathcal{M} \vert^{2}=L^{\mu \nu} X_{\mu \nu} \enskip.
\label{eq:squared_amp-master}
\end{equation}
The lepton tensor given by
\begin{equation}
L^{\mu \nu}=\frac{e^{2}}{2 Q^{4}} \,  \text{tr} \, [ \tilde{\slashed{l}} \mathcal{G}^{\mu}_{q} {\slashed{\tilde{l}}}^\prime \mathcal{G}^{\nu}_{q} ]  \, ,
\label{eq:L-tensor}
\end{equation} 
is identical to the one obtained in fully inclusive DIS.

In the following, we will concentrate on obtaining the NLO contributions to the hadron tensor which is defined as 
\begin{equation}
X_{\mu \nu}= - \sum_{\text{spins}} \left \langle  \mathcal{M}_{\mu \alpha}^{*} (\bm{q},\bm{k},\bm{p},\bm{k}_{\gamma}) {\mathcal{M}_{\nu}}^{\alpha} (\bm{q},\bm{k},\bm{p},\bm{k}_{\gamma}) \right \rangle \, .
\label{eq:H-tensor}
\end{equation}
The $\langle \ldots \rangle $ in the above equation refers to the CGC  averaging over all possible source charge configurations $\rho_{A}$. From a first principles Quantum Field Theory  perspective, this corresponds to the systematic computation of Feynman diagrams in the presence of static sources, and subsequently performing averages over the source distribution, as spelled out in \cite{Gelis:2006yv,Gelis:2006cr,Gelis:2007kn}, and references therein. 

For a generic operator $\hat{\mathcal{O}}$ this is quantified as~\cite{Iancu:2000hn,Ferreiro:2001qy}
 \begin{equation}
\langle \hat{\mathcal{O}} \rangle = \int [ \mathcal{D} \rho_{A} ] \, W_{Y} [\rho_{A}] \, \hat{\mathcal{O}} [\rho_{A}] \, .
\label{eq:expectation-value-cgc}
 \end{equation}
 In this equation,  $\hat{\mathcal{O}} [\rho_A]$ is the quantum expectation value of the operator for a given charge configuration $\rho_A$. One then performs the classical-statistical average of $\hat{\mathcal{O}}$ over all possible color charge configurations with the gauge invariant weight functional $W_Y[\rho_A]$ representing the distribution of the color charge configurations at a rapidity $Y=\ln(x/x_0)$ in the target. This double average is justified because the color charges $\rho_A$ are long-lived (or static) on the time scales corresponding to the (quantum) dynamics of the gauge fields.  The functional dependence on $\rho_{A}$ enters the amplitude $\mathcal{M}_{\mu \alpha}$ through Wilson lines which are also the phase rotations in color space obtained by the quark and antiquark during their eikonal propagation along the light cone. We will see this more clearly when we present the structure of the amplitudes at LO and NLO in the upcoming discussion. 
 
Since we wish our presentation to be self-contained, we will sketch here the LO contributions to the amplitude derived in \cite{Roy:2018jxq}.
% albeit with minor redefinitions of the momenta suited for the present calculation.  
 At LO in the CGC power counting, there are four contributions to the amplitude; two of these are shown below in Fig.~\ref{fig:LO-diagrams} and the other two obtained by interchanging the quark and antiquark lines.
\begin{figure}[!htbp]
\begin{minipage}[b]{0.7\textwidth}
\includegraphics[width=\textwidth]{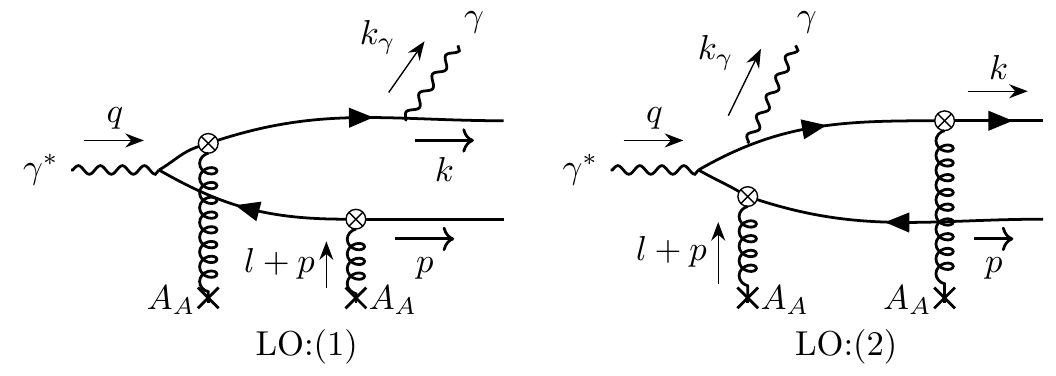}
\end{minipage}
\caption{Leading order contributions to the amplitude for photon production. The other two contributions (LO:(3) and LO:(4)) are obtained by interchanging the quark and antiquark lines in LO:(1) and LO:(2) respectively. The blobs with the crosses mark represent all possible eikonal scatterings with the background classical field of the nucleus including the possibility of ``no scattering''.
\label{fig:LO-diagrams}}
\end{figure}
As noted previously, an important ingredient in the computation is the simple form of the dressed quark propagator in the classical background field of the target nucleus~\cite{McLerran:1994vd,Ayala:1995hx,Ayala:1995kg,Balitsky:1995ub,Balitsky:2001mr}. In $A^-=0$ gauge, this can be expressed as~\cite{McLerran:1998nk}
\begin{equation}
S_{ij}(p,q) =  S_0(p)\,\mathcal{T}_{q;ij}(p,q)\,S_0 (q)
\label{eq:dressed-quark-mom-prop}
\end{equation}
where 
\begin{equation}
S_0 (p) = \frac{i \slashed{p}}{p^{2}+i \varepsilon}\,,
\end{equation}
is the free fermion propagator, and 
\begin{equation}
\mathcal{T}_{q;ji}(q,p)= (2 \pi)\delta(p^{-}-q^{-}) \gamma^{-} {\text{sign} (q^-)}\int \mathrm{d}^{2} \bm{z}_{\perp} \enskip e^{-i(\bm{q}_{\perp} - \bm{p}_{\perp})\cdot \bm{z}_{\perp}} \tilde{U}_{ji}^{\text{sign}(q^-)}(\bm{z}_{\perp}) \enskip ,
\label{eq:effective-vertex}
\end{equation}
is the effective vertex corresponding to the multiple scattering of the quark (or antiquark) off the shock wave background field. The dependence on the latter is given by the Wilson line defined in Eq.~(\ref{eq:Wilson-line}); here $i$ and $j$ label the colors of the incoming and outgoing quarks. Because we are including the possibility of ``no scattering'' within the definition of the effective vertex, the dressed propagator in Eq.~\ref{eq:dressed-quark-mom-prop} also contains a free part given by $(2\pi)^{4} \delta^{(4)}(p-q) S_{0}(p)$ and an interacting part which contains all possible scattering with the nuclear shock wave.

The diagram labeled as LO:(1) can be written as 
\begin{align}
\mathcal{M}^{\rm LO:(1)}_{\mu \alpha;ij}(\bm{q},\bm{k},\bm{p},\bm{k}_{\gamma})  =  - (eq_{f})^{2} \int \frac{\mathrm{d}^{4}l}{(2\pi)^{4}} \enskip \overline{u}(\bm{k}) \gamma_{\alpha} S_{0}(k+k_{\gamma})  \mathcal{T}_{q;ik}(k+k_{\gamma},q+l) S_{0}(q+l)  \gamma_{\mu}S_{0}(l) \mathcal{T}_{q;kj}(-p,l) v(\bm{p}) \enskip ,
\end{align}
The integration over $l^{-}$ is trivial because of the $\delta(l^{-}+p^{-})$ factor from one of the effective vertices. The integration over $l^{+}$ is performed using the theorem of residues. After subtracting the ``no scattering $\mathds{1}$" contribution in which neither the quark or antiquark cross the nuclear shock wave, we can write the result compactly\footnote{In the following, we will use the shorthand notations: $\int_{l}=\int \mathrm{d}^{4} l /(2\pi)^{4}$, $\int_{\bm{l}_{\perp}}=\int \mathrm{d}^{2} \bm{l}_{\perp}/(2\pi)^{2}$, $\int_{l^{\pm}}=\int \mathrm{d}l^{\pm}/2\pi$ and $\int_{\bm{x}_{\perp}}=\int \mathrm{d}^{2} \bm{x}_{\perp}$.}  as

\begin{align}
 \mathcal{M}^{\textrm{LO}:(1)}_{\mu \alpha}( \bm{q},\bm{k},\bm{p},\bm{k}_{\gamma}) = 2\pi (eq_{f})^{2}\delta(P^{-}-q^{-}) \int_{\bm{x}_{\perp} , \bm{y}_{\perp}} \int_{\bm{l}_{\perp}} & e^{i \bm{l}_{\perp}.(\bm{x}_{\perp}-\bm{y}_{\perp})-i(\bm{k}_{\perp}+\bm{k}_{\gamma \perp}).\bm{x}_{\perp}-i\bm{p}_{\perp}.\bm{y}_{\perp}}   \enskip \overline{u}(\bm{k}) R^{\text{LO}:(1)}_{\mu \alpha}(\bm{l}_{\perp}) \nonumber \\
 &\times\Big[\big( \tilde{U}(\bm{x}_{\perp})  \tilde{U}^{\dagger}(\bm{y}_{\perp})- \mathds{1} \big)   \Big]_{ij} v(\bm{p}) \enskip ,
\end{align}
with 
\begin{equation}
R^{\text{LO}:(1)}_{\mu \alpha}(\bm{l}_{\perp})= -\frac{1}{2(q^{-})^{2}} \gamma_{\alpha} \frac{\slashed{k}+\slashed{k}_{\gamma}}{2k.k_{\gamma}}  \gamma^{-} \frac{[\gamma^{+}(1-z_{\bar{q}})q^{-}-\bm{\gamma}_{\perp}. \, \bm{l}_{\perp}]\gamma_{\mu} [\gamma^{+}z_{\bar{q}}q^{-}+\bm{\gamma}_{\perp}. \, \bm{l}_{\perp}]}{\bm{l}_{\perp}^{2}+\Delta^{\text{LO}:(1)}} \gamma^{-} \, .
\label{eq:R-LO-1}
\end{equation}
Here $\Delta^{\rm LO:(1)}=Q^{2}z_{\bar{q}}(1-z_{\bar{q}})-i\varepsilon$. 

The other diagram in Fig.~\ref{fig:LO-diagrams} can be expressed similarly with $R^{\text{LO}:(1)}_{\mu \alpha}(\bm{l}_{\perp})\rightarrow R^{\text{LO}:(2)}_{\mu \alpha}(\bm{l}_{\perp})$, with 
\begin{align}
R^{\text{LO}:(2)}_{\mu \alpha}(\bm{l}_{\perp})&= \frac{1}{2(q^{-})^{2}} \gamma^{-} \frac{\gamma^{+}z_{q}q^{-} -\bm{\gamma}_{\perp}.(\bm{l}_{\perp}-\bm{k}_{\gamma \perp})}{(z_{q}+z_{\bar{q}})/z_{\bar{q}}}  \gamma_{\alpha} \frac{\gamma^{+}(1-z_{\bar{q}})q^{-}-\gamma^{-} \Big( (Q^{2} z_{\bar{q}}+\bm{l}_{\perp}^{2})/2z_{\bar{q}}q^{-} \Big) -\bm{\gamma}_{\perp}. \, \bm{l}_{\perp}}{\bm{l}_{\perp}^{2}+\Delta^{\text{LO}:(1)}} \gamma_{\mu} \nonumber \\
& \times \frac{\gamma^{+}z_{\bar{q}}q^{-}+\bm{\gamma}_{\perp}.\, \bm{l}_{\perp}}{\Big(\bm{l}_{\perp}+\bm{v}_{\perp}^{\text{LO}:(2)} \Big)^{2}+\Delta^{\text{LO}:(2)}    } \gamma^{-} \, ,
\label{eq:R-LO-2}
\end{align}
where
\begin{equation}
\bm{v}_{\perp}^{\text{LO}:(2)} =- \frac{z_{\bar{q}}}{1-z_{\gamma}}  \, \bm{k}_{\gamma \perp} \,\,;\,\,
\Delta^{\text{LO}:(2)} = \frac{z_{q}z_{\bar{q}}}{z_{\gamma} \, (1-z_{\gamma})^{2}} \bm{k}_{\gamma \perp}^{2}+\frac{Q^{2}z_{q}z_{\bar{q}}}{1-z_{\gamma}} -i\varepsilon \, .
\label{eq:denominator-factors-LO-amplitude}
\end{equation}
The remaining R-functions are related to those in Eqs.~\ref{eq:R-LO-1} and \ref{eq:R-LO-2} by the following replacements:  $ \bm{x}_{\perp} \leftrightarrow \bm{y}_{\perp}$, $ k \leftrightarrow p$ and $\bar{u}(\bm{k}) \leftrightarrow v(\bm{p})$. If we keep the internal momentum  labels identical to that in Fig.~\ref{fig:LO-diagrams}, this also results in an overall change in sign. Finally, one needs to redefine $\bm{l}_{\perp} \rightarrow -\bm{l}_{\perp}+\bm{k}_{\gamma \perp}$ in order to make the transverse phases in all four contributions identical.

The sum of the four contributions to the LO amplitude can therefore be compactly written as
\begin{align}
\mathcal{M}_{\mu \alpha}^{\rm LO}& =2\pi (eq_{f})^{2} \delta\Big( 1-z_{q}-z_{\bar{q}}-z_{\gamma} \Big) \int_{\bm{l}_{\perp}} \int_{\bm{x}_{\perp},\bm{y}_{\perp}} e^{i \bm{l}_{\perp}.(\bm{x}_{\perp}-\bm{y}_{\perp})-i(\bm{k}_{\perp}+\bm{k}_{\gamma \perp}).\bm{x}_{\perp}-i\bm{p}_{\perp}.\bm{y}_{\perp}} \nonumber \\
& \times \bar{u} (\bm{k}) T_{\mu \alpha}^{\rm LO}(\bm{l}_{\perp})\, \Big( \tilde{U} (\bm{x}_{\perp}) \tilde{U}^{\dagger} (\bm{y}_{\perp})- \mathds{1} \Big) v(\bm{p}) \, ,
\label{eq:LO-amp-master}
\end{align}
where $q_{f}$ is the charge of a quark or antiquark of a certain flavor $f$ and 
\begin{equation}
T^{\text{LO}}_{\mu \alpha} (\bm{l}_{\perp}) =\sum_{i=1}^{4} R^{\text{LO}:(i)}_{\mu \alpha}(\bm{l}_{\perp})  \, ,
\label{eq:T-LO}
\end{equation}
is the sum of the contributions from the four processes, whose individual contributions are given by $R^{\rm LO:(i)}_{\mu \alpha}$. 
Plugging this expression for the amplitude (and its complex conjugate) back into Eq.~(\ref{eq:H-tensor}), we obtain the LO triple differential inclusive cross-section for the production of a prompt photon in association with a dijet as~\cite{Roy:2018jxq}
\begin{equation}
\frac{\mathrm{d}^{3} \sigma^{\rm LO}}{\mathrm{d}x \,  \mathrm{d}Q^{2} \mathrm{d}^{6} K_{\perp} \mathrm{d}^{3} \eta_{K} }= \frac{\alpha_{em}^{2}q_{f}^{4}y^{2}N_{c}}{512 \pi^{5} Q^{2}} \, \frac{1}{(2\pi)^{4}} \,  \frac{1}{2} \,  L^{\mu \nu} \tilde{X}_{\mu \nu}^{\text{LO}} \, ,
\label{eq:triple-differential-CS-LO}
\end{equation}
where $\alpha_{em}=e^{2}/4\pi$ is the electromagnetic fine structure constant, $y=q\cdot P_{N}/\tilde{l}\cdot P_{N}$ is the familiar inelasticity variable of DIS and $L^{\mu \nu}$ is the lepton tensor in Eq.~\ref{eq:L-tensor}. We also introduced the differential phase space measures, $\mathrm{d}^{6}K_{\perp}=\mathrm{d}^{2}\bm{k}_{\perp} \mathrm{d}^{2}\bm{p}_{\perp} \mathrm{d}^{2}\bm{k}_{\gamma \perp} $ and $\mathrm{d}^{3} \eta_{K}=\mathrm{d}\eta_{k} \mathrm{d}\eta_{p} \mathrm{d}\eta_{k_{\gamma}}$.
In deriving the triple differential cross-section, we also isolated the prefactors $(eq_{f})^{4} N_{c}$ of the hadron tensor in Eq.~\ref{eq:H-tensor} and used a properly normalized wave packet description for the incoming virtual photon~\cite{Gelis:2002ki,Roy:2018jxq}. 

The leading order hadron tensor is given by
\begin{equation}
\tilde{X}^{\text{LO}}_{\mu \nu}=  2\pi  \, \delta(1-z_{q}-z_{\bar{q}}-z_{\gamma}) \int \mathrm{d} \Pi_{\perp}^{\text{LO}} \int \mathrm{d} {{\Pi_{\perp}^{\prime}}^{\text{LO}}}^{\star} \, \tau^{q\bar{q},q\bar{q}}_{\mu \nu}(\bm{l}_{\perp},\bm{l'}_{\perp})\, \Xi(\bm{x}_{\perp},\bm{y}_{\perp};\bm{y'}_{\perp},\bm{x'}_{\perp}) \, ,
\label{eq:H-tensor-LO}
\end{equation}
where we introduced a compact notation for the integrals over the phases appearing in the amplitude expression in Eq.~\ref{eq:LO-amp-master}
\begin{equation}
\int \mathrm{d} \Pi^{\text{LO}}_{\perp} = \int_{\bm{l}_{\perp}} \int_{\bm{x}_{\perp},\bm{y}_{\perp}} e^{i \bm{l}_{\perp}.(\bm{x}_{\perp}-\bm{y}_{\perp})-i(\bm{k}_{\perp}+\bm{k}_{\gamma \perp}).\bm{x}_{\perp}-i\bm{p}_{\perp}.\bm{y}_{\perp}} \, .
\label{eq:transverse-phases-LO}
\end{equation}
The second such term appearing in Eq.~\ref{eq:H-tensor-LO} results from the complex conjugate of Eq.~\ref{eq:transverse-phases-LO} and corresponds to replacing all transverse coordinates and internal momenta therein by their primed counterparts. The function 
\begin{equation}
\tau^{q\bar{q},q\bar{q}}_{\mu \nu}(\bm{l}_{\perp},\bm{l'}_{\perp})= \text{Tr} \Big[ \slashed{k} \, {T^{\text{LO}}_{\nu}}^{\alpha} (\bm{l}_{\perp}) \, (-\slashed{p}) \, \hat{\gamma}^{0} \, (T^{\text{LO}}_{\mu \alpha})^\dagger (\bm{l'}_{\perp}) \, \hat{\gamma}^{0} \Big] \, ,
\end{equation}
represents the spinor trace in the cross-section. 

\begin{figure}[!htbp]
    \centering
    \begin{minipage}{0.5\textwidth}
        \centering
        \includegraphics[width=0.75\linewidth, height=0.10\textheight]{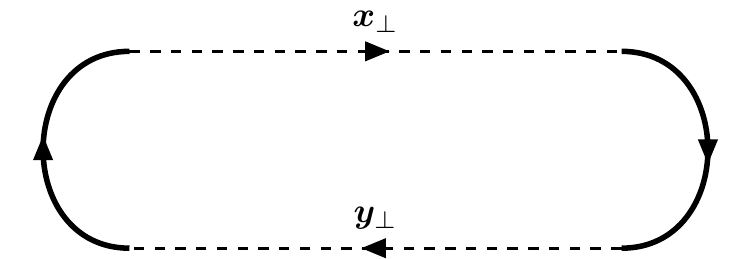}
    \end{minipage}%
    \begin{minipage}{0.5\textwidth}
        \centering
        \includegraphics[width=0.75\linewidth, height=0.20\textheight]{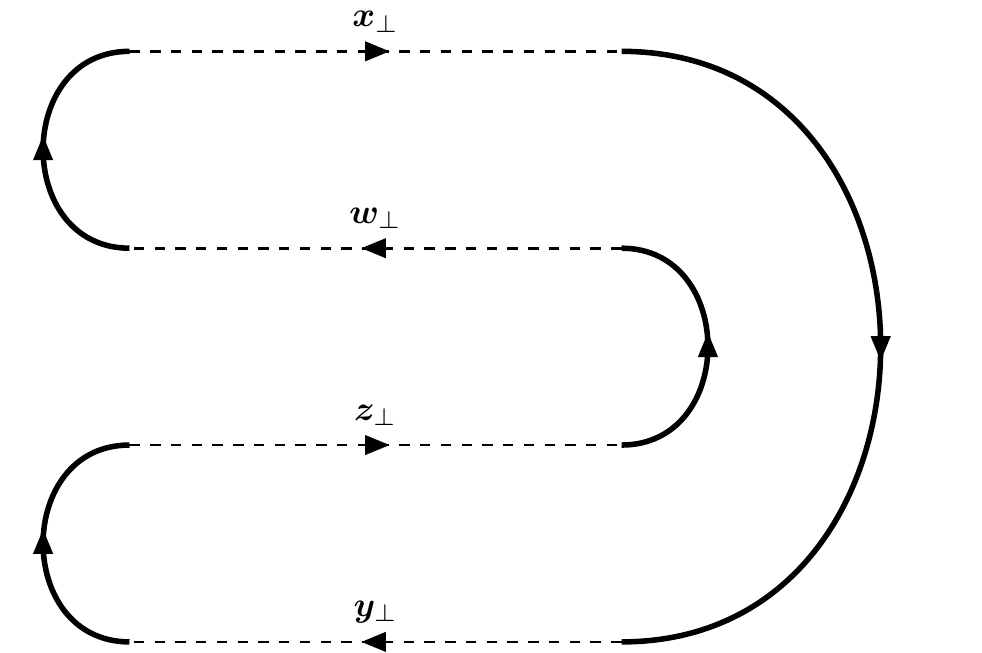}
    \end{minipage}
    \caption{Diagrammatic representations of dipole (left) and quadrupole (right) Wilson line correlators. The dotted lines represent Wilson lines at the given transverse spatial position in the fundamental representation of $SU(N_{c})$ and the horizontal axis  represents the $x^{-}$ from $x^{-}= -\infty$ to $\infty$. The arrows indicate the direction of path-ordering--whether we have a $\tilde{U}$ or $\tilde{U}^{\dagger}$. The Wilson lines are connected at $x^{-}=\pm \infty$ by solid lines which show the order in which the Wilson lines are multiplied and the closed loop represents the trace over the product.  
   \label{fig:dipole-quadrupole-correlators}}
\end{figure}

Finally, the nonperturbative input from the dynamics of saturated gluons in the nuclear target is contained in $\Xi(\bm{x}_{\perp},\bm{y}_{\perp};\bm{y'}_{\perp},\bm{x'}_{\perp})$ which can be decomposed as
\begin{equation}
\Xi(\bm{x}_{\perp},\bm{y}_{\perp};\bm{y'}_{\perp},\bm{x'}_{\perp})=1-D_{xy}-D_{y'x'}+Q_{y'x';xy} \, ,
\label{eq:LO-cross-section-color-structure}
\end{equation}
where 
\begin{align}
D_{xy}& =\frac{1}{N_{c}} \left \langle \text{Tr}\Big( \tilde{U}(\bm{x}_{\perp}) \tilde{U}^{\dagger}(\bm{y}_{\perp}) \Big) \right \rangle \, , \nonumber \\
Q_{xy;zw} & =\frac{1}{N_{c}} \left \langle \text{Tr} \Big( \tilde{U}(\bm{x}_{\perp}) \tilde{U}^{\dagger}(\bm{y}_{\perp})  \tilde{U}(\bm{z}_{\perp}) \tilde{U}^{\dagger}(\bm{w}_{\perp}) \Big)\right \rangle =Q_{zw;xy} \, ,
\label{eq:dipole-quadrupole-Wilson-line-correlators}
\end{align} 
represent respectively dipole and quadrupole Wilson line correlators. A pictorial representation of these correlators is given in Fig.~\ref{fig:dipole-quadrupole-correlators}.  These gauge invariant quantities appear in a variety of processes in both $p+A$ and $e+A$ collisions. Explicit expressions for these correlators are available~\cite{Kovchegov:2012mbw,Blaizot:2004wv,Dominguez:2012ad} in the McLerran-Venugopalan model~\cite{McLerran:1993ni,McLerran:1993ka,McLerran:1994vd}, where the distribution of sources $W[\rho_A]$ is Gaussian distributed with 
\begin{equation}
\left \langle \rho_{A}^{a} (x^{-},\bm{x}_{\perp}) \rho_{A}^{b} (y^{-},\bm{y}_{\perp}) \right \rangle = \delta^{ab} \delta(x^{-}-y^{-}) \delta^{(2)} (\bm{x}_{\perp}-\bm{y}_{\perp}) \lambda_{A}(x^{-}) \enskip .
\end{equation}
Here 
\begin{equation}
\int \mathrm{d} x^{-} \lambda_{A}(x^{-}) =\mu^{2}_{A}  \enskip,
\end{equation}
where $\mu^{2}_{A}=A/2\pi R^{2} \sim A^{1/3}$ is the average color charge squared of the valence quarks per color and per unit transverse area of a nucleus with mass number A.
As we will discuss later, the JIMWLK evolution equation can be reexpressed as evolution equations for these gauge invariant quantities.

\section{Outline of the NLO computation} \label{sec:contributions-NLO}

Before we dive into the rather involved computations (which, as articulated briefly in the introduction, have much of the complexity of two-loop computations in standard pQCD) it is useful to outline the structure of various contributions to the computation at NLO. These can be classified into real and virtual contributions; the latter can be further subdivided into self-energy and vertex corrections. An important simplification in the Regge limit is that the shock wave interaction is instantaneous, which eliminates more than one insertion from the effective vertex on any given line in a Feynman diagram. In addition to outlining the structures comprising the different contributions, we will also provide in this section a flow chart which points to the different contributions, and links that take the reader to specific terms in the computation, without having to wade through the entire detailed computation in the next section. 
 
 \subsection{Structure of contributing processes} \label{sec:structure-NLO-contributions}

%The addition of a final state photon increases the tediousness of the present problem manifold compared to the case of fully inclusive DIS. This is clear from the LO contributions which are four times in number that of inclusive DIS.
 There are both real gluon radiation and virtual gluon exchange processes that contribute at O($\alpha_{S}$) to inclusive photon$+$dijet production. For the computation of the NLO differential cross-section, we need to take the modulus squared of the amplitudes for gluon emission for fixed static color sources, perform the CGC averaging over the distribution of these color charge configurations, and finally integrate over the phase space of the emitted gluon. In the case of the amplitudes of graphs containing virtual gluons, we need to include the interference of these with the leading order amplitude given by Eq.~\ref{eq:LO-amp-master}, before  performing the CGC average over static color charge distributions. 
% Since the lepton sub-process remains the same, we have the same expression for the lepton tensor as given by Eq.~\ref{eq:L-tensor}. 
 
 The NLO hadron tensor is then given by
 \begin{align}
 X^{\rm NLO}_{\mu \nu}& = \Big\{ \int_{\mathrm{d} \Omega(\bm{k}_{g})} \Bigg \langle \Big( \mathcal{M}^{\rm NLO;Real}_{\mu \alpha} (\bm{q},\bm{k},\bm{p},\bm{k}_{\gamma},\bm{k}_{g} ) \Big)^{*}  \Big( \mathcal{M}^{\rm NLO;Real} (\bm{q},\bm{k},\bm{p},\bm{k}_{\gamma},\bm{k}_{g} )  \Big)_{\nu}^{\alpha} \Bigg \rangle \Big\} \nonumber \\
 & + \Big\{ \Bigg \langle \Big( \mathcal{M}^{\rm LO}_{\mu \alpha} (\bm{q},\bm{k},\bm{p},\bm{k}_{\gamma} ) \Big)^{*} \Big( \mathcal{M}^{\rm NLO;SE} (\bm{q},\bm{k},\bm{p},\bm{k}_{\gamma} )+  \mathcal{M}^{\rm NLO;Vert} (\bm{q},\bm{k},\bm{p},\bm{k}_{\gamma} ) \Big)_{\nu}^{\alpha} + c.c \Bigg \rangle \Big\}  \, ,
 \label{eq:H-tensor-NLO-generic}
 \end{align}
where
\begin{equation}
\int_{\mathrm{d} \Omega(\bm{k}_{g})} =  \int \frac{\mathrm{d}^{2} \bm{k}_{g \perp}}{(2\pi)^{2}} \int \frac{\mathrm{d}z_{g}}{(2\pi) \, 2z_{g}} \, , 
\end{equation}
is shorthand notation for integration over the phase space of the emitted gluon and $c.c$ denotes the complex conjugate. 
%The dependence of the amplitudes on the various 3-momenta are clearly shown. 
For virtual exchange graphs, we can broadly classify the two topologies of diagrams as self-energy and vertex contributions, which we have denoted above with the superscripts `$\rm SE$' and `$\rm Vert$' respectively. We will now describe the further systematic classification of the contributions to the amplitudes in each category in terms of their color structure.
 
 \subsubsection{Real emissions}
There are 20 Feynman graphs that describe the radiation of a gluon in addition to the photon radiated in the final state. Further, there are distinct topologies of these graphs depending on whether
\begin{enumerate}
\item the gluon is emitted prior to scattering of the quark \textbf{and} antiquark, or 
\item 
emitted by the quark \textbf{or} antiquark after they scatter off the nucleus.
\end{enumerate}
 In the former case, the gluon has the possibility of scattering off the background classical field whereas in the latter case it does not. For each of these diagrams, we need to subtract the ``no-scattering" contribution to the amplitude, which is obtained by setting $U$ and $\tilde{U}$'s to unity.
  
As in the case of the LO amplitude, we can write the amplitude for real emissions as 
\begin{align}
\mathcal{M}^{\text{NLO; Real}}_{\mu \alpha;b}&=2\pi (eq_{f})^{2} g \, \delta(q^{-}-P^{-}_{tot}) \,\, \int \mathrm{d} \Pi_{R} \,\,  \bar{u}(\bm{k}) \Bigg( T^{(1)}_{R;\mu \alpha} \Big( (\tilde{U}(\bm{x}_{\perp})t^{a}\tilde{U}^{\dagger}(\bm{y}_{\perp})) \, U_{ba}(\bm{z}_{\perp}) -t_{b} \Big)  \nonumber \\
& + T^{(2)}_{R;\mu \alpha} \Big( (t_{b} \tilde{U}(\bm{x}_{\perp})\tilde{U}^{\dagger} (\bm{y}_{\perp})) - t_{b}\Big)+ T^{(3)}_{R;\mu \alpha} \Big( ( \tilde{U}(\bm{x}_{\perp})\tilde{U}^{\dagger} (\bm{y}_{\perp})t_{b} ) - t_{b}\Big) \Bigg) v(\bm{p}) \, ,
\label{eq:real-emission-master-amplitude}
\end{align}
where $\int \mathrm{d} \Pi_{R}$ represents the integrals over the transverse Fourier phases associated with the effective vertices. By an appropriate redefinition of momenta, these can be made identical for all the contributions. Their exact form is not important for the present discussion but will be delineated in the upcoming sections which contain the detailed computation of the amplitudes for the various processes. The essential features of $T^{(1)}_{R}$, $T^{(2)}_{R}$, and $T^{(3)}_{R}$ are as follows:

\begin{itemize}
\item There are a set of 10 diagrams that contribute to the factor $T^{(1)}_{R}$. These are the processes where the emitted gluon may simultaneously scatter off the background classical field in addition to scattering of the quark-antiquark dipole. A representative diagram is shown in Fig.~\ref{fig:NLO-real-allscatter-representative}, with the other diagrams  obtained simply both by permutations of the emission vertex for the final state photon and by interchanging the quark-antiquark lines. 

\begin{figure}[!htbp]
\begin{center}
\includegraphics[scale=1]{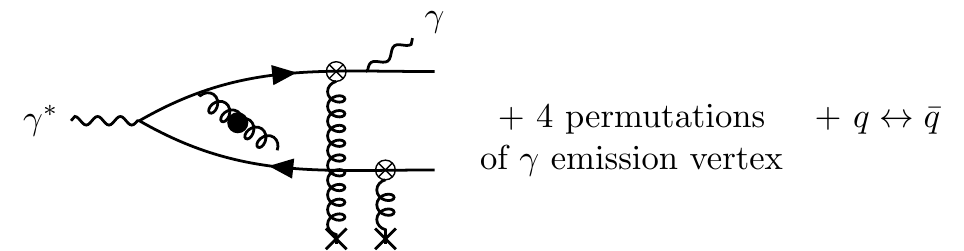}
\caption{Feynman diagram for gluon emission with the quark-antiquark dipole as well as the gluon scattering off of the background classical field. The other such diagrams are obtained by permutations of the photon emission vertex and their quark$\leftrightarrow$antiquark interchanged counterparts.\label{fig:NLO-real-allscatter-representative}}
\end{center}
\end{figure}

 \item There are  5 contributions that constitute $T^{(2)}_{R}$ in Eq.~\ref{eq:real-emission-master-amplitude}. These correspond to gluon emission from the quark after it scatters off the nucleus. A representative graph is shown in Fig.~\ref{fig:NLO-real-nogluonscatter-representative}; the others are obtained by permutations of the vertex for the final state photon.
 
\begin{figure}[!htbp]
\begin{center}
\includegraphics[scale=1]{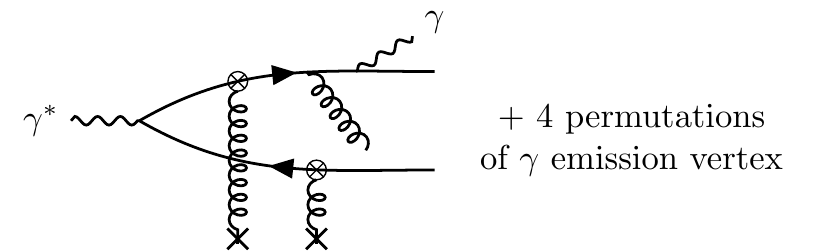}
\caption{Representative diagram for the NLO process involving real gluon emission from a quark after the quark-antiquark dipole gets scattered off the background classical field. The gluon does not get scattered in this scenario. The remaining diagrams are obtained simply by permutations of the photon emission vertex.\label{fig:NLO-real-nogluonscatter-representative}}
\end{center}
\end{figure}
 
\item Finally, there are 5 contributions constituting $T^{(3)}_{R}$ which are obtained by interchanging the quark and antiquark lines in the Feynman graphs of Fig.~\ref{fig:NLO-real-nogluonscatter-representative}. These are identified separately because they have a different color structure from the diagrams comprising $T^{(2)}_{R}$.
\end{itemize}
Thus at the level of the squared amplitude, 400 diagrams contribute to the NLO photon+dijet cross-section.

\subsubsection{Virtual contributions}
Broadly speaking, virtual contributions can be classified into vertex and self-energy graphs. In addition, there are diagrams in which the emitted gluon scatters off the shock wave before being reabsorbed by the quark/antiquark. To add to the complexity of such computations, the photon can be emitted either before or after these scatterings from the quark or antiquark. Thus the total number of diagrams to compute is significantly more than fully inclusive DIS at NLO. These can however be classed into distinct categories based on their Wilson lines structures. 
\begin{enumerate}
\item Self-energy contributions: We can write the amplitude of the self-energy contributions as
\begin{align}
\mathcal{M}^{\text{NLO; SE}}_{\mu \alpha}&=2\pi (eq_{f}g)^{2}  \, \delta(q^{-}-P^{-}) \,\, \int \mathrm{d} \Pi_{S} \,\,  \bar{u}(\bm{k}) \Bigg( T^{(1)}_{S;\mu \alpha} \Big( (t^{a}\tilde{U}(\bm{x}_{\perp})t^{b}\tilde{U}^{\dagger}(\bm{y}_{\perp})) \, U_{ab}(\bm{z}_{\perp}) -C_{F} \mathds{1} \Big)  \nonumber \\
& + T^{(2)}_{S;\mu \alpha} \Big( ( \tilde{U}(\bm{x}_{\perp})t^{a}\tilde{U}^{\dagger} (\bm{y}_{\perp}) t^{b}   ) \, U_{ba}(\bm{z}_{\perp})-C_{F} \mathds{1} \Big)+ T^{(3)}_{S;\mu \alpha}\Big( C_{F}( \tilde{U}(\bm{x}_{\perp})\tilde{U}^{\dagger} (\bm{y}_{\perp}) - \mathds{1}) \Big)  \Bigg) v(\bm{p})
 \, ,
\label{eq:selfenergy-correction-master-amplitude}
\end{align}
where $\int \mathrm{d}\Pi_{S}$ denotes the phase space factor corresponding to the self-energy contribution. 
\begin{itemize}
\item There are 6 contributions proportional to $T^{(1)}_{S}$ in the expression above; one such diagram is shown in Fig.~\ref{fig:NLO-self-energy-allscatter}. The topology of these diagrams corresponds to that of a gluon emitted by the quark prior to scattering and then reabsorbed by the quark after the $q\bar{q}g$ state scatters off the shock wave.
\begin{figure}[!htbp]
\begin{center}
\includegraphics[scale=1]{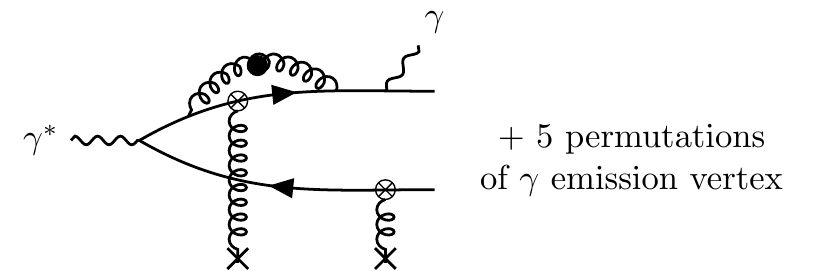}
\caption{Representative diagram involving a gluon loop where the gluon is emitted and reabsorbed by the quark with the possibility of scattering from the background field. The remaining 5 diagrams are obtained simply by permutations of the final state photon emission vertex.\label{fig:NLO-self-energy-allscatter}}
\end{center}
\end{figure}
\item The 6 contributions that constitute  $T^{(2)}_{S}$ in Eq.~\ref{eq:selfenergy-correction-master-amplitude} are obtained by interchanging the quark and antiquark lines in Fig.~\ref{fig:NLO-self-energy-allscatter}. They are classified separately because the interchange modifies the  color structure of the diagrams.
\item There are 24 other contributions in which the emission and absorption of the gluon occurs either prior to or subsequent to scattering with the nucleus; these are proportional to $T^{(3)}_{S}$. Two examples of such diagrams are shown in Fig.~\ref{fig:NLO-self-energy-noscatter}. The multitude of diagrams is primarily because of the various possibilities associated with the emission of the final state photon.
\begin{figure}[!htbp]
\centering
\begin{minipage}[b]{0.55\textwidth}
\includegraphics[width=\textwidth]{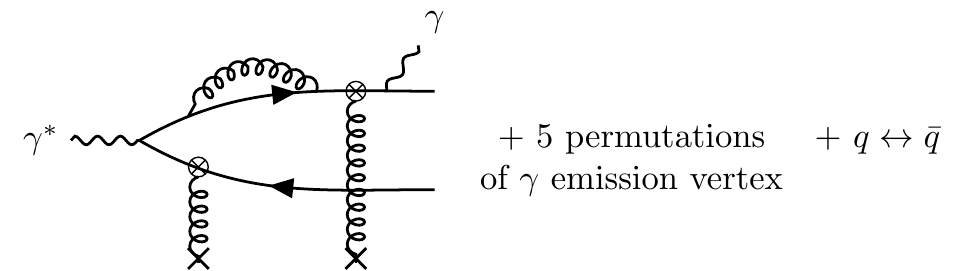}
\end{minipage}
\begin{minipage}[b]{0.55\textwidth}
\includegraphics[width=\textwidth]{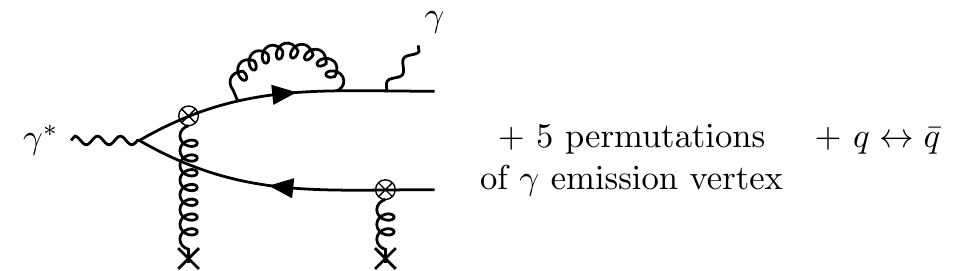}
\end{minipage}
\caption{Representative Feynman diagrams for self-energy graphs with no scattering by the virtual gluon. The gluon loop can be present either before or after the shock wave.  \label{fig:NLO-self-energy-noscatter} }
\end{figure}
\end{itemize}

\item Vertex contributions: Similarly to the self-energy contributions, the general expression for the amplitude of vertex contributions can be written as 
\begin{align}
\mathcal{M}^{\text{NLO; Vert}}_{\mu \alpha}&=2\pi (eq_{f}g)^{2}  \, \delta(q^{-}-P^{-}) \,\, \int \mathrm{d} \Pi_{V} \,\,  \bar{u}(\bm{k}) \Bigg( T^{(1)}_{V;\mu \alpha} \Big( (t^{a}\tilde{U}(\bm{x}_{\perp})t^{b}\tilde{U}^{\dagger}(\bm{y}_{\perp})) \, U_{ab}(\bm{z}_{\perp}) -C_{F} \mathds{1} \Big)  \nonumber \\
& + T^{(2)}_{V;\mu \alpha} \Big( ( \tilde{U}(\bm{x}_{\perp})t^{a}\tilde{U}^{\dagger} (\bm{y}_{\perp}) t^{b}   ) \, U_{ba}(\bm{z}_{\perp})-C_{F} \mathds{1} \Big)+ T^{(3)}_{V;\mu \alpha} \Big( C_{F}( \tilde{U}(\bm{x}_{\perp})\tilde{U}^{\dagger} (\bm{y}_{\perp}) - \mathds{1}) \Big) \nonumber \\
&+ T^{(4)}_{V;\mu \alpha}\Big(t^{a} \tilde{U}(\bm{x}_{\perp})\tilde{U}^{\dagger} (\bm{y}_{\perp}) t_{a} - C_{F} \mathds{1} \Big) \Bigg) v(\bm{p}) \, ,
\label{eq:vertex-correction-master-amplitude}
\end{align}
where $\int \mathrm{d} \Pi_{V}$ represents the phase space factor for vertex-like corrections and $C_{F}=(N^{2}_{c}-1)/2N_{c}$ is the quadratic Casimir for the fundamental representation of $SU(N_{c})$.

\begin{itemize}
\item There are 6 contributions to  $T^{(1)}_{V}$. A typical diagram is shown in Fig.~\ref{fig:NLO-vertex-allscatter};  the rest are obtained by permutations of the photon emission vertex. These correspond to the virtual gluon emitted by the antiquark following which it crosses the shock wave before being absorbed by the quark.

\begin{figure}[!htbp]
\begin{center}
\includegraphics[scale=1]{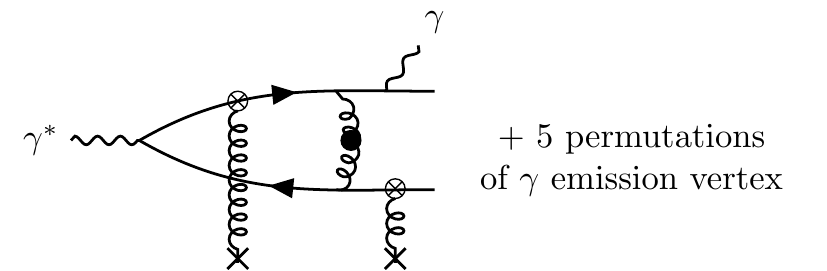}
\caption{Representative Feynman diagrams for the vertex corrections in which the exchanged gluon crosses the shock wave. \label{fig:NLO-vertex-allscatter}}
\end{center}
\end{figure}

\item The  $T^{(2)}_{V}$ are obtained by interchanging quark and antiquark lines in Fig.~\ref{fig:NLO-vertex-allscatter}.

\item There are 6 contributions proportional to $T^{(3)}_{V}$; one such graph is shown in Fig.~\ref{fig:NLO-vertex-noscatter}. These are part of the radiative corrections to the virtual photon wavefunction fluctuating into a quark-antiquark dipole with the addition of a final state photon. Consequently, the Wilson line factor is identical to that in the LO amplitude times the color factor $C_{F}$.

\begin{figure}[!htbp]
\begin{center}
\includegraphics[scale=1]{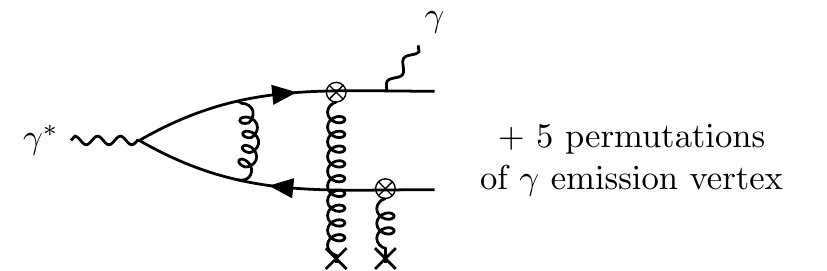}
\caption{Vertex corrections to dijet+photon production where the gluon does not scatter off the background classical field. The 5 other permutations are those of the photon emission vertex. Half of them are connected to the other half by quark$\leftrightarrow$antiquark interchange.  
 \label{fig:NLO-vertex-noscatter}}
\end{center}
\end{figure}

\item Finally, we have 6 contributions proportional to $T^{(4)}_{V}$, representing final state gluon interactions between the quark and antiquark after the latter cross the shock wave. An example of this process is shown below in Fig.~\ref{fig:NLO-vertex-final-state-interactions} and the remaining ones can be obtained via permutations of the final state photon vertex.  
Half of the diagrams are connected to the other half by quark$\leftrightarrow$antiquark interchange.

\begin{figure}[!htbp]
\begin{center}
\includegraphics[scale=1]{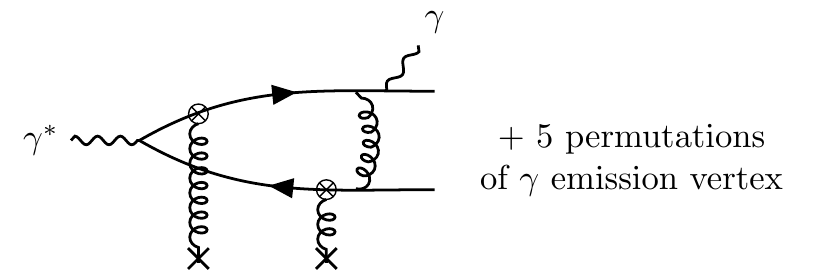}
\caption{Representative diagram for final state interactions between the quark and antiquark. \label{fig:NLO-vertex-final-state-interactions}}
\end{center}
\end{figure}
\end{itemize}

\end{enumerate}

\begin{figure}[!htbp]
\begin{center}
\begin{tikzpicture}
\draw (-6,0) rectangle (-2,1);
\node at (-4, 0.5) {\hyperref[eq:H-tensor-NLO-generic]{$X^{\rm NLO}$}};
 \draw  (-0.25, -0.5)-- (-0.25,1);
\draw (-1, 1) rectangle (0.5, 2);
\node at (-0.25, 1.5) {\hyperref[eq:LO-amp-master]{$\mathcal{M}^{\rm LO}$}};
\draw (0.5, 1.5) -- (1,1.5);
\draw (1, 1) rectangle (2.5,2);
\node at (1.7,1.5) {\hyperref[eq:T-LO]{$T^{\rm LO}$}};
\draw (2.5, 1.5) -- (3,1.5);
\draw (3, 0.2) rectangle (4.5, 2.8);
\node at (3.7,2.3) {\hyperref[eq:R-LO-1]{$R^{\rm LO:(1)}$}};
\node at (3.7,1.8) {\hyperref[eq:R-LO-2]{$R^{\rm LO:(2)}$}};
\node at (3.7,1.3) {$R^{\rm LO:(3)}$};
\node at (3.7,0.8) {$R^{\rm LO:(4)}$};
\draw [rounded corners, blue, ultra thick] (2.9, 0.4) rectangle (4.6, 1.55);
\draw (-10,-0.5)--(2, -0.5);
\draw (-10,-0.5) -- (-10,-1);
\draw (-4,0) -- (-4,-1);
\draw (2,-0.5) -- (2,-1);
\draw (-11,-2) rectangle (-9,-1);
\draw (-10,-2) -- (-10,-2.5);
\node at (-10, -1.5) {\hyperref[eq:master-NLOamp-real-emission]{$\mathcal{M}^{\rm NLO; Real}$}};
\draw (-11.5,-2.5) -- (-8.5, -2.5);
\draw (-11.5, -2.5) -- (-11.5, -3);
\draw (-12,-4) rectangle (-11, -3);
\node at (-11.5, -3.5) {\hyperref[eq:T-R1]{$T^{(1)}_{R}$}};
\draw (-11.5,-4) -- (-11.5,-4.5);
\draw (-12.5,-9) rectangle (-11, -4.5);
\node at (-11.8, -5) {\hyperref[eq:R-R1]{$R^{(R1)}$}};
\node at (-11.8, -5.5) {\hyperref[eq:R-R2]{$R^{(R2)}$}};
\node at (-11.8, -6) {\hyperref[eq:R-R3]{$R^{(R3)}$}};
\node at (-11.8, -6.5) {\hyperref[eq:R-R4]{$R^{(R4)}$}};
\node at (-11.8, -7) {\hyperref[eq:R-R5]{$R^{(R5)}$}};
\node at (-11.8, -7.5) {\hyperref[eq:R-R6]{$R^{(R6)}$}};
\node at (-11.8, -7.9) {\vdots};
\node at (-11.8, -8.5) {$R^{(R10)}$};
\draw [rounded corners,blue,ultra thick] (-12.7,-8.8) rectangle (-10.8,-7.2);
%\draw [->] (-11.8, -9) -- (-11.8, -9.5);
%\draw (-12.5, -12) rectangle (-7.5, -9.8);
\draw (-10,-2.5) -- (-10, -3);
\draw (-10.5, -4) rectangle (-9.5, -3);
\node at (-10, -3.5) {\hyperref[eq:T-R2]{$T^{(2)}_{R}$}};
\draw (-10,-4) -- (-10,-4.5);
\draw (-10.7,-7.5) rectangle (-9.3, -4.5);
\node at (-10, -5) {\hyperref[eq:R-R11]{$R^{(R11)}$}};
\node at (-10, -5.5) {\hyperref[eq:R-R12]{$R^{(R12)}$}};
\node at (-10, -6) {\hyperref[eq:R-R13]{$R^{(R13)}$}};
\node at (-10, -6.5) {\hyperref[eq:R-R14]{$R^{(R14)}$}};
\node at (-10, -7) {\hyperref[eq:R-R15]{$R^{(R15)}$}};
\draw (-8.5,-2.5) -- (-8.5, -3);
\node at (-8.5, -3.5) {\hyperref[eq:T-R3]{$T^{(3)}_{R}$}};
\draw (-8.5,-4) -- (-8.5,-4.5);
\draw (-9,-6.5) rectangle (-7.5, -4.5);
\node at (-8.3, -5) {$R^{(R16)}$};
\node at (-8.3, -5.4) {\vdots};
\node at (-8.3, -6) {$R^{(R20)}$};
\draw [<->, ultra thick] (-10,-7.7) -- (-10,-8)--(-8.3,-8) -- (-8.3,-7.7);
\node at (-9, -8.5) {\rm  \hyperref[eq:replacements-qqbar-exchange]{$q\leftrightarrow \bar{q}$  interchange}};
\draw (-9,-4) rectangle (-8,-3);
\draw (-5,-2) rectangle (-3,-1); %Second block
\draw (-4, -2) -- (-4, -3);
\draw (-5.5,-3) -- (-5.5,-2.5) -- (-2.5,-2.5) -- (-2.5,-3);
\draw (-6,-4) rectangle (-5,-3);
\node at (-5.5, -3.5) {\hyperref[eq:T-S1]{$T^{(1)}_{S}$}};
\draw (-4.5, -4) rectangle (-3.5, -3);
\node at (-4, -3.5) {\hyperref[eq:T-S2]{$T^{(2)}_{S}$}};
\draw (-3,-4) rectangle (-2,-3);
\node at (-2.5, -3.5) {\hyperref[eq:T-S3]{$T^{(3)}_{S}$}};
\node at (-2.3, -5) {\hyperref[eq:R-S13]{$R^{(S13)}$}};
\node at (-2.3, -5.5) { \hyperref[eq:R-S14-S15-S16]{$R^{(S14)}$}};
\node at (-2.3, -6) {\hyperref[eq:R-S14-S15-S16] {$R^{(S15)}$}};
\node at (-2.3, -6.5) {\hyperref[eq:R-S14-S15-S16]{$R^{(S16)}$}};
\node at (-2.3, -7) {$R^{(S17)}$};
\draw [rounded corners, ultra thick, red] (-3.2,-8.2) rectangle (-1.4,-6.8);
\node at (-1,-7.5) {$=0$};
\node at (-2.3, -7.4) {\vdots};
\node at (-2.3, -8) {$R^{(S20)}$};
\node at (-2.3, -8.5) {\hyperref[eq:R-S21-S23]{$R^{(S21)}$}};
\node at (-2.3, -9) {\hyperref[eq:R-S21-S23]{$R^{(S22)}$}};
\node at (-2.3, -9.5) {\hyperref[eq:R-S23]{$R^{(S23)}$}};
\node at (-2.3, -10) {\hyperref[eq:R-S24]{$R^{(S24)}$}};
\node at (-2.3, -10.5) {$R^{(S25)}$};
\node at (-2.3, -10.9) {\vdots};
\node at (-2.3, -11.5) {$R^{(S36)}$};
\node at (-11.5, -10.5) {\large \textbf{Legend:}};
\draw [rounded corners, ultra thick, blue] (-3.2, -11.8) rectangle (-1.4, -10.3);
\draw (-11.8,-14) rectangle (-10.7, -11);
\node at (-11.2, -11.5) {$R^{(a)}$};
\node at (-11.2, -12) {$R^{(b)}$};
\node at (-11.2, -12.8) {$R^{(c)}$};
\draw [ultra thick,<-] (-10.8, -11.5)  arc[radius=1, start angle= 45, end angle = -45] (-10.8,-12.8);
\draw [ultra thick,<-] (-10.8, -12)  arc[radius=1, start angle= 45, end angle = -45] (-10.8,-13.3);
\node at (-11.2, -13.3) {$R^{(d)}$};
\draw [rounded corners, ultra thick, blue]  (-12,-13.8) rectangle (-10.5,-12.5);
\node at (-8.5,-12.5) [align=left] {related by \\
 \hyperref[eq:replacements-qqbar-exchange]{$q \leftrightarrow \bar{q}$ interchange}};
\draw [rounded corners, ultra thick, blue] (-3.2, -11.8) rectangle (-1.4, -10.3);
%\node at (-4,-13) [align=left] {\rm Quantities within blue boxes are %obtained by imposing $q\leftrightarrow \bar{q}$ replacements \\
%to their counterparts that appear above them from top to bottom %within the same column};
\draw (-5.5,-4) -- (-5.5,-4.5);
\draw (-4,-4) -- (-4, -4.5);
\draw (-2.5, -4) -- (-2.5, -4.5);
\draw (-6.2, -8) rectangle (-4.8,-4.5);
\draw (-4.6, -6.5) rectangle (-3.2,-4.5);
\draw (-3, -12) rectangle (-1.6, -4.5);
\node at (-5.5, -5) {\hyperref[eq:R-S1]{$R^{(S1)}$}};
\node at (-5.5,-5.5) {\hyperref[eq:R-S2]{$R^{(S2)}$}};
\node at (-5.5, -6) {\hyperref[eq:R-S3]{$R^{(S3)}$}};
\node at (-5.5, -6.5) {\hyperref[eq:R-S4-R-S6]{$R^{(S4)}$}};
\node at (-5.5, -7) {\hyperref[eq:R-S4-R-S6]{$R^{(S5)}$}};
\node at (-5.5, -7.5) {\hyperref[eq:R-S4-R-S6]{$R^{(S6)}$}};
\node at (-4, -5) {$R^{(S7)}$};
\node at (-4,-5.4) {\vdots};
\node at (-4, -6) {$R^{(S12)}$};
\draw [<->,ultra thick] (-5.5,-8.2) -- (-5.5, -8.5) -- (-3.7,-8.5) -- (-3.7,-8.2);
\node at (-5, -9) { \hyperref[eq:replacements-qqbar-exchange]{$q \leftrightarrow \bar{q}$ \rm interchange}};
\node at (-4, -1.5) {\hyperref[eq:selfenergy-correction-master-amplitude]{$\mathcal{M}^{\rm NLO; SE}$}};
\draw (1,-2) rectangle (3,-1); %Third block 
\draw (2,-2)-- (2,-2.5);
\draw (0,-2.5) -- (4.5, -2.5);
\draw (0,-2.5) -- (0,-3);
\node at (0,-3.5) {\hyperref[eq:T-V1]{$T^{(1)}_{V}$}};
\draw (0,-4) -- (0,-4.5);
\draw (-0.5, -4) rectangle (0.5, -3);
\draw (-0.5,-8) rectangle (0.7, -4.5);
\node at (0.1, -5) {\hyperref[eq:R-V1]{$R^{(V1)}$}};
\node at (0.1, -5.5) {\hyperref[eq:R-V2-R-V5]{$R^{(V2)}$}};
\node at (0.1, -6) {\hyperref[eq:R-V3]{$R^{(V3)}$}};
\node at (0.1, -6.5) {\hyperref[eq:R-V4]{$R^{(V4)}$}};
\node at (0.1, -7) {\hyperref[eq:R-V2-R-V5]{$R^{(V5)}$}};
\node at (0.1, -7.5) {\hyperref[eq:R-V6]{$R^{(V6)}$}};
\draw (1.5,-2.5) -- (1.5,-3);
\node at (1.5,-3.5) {\hyperref[eq:T-V2]{$T^{(2)}_{V}$}};
\draw [<->,ultra thick] (0.1,-8.2)-- (0.1,-8.5) -- (1.8,-8.5) -- (1.8,-8.2);
\node at (1,-8.8) { \hyperref[eq:replacements-qqbar-exchange]{$q \leftrightarrow \bar{q}$ \rm interchange}};
\draw (1.5, -4) -- (1.5, -4.5);
\draw (0.9,-6.5) rectangle (2.1, -4.5);
\node at (1.5, -5) {$R^{(V7)}$};
\node at (1.5, -5.3) {\vdots};
\node at (1.5, -6) {$R^{(V12)}$};
\draw (1, -4) rectangle (2,-3);
\draw (3,-2.5) -- (3,-3);
\node at (3,-3.5) {\hyperref[eq:T-V3]{$T^{(3)}_{V}$}};
\draw (3,-4) -- (3,-4.5);
\draw (2.5, -4) rectangle (3.5, -3);
\draw (2.3,-8) rectangle (3.6,-4.5);
\node at (3,-5) {\hyperref[eq:R-V13]{$R^{(V13)}$}};
\node at (3,-5.5) {\hyperref[eq:R-V14]{$R^{(V14)}$}};
\node at (3,-6) {\hyperref[eq:R-V15]{$R^{(V15)}$}};
\node at (3,-6.6) {$R^{(V16)}$};
\node at (3,-7.1) {$R^{(V17)}$};
\node at (3,-7.6) {$R^{(V18)}$};
\draw (4.5, -2.5) -- (4.5, -3);
\node at (4.5, -3.5) {\hyperref[eq:T-V4]{$T^{(4)}_{V}$}};
\draw (4.5, -4) -- (4.5,-4.5);
\node at (4.5,-5) {\hyperref[eq:R-V19]{$R^{(V19)}$}};
\node at (4.5,-5.5) {$R^{(V20)}$};
\node at (4.5,-6) {\hyperref[eq:R-V21]{$R^{(V21)}$}};
\node at (4.5,-6.6) {$R^{(V22)}$};
\node at (4.5,-7.1) {$R^{(V23)}$};
\node at (4.5,-7.6) {$R^{(V24)}$};
\draw (3.8, -8) rectangle (5.2,-4.5);
\draw [rounded corners, blue, ultra thick] (2.2,-8.1) rectangle (5.3,-6.3);
\draw (4, -4) rectangle (5,-3);
\node at (2, -1.5) {\hyperref[eq:vertex-correction-master-amplitude]{$\mathcal{M}^{\rm NLO; Vert}$}};
\end{tikzpicture}
\end{center}
\caption{Computational tree for the NLO computation of the hadron tensor. The different branches correspond to the different components constituting $X^{\rm NLO}_{\mu \nu}$. The first of these nodes represent the amplitude contributions that comprise the NLO hadron tensor. The sub-branches contain the combined result for the hard parts of the amplitude for the different contributing processes. These  individual contributions from diagrams  which are categorized based on the color structure are provided in the long columns. Quark-antiquark interchanged counterparts of quantities appearing in the same class of diagrams and hence the same column are grouped within blue rectangular boxes. As an example, $R^{(R6),\ldots,(R10)}$ are obtained respectively by exchanging $q\leftrightarrow \bar{q}$ in $R^{(R1),\ldots, (R5)}$.  The labeling follows the same ordering for the other quantities grouped in blue boxes. Terms grouped in the red box are all zero and do not contribute to the amplitude. \label{fig:computational-tree}}
\end{figure}
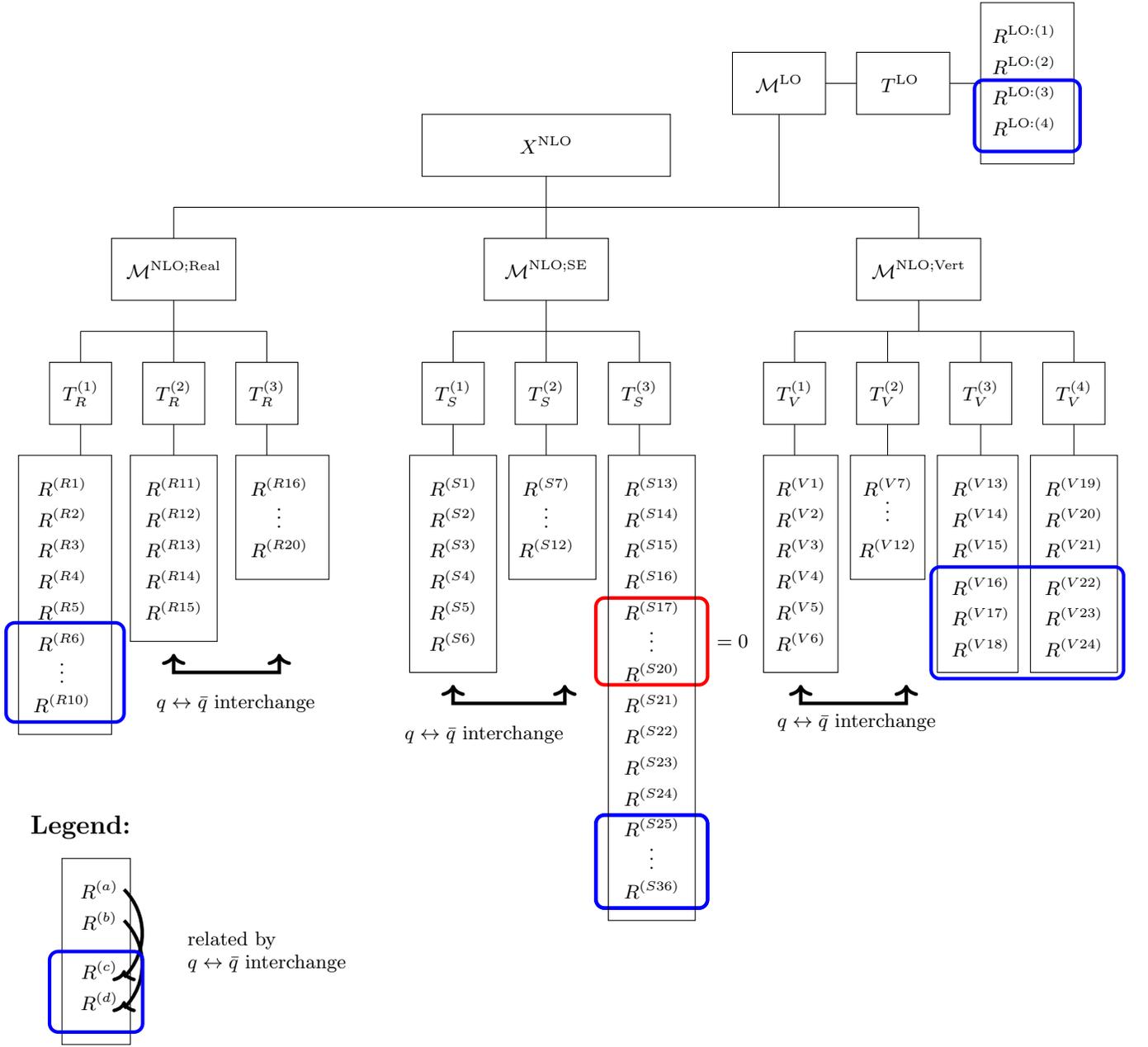

For the convenience of the reader, the computational tree depicted in Figure~\ref{fig:computational-tree} shows the components and sub-components building up the NLO hadron tensor in Eq.~\ref{eq:H-tensor-NLO-generic}.  Clicking on each of these will take him or her to the particular expression desired. This computational tree is therefore also a template for numerical evaluation of the NLO photon+dijet cross-section that will be the subject of future work. 

As discussed above,  perturbative contributions from kinematically allowed diagrams with similar color structure are contained in the various $T_{R,S,V}$ functions. Each of these is the sum of the contributions of the different Feynman diagrams denoted by $R^{i}_{\mu \alpha}$ and are presented in columns under the $T_{R,S,V}$  functions in Fig.~\ref{fig:computational-tree}. Within a certain column, there may be diagrams that are connected to one another by quark-antiquark interchange. We have put these together within blue rectangular boxes in Fig.~\ref{fig:computational-tree} . These $R$-functions can be obtained in sequence by imposing the $q \leftrightarrow \bar{q}$ replacements given by  Eq.~\ref{eq:replacements-qqbar-exchange} (later in the text) in the functions appearing in the columns above the blue boxes. Moreover, there are entire categories of processes related by interchange of the quark and antiquark lines. These are also shown in Figure~\ref{fig:computational-tree}. 

\subsection{Assembling the different contributions in the amplitude squared} \label{sec:assembly}

For the computation of the differential cross-section, we need to take the modulus squared of the amplitude for the real emission processes and the interference of the virtual graphs with LO processes. The general expressions for the NLO amplitudes are given by Eqs.~\ref{eq:real-emission-master-amplitude}, \ref{eq:selfenergy-correction-master-amplitude} and \ref{eq:vertex-correction-master-amplitude} respectively while Eq.~\ref{eq:LO-amp-master} denotes the same for the LO amplitude. The squared amplitude, a functional of the stochastic source charge density $\rho_{A}$, then needs to be averaged over all possible charge configurations weighted by the distribution $W[\rho_{A}]$. Following  extensive use of the Fierz identity
\begin{equation}
(t^{a})_{ij} (t^{a})_{kl}=\frac{1}{2} \Big( \delta_{il} \delta_{jk} -\frac{1}{N_{c}} \delta_{ij} \delta_{kl} \Big) \, ,
\label{eq:Fierz-identity}
\end{equation}
and the relation
\begin{equation}
\tilde{U} (\bm{x}_{\perp}) t_{a} \tilde{U}^{\dagger} (\bm{x}_{\perp}) = t^{b} U_{ba} (\bm{x}_{\perp}) \, ,
\label{eq:adjoint-to-fundamental-connection}
\end{equation}
connecting adjoint Wilson lines to fundamental ones,  we get non-trivial combinations of dipole and quadrupole Wilson line correlators (see Eqs.~\ref{eq:dipole-quadrupole-Wilson-line-correlators}). These are summarized clearly in Table~\ref{tab:NLO-assembly} below.
\begin{table}[!htbp]
\begin{center}
    \begin{tabular}{ | p{5cm} | p{3cm} | p{3cm} | p{3cm} | }
    \hline
    Wilson line factor & Real emission & Virtual: Vertex& Virtual: Self-energy\\ \hline
     $\frac{N^{2}_{c}}{2}\Big( 1-D_{xz}D_{zy}-D_{y'z}D_{zx'}+D_{y'y}D_{xx'} \Big) -\frac{1}{2}\Xi(\bm{x}_{\perp},\bm{y}_{\perp};\bm{y'}_{\perp},\bm{x'}_{\perp}) $ & ${T^{(1)}_{R}}^{*}T^{(1)}_{R}$ &  &  \\
    \hline
    $\frac{N^{2}_{c}}{2} \Big( 1- D_{xy} -D_{y'x'} +Q_{xy;y'x'} \Big) -\frac{1}{2} \Xi(\bm{x}_{\perp},\bm{y}_{\perp};\bm{y'}_{\perp},\bm{x'}_{\perp})$ & ${T^{(2)}_{R}}^{*} T^{(2)}_{R}+ {T^{(3)}_{R}}^{*} T^{(3)}_{R}$  &$T^{*}_{LO}T^{(3)}_{V}+c.c$ & $ T^{*}_{LO}T^{(3)}_{S}+c.c$ \\ \hline
   $\frac{N^{2}_{c}}{2}[(1-D_{xy})(1-D_{y'x'})] -\frac{1}{2}\Xi(\bm{x}_{\perp},\bm{y}_{\perp};\bm{y'}_{\perp},\bm{x'}_{\perp})$ & ${T^{(2)}_{R}}^{*}T^{(3)}_{R}+c.c$  & $T^{*}_{LO} T^{(4)}_{V}+c.c$ & \\ \hline
    $\frac{N^{2}_{c}}{2}\Big( 1+(Q_{zy;y'x'}-D_{zy}) D_{xz}-D_{y'x'} \Big) -\frac{1}{2}\Xi(\bm{x}_{\perp},\bm{y}_{\perp};\bm{y'}_{\perp},\bm{x'}_{\perp}) $ & ${T^{(2)}_{R}}^{*}T^{(1)}_{R}$ & $T^{*}_{LO} T^{(1)}_{V}$ & $T^{*}_{LO}T^{(1)}_{S}$ \\
    \hline
     %$\frac{N^{2}_{c}}{2}\Big( 1+(Q_{xy;y'z}-D_{y'z}) D_{zx'}-D_{xy} \Big) -\frac{1}{2}\Xi(\bm{x}_{\perp},\bm{y}_{\perp};\bm{y'}_{\perp},\bm{x'}_{\perp}) $ & ${T^{(1)}_{R}}^{*}T^{(2)}_{R}$ & ${T^{(1)}_{V}}^{*} T_{LO} $ & $ {T^{(1)}_{S}}^{*} T_{LO} $ \\
    %\hline
     $\frac{N^{2}_{c}}{2}\Big( 1+(Q_{y'x';xz}-D_{xz}) D_{zy}-D_{y'x'} \Big) -\frac{1}{2}\Xi(\bm{x}_{\perp},\bm{y}_{\perp};\bm{y'}_{\perp},\bm{x'}_{\perp}) $ & ${T^{(3)}_{R}}^{*}T^{(1)}_{R}$ & $T^{*}_{LO} T^{(2)}_{V}  $ & $ T^{*}_{LO} T^{(2)}_{S}  $ \\
    \hline
    % $\frac{N^{2}_{c}}{2}\Big( 1+(Q_{xy;zx'}-D_{zx'}) D_{y'z}-D_{xy} \Big) -\frac{1}{2}\Xi(\bm{x}_{\perp},\bm{y}_{\perp};\bm{y'}_{\perp},\bm{x'}_{\perp}) $ & ${T^{(1)}_{R}}^{*}T^{(3)}_{R}$ & ${T^{(2)}_{V}}^{*} T_{LO} $ & $ {T^{(2)}_{S}}^{*} T_{LO} $ \\
    % \hline
    
    \end{tabular}
\end{center}
\caption{Classification of the dipole and quadrupole Wilson line structures from different contributions to the amplitude squared. The Wilson lines for the conjugates of the terms in row 4 (5) are obtained by replacing $(\bm{x}_{\perp}, \bm{y}_{\perp}) \rightarrow (\bm{y'}_{\perp}, \bm{x'}_{\perp})$ on the factor in row 5 (4). The expression for $\Xi$ is given in 
Eq.~\ref{eq:LO-cross-section-color-structure}. }
\label{tab:NLO-assembly}
\end{table}
 The terms  proportional to products of T's represent Dirac traces obtained from expressions for the squared amplitudes. The corresponding color trace over products of Wilson lines  corresponding to each row is given in the leftmost column. To obtain the color factors for the conjugates of the terms in rows 4 (5), we need to replace $(\bm{x}_{\perp}, \bm{y}_{\perp}) \rightarrow (\bm{y'}_{\perp}, \bm{x'}_{\perp})$ in the transverse coordinates of the corresponding color factors of rows 5 (4). As evident from Table~\ref{tab:NLO-assembly}, the fundamental building blocks which span the entire high energy computation have the structures $D$, $Q$, $DD$ and $DQ$, albeit with different dependence on the transverse coordinates. In the sections that follow, we will carry out detailed computations of the various entities in Table~\ref{tab:NLO-assembly}. The organization of the NLO computation in the manner described here will provide a transparent guide to the identification of soft, collinear and ultraviolet divergences in the computations.

\section{NLO contributions to the amplitude from real emissions: Detailed calculations} \label{sec:real-emission-details}

In this section, we will compute in detail the amplitudes for the various real emission graphs presented in Sec.~\ref{sec:contributions-NLO}. As discussed there, there are two distinct topologies based on gluon emission before and after scattering of the dipole off the background classical field. We shall now systematically illustrate how to calculate the various diagrams contributing to each of the three terms in the general amplitude expression in Eq.~\ref{eq:real-emission-master-amplitude}. Readers uninterested in these details can 
proceed directly to Section~\ref{sec:jet-cross-section}.

\begin{enumerate}
\item \textbf{Contributions to $T^{(1)}_{R}$}:  The processes that contribute to $T^{(1)}_{R}$ in the amplitude (see Eq.~\ref{eq:real-emission-master-amplitude}) for real gluon emission are shown below in Fig.~\ref{fig:NLO-real-allscatter}. Only half of them (labeled $(R1)-(R5)$) are presented here. The quark$\leftrightarrow$antiquark interchanged counterparts of these 5 diagrams are respectively labeled $(R6)-(R10)$.

 \begin{figure}[!htbp]
\centering
\begin{minipage}[b]{0.75\textwidth}
\includegraphics[width=\textwidth]{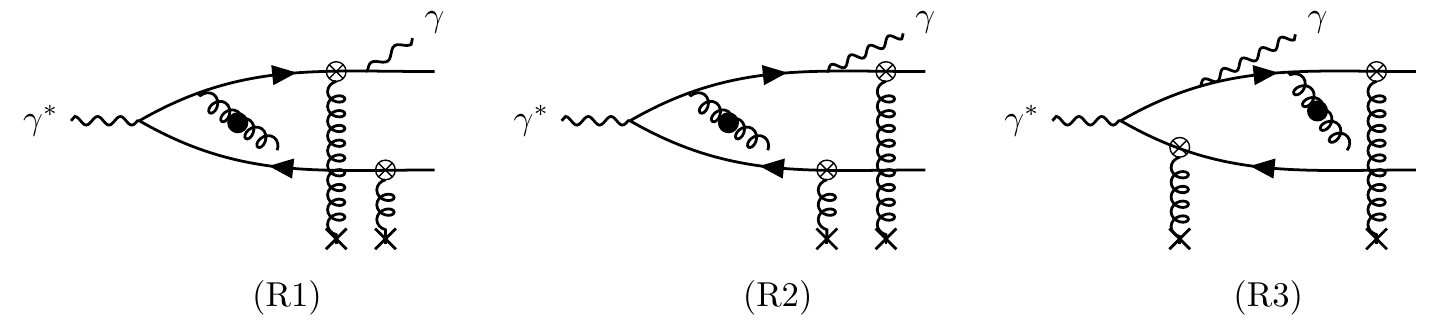}
\end{minipage}

\begin{minipage}[b]{0.5\textwidth}
\includegraphics[width=\textwidth]{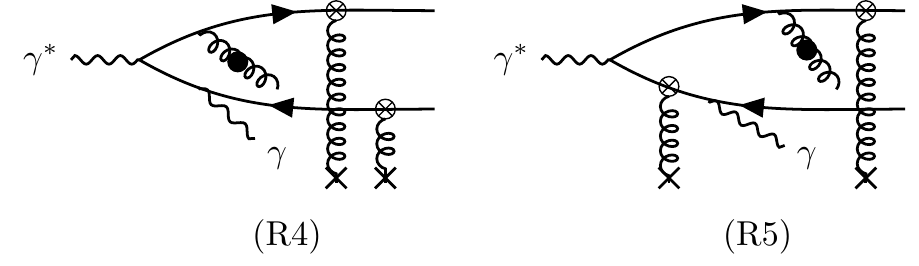}
\end{minipage}
 \caption{Real emission diagrams contributing at NLO to the ``impact factor" with the gluon emitted prior to the scattering of the quark and antiquark off the nucleus. The other five diagrams are labeled $(R6)-(R10)$ and can be obtained respectively by quark-antiquark interchange of $(R1)-(R5)$. These 10 contributions constitute in total the coefficient $T^{(1)}_{R}$ in Eq.~\ref{eq:real-emission-master-amplitude}. \label{fig:NLO-real-allscatter}}
 \end{figure}

 Before delving into the details of the computation, we will write down the general form of the contribution from these 10 processes to the total amplitude. After subtracting the ``no scattering'' contribution from each of them, this is given by 
\begin{align}
\mathcal{M}_{\mu \alpha;b}^{\rm NLO:Real(1)}&=\sum_{\beta=1}^{10} \mathcal{M}^{(R\beta)}_{\mu \alpha;b}\nonumber \\
& =2\pi (eq_{f})^{2}g \, \delta(1-z_{\text{tot}}^{r})  \int \mathrm{d}  \Pi_{\perp}^{\rm LO} \,\, \overline{u}(\bm{k})  \Bigg\{  \int_{\bm{z}_{\perp}} \!\! e^{-i \bm{k}_{g\perp}.\bm{z}_{\perp}} \,  T^{(1)}_{R;\mu \alpha}  (\bm{l}_{1\perp})   \Big[ \Big( \tilde{U}(\bm{x}_{\perp}) t^{a} \tilde{U}^{\dagger}(\bm{y}_{\perp}) \Big) U_{ab}(\bm{z}_{\perp})-t_{b} \Big] \Bigg\}  v(\bm{p}) \, .
\label{eq:amplitude-R1-R10}
\end{align}
Here $z_{\text{tot}}^{r}=z_{q}+z_{\bar{q}}+z_{\gamma}+z_{g}$ is the total momentum fraction for real emission and $\int \mathrm{d} \Pi_{\perp}^{\rm LO}$ is a shorthand 
for the integrals 
\begin{equation}
\int \mathrm{d} \Pi_{\perp}^{\rm LO}= \int_{\bm{l}_{1\perp}} \int_{\bm{x}_{\perp},\bm{y}_{\perp}} \!\!\!\!\!\!\!\!\!\!\! e^{i \bm{l}_{1\perp}.(\bm{x}_{\perp}-\bm{y}_{\perp})-i(\bm{k}_{\perp}+\bm{k}_{\gamma \perp}).\bm{x}_{\perp}-i\bm{p}_{\perp}.\bm{y}_{\perp} } \, .
\label{eq:phase-transverse}
\end{equation}

We will now discuss the computation of the $R^{(R\beta)}$'s that constitute $T^{(1)}_{R}$ given by
\begin{equation}
T_{R,\mu \alpha}^{(1)} (\bm{l}_{1\perp}) =\sum_{\beta=1}^{10} R^{(R\beta)}_{\mu \alpha}(\bm{l}_{1\perp} ) \, .
\label{eq:T-R1}
\end{equation}
Note that in this discussion, and all subsequent discussions, we will only explicitly show the dependence (if any) of these functions on internal momenta (that are integrated over) albeit they are of course functions of the external momenta as well. 

The contribution to the amplitude for the processes labeled $(R1)$ in Fig.~\ref{fig:NLO-real-R1} is given by
\begin{figure}[!htbp]
\begin{minipage}[b]{0.5\textwidth}
\includegraphics[width=\textwidth]{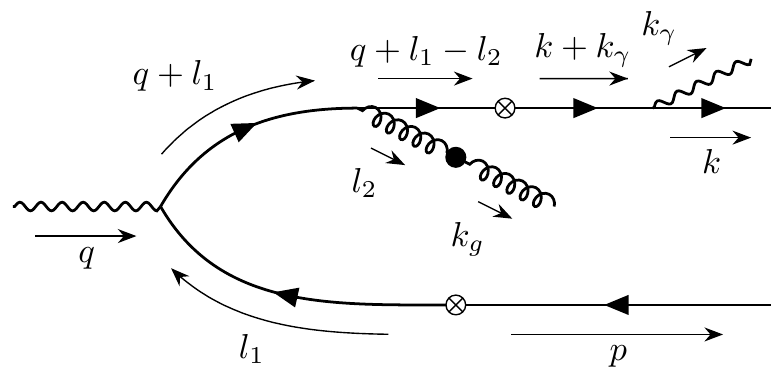}
\end{minipage}
\caption{The NLO process labeled $(R1)$ in Fig.~\ref{fig:NLO-real-allscatter} with all momentum assignments and directions shown. The effective vertices are clearly shown in Fig.~\ref{fig:effective-vertices}.  \label{fig:NLO-real-R1}}
\end{figure}
\begin{align}
\mathcal{M}^{(R1)}_{\mu \alpha;b}& =-i(eq_{f})^{2}g \int_{l_{1},l_{2}} \overline{u}(\bm{k}) \gamma_{\alpha} S_{0}(k+k_{\gamma}) \mathcal{T}_{q}(k+k_{\gamma},q+l_{1}-l_{2})S_{0}(q+l_{1}-l_{2})  \, t^{a} \, \gamma^{\beta}  \nonumber \\
& \times S_{0}(q+l_{1})  \gamma_{\mu}  S_{0}(l_{1})  \mathcal{T}_{q}(l_{1},-p) v(\bm{p}) \times \mathcal{T}^{\sigma \rho;bc}_{g} (k_{g},l_{2}) \, G^{0}_{\rho \beta;ca}(l_{2}) \, \epsilon^{*}_{\sigma}(\bm{k}_{g}) \enskip,
\label{eq:amplitude-A1}
\end{align}
where the free fermion and gluon propagators in $A^{-}=0$ gauge are respectively given by
\begin{align}
S_{0}(p)&= \frac{i\, \slashed{p}}{p^{2}+i\varepsilon} \, , \nonumber \\
G^{0}_{\mu \nu;ab}(p)&=\frac{i}{p^{2}+i\varepsilon}\Big( -g_{\mu \nu}+\frac{p_{\mu}n_{\nu}+n_{\mu}p_{\nu}}{n.p} \Big)\delta_{ab} ,\quad n^{\mu}=\delta^{\mu +} \enskip.
\label{eq:fermion-gluon-propagator}
\end{align}
The effective vertices for the quark and gluon  are contained in the expressions for their dressed propagators in the background of the strong classical color field of the nucleus. Recall that the  expression for the quark propagator is (given previously in Eq.~\ref{eq:dressed-quark-mom-prop})
\begin{equation}
S_{ij}(p,q) =  S_0(p)\,\mathcal{T}_{q;ij}(p,q)\,S_0 (q) \, .
\end{equation}

In the ``wrong'' light cone gauge $A^{-}=0$, we conveniently obtain an analogous form for the dressed gluon propagator which can be written as~\cite{McLerran:1994vd,Ayala:1995hx,Ayala:1995kg,Balitsky:1995ub,McLerran:1998nk,Balitsky:2001mr}
\begin{equation}
G_{\mu \nu;ab}(p,q) =  G^{0}_{\mu \rho;ac}(p)\,\mathcal{T}_{g}^{\rho \sigma;cd}(p,q)\,G^{0}_{\sigma \nu;db} (q) \, ,
\label{eq:dressed-gluon-mom-prop}
\end{equation}
where $\mu , \nu $ and $a,b$ are the Lorentz and adjoint color indices for the outgoing and incoming gluon which respectively carry momenta $p$ and $q$. Again, to recapitulate, the expressions for the effective vertices (introduced in Fig.~\ref{fig:effective-vertices}) are,
\begin{align}
\mathcal{T}_{q}(p,p')&=  (2 \pi)\delta(p^{-}-p'^{-}) \gamma^{-} \text{sign}(p^{-}) \int \mathrm{d}^{2} \bm{z}_{\perp} \, e^{-i(\bm{p}_{\perp} - \bm{p'}_{\perp}).\bm{z}_{\perp}}\,\, \tilde{U}^{\text{sign}(p^{-})} (\bm{z}_{\perp}) \, , \nonumber \\
\mathcal{T}^{\mu \nu;ab}_{g}(p,p') &=-2\pi \delta(p^{-}-p'^{-}) \times (2p^{-}) g^{\mu \nu} \, \text{sign}(p^{-}) \int \mathrm{d}^{2} \bm{z}_{\perp} \, e^{-i(\bm{p}_{\perp}-\bm{p'}_{\perp}).\bm{z}_{\perp} }\,\, \Big( U^{ab} \Big)^{\text{sign}(p^{-})} (\bm{z}_{\perp}) \, .
\label{eq:vertex-factors-quark-gluon}
\end{align}

Since $n.\epsilon^{*}(\bm{k}_{g})=0$  in $A^-=0$ gauge, the polarization vector for the outgoing gluon has the form  $\epsilon(\bm{k}_{g})=\Big( \frac{\bm{k}_{g\perp}.\bm{\epsilon}_{\perp}}{k_{g}^{-}}, 0, \bm{\epsilon}_{\perp} \Big)$. Using this, we can derive the following useful relation
\begin{equation}
\gamma^{\beta} \Big( -g_{ \rho \beta}+\frac{l_{2 \rho}n_{\beta}+n_{\rho}l_{2\beta}}{n.l_{2}} \Big) g^{ \sigma \rho} \epsilon^{*}_{\sigma}(\bm{k}_{g})= \Big(\bm{\gamma}_{\perp}-\frac{\bm{l}_{2\perp}}{k^{-}_{g}}\gamma^{-} \Big) . \bm{\epsilon}_{\perp}^{*}(\bm{k}_{g}) \enskip ,
\label{eq:gamma-relation}
\end{equation}
where we have used the eikonal approximation $l_{2}^{-}=k_{g}^{-}$ contained in the expression for the effective gluon vertex. Now integrating over $l_{1}^{-}$ and $l_{2}^{-}$ using the $\delta$-functions appearing in the effective vertex factors, we can rewrite Eq.~\ref{eq:amplitude-A1} as
\begin{align}
\mathcal{M}^{(R1)}_{\mu \alpha;b} =2\pi (eq_{f})^{2}g \, \delta(q^{-}-P^{-}_{\text{tot}} ) \int \mathrm{d} \Pi_{\perp}^{\rm LO}  \int_{\bm{z}_{\perp}} e^{-i \bm{k}_{g\perp}.\bm{z}_{\perp}} \, \Big( \tilde{U}(\bm{x}_{\perp}) t^{a} \tilde{U}^{\dagger}(\bm{y}_{\perp}) \Big) U_{ba}(\bm{z}_{\perp})   \, \int_{\bm{l}_{2\perp}} e^{i\bm{l}_{2\perp}.\bm{r}_{zx}} \, \int_{l_{1}^{+},l_{2}^{+}}  \frac{N}{D} \, ,
\
\end{align}
where $\bm{r}_{zx}=\bm{z}_{\perp}-\bm{x}_{\perp}$ and the numerator and denominator are respectively given by
\begin{align}
N&=\overline{u}(\bm{k}) \gamma_{\alpha}(\slashed{k}+\slashed{k}_{\gamma}) \gamma^{-} [\gamma^{+}(k^{-}+k^{-}_{\gamma})-\bm{\gamma}_{\perp}.(\bm{l}_{1\perp}-\bm{l}_{2\perp})] \,  \Big[ \Big(\bm{\gamma}_{\perp}-\frac{\bm{l}_{2\perp}}{k^{-}_{g}}\gamma^{-} \Big) . \bm{\epsilon}_{\perp}^{*}(\bm{k}_{g}) \Big]  \nonumber \\
& \times[\gamma^{+}(k^{-}+k^{-}_{\gamma}+k^{-}_{g})+\gamma^{-}(q^{+}+ l^{+}_{1})-\bm{\gamma}_{\perp}.\bm{l}_{1\perp}]   \gamma_{\mu} [\gamma^{+}p^{-}+\bm{\gamma}_{\perp}.\bm{l}_{1\perp}]\gamma^{-} v(\bm{p}) \, ,
\end{align}
and
\begin{align}
D&=8p^{-}(k^{-}+k^{-}_{\gamma})(k^{-}+k^{-}_{\gamma}+k^{-}_{g}) (2k.k_{\gamma})\Big( l^{+}_{2}-\frac{\bm{l}_{2\perp}^{2}}{2k^{-}_{g}}
+\frac{i\varepsilon}{2k^{-}_{g}} \Big) \Big(l^{+}_{1}+\frac{\bm{l}^{2}_{1\perp}}{2p^{-}} -\frac{i\varepsilon}{2p^{-}} \Big) \nonumber \\
& \times  \Bigg(q^{+}+ l^{+}_{1}-\frac{\bm{l}_{1\perp}^{2}}{2(k^{-}+k^{-}_{\gamma}+k_{g}^{-})}+\frac{i\varepsilon}{2(k^{-}+k^{-}_{\gamma}+k_{g}^{-})} \Bigg)  \nonumber \\
& \times \Bigg(q^{+}+ l^{+}_{1}-l_{2}^{+}-\frac{(\bm{l}_{1\perp}-\bm{l}_{2\perp})^{2}}{2(k^{-}+k^{-}_{\gamma})}+\frac{i\varepsilon}{2(k^{-}+k^{-}_{\gamma})} \Bigg) \, .
\end{align}
As expected, we have an overall longitudinal momentum conserving $\delta$-function where $P^{-}_{\text{tot}}=k^{-}+p^{-}+k^{-}_{\gamma}+k^{-}_{g}$, which is a reflection of the eikonal approximation ingrained in our analyses. 

When we examine the above equations, it is clear that the numerator structure allows for the use of Cauchy's residue theorem to evaluate the $+$  -integrals by complex contour integration. There are two $l_{2}^{+}$ poles on either side of the real axis. We deform the contour clockwise so as to enclose the pole below and subsequently perform the $l_{1}^{+}$ integration by an anticlockwise contour deformation. We next perform the $\bm{l}_{2\perp}$ integration using the expressions for the relevant integrals tabulated in Appendix~\ref{sec:constituent-integrals-real-emission} (see Eqs.~\ref{eq:I-r-10} and \ref{eq:I-r-11} for the expressions in $d$ dimensions). Finally, subtracting the ``no scattering'' contribution by setting the Wilson lines to unity, we write the resulting amplitude as
\begin{equation}
\mathcal{M}^{(R1)}_{\mu \alpha;b}= 2\pi (eq_{f})^{2}g \, \delta(1-z_{\text{tot}}^{r}) \int \mathrm{d}  \Pi_{\perp}^{\rm LO} \,\, \overline{u}(\bm{k})  \Bigg\{  \int_{\bm{z}_{\perp}} \!\!\!\! e^{-i \bm{k}_{g\perp}.\bm{z}_{\perp}} \,   R^{(R1)}_{\mu \alpha}(\bm{l}_{1\perp})  \Big[ \Big( \tilde{U}(\bm{x}_{\perp}) t^{a} \tilde{U}^{\dagger}(\bm{y}_{\perp}) \Big) U_{ab}(\bm{z}_{\perp})-t_{b} \Big] \Bigg\}  v(\bm{p}) \, ,
\end{equation}
where
\begin{align}
R^{(R1)}_{\mu \alpha} (\bm{l}_{1\perp}) = &  \gamma_{\alpha}\frac{\slashed{k}+\slashed{k}_{\gamma}}{2k.k_{\gamma}} \gamma^{-} \Big[ \Big(  \{ \gamma^{+}(1-z_{\bar{q}}-z_{g})q^{-}-\bm{\gamma}_{\perp}.\bm{l}_{1\perp}\} \, \mathcal{I}^{(1,0)}_{r}(\bm{v}_{\perp}^{(R1)},\Delta^{(R1)};\bm{r}_{zx}) + \gamma^{j} \mathcal{I}^{(1,j)}_{r} (\bm{v}_{\perp}^{(R1)},\Delta^{(R1)}; \bm{r}_{zx}) \Big)    \nonumber \\
& \times \bm{\gamma}_{\perp}.\bm{\epsilon}_{\perp}^{*} (\bm{k}_{g})- 2 \frac{(1-z_{\bar{q}}-z_{g})}{z_{g}} \, \mathcal{I}^{(1,i)}_{r} (\bm{v}_{\perp}^{(R1)},\Delta^{(R1)};\bm{r}_{zx}) \, {\epsilon^{i}}^{*}(\bm{k}_{g}) \Big] \,   \nonumber \\
&  \times \frac{\gamma^{+}(1-z_{\bar{q}})q^{-}-\gamma^{-}\Big( Q^{2}z_{\bar{q}}+\bm{l}_{1\perp}^{2} \Big)/2z_{\bar{q}}q^{-} -\bm{\gamma}_{\perp}.\bm{l}_{1\perp}  }{\bm{l}_{1\perp}^{2}+Q^{2}z_{\bar{q}}(1- z_{\bar{q}}) -i\varepsilon} \gamma_{\mu}   \frac{\gamma^{+}z_{\bar{q}}q^{-}+\bm{\gamma}_{\perp}.\bm{l}_{1\perp}}{2(q^{-})^{2} (1-z_{\bar{q}})/z_{g}} \gamma^{-}  \, ,
\label{eq:R-R1}
\end{align}
is independent of $\bm{k}_{g\perp}$ but depends on the other external momenta which are not shown explicitly in its argument. In the above equation, the functions $\mathcal{I}$ are proportional to modified Bessel functions of the second kind (or Macdonald functions). In $d=2$ dimensions, and for the process $(R1)$, these can be written as
\begin{align}
\mathcal{I}^{(1,0)}_{r}(\bm{v}_{\perp}^{(R1)},\Delta^{(R1)};\bm{r}_{zx})& = \frac{1}{2\pi} \, e^{-i \bm{v}_{\perp}^{(R1)}.\bm{r}_{zx}} \, K_{0} \Big( \sqrt{\bm{r}_{zx}^{2} \, \Delta^{(R1)}} \Big) \, , \nonumber \\
\mathcal{I}^{(1,i)}_{r}(\bm{v}_{\perp}^{(R1)},\Delta^{(R1)};\bm{r}_{zx})&=\frac{1}{2\pi} \, e^{-i \bm{v}_{\perp}^{(R1)}.\bm{r}_{zx}} \, \Bigg\{ \frac{ir^{i}_{zx}}{2} \Big(\frac{\bm{r}_{zx}^{2}}{4\Delta^{(R1)}} \Big)^{-1/2} \, K_{1}\Big( \sqrt{\bm{r}_{zx}^{2} \, \Delta^{(R1)}} \Big) - (v^{(R1)})^{i}\, K_{0} \Big( \sqrt{\bm{r}_{zx}^{2} \,  \Delta^{(R1)}} \Big) \Bigg\}  \, ,
\label{eq:I-factors-R1}
\end{align}
with the arguments of the functions given by 
\begin{align}
\bm{v}_{\perp}^{(R1)}& =- \frac{z_{g}}{1-z_{\bar{q}}} \, \bm{l}_{1\perp}  \, , \qquad \bm{r}_{zx}=\bm{z}_{\perp}-\bm{x}_{\perp} \, , \nonumber\\ 
\Delta^{(R1)}& = \frac{z_{g}}{(1-z_{\bar{q}})^{2}} \, \frac{1-z_{\bar{q}}-z_{g}}{z_{\bar{q}}} \, \bm{l}_{1\perp}^{2} +\frac{Q^{2}z_{g} (1-z_{\bar{q}}-z_{g})}{1-z_{\bar{q}}}-i\varepsilon \, .
\label{eq:v-perp-R1-Delta-R1}
\end{align}

At the level of the differential cross-section, we will integrate over the phase space of the real gluon which includes an integration over $k_{g}^{-}$ from $0$ to $+\infty$. If we examine closely Eqs.~\ref{eq:I-factors-R1} and~\ref{eq:v-perp-R1-Delta-R1} above, we observe a logarithmic singularity in the limit $k_{g}^{-} (z_{g}) \rightarrow 0$. The other limit ($k_{g}^{-} \rightarrow +\infty$) converges because of the oscillatory nature of the exponentials in the $\mathcal{I}$-functions. We will show later that this slow gluon limit ($z_{g} \rightarrow 0$) is what generates the large logs in $x$ -- the net contribution of terms with these large logs multiplies the JIMWLK kernel. This aspect of the computation will be discussed at length in Sec.~\ref{sec:JIMWLK-evolution}.

The remaining four diagrams in Fig.~\ref{fig:NLO-real-allscatter} can also be computed in a similar fashion and the combined contribution is given by
\begin{equation}
\sum_{\beta=1}^{5} \mathcal{M}^{(R\beta)}_{\mu \alpha;b}=\sum_{\beta=1}^{5} 2\pi (eq_{f})^{2}g \, \delta(1-z_{\text{tot}}^{r}) \int \mathrm{d}  \Pi_{\perp}^{\rm LO} \,\, \overline{u}(\bm{k})  \Bigg\{  \int_{\bm{z}_{\perp}} \!\!\!\! e^{-i \bm{k}_{g\perp}.\bm{z}_{\perp}} \,   R^{(R\beta)}_{\mu \alpha}(\bm{l}_{1\perp})  \Big[ \Big( \tilde{U}(\bm{x}_{\perp}) t^{a} \tilde{U}^{\dagger}(\bm{y}_{\perp}) \Big) U_{ab}(\bm{z}_{\perp})-t_{b} \Big] \Bigg\}  v(\bm{p}) \, ,
\label{eq:amplitude-A1-A5}
\end{equation}
where the R-functions are given in Appendix~\ref{sec:R-factors-real-emission}.

In order to find the corresponding contributions of Fig.~\ref{fig:NLO-real-allscatter} (with the quark and antiquark lines interchanged) which we call $(R6)-(R10)$, we need to impose the following replacements in the R-functions in Eqs.~\ref{eq:R-R1} and \ref{eq:R-R2}-\ref{eq:R-R5} 
\begin{equation}
k\leftrightarrow p \, (z_{q} \leftrightarrow z_{\bar{q}}), \enskip, \bar{u}(\bm{k}) \leftrightarrow v(\bm{p}) \enskip, \bm{x}_{\perp} \leftrightarrow \bm{y}_{\perp} \enskip  \text{and} \enskip \bm{l}_{1\perp} \rightarrow -\bm{l}_{1\perp}+\bm{k}_{\gamma \perp} \, .
\label{eq:replacements-qqbar-exchange}
\end{equation}
As discussed at the end of Sec.~\ref{sec:LO-review}, the last redefinition is only to ensure that the transverse phases defined by Eq.~\ref{eq:phase-transverse} remain the same so that the net contribution to the amplitude from the 10 diagrams can be compactly written as in Eq.~\ref{eq:amplitude-R1-R10}.

\item \textbf{Contributions to $T^{(2)}_{R}$}: 

 \begin{figure}[!htbp]
\centering
\begin{minipage}[b]{0.75\textwidth}
\includegraphics[width=\textwidth]{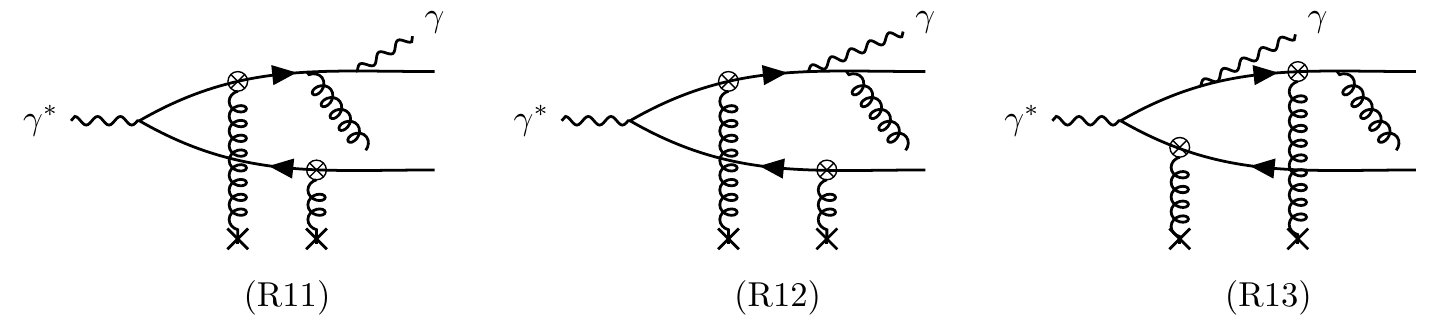}
\end{minipage}

\begin{minipage}[b]{0.5\textwidth}
\includegraphics[width=\textwidth]{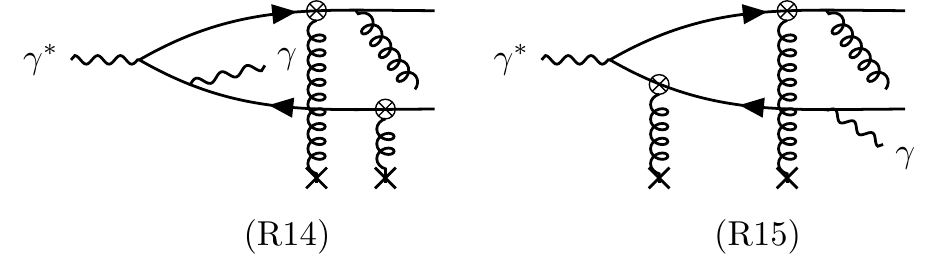}
\end{minipage}
\caption{Real emission graphs with the gluon emitted by the quark subsequent to its scattering off the nucleus. The graphs obtained by interchanging the quark and antiquark lines in the above diagrams are respectively labeled $(R16)-(R20)$ and they constitute $T^{(3)}_{R}$ in Eq.~\ref{eq:real-emission-master-amplitude}.\label{fig:NLO-real-nogluonscatter}}
\end{figure}

We will now compute the contributions from diagrams shown in Fig.~\ref{fig:NLO-real-nogluonscatter} in which the gluon is emitted by the quark after it scatters off the background classical field. The combined amplitude from these 5 processes can be written as
\begin{align}
\mathcal{M}_{\mu \alpha;b}^{\rm NLO:Real(2)}&=\sum_{\beta=11}^{15} \mathcal{M}^{(R\beta)}_{\mu \alpha;b} \nonumber \\
& =2\pi (eq_{f})^{2}g \, \delta(1-z_{\text{tot}}^{r}) \, \int \mathrm{d}  \Pi_{\perp}^{\rm LO} \,\, \overline{u}(\bm{k}) \Bigg\{ e^{-i\bm{k}_{g\perp}.\bm{x}_{\perp}} \, T^{(2)}_{R;\mu \alpha} (\bm{k}_{g\perp},\bm{l}_{1\perp})\Big[ \Big( t_{b} \tilde{U}(\bm{x}_{\perp}) \tilde{U}^{\dagger}(\bm{y}_{\perp})  \Big) -t_{b} \Big] \Bigg\} v(\bm{p}) \, ,
\label{eq:amplitude-R11-R15}
\end{align}
where
\begin{equation}
T_{R;\mu \alpha}^{(2)} (\bm{k}_{g\perp},\bm{l}_{1\perp})=\sum_{\beta=11}^{15} R^{(R\beta)}_{\mu \alpha}(\bm{k}_{g\perp},\bm{l}_{1\perp}) \, ,
\label{eq:T-R2}
\end{equation}
has an explicit $\bm{k}_{g\perp}$ dependence. 

In the following, we will show how to compute one such contribution. The rest can be computed using the same techniques. The contribution to the amplitude from the diagram labeled $(R11)$ , with the detailed momentum assignments and directions shown in Fig.~\ref{fig:NLO-real-R11}, is given by
\begin{figure}[!htbp]
\begin{minipage}[b]{0.5\textwidth}
\includegraphics[width=\textwidth]{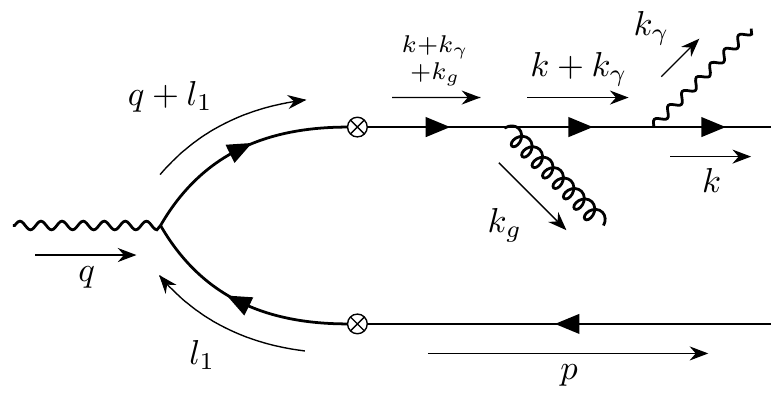}
\end{minipage}
\caption{NLO process labeled $(R11)$ in Fig.~\ref{fig:NLO-real-nogluonscatter} with the momentum assignments and directions shown. The gluon does not suffer scattering off the nucleus in the above scenario.\label{fig:NLO-real-R11}}
\end{figure}
\begin{align}
\mathcal{M}^{(R11)}_{\mu \alpha;b}& =2\pi (eq_{f})^{2}g \, \delta(q^{-}-P^{-}_{\text{tot}}) \, \int \mathrm{d} \Pi_{\perp}^{\rm LO} \,  e^{-i\bm{k}_{g\perp}.\bm{x}_{\perp}}  \Big( t_{b} \tilde{U}(\bm{x}_{\perp}) \tilde{U}^{\dagger}(\bm{y}_{\perp}) \Big) \, \int_{l^{+}_{1}} (i) \, \overline{u}(\bm{k}) \gamma_{\alpha} \frac{\slashed{k}+\slashed{k}_{\gamma}}{2k.k_{\gamma}} \frac{(\gamma.\epsilon^{*}(\bm{k}_{g}))}{4p^{-}(k^{-}+k_{\gamma}^{-}+k_{g}^{-})} \nonumber \\
& \times \frac{\slashed{k}+\slashed{k}_{\gamma}+\slashed{k}_{g}}{2k.k_{\gamma}+2k_{g}.(k+k_{\gamma})}  \gamma^{-} \frac{\gamma^{+}(k^{-}+k^{-}_{\gamma}+k_{g}^{-})-\bm{\gamma}_{\perp}.\bm{l}_{1\perp}}{q^{+}+l^{+}_{1}-\frac{\bm{l}_{1\perp}^{2}}{2(k^{-}+k^{-}_{\gamma}+k_{g}^{-})}+\frac{i\varepsilon}{2(k^{-}+k^{-}_{\gamma}+k^{-}_{g})} }   \gamma_{\mu} \frac{\gamma^{+}p^{-}+\bm{\gamma}_{\perp}.\bm{l}_{1\perp}}{l^{+}_{1}+\frac{\bm{l}_{1\perp}^{2}}{2p^{-}}-\frac{i\varepsilon}{2p^{-}}} \gamma^{-}  v(\bm{p}) \enskip,
\end{align}
where $l^{-}_{1}$ has been integrated out using the $\delta$-functions contained in the expressions for the effective vertices in Eqs.~\ref{eq:vertex-factors-quark-gluon}. We perform the integration over $l^{+}_{1}$ by a clockwise deformation of the contour. Finally subtracting the ``no scattering" contribution, we get the amplitude as
\begin{align}
\mathcal{M}^{(R11)}_{\mu \alpha;b}=2\pi (eq_{f})^{2}g \, \delta(1-z_{\text{tot}}^{r}) \, \int \mathrm{d}  \Pi_{\perp}^{\rm LO} \,\, \overline{u}(\bm{k}) \Bigg\{ e^{-i\bm{k}_{g\perp}.\bm{x}_{\perp}} \, R^{(R11)}_{\mu \alpha} (\bm{k}_{g\perp},\bm{l}_{1\perp}) \Big[ \Big( t_{b} \tilde{U}(\bm{x}_{\perp}) \tilde{U}^{\dagger}(\bm{y}_{\perp})  \Big) -t_{b} \Big] \Bigg\} v(\bm{p}) \, ,
\label{eq:amplitude-R11}
\end{align} 
where 
\begin{align}
R^{(R11)}_{\mu \alpha} (\bm{k}_{g\perp},\bm{l}_{1\perp})& = -  \gamma_{\alpha} \frac{\slashed{k}+\slashed{k}_{\gamma}}{2k.k_{\gamma}} \Bigg[ \bm{\gamma}_{\perp}.\bm{\epsilon}_{\perp}^{*}(\bm{k}_{g})\, \Big( \{\gamma^{+}(1-z_{\bar{q}})q^{-}-\bm{\gamma}_{\perp}.(\bm{k}_{\perp}+\bm{k}_{\gamma \perp}) \} -\bm{\gamma}_{\perp}.\bm{k}_{g\perp} \Big) - \frac{2(1-z_{\bar{q}})}{z_{g}} \, \bm{k}_{g\perp}.\bm{\epsilon}^{*}_{\perp}(\bm{k}_{g}) \Bigg]  \nonumber \\
& \times \gamma^{-} \frac{\gamma^{+}(1-z_{\bar{q}})q^{-}-\bm{\gamma}_{\perp}.\bm{l}_{1\perp}}{2(q^{-})^{2}(1-z_{\bar{q}}-z_{g})/z_{g} \Big[(\bm{k}_{g\perp}+\bm{v}^{(R11)}_{\perp})^{2}+\Delta^{(R11)}\Big]    } \, \gamma_{\mu} \, \frac{\gamma^{+}z_{\bar{q}}q^{-}+\bm{\gamma}_{\perp}.\bm{l}_{1\perp}}{\bm{l}^{2}_{1\perp}+Q^{2}z_{\bar{q}}(1-z_{\bar{q}})-i\varepsilon} \, \gamma^{-} \, ,
\label{eq:R-R11}
\end{align}
and 
\begin{align}
\bm{v}_{\perp}^{(R11)}&= -\frac{z_{g}}{z_{q}+z_{\gamma}} \, (\bm{k}_{\perp}+\bm{k}_{\gamma \perp} ) \, , \qquad  \Delta^{(R11)}=\frac{z_{g}}{z_{q}+z_{\gamma}} \,  (1-z_{\bar{q}}) \, (2k.k_{\gamma}) \, .
\label{eq:v-R11-Delta-R11}
\end{align}
At the level of the inclusive cross-section when we integrate over $\bm{k}_{g\perp}$, it is evident from Eq.~\ref{eq:R-R11} that we will once again encounter tensor integrals of the kind given by Eq.~\ref{eq:generic-tensor-integral-real}. The other contributions arising from the emission of the gluon after the scattering off the shock wave can be similarly computed and their combined contribution is represented by Eq.~\ref{eq:amplitude-R11-R15}. Expressions for the remaining R-functions are provided in Appendix~\ref{sec:R-factors-real-emission}.

\item \textbf{Contributions to $T^{(3)}_{R}$}: These are the processes obtained by interchanging the quark and antiquark lines in $(R11)-(R15)$. They are respectively labeled $(R16)-(R20)$ and their contributions to the amplitude are obtained by imposing the replacements given by Eq.~\ref{eq:replacements-qqbar-exchange} in Eq.~\ref{eq:T-R2}. This ensures that the transverse phases remain the same throughout. The total contribution from this final sub-category of diagrams can then be written as
\begin{align}
\mathcal{M}_{\mu \alpha;b}^{\rm NLO:Real(3)}&=\sum_{\beta=16}^{20} \mathcal{M}^{(R\beta)}_{\mu \alpha;b} \nonumber \\
& =2\pi (eq_{f})^{2}g \, \delta(1-z_{\text{tot}}^{r}) \, \int \mathrm{d}  \Pi_{\perp}^{\rm LO} \,\, \overline{u}(\bm{k}) \Bigg\{ e^{-i\bm{k}_{g\perp}.\bm{y}_{\perp}} \, T^{(3)}_{R;\mu \alpha} (\bm{k}_{g\perp},\bm{l}_{1\perp})\Big[ \Big( \tilde{U}(\bm{x}_{\perp}) \tilde{U}^{\dagger}(\bm{y}_{\perp}) t_{b}   \Big) -t_{b} \Big] \Bigg\} v(\bm{p}) \, ,
\label{eq:amplitude-R16-R20}
\end{align}
with 
\begin{equation}
T_{R;\mu \alpha}^{(3)}(\bm{k}_{g\perp},\bm{l}_{1\perp})=\sum_{\beta=16}^{20} R^{(R\beta)}_{\mu \alpha}(\bm{k}_{g\perp},\bm{l}_{1\perp}) \, .
\label{eq:T-R3}
\end{equation}

\end{enumerate}
Finally, we can write the combined contribution from all the allowed 20 real emission diagrams as
\begin{align}
\mathcal{M}_{\mu \alpha;b}^{\text{NLO;Real}}&=  2\pi (eq_{f})^{2}g \, \delta(1-z_{\text{tot}}^{r}) \, \int \mathrm{d}  \Pi_{\perp}^{\rm LO}  \,\, \overline{u}(\bm{k})  \Bigg\{  \int_{\bm{z}_{\perp}} \!\!\!\! e^{-i \bm{k}_{g\perp}.\bm{z}_{\perp}} \,  T^{(1)}_{R;\mu \alpha}  (\bm{l}_{1\perp})   \Big[ \Big( \tilde{U}(\bm{x}_{\perp}) t^{a} \tilde{U}^{\dagger}(\bm{y}_{\perp}) \Big) U_{ab}(\bm{z}_{\perp})-t_{b} \Big] \nonumber \\
&+ e^{-i\bm{k}_{g\perp}.\bm{x}_{\perp}} \, T^{(2)}_{R;\mu \alpha} (\bm{k}_{g\perp},\bm{l}_{1\perp})\Big[ \Big( t_{b} \tilde{U}(\bm{x}_{\perp}) \tilde{U}^{\dagger}(\bm{y}_{\perp})  \Big) -t_{b} \Big] \nonumber \\
& + e^{-i\bm{k}_{g\perp}.\bm{y}_{\perp}} \, T^{(3)}_{R;\mu \alpha} (\bm{k}_{g\perp},\bm{l}_{1\perp})  \Big[ \Big( \tilde{U}(\bm{x}_{\perp}) \tilde{U}^{\dagger}(\bm{y}_{\perp}) t_{b} \Big) -t_{b} \Big]  \Bigg\}  v(\bm{p}) \, .
\label{eq:master-NLOamp-real-emission}
\end{align}
To compute the contributions of these real graphs to the differential cross-section, we need to take the modulus squared of the amplitude in Eq.~\ref{eq:master-NLOamp-real-emission} and then integrate over the phase space of the real gluon. From the discussion here, and the expressions given in Appendix~\ref{sec:R-factors-real-emission}, it is clear that for the 10 diagrams (see Fig.~\ref{fig:NLO-real-allscatter}) in which the gluon gets scattered off the nucleus, the amplitudes can be written in terms of MacDonald functions whose arguments in general depend on the gluon momenta. As such, it is difficult to isolate analytically the rapidity divergent pieces and the finite contributions from the squared amplitudes for these graphs. In Sec.~\ref{sec:JIMWLK-evolution}, we will obtain the rapidity divergent pieces from these amplitudes by explicitly taking the $k_{g}^{-} \rightarrow 0$ limit and show that these pieces contribute towards small $x$ evolution. To compute the finite pieces from this class of diagrams, one however needs to perform the integration over the gluon phase space numerically by imposing a cutoff for the gluon momentum fraction $z_{g}$. Because of the interaction with the nuclear shock wave, there are no collinear divergences associated with these diagrams. 
%The only divergences would be in rapidity which have logarithmic and not power law behavior and therefore should be numerically easy to regulate.

For the processes shown in Fig.~\ref{fig:NLO-real-nogluonscatter} (and their $q\leftrightarrow \bar{q}$ counterparts) in which the gluon does not scatter off the nucleus, there are divergences from the region of phase space where the gluon is soft ($k_{g} \rightarrow 0$) and/or collinear ($\bm{k}_{g\perp} \propto \bm{k}_{\perp}, 
\bm{p}_{\perp}$) respectively to the antiquark and quark. In Sec.~\ref{sec:jet-cross-section}, we will promote  partons to jets and explicitly extract these divergent structures by using a jet cone algorithm. This will allow us to show the cancellation between residual collinear divergences from the virtual graphs with those in the real gluon amplitude squared and therefore obtain an IR safe cross-section.

\section{NLO contributions to the amplitude from virtual graphs: Detailed calculations} \label{sec:virtual-emission-details}

In this section, we will illustrate the details of the computation of the amplitudes corresponding to the virtual diagrams shown in Sec.~\ref{sec:contributions-NLO}. We will start with the self-energy diagrams and follow this with the computation of the vertex correction graphs. An additional feature of these processes relative to the usual Feynman diagram computations is that the emitted gluon can scatter off the background shock wave classical field before being absorbed by the quark or antiquark. 
%So when we write self-energy or vertex corrections, we mean in the sense of topology of the Feynman graphs.

\subsection{Self-energy graphs with dressed gluon propagator} \label{sec:virtual-self-energy-dressed}
As discussed previously, there are three distinct topologies of the Feynman graphs describing self-energy contributions. These are discussed individually below. 
%The bulk of the contributions comes from processes in which the gluon in the loop does not suffer scattering off the shock wave. However, the more interesting cases with novel Wilson line structures are the ones in which there is indeed a possibility that the gluon crosses the shock wave before being absorbed. We initiate our discussion with these two broad categories of diagrams each related to the other by quark-antiquark interchange. The results for the multitude of processes in which the gluon does not cross the shock wave will be presented in the next subsection.
\begin{enumerate}
\item \textbf{Contributions to $T^{(1)}_{S}$}: The diagrams contributing to $T^{(1)}_{S}$ in the general expression for the amplitude given by Eq.~\ref{eq:selfenergy-correction-master-amplitude} are presented in Fig.~\ref{fig:NLO-self-1}. These are the processes which allow for a virtual gluon emitted from the quark line to scatter off the shock wave before being reabsorbed. We will first present the combined result for the amplitude from all such processes and then demonstrate the details of the computation using a representative diagram. 
 
 \begin{figure}[!htbp]
\centering
\begin{minipage}[b]{0.75\textwidth}
\includegraphics[width=\textwidth]{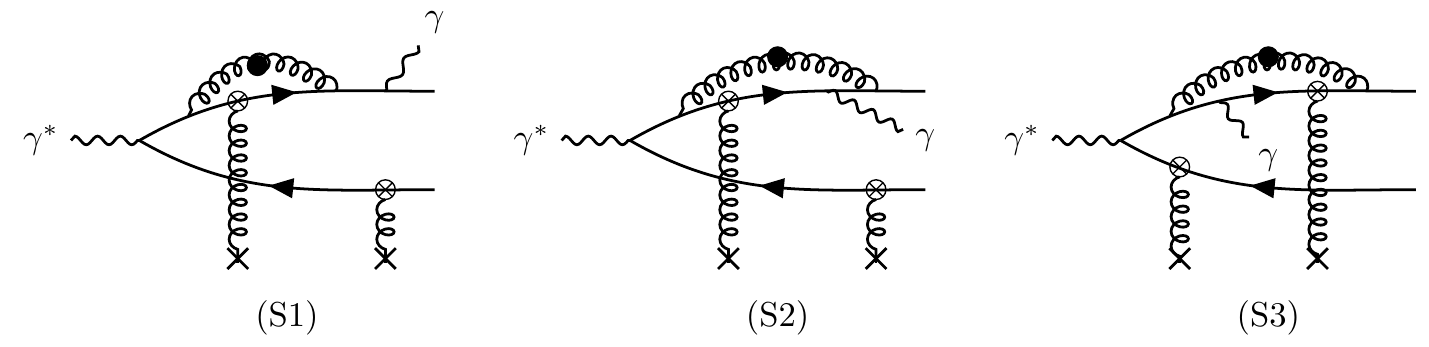}
\end{minipage}

\begin{minipage}[b]{0.75\textwidth}
\includegraphics[width=\textwidth]{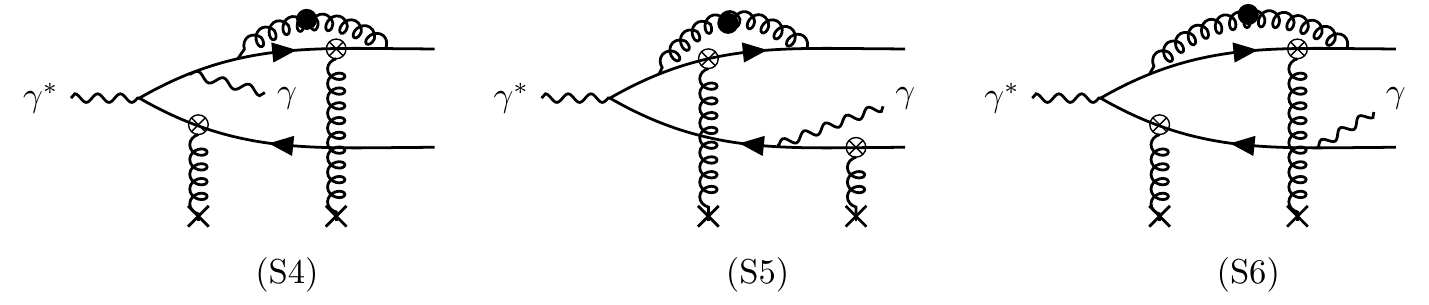}
\end{minipage}
\caption{Self-energy corrections with the gluon emitted and absorbed by the quark after it crosses the shock wave. These contributions constitute $T^{(1)}_{S}$. By interchanging the quark and antiquark lines in the above diagrams, we get the contributions composing $T^{(2)}_{S}$ which are labeled (S7)-(S12). \label{fig:NLO-self-1}}
\end{figure}
The combined amplitude from these 6 processes has the structure
\begin{align}
\mathcal{M}^{\text{NLO};\text{SE}(1)}_{\mu \alpha}& = 2\pi \, \delta(1-z^{v}_{\text{tot}}) \, (eq_{f} g)^{2} \,  \int \mathrm{d} \Pi_{\perp}^{v} \, \overline{u} (\bm{k}) \Bigg\{ T^{(1)}_{S;\mu \alpha} (\bm{l}_{1\perp})  \, \Big[ \Big( t^{b} \tilde{U} (\bm{x}_{\perp}) t^{a} \tilde{U}^{\dagger} (\bm{y}_{\perp}) \Big) U_{ba}(\bm{z}_{\perp}) - C_{F} \mathds{1} \Big] \Bigg\} v(\bm{p}) \, ,
\label{eq:amplitude-SE1-generic}
\end{align}
where $z^{v}_{\text{tot}}=z_{q}+z_{\bar{q}}+z_{\gamma}$ is the total momentum fraction for virtual emission and we have bundled together the transverse Fourier phases using the short-hand convention
\begin{equation}
\int \mathrm{d} \Pi_{\perp}^{v}= \int_{\bm{l}_{1\perp}} \int_{\bm{x}_{\perp},\bm{y}_{\perp},\bm{z}_{\perp}} \!\!\!\!\!\!\!\!\!\!\! e^{i \bm{l}_{1\perp}.(\bm{x}_{\perp}-\bm{y}_{\perp})-i(\bm{k}_{\perp}+\bm{k}_{\gamma \perp}).\bm{x}_{\perp}-i\bm{p}_{\perp}.\bm{y}_{\perp} } =\int_{\bm{z}_{\perp}}  \mathrm{d} \Pi_{\perp}^{\rm LO
}  \, ,
\label{eq:phase-transverse-virtual}
\end{equation}
which contains an additional integration over $\bm{z}_{\perp}$ relative to the similar expression given in Eq.~\ref{eq:phase-transverse}. Finally, $T^{(1)}_{S;\mu \alpha}$ can be written as the sum  of a piece that includes UV and rapidity divergences and a finite part,
\begin{align}
& T^{(1)}_{S;\mu \alpha} (\bm{l}_{1\perp})  = \sum_{\beta=1}^{6} R^{(S\beta)}_{\mu \alpha} (\bm{l}_{1\perp})  \nonumber \\
&= \frac{1}{2\pi^{2}} \int_{\bm{l}_{2\perp}} \!\!\!\! e^{i\bm{l}_{2\perp}.\bm{r}_{zx}} \, T^{\text{LO}}_{\mu \alpha} (\bm{l}_{1\perp}) \,  \Bigg\{ \ln \Big(\frac{1}{z_{0}} \Big) \, \Bigg(\frac{1}{\epsilon} +\frac{1}{2} \, \ln \Big(\frac{\tilde{\mu}^{2}}{Q^{2}} \Big)  \Bigg)   -\frac{3}{4} \Bigg( \frac{1}{\epsilon} +\frac{1}{2} \, \ln \Big(\frac{\tilde{\mu}^{2}}{Q^{2}} \Big) \Bigg) +\frac{1}{2} \ln \Big( \frac{1}{z_{0}} \Big) \ln \Big( \frac{2Q^{2}}{\bm{l}_{2\perp}^{2}} \Big)  \Bigg\}\nonumber \\
& +R^{\text{SE}(1)}_{\text{finite};\mu \alpha} (\bm{l}_{1\perp})   \, .
\label{eq:T-S1}
\end{align} 
Once again, we are including the dependence on momenta that are integrated over. Also note that as previously, $\bm{r}_{zx}=\bm{z}_{\perp}-\bm{x}_{\perp}$.  In the above equation, $T^{\text{LO}}$ (given previously in Eq.~\ref{eq:T-LO}) contains contributions to the amplitude from the four allowed LO processes. The pieces that are independent of $\bm{l}_{2\perp}$ will produce a $\delta^{(2)}( \bm{r}_{zx})$ from the $\bm{l}_{2\perp}$ integration. Once the $\bm{z}_{\perp}$ integration is done, the color structure corresponding to these pieces then reduce to that for the LO amplitudes times the quadratic Casimir $C_{F}$. We can therefore write the amplitude in Eq.~\ref{eq:amplitude-SE1-generic} as
\begin{align}
\mathcal{M}^{\text{NLO};\text{SE}(1)}_{\mu \alpha}& = \frac{2\alpha_{S}C_{F}}{\pi} \, \mathcal{M}^{\text{LO}}_{\mu \alpha} \,  \Bigg\{\ln \Big(\frac{1}{z_{0}} \Big) \, \Bigg(\frac{1}{\epsilon} +\frac{1}{2} \, \ln \Big(\frac{\tilde{\mu}^{2}}{Q^{2}} \Big) \Bigg)   -\frac{3}{4} \Bigg( \frac{1}{\epsilon} +\frac{1}{2} \, \ln \Big(\frac{\tilde{\mu}^{2}}{Q^{2}} \Big) \Bigg) \Bigg\} \nonumber \\
& +\frac{2\alpha_{S}}{\pi} \, \Big[ 2\pi   \, \delta(1-z^{v}_{\text{tot}}) \, (eq_{f})^{2} \,  \int \mathrm{d} \Pi_{\perp}^{v} \int_{\bm{l}_{2\perp}}  \!\!\!\! e^{i\bm{l}_{2\perp}.\bm{r}_{zx}} \, \frac{1}{2}\ln \Big( \frac{1}{z_{0}} \Big) \ln \Big( \frac{2Q^{2}}{\bm{l}_{2\perp}^{2}} \Big)  \overline{u} (\bm{k}) \Bigg\{ T^{\rm LO}_{\mu \alpha} (\bm{l}_{1\perp})  \nonumber \\
& \times  \Big[ \Big( t^{b} \tilde{U} (\bm{x}_{\perp}) t^{a} \tilde{U}^{\dagger} (\bm{y}_{\perp}) \Big) U_{ba}(\bm{z}_{\perp}) - C_{F} \mathds{1} \Big] \Bigg\} v(\bm{p})  + \mathcal{M}^{\text{SE}(1)}_{\text{finite};\mu \alpha} \, ,
\label{eq:amplitude-SE1}
\end{align}
where $\mathcal{M}^{\text{LO}}$ is the leading order amplitude given by Eq.~\ref{eq:LO-amp-master}. 

The divergence free pieces of the various contributions are combined to constitute $\mathcal{M}^{\text{SE}(1)}_{\text{finite}}$. The logarithmic divergence $\ln (1/z_{0})= \ln(q^-/\Lambda_0^-)$  in the above expression, where $\Lambda_{0}^{-}$ is given by Eq.~\ref{eq:initial-energy-scale}, arises from the integration over the momentum fraction $z_{l}$ of the gluon in the loop.  Its upper limit, up to logarithmic accuracy at this order, is controlled by the $q^-$ momentum component of the photon and its lower limit by the longitudinal width $1/\Lambda_0^-$ (or equivalently $1/P_N^+$) of the target nucleus. The latter is a cutoff that we are imposing to regulate the gauge pole $l^{-}=0$ in the LC gauge gluon propagator.  

Further, from the expression for the amplitude in Eq.~\ref{eq:amplitude-SE1}, we can see that there are two kinds of singular logarithms multiplying $\ln(1/z_{0})$. The one that appears in the first line of Eq.~\ref{eq:amplitude-SE1} arises from the collinear limit $\bm{z}_{\perp} \rightarrow \bm{x}_{\perp}$ and are not part of the small $x$ logarithms contributing to JIMWLK evolution. Some of these divergences will cancel out already at the amplitude level between different classes of diagrams and the rest will cancel between real and virtual graphs. 

The  limit $z_{\perp} \gg x_{\perp},y_{\perp}$ of the evolution kernels is captured by the logarithms appearing in the second line of Eq.~\ref{eq:amplitude-SE1}. At this order of accuracy, the $\ln(1/z_{0})$ log can also be expressed as the $\ln(x_0/x_{\rm Bj})$ log giving rise to small $x$ JIMWLK evolution. We will discuss this point in greater detail in Section~\ref{sec:JIMWLK-evolution}. {\it In the limit of large $Q^2$, these will give the double log limit of the DGLAP evolution equation~\cite{Gribov:1972ri,Lipatov:1974qm,Altarelli:1977zs,Dokshitzer:1977sg} or equivalently the large $Q^{2}$ limit of BFKL equation.}

The $1/\epsilon$ singularities for $\epsilon \rightarrow 0$ arise from regulating the UV divergences in the integrations over the transverse loop momentum of the gluon using dimensional regularization in $d=2-\epsilon$ dimensions. In our expressions, $\tilde{\mu}^{2}=4\pi \mu^{2}e^{-\gamma_{E}}$, where $\gamma_{E}$ is the Euler-Mascheroni constant, is the reference scale used in the $\overline{MS}$-scheme. 

We will now use one representative process to explicitly demonstrate the steps leading to the above result, in particular, the computation of the divergent pieces. The calculation of the finite pieces, while absolutely essential for precision computations, are not particularly illuminating; they are discussed in Appendix~\ref{sec:finite-pieces-virtual-graphs}.
The amplitude for the process labeled $(S1)$ with momentum assignments shown in Fig.~\ref{fig:S1} is given by 
\begin{figure}[!htbp]
\begin{minipage}[b]{0.5\textwidth}
\includegraphics[width=\textwidth]{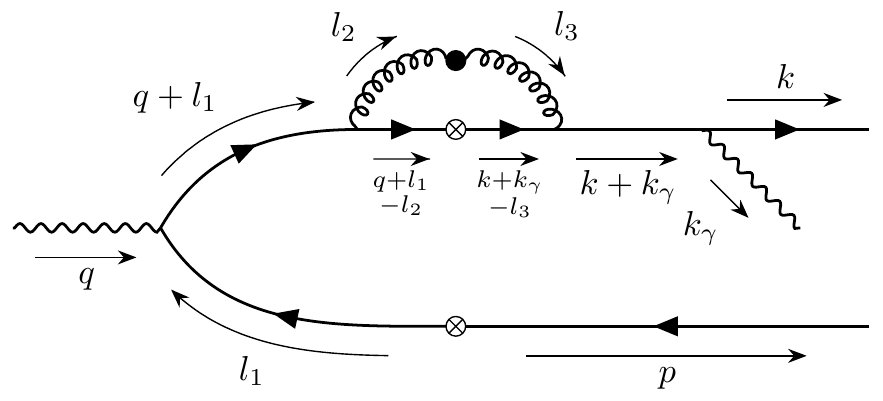}
\end{minipage}
\caption{The process labeled $(S1)$ in Fig.~\ref{fig:NLO-self-1} with momenta and their directions shown. We must subtract the ``no scattering" case (corresponding to the same amplitude without crossed or filled blobs) from the above contribution to obtain the physical amplitude. \label{fig:S1}}
\end{figure}
\begin{align}
\mathcal{M}^{(S1)}_{\mu \alpha} & = (eg q_{f})^{2}  \int_{l_{1},l_{2},l_{3}} \bar{u}(\bm{k}) \gamma_{\alpha} S_{0}(k+k_{\gamma}) t^{b} \gamma^{\beta}  S_{0}(k+k_{\gamma}-l_{3}) \mathcal{T}_{q}(k+k_{\gamma}-l_{3},q+l_{1}-l_{2}) S_{0}(q+l_{1}-l_{2})  \nonumber \\
& \times t^{a} \gamma^{\nu} S_{0}(q+l_{1}) \gamma_{\mu} S_{0}(l_{1}) \mathcal{T}_{q} (l_{1},-p) v(\bm{p}) \times G^{0}_{\beta \sigma;bd} (l_{3})\mathcal{T}^{\sigma \rho;dc}_{g}(l_{3},l_{2}) G^{0}_{\rho \nu;ca} (l_{2}) \, ,
\label{eq:amplitude-S1-general}
\end{align}
where the free fermion and gluon propagators are given respectively in Eq.~\ref{eq:fermion-gluon-propagator} and the corresponding effective vertices are in Eqs.~\ref{eq:vertex-factors-quark-gluon}.

 Although the structure of the free gluon propagator in $A^-=0$  gauge is in general more complicated compared to that in covariant gauges, we will use the light cone condition to derive an identity that allows one to simplify the amplitude above. To show this, we start with the expression
\begin{equation}
\gamma^{\beta} (\ldots)\gamma^{\nu} \times \Big( -g_{\beta \sigma}+\frac{l_{3\beta }n_{\sigma}+n_{\beta}l_{3\sigma}}{l_{3}^{-}} 
\Big) \, g^{\sigma \rho} \, \Big( -g_{\rho \nu}+\frac{l_{2\rho }n_{\nu}+n_{\rho}l_{2\nu}}{l_{2}^{-}} 
\Big) \, ,
\label{eq:vertex-contraction-LCgauge}
\end{equation}
where $(\ldots)$ denotes the terms between the two gamma matrices. The terms in parentheses on the right denote the Lorentz structure of the free gluon propagators while the metric $g^{\sigma \rho}$ in between these is from the effective gluon vertex. Using $n^{\mu}=(1,0,\bm{0}_{\perp})$ and some algebra with the indices, it is possible to reduce the above expression to 
\begin{equation}
\Big( \gamma^{k} -\frac{l^{k}_{3}}{l^{-}_{3}} \gamma^{-} \Big) (\ldots) \Big( \gamma_{k} -\frac{l_{2k}}{l^{-}_{2}} \gamma^{-} \Big) \, ,
\label{eq:gamma-identity}
\end{equation}
where $k={1,2}$ is summed over. Using the above identity, analogous to the one in Eq.~\ref{eq:gamma-relation}, and integrating out $l_{1}^{-}, l^{-}_{2}$ using the $\delta$-functions in the effective vertices, Eq.~\ref{eq:amplitude-S1-general} can be written as
\begin{align}
\mathcal{M}^{(S1)}_{\mu \alpha} & = 2\pi \delta(q^{-}-P^{-}) (egq_{f})^{2}  \int \mathrm{d}\Pi_{\perp}^{v} \, \int_{\bm{l}_{2\perp},\bm{l}_{3\perp}} \!\!\!\!e^{i(\bm{l}_{2\perp}-\bm{l}_{3\perp}).\bm{r}_{zx}}  \int_{l_{3}^{-}}  \text{sign}(l_{3}^{-}) \, \text{sign}(k^{-}+k^{-}_{\gamma}-l^{-}_{3}) \nonumber \\
& \times  \Big( t^{b} \tilde{U}^{\text{sign}(k^{-}+k^{-}_{\gamma}-l^{-}_{3})} (\bm{x}_{\perp}) t^{a} \tilde{U}^{\dagger} (\bm{y}_{\perp}) \Big) U^{\text{sign}(l^{-}_{3})}_{ba}(\bm{z}_{\perp})  \int_{l_{1}^{+},l_{2}^{+},l_{3}^{+}} \, i \, \frac{N^{(S1)}}{D^{(S1)}} \, ,
\end{align}
with 
\begin{align}
N^{(S1)}& = \bar{u}(\bm{k}) \gamma_{\alpha} \frac{\slashed{k}+\slashed{k}_{\gamma}}{2k.k_{\gamma}} \Big( \gamma^{k}-\frac{l_{3}^{k}}{l^{-}_{3}} \gamma^{-} \Big) [\gamma^{+} (k^{-}+k^{-}_{\gamma}-l_{3}^{-}) -\bm{\gamma}_{\perp}.(\bm{k}_{\perp}+\bm{k}_{\gamma \perp}-\bm{l}_{3\perp} )  ] \gamma^{-}  \nonumber \\
& \times [\gamma^{+} (k^{-}+k^{-}_{\gamma}-l_{3}^{-}) -\bm{\gamma}_{\perp}.(\bm{l}_{1\perp}-\bm{l}_{2\perp})]  \Big( \gamma_{k}-\frac{l_{2k}}{l^{-}_{3}} \gamma^{-} \Big)[\gamma^{+}(k^{-}+k^{-}_{\gamma}) +\gamma^{-}(q^{+}+l^{+}_{1}) -\bm{\gamma}_{\perp}.\bm{l}_{1\perp} ] \nonumber \\
& \times \gamma_{\mu} [\gamma^{+}p^{-} +\bm{\gamma}_{\perp}.\bm{l}_{1\perp}] \gamma^{-} v(\bm{p}) \,  ,
\label{eq:numerator-S1}
\end{align}
and 
\begin{align}
D^{(S1)}&= 32p^{-}  (k^{-}+k^{-}_{\gamma}) \, l^{-}_{3} (k^{-}+k^{-}_{\gamma}-l_{3}^{-})^{2} \Bigg( l_{1}^{+}+\frac{\bm{l}_{1\perp}^{2}}{2p^{-}} -\frac{i\varepsilon}{2p^{-}} \Bigg) \Bigg( l^{+}_{1}+q^{+}-\frac{\bm{l}_{1\perp}^{2}}{2(k^{-}+k^{-}_{\gamma})} +\frac{i\varepsilon}{2(k^{-}+k^{-}_{\gamma})}    \Bigg)         \nonumber \\
& \times \Bigg( l_{1}^{+}-l_{2}^{+}+q^{+}-\frac{(\bm{l}_{1\perp}-\bm{l}_{2\perp})^{2}}{2(k^{-}+k^{-}_{\gamma}-l^{-}_{3})} +\frac{i\varepsilon}{2(k^{-}+k^{-}_{\gamma}-l^{-}_{3})} \Bigg) \Bigg(l^{+}_{2}-\frac{\bm{l}^{2}_{2\perp}}{2l^{-}_{3}}+\frac{i\varepsilon}{2l^{-}_{3}}\Bigg) \nonumber \\
&\times  \Bigg( l^{+}_{3}-\frac{\bm{l}_{3\perp}^{2}}{2l_{3}^{-}}+\frac{i\varepsilon}{2l^{-}_{3}} \Bigg) \Bigg( l_{3}^{+}-k^{+}-k^{+}_{\gamma} +\frac{(\bm{k}_{\perp}+\bm{k}_{\gamma \perp} -\bm{l}_{3\perp})^{2}}{2(k^{-}+k^{-}_{\gamma}-l^{-}_{3})} -\frac{i\varepsilon}{2(k^{-}+k^{-}_{\gamma}-l^{-}_{3})}\Bigg) \, .
\label{eq:denominator-S1}
\end{align}
Note that in these expressions we have canceled a factor of $2l_3^-$ from the gluon effective vertex in the numerator with a corresponding factor from one of the propagators in the denominator. 

From the above expression for the denominator, it is evident that there are two $l_{3}^{+}$ poles which are located at 
\begin{align}
l^{+}_{3} \vert_{A}  = k^{+}+k^{+}_{\gamma} -\frac{(\bm{k}_{\perp}+\bm{k}_{\gamma \perp} -\bm{l}_{3\perp})^{2}}{2(k^{-}+k^{-}_{\gamma}-l^{-}_{3})} + \frac{i\varepsilon}{2(k^{-}+k^{-}_{\gamma}-l^{-}_{3})} \,\,\,;\,\,\,
l^{+}_{3} \vert_{B} = \frac{\bm{l}_{3\perp}^{2}}{2l_{3}^{-}}-\frac{i\varepsilon}{2l^{-}_{3}} \, .
\end{align}
 For $l^{-}_{3} < 0$, we have $k^{-}+k^{-}_{\gamma}-l^{-}_{3} > 0$ which implies that the poles are on the same side (above) of the real axis. Since the numerator is independent of $l_{3}^{+}$, the contour can always be deformed such that none of the poles are enclosed, thereby giving a null result. Hence we must have $l_{3}^{-} > 0$ as well as $k^{-}+k^{-}_{\gamma}-l^{-}_{3} > 0 $ for a non-zero contribution. As discussed earlier, logarithmically divergent integrals in $l^{-}_{3}$ are regulated by introducing a lower cutoff $\Lambda_{0}^{-}$ given by Eq.~\ref{eq:initial-energy-scale}. 
 
We  observe that there is an equivalence of this picture to analyses in light cone perturbation theory, where the positivity of $l^{-}_{3}$ here correspond to forward propagation (in light cone time) of the exchanged gluon. From inspection, an identical argument holds for the  $l_{2}^{+}$ pole.

We will enclose $l_{3}^{+} \vert_{B}$ and the following poles for the contour integration over $l_{1}^{+}$ and $l_{2}^{+}$ respectively,
\begin{align}
l_{1}^{+} \vert_{pole} =- \frac{\bm{l}_{1\perp}^{2}}{2p^{-}} +\frac{i\varepsilon}{2p^{-}} \,\,\,;\,\,\, l_{2}^{+} \vert_{pole}  = \frac{\bm{l}^{2}_{2\perp}}{2l^{-}_{3}}- \frac{i\varepsilon}{2l^{-}_{3}} \, .
\end{align}
Finally subtracting the ``no scattering'' contribution we arrive at the following expression for the amplitude,
\begin{align}
\mathcal{M}^{(S1)}_{\mu \alpha}&= 2\pi \delta(1-z_{\text{tot}}^{v}) \, (eq_{f} g)^{2} \,  \int \mathrm{d} \Pi_{\perp}^{v} \, \Bigg[ \Big( t^{b} \tilde{U} (\bm{x}_{\perp}) t^{a} \tilde{U}^{\dagger} (\bm{y}_{\perp}) \Big) U_{ba}(\bm{z}_{\perp}) - C_{F} \mathds{1} \Bigg] \nonumber \\
& \times \int \frac{\mathrm{d}z_{l}}{2\pi} \int_{\bm{l}_{2\perp},\bm{l}_{3\perp}} \!\!\!\!\!\!\!\!\!\!\!\! e^{i\bm{l}_{2\perp}.\bm{r}_{zx}+i\bm{l}_{3\perp}.\bm{r}_{xz}} \,\,  \overline{u} (\bm{k}) \gamma_{\alpha}\frac{\slashed{k}+\slashed{k}_{\gamma}}{2k.k_{\gamma}} \Bigg[ \gamma^{i} \Big( \{\gamma^{+}(1-z_{\bar{q}}-z_{l})q^{-}-\bm{\gamma}_{\perp}.(\bm{k}_{\perp}+\bm{k}_{\gamma \perp}) \} +\gamma^{j}l^{j}_{3} \Big) \nonumber \\
& -\frac{2(1-z_{\bar{q}}-z_{l})}{z_{l}} \, l^{i}_{3} \Bigg] \gamma^{-} \Bigg[ \Big( \{ \gamma^{+}(1-z_{\bar{q}}-z_{l})q^{-} -\bm{\gamma}_{\perp}.\bm{l}_{1\perp} \} +\gamma^{k}l^{k}_{2} \Big) \, \gamma_{i} +\frac{2(1-z_{\bar{q}}-z_{l})}{z_{l}} \, l^{i}_{2} \Bigg]   \nonumber \\
& \times \frac{\gamma^{+}(1-z_{\bar{q}})q^{-}-\gamma^{-}\Big(Q^{2}z_{\bar{q}}+\bm{l}_{1\perp}^{2}\Big)/2z_{\bar{q}}q^{-}-\bm{\gamma}_{\perp}.\bm{l}_{1\perp}}{\Big[(\bm{l}_{2\perp}+\bm{v}_{1\perp}^{(S1)})^{2}+\Delta_{1}^{(S1)} \Big] \, \Big[(\bm{l}_{3\perp}+\bm{v}_{2\perp}^{(S1)})^{2}+\Delta_{2}^{(S1)}\Big]} \gamma_{\mu} \frac{\gamma^{+}z_{\bar{q}}q^{-}+\bm{\gamma}_{\perp}.\bm{l}_{1\perp}}{4(q^{-})^{2} \, (1-z_{\bar{q}})^{2}/z_{l} [\bm{l}_{1\perp}^{2}+Q^{2}z_{\bar{q}}(1-z_{\bar{q}})-i\varepsilon]} \gamma^{-} v(\bm{p}) \, ,
\label{eq:M-S1-form-1}
\end{align}
where $\int \mathrm{d} \Pi_{\perp}^{v}$ is defined by Eq.~\ref{eq:phase-transverse-virtual}, $z_{l}=l^{-}_{3}/q^{-}$ is the gluon momentum fraction in the loop and  
\begin{align}
\bm{v}_{1\perp}^{(S1)}&= -\frac{z_{l}}{1-z_{\bar{q}}} \, \bm{l}_{1\perp} \, , \, \Delta_{1}^{(S1)}= \frac{z_{l}}{1-z_{\bar{q}}} \Big( 1-\frac{z_{l}}{1-z_{\bar{q}}} \Big) \{ \bm{l}_{1\perp}^{2}+z_{\bar{q}}(1-z_{\bar{q}}) Q^{2} \}/z_{\bar{q}}  -i\varepsilon \, ,\nonumber \\
\bm{v}_{2\perp}^{(S1)}&= -\frac{z_{l}}{1-z_{\bar{q}}}  \, (\bm{k}_{\perp}+\bm{k}_{\gamma \perp} ) \, , \, \Delta_{2}^{(S1)}= - \frac{z_{l}}{1-z_{\bar{q}}} \Big( 1-\frac{z_{l}}{1-z_{\bar{q}}} \Big) (2k.k_{\gamma}) -i\varepsilon \, .
\label{eq:S1-integration-parameters}
\end{align}

As previously for the real contributions, we have here too the familiar integrals over transverse momenta $\bm{l}_{2\perp}$ and $\bm{l}_{3\perp}$, that we encountered in the computation of the real emission amplitude, which contain the exponential terms $\text{exp}(\pm i \bm{v}_{i \perp}^{(S1)}.\bm{r}_{zx})$ ($i=1,2$) and are proportional to Macdonald functions. We also learnt from our  previous discussion that these can give singular logarithms from the $z_{l}$ integration in the limit $z_{l} \rightarrow 0$ when these phases become unity. We will see later that this is indeed the case. Another interesting possibility is the limit $\bm{r}_{zx} \rightarrow 0$ which corresponds to the UV limit for the momenta, $\bm{l}_{i \perp}$ ($i=2,3$) conjugate to this transverse coordinate vector. The manifestation of this UV divergence is explicit if we perform the following momentum redefinition, $\bm{l}_{2\perp} -\bm{l}_{3\perp} \rightarrow \bm{l}_{2\perp}$ in Eq.~\ref{eq:M-S1-form-1} with $\bm{l}_{3\perp}$ remaining unchanged.

 If we now do a naive power counting in $\bm{l}_{3\perp}$, it is clear that the numerator ($N^{(S1)}$) and denominator ($D^{(S1)}$) are both proportional to $\bm{l}_{3\perp}^{4}$ for large $\bm{l}_{3\perp}$. By a careful use of Dirac algebra, it can be shown that the terms proportional to $l_{3i}^{4}$ and $l_{3i}^{3}$ vanish and hence the transverse momentum integral is at most logarithmically divergent. We will use dimensional regularization\footnote{This implies the following replacement: $\int \mathrm{d}^{2} \bm{l}_{3\perp} / (2\pi)^{2} \rightarrow \mu^{\epsilon} \int \mathrm{d}^{2-\epsilon} \bm{l}_{3\perp} /(2\pi)^{2-\epsilon}$, where $\mu $ has mass dimension of one.} in $d=2-\epsilon$ dimensions for terms in the integrand proportional to $\bm{l}_{3\perp}^{2}$ and use the $d=2$ results for the convergent pieces. 
 
 However we also have to account for rapidity logarithms from the $z_{l}$ integration which as one can see arises from terms proportional to $1/z_{l}$. We will therefore separate our amplitude into two contributions. The first contribution is constituted of terms in the integrand proportional to $\bm{l}_{3\perp}^{2}$ and also includes terms which will be finite from the $\bm{l}_{3\perp}$ integration but contains $1/z_{l}$ pieces . The second contribution is comprised of terms proportional to $ l^{p}_{3\perp} \, (p<2)$ and does not contain any piece proportional to $1/z_{l}$. This will then be a genuinely finite part of the amplitude. 
 
 With this in mind, we can rewrite the amplitude in Eq.~\ref{eq:M-S1-form-1} as
\begin{align}
\mathcal{M}_{\mu \alpha}^{(S1)} =2\pi \, \delta(1-z_{\text{tot}}^{v}) \, (egq_{f})^{2} \int \mathrm{d} \Pi_{\perp}^{v}  \, \overline{u}(\bm{k}) \, R^{(S1)}_{\mu \alpha} (\bm{l}_{1\perp}) \,  \Bigg[ \Big( t^{b} \tilde{U} (\bm{x}_{\perp}) t^{a} \tilde{U}^{\dagger} (\bm{y}_{\perp}) \Big) U_{ba}(\bm{z}_{\perp}) - C_{F} \mathds{1} \Bigg] \, v(\bm{p}) \, ,
\label{eq:amplitude-S1-conv-div}
\end{align}
where $R^{(S1)}_{\mu \alpha} = R^{(S1)}_{(\textrm{I}); \mu \alpha} + R^{(S1)}_{(\textrm{II}); \mu \alpha}$. Here  $R^{(S1)}_{(\textrm{I})}$ contains the UV and rapidity divergent pieces and some finite terms. We will isolate these remaining finite pieces and combine them with the genuinely finite contribution $R^{(S1)}_{(\textrm{II})}$ to obtain the net finite contribution from $(S1)$.

Carefully isolating these pieces from the numerator in Eq.~\ref{eq:M-S1-form-1} and after an extensive use of Dirac algebra in $d$-dimensions and the identity $\gamma^{i}\gamma^{j} l^{3i}l^{3j}= -\bm{l}_{3\perp}^{2}$ we can simplify the first contribution to read 
 \begin{align}
 R^{(S1)}_{(\textrm{I}); \mu \alpha} (\bm{l}_{1\perp}) & =  \int_{\bm{l}_{2\perp}} \, \!\!\!\! e^{i\bm{l}_{2\perp}.\bm{r}_{zx}} \,\, \int \frac{\mathrm{d}z_l}{\pi} \Bigg\{ \Bigg(  \frac{1}{z_{l}}  -\frac{1}{1-z_{\bar{q}}} + \frac{(2-\epsilon)\, z_{l}}{4 \, (1-z_{\bar{q}})^{2}}  \Bigg) \, \mu^{\epsilon} \, \mathcal{I}^{(2,ii)}_{v} (\bm{V}_{\perp}^{(S1)}, \Delta^{(S1)}) \nonumber \\
 & +\frac{1}{z_{l}} \, l_{2}^{i} \, \mathcal{I}_{v}^{(2,i)} (\bm{V}_{\perp}^{(S1)}, \Delta^{(S1)}) \Bigg\} \, R^{\rm LO:(1)}_{\mu \alpha}(\bm{l}_{1\perp})    \, ,
 \label{eq:R-S1-divergent}
 \end{align}
where the constituent integrals appearing above are given by 
\begin{align}
\mu^{\epsilon} \mathcal{I}^{(2,ii)}_{v} (\bm{V}_{\perp}^{(S1)}, \Delta^{(S1)})& =  \mu^{\epsilon} \int \frac{\mathrm{d}^{2-\epsilon} \bm{l}_{3\perp}}{(2\pi)^{2-\epsilon}} \, \frac{\bm{l}_{3\perp}^{2}}{\Big[ (\bm{l}_{3\perp}+\bm{\tilde{v}}_{1 \perp}^{(S1)})^{2}+\Delta_{1}^{(S1)} \Big] \, \Big[ (\bm{l}_{3\perp}+\bm{v}_{2 \perp}^{(S1)})^{2}+\Delta_{2}^{(S1)} \Big]} \nonumber \\
& = \int_{0}^{1} \frac{\mathrm{d} \alpha}{4\pi} \Bigg(\frac{4\pi \mu^{2}}{\Delta^{(S1)}} \Bigg)^{\epsilon/2} \, \Gamma\Big(\frac{\epsilon}{2}\Big) \, \Bigg\{ 1+\frac{\epsilon}{2} \Big(\frac{(\bm{V}_{\perp}^{(S1)})^{2}}{\Delta^{(S1)}} -1 \Big) \Bigg\}  \, ,
\end{align}
and
\begin{equation}
\mathcal{I}^{(2,i)}(\bm{V}_{\perp}^{(S1)},\Delta^{(S1)})= -\frac{1}{4\pi} \int_{0}^{1} \mathrm{d} \alpha \, \frac{(V^{(S1)})^{i} }{\Delta^{(S1)}} \, .
\end{equation}
We have also introduced here the short-hand notations
\begin{align}
\bm{V}_{\perp}^{(S1)} = \alpha \, \bm{\tilde{v}}_{1 \perp}^{(S1)} + (1-\alpha) \,  \bm{v}_{2 \perp}^{(S1)} \,\,;\,\,
\bm{\tilde{v}}_{1\perp}^{(S1)} = \bm{l}_{2\perp}+\bm{v}_{1\perp}^{(S1)} \,\,;\,\,
\Delta^{(S1)} = \alpha(1-\alpha) \, (\bm{\tilde{v}}_{1\perp}^{(S1)} -\bm{v}_{2\perp}^{(S1)})^{2} +\alpha \, \Delta_{1}^{(S1)} +(1-\alpha) \, \Delta_{2}^{(S1)} \,,
\end{align}
where the relevant terms $\bm{v}_{1,2\perp}^{(S1)}$ and $\Delta_{1,2}^{(S1)}$ were previously defined in Eqs.~\ref{eq:S1-integration-parameters}. Using the standard identities
\begin{align}
\Gamma(\epsilon/2 )  = \frac{2}{\epsilon} - \gamma_{E}+\text{O}(\epsilon) +\ldots \,\,\,;\,\,\,
A^{\epsilon/2}= 1+\frac{\epsilon}{2}\ln A+\ldots  \, , 
\end{align}
for $\epsilon \rightarrow 0$ in Eq.~\ref{eq:R-S1-divergent} and defining $\tilde{\mu}^{2}=4\pi \mu^{2} e^{-\gamma_{E}}$, we can finally write
\begin{align}
R^{(S1)}_{(\textrm{I}); \mu \alpha} (\bm{l}_{1\perp}) & = \frac{1}{2\pi^{2}} \, \int_{\bm{l}_{2\perp}} \!\!\!\!e^{i\bm{l}_{2\perp}.\bm{r}_{zx}} \, R^{\text{LO}:(1)}_{\mu \alpha}(\bm{l}_{1\perp}) \Bigg\{  \ln \Big(\frac{1}{z_{0}} \Big) \, \Bigg(\frac{1}{\epsilon} +\frac{1}{2} \, \ln \Big(\frac{\tilde{\mu}^{2}}{Q^{2}} \Big) \Bigg)   -\frac{3}{4} \Bigg( \frac{1}{\epsilon} +\frac{1}{2} \, \ln \Big(\frac{\tilde{\mu}^{2}}{Q^{2}} \Big)\Bigg) \nonumber \\
&+\frac{1}{2} \, \ln \Big( \frac{1}{z_{0}} \Big) \ln \Big( \frac{2Q^{2}}{\bm{l}_{2\perp}^{2}} \Big)  \Bigg\} +\Re^{(S1)}_{\mu \alpha} (\bm{l}_{1\perp})   \, ,
\label{eq:R-S1-div-pieces}
\end{align}
where we have deliberately separated the logarithm containing  $l_{2\perp}$ by introducing the resolution scale $Q^{2}$. We will show later that divergent terms of the kind appearing in the first line of the above equation cancel and do not appear in the final cross-section. This in turn also shows the independence of our final observable on the scale $\tilde{\mu}$. 

The finite pieces from $R_{(\rm I)}^{(S1)}$ that do not contain any logarithms in $z_{0}$ nor UV logs are contained in the remainder term defined as\footnote{Even though there is an apparent log divergence in $z_0$ in the first line of the r.h.s, this is not the case because of cancelations between the four individual terms.}
\begin{align}
\Re^{(S1)}_{\mu \alpha} (\bm{l}_{1\perp}) & = \frac{1}{4\pi^{2}} \, \int_{\bm{l}_{2\perp}} \!\!\!\! e^{i\bm{l}_{2\perp}.\bm{r}_{zx}} \, R^{\text{LO}:(1)}_{\mu \alpha}(\bm{l}_{1\perp})   \int \mathrm{d}z_{l} \int \mathrm{d} \alpha \Bigg[ \frac{1}{z_{l}} \Bigg\{ \frac{(\bm{V}_{\perp}^{(S1)})^{2}}{\Delta^{(S1)}} +\ln \Big( \frac{\bm{l}_{2\perp}^{2}}{2\Delta^{(S1)}} \Big) -\frac{\bm{l}_{2\perp}.\bm{V}_{\perp}^{(S1)}}{\Delta^{(S1)}} -1 \Bigg\}  \nonumber \\
&+\Big(\frac{z_{l}}{2(1-z_{\bar{q}})^{2}} - \frac{1}{1-z_{\bar{q}}} \Big) \,  \Bigg( \frac{(\bm{V}_{\perp}^{(S1)})^{2}}{\Delta^{(S1)}} +\ln \Big( \frac{Q^{2}}{\Delta^{(S1)}} \Big)  \Bigg) + \frac{1}{2 (1-z_{\bar{q}})} \, \Bigg]  
 \, .
\label{eq:remainder-divergent-part-S1}
\end{align}
In Appendix~\ref{sec:finite-pieces-virtual-graphs}, we show the computation of the above term in detail. The $R$-function appearing in Eq.~\ref{eq:amplitude-S1-conv-div} can finally be written as
\begin{align}
R^{(S1)}_{ \mu \alpha} (\bm{l}_{1\perp}) & =\frac{1}{2\pi^{2}} \, \int_{\bm{l}_{2\perp}} \!\!\!\!e^{i\bm{l}_{2\perp}.\bm{r}_{zx}} \, R^{\text{LO}:(1)}_{\mu \alpha}(\bm{l}_{1\perp}) \,  \Bigg\{  \ln \Big(\frac{1}{z_{0}} \Big) \, \Bigg(\frac{1}{\epsilon} +\frac{1}{2} \, \ln \Big(\frac{\tilde{\mu}^{2}}{Q^{2}} \Big) \Bigg)   -\frac{3}{4} \Bigg( \frac{1}{\epsilon} +\frac{1}{2} \, \ln \Big(\frac{\tilde{\mu}^{2}}{Q^{2}} \Big)\Bigg) \nonumber \\
&+\frac{1}{2} \, \ln \Big( \frac{1}{z_{0}} \Big) \ln \Big( \frac{2Q^{2}}{\bm{l}_{2\perp}^{2}} \Big)  \Bigg\} +\Big( \hyperref[eq:remainder-divergent-part-S1]{\Re^{(S1)}_{\mu \alpha} (\bm{l}_{1\perp}) }+\hyperref[eq:convergent-piece-S1]{R^{(S1)}_{(\textrm{II});\mu \alpha} (\bm{l}_{1\perp})  } \Big) \, .
\label{eq:R-S1}
\end{align}
In case of the diagrams $(S2)$ and $(S3)$, we get two independent contributions to the amplitude because we have different choices of contours for the integration over $l_{2,3}^{+}$ depending on whether $0<l_{3}^{-} <k^{-}$ or $k^{-} < l_{3}^{-} <k^{-}+k^{-}_{\gamma}$. The net amplitude is therefore obtained by summing these individual contributions. We can show that for either range of $l^{-}_{3}$ the divergent pieces in $R^{(S\beta)}_{(\textrm{I});\mu \alpha}$ ($\beta=2,3$), are proportional to $g_{\alpha +}$ which will yield zero when contracted with the polarization vector for the outgoing photon because ${\epsilon^{-}}^{*}(\bm{k}_{\gamma})=0$ in our choice of gauge. Therefore, for these diagrams, only the finite pieces survive; we will present the expressions for these in Appendix~\ref{sec:finite-pieces-virtual-graphs}.
 \begin{align}
 R^{(S2)}_{\mu \alpha} (\bm{l}_{1\perp})  &=\hyperref[eq:R-II-S2]{ R^{(S2)}_{\rm (II); \mu \alpha} (\bm{l}_{1\perp}) } \, , \label{eq:R-S2} \\
R^{(S3)}_{\mu \alpha} (\bm{l}_{1\perp})  &= \hyperref[eq:R-II-S3]{   R^{(S3)}_{\rm (II); \mu \alpha} (\bm{l}_{1\perp})} \, .
\label{eq:R-S3}
\end{align}

Finally, for diagrams $(S4)-(S6)$, we can show that the divergent parts of the amplitudes have exactly the same structure as Eq.~\ref{eq:R-S1-div-pieces} albeit with the (previously specified) different $R^{\text{LO}}$'s depending on the topology of the diagram. 
\begin{align}
\begin{pmatrix}
R^{(S4)}_{\mu \alpha} (\bm{l}_{1\perp})  \\
R^{(S5)}_{\mu \alpha} (\bm{l}_{1\perp})  \\
R^{(S6)}_{\mu \alpha} (\bm{l}_{1\perp}) 
\end{pmatrix}
& = \frac{1}{2\pi^{2}} \, \delta^{(2)}(\bm{r}_{zx}) \, \begin{pmatrix}
R^{\text{LO}:(2)}_{\mu \alpha}(\bm{l}_{1\perp}) \\
R^{\text{LO}:(4)}_{\mu \alpha}(\bm{l}_{1\perp}) \\
R^{\text{LO}:(3)}_{\mu \alpha}(\bm{l}_{1\perp})
\end{pmatrix}
  \times   \Bigg\{  \ln \Big(\frac{1}{z_{0}} \Big) \, \Bigg(\frac{1}{\epsilon} +\frac{1}{2} \, \ln \Big(\frac{\tilde{\mu}^{2}}{Q^{2}} \Big) \Bigg)   -\frac{3}{4} \Bigg( \frac{1}{\epsilon} +\frac{1}{2} \, \ln \Big(\frac{\tilde{\mu}^{2}}{Q^{2}} \Big)\Bigg) \nonumber \\
&+\frac{1}{2} \, \ln \Big( \frac{1}{z_{0}} \Big) \ln \Big( \frac{2Q^{2}}{\bm{l}_{2\perp}^{2}} \Big)  \Bigg\} + \begin{pmatrix}
 \hyperref[eq:remainder-generic-S1-S4-S6]{\Re^{(S4)}_{\mu \alpha}(\bm{l}_{1\perp}) }+\hyperref[eq:R-II-S4]{R^{(S4)}_{(\textrm{II});\mu \alpha}(\bm{l}_{1\perp})}   \\
 \hyperref[eq:remainder-S5]{\Re^{(S5)}_{\mu \alpha} (\bm{l}_{1\perp}) }+\hyperref[eq:R-II-S5]{R^{(S5)}_{(\textrm{II});\mu \alpha}(\bm{l}_{1\perp})}   \\
  \hyperref[eq:remainder-generic-S1-S4-S6]{\Re^{(S6)}_{\mu \alpha}(\bm{l}_{1\perp}) }+\hyperref[eq:R-II-S6]{R^{(S6)}_{(\textrm{II});\mu \alpha} (\bm{l}_{1\perp}) }
\end{pmatrix} \, .
\label{eq:R-S4-R-S6}
\end{align}
We would like to remind the reader that in our chosen convention the $q\leftrightarrow\bar{q}$ interchanged counterparts of the LO processes labeled LO:(1) and LO:(2) as shown in Fig.~\ref{fig:LO-diagrams} are respectively labeled LO:(3) and LO:(4). We can now finally express the sum of the contributions to the amplitude from the six processes (in Fig.~\ref{fig:NLO-self-1}) in the form given by Eq.~\ref{eq:amplitude-SE1}.

The finite pieces of the amplitude are contained in 
\begin{align}
\mathcal{M}^{\rm SE(1)}_{\text{finite};\mu \alpha}= 2\pi \, \delta(1-z_{\text{tot}}^{v}) \, (egq_{f})^{2} \int \mathrm{d} \Pi_{\perp}^{v} \, \overline{u}(\bm{k}) \,  R^{\rm SE(1)}_{\rm finite;\mu \alpha} (\bm{l}_{1\perp}) \, \Bigg[ \Big( t^{b} \tilde{U} (\bm{x}_{\perp}) t^{a} \tilde{U}^{\dagger} (\bm{y}_{\perp}) \Big) U_{ba}(\bm{z}_{\perp}) - C_{F} \mathds{1} \Bigg] \,  v(\bm{p}) \, ,
\label{eq:amplitude-SE1-finite}
\end{align}
where
\begin{equation}
R^{\rm SE(1)}_{\rm finite;\mu \alpha} (\bm{l}_{1\perp}) = \sum_{\beta=1}^{6}  \Big( \Re^{(S\beta)}_{\mu \alpha} (\bm{l}_{1\perp})   + R^{(S\beta)}_{(\textrm{II}); \mu \alpha} (\bm{l}_{1\perp}) \, \Big)  \, .
\label{eq:R-SE1-finite}
\end{equation}
These are presented in detail in Appendix~\ref{sec:finite-pieces-virtual-graphs}. 

The fact that the divergent part of the amplitude in Eq.~\ref{eq:amplitude-SE1} is proportional to the LO amplitude shows that scattering off of the background classical field does not affect the UV structure of these processes. {\it This is to be expected because the gluon in the loop only  experiences transverse momentum kicks while propagating through the nuclear shock wave; in the limit of large loop momentum, this scattering should have no effect on the short distance structure of the theory.}

\item \textbf{Contributions to $T^{(2)}_{S}$}: These are the processes labeled $(S7)-(S12)$ which are obtained by interchanging the quark and antiquark lines in the diagrams shown in Fig.~\ref{fig:NLO-self-1}. The combined contribution to the amplitude from these processes can be written as
\begin{align}
\mathcal{M}^{\text{NLO};\text{SE}(2)}_{\mu \alpha}& = 2\pi \delta(1-z^{v}_{\text{tot}}) \, (eq_{f} g)^{2} \,  \int \mathrm{d} \Pi_{\perp}^{v} \, \overline{u} (\bm{k}) \Bigg\{ T^{(2)}_{S;\mu \alpha} (\bm{l}_{1\perp})  \, \Big[ \Big(  \tilde{U} (\bm{x}_{\perp}) t^{a} \tilde{U}^{\dagger} (\bm{y}_{\perp}) t^{b} \Big) U_{ba}(\bm{z}_{\perp}) - C_{F} \mathds{1} \Big] \Bigg\} v(\bm{p}) \, ,
\label{eq:amplitude-SE2-generic}
\end{align}
where $T^{(2)}_{S}$ is obtained by imposing the replacements in Eq.~\ref{eq:replacements-qqbar-exchange} to the corresponding expression in Eq.~\ref{eq:T-S1} of  $T^{(1)}_{S}$. The resulting expression is given by 
\begin{align}
T^{(2)}_{S;\mu \alpha} (\bm{l}_{1\perp}) & = \sum_{\beta=1}^{6} R^{(S\beta)}_{\mu \alpha} (\bm{l}_{1\perp})  \nonumber \\
&= \frac{1}{2\pi^{2}} \int_{\bm{l}_{2\perp}} \!\!\!\! e^{i\bm{l}_{2\perp}.\bm{r}_{zx}} \, T^{\text{LO}}_{\mu \alpha} (\bm{l}_{1\perp}) \,  \Bigg\{ \ln \Big(\frac{1}{z_{0}} \Big) \, \Bigg(\frac{1}{\epsilon} +\frac{1}{2} \, \ln \Big(\frac{\tilde{\mu}^{2}}{Q^{2}} \Big)  \Bigg)   -\frac{3}{4} \Bigg( \frac{1}{\epsilon} +\frac{1}{2} \, \ln \Big(\frac{\tilde{\mu}^{2}}{Q^{2}} \Big) \Bigg) +\frac{1}{2} \ln \Big( \frac{1}{z_{0}} \Big) \ln \Big( \frac{2Q^{2}}{\bm{l}_{2\perp}^{2}} \Big)  \Bigg\}\nonumber \\
& +R^{\text{SE}(2)}_{\text{finite};\mu \alpha} (\bm{l}_{1\perp})   \, .
\label{eq:T-S2}
\end{align} 
The amplitude in Eq.~\ref{eq:amplitude-SE2-generic} can therefore be rewritten as 
\begin{align}
\mathcal{M}^{\text{NLO};\text{SE}(2)}_{\mu \alpha}& = \frac{2\alpha_{S}C_{F}}{\pi} \, \mathcal{M}^{\text{LO}}_{\mu \alpha} \,  \Bigg\{\ln \Big(\frac{1}{z_{0}} \Big) \, \Bigg(\frac{1}{\epsilon} +\frac{1}{2} \, \ln \Big(\frac{\tilde{\mu}^{2}}{Q^{2}} \Big) \Bigg)   -\frac{3}{4} \Bigg( \frac{1}{\epsilon} +\frac{1}{2} \, \ln \Big(\frac{\tilde{\mu}^{2}}{Q^{2}} \Big) \Bigg) \Bigg\} \nonumber \\
& +\frac{2\alpha_{S}}{\pi} \, \Big[ 2\pi   \, \delta(1-z^{v}_{\text{tot}}) \, (eq_{f})^{2} \,  \int \mathrm{d} \Pi_{\perp}^{v} \int_{\bm{l}_{2\perp}}  \!\!\!\! e^{i\bm{l}_{2\perp}.\bm{r}_{zx}} \, \frac{1}{2}\ln \Big( \frac{1}{z_{0}} \Big) \ln \Big( \frac{2Q^{2}}{\bm{l}_{2\perp}^{2}} \Big)  \overline{u} (\bm{k}) \Bigg\{ T^{\rm LO}_{\mu \alpha} (\bm{l}_{1\perp})  \nonumber \\
& \times  \Big[ \Big( t^{b} \tilde{U} (\bm{x}_{\perp}) t^{a} \tilde{U}^{\dagger} (\bm{y}_{\perp}) \Big) U_{ba}(\bm{z}_{\perp}) - C_{F} \mathds{1} \Big] \Bigg\} v(\bm{p})  + \mathcal{M}^{\text{SE}(2)}_{\text{finite};\mu \alpha} \, ,
\label{eq:amplitude-SE2}
\end{align}
where the finite piece is
\begin{align}
\mathcal{M}^{\text{SE}(2)}_{\text{finite};\mu \alpha}= 2\pi \, \delta(1-z_{\text{tot}}^{v}) \, (egq_{f})^{2} \int \mathrm{d} \Pi_{\perp}^{v} \, \overline{u}(\bm{k}) \, R^{\rm SE(2)}_{ \rm finite;\mu \alpha} (\bm{l}_{1\perp}) \Big[ \Big(  \tilde{U} (\bm{x}_{\perp}) t^{a} \tilde{U}^{\dagger} (\bm{y}_{\perp}) t^{b} \Big) U_{ba}(\bm{z}_{\perp}) - C_{F} \mathds{1} \Big] \,   v(\bm{p}) \, ,
\label{eq:amplitude-SE2-finite}
\end{align}
with 
\begin{equation}
R^{\rm SE(2)}_{ \rm finite;\mu \alpha} (\bm{l}_{1\perp})= \sum_{\beta=7}^{12} \,  \Big( \Re^{(S\beta)}_{\mu \alpha} (\bm{l}_{1\perp})   + R^{(S\beta)}_{(\textrm{II}); \mu \alpha} (\bm{l}_{1\perp}) \, \Big) \, .
\label{eq:R-SE2-finite} 
\end{equation}
As noted earlier, the two functions appearing above are obtained from their quark$\leftrightarrow$antiquark interchanged  counterparts by imposing the replacements in Eq.~\ref{eq:replacements-qqbar-exchange}.
\end{enumerate}

\subsection{Self-energy graphs with free gluon propagator} \label{sec:virtual-self-energy}

\begin{figure}[!htbp]
\centering
\begin{minipage}[b]{0.9\textwidth}
\includegraphics[width=\textwidth]{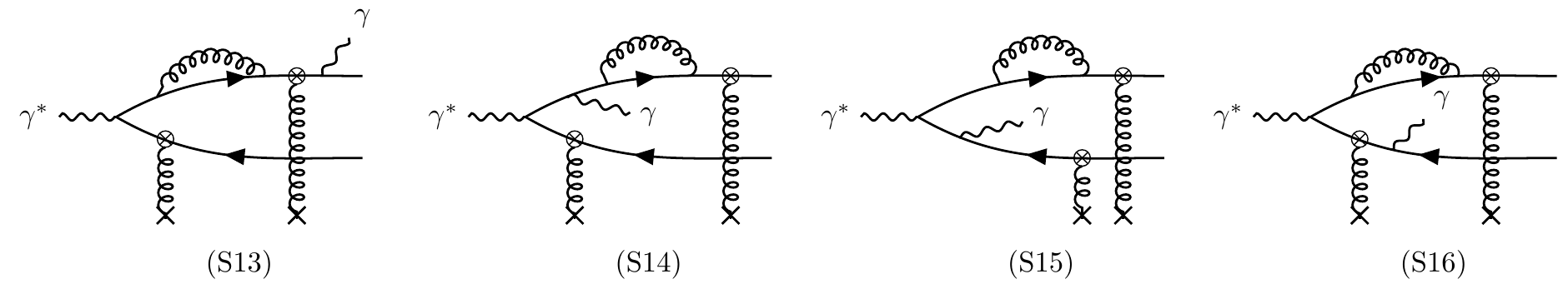}
\end{minipage}

\begin{minipage}[b]{0.9\textwidth}
\includegraphics[width=\textwidth]{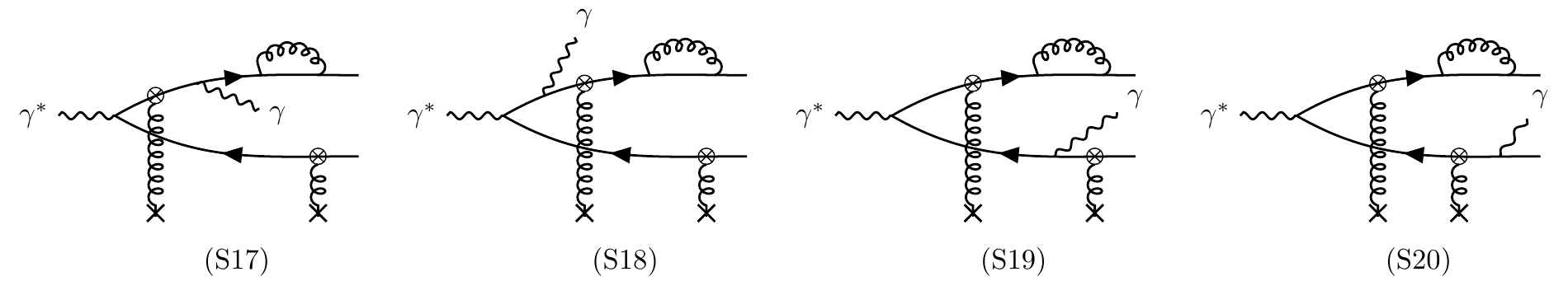}
\end{minipage}

\begin{minipage}[b]{0.9\textwidth}
\includegraphics[width=\textwidth]{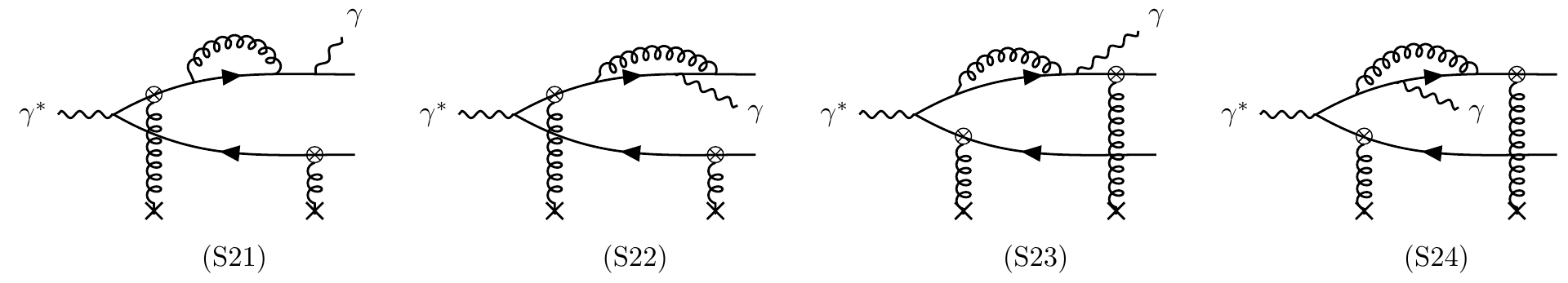}
\end{minipage}
\caption{Self-energy corrections to the quark and antiquark propagator with the additional complication brought in by the final state photon. These contributions constitute $T^{(3)}_{S}$ in Eq.\ref{eq:selfenergy-correction-master-amplitude}.  The remaining 12 diagrams labeled (S25)-(S36) are obtained by interchanging the quark and antiquark lines. \label{fig:NLO-self-3}       }
\end{figure}

\textbf{Contributions to $T^{(3)}_{S}$:} There are a total of 24 diagrams which contribute to the quark self-energy corrections at NLO. Half of them are shown in Fig.~\ref{fig:NLO-self-3} and the other half are obtained simply by interchanging the quark and antiquark lines. We have grouped them in three rows depending on their divergence structure. As we will show, the first four processes labeled $(S13)-(S16)$ inherit a UV divergence structure which is identical to the one described in the previous section.

Because we are working in the limit of massless quarks, we do not need to renormalize the quark mass. So diagrams in which the gluon loop is on an external on-shell quark or antiquark line will contribute zero to the amplitude. This is true for the four diagrams labeled $(S17)-(S20)$ appearing in the second row of Fig.~\ref{fig:NLO-self-3}. A detailed computation of the quark self-energy in Appendix~\ref{sec:quark-self-energy} in the ``wrong'' LC gauge $A^{-}=0$ demonstrates this result. 

We will also use the  expressions obtained in this Appendix for the self-energy loop to compute the diagrams which have a similar topology albeit different locations of the emission of the photon. Finally we will compute here the amplitude for the two processes $(S22)$ and $(S24)$ which represent $\alpha_{S}$ corrections to the quark-photon-quark vertex. Although these diagrams will yield a different UV divergence structure from the rest, we will see that there are cancellations amongst the divergences of the four processes in the third row (which explains why they are grouped together). A general  expression for the gluon loop correction to the quark-photon-quark vertex is derived in Appendix~\ref{sec:quark-real photon-quark-vertex-gluon-correction}.

We will begin our discussion by considering the amplitudes for $(S13)-(S16)$. Subtracting the ``no scattering'' contribution from each of these diagrams, we can write the sum of the amplitudes from these processes as
\begin{align}
\sum_{\beta=13}^{16} \mathcal{M}^{(S\beta)}_{\mu \alpha}& = 2\pi \, (eq_{f}g)^{2} \, \delta(1-z_{\rm tot}^{v}) \, \int \mathrm{d} \Pi_{\perp}^{\rm LO} \, \overline{u}(\bm{k}) \, \sum_{\beta=13}^{16}  R^{(S\beta)}_{\mu \alpha} (\bm{l}_{1\perp})  \, \Bigg( C_{F} \Big( \tilde{U}(\bm{x}_{\perp}) \tilde{U}^{\dagger} (\bm{y}_{\perp}) -\mathds{1} \Big) \Bigg) v(\bm{p}) \, ,
\label{eq:amplitude-S13-S16}
\end{align}
where 
\begin{align}
 \sum_{\beta=13}^{16}  R^{(S\beta)}_{\mu \alpha} (\bm{l}_{1\perp})  &= -\frac{1}{2\pi^{2}} \, T^{\rm LO}_{\mu \alpha} (\bm{l}_{1\perp}) \Bigg\{ \ln \Big(\frac{1}{z_{0}} \Big) \, \Bigg( \frac{1}{\epsilon}+\frac{1}{2} \ln \Big( \frac{\tilde{\mu}^{2}}{Q^{2}} \Big) \Bigg) -\frac{3}{4} \, \Bigg( \frac{1}{\epsilon}+\frac{1}{2} \ln \Big( \frac{\tilde{\mu}^{2}}{Q^{2}} \Big) \Bigg) \Bigg\} \nonumber \\
 & - \frac{1}{4\pi^{2}} \, \ln  \Big(\frac{1}{z_{0}} \Big) \Bigg\{ \sum_{\beta=13}^{16} \, A^{(S\beta)}_{\mu \alpha} (\bm{l}_{1\perp})  - \frac{1}{2} \, \ln \Big(\frac{1}{z_{0}} \Big) \sum_{\beta=13}^{16} \, B^{(S\beta)}_{\mu \alpha} (\bm{l}_{1\perp})   \Bigg\} + \sum_{\beta=13}^{16}  R^{(S\beta)}_{\text{finite};\mu \alpha} (\bm{l}_{1\perp})  \, .
\label{eq:R-S13-S16}
\end{align}
In the above equation, $T^{\rm LO}$ is given by Eq.~\ref{eq:T-LO} and the coefficients $A^{(S\beta)}$ and $B^{(S\beta)}$ will be given later. The finite pieces of each process are contained in $R^{(S\beta)}_{\text{finite}}$ and expressions for these are provided in Appendix~\ref{sec:finite-pieces-SE-3}.

We shall now discuss in detail the computation for $(S13)$. The results for the other three can be obtained using the same methods.
\begin{figure}[!htbp]
\includegraphics[scale=1]{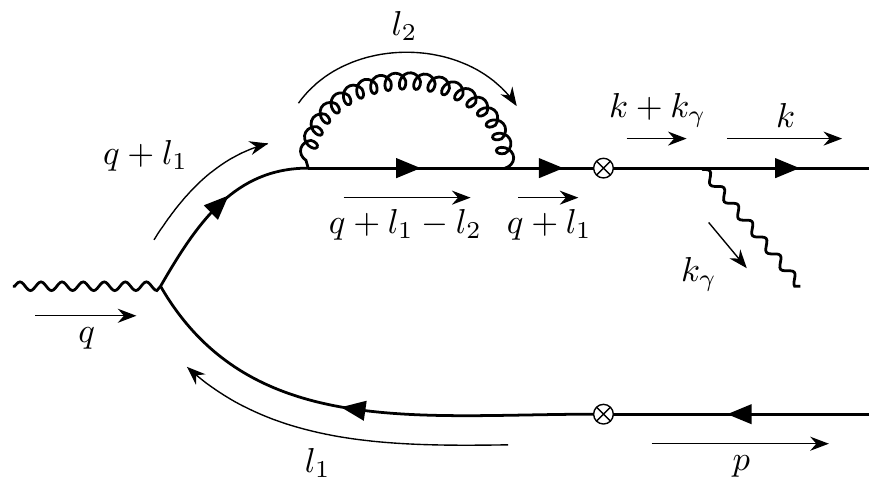}
\caption{The process labeled $(S13)$ in Fig.~\ref{fig:NLO-self-3} with the momentum assignments and directions shown. The gluon does not scatter off of the background classical field in this picture. Only the quark and antiquark do which are represented by the crossed blobs.  \label{fig:S13}}
\end{figure}
The amplitude for this diagram (shown in Fig.~\ref{fig:S13}) is 
\begin{align}
\mathcal{M}_{\mu \alpha}^{(S13)}&= \int_{l_{1}} \overline{u}(\bm{k}) \, (-ieq_{f}) \, \gamma_{\alpha} \, S_{0}(k+k_{\gamma}) \, \mathcal{T}_{q}(k+k_{\gamma}, q+l_{1}) \, S_{0}(q+l_{1}) \Sigma(q+l_{1}) \, S_{0}(q+l_{1})  \nonumber \\
& \times (-ieq_{f}) \, \gamma_{\mu } \, S_{0}(l_{1}) \, \mathcal{T}_{q}(l_{1},-p) \, v(\bm{p}) \, ,
\label{eq:amplitude-S13-generic}
\end{align}
where the free quark propagator and effective vertices are given respectively in Eqs.~\ref{eq:fermion-gluon-propagator} and \ref{eq:vertex-factors-quark-gluon}. The self-energy contribution $\Sigma$ computed in Eq.~\ref{eq:loop-contribution-final-expression} of Appendix~\ref{sec:quark-real photon-quark-vertex-gluon-correction}  is 
\begin{align}
\Sigma(k_{f})&= ig^{2} C_{F}  \int_{l^{-}_{2}} \, \frac{1}{2k^{-}_{f}} \, \Bigg\{  \Bigg(\frac{2+\epsilon}{2l^{-}_{2}} +\frac{2(k^{-}_{f}-l^{-}_{2})}{(l^{-}_{2})^{2}} \Bigg) \gamma^{-} \, \mu^{\epsilon}  \, \mathcal{I}_{v}^{(1,ii)}(\bm{0}_{\perp},\Delta_{s}) \nonumber \\
&+ (2-\epsilon) \, \Bigg[ \slashed{k}_{f} - \Big( \gamma^{+}k^{-}_{f} + \gamma^{-} \frac{\bm{k}_{f\perp}^{2}}{2k^{-}_{f}} - \bm{\gamma}_{\perp}.\bm{k}_{f\perp} \Big) \, \frac{l^{-}_{2}}{k^{-}_{f}} \Bigg] \, \mu^{\epsilon}  \, \mathcal{I}_{v}^{(1,0)}(\bm{0}_{\perp},\Delta_{s})  \Bigg\}  \, ,
\end{align}
with the four-momentum $k_{f}=(q^{+}+l_{1}^{+},k^{-}+k^{-}_{\gamma},\bm{l}_{1\perp})$. The constituent integrals are given by
\begin{align}
\mu^{\epsilon}  \, \mathcal{I}_{v}^{(1,ii)}(\bm{0}_{\perp},\Delta_{s})&= \mu^{\epsilon}  \, \int \frac{\mathrm{d}^{2-\epsilon} \bm{l}_{2\perp}}{(2\pi)^{2-\epsilon}} \, \frac{\bm{l}_{2\perp}^{2}}{\bm{l}_{2\perp}^{2}+\Delta_{s}}= -\frac{\Delta_{s}}{4\pi} \, \Bigg( \frac{2}{\epsilon} +\ln \Big(\frac{\tilde{\mu}^{2}}{\Delta_{s}} \Big)+O(\epsilon) \Bigg)  \, ,  \enskip \text{for} \, \, \, \epsilon \rightarrow 0 \, , \nonumber \\
\mu^{\epsilon}  \, \mathcal{I}_{v}^{(1,0)}(\bm{0}_{\perp},\Delta_{s})&=\mu^{\epsilon}  \,  \int \frac{\mathrm{d}^{2-\epsilon} \bm{l}_{2\perp}}{(2\pi)^{2-\epsilon}} \, \frac{1}{\bm{l}_{2\perp}^{2}+\Delta_{s}}= \frac{1}{4\pi} \, \Bigg( \frac{2}{\epsilon} +\ln \Big(\frac{\tilde{\mu}^{2}}{\Delta_{s}} \Big) +O(\epsilon) \Bigg)  \, , \enskip \text{for} \,\, \,  \epsilon \rightarrow 0 \, ,
 \end{align}
where $\Delta_{s}= - \frac{l^{-}_{2}}{k^{-}_{f}} \Big( 1-\frac{l^{-}_{2}}{k^{-}_{f}} \Big) \, k_{f}^{2}$. 

We have used here the conservation of momentum across the eikonal quark vertex to write $q^{-}+l_{1}^{-}=k^{-}+k^{-}_{\gamma}$. The integration over $l_{1}^{-}$ follows trivially from the Dirac delta functions contributed by the effective vertices and we will obtain an overall momentum conserving delta function. Since $k_{f}^{2}$ depends on $l_{1}^{+}$ the value of $\Delta_{s}$ will be determined by the pole enclosed in the contour integration over $l_{1}^{+}$. We will enclose the pole at
\begin{equation*}
l_{1}^{+} \vert_{\text{pole}} = -\frac{\bm{l}_{1\perp}^{2}}{2p^{-}}+\frac{i\varepsilon}{2p^{-}} \, ,
\end{equation*}
which is above the real $l_{1}^{+}$ axis. The $\Delta_{s}$ factor, appearing in the constituent integrals in Eqs.~\ref{eq:constituent-integrals-self-energy}, for $(S13)$  therefore becomes 
\begin{equation}
\Delta_{s}^{(S13)}= \frac{z_{l}}{1-z_{\bar{q}}} \, \Big( 1- \frac{z_{l}}{1-z_{\bar{q}}} \Big) \frac{\bm{l}_{1\perp}^{2}+\Delta^{\text{LO}:(1)}}{z_{\bar{q}}} \, ,
\end{equation}
where $z_{l}= l_{2}^{-}/q^{-}$ is the gluon momentum fraction in the loop and $\Delta^{\text{LO}:(1)}=Q^{2}z_{\bar{q}}(1-z_{\bar{q}})-i\varepsilon$. Using the identities given by Eqs.~\ref{eq:self-energy-loop-internal-shockwave} and \ref{eq:constituent-integrals-self-energy} in Eq.~\ref{eq:amplitude-S13-generic}, we can finally write the expression for the expression in Eq.~\ref{eq:amplitude-S13-generic} as
\begin{align}
\mathcal{M}^{(S13)}_{\mu \alpha}&= 2\pi \, (egq_{f})^{2} \, \delta(1-z_{\text{tot}}^{v}) \int \mathrm{d} \Pi^{\text{LO}}_{\perp} \, \overline{u} (\bm{k}) \, R^{(S13)}_{\mu \alpha} (\bm{l}_{1\perp}) \, \Big[  C_{F} \, \Big( \tilde{U} (\bm{x}_{\perp}) \tilde{U}^{\dagger} (\bm{y}_{\perp}) -\mathds{1} \Big) \Big] \,   v(\bm{p}) \, ,
\end{align}
where we employ the compact notation for the transverse Fourier phases appearing in the LO amplitude introduced in Eq.~\ref{eq:phase-transverse}.

 For clarity, as previously, we split $R^{(S13)}$ as
\begin{equation}
R^{(S13)}_{\mu \alpha} (\bm{l}_{1\perp})  = \hyperref[eq:R-S13-div]{R^{(S13)}_{\text{div.};\mu \alpha} (\bm{l}_{1\perp})  }+\hyperref[eq:R-S13-finite]{R^{(S13)}_{\text{finite};\mu \alpha} (\bm{l}_{1\perp})  }  \, ,
\label{eq:R-S13}
\end{equation}
where the first part is constituted of logarithms in $z_{0}$ (equivalent to logarithms in rapidity) and UV singularities and the second part is free of such terms. For $(S13)$ these are obtained respectively to be 
\begin{align}
 R^{(S13)}_{\text{div.};\mu \alpha} (\bm{l}_{1\perp}) &=  -\frac{1}{2\pi^{2}} \, R^{\text{LO}:(1)}_{\mu \alpha}(\bm{l}_{1\perp} ) \, \Bigg\{ \ln \Big(\frac{1}{z_{0} } \Big) \Bigg(\frac{1}{\epsilon} +\frac{1}{2} \ln \Big( \frac{\tilde{\mu}^{2}}{Q^{2}}  \Big)  \, \Bigg) -\frac{3}{4}  \Bigg(\frac{1}{\epsilon} +\frac{1}{2} \ln \Big( \frac{\tilde{\mu}^{2}}{Q^{2}}  \Big)  \, \Bigg)   \nonumber \\
& +\frac{1}{2} \, \ln \Big(\frac{1}{z_{0} } \Big) \, \Bigg[  \ln \Bigg(\frac{Q^{2}z_{\bar{q}}}{\bm{l}_{1\perp}^{2}+\Delta^{\text{LO}:(1)}} \Bigg) - \frac{1}{2} \, \ln \Big(\frac{1}{z_{0} } \Big) \Bigg] \Bigg\} \, ,
\label{eq:R-S13-div}
\end{align}
and 
\begin{align}
 R^{(S13)}_{\text{finite};\mu \alpha} (\bm{l}_{1\perp}) &= \frac{1}{2\pi^{2}} \, R^{\text{LO}:(1)}_{\mu \alpha}(\bm{l}_{1\perp} ) \, \Bigg\{ \frac{7}{8} +\frac{3}{8} \, \ln \Bigg( \frac{Q^{2} z_{\bar{q}}  }{\bm{l}_{1\perp}^{2}+\Delta^{\rm LO:(1)}} \Bigg) -\frac{\pi^{2}}{12} \Bigg\} \, .
 \label{eq:R-S13-finite}
\end{align}

Following the structure of the amplitude for these four processes given by Eqs.~\ref{eq:amplitude-S13-S16} and \ref{eq:R-S13-S16}, the coefficients $A$ and $B$ for $(S13)$ are respectively 
\begin{equation}
A^{(S13)}_{\mu \alpha} (\bm{l}_{1\perp}) = \ln \Bigg(\frac{Q^{2} z_{\bar{q}}}{\bm{l}_{1\perp}^{2}+\Delta^{\text{LO}:(1)}} \Bigg) \, R^{\rm LO:(1)}_{\mu \alpha} (\bm{l}_{1\perp})  \, , \quad B^{(S13)}_{\mu \alpha} (\bm{l}_{1\perp}) =  R^{\rm LO:(1)}_{\mu \alpha}(\bm{l}_{1\perp})   \, .
\label{eq:A-S13-B-S13}
\end{equation}

We can now employ these techniques to compute the contributions $(S14)-(S16)$: 
\begin{align}
\begin{pmatrix}
R^{(S14)}_{\mu \alpha} (\bm{l}_{1\perp})  \\
R^{(S15)}_{\mu \alpha} (\bm{l}_{1\perp})  \\
R^{(S16)}_{\mu \alpha} (\bm{l}_{1\perp}) 
\end{pmatrix}
 & =-\frac{1}{2\pi^{2}} \, \begin{pmatrix} 
 R^{\rm LO:(2)}_{\mu \alpha} (\bm{l}_{1\perp}) \\
  R^{\rm LO:(4)}_{\mu \alpha} (\bm{l}_{1\perp}) \\
   R^{\rm LO:(3)}_{\mu \alpha} (\bm{l}_{1\perp})
   \end{pmatrix}
    \Bigg\{ \ln \Big(\frac{1}{z_{0}} \Big) \, \Bigg( \frac{1}{\epsilon}+\frac{1}{2} \ln \Big( \frac{\tilde{\mu}^{2}}{Q^{2}} \Big) \Bigg) -\frac{3}{4} \, \Bigg( \frac{1}{\epsilon}+\frac{1}{2} \ln \Big( \frac{\tilde{\mu}^{2}}{Q^{2}} \Big) \Bigg) \Bigg\} \nonumber \\
 & -\frac{1}{4\pi^{2}} \, \ln  \Big(\frac{1}{z_{0}} \Big) \Bigg\{ \begin{pmatrix}
 \hyperref[eq:A-S14-B-S14]{A^{(S14)}_{\mu \alpha} (\bm{l}_{1\perp}) } \\
 \hyperref[eq:A-S15-B-S15]{A^{(S15)}_{\mu \alpha} (\bm{l}_{1\perp})  } \\
 \hyperref[eq:A-S16-B-S16]{A^{(S16)}_{\mu \alpha} (\bm{l}_{1\perp}) }
\end{pmatrix} 
 - \frac{1}{2} \, \ln \Big(\frac{1}{z_{0}} \Big) \begin{pmatrix}
 \hyperref[eq:A-S14-B-S14]{B^{(S14)}_{\mu \alpha} (\bm{l}_{1\perp}) } \\
\hyperref[eq:A-S15-B-S15]{ B^{(S15)}_{\mu \alpha} (\bm{l}_{1\perp})  } \\
\hyperref[eq:A-S16-B-S16]{ B^{(S16)}_{\mu \alpha}(\bm{l}_{1\perp})  }
\end{pmatrix} \Bigg\} + 
\begin{pmatrix}
\hyperref[eq:R-S14-finite]{R^{(S14)}_{\text{finite};\mu \alpha}(\bm{l}_{1\perp})}  \\
\hyperref[eq:R-S15-finite]{R^{(S15)}_{\text{finite};\mu \alpha} (\bm{l}_{1\perp})} \\
\hyperref[eq:R-S16-finite]{R^{(S16)}_{\text{finite};\mu \alpha}(\bm{l}_{1\perp})} 
\end{pmatrix}  \, .
\label{eq:R-S14-S15-S16}
\end{align}
 The $A$ and $B$ coefficients above are 
\begin{align}
A^{(S14)}_{\mu \alpha} (\bm{l}_{1\perp}) & = \ln \Bigg(\frac{z_{\bar{q}}Q^{2}}{(z_{q}+z_{\bar{q}}) \, \Big[ (\bm{l}_{1\perp}+\bm{v}_{\perp}^{\rm LO:(2)})^{2}+\Delta^{\rm LO:(2)} \Big]  } \Bigg) \, R^{\rm LO:(2)}_{\mu \alpha} (\bm{l}_{1\perp}) \, , \quad B^{(S14)}_{\mu \alpha} (\bm{l}_{1\perp}) =  R^{\rm LO:(2)}_{\mu \alpha} (\bm{l}_{1\perp}) \, ,
\label{eq:A-S14-B-S14}
\end{align} 
\begin{align}
A^{(S15)}_{\mu \alpha} (\bm{l}_{1\perp})& = \ln \Bigg(\frac{Q^{2} z_{\bar{q}}}{(z_{q}+z_{\bar{q}}) \, \Big[ (\bm{l}_{1\perp}+\bm{v}_{\perp}^{\rm LO:(2)})^{2}+\Delta^{\rm LO:(2)} \Big]  } \Bigg) \, R^{\rm LO:(4)}_{\mu \alpha} (\bm{l}_{1\perp}) + \overline{A}_{\mu \alpha} (\bm{l}_{1\perp}) \, \ln \Big(\frac{a}{b} \Big) \, , \quad B^{(S15)}_{\mu \alpha} (\bm{l}_{1\perp}) =  R^{\rm LO:(4)}_{\mu \alpha} (\bm{l}_{1\perp}) \, ,
\label{eq:A-S15-B-S15}
\end{align}
where 
\begin{align}
\overline{A}_{\mu \alpha} (\bm{l}_{1\perp})&= \gamma^{-} \frac{\gamma^{+}z_{q}q^{-}-\bm{\gamma}_{\perp}.(\bm{l}_{1\perp}-\bm{k}_{\gamma \perp})}{ (\bm{l}_{1\perp} +z_{\bar{q}}/z_{\gamma} \, \bm{k}_{\gamma \perp})^{2}-i\varepsilon } \gamma_{\mu} \frac{\gamma^{+}(1-z_{\gamma})q^{-}+\gamma^{-} \, (\bm{l}_{1\perp}-\bm{k}_{\gamma \perp})^{2} / 2(1-z_{\gamma})q^{-} +\bm{\gamma}_{\perp}.(\bm{l}_{1\perp}-\bm{k}_{\gamma \perp})}{(\bm{l}_{1\perp}-\bm{k}_{\gamma \perp})^{2}+Q^{2}z_{q}(1-z_{q}) -i\varepsilon}  \nonumber \\
& \times \gamma_{\alpha} \, \frac{\gamma^{+}z_{\bar{q}}q^{-}+\bm{\gamma}_{\perp}.\bm{l}_{1\perp}  }{2(q^{-})^{2} z_{\gamma}/(1-z_{q}) } \gamma^{-} \, ,
\end{align}
and 
\begin{align}
a=\frac{(\bm{l}_{1\perp}-\bm{k}_{\gamma \perp})^{2}+\Delta^{\rm LO:(3)}}{1-z_{q}} \, , b=\frac{1-z_{\gamma}}{z_{\bar{q}}} \, \Big[ (\bm{l}_{1\perp}+\bm{v}_{\perp}^{\rm LO:(2)})^{2}+\Delta^{\rm LO:(2)} \Big] \, .
\end{align}
Finally we have 
\begin{align}
A^{(S16)}_{\mu \alpha} (\bm{l}_{1\perp})&= \ln \Bigg(\frac{Q^{2}(1-z_{q})}{(\bm{l}_{1\perp}-\bm{k}_{\gamma \perp})^{2}+\Delta^{\text{LO}:(3)}} \Bigg) \, R^{\rm LO:(3)}_{\mu \alpha} (\bm{l}_{1\perp}) \, , \quad B^{(S16)}_{\mu \alpha} (\bm{l}_{1\perp}) =  R^{\rm LO:(3)}_{\mu \alpha} (\bm{l}_{1\perp}) \, .
\label{eq:A-S16-B-S16}
\end{align}
The terms $\bm{v}_{\perp}^{\rm LO:(1,2)}$ and $\Delta^{\rm LO:(1,2)}$ are given in Eq.~\ref{eq:denominator-factors-LO-amplitude} and $\Delta^{\rm LO:(3)}=z_{q}(1-z_{q})-i\varepsilon $. The finite pieces for these processes are given in Appendix~\ref{sec:finite-pieces-virtual-graphs}.

Moving now to the third line of Fig.~\ref{fig:NLO-self-3}, we can use the self-energy function in Eq.~\ref{eq:loop-contribution-final-expression} to compute the contributions to the amplitude from diagrams $(S21)$ and $(S23)$. As for the other diagrams, the expression for the amplitude can be written as
\begin{align}
\mathcal{M}^{(S \beta)}_{\mu \alpha}&= 2\pi \, (egq_{f})^{2} \, \delta(1-z_{\text{tot}}^{v}) \int \mathrm{d} \Pi^{\text{LO}}_{\perp} \, \overline{u} (\bm{k}) \, R^{(S\beta)}_{\mu \alpha} (\bm{l}_{1\perp}) \, \Big[ C_{F} \, \Big( \tilde{U} (\bm{x}_{\perp} \tilde{U}^{\dagger} (\bm{y}_{\perp}) -\mathds{1} \Big)\Big]  \,  v(\bm{p}) \, , \quad \beta=21, 23 \,  ,
\end{align}
where 
\begin{align}
\begin{pmatrix}
R^{(S21)}_{\mu \alpha} (\bm{l}_{1\perp}) \\
R^{(S23)}_{\mu \alpha} (\bm{l}_{1\perp})
\end{pmatrix}
& = \begin{pmatrix}
\hyperref[eq:R-S21-div]{R^{(S21)}_{\rm div.;\mu \alpha} (\bm{l}_{1\perp})} \\
\hyperref[eq:R-S23-div]{R^{(S23)}_{\rm div.;\mu \alpha} (\bm{l}_{1\perp})}
\end{pmatrix}
+
 \begin{pmatrix}
\hyperref[eq:R-S21-finite]{R^{(S21)}_{\rm finite;\mu \alpha} (\bm{l}_{1\perp})} \\
\hyperref[eq:R-S23-finite]{R^{(S23)}_{\rm finite;\mu \alpha} (\bm{l}_{1\perp})}
\end{pmatrix} \, .
\label{eq:R-S21-S23}
\end{align}
We will provide expressions only for the divergent parts of these amplitudes in this section. The finite pieces are given in the Appendix~\ref{sec:finite-pieces-SE-3}. For $(S21)$ and $(S23)$ the divergent pieces are respectively given by
\begin{align}
 R^{(S21)}_{\text{div.};\mu \alpha} (\bm{l}_{1\perp}) &=  -\frac{1}{2\pi^{2}} \, R^{\text{LO}:(1)}_{\mu \alpha}(\bm{l}_{1\perp} ) \, \Bigg\{ \ln \Big(\frac{1}{z_{0} } \Big) \Bigg(\frac{1}{\epsilon} +\frac{1}{2} \ln \Big( \frac{\tilde{\mu}^{2}}{Q^{2}}  \Big)  \, \Bigg) -\frac{3}{4}  \Bigg(\frac{1}{\epsilon} +\frac{1}{2} \ln \Big( \frac{\tilde{\mu}^{2}}{Q^{2}}  \Big)  \, \Bigg)   \nonumber \\
& +\frac{1}{2} \, \ln \Big(\frac{1}{z_{0} } \Big) \, \Bigg[  \ln \Bigg(\frac{Q^{2}(1-z_{\bar{q}}) }{z_{q} \, (-2k.k_{\gamma})} \Bigg) +\frac{1}{2} \, \ln \Big(\frac{1}{z_{0} } \Big) \Bigg] \Bigg\} \, ,
\label{eq:R-S21-div}
\end{align}
and 
\begin{align}
R^{(S23)}_{\rm div.;\mu \alpha} (\bm{l}_{1\perp})&=  -\gamma^{-} \, \frac{\gamma^{+}z_{q}q^{-}-\bm{\gamma}_{\perp}.(\bm{l}_{1\perp}-\bm{k}_{\gamma \perp})}{\Big(\bm{l}_{1\perp}  + \bm{v}_{\perp}^{\rm LO:(2)} \Big)^{2}+\Delta^{\rm LO:(2)}    } \gamma_{\alpha} \, \frac{\gamma^{+}(1-z_{\bar{q}})q^{-}+\gamma^{-} \frac{\bm{l}_{1\perp}^{2}}{2(1-z_{\bar{q}})q^{-}}  -\bm{\gamma}_{\perp}.\bm{l}_{1\perp}    }{\bm{l}_{1\perp}^{2}+\Delta^{\rm LO:(1)} }  \gamma_{\mu } \frac{\gamma^{+}z_{\bar{q}}q^{-}+\bm{\gamma}_{\perp}.\bm{l}_{1\perp}}{2(1-z_{\gamma})/z_{\bar{q}} \, (q^{-})^{2}} \, \gamma^{-}  \nonumber \\
& \times \frac{1}{2\pi^{2}} \Bigg\{ \ln \Big(\frac{1}{z_{0}} \Big) \Bigg( \frac{1}{\epsilon} +\frac{1}{2} \ln \Big( \frac{\tilde{\mu}^{2}}{Q^{2}} \Big)  \Bigg)  -\frac{3}{4} \Bigg( \frac{1}{\epsilon} +\frac{1}{2} \ln \Big( \frac{\tilde{\mu}^{2}}{Q^{2}}  \Big) \Bigg) +\frac{1}{2} \ln \Big(\frac{1}{z_{0}} \Big)  \Bigg( \ln \frac{Q^{2}z_{\bar{q}}}{\bm{l}_{1\perp}^{2}+\Delta^{\rm LO:(1)}} +\frac{1}{2}  \ln \Big(\frac{1}{z_{0}} \Big)  \Bigg) \Bigg\}  \nonumber \\
& - \frac{\gamma^{-}\gamma_{\alpha} \gamma_{\mu} }{2(1-z_{\gamma})/z_{\bar{q}} \, (q^{-})^{2}  \,\Big[ \Big(\bm{l}_{1\perp}  + \bm{v}_{\perp}^{\rm LO:(2)} \Big)^{2}+\Delta^{\rm LO:(2)} \Big] } \, \frac{z_{q}q^{-}}{2(1-z_{\bar{q}})} \, \frac{1}{2\pi^{2}}  \Bigg( \frac{1}{\epsilon} +\frac{1}{2} \ln \Big( \frac{\tilde{\mu}^{2}}{Q^{2}}  \Big) \Bigg) \, ,
\label{eq:R-S23-div}
\end{align}
We conclude this section by detailing the computation of the amplitude for the remaining two diagrams labeled $(S22)$ and $(S24)$ that appear in the third row of Fig.~\ref{fig:NLO-self-3}. For this we will use the general results in Appendix \ref{sec:quark-real photon-quark-vertex-gluon-correction} for the gluon loop contribution. 
For the diagram $(S22)$, we write the amplitude as 
\begin{align}
\mathcal{M}_{\mu \alpha}^{(S22)}= \int_{l_{1}} \overline{u}(\bm{k}) \, \tilde{\Sigma}_{\alpha} (k,k_{\gamma}) \, S_{0}(k+k_{\gamma}) \, \mathcal{T}_{q}(k+k_{\gamma},q+l_{1}) \, S_{0}(q+l_{1}) \, (-ieq_{f}) \gamma_{\mu} \, S_{0}(l_{1}) \, \mathcal{T}_{q}(l_{1},-p) v(\bm{p})  \, ,
\label{eq:amplitude-S23-generic}
\end{align}
In Appendix~\ref{sec:quark-real photon-quark-vertex-gluon-correction}, we wrote $\tilde{\Sigma}_{\alpha}$ as the sum of two contributions $\tilde{\Sigma}=\tilde{\Sigma}_{\alpha}^{A}+\tilde{\Sigma}_{\alpha}^{B}$ where $A$ and $B$ denote respectively the cases where the gluon loop momentum $l^{-}$ is in the range $(\Lambda^{-}_{0}, k^{-})$ and $(k^{-}, k^{-}+k^{-}_{\gamma})$. The divergent pieces of each of these contributions are given respectively in Eqs.~\ref{eq:sigma-tilde-divergent-case-A} and \ref{eq:sigma-tilde-divergent-case-B}. Using the Dirac equation $\overline{u}(\bm{k}) \slashed{k}=0$ for massless quarks, and a bit of algebra involving gamma matrices, we can rewrite Eq.~\ref{eq:amplitude-S23-generic} as
\begin{align}
\mathcal{M}^{(S22)}_{\mu \alpha}&= 2\pi \, (egq_{f})^{2} \, \delta(1-z_{\text{tot}}^{v}) \int \mathrm{d} \Pi^{\text{LO}}_{\perp} \, \overline{u} (\bm{k}) \, R^{(S22)}_{\mu \alpha} (\bm{l}_{1\perp}) \, \Big[ C_{F} \, \Big( \tilde{U} (\bm{x}_{\perp} \tilde{U}^{\dagger} (\bm{y}_{\perp}) -\mathds{1} \Big) \Big]  \,  v(\bm{p}) \, ,
\end{align}
where 
\begin{equation}
R^{(S22)}_{\mu \alpha}(\bm{l}_{1\perp})= \hyperref[eq:R-S22-div]{R^{(S22)}_{\text{div.};\mu \alpha}(\bm{l}_{1\perp})}+\hyperref[eq:R-S22-finite]{R^{(S22)}_{\text{finite};\mu \alpha}(\bm{l}_{1\perp})} \, .
\label{eq:R-S22}
\end{equation}
The divergent piece is 
\begin{align}
R^{(S22)}_{\text{div.};\mu \alpha} (\bm{l}_{1\perp})&= \frac{1}{2\pi^{2}} \,  R^{\rm LO:(1)}_{\mu \alpha} (\bm{l}_{1\perp} ) \, \Bigg\{ \ln \Big(\frac{1}{z_{0} } \Big) \Bigg(\frac{1}{\epsilon} +\frac{1}{2} \ln \Big( \frac{\tilde{\mu}^{2}}{Q^{2}}  \Big)  \, \Bigg) -\frac{3}{4}  \Bigg(\frac{1}{\epsilon} +\frac{1}{2} \ln \Big( \frac{\tilde{\mu}^{2}}{Q^{2}}  \Big)  \, \Bigg) +\frac{1}{2} \, \ln \Big(\frac{1}{z_{0} } \Big) \, \Bigg[  \ln \Bigg(\frac{Q^{2}(1-z_{\bar{q}}) }{z_{q} \, (-2k.k_{\gamma})} \Bigg)   \nonumber \\
& +\frac{1}{2} \, \ln \Big(\frac{1}{z_{0} } \Big) \Bigg] \Bigg\}  \, .
\label{eq:R-S22-div}
\end{align}
We can clearly see from Eqs.~\ref{eq:R-S21-div} and \ref{eq:R-S22-div} that the divergent pieces for $(S21)$ and $(S22)$ exactly cancel leaving only the finite pieces from these graphs towards their contribution to the amplitude. 

In a similar fashion, we write the amplitude for $(S24)$ as 
\begin{align}
\mathcal{M}^{(S24)}_{\mu \alpha}&= 2\pi \, (egq_{f})^{2} \, \delta(1-z_{\text{tot}}^{v}) \int \mathrm{d} \Pi^{\text{LO}}_{\perp} \, \overline{u} (\bm{k}) \, R^{(S24)}_{\mu \alpha} (\bm{l}_{1\perp}) \, \Big[ C_{F} \, \Big( \tilde{U} (\bm{x}_{\perp} \tilde{U}^{\dagger} (\bm{y}_{\perp}) -\mathds{1} \Big) \Big] \,  v(\bm{p}) \, ,
\end{align}
where 
\begin{equation}
R^{(S24)}_{\mu \alpha} (\bm{l}_{1\perp}) = \hyperref[eq:R-S24-div]{R^{(S24)}_{\text{div.};\mu \alpha} (\bm{l}_{1\perp})}+\hyperref[eq:R-S24-finite]{R^{(S24)}_{\text{finite};\mu \alpha} (\bm{l}_{1\perp})} \, .
\label{eq:R-S24}
\end{equation} 
The divergent piece for this diagram is provided in Eq.~\ref{eq:R-S24-div} of Appendix~\ref{sec:quark-real photon-quark-vertex-gluon-correction}. 

Comparing this expression with the divergent part of $(S23)$ (see Eq.~\ref{eq:R-S23-div}), we can check that there are indeed UV and rapidity divergence cancellations occurring in the net contribution from these processes. The only UV divergent term that survives is the one whose Dirac gamma matrix structure is $\gamma^{-} \gamma_{\alpha} \gamma_{\mu}$. As we will show in the next section, a similarly divergent piece is obtained from the vertex correction graph labeled $(V15)$ in Fig.~\ref{fig:NLO-vertex-V15} that cancels this UV divergence. The result for the finite piece of the loop diagram in Fig.~\ref{fig:self-energy-nested-photon} is given in Appendix~\ref{sec:quark-real photon-quark-vertex-gluon-correction}. This result has been used to compute the finite pieces of the amplitude for $(S22)$ and $(S24)$ in Appendix~\ref{sec:finite-pieces-SE-3}.

We can now write down the final expression for the amplitude from the 12 processes shown in Fig.~\ref{fig:NLO-self-3} (and their quark$\leftrightarrow$antiquark interchanged counterparts) that contribute to the ``free gluon" self-energy amplitude as
\begin{align}
\mathcal{M}_{\mu \alpha}^{\rm NLO:SE(3)}&= 2\pi \, (egq_{f})^{2} \, \delta(1-z_{\rm tot}^{v}) \int \mathrm{d} \Pi_{\perp}^{\rm LO} \, \overline{u} (\bm{k}) \Bigg\{ T^{(3)}_{S;\mu \alpha} (\bm{l}_{1\perp}) \, \Big[ C_{F} \Big(\tilde{U}(\bm{x}_{\perp}) \tilde{U}^{\dagger}(\bm{y}_{\perp})-\mathds{1} \Big) \Big] \Bigg\} \, v(\bm{p}) \, ,
\label{eq:amplitude-self-energy-SE3}
\end{align}
where
\begin{align}
T^{(3)}_{S;\mu \alpha} (\bm{l}_{1\perp})= \sum_{\beta=13}^{36} R^{(S\beta)}_{\mu \alpha} (\bm{l}_{1\perp})=\Big\{ \sum_{\beta=13}^{24} \Big( R^{(S\beta)}_{\rm div.;\mu \alpha} (\bm{l}_{1\perp})+R^{(S\beta)}_{\rm finite;\mu \alpha} (\bm{l}_{1\perp}) \Big) \Big\} + (q\leftrightarrow\bar{q}) \, .
\label{eq:T-S3}
\end{align}
The $(q\leftrightarrow \bar{q})$ interchange in the above equation represents the contribution to the amplitude from the processes labeled $(S25)-(S36)$. These are obtained by imposing the replacements in Eq.~\ref{eq:replacements-qqbar-exchange} in the result we computed for $(S13)-(S24)$. 

The divergent part of the amplitude coming from the 12 diagrams in Fig.~\ref{fig:NLO-self-3} is given by 
\begin{align}
\sum_{\beta=13}^{24} R^{(S\beta)}_{\rm div.;\mu \alpha} &  (\bm{l}_{1\perp})= -\frac{1}{2\pi^{2}} \, T^{\rm LO}_{\mu \alpha} (\bm{l}_{1\perp}) \Bigg\{ \ln \Big(\frac{1}{z_{0}} \Big) \, \Bigg( \frac{1}{\epsilon}+\frac{1}{2} \ln \Big( \frac{\tilde{\mu}^{2}}{Q^{2}} \Big) \Bigg) -\frac{3}{4} \, \Bigg( \frac{1}{\epsilon}+\frac{1}{2} \ln \Big( \frac{\tilde{\mu}^{2}}{Q^{2}} \Big) \Bigg) \Bigg\} \nonumber \\
& -\frac{1}{4\pi^{2}} \, \ln  \Big(\frac{1}{z_{0}} \Big) \Bigg\{  \, \sum_{\beta=13}^{16} A^{(S\beta)}_{\mu \alpha} (\bm{l}_{1\perp})  - \frac{1}{2} \, \ln \Big(\frac{1}{z_{0}} \Big) \, T^{\rm LO}_{\mu \alpha}(\bm{l}_{1\perp})  \Bigg\} \nonumber \\
& +\frac{1}{4\pi^{2}} \,  \Bigg( \frac{1}{\epsilon}+\frac{1}{2} \ln \Big( \frac{ \tilde{\mu}^{2}   }{Q^{2}} \Big) \Bigg) \, \frac{z_{q}^{2} \, z_{\bar{q}}}{2 \, (1-z_{\bar{q}})^{2} \, (1-z_{\gamma}) \, q^{-}}  \, \frac{ \gamma_{\alpha}\gamma_{\mu} \gamma^{-} }{  \Big[(\bm{l}_{1\perp}+\bm{v}_{\perp}^{\rm LO:(2)}  )^{2}+\Delta^{\rm LO:(2)}  \Big]  }  \, .
 \label{eq:R-S13-S24}
\end{align}
The terms appearing in the first line of the above equation are equal in magnitude but opposite in sign with respect to the amplitudes from the six diagrams $(S1)-(S6)$ given by Eq.~\ref{eq:T-S1}. They will therefore cancel out when we add the contributions from all possible self-energy corrections. 

There are two kinds of rapidity divergent pieces in the second line of the above equation. (We have used $\sum_{\beta=13}^{16} B^{(S\beta)}_{\mu \alpha} (\bm{l}_{1\perp})= T^{\rm LO}_{\mu \alpha}(\bm{l}_{1\perp})$ in writing the second term.) The terms proportional to $\ln(1/z_{0})$, with the coefficients $A^{(S\beta)}$ given in Eqs.~\ref{eq:A-S13-B-S13}, \ref{eq:A-S14-B-S14}, \ref{eq:A-S15-B-S15} and \ref{eq:A-S16-B-S16} contribute to the LO JIMWLK Hamiltonian. Indeed, these are precisely the terms that will give the double log limit of the DGLAP/BFKL equations in the limit of large $Q^{2}$. On the other hand, the terms proportional to $\ln^{2}(1/z_{0})$ cancel between real and virtual graphs; we will demonstrate this explicitly in section~\ref{sec:jet-cross-section}. 

Finally, the term in the last line of Eq.~\ref{eq:R-S13-S24} is  the UV divergent piece that remains after cancellations between the divergent pieces of graphs $(S23)$ and $(S24)$. This divergence will be canceled by a similar contribution from the vertex correction graph $(V15)$ which also has a photon nested in the gluon loop.

\subsection{Vertex graphs with dressed gluon propagator } \label{sec:virtual-corrections-vertex}

In this section, we will compute the vertex corrections to the LO amplitude in which the gluon crosses the nuclear shock wave. As discussed in Sec.~\ref{sec:contributions-NLO}, there are two distinct topologies with six contributions in each class. These two sets of contributions are related to each other by $q \leftrightarrow \bar{q}$ interchange. Following a logic identical to the discussion of self-energy graphs, we will detail the computation of the amplitude for one such representative process. The remaining processes can then be computed following similar techniques.

\begin{enumerate} 
\item \textbf{Contributions to $T^{(1)}_{V}$:} 

\begin{figure}[!htbp]
\centering
\begin{minipage}[b]{0.75\textwidth}
\includegraphics[width=\textwidth]{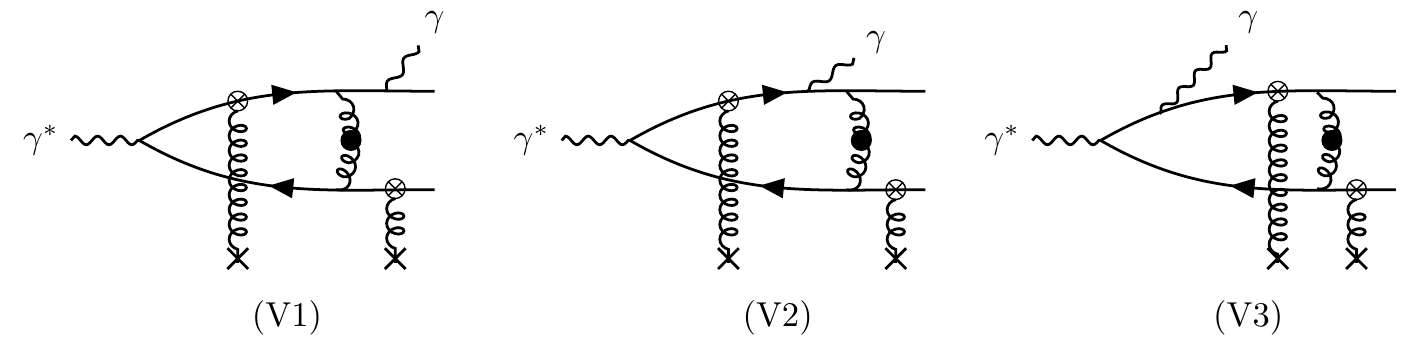}
\end{minipage}

\begin{minipage}[b]{0.75\textwidth}
\includegraphics[width=\textwidth]{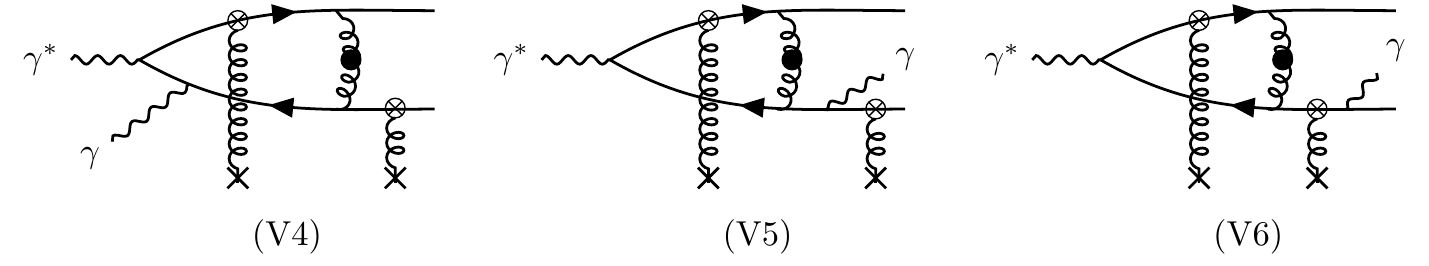}
\end{minipage}
\caption{Vertex corrections proportional to $T^{(1)}_{V}$ in Eq.~\ref{eq:vertex-correction-master-amplitude}.  The  topology corresponds to that of a gluon emitted by the antiquark which propagates forward and is absorbed by the quark after the $q\bar{q}g$ state scatters off the nucleus. The different diagrams correspond then to the possible locations from where the final state photon might be emitted. By interchanging the quark and antiquark lines, we get the diagrams constituting $T^{(2)}_{V}$. Those are labeled as (V7)-(V12).\label{fig:NLO-vertex-1} }
\end{figure}

The six diagrams contributing to $T^{(1)}_{V}$ in the general structure of the amplitude given in Eq.~\ref{eq:vertex-correction-master-amplitude} are shown in Fig.~\ref{fig:NLO-vertex-1}. Their combined contribution to the amplitude can be written as
\begin{align}
\mathcal{M}^{\text{NLO};\text{Vert.}(1)}_{\mu \alpha}& = 2\pi \delta(1-z^{v}_{\text{tot}}) \, (eq_{f} g)^{2} \,  \int \mathrm{d} \Pi_{\perp}^{v} \, \overline{u} (\bm{k}) \Bigg\{ T^{(1)}_{V;\mu \alpha} (\bm{l}_{1\perp}) \, \Big[ \Big( t^{b} \tilde{U} (\bm{x}_{\perp}) t^{a} \tilde{U}^{\dagger} (\bm{y}_{\perp}) \Big) U_{ba}(\bm{z}_{\perp}) - C_{F} \mathds{1} \Big] \Bigg\} v(\bm{p}) \, ,
\label{eq:amplitude-V1-generic}
\end{align}
where $z^{v}_{\text{tot}}=z_{q}+z_{\bar{q}}+z_{\gamma}$ and 
\begin{align}
T^{(1)}_{V;\mu \alpha}(\bm{l}_{1\perp}) &=\sum_{\beta=7}^{12} R^{(V\beta)}_{\mu \alpha}(\bm{l}_{1\perp})=  \hyperref[eq:R-Ver1-div-finite]{R^{\rm Vert.(1)}_{\rm div.;\mu \alpha} (\bm{l}_{1\perp})} + \hyperref[eq:R-Ver1-div-finite]{R^{\rm Vert.(1)}_{\rm finite;\mu \alpha} (\bm{l}_{1\perp})}  \, ,
\label{eq:T-V1}
\end{align}
is the net perturbative contribution from the six processes which can be expressed as the combination of a divergent and a finite part. We will see that there are no UV divergent pieces for these diagrams and the only singularities are those arising from the $l^{-}=0$ pole in the free gluon propagator. In the following, we will outline the steps leading to the above conclusion by considering the representative process $(V1)$. 

The amplitude for $(V1)$ with the momentum assignments shown in Fig.~\ref{fig:V1} is given by 
\begin{figure}[!htbp]
\begin{minipage}[b]{0.5\textwidth}
\includegraphics[width=\textwidth]{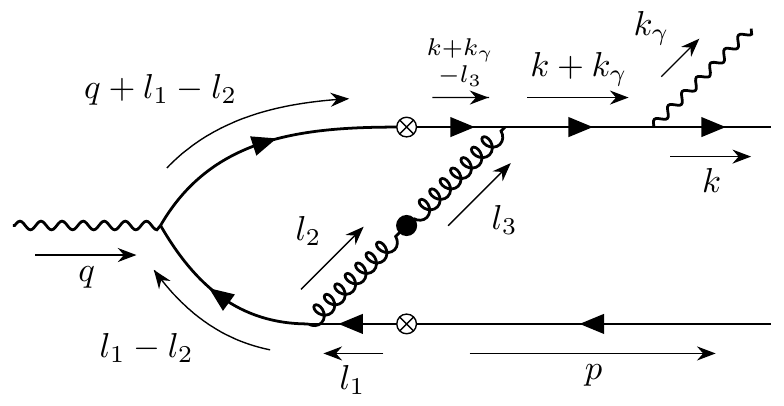}
\end{minipage}
\caption{The process labeled $(V1)$ in Fig.~\ref{fig:NLO-vertex-1} with momenta and their directions shown. We must subtract the ``no scattering" case from the above contribution to obtain the physical amplitude. This will correspond to the diagram above but with no crossed and filled blobs. \label{fig:V1}}
\end{figure}
\begin{align}
\mathcal{M}^{(V1)}_{\mu \alpha}&= (egq_{f})^{2} \int_{l_{1},l_{2},l_{3}} \!\!\!\! \overline{u}(\bm{k}) \, \gamma_{\alpha}  S_{0}(k+k_{\gamma}) t^{b} \, \gamma^{\nu} S_{0}(k+k_{\gamma}-l_{3}) \, \mathcal{T}_{q}(k+k_{\gamma}-l_{3},q+l_{1}-l_{2}) S_{0}(q+l_{1}-l_{2}) \nonumber \\
& \times  \gamma_{\mu} \, S_{0}(l_{1}-l_{2}) t^{a} \, \gamma^{\beta} S_{0}(l_{1}) \mathcal{T}_{q}(l_{1},-p) v(\bm{p}) \times G^{0}_{\nu \rho;ac}(l_{3})\,  \mathcal{T}^{\rho \sigma;cd}_{g}(l_{3},l_{2}) \, G^{0}_{\sigma \beta;db}(l_{2}) \, ,
\label{eq:amplitude-V1-general}
\end{align}
where we remind the reader that the free fermion and gluon propagators are given respectively in Eq.~\ref{eq:fermion-gluon-propagator} and the corresponding effective vertices in Fig.~\ref{fig:effective-vertices}. The various Lorentz indices appearing above can be contracted to obtain the simplified expression given by Eq.~\ref{eq:gamma-identity} for the gamma matrices. 

The integrals over $l_{1}^{-}$ and $l_{2}^{-}$ can be trivially performed using the delta functions appearing in the effective vertices. The integrations over $l_{1}^{+}$, $l_{2}^{+}$ and $l_{3}^{+}$ can then be systematically performed using Cauchy's theorem of residues. These contour integrations are non-zero for $l_{3}^{-}$ in the range $0< l_{3}^{-} < (k^{-}+k^{-}_{\gamma})$. Finally introducing the gluon loop momentum fraction, $z_{l}=l_{3}^{-}/q^{-}$ and subtracting the ``no scattering'' contribution we can write the $(V1)$ amplitude as
\begin{align}
\mathcal{M}^{(V1)}_{\mu \alpha}&= 2\pi (egq_{f})^{2} \delta(1-z^{v}_{\text{tot}}) \int \mathrm{d} \Pi_{\perp}^{v} \int_{\bm{l}_{2\perp}, 
\bm{l}_{3\perp}} \!\!\!\!\!\!\! e^{i(\bm{l}_{2\perp}-\bm{l}_{3\perp}).\bm{r}_{zx}} \, \Big[ \Big( t^{b} \tilde{U} (\bm{x}_{\perp}) t^{a} \tilde{U}^{\dagger} (\bm{y}_{\perp}) \Big) U_{ba}(\bm{z}_{\perp}) -C_{F} \mathds{1} \Big]\nonumber \\
& \times \int \frac{\mathrm{d}z_{l}}{2\pi}  \, \frac{N^{(V1)}}{D^{(V1)}} \, ,
\label{eq:amplitude-diagV1-generic-form}
\end{align}
where the numerator and denominator are respectively given by
\begin{align}
N^{(V1)}&= \overline{u}(\bm{k}) \, \gamma_{\alpha}\frac{\slashed{k}+\slashed{k}_{\gamma}}{2k.k_{\gamma}} \Big( \gamma^{i}-\frac{l^{i}_{3}}{z_{l}q^{-}}\gamma^{-}\Big) \, [\gamma^{+}(1-z_{\bar{q}}-z_{l})q^{-}-\bm{\gamma}_{\perp}.(\bm{k}_{\perp}+\bm{k}_{\gamma \perp}) +\bm{\gamma}_{\perp}.\bm{l}_{3\perp}] \, \gamma^{-} \nonumber \\
& \times [\gamma^{+}(1-z_{\bar{q}}-z_{l})q^{-}-\bm{\gamma}_{\perp}.(\bm{l}_{1\perp}-\bm{l}_{2\perp}) ] \gamma_{\mu} \Big[\gamma^{+}(z_{\bar{q}}+z_{l})q^{-}-\gamma^{-} \Big( \frac{Q^{2}}{2q^{-}}+ \frac{(\bm{l}_{1\perp}-\bm{l}_{2\perp})^{2}}{2(1-z_{\bar{q}}-z_{l})q^{-}}  \Big)+\bm{\gamma}_{\perp}.(\bm{l}_{1\perp}-\bm{l}_{2\perp}) \Big] \nonumber \\
&\times \Big( \gamma_{i}-\frac{l_{2i}}{z_{l}q^{-}}\gamma^{-}  \Big) \, [\gamma^{+}z_{\bar{q}}q^{-}+\bm{\gamma}_{\perp}.\bm{l}_{1\perp}]\gamma^{-} v(\bm{p}) \, ,
\label{eq:numerator-V1}
\end{align}
and 
\begin{align}
D^{(V1)}& =32z_{\bar{q}}z_{l}(z_{\bar{q}}+z_{l}) (1-z_{\bar{q}}-z_{l})^{2} \, (q^{-})^{5} \Big( \frac{\bm{k}_{\perp}^{2}}{2z_{q}q^{-}}+ \frac{\bm{k}_{\gamma \perp}^{2}}{2z_{q}q^{-}} -\frac{\bm{l}_{3\perp}^{2}}{2z_{l}q^{-}} -\frac{(\bm{k}_{\perp}+\bm{k}_{\gamma \perp}-\bm{l}_{3\perp})^{2}}{2(1-z_{\bar{q}}-z_{l})q^{-}} +\frac{i\varepsilon (1-z_{\bar{q}})}{2z_{l}(1-z_{\bar{q}}-z_{l})q^{-}} \Big) \nonumber \\
& \times \Big( \frac{Q^{2}}{2q^{-}} +\frac{\bm{l}_{1\perp}^{2}}{2z_{\bar{q}}q^{-}} +\frac{\bm{l}_{2\perp}^{2}}{2z_{l}q^{-}}+\frac{(\bm{l}_{1\perp}-\bm{l}_{2\perp})^{2}}{ 2(1-z_{\bar{q}}-z_{l})q^{-}} -\frac{i\varepsilon (1-z_{\bar{q}})}{2z_{l}(1-z_{\bar{q}}-z_{l})q^{-}} \Big) \, \Big( \frac{Q^{2}}{2q^{-}}+\frac{(\bm{l}_{1\perp}-\bm{l}_{2\perp})^{2}}{2(z_{\bar{q}}+z_{l})(1-z_{\bar{q}}-z_{l})q^{-}} \nonumber \\
&  -\frac{i\varepsilon}{2(z_{\bar{q}}+z_{l})(1-z_{\bar{q}}-z_{l})q^{-}} \Big) \, .
\label{eq:denominator-V1}
\end{align}
Note that in these expressions we have canceled a factor of $2z_{l}$ from the gluon effective vertex in the numerator with a corresponding factor from one of the propagators in the denominator. This particular form of the amplitude given by Eqs.~\ref{eq:amplitude-diagV1-generic-form}, \ref{eq:numerator-V1} and \ref{eq:denominator-V1} will be useful later in Sec.~\ref{sec:JIMWLK-evolution}. For extracting the UV divergent pieces we will redefine $\bm{l}_{2\perp}-\bm{l}_{3\perp} \rightarrow \bm{l}_{2\perp}$ and $\bm{l}_{3\perp}-z_{l} / (1-z_{\bar{q}} ) \, (\bm{k}_{\perp}+\bm{k}_{\gamma \perp} ) \rightarrow \bm{l}_{3\perp}$ to rewrite Eq.~\ref{eq:amplitude-diagV1-generic-form} as
\begin{align}
\mathcal{M}^{(V1)}_{\mu \alpha}&= - 2\pi (egq_{f})^{2} \delta(1-z^{v}_{\text{tot}}) \int \mathrm{d} \Pi_{\perp}^{v} \int_{\bm{l}_{2\perp}} \!\!\!\! e^{i\bm{l}_{2\perp}.\bm{r}_{zx}} \,\Big[ \Big( t^{b} \tilde{U} (\bm{x}_{\perp}) t^{a} \tilde{U}^{\dagger} (\bm{y}_{\perp}) \Big) U_{ba}(\bm{z}_{\perp}) -C_{F} \mathds{1} \Big] \nonumber \\
& \times \int \frac{\mathrm{d}z_{l}}{2\pi} \int_{\bm{l}_{3\perp}} \frac{\tilde{N}^{(V1)}}{\tilde{D}^{(V1)}} \, ,
\label{eq:amplitude-diagV1-modified}
\end{align}
where the modified numerator and denominator are 
\begin{align}
\tilde{N}^{(V1)}&= \overline{u}(\bm{k}) \gamma_{\alpha}\frac{\slashed{k}+\slashed{k}_{\gamma}}{2k.k_{\gamma}} \Big( \gamma^{i}-\frac{k^{i}+k^{i}_{\gamma}}{(1-z_{\bar{q}})q^{-}}  \gamma^{-}-\frac{l_{3}^{i}}{z_{l}q^{-}} \gamma^{-}  \Big) \, [\gamma^{+}(1-z_{\bar{q}}-z_{l})q^{-}-\bm{\gamma}_{\perp}.(\bm{k}_{\perp}+\bm{k}_{\gamma \perp}) \, \frac{1-z_{\bar{q}}-z_{l}}{1-z_{\bar{q}}} +\bm{\gamma}_{\perp}.\bm{l}_{3\perp}] \nonumber \\
& \times \gamma^{-} [\gamma^{+}(1-z_{\bar{q}}-z_{l})q^{-} -\bm{\gamma}_{\perp}.\Big(\bm{l}_{1\perp}-\bm{l}_{2\perp}-\frac{z_{l}}{1-z_{\bar{q}}} \, (\bm{k}_{\perp}+\bm{k}_{\gamma \perp}) \Big)   +\bm{\gamma}_{\perp}.\bm{l}_{3\perp}] \gamma_{\mu} \Big[ \Big\{ \gamma^{+}(z_{\bar{q}}+z_{l})q^{-}-\gamma^{-} \Big( \frac{Q^{2}}{2q^{-}} \nonumber \\
& + \frac{\Big(\bm{l}_{1\perp}-\bm{l}_{2\perp}-\frac{z_{l}}{1-z_{\bar{q}}} \, (\bm{k}_{\perp}+\bm{k}_{\gamma \perp}) \Big)^{2}}{2(1-z_{\bar{q}}-z_{l})q^{-}} \Big) +\bm{\gamma}_{\perp}.\Big(\bm{l}_{1\perp}-\bm{l}_{2\perp}-\frac{z_{l}}{1-z_{\bar{q}}} \, (\bm{k}_{\perp}+\bm{k}_{\gamma \perp}) \Big) \Big\} -\gamma^{-} \frac{\bm{l}_{3\perp}^{2}}{2(1-z_{\bar{q}}-z_{l})q^{-}} \nonumber \\
&- \Big( \bm{\gamma}_{\perp}-\frac{\bm{l}_{1\perp}-\bm{l}_{2\perp}-\frac{z_{l}}{1-z_{\bar{q}}} \, (\bm{k}_{\perp}+\bm{k}_{\gamma \perp}) }{2(1-z_{\bar{q}}-z_{l})q^{-}} \, \gamma^{-} \Big).\bm{l}_{3\perp} \Big]  \Big( \gamma_{i}-  \frac{k_{i}+k_{\gamma i}}{(1-z_{\bar{q}})q^{-}}  \gamma^{-} +\frac{l^{i}_{2}+l^{i}_{3}}{z_{l}q^{-}}\gamma^{-} \Big) [\gamma^{+}z_{\bar{q}}q^{-}+\bm{\gamma}_{\perp}.\bm{l}_{1\perp}]\gamma^{-} v(\bm{p}) \, , 
\label{eq:numerator-V1-modified}
\end{align}
and 
\begin{align}
\tilde{D}^{(V1)}& = \frac{4z_{\bar{q}}(1-z_{\bar{q}})^{2} \, (q^{-})^{2}}{z_{l}(1-z_{\bar{q}}-z_{l})}  \, \Big[(\bm{l}_{3\perp}+\bm{v}_{1\perp}^{(V1)})^{2}+\Delta_{1}^{(V1)}\Big] \,  \Big[(\bm{l}_{3\perp}+\bm{v}_{2\perp}^{(V1)})^{2}+\Delta_{2}^{(V1)} \Big] \, \Big[\bm{l}_{3\perp}^{2}+\Delta_{3}^{(V1)}\Big]\, .
\label{eq:denominator-V1-modified}
\end{align}
To write this denominator compactly, we have introduced the variables 
\begin{align}
\bm{v}_{1\perp}^{(V1)}&= \bm{l}_{2\perp}+\frac{z_{l}}{1-z_{\bar{q}}} (\bm{k}_{\perp}+\bm{k}_{\gamma \perp}-\bm{l}_{1\perp})    \, , \enskip \bm{v}_{2\perp}^{(V1)} = \bm{l}_{2\perp}-\bm{l}_{1\perp}+\frac{z_{l}}{1-z_{\bar{q}}} \, (\bm{k}_{\perp}+\bm{k}_{\gamma \perp}) \, , \nonumber \\
\Delta_{1}^{(V1)}&= \frac{z_{l}}{1-z_{\bar{q}}} \Big(1-\frac{z_{l}}{1-z_{\bar{q}}} \Big) \, [\bm{l}_{1\perp}^{2}+Q^{2}z_{\bar{q}}(1-z_{\bar{q}}) ]/z_{\bar{q}} -i\varepsilon \, , \quad \Delta_{2}^{(V1)}= Q^{2}(z_{\bar{q}}+z_{l})(1-z_{\bar{q}}-z_{l}) -i\varepsilon \, , \nonumber \\
\Delta_{3}^{(V1)}&= -\frac{z_{l}}{1-z_{\bar{q}}} \Big(1-\frac{z_{l}}{1-z_{\bar{q}}} \Big) \, (2k.k_{\gamma}) -i\varepsilon \, .
\label{eq:denominator-factors-V1}
\end{align}
The denominator in Eq.~\ref{eq:denominator-V1-modified} grows as $\bm{l}_{3\perp}^{6}$ for very large $\bm{l}_{3\perp}$. As we argued in the computation of $(S1)$ in Sec.~\ref{sec:virtual-self-energy-dressed},  terms in the numerator in Eq.~\ref{eq:numerator-V1} proportional to $\bm{l}_{3\perp}^{5}$ and above vanish. This implies that the integral over $\bm{l}_{3\perp}$ is at most logarithmically divergent in the limit of large $\bm{l}_{3\perp}$. Collecting terms in the numerator proportional to $\bm{l}_{3\perp}^{4}$ (and using $\gamma^{i}\gamma^{j} l_{3}^{i} l_{3}^{j} = - \bm{l}_{3\perp}^{2}$, $\gamma^{-}\gamma_{\mu} \gamma^{-}= 2g_{\mu +}=2\delta_{\mu -}$), we can write the piece of the numerator (in $d=2-\epsilon$ dimensions) contributing to the UV divergence as 
\begin{equation}
\tilde{N}_{\rm UV \, div.}^{(V1)}= \overline{u}(\bm{k}) \gamma_{\alpha}\frac{\slashed{k}+\slashed{k}_{\gamma}}{2k.k_{\gamma}} \gamma^{-} v(\bm{p}) \times \Bigg\{ \frac{8z_{\bar{q}}(1-z_{\bar{q}}-z_{l})}{z_{l}^{2}}+\frac{8z_{\bar{q}}}{z_{l}}+\frac{(4-2\epsilon) \, z_{\bar{q}}}{1-z_{\bar{q}}-z_{l}} \Bigg\} \, \delta_{\mu -} \, .
\end{equation}
However from the general expression for the amplitude of our photon$+$dijet production process in Eq.~\ref{eq:amplitude-master}, it is apparent that contractions of the virtual photon effective vertex given by Eq.~\ref{eq:gamma-contraction-with-virtual-photon-propagator} with $\delta_{\mu -}$ will give the result to be zero. This is a consequence of our choice of gauge. We therefore do not have any UV divergences in the $(V1)$ amplitude. 
There are however singularities arising from the $l^{-}_{3}=0$ pole in the gluon propagator. They will be regulated by imposing a lower cutoff in the integral over $z_{l}$ at $z_{0}=\Lambda_{0}^{-}/q^{-}$, where $\Lambda_{0}^{-}$ was specified in Eq.~\ref{eq:initial-energy-scale}. To compute the terms proportional to logarithms in $z_{0}$ we will extract the contributions in Eq.~\ref{eq:amplitude-diagV1-modified} that are proportional to $1/z_{l}$. 

In terms of the constituent integrals given in Appendix~\ref{sec:constituent-integrals-real-emission} that appear in the computation of virtual graphs,
\begin{align}
\mathcal{M}_{\mu \alpha}^{(V1)} \vert_{\propto 1/z_{l}}&= 2\pi (eq_{f}g)^{2} \delta(1-z^{v}_{\rm tot}) \int \mathrm{d} \Pi_{\perp}^{v} \, \overline{u} (\bm{k}) R_{\rm (I);\mu \alpha}^{(V1)} (\bm{l}_{1\perp})  \,  \Big[ \Big( t^{b} \tilde{U} (\bm{x}_{\perp}) t^{a} \tilde{U}^{\dagger} (\bm{y}_{\perp}) \Big) U_{ba}(\bm{z}_{\perp}) -C_{F} \mathds{1} \Big]  v(\bm{p}) \, ,
\label{eq:log-generating-piece-V1}
\end{align}
where
\begin{align}
R_{\rm (I);\mu \alpha}^{(V1)} (\bm{l}_{1\perp})  &= \int  \frac{\mathrm{d}z_{l}}{(2\pi) z_{l}} \int_{\bm{l}_{2\perp}} \!\!\!\! e^{i\bm{l}_{2\perp}.\bm{r}_{zy}} \,\frac{1}{(q^{-})^{2}} \,   \gamma_{\alpha} \frac{\slashed{k}+\slashed{k}_{\gamma}}{2k.k_{\gamma}} \Big[ \big\{4z_{\bar{q}}q^{-} \delta_{i \mu} \gamma^{-} -2q^{-} \gamma_{\mu} \gamma^{i} \gamma^{-} \big\} \Big( \mathcal{I}_{v}^{(3,ijj)}(\bm{V}_{\perp}^{(V1)},\Delta^{(V1)}) \nonumber \\
&+l_{2}^{j} \, \mathcal{I}_{v}^{(3,ij)}(\bm{V}_{\perp}^{(V1)},\Delta^{(V1)}) \Big)  +\gamma^{-} [\gamma^{+}(1-z_{\bar{q}})q^{-}-\bm{\gamma}_{\perp}.\bm{l}_{1\perp} ]\gamma_{\mu} [\gamma^{+}z_{\bar{q}}q^{-}+\bm{\gamma}_{\perp}.\bm{l}_{1\perp} ]\gamma^{-} \, \Big( \mathcal{I}_{v}^{(3,ii)} (\bm{V}_{\perp}^{(V1)},\Delta^{(V1)})\nonumber \\
&+ l_{2}^{i} \, \mathcal{I}_{v}^{(3,i)} (\bm{V}_{\perp}^{(V1)},\Delta^{(V1)}) \Big) \Big]  \, .
\label{eq:R-V1-propto-zl-inverse}
\end{align}
To obtain this expression, we redefined $\bm{l}_{1\perp}-\bm{l}_{2\perp} \rightarrow \bm{l}_{1\perp}$ in Eq.~\ref{eq:amplitude-diagV1-modified}. The constituent integrals here have the  (finite) expressions in $d=2$ dimensions,
\begin{align}
\mathcal{I}_{v}^{(3,ijj)}(\bm{V}_{\perp},\Delta)&= -\frac{1}{4\pi} \int_{0}^{1} \mathrm{d} \alpha_{1} \int_{0}^{1-\alpha_{1}} \mathrm{d} \alpha_{2} \, \Big( \frac{2}{\Delta} +\frac{\bm{V}_{\perp}^{2}}{\Delta^{2}} \Big) \, V^{i} \, ,  \nonumber \\
\mathcal{I}^{(3,ij)}_{v} (\bm{V}_{\perp},\Delta)&=\frac{1}{4\pi} \int_{0}^{1} \mathrm{d} \alpha_{1} \int_{0}^{1-\alpha_{1}} \mathrm{d} \alpha_{2} \, \Big( \frac{V^{i}V^{j}}{\Delta^{2}} +\frac{\delta^{ij}}{\Delta} \Big) \, , \nonumber \\
\mathcal{I}^{(3,i)}_{v} (\bm{V}_{\perp},\Delta)&= -\frac{1}{4\pi} \int_{0}^{1} \mathrm{d} \alpha_{1} \int_{0}^{1-\alpha_{1}} \mathrm{d} \alpha_{2} \, \frac{V^{i}}{\Delta^{2}} \, ,
\label{eq:constituent-integrals-V1}
\end{align}
where $\alpha_{1}$ and $\alpha_{2}$ are Feynman parameters. For the process $(V1)$ the arguments $\bm{V}_{\perp}$ and $\Delta$ are defined in terms of the factors (after imposing $\bm{l}_{1\perp}-\bm{l}_{2\perp} \rightarrow \bm{l}_{1\perp}$) in Eq.~\ref{eq:denominator-factors-V1} as
\begin{align}
\bm{V}_{\perp}^{(V1)}& =\alpha_{1} \, \bm{v}_{1\perp}^{(V1)} +\alpha_{2} \, \bm{v}_{2\perp}^{(V1)} \, , \nonumber \\
\Delta^{(V1)}&= \alpha_{1} (1-\alpha_{1}) \, (\bm{v}_{1\perp}^{(V1)})^{2}+\alpha_{2} (1-\alpha_{2}) \, (\bm{v}_{2\perp}^{(V1)})^{2} -2\alpha_{1} \alpha_{2} \bm{v}_{1\perp}^{(V1)}.\bm{v}_{2\perp}^{(V1)} \nonumber \\
& +\alpha_{1} \Delta_{1}^{(V1)} +\alpha_{2} \Delta_{2}^{(V1)} +(1-\alpha_{1}-\alpha_{2}) \Delta_{3}^{(V1)} \, .
\end{align}
Interestingly, these arguments $\bm{V}_{\perp}$ and $\Delta$ appearing in the constituent integrals too can be expanded in terms of the gluon loop momentum fraction $z_{l}$ as 
\begin{align}
V_{\perp}^{i}& =c_{1}^{i} +z_{l} \, c_{2}^{i} \, , \enskip \Delta=c_{3}+c_{4} \, z_{l}+c_{5} \, z_{l}^{2}  \, ,
\label{eq:V-and-Delta-in-terms-of-zl}
\end{align}
where the coefficients $c_{j}$ ($j=1,\ldots,5$
) are different for each process. Using the above forms of the arguments one can perform the integration over $z_{l}$ in Eq.~\ref{eq:log-generating-piece-V1} to extract the rapidity divergent term and a remainder piece which is finite. 

In general, we can write the  amplitude for each processes as
\begin{align}
\mathcal{M}^{(V\beta)}_{\mu \alpha}&=2\pi (eq_{f}g)^{2} \delta(1-z^{v}_{\rm tot}) \, \int \mathrm{d} \Pi_{\perp}^{v} \, \overline{u}(\bm{k}) \, R^{(V\beta)}_{\mu \alpha} (\bm{l}_{1\perp})  \, \Big[ \Big( t^{b} \tilde{U} (\bm{x}_{\perp}) t^{a} \tilde{U}^{\dagger} (\bm{y}_{\perp}) \Big) U_{ba}(\bm{z}_{\perp}) -C_{F} \mathds{1} \Big] \, v(\bm{p})  \enskip , \, \beta=1,\ldots,6 \, ,
\label{eq:amplitude-generic-V1-V6}
\end{align}
where 
\begin{equation}
R^{(V\beta)}_{\mu \alpha} (\bm{l}_{1\perp})=R^{(V\beta)}_{(\rm I);\mu \alpha}  (\bm{l}_{1\perp})+R^{(V\beta)}_{(\rm II);\mu \alpha}  (\bm{l}_{1\perp}) \, .
\label{eq:R-V1-V6}
\end{equation}
The first piece is obtained from the contributions proportional to $1/z_{l}$. For the process $(V1)$, this contribution is the one given by Eq.~\ref{eq:R-V1-propto-zl-inverse}. Following the same logic as articulated for the self-energy computations, it can be written (after extracting the rapidity logarithms) as the sum of a divergent contribution and a remainder term: $R^{(V\beta)}_{(\rm I);\mu \alpha}=R^{(V\beta)}_{\mu \alpha;\rm div.}+\Re^{(V\beta)}_{\mu \alpha}$. The remainder term is comprised of terms in the amplitude which are proportional to $1/z_{l}$ but are devoid of logarithms in $z_{0}$. We can combine this piece with other finite terms in the amplitude that are not proportional to $1/z_{l}$. The contribution of the latter is denoted above by $R^{(V\beta)}_{(\rm II);\mu \alpha}$ and is computed in Appendix~\ref{sec:finite-pieces-Ver1} for the different processes.

With this decomposition, we can finally write the net amplitude from the processes in Fig.~\ref{fig:NLO-vertex-1} as
\begin{align}
\mathcal{M}^{\text{NLO};\text{Vert.}(1)}_{\mu \alpha}&=2\pi (eq_{f}g)^{2} \delta(1-z^{v}_{\rm tot}) \, \int \mathrm{d} \Pi_{\perp}^{v} \,   \overline{u}(\bm{k}) \,  \Big( R^{\rm Vert.(1)}_{\rm div.;\mu \alpha}(\bm{l}_{1\perp}) +R^{\rm Vert.(1)}_{\rm finite.;\mu \alpha}(\bm{l}_{1\perp})  \Big) \nonumber \\
& \times \Big[ \Big( t^{b} \tilde{U} (\bm{x}_{\perp}) t^{a} \tilde{U}^{\dagger} (\bm{y}_{\perp}) \Big) U_{ba}(\bm{z}_{\perp}) -C_{F} \mathds{1} \Big]  \, v(\bm{p}) \, ,
\end{align}
where
\begin{align}
  R^{\rm Vert.(1)}_{\rm div.;\mu \alpha}(\bm{l}_{1\perp}) & = \sum_{\beta=1}^{6}  R^{(V\beta)}_{\rm div.;\mu \alpha}  (\bm{l}_{1\perp}) \, , \quad
 R^{\rm Vert.(1)}_{\rm finite.;\mu \alpha}(\bm{l}_{1\perp})  = \sum_{\beta=1}^{6}  \big\{\Re^{(V\beta)}_{\mu \alpha}  (\bm{l}_{1\perp})+R^{(V\beta)}_{(\rm II);\mu \alpha}  (\bm{l}_{1\perp}) \big\}  \, .
 \label{eq:R-Ver1-div-finite} 
\end{align}
In particular, for $(V1)$ we have
\begin{equation}
R^{(V1)}_{\mu \alpha}  (\bm{l}_{1\perp})= \hyperref[eq:R-div-V1]{R^{(V1)}_{\rm div.;\mu \alpha} (\bm{l}_{1\perp})} + \big\{ \hyperref[eq:remainder-V1]{\Re^{(V1)}_{\mu \alpha}  (\bm{l}_{1\perp})}+\hyperref[eq:R-finite-V1]{R^{(V1)}_{(\rm II);\mu \alpha}  (\bm{l}_{1\perp})} \big\}  \, ,
\label{eq:R-V1}
\end{equation}
where the divergent term is 
\begin{align}
R^{(V1)}_{\rm div.;\mu \alpha}(\bm{l}_{1\perp}) &= \ln \Big(\frac{1}{z_{0}} \Big) \times \int_{\bm{l}_{2\perp}} \!\!\!\! e^{i\bm{l}_{2\perp}.\bm{r}_{zy}} \, \frac{1}{(q^{-})^{2}}  \, \gamma_{\alpha} \frac{\slashed{k}+\slashed{k}_{\gamma}}{2k.k_{\gamma}} \Bigg[ \big\{4z_{\bar{q}}q^{-} \delta_{i\mu} \gamma^{-}-2q^{-}\gamma_{\mu} \gamma^{i} \gamma^{-} \big\}  \big\{ \mathcal{I}_{v;\rm log}^{(3,ijj)} (\bm{c}_{1\perp}^{(V1)},c_{3}^{(V1)}) \nonumber \\
& + l_{2}^{j} \,  \mathcal{I}_{v;\rm log}^{(3,ij)} (\bm{c}_{1\perp}^{(V1)},c_{3}^{(V1)}) \big\} +\big\{ \gamma^{-} [\gamma^{+}(1-z_{\bar{q}})q^{-}-\bm{\gamma}_{\perp}.\bm{l}_{1\perp}]\gamma_{\mu}  [\gamma^{+}z_{\bar{q}}q^{-}+\bm{\gamma}_{\perp}.\bm{l}_{1\perp}]\gamma^{-} \big\}  \nonumber \\
& \times \big\{ \mathcal{I}_{v;\rm log}^{(3,ii)} (\bm{c}_{1\perp}^{(V1)},c_{3}^{(V1)}) + l_{2}^{i} \,  \mathcal{I}_{v;\rm log}^{(3,i)} (\bm{c}_{1\perp}^{(V1)},c_{3}^{(V1)}) \big\} \Bigg] \, .
\label{eq:R-div-V1}
\end{align}
The integrals $\mathcal{I}_{v;\rm log}$ appearing in the above equation are the components of the constituent integrals in Eq.~\ref{eq:constituent-integrals-V1} that are proportional to logarithms in $z_{0}$. These depend only on the coefficients $\bm{c}_{1\perp}$ and $c_{3}$ of the arguments $\bm{V}_{\perp}$ and $\Delta$ of the constituent integrals when they are decomposed in the form shown in Eq.~\ref{eq:V-and-Delta-in-terms-of-zl}. This is a very general feature of our computation and we will express the divergent pieces for the other processes in terms of these integrals. Their expressions can be obtained as
\begingroup
\allowdisplaybreaks
\begin{align}
\mathcal{I}_{v;\rm log}^{(3,ijj)} (\bm{c}_{1\perp},c_{3}) &= -\frac{1}{8 \pi^{2}} \int_{0}^{1} \mathrm{d} \alpha_{1} \int_{0}^{1-\alpha_{1}} \mathrm{d} \alpha_{2} \, \Big( \frac{2 \, c_{1}^{i}}{c_{3}}+c_{1}^{i} \, \frac{ \bm{c}_{1\perp}^{2}}{c_{3}^{2}} \Big)  \, , \nonumber \\
\mathcal{I}_{v;\rm log}^{(3,ij)} (\bm{c}_{1\perp},c_{3}) &= \frac{1}{8 \pi^{2}} \int_{0}^{1} \mathrm{d} \alpha_{1} \int_{0}^{1-\alpha_{1}} \mathrm{d} \alpha_{2} \, \Big( \frac{\delta^{ij}}{c_{3}}+\frac{c_{1}^{i} \, c_{1}^{j}}{c_{3}^{2}} \Big) \, , \nonumber \\
\mathcal{I}_{v;\rm log}^{(3,i)} (\bm{c}_{1\perp},c_{3}) &=-  \frac{1}{8 \pi^{2}} \int_{0}^{1} \mathrm{d} \alpha_{1} \int_{0}^{1-\alpha_{1}} \mathrm{d} \alpha_{2} \,  \frac{c_{1}^{i}}{c_{3}^{2}} \, ,
\label{eq:constituent-integrals-V1-log}
\end{align}
\endgroup
where repeated indices are summed over. For $(V1)$, the coefficients $c_{1}$ and $c_{3}$ appearing in these integrals are respectively
\begin{align}
\bm{c}_{1\perp}^{(V1)}& =\alpha_{1} \, \bm{l}_{2\perp}-\alpha_{2} \,  \bm{l}_{1\perp} \, , \nonumber \\
c_{3}^{(V1)}&= \alpha_{1}(1-\alpha_{1}) \, \bm{l}_{2\perp}^{2} +\alpha_{2}(1-\alpha_{2}) \, \bm{l}_{1\perp}^{2}+2\alpha_{1} \alpha_{2} \, \bm{l}_{1\perp}.\bm{l}_{2\perp}+\alpha_{2} \, Q^{2}z_{\bar{q}}(1-z_{\bar{q}}) \, .
\label{eq:c1-c3-V1}
\end{align}
Using the same techniques as described previously, it is straightforward to show that the UV divergent parts of the amplitude for the remaining processes $(V2),\ldots,(V6)$ are proportional to $\delta_{\mu -}$ and hence yield zero. In addition, there are no rapidity divergent contributions from the processes $(V2)$ and $(V5)$. Therefore we need to compute only the finite pieces $R^{(V\beta)}_{(\rm II);\mu \alpha}$ ($\beta=2,5$) for these diagrams. 
 
The  divergent term for $(V6)$ is similar to the one for $(V1)$. The computation of the divergent terms for $(V3)$ and $(V4)$ is especially tedious but can be carried out along the same lines as discussed here. These computations are cumbersome because we encounter more denominators whose evaluation, in turn, requires additional Feynman parameters in the transverse momentum integrations. However the arguments in these constituent integrals can again be written generically in terms of the gluon loop momentum fraction $z_{l}$, as in Eq.~\ref{eq:V-and-Delta-in-terms-of-zl}. We will provide the results for the divergent pieces of $(V4)$ in Appendix~\ref{sec:T-V1-div-parts}. The expressions for $(V3)$ can be obtained following similar methods. 

Finally, the finite contributions from these 6 graphs ($(V1)-(V6)$) can be written compactly as
\begin{align}
\mathcal{M}^{\text{Vert.}(1)}_{\mu \alpha;\rm finite}&=2\pi (eq_{f}g)^{2} \delta(1-z^{v}_{\rm tot}) \, \int \mathrm{d} \Pi_{\perp}^{v} \,  \overline{u}(\bm{k}) \,  \sum_{\beta=1}^{6}   \Big( \Re^{(V\beta)}_{\mu \alpha} (\bm{l}_{1\perp}) +R^{(V\beta)}_{(\rm II);\mu \alpha} (\bm{l}_{1\perp})  \Big) \nonumber \\
& \times  \Big[ \Big( t^{b} \tilde{U} (\bm{x}_{\perp}) t^{a} \tilde{U}^{\dagger} (\bm{y}_{\perp}) \Big) U_{ba}(\bm{z}_{\perp}) -C_{F} \mathds{1} \Big] \, v(\bm{p}) \, .
\label{eq:amplitude-finite-V1-generic}
\end{align}
The detailed computation of each of these terms is provided in Appendix~\ref{sec:finite-pieces-Ver1}.

\item  \textbf{Contributions to $T^{(2)}_{V}$:} These are the processes labeled $(V7)-(V12)$ which are obtained by interchanging the quark and antiquark lines in the diagrams shown in Fig.~\ref{fig:NLO-vertex-1}. As in Eq.~\ref{eq:amplitude-V1-generic} their net amplitude can be written as 
\begin{align}
\mathcal{M}^{\text{NLO};\text{Vert.}(2)}_{\mu \alpha}& = 2\pi \delta(1-z^{v}_{\text{tot}}) \, (eq_{f} g)^{2} \,  \int \mathrm{d} \Pi_{\perp}^{v} \, \overline{u} (\bm{k}) \Bigg\{ T^{(2)}_{V;\mu \alpha} (\bm{l}_{1\perp}) \, \Big[ \Big(  \tilde{U} (\bm{x}_{\perp}) t^{a} \tilde{U}^{\dagger} (\bm{y}_{\perp})t^{b} \Big) U_{ba}(\bm{z}_{\perp}) - C_{F} \mathds{1} \Big] \Bigg\} \, v(\bm{p}) \, ,
\label{eq:amplitude-V2-generic}
\end{align}
where 
\begin{align}
T^{(2)}_{V;\mu \alpha}(\bm{l}_{1\perp}) &=  \sum_{\beta=7}^{12} R^{(V\beta)}_{\mu \alpha}(\bm{l}_{1\perp})=\hyperref[eq:R-Ver2-div-finite]{R^{\rm Vert.(2)}_{\rm div.;\mu \alpha} (\bm{l}_{1\perp})} + \hyperref[eq:R-Ver2-div-finite]{R^{\rm Vert.(2)}_{\rm finite;\mu \alpha} (\bm{l}_{1\perp})} \, , 
\label{eq:T-V2}
\end{align}
is composed of divergent and finite contributions defined by
\begin{align}
 R^{\rm Vert.(2)}_{\rm div.;\mu \alpha}(\bm{l}_{1\perp})  = \sum_{\beta=7}^{12}  R^{(V\beta)}_{\rm div.;\mu \alpha}  (\bm{l}_{1\perp}) \, , \quad 
 R^{\rm Vert.(2)}_{\rm finite.;\mu \alpha}(\bm{l}_{1\perp})  = \sum_{\beta=7}^{12}  \big\{\Re^{(V\beta)}_{\mu \alpha}  (\bm{l}_{1\perp})+R^{(V\beta)}_{(\rm II);\mu \alpha}  (\bm{l}_{1\perp}) \big\}  \, .
\label{eq:R-Ver2-div-finite} 
\end{align}
In order to obtain $T^{(2)}_{V}$, we have to impose the replacements defined in Eq.~\ref{eq:replacements-qqbar-exchange} along with a change of sign in the various terms constituting $T^{(1)}_{V}$ in Eq.~\ref{eq:T-V1}.
\end{enumerate}

\subsection{Vertex graphs with free gluon propagator} \label{sec:vertex-corrections-free-gluon}

There are two broad categories of processes with six diagrams each that have the topology of vertex corrections with the gluon not crossing the shockwave. Three representative diagrams belonging to each of these classes are shown respectively in Figs.~\ref{fig:NLO-vertex-3} and \ref{fig:NLO-vertex-4}. The other half can be obtained by interchanging the quark and antiquark lines. The second class of processes labeled $(V19)-(V24)$ actually resemble final state interactions between the quark and antiquark but we will also include them in this section. In the following sections, we will demonstrate the computation of such amplitudes by considering one representative diagram from each class, with the full result given in the Appendices. 

\begin{enumerate}

\item \textbf{Contributions to $T^{(3)}_{V}$}: The diagrams contributing to $T^{(3)}_{V}$ in the general structure in Eq.~\ref{eq:vertex-correction-master-amplitude} of the amplitudes are shown in Fig.~\ref{fig:NLO-vertex-3}. 
\begin{figure}[!htbp]
\begin{minipage}[b]{0.8\textwidth}
\includegraphics[width=\textwidth]{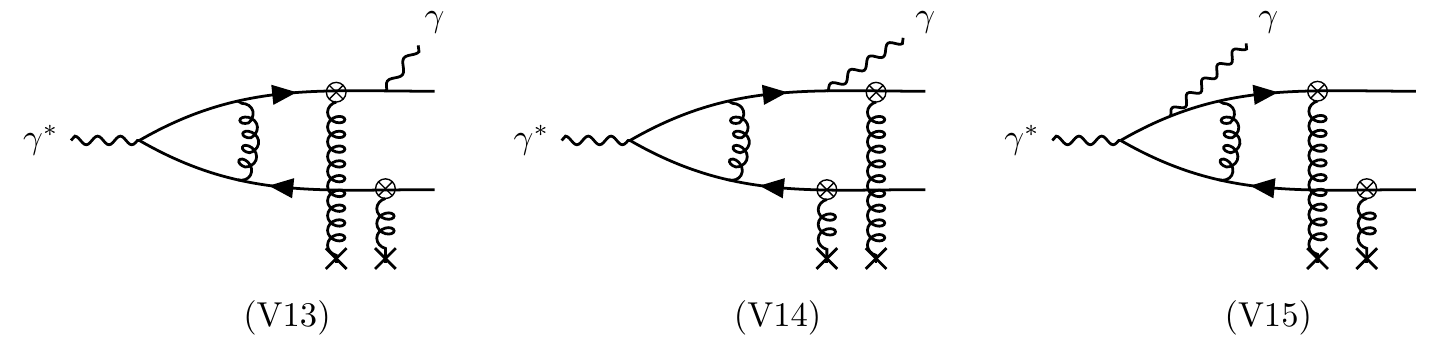}
\end{minipage}
\caption{Feynman diagrams that give $T^{(3)}_{V}$. Only half of the contributions are shown. The other half can be obtained by quark-antiquark interchange and are labeled (V16)-(V18) respectively. \label{fig:NLO-vertex-3}}
\end{figure}
The amplitude for these six processes can be written as
\begin{align}
\mathcal{M}^{\rm NLO;Vert.(3)}_{\mu \alpha}= 2\pi \delta(1-z_{\rm tot}^{v}) \, (eq_{f}g)^{2} \, \int \mathrm{d} \Pi_{\perp}^{\rm LO} \, \overline{u}(\bm{k}) \, \Bigg\{ T^{(3)}_{V;\mu \alpha} (\bm{l}_{1\perp}) \, \Big[ C_{F} \, \Big( \tilde{U}(\bm{x}_{\perp}) \tilde{U}^{\dagger} (\bm{y}_{\perp}) -\mathds{1} \Big) \Big] \Bigg\} \, v(\bm{p}) \, , 
\label{eq:amplitude-V3-generic}
\end{align}
where 
\begin{align}
T^{(3)}_{V;\mu \alpha} (\bm{l}_{1\perp}) &= \sum_{\beta=13}^{18} R^{(V\beta)}_{\mu \alpha}(\bm{l}_{1\perp}) = R^{\rm Vert.(3)}_{\rm div.;\mu \alpha} (\bm{l}_{1\perp}) +R^{\rm Vert.(3)}_{\rm finite;\mu \alpha} (\bm{l}_{1\perp}) \, .
\label{eq:T-V3}
\end{align}
is the sum of the perturbative contributions from each process which can be decomposed into a divergent and a finite piece. We show below the computation of the divergent pieces constituting 
\begin{align}
R^{\rm Vert.(3)}_{\rm div.;\mu \alpha} (\bm{l}_{1\perp})= \sum_{\beta=13}^{18} R^{(V\beta)}_{\rm div.;\mu \alpha} (\bm{l}_{1\perp}) \, ,
\end{align}
by considering the representative diagrams $(V13)$ and $(V15)$. The latter has a nested photon in the loop and hence has a topological similarity to the graph $(S24)$ in Fig.~\ref{fig:NLO-self-3}. We will see that there are indeed divergent pieces that arise from this graph which cancel the divergences appearing in the third line of Eq.~\ref{eq:R-S13-S24}.

The diagrams labeled $(V16)-(V18)$ are related to $(V13)-(V15)$ respectively by quark-antiquark interchange; we can therefore obtain $R^{(V\beta)}$  $(\beta=16,17,18)$ for these processes with the redefinitions given in Eq.~\ref{eq:replacements-qqbar-exchange}.

Fig.~\ref{fig:NLO-vertex-V13} shows the Feynman diagram for the process $(V13)$ with detailed momentum assignments.
\begin{figure}[!htbp]
\begin{center}
\includegraphics[scale=1]{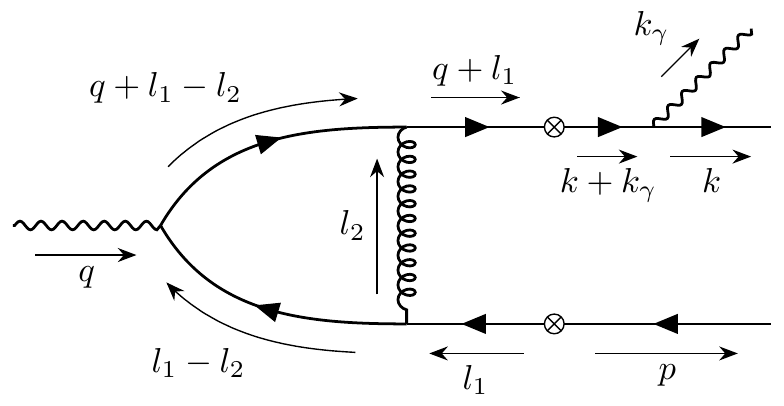}
\caption{The NLO process labeled $(V13)$ in Fig.~\ref{fig:NLO-vertex-3} with momenta and directions shown. Only the quark and antiquark scatter off the background classical field in this scenario; as previously, multiple scattering off this shock wave is represented by blobs with cross marks in the figure. \label{fig:NLO-vertex-V13}}
\end{center}
\end{figure}

The amplitude for this process is given by 
\begin{align}
\mathcal{M}^{(V13)}_{\mu \alpha}&= \int_{l_{1}} \overline{u} (\bm{k}) \,  (-ieq_{f}) \, \gamma_{\alpha} \, S_{0}(k+k_{\gamma}) \, \mathcal{T}_{q}(k+k_{\gamma},q+l_{1}) \, S_{0}(q+l_{1}) \, \Gamma_{\mu}(l_{1}^{+},l_{1}^{-}) \, S_{0}(l_{1}) \, \mathcal{T}_{q}(l_{1},-p) \, v(\bm{p}) \, .
\label{eq:amplitude-V13-master-expression}
\end{align}
The function $\Gamma_{\mu}(l_{1}^{+},l_{1}^{-})$ represents the gluon loop contribution to the $\gamma^{*} q\bar{q}$ vertex. Unlike the processes discussed in the previous sections, where the sign of the loop longitudinal momentum $l_{2}^{-}$ is determined by the location of the shock wave, there is nothing preventing $l_{2}^{-}$ from taking both positive and negative signs. The expressions for the loop contribution for these two cases are derived in Appendix~\ref{sec:vertex-correction-computation} and are given respectively by Eqs.~\ref{eq:vertex-correction-caseA} and \ref{eq:vertex-correction-caseB}.

Using the delta functions present in the expressions for the effective quark vertices in Eq.~\ref{eq:effective-vertex}, it is trivial to perform the $l_{1}^{-}$ integration. We get $l_{1}^{-}=-p^{-}$ and the overall momentum conserving delta function $\delta(q^{-}-p^{-}-k^{-}-k^{-}_{\gamma})$. In order to perform the $l_{1}^{+}$ integration, we need to separately analyze the two cases discussed above because the location of the $l_{1}^{+}$ poles are dependent on the sign of $l_{2}^{-}$. The contribution to the amplitude for the process will then be the sum of the individual results obtained for these two cases.  
\begin{itemize}
\item Case A: For $0<l_{2}^{-}<k^{-}+k^{-}_{\gamma}$ we have the following locations of the $l_{1}^{+}$ poles,
\begin{align}
l_{1}^{+} \vert_{a}&= -\frac{\bm{l}_{1\perp}^{2}}{2p^{-}}+\frac{i\varepsilon}{2p^{-}} \enskip (\rm above) \, , \nonumber \\
l_{1}^{+} \vert_{b}&= -q^{+}+\frac{\bm{l}_{2\perp}^{2}}{2(k^{-}+k_{\gamma}^{-}-l_{2}^{-})} +\frac{(\bm{l}_{2\perp} +\bm{l}_{1\perp})^{2}}{2l^{-}_{2}} -\frac{i\varepsilon(k^{-}+k^{-}_{\gamma})}{2l^{-}_{2} (k^{-}+k^{-}_{\gamma}-l_{2}^{-}) } \enskip (\rm below) \, , \nonumber \\
l_{1}^{+} \vert_{c}&= -q^{+}+\frac{\bm{l}_{1\perp}^{2}}{2(k^{-}+k^{-}_{\gamma})} -\frac{i\varepsilon}{2(k^{-}+k^{-}_{\gamma})} \enskip (\rm below) \, .
\end{align}
The second pole comes from one of the denominators in Eq.~\ref{eq:vertex-correction-caseA}. We will enclose the pole at $l_{1}^{+} \vert_{a}$ for convenience. 
\item Case B: For $0>l_{2}^{-}>-p^{-}$ we have the following locations of the $l_{1}^{+}$ poles,
\begin{align}
l_{1}^{+} \vert_{a}&= -\frac{\bm{l}_{1\perp}^{2}}{2p^{-}}+\frac{i\varepsilon}{2p^{-}} \enskip (\rm above) \, , \nonumber \\
l_{1}^{+} \vert_{b}&= -\frac{\bm{l}_{2\perp}^{2}}{2(l_{2}^{-}+p^{-})}+\frac{(\bm{l}_{2\perp}+\bm{l}_{1\perp})^{2}}{2l^{-}_{2}} -\frac{i\varepsilon p^{-}}{2l^{-}_{2}(l_{2}^{-}+p^{-})}    \enskip (\rm above) \, , \nonumber \\
l_{1}^{+} \vert_{c}&= -q^{+}+\frac{\bm{l}_{1\perp}^{2}}{2(k^{-}+k^{-}_{\gamma})} -\frac{i\varepsilon}{2(k^{-}+k^{-}_{\gamma})} \enskip (\rm below) \, .
\end{align}
In this case, we will enclose the pole at $l_{1}^{+} \vert_{c}$.
\end{itemize}

Using these results in Eq.~\ref{eq:amplitude-V13-master-expression} and finally subtracting the ``no scattering'' contribution we can write the amplitude for $(V13)$ as 
\begin{align}
\mathcal{M}_{\mu \alpha}^{(V13)}&= -ieq_{f} \, 2\pi \delta(q^{-}-P^{-}) \int \mathrm{d} \Pi_{\perp}^{\rm LO} \Big( \tilde{U} (\bm{x}_{\perp} ) \tilde{U}^{\dagger}(\bm{y}_{\perp}) -\mathds{1} \Big) \, \overline{u} (\bm{k}) \gamma_{\alpha} \frac{\slashed{k}+\slashed{k}_{\gamma}}{2k.k_{\gamma}} \gamma^{-} \frac{\gamma^{+}(k^{-}+k^{-}_{\gamma})-\bm{\gamma}_{\perp}.\bm{l}_{1\perp}}{2q^{-} \, \Big[\bm{l}_{1\perp}^{2}+\Delta^{\rm LO:(1)} \Big]} \nonumber \\
& \times \Big\{ \Gamma_{\mu}^{A}(l_{1}^{+} \vert_{a},-p^{-}) + \Gamma_{\mu}^{B}(l_{1}^{+} \vert_{c},-p^{-}) \Big\} \, [\gamma^{+}p^{-}+\bm{\gamma}_{\perp}.\bm{l}_{1\perp} ] \gamma^{-} v(\bm{p}) \, ,
\end{align} 
where 
\begin{align}
\Gamma_{\mu}^{A}(l_{1}^{+} \vert_{a},-p^{-})&= ieq_{f}g^{2} \, C_{F} \int \frac{\mathrm{d}z_{l}}{2\pi} \int_{\bm{l}_{2\perp}} \, \gamma^{\beta} \frac{\gamma^{+}(1-z_{\bar{q}}-z_{l})q^{-}+\gamma^{-} \, \frac{\bm{l}_{2\perp}^{2}}{2(1-z_{\bar{q}}-z_{l})q^{-}} +\bm{\gamma}_{\perp}.\bm{l}_{2\perp}   }{\Big(\bm{l}_{2\perp}+\bm{v}^{V}_{1\perp} \Big)^{2}+\Delta^{V}_{1} } \gamma_{\mu} \nonumber \\
 & \times \frac{\gamma^{+}(z_{\bar{q}}+z_{l})q^{-} -\gamma^{-} \Bigg( \frac{Q^{2}}{2q^{-}}+\frac{\bm{l}_{2\perp}^{2}}{2(1-z_{\bar{q}}-z_{l})q^{-}} \Bigg) -\bm{\gamma}_{\perp}.\bm{l}_{2\perp}              }{     2\frac{1-z_{\bar{q}}}{1-z_{\bar{q}}-z_{l}} \, \Big[  \bm{l}_{2\perp}^{2}+Q^{2}(z_{\bar{q}}+z_{l}) (1-z_{\bar{q}}-z_{l})-i\varepsilon \Big]}   \gamma^{\nu}  \Bigg( -g_{\beta \nu}+\frac{(l_{2}+l_{1})_{\beta}n_{\nu}+(l_{2}+l_{1})_{\nu}n_{\beta}}{z_{l} q^{-}} \Bigg|_{l_{2}^{+}\vert_{c}} \Bigg) \, , 
 \label{eq:gamma-A-V13}
\end{align}
and 
\begin{align}
\Gamma_{\mu}^{B}(l_{1}^{+} \vert_{c},-p^{-})&= ieq_{f}g^{2} \, C_{F} \int \frac{\mathrm{d}z_{l}}{2\pi} \int_{\bm{l}_{2\perp}} \, \gamma^{\beta} \frac{\gamma^{+}(1-z_{\bar{q}}-z_{l})q^{-}-\gamma^{-} \Bigg( \frac{Q^{2}}{2q^{-}}+\frac{\bm{l}_{2\perp}^{2}}{2(z_{\bar{q}}+z_{l})q^{-}} \Bigg) +\bm{\gamma}_{\perp}.\bm{l}_{2\perp}   }{\Big(\bm{l}_{2\perp}+\bm{v}^{V}_{2\perp} \Big)^{2}+\Delta^{V}_{2} } \gamma_{\mu} \nonumber \\
 & \times \frac{\gamma^{+}(z_{\bar{q}}+z_{l})q^{-} + \gamma^{-}  \, \frac{\bm{l}_{2\perp}^{2}}{2(z_{\bar{q}}+z_{l})q^{-}}  -\bm{\gamma}_{\perp}.\bm{l}_{2\perp}              }{2\frac{z_{\bar{q}}}{z_{\bar{q}}+z_{l}} \, \Big[ \bm{l}_{2\perp}^{2}+Q^{2}(z_{\bar{q}}+z_{l}) (1-z_{\bar{q}}-z_{l})-i\varepsilon \Big]} \gamma^{\nu}  \Bigg( -g_{\beta \nu}+\frac{(l_{2}+l_{1})_{\beta}n_{\nu}+(l_{2}+l_{1})_{\nu}n_{\beta}}{z_{l}q^{-}} \Bigg|_{l_{2}^{+}\vert_{b}} \Bigg) \, .
 \label{eq:gamma-B-V13}
\end{align}
In this expression, we defined $z_{l}=l_{2}^{-}/q^{-}$ and introduced  
\begin{align}
\bm{v}_{1\perp}^{V}& = \frac{1-z_{\bar{q}}-z_{l}}{1-z_{\bar{q}}} \, \bm{l}_{1\perp} \, , \quad \Delta_{1}^{V}= \frac{z_{l}}{1-z_{\bar{q}}} \Big(1-\frac{z_{l}}{1-z_{\bar{q}}} \Big) \, \frac{\bm{l}_{1\perp}^{2}+Q^{2}z_{\bar{q}}(1-z_{\bar{q}})}{z_{\bar{q}}}  \, , \nonumber \\
\bm{v}_{2\perp}^{V}&= \frac{z_{\bar{q}}+z_{l}}{z_{\bar{q}}} \, \bm{l}_{1\perp} \, , \quad \Delta^{V}_{2}= -\frac{z_{l}}{z_{\bar{q}}} \Big(1+\frac{z_{l}}{z_{\bar{q}}} \Big) \, \frac{\bm{l}_{1\perp}^{2}+Q^{2}z_{\bar{q}}(1-z_{\bar{q}})}{1-z_{\bar{q}}} \,  ,
\end{align}
and
\begin{align}
l_{2}^{+} \vert_{c}=- \frac{Q^{2}}{2q^{-}} -\frac{\bm{l}_{1\perp}^{2}}{2z_{\bar{q}}q^{-}} - \frac{\bm{l}_{2\perp}^{2}}{2(1-z_{\bar{q}}-z_{l})q^{-}} \,\,;\,\,
l_{2}^{+} \vert_{b}= \frac{Q^{2}}{2q^{-}} +\frac{\bm{l}_{1\perp}^{2}}{2(1-z_{\bar{q}})q^{-}} +\frac{\bm{l}_{2\perp}^{2}}{2(z_{\bar{q}}+z_{l})q^{-}} \,.
\end{align}
If we use the identity in Eq.~\ref{eq:gamma-identity-2} to expand the terms in the r.h.s of Eqs.~\ref{eq:gamma-A-V13} and \ref{eq:gamma-B-V13} we can infer that the singular logarithms in rapidity will  come from terms proportional to $1/z_{l}$ whereas the UV divergent pieces will come from terms proportional to $\bm{l}_{2\perp}^{2} l^{i}$, $l_{2}^{i} l_{2}^{j}$ and $\bm{l}_{2\perp}^{2}$. We regulate these divergences using dimensional regularization in $d=2-\epsilon$ dimensions. We will therefore require the following constituent integrals to compute the divergent pieces of the vertex correction terms in Eqs.~\ref{eq:gamma-A-V13} and \ref{eq:gamma-B-V13},
\begin{align}
\mu^{\epsilon} \mathcal{I}^{(2,ijj)}_{v} (\bm{V}_{\perp},\Delta ) & = \mu^{\epsilon} \int \, \frac{\mathrm{d}^{2-\epsilon} \bm{l}_{2\perp}}{(2\pi)^{2-\epsilon}} \frac{\bm{l}_{2\perp}^{2} l^{i}}{\Big[ \bm{l}_{2\perp}^{2}+\Delta_{1} \Big] \, \Big[ (\bm{l}_{2\perp}+\bm{v}_{\perp})^{2}+\Delta_{2} \Big] } \nonumber \\
&= -\frac{1}{\pi} \int_{0}^{1} \mathrm{d}\alpha \Bigg\{ \Bigg(\frac{1}{\epsilon} +\frac{1}{2} \ln \Big( \frac{\tilde{\mu}^{2}}{Q^{2}} \Big) \Bigg)+\frac{1}{2} \Bigg( \ln \Big( \frac{Q^{2}}{\Delta} \Big) +\frac{\bm{V}_{\perp}^{2}}{2\Delta} -\frac{1}{2} \Bigg) \Bigg\} \, V^{i} \, , \nonumber \\
\mu^{\epsilon} \mathcal{I}^{(2,ij)}_{v} (\bm{V}_{\perp},\Delta ) & = \mu^{\epsilon} \int \, \frac{\mathrm{d}^{2-\epsilon} \bm{l}_{2\perp}}{(2\pi)^{2-\epsilon}} \frac{l_{2}^{i}l_{2}^{j}}{\Big[ \bm{l}_{2\perp}^{2}+\Delta_{1} \Big] \, \Big[ (\bm{l}_{2\perp}+\bm{v}_{\perp})^{2}+\Delta_{2} \Big] } \nonumber \\
&= \frac{1}{4\pi} \Bigg\{  \Bigg(\frac{1}{\epsilon} +\frac{1}{2} \ln \Big( \frac{\tilde{\mu}^{2}}{Q^{2}} \Big) \Bigg) \delta^{ij} +\int_{0}^{1} \mathrm{d}\alpha \Bigg( \frac{1}{2} \ln \Big( \frac{Q^{2}}{\Delta} \Big) \delta^{ij} +\frac{V^{i}V^{j}}{\Delta} \Bigg) \Bigg\}  \, ,
\label{eq:constituent-integrals-V13-V14}
 \end{align}
where 
\begin{equation}
\bm{V}_{\perp}= \alpha \bm{v}_{\perp} \, , \quad \Delta= \alpha \bm{v}_{\perp}^{2}+\alpha \Delta_{2}+(1-\alpha) \Delta_{1} \, ,
\end{equation}
and $\alpha$ is a Feynman parameter. The forms of these arguments are different for the two cases. Employing all the above results, and a fair amount of algebra involving gamma matrices, we finally arrive at the following result for the amplitude $(V13)$:
\begin{align}
\mathcal{M}^{(V13)}_{\mu \alpha}&= 2\pi (egq_{f})^{2} \delta(1-z_{\rm tot}^{v}) \int \mathrm{d} \Pi_{\perp}^{\rm LO} \, \overline{u} (\bm{k}) \, R^{(V13)}_{\mu \alpha} (\bm{l}_{1\perp}) \,  \Big[ C_{F} \Big( \tilde{U} (\bm{x}_{\perp}) \tilde{U}^{\dagger} (\bm{y}_{\perp}) -\mathds{1} \Big) \Big] \, v(\bm{p}) \, ,
\label{eq:amplitude-expression-V13}
\end{align}
where
\begin{align}
R^{(V13)}_{\mu \alpha} (\bm{l}_{1\perp}) &= \frac{1}{4\pi^{2}} \Bigg\{ \ln \Big( \frac{1}{z_{0}^{2}} \Big)  \Bigg(\frac{1}{\epsilon} +\frac{1}{2} \ln \Big( \frac{\tilde{\mu}^{2}}{Q^{2}} \Big) \Bigg) - \frac{3}{2}  \Bigg(\frac{1}{\epsilon} +\frac{1}{2} \ln \Big( \frac{\tilde{\mu}^{2}}{Q^{2}} \Big) \Bigg)   \Bigg\} \, R^{\rm LO:(1)}_{\mu \alpha} (\bm{l}_{1\perp})  +R^{(V13)}_{\rm finite;\mu \alpha}  (\bm{l}_{1\perp})  \, .
\label{eq:R-V13}
\end{align}
In an identical manner, the amplitude for $(V14)$ is computed to be 
\begin{align}
R^{(V14)}_{\mu \alpha}  (\bm{l}_{1\perp})  &= \frac{1}{4\pi^{2}} \Bigg\{ \ln \Big( \frac{1}{z_{0}^{2}} \Big)  \Bigg(\frac{1}{\epsilon} +\frac{1}{2} \ln \Big( \frac{\tilde{\mu}^{2}}{Q^{2}} \Big) \Bigg) - \frac{3}{2}  \Bigg(\frac{1}{\epsilon} +\frac{1}{2} \ln \Big( \frac{\tilde{\mu}^{2}}{Q^{2}} \Big) \Bigg)   \Bigg\} \, R^{\rm LO:(2)}_{\mu \alpha} (\bm{l}_{1\perp})  +R^{(V14)}_{\rm finite;\mu \alpha}  (\bm{l}_{1\perp})  \, .
\label{eq:R-V14}
\end{align}
One can therefore rewrite the amplitudes for $(V13)$ and $(V14)$ in terms of the amplitude for the leading order processes (labeled $\rm LO:(1,2)$ in Fig.~\ref{fig:LO-diagrams}) as, 
\begin{align}
\begin{pmatrix}
\mathcal{M}^{(V13)}_{\mu \alpha} \\
\mathcal{M}^{(V14)}_{\mu \alpha} 
\end{pmatrix}   &= \Big( \frac{2\alpha_{S}C_{F}}{\pi } \Big) \Bigg\{ \frac{1}{2}   \ln \Big( \frac{1}{z_{0}^{2}} \Big)  \Bigg(\frac{1}{\epsilon} +\frac{1}{2} \ln \Big( \frac{\tilde{\mu}^{2}}{Q^{2}} \Big) \Bigg) -\frac{3}{4}  \Bigg(\frac{1}{\epsilon} +\frac{1}{2} \ln \Big( \frac{\tilde{\mu}^{2}}{Q^{2}} \Big) \Bigg) \Bigg\}     \, \begin{pmatrix}
\mathcal{M}^{\rm LO:(1)}_{\mu \alpha} \\
\mathcal{M}^{\rm LO:(2)}_{\mu \alpha}
\end{pmatrix} + \begin{pmatrix}
 \mathcal{M}_{\rm finite;\mu \alpha}^{(V13)} \\
 \mathcal{M}_{\rm finite;\mu \alpha}^{(V14)}
\end{pmatrix}   \, .
\label{eq:amplitude-final-V13-V14}
\end{align}
We will now outline the computation of the amplitude of the graph $(V15)$. This is considerably more tedious than the previous two diagrams because the presence of the real photon in the loop allows for two possible regimes for the integration over $l_{2}^{-}$ when $l_{2}^{-}>0$, namely $\Lambda^{-}_{0} <l_{2}^{-}<k^{-}$ and $k^{-}<l_{2}^{-} <k^{-}+k^{-}_{\gamma}$. To see this explicitly, we will start with the amplitude  for $(V15)$ (see Fig.~\ref{fig:NLO-vertex-V15} for the Feynman diagram with momentum assignments) given by
\begin{align}
\mathcal{M}_{\mu \alpha}^{(V15)}& = \int_{l_{1},l_{2}} \overline{u}(\bm{k}) \, \mathcal{T}_{q}(k,q+l_{1}+l_{2}-k_{\gamma}) \, S_{0} (q+l_{1}+l_{2}-k_{\gamma}) \, (igt^{a}) \, \gamma^{\beta} \, S_{0} (q+l_{1}-k_{\gamma}) \, (-ieq_{f}) \gamma_{\alpha} \, S_{0}(q+l_{1}) \nonumber \\
& \times  (-ieq_{f}) \gamma_{\mu} \, S_{0}(l_{1}) \, (igt^{b}) \gamma^{\nu} \, S_{0}(l_{1}+l_{2}) \, \mathcal{T}_{q} (l_{1}+l_{2},-p) \, v(\bm{p}) \times G_{\beta \nu;ab}^{0} (l_{2}) \, .
\end{align}
\begin{figure}[!htbp]
\begin{center}
\includegraphics[scale=1]{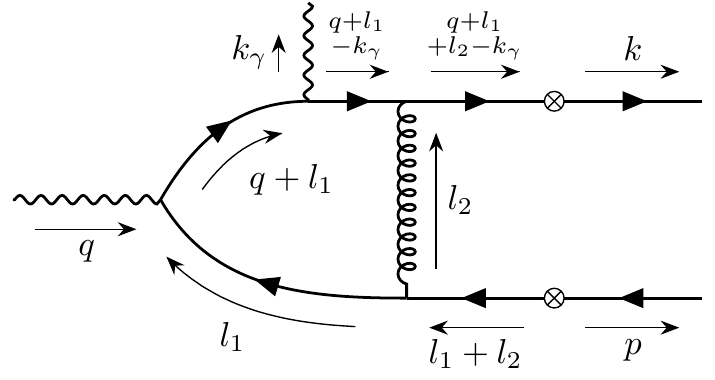}
\caption{The NLO process labeled $(V15)$ in Fig.~\ref{fig:NLO-vertex-3} with momenta and directions shown.  \label{fig:NLO-vertex-V15}}
\end{center}
\end{figure}
As usual, the integration over $l_{1}^{-}$ can be trivially performed using the Dirac delta functions contained in the effective vertices. The integration over $l_{1}^{+}$ and $l_{2}^{+}$ will be performed using the theorem of residues where the choice of contours is decided by the sign and magnitude of $l_{2}^{-}$. Expanding the denominators of the propagators we can see that there are three $l_{2}^{+}$ poles located at 
\begin{align}
l_{2}^{+} \vert_{a}&= -q^{+}+k^{+}_{\gamma}-l_{1}^{+}+\frac{(\bm{l}_{1\perp}+\bm{l}_{2\perp}-\bm{k}_{\gamma \perp})^{2}}{2k^{-}} - \frac{i\varepsilon}{2k^{-}} \, ; \quad l_{2}^{+} \vert_{b}= -l_{1}^{+}- \frac{(\bm{l}_{1\perp}+\bm{l}_{2\perp})^{2}}{2p^{-}}+ \frac{i\varepsilon}{2p^{-}} \, ; \quad l_{2}^{+} \vert_{c}= \frac{\bm{l}_{2\perp}^{2}}{2l^{-}_{2}} -\frac{i\varepsilon}{2l^{-}_{2}} \, . 
\end{align}
Clearly the locations of $l_{2}^{+} \vert_{a}$ and $l_{2}^{+} \vert_{b}$ are well defined by the signs of the external longitudinal momenta but the pole at $l_{2}^{+} \vert_{c}$ can be below or above the real $l_{2}^{+}$ axis respectively depending on $l_{2}^{-}$ being positive or negative. We therefore have two cases similar to the computations of the graphs $(V13)$ and $(V14)$ in Fig.~\ref{fig:NLO-vertex-3}. However for the case $l_{2}>0$ we can have two separate contributions depending on the magnitude of $l_{2}^{-}$ relative to $k^{-}$. 

For $l_{2}^{-}>0$, if we enclose the pole at $l_{2}^{+} \vert_{b}$ in the contour integration over $l_{2}^{+}$ we have four $l_{1}^{+}$ poles located at 
\begin{align}
l_{1}^{+} \vert_{a} &= -q^{+}+k^{+}_{\gamma} + \frac{(\bm{l}_{1\perp}-\bm{k}_{\gamma \perp})^{2}}{2 \, (k^{-}-l^{-}_{2}) } -\frac{i \varepsilon}{2 \, (k^{-}-l^{-}_{2}) } \, ; \quad l_{1}^{+} \vert_{b}= -q^{+}+\frac{\bm{l}_{1\perp}^{2}}{2 \, (k^{-}+k^{-}_{\gamma}-l^{-}_{2}) } - \frac{i\varepsilon}{2 \,  (k^{-}+k^{-}_{\gamma}-l^{-}_{2}) } \, , \nonumber \\
l_{1}^{+} \vert_{c} &= - \frac{\bm{l}_{1\perp}^{2}}{2 \, (l_{2}^{-}+p^{-})} + \frac{i \varepsilon}{2 (l_{2}^{-}+p^{-}) } \, ; \quad l_{1}^{+} \vert_{d} = - \frac{(\bm{l}_{1\perp}+\bm{l}_{2\perp})^{2}}{2 p^{-}} - \frac{\bm{l}_{2\perp}^{2}}{2l^{-}_{2}} +\frac{i\varepsilon}{2l^{-}_{2}} + \frac{i\varepsilon}{2p^{-}} \, .
\end{align}
For $0<l_{2}^{-}<k^{-}$ we have the poles at $l_{1}^{+} \vert_{a}$ and $l_{1}^{+} \vert_{b}$ located below the real $l_{1}^{+}$ axis whereas the remaining two are located above. We will deform the contour clockwise in our computation. For $k^{-}<l_{2}^{-}<k^{-}+k^{-}_{\gamma}$ we have the poles at $l_{1}^{+} \vert_{a,c,d} $ located above the real axis and that at $l_{1}^{+} \vert_{b} $ located below which we will enclose through a clockwise deformation of the contour. The total contribution to the amplitude for the case $l_{2}^{-}>0$ is therefore obtained by summing these individual contributions. The case $l_{2}^{-}<0$ is simpler because we have a single pole located above the real $l_{1}^{+}$ axis whose contribution obtained through an anticlockwise deformation can be computed using Cauchy's residue theorem. 

The UV divergent terms are obtained from pieces proportional to the constituent integrals $\mathcal{I}_{v}^{(3,iijj)}$ and $\mathcal{I}_{v}^{(3,ijkk)}$. There are no rapidity divergent pieces from this graph. We can finally write the amplitude for $(V15)$ as
\begin{align}
\mathcal{M}^{(V15)}_{\mu \alpha}&= 2\pi (egq_{f})^{2} \delta(1-z_{\rm tot}^{v}) \int \mathrm{d} \Pi_{\perp}^{\rm LO} \, \overline{u} (\bm{k}) \, R^{(V15)}_{\mu \alpha} (\bm{l}_{1\perp}) \,  \Big[ C_{F} \Big( \tilde{U} (\bm{x}_{\perp}) \tilde{U}^{\dagger} (\bm{y}_{\perp}) -\mathds{1} \Big) \Big] \, v(\bm{p}) \, ,
\label{eq:amplitude-expression-V15}
\end{align}
where 
\begin{align}
R^{(V15)}_{\mu \alpha} (\bm{l}_{1\perp})&=- \frac{1}{4\pi^{2}} \,  \Bigg( \frac{1}{\epsilon}+\frac{1}{2} \ln \Big( \frac{ \tilde{\mu}^{2}   }{Q^{2}} \Big) \Bigg) \, \frac{z_{q}^{2} \, z_{\bar{q}}}{2 \, (1-z_{\bar{q}})^{2} \, (1-z_{\gamma}) \, q^{-}}  \, \frac{ \gamma_{\alpha}\gamma_{\mu} \gamma^{-} }{  \Big[(\bm{l}_{1\perp}+\bm{v}_{\perp}^{\rm LO:(2)}  )^{2}+\Delta^{\rm LO:(2)}  \Big]  } +R^{(V15)}_{\rm finite; \mu \alpha} (\bm{l}_{1\perp}) \, ,
\label{eq:R-V15}
\end{align}
 contains the divergent piece that exactly cancels the residual divergence in the sum of the contributions from $(S23)$ and $(S24)$ in Fig.~\ref{fig:NLO-self-3}.

The finite contributions to the amplitudes for these processes are part of 
\begin{align}
\mathcal{M}^{\rm Vert.(3)}_{\rm finite;\mu \alpha}& = 2\pi \, \delta(1-z_{\rm tot}^{v})  (eq_{f}g)^{2}  \int \mathrm{d} \Pi_{\perp}^{\rm LO}  \, \overline{u}(\bm{k}) \,  \sum_{\beta=13}^{18} R^{(V\beta)}_{\rm finite;\mu \alpha} (\bm{l}_{1\perp})   \, \Big[ C_{F} \Big( \tilde{U}(\bm{x}_{\perp}) \tilde{U}^{\dagger}(\bm{y}_{\perp}) -\mathds{1} \Big) \Big] \, v(\bm{p}) \, ,
\label{eq:amplitude-finite-V3-generic}
\end{align}
where the expressions for $R^{(V\beta)}_{\rm finite}$ $(\beta=13,14,15)$ are provided in Appendix~\ref{sec:finite-pieces-Ver3}. As in the previous cases considered, Eq.~\ref{eq:replacements-qqbar-exchange} allows us to obtain the corresponding expressions for the finite contributions of the $q\leftrightarrow\bar{q}$ interchanged processes $(V16)-(V18)$.

\item \textbf{Contributions to $T^{(4)}_{V}$:} The final set of virtual gluon exchange diagrams are those with final state interactions between the outgoing quark and antiquark. These are shown in Fig.~\ref{fig:NLO-vertex-4}.
\begin{figure}[!htbp]
\begin{minipage}[b]{0.8\textwidth}
\includegraphics[width=\textwidth]{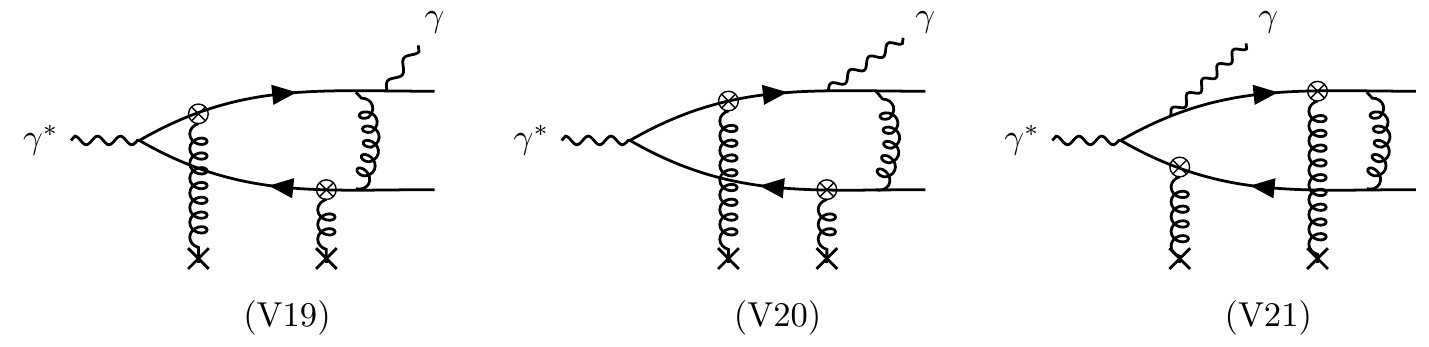}
\end{minipage}
\caption{Contributions that constitute $T^{(4)}_{V}$. The other three diagrams labeled (V22)-(V24) are obtained by swapping the quark and antiquark lines. \label{fig:NLO-vertex-4}}
\end{figure}

The amplitude for these processes is given by  
\begin{align}
\mathcal{M}^{\rm NLO;Vert.(4)}_{\mu \alpha}= 2\pi \delta(1-z_{\rm tot}^{v}) \, (eq_{f}g)^{2} \, \int \mathrm{d} \Pi_{\perp}^{\rm LO} \, \overline{u}(\bm{k}) \, \Bigg\{ T^{(4)}_{V;\mu \alpha} (\bm{l}_{1\perp}) \, \Big[  t^{a} \tilde{U}(\bm{x}_{\perp}) \tilde{U}^{\dagger} (\bm{y}_{\perp})t_{a} -C_{F} \, \mathds{1}  \Big] \Bigg\} \, v(\bm{p}) \, , 
\label{eq:amplitude-V4-generic}
\end{align}
where 
\begin{align}
T^{(4)}_{V;\mu \alpha} (\bm{l}_{1\perp}) &= \sum_{\beta=19}^{24} R^{(V\beta)}_{\mu \alpha}(\bm{l}_{1\perp}) = R^{\rm Vert.(4)}_{\rm div.;\mu \alpha} (\bm{l}_{1\perp}) +R^{\rm Vert.(4)}_{\rm finite;\mu \alpha} (\bm{l}_{1\perp}) \, .
\label{eq:T-V4}
\end{align}
We have again expressed the perturbative contributions from each process as the sum of a divergent and a finite part. We will see that there are only rapidity singularities associated with these processes. We will detail below the computation for one such representative diagram $(V19)$. The amplitude contributions for the other two diagrams in Fig.~\ref{fig:NLO-vertex-4} can be obtained following similar methods. In order to obtain the expressions for the $q\leftrightarrow \bar{q}$ interchanged counterparts of diagrams $(V19)-(V21)$ we will use Eq.~\ref{eq:replacements-qqbar-exchange} in $R^{(V19)}-R^{(V21)}$.

Fig.~\ref{fig:NLO-vertex-V19} shows the Feynman graph for $(V19)$ with the momentum assignments and directions.
\begin{figure}[!htbp]
\begin{center}
\includegraphics[scale=1]{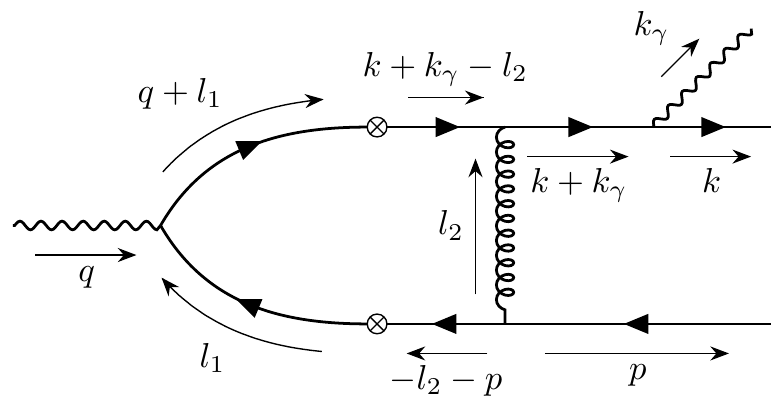}
\caption{The NLO process labeled $(V19)$ in Fig.~\ref{fig:NLO-vertex-4} with momenta and directions shown. This is representative of final state interactions between the quark and antiquark after they scatter off the shock wave.  \label{fig:NLO-vertex-V19}}
\end{center}
\end{figure}
The amplitude for this process is given by 
\begin{align}
\mathcal{M}^{(V19)}_{\mu \alpha}&= (eq_{f}g)^{2} \int_{l_{1},l_{2}} \overline{u}(\bm{k}) \, \gamma_{\alpha} \, S_{0}(k+k_{\gamma}) t^{a} \gamma^{\beta} \, S_{0}(k+k_{\gamma}-l_{2}) \, \mathcal{T}_{q} (k+k_{\gamma}-l_{2},q+l_{1}) \, S_{0}(q+l_{1}) \nonumber \\
& \times \gamma_{\mu} \, S_{0}(l_{1}) \, \mathcal{T}_{q}(l_{1},-l_{2}-p) \, S_{0}(-l_{2}-p) \, t^{b} \gamma^{\nu} \, v(\bm{p}) \, G^{0}_{\beta \nu;ab}(l_{2}) \, ,
\end{align}
where the free quark and gluon propagators are given by Eqs.~\ref{eq:fermion-gluon-propagator} and the effective vertex for the dressed quark propagator is given by Eq.~\ref{eq:effective-vertex}.

The integration over $l_{1}^{-}$ can be trivially performed using the delta functions contained in the effective vertex factors resulting in an overall longitudinal momentum conserving delta function $\delta(q^{-}-p^{-}-k^{-}-k^{-}_{\gamma})$. We next perform the contour integration over $l_{1}^{+}$ by enclosing the single pole situated below the real $l_{1}^{+}$ axis. This results in the following expression for the amplitude:
\begin{align}
\mathcal{M}^{(V19)}_{\mu \alpha}= 2\pi \, (eq_{f}g)^{2} \, \delta(q^{-}-p^{-}-k^{-}-k^{-}_{\gamma}) \int \mathrm{d}\Pi_{\perp}^{\rm LO} \, \Big[  t^{a} \tilde{U}(\bm{x}_{\perp}) \tilde{U}^{\dagger} (\bm{y}_{\perp})t_{a} -C_{F} \, \mathds{1}  \Big] \, \int_{\bm{l}_{2\perp}} \!\!\!\!e^{i\bm{l}_{2\perp}.\bm{r}_{xy}} \int_{l_{2}^{-},l_{2}^{+}} i \, \frac{N^{(V19)}}{D^{(V19)}} \, ,
\label{eq:amplitude-V19-generic}
\end{align} 
where the numerator and denominator are respectively given by
\begin{align}
N^{(V19)}&= \overline{u}(\bm{k}) \, \gamma_{\alpha} \, \frac{\slashed{k}+\slashed{k}_{\gamma}}{2k.k_{\gamma}} \, \gamma^{\beta} \, [\gamma^{+}(k^{-}+k^{-}_{\gamma}-l_{2}^{-})-\bm{\gamma}_{\perp}.(\bm{k}_{\perp}+\bm{k}_{\gamma \perp}-\bm{l}_{2\perp}) ] \, \gamma^{-} \, [\gamma^{+}(k^{-}+k^{-}_{\gamma}-l_{2}^{-})-\bm{\gamma}_{\perp}.\bm{l}_{1\perp} ] \nonumber \\
& \times \gamma_{\mu} \, [\gamma^{+}(l_{2}^{-}+p^{-}) +\bm{\gamma}_{\perp}.\bm{l}_{1\perp}] \, \gamma^{-} \, [\gamma^{+}(l_{2}^{-}+p^{-})-\bm{\gamma}_{\perp}.(\bm{l}_{2\perp}+\bm{p}_{\perp})] \, \gamma^{\nu} \, v(\bm{p}) \, \Big( -g_{\beta \nu}+\frac{l_{2\beta}n_{\nu}+l_{2\nu}n_{\beta}}{l_{2}^{-}} \Big) \, ,
\end{align}
and 
\begin{align}
D^{(V19)}&= 16 \, l_{2}^{-} (l_{2}^{-}+p^{-}) (k^{-}+k^{-}_{\gamma}-l_{2}^{-}) q^{-} \, [\bm{l}_{1\perp}^{2}+Q^{2}(z_{\bar{q}}+z_{l}) (1-z_{\bar{q}}-z_{l}) -i\varepsilon ] \, \Big( l_{2}^{+}-\frac{\bm{l}_{2\perp}^{2}}{2l_{2}^{-}} +\frac{i\varepsilon}{2l_{2}^{-}} \Big) \nonumber \\
& \times \Big( l_{2}^{+}+p^{+}-\frac{(\bm{l}_{2\perp}+\bm{p}_{\perp})^{2}}{2(l_{2}^{-}+p^{-})} +\frac{i\varepsilon}{2(l_{2}^{-}+p^{-})} \Big) \, \Big( l_{2}^{+}-k^{+}-k^{+}_{\gamma}+\frac{(\bm{k}_{\perp}+\bm{k}_{\gamma \perp}-\bm{l}_{2\perp})^{2}}{2(k^{-}+k^{-}_{\gamma}-l_{2}^{-})} -\frac{i\varepsilon}{2(k^{-}+k^{-}_{\gamma}-l_{2}^{-})} \Big) \, .
\end{align}
If we use the identity in Eq.~\ref{eq:gamma-identity-2} for the numerator, we see that it is at most proportional to $l_{2}^{+}$ and hence the contour integration is well defined. There are three $l_{2}^{+}$ poles located at 
\begin{align}
l_{2}^{+} \vert_{a}&=\frac{\bm{l}_{2\perp}^{2}}{2l_{2}^{-}}-\frac{i\varepsilon}{2l_{2}^{-}} \, ; \quad
 l_{2}^{+} \vert_{b}= -p^{+}+\frac{(\bm{l}_{2\perp}+\bm{p}_{\perp})^{2}}{2(l_{2}^{-}+p^{-})} -\frac{i\varepsilon}{2(l_{2}^{-}+p^{-})}  \, , \nonumber \\
l_{2}^{+} \vert_{c}&=k^{+}+k^{+}_{\gamma}-\frac{(\bm{k}_{\perp}+\bm{k}_{\gamma \perp}-\bm{l}_{2\perp})^{2}}{2(k^{-}+k^{-}_{\gamma}-l_{2}^{-})} +\frac{i\varepsilon}{2(k^{-}+k^{-}_{\gamma}-l_{2}^{-})} \, .
\end{align}
The location of the first pole clearly depends on the sign of $l_{2}^{-}$. We therefore have the following cases:
\begin{itemize}
\item Case A: For $0<l_{2}^{-} <k^{-}+k^{-}_{\gamma}$ we have $l_{2}^{+}\vert_{a}$ and $l_{2}^{+}\vert_{b}$ located below the real $l_{2}^{+}$ axis whereas $l_{2}^{+}\vert_{c}$ is above. We will deform the contour anticlockwise to enclose this pole.
\item Case B: For $0>l_{2}^{-}> -p^{-}$ we will deform the contour clockwise to enclose the pole at $l_{2}^{+}\vert_{b}$.
\end{itemize}

We are only interested in extracting the divergent pieces for these two cases. Redefining $\bm{l}_{1\perp}+\bm{l}_{2\perp} \rightarrow \bm{l}_{1\perp}$ we can get rid of the transverse phase containing $\bm{l}_{2\perp}$ in Eq.~\ref{eq:amplitude-V19-generic}. The denominator is now proportional to $\bm{l}_{2\perp}^{6}$ in the limit of large $l_{2\perp}$. Further, using the fact that terms proportional to $\delta_{\mu -}$ contributes zero to the cross-section, we can easily see that $(V19)$ is UV finite. 

There are however rapidity singularities and to extract them we need to isolate the terms proportional to $1/z_{l}$. Following the conventions used in Sec.~\ref{sec:virtual-corrections-vertex} we can write the amplitude for $(V19)$ as 
\begin{align}
\mathcal{M}_{\mu \alpha}^{(V19)}& = 2\pi \, (eq_{f}g)^{2} \, \delta(1-z_{\rm tot}^{v}) \int \mathrm{d}\Pi_{\perp}^{\rm LO} \, \overline{u}(\bm{k}) \, R^{(V19)}_{\mu \alpha} (\bm{l}_{1\perp}) \,  \Big[  t^{a} \tilde{U}(\bm{x}_{\perp}) \tilde{U}^{\dagger} (\bm{y}_{\perp})t_{a} -C_{F} \, \mathds{1}  \Big] \, v(\bm{p})  \, ,
\end{align}
where 
\begin{align}
R^{(V19)}_{\mu \alpha}(\bm{l}_{1\perp}) &= \Big\{ R^{(V19);A}_{(\rm I);\mu \alpha} (\bm{l}_{1\perp})+  R^{(V19);B}_{(\rm I);\mu \alpha} (\bm{l}_{1\perp}) \Big\}+\Big\{ R^{(V19);A}_{(\rm II);\mu \alpha} (\bm{l}_{1\perp})+  R^{(V19);B}_{(\rm II);\mu \alpha} (\bm{l}_{1\perp}) \Big\}   \nonumber \\
&= \hyperref[eq:R-div-V19]{R^{(V19)}_{\rm div.;\mu \alpha} (\bm{l}_{1\perp}) }+ R^{(V19)}_{\rm finite;\mu \alpha} (\bm{l}_{1\perp}) \, .
\label{eq:R-V19}
\end{align}
We have here terms proportional to $1/z_{l}$ for the two cases A and B denoted respectively by $R^{(V19);A}_{(\rm I)} $ and $R^{(V19);B}_{(\rm I)}$. Contributions from terms not proportional to $1/z_{l}$ and hence yielding finite pieces are denoted respectively for the two cases by $R^{(V19);A}_{(\rm II)} $ and $R^{(V19);B}_{(\rm II)} $. From the first set of terms, we will get a logarithmically divergent contribution leaving behind a finite remainder. Adding these remainders with the other finite contributions $R^{(V19);A,B}_{(\rm II)}$ results in the net finite contribution from this process:
\begin{align}
R^{(V19)}_{\rm finite;\mu \alpha} (\bm{l}_{1\perp})&= \Big\{ \Re^{(V19);A}_{\mu \alpha} (\bm{l}_{1\perp})  + R^{(V19);A}_{(\rm II);\mu \alpha} (\bm{l}_{1\perp}) \Big\}  +A \rightarrow  B \, ,
\label{eq:R-finite-V19}
\end{align}
where the remainders are given by
\begin{align}
\Re^{(V19);A,B}_{\mu \alpha} (\bm{l}_{1\perp}) =R^{(V19);A,B}_{(\rm I);\mu \alpha} (\bm{l}_{1\perp}) -  R^{(V19);A,B}_{\rm div.;\mu \alpha} (\bm{l}_{1\perp}) \, .
\label{eq:remainder-V19}
\end{align}
With these definitions in mind, we will now write down the expressions for the contributions proportional to $1/z_{l}$ for the two cases considered above.

\begin{itemize}
\item Case A: $0<l_{2}^{-} <k^{-}+k^{-}_{\gamma}$
\begin{align}
R^{(V19);A}_{(\rm I);\mu \alpha} (\bm{l}_{1\perp}) &= \frac{1}{2 \, (q^{-})^{2}} \!\!  \int \!\! \frac{\mathrm{d}z_{l}}{(2\pi) \, z_{l}}  \,  \gamma_{\alpha} \, \frac{\slashed{k}+\slashed{k}_{\gamma}}{2k.k_{\gamma}}  \Big[\mathcal{R}_{(1);\mu}^{i}  \, \mathcal{I}_{v}^{(3,ijj)}( \bm{V}_{\perp}^{(V19);A},\Delta^{(V19);A} )  + \mathcal{R}_{(2);\mu } \, \mathcal{I}_{v}^{(3,ii)} ( \bm{V}_{\perp}^{(V19);A},\Delta^{(V19);A} ) \nonumber \\
& -\mathcal{R}_{(3)}^{i} \, \mathcal{R}_{(1);\mu }^{j} \, \mathcal{I}_{v}^{(3,ij)} ( \bm{V}_{\perp}^{(V19);A},\Delta^{(V19);A} ) - (\mathcal{R}_{(3)}^{i} \, \mathcal{R}_{(2);\mu }+\mathcal{R}_{(4)} \, \mathcal{R}_{(1);\mu}^{i} ) \, \mathcal{I}_{v}^{(3,i)} ( \bm{V}_{\perp}^{(V19);A},\Delta^{(V19);A} )  \nonumber  \\
& - \mathcal{R}_{(4)} \, \mathcal{R}_{(2);\mu} \, \mathcal{I}_{v}^{(3,0)} ( \bm{V}_{\perp}^{(V19);A},\Delta^{(V19);A} ) \Big]  \, , 
\label{eq:R-V19-propto-zl-inverse-A}
\end{align}
where the coefficients multiplying the constituent integrals are given by 
\begin{align}
\mathcal{R}^{i}_{(1);\mu} & = 4 \, z_{\bar{q}}q^{-} \, \delta^{i}_{\mu} \gamma^{-} -2q^{-} \gamma^{-} \gamma_{\mu} \gamma^{i} \, , \quad
\mathcal{R}_{(2);\mu} = \gamma^{-} [\gamma^{+}(1-z_{\bar{q}})q^{-}-\bm{\gamma}_{\perp}.\bm{l}_{1\perp} ] \gamma_{\mu} [\gamma^{+}z_{\bar{q}}q^{-}+\bm{\gamma}_{\perp}.\bm{l}_{1\perp}] \gamma^{-} \, , \nonumber \\
\mathcal{R}_{(3)}^{i}& = 2z_{\bar{q}} \, (k^{i}+k^{i}_{\gamma}) -2(1-z_{\bar{q}}) \, p^{i}+(\slashed{k}+\slashed{k}_{\gamma}) \, \gamma^{i} z_{\bar{q}}  \, , \quad \mathcal{R}_{(4)}= 2z_{\bar{q}} \, (2k.k_{\gamma}) \, .
\label{eq:coefficients-V19-A}
\end{align}
The expressions for the constituent integrals are given in Eq.~\ref{eq:constituent-integrals-V1}. We can always express the arguments $\bm{V}_{\perp}$ and $\Delta$ of these integrals in terms of the gluon loop momentum fraction $z_{l}$, as in Eq.~\ref{eq:V-and-Delta-in-terms-of-zl}. For the purposes of the discussion in this section where we extract the logarithmic singularity from Eq.~\ref{eq:R-V19-propto-zl-inverse-A} we will only require the expressions for the coefficients $\bm{c}_{1\perp}^{(V19);A}$ and $c_{3}^{(V19);A}$. 

We can now finally write the rapidity divergent piece for Case A as 
\begin{align}
R^{(V19);A}_{\rm div.;\mu \alpha} (\bm{l}_{1\perp}) &= \ln \Big( \frac{1}{z_{0}} \Big)  \, \frac{\slashed{k}+\slashed{k}_{\gamma}}{2k.k_{\gamma}} \Bigg[  \mathcal{R}_{(1);\mu}^{i} \,  \mathcal{I}_{v;\rm log}^{(3,ijj)}\big(\bm{c}_{1\perp}^{(V19);A},\Delta^{(V19);A} \big)  + \mathcal{R}_{(2);\mu } \,  \mathcal{I}_{v;\rm log}^{(3,ii)}\big(\bm{c}_{1\perp}^{(V19);A},\Delta^{(V19);A} \big)  \nonumber \\
& -\mathcal{R}_{(3)}^{i} \, \mathcal{R}_{(1);\mu }^{j} \,   \mathcal{I}_{v;\rm log}^{(3,ij)}\big(\bm{c}_{1\perp}^{(V19);A},\Delta^{(V19);A} \big) - (\mathcal{R}_{(3)}^{i} \, \mathcal{R}_{(2);\mu }+\mathcal{R}_{(4)} \, \mathcal{R}_{(1);\mu}^{i} ) \,   \mathcal{I}_{v;\rm log}^{(3,i)}\big(\bm{c}_{1\perp}^{(V19);A},\Delta^{(V19);A} \big)  \nonumber \\
& - \mathcal{R}_{(4)} \, \mathcal{R}_{(2);\mu} \,  \mathcal{I}_{v;\rm log}^{(3,0)}\big(\bm{c}_{1\perp}^{(V19);A},\Delta^{(V19);A} \big)  \Bigg]  \, , 
\label{eq:R-div-V19-A}
\end{align}
where all but one of the integrals\footnote{The only missing expression is for the integral $\mathcal{I}_{v;\rm log}^{(3,0)}(\bm{c}_{1\perp},c_{3})=\frac{1}{8\pi^{2}} \int_{0}^{1} \mathrm{d} \alpha_{1} \int_{0}^{1-\alpha_{1}}  \mathrm{d} \alpha_{2} \, \big( 1/c_{3}^{2} \big) \, . $ } appearing above are given in Eqs.~\ref{eq:constituent-integrals-V1-log}. The arguments appearing in these can be obtained for case A as 
\begin{align}
\bm{c}_{1\perp}^{(V19);A}&= -\alpha_{1} \, \bm{l}_{1\perp} +\alpha_{2} \, \{ (1-z_{\bar{q}}) \, \bm{p}_{\perp} -z_{\bar{q}} \, (\bm{k}_{\perp}+\bm{k}_{\gamma \perp} ) \} \, , \nonumber \\
c_{3}^{(V19);A}& = \alpha_{1} (1-\alpha_{1}) \, \bm{l}_{1\perp}^{2} +\alpha_{2} (1-\alpha_{2}) \, \{ (1-z_{\bar{q}}) \, \bm{p}_{\perp} -z_{\bar{q}} \, (\bm{k}_{\perp}+\bm{k}_{\gamma \perp} ) \}^{2}  \nonumber \\
& +2\alpha_{1} \alpha_{2} \, \bm{l}_{1\perp}. \{ (1-z_{\bar{q}}) \, \bm{p}_{\perp} -z_{\bar{q}} \, (\bm{k}_{\perp}+\bm{k}_{\gamma \perp} ) \} +\alpha_{1} \, z_{\bar{q}} (1-z_{\bar{q}}) \, Q^{2} \nonumber \\
& - \alpha_{2} \, z_{\bar{q}} (1-z_{\bar{q}}) \, (2p.k+2p.k_{\gamma}+2k.k_{\gamma}) \, .
\label{eq:c1-c3-V19-A}
\end{align}

\item Case B: $0>l_{2}^{-}> -p^{-}$ 
\begin{align}
R^{(V19);B}_{(\rm I);\mu \alpha} (\bm{l}_{1\perp}) &= \frac{1}{2 \, (q^{-})^{2}} \!\!  \int \!\! \frac{\mathrm{d}z_{l}}{(2\pi) \, z_{l}}  \,  \gamma_{\alpha} \, \frac{\slashed{k}+\slashed{k}_{\gamma}}{2k.k_{\gamma}}  \Big[\mathcal{R}_{(1);\mu}^{i}  \, \mathcal{I}_{v}^{(3,ijj)}( \bm{V}_{\perp}^{(V19);B},\Delta^{(V19);B} )  + \mathcal{R}_{(2);\mu } \, \mathcal{I}_{v}^{(3,ii)} ( \bm{V}_{\perp}^{(V19);B},\Delta^{(V19);B} ) \nonumber \\
& -\overline{\mathcal{R}}_{(3)}^{i} \, \mathcal{R}_{(1);\mu }^{j} \, \mathcal{I}_{v}^{(3,ij)} ( \bm{V}_{\perp}^{(V19);B},\Delta^{(V19);B} ) - \overline{\mathcal{R}}_{(3)}^{i}  \, \mathcal{R}_{(2);\mu } \, \mathcal{I}_{v}^{(3,i)} ( \bm{V}_{\perp}^{(V19);B},\Delta^{(V19);B} ) \Big]   \, , 
\label{eq:R-V19-propto-zl-inverse-B}
\end{align}
where the coefficients $\mathcal{R}_{(1,2,4)}$ are the same as in Eq.~\ref{eq:coefficients-V19-A} and 
\begin{align}
\overline{\mathcal{R}}_{(3)}^{i}& = 2z_{\bar{q}} \, (k^{i}+k^{i}_{\gamma}) -2(1-z_{\bar{q}}) \, p^{i}-(\slashed{k}+\slashed{k}_{\gamma}) \, \gamma^{i} z_{\bar{q}}   \, .
\end{align}
The divergent piece for  Case B can now be obtained as 
\begin{align}
R^{(V19);B}_{\rm div.;\mu \alpha} (\bm{l}_{1\perp}) &=\ln \Big( \frac{1}{z_{0}} \Big) \, \gamma_{\alpha} \, \frac{\slashed{k}+\slashed{k}_{\gamma}}{2k.k_{\gamma}} \Bigg[  \mathcal{R}_{(1);\mu}^{i} \,  \mathcal{I}_{v;\rm log}^{(3,ijj)}\big(\bm{c}_{1\perp}^{(V19);A},\Delta^{(V19);A} \big)  + \mathcal{R}_{(2);\mu } \,  \mathcal{I}_{v;\rm log}^{(3,ii)}\big(\bm{c}_{1\perp}^{(V19);A},\Delta^{(V19);A} \big) \nonumber \\
& -\overline{\mathcal{R}}_{(3)}^{i} \, \mathcal{R}_{(1);\mu }^{j} \,   \mathcal{I}_{v;\rm log}^{(3,ij)}\big(\bm{c}_{1\perp}^{(V19);A},\Delta^{(V19);A} \big)    - \overline{\mathcal{R}}_{(3)}^{i} \, \mathcal{R}_{(2);\mu }\,   \mathcal{I}_{v;\rm log}^{(3,ijj)}\big(\bm{c}_{1\perp}^{(V19);A},\Delta^{(V19);A} \big)   \, \Bigg]  \, , 
\label{eq:R-div-V19-B}
\end{align}
with $\bm{c}_{1\perp}$ and $c_{3}$ defined in Eq.~\ref{eq:c1-c3-V19-A}.
\end{itemize}

Combining the results for the two cases, we obtain the net divergent contribution to $(V19)$ as 
\begin{equation}
R^{(V19)}_{\rm div.;\mu \alpha} (\bm{l}_{1\perp}) = R^{(V19);A}_{\rm div.;\mu \alpha} (\bm{l}_{1\perp})+R^{(V19);B}_{\rm div.;\mu \alpha} (\bm{l}_{1\perp}) \, , 
\label{eq:R-div-V19}
\end{equation}
where $R^{(V19);A}_{\rm div.}$ and $R^{(V19);B}_{\rm div.}$ are given respectively by Eqs.~\ref{eq:R-div-V19-A} and \ref{eq:R-div-V19-B}.
The expressions for the divergent pieces of the other contributions are provided in Appendix~\ref{sec:T-V4-div-parts}.
\end{enumerate}

\section{Constructing the inclusive photon$+$2 jet cross-section}\label{sec:jet-cross-section}

In this section, we will define jets\footnote{Although the underlying principle is identical to prior discussions in the literature~\cite{Sterman:1977wj,Furman:1981kf,Aversa:1988vb,Aversa:1989xw,Jager:2004jh,Kang:2017mda} that utilize the small cone condition, the framework in~\cite{Ivanov:2012ms} is best suited for our computation.} using the small cone algorithm of Ivanov and Papa~\cite{Ivanov:2012ms} to extract the collinear and soft contributions from the expressions for the squared amplitude in the relevant diagrams of real gluon emission processes. These jet definitions can then be employed for amplitudes with virtual loops. When their cross-sections are combined with gluon emission contributions  including collinear and soft divergences, we will obtain a finite cross-section for  inclusive photon production in association with a quark-antiquark dijet.

Following the general definitions in Sec.~\ref{sec:LO-review}, the differential cross-section for inclusive $\gamma+q\bar{q}$ production at NLO can be written as
\begin{equation}
\frac{\mathrm{d}^{3} \sigma^{\rm NLO; parton}}{\mathrm{d}x \,  \mathrm{d}Q^{2} \mathrm{d}^{6} K_{\perp} \mathrm{d}^{3} \eta_{K} }= \frac{\alpha_{em}^{2}q_{f}^{4}y^{2}N_{c}}{512 \pi^{5} Q^{2}} \, \frac{1}{(2\pi)^{4}} \,  \frac{1}{2} \,  L^{\mu \nu} \tilde{X}_{\mu \nu}^{\rm NLO; parton} \, ,
\label{eq:triple-differential-CS-NLO}
\end{equation}
where $\tilde{X}_{\mu \nu}^{\rm NLO; parton}$ represents the NLO contributions to the hadron tensor, in analogy to its LO counterpart in Eq.~\ref{eq:triple-differential-CS-LO}.
Its computation was outlined in Eq.~\ref{eq:H-tensor-NLO-generic} and carried out in the previous two sections.
% The lepton tensor is unaffected and has the same expression as Eq.~\ref{eq:L-tensor}. 
The superscript ``parton'' indicates that the various components that build up the hadron tensor in Eq.~\ref{eq:triple-differential-CS-NLO} are all calculated at the parton level. We will now 
discuss how to promote these quantities to the level of jets and shall construct the inclusive cross-section for $\gamma+2$ jets. 

As a first step towards writing down our final result, we refer the reader back to Table~\ref{tab:NLO-assembly}. This table contains all the elements to construct the cross-section organized in terms of their color structures.  Those with identical structures are placed in the same row, which makes the cancellations of divergences transparent. We begin with the virtual graphs discussed at length in the previous section; the structure of their divergences is apparent at the amplitude level. For most of these graphs, the divergent pieces are proportional to the LO amplitude; hence their interference contributions with the LO amplitudes are proportional to the LO cross-section. 

We observed that the UV divergences arising from the self-energy graphs with dressed gluon propagators cancel with the divergent contributions from the self-energy graphs $(S13)-(S16)$ where the gluon does not scatter off the shock wave. A similar cancellation of divergences that are proportional to the LO cross-section also occurs between the graphs $(S21)$ and $(S22)$. In addition, there are a few virtual gluon exchange processes, unique to photon+dijet production, which do not have a divergence structure proportional to the LO cross-section. In these graphs, the photon is nested in the gluon loop for both self-energy and vertex contributions. These correspond to the processes $(S24)$ in  Fig.~\ref{fig:NLO-self-3} and $(V15)$ in Fig.~\ref{fig:NLO-vertex-3} (and their $q\leftrightarrow \bar{q}$ counterparts). However as shown in section~\ref{sec:virtual-self-energy}, there is an intricate cancellation of divergences that takes place between the net contribution of self-energy graphs $(S23)$ and $(S24)$ in Fig.~\ref{fig:NLO-self-3} and the vertex contribution from $(V15)$ in Fig.~\ref{fig:NLO-vertex-3}. The rapidity  divergences also cancel between the diagrams $\{ (S21), (S22) \}$ and $\{ (S23), (S24) \}$; they therefore do not contribute to the JIMWLK kernel. From the net UV contributions of virtual graphs, we are therefore only left with the divergences from vertex contributions with free gluon namely $(V13)$ and $(V14)$ (and their $q\leftrightarrow \bar{q}$ counterparts).

We can now add the amplitudes for all the allowed virtual graphs and obtain the following result for their contribution to the NLO hadron tensor:
\begin{align}
\tilde{X}_{\mu \nu}^{\rm NLO; parton} \Big \vert_{\rm Virtual} & \propto  \Bigg \langle \Big( \mathcal{M}^{\rm LO}_{\mu \alpha} (\bm{q},\bm{k},\bm{p},\bm{k}_{\gamma} ) \Big)^{*} \Big( \mathcal{M}^{\rm NLO;SE} (\bm{q},\bm{k},\bm{p},\bm{k}_{\gamma} )+  \mathcal{M}^{\rm NLO;Vert} (\bm{q},\bm{k},\bm{p},\bm{k}_{\gamma} ) \Big)_{\nu}^{\alpha} + c.c \Bigg \rangle \nonumber \\
&=\Bigg\{ \frac{2\alpha_{S}C_{F}}{\pi} \,  \ln \Big( \frac{1}{z_{0}^{2}} \Big) \, \Bigg( \frac{1}{\epsilon}+\frac{1}{2} \, \ln \Big( \frac{\tilde{\mu}^{2}}{Q^{2}} \Big)  \Bigg)-\frac{2\alpha_{S}C_{F}}{\pi} \, \frac{3}{2} \, \Bigg( \frac{1}{\epsilon}+\frac{1}{2} \, \ln \Big( \frac{\tilde{\mu}^{2}}{Q^{2}} \Big)  \Bigg)  \Bigg\} \,  \tilde{X}_{\mu \nu}^{\rm LO; parton} \nonumber \\
&+\frac{2\alpha_{S}C_{F}}{\pi} \, \ln^{2} \Big( \frac{1}{z_{0}} \Big) \,  \tilde{X}_{\mu \nu}^{\rm LO; parton} +\alpha_{S} \, \ln \Big( \frac{1}{z_{0}} \Big) \, \mathcal{H}_{V} \otimes  \tilde{X}_{\mu \nu}^{\rm LO; parton} +\tilde{X}^{\rm NLO; parton}_{\mu \nu;\rm finite} \Big  \vert_{\rm virtual}  \, .
\label{eq:interference-contributions-1}
\end{align}
The parton level LO hadron tensor is defined in Eq.~\ref{eq:H-tensor-LO}. Recall that $\langle \cdots\rangle$ corresponds to the CGC averaging in Eq.~\ref{eq:expectation-value-cgc}. The operator $\mathcal{H}_{V}$ here contains bilinear functional derivatives in the classical gauge field $A_{\rm cl}^+$; as we will discuss in the next section, the action of these on the color structure $\Xi$ of the leading order cross-section given in section~\ref{sec:LO-review} generates the leading log color structures of the NLO cross-section. The computation of the various pieces that constitute the finite contribution from the virtual graphs (denoted above by $\tilde{X}^{\rm NLO; parton}_{\mu \nu;\rm finite} \Big  \vert_{\rm virtual}$) are spelled out over seven subsections in Appendix~\ref{sec:finite-pieces-virtual-graphs}.

In this section, we will focus on the divergent structures shown in the first line on the r.h.s. These are the residual collinear singularities that survive after cancellations of UV divergences between different individual graphs in the virtual amplitude. One must therefore combine the cross-section containing these divergences with those of the real $q\bar{q}\gamma+g$ cross-section to obtain further cancellations.
 
For real gluon emission processes, it is difficult to extract their soft and collinear structures at the amplitude level. One needs to evaluate the squared amplitude and then integrate over the phase space of the outgoing gluon to recover these. In the notation of Table~\ref{tab:NLO-assembly}, contributions from processes in which the gluon interacts with the shock wave (denoted as $T_{R}^{(1)}$) do not contain soft or collinear singularities. This is because the gluon gains a net transverse momentum from the cumulative effect of successive ``kicks'' received during multiple scattering off the shock wave. Likewise, any soft gluon emitted before the shock wave is not energetic enough to cross it without being reabsorbed. This is true for the squared amplitude proportional to ${T_{R}^{(1)} }^{*} T_{R}^{(1)}$ as well as the interference contributions of such graphs with those in which the gluon does not scatter off the shock wave; the latter are proportional to ${T^{(2)}_{R}}^{*} T_{R}^{(1)} $ and ${T^{(3)}_{R}}^{*} T_{R}^{(1)}$ in Table~\ref{tab:NLO-assembly}. The only kind of divergences present in these contributions are the small $x$ logarithmic singularities which can be isolated by taking the slow gluon limit, $k_{g}^{-} \rightarrow 0$ in the results obtained in Sec.~\ref{sec:real-emission-details}. We will demonstrate these in detail in the next section. One can then obtain the finite contribution to the cross-section from $T_{R}^{(1)}$ and its interference contributions with $T^{(2,3)}_{R} $ by numerically evaluating the gluon phase space integrated results for the appropriate squared amplitudes in Eqs.~\ref{eq:amplitude-R1-R10}, \ref{eq:amplitude-R11-R15} and \ref{eq:amplitude-R16-R20} and then subtracting the pieces that contribute to leading log JIMWLK evolution. 

Such rescattering contributions can therefore be written generically as 
\begin{align}
\tilde{X}^{\rm NLO; parton}_{\mu \nu} \Big \vert_{\rm Real;scatter} &  \propto \Big \langle \Big( \mathcal{M}^{\rm NLO;Real(1)}_{\mu \alpha} \Big)^{*} \Big( \mathcal{M}^{\rm NLO;Real(1)}_{\mu \alpha} \Big)
%_{\nu}^{\alpha} 
+ \Big\{ \Big(  \mathcal{M}^{\rm NLO;Real(2)}_{\mu \alpha} +  \mathcal{M}^{\rm NLO;Real(3)}_{\mu \alpha} \Big)^{*}       \Big( \mathcal{M}^{\rm NLO;Real(1)}_{\mu \alpha} \Big)
%_{\nu}^{\alpha} 
+ c.c \Big\} \Big \rangle \nonumber \\
& = \alpha_{S} \, \ln \Big( \frac{1}{z_{0}} \Big) \, \mathcal{H}_{R}^{(1)} \otimes  \tilde{X}_{\mu \nu}^{\rm LO; parton}  + \tilde{X}^{\rm NLO; parton}_{\mu \nu; \rm finite} \Big \vert_{\text{real scatter}}  \, ,
\label{eq:real-emission-scattered-gluon-generic-CS}
\end{align}
where the term proportional to the small $x$ logs can be expressed as the convolution of an operator $ \mathcal{H}_{R}^{(1)}$ (analogous to $\mathcal{H}_{V}$ in Eq.~\ref{eq:interference-contributions-1} above) acting on the leading order hadron tensor. This too will be discussed in the next section. The finite pieces can be evaluated numerically; doing so will be a topic for future work.

The gluon emission processes of interest in this section will be the ones in which the gluon does not cross the shock wave and we have a region of phase space where it can be soft or collinear with respect to the quark or antiquark. Depending on whether the gluon is emitted from the quark or antiquark, we have denoted the contributions from these kinds of processes respectively by $T_{R}^{(2)}$ and $T_{R}^{(3)}$. It is clear from the first column of Table~\ref{tab:NLO-assembly} that we get different results for the respective color structures in the NLO cross-section depending on whether the processes constituting $T_{R}^{(2,3)}$ interfere between themselves or with their $q\leftrightarrow \bar{q}$ interchanged counterparts. This dependence on color flow at the loop level is absent for the case discussed in~\cite{Boussarie:2014lxa,Boussarie:2016bkq,Boussarie:2016ogo} where the $q\bar{q}$ is projected on to a singlet final state. We will show that these differing color structures have  interesting implications for soft gluon factorization.  

%We therefore have to investigate the structure of the squared amplitudes for processes in which the gluon is emitted by the quark or antiquark after the shock wave interaction. 

Since we are not integrating over the phase space of the quark and antiquark, combining the NLO cross-sections for $q\bar{q}\gamma$ and $q\bar{q}\gamma+g$ will not in general cure all infrared (IR) singularities. We will still have a remnant collinear divergence which can be absorbed in the jet fragmentation function of the quark or antiquark and this can be interpreted as contributing to the evolution in energy of the fragmentation function. Conversely, we can construct an IR-safe cross-section for photon$+$dijet production by using a jet algorithm that restricts the phase space for the final state gluon.

Following~\cite{Ivanov:2012ms,Boussarie:2016ogo}, we will work in the small cone approximation (SCA) in which the extent of the jets in the rapidity-azimuthal angle $(Y,\phi)$ plane is small, or more quantitatively, the jet cone radius $R$ is not too large ($R^{2} \ll 1$). In this approximation, one then systematically expands the partonic cross-sections around $R=0$. The dependence on $R$ is of the form $A \, \ln (R)+B+ O(R^{2})$, where the coefficients $A$ and $B$ can be evaluated analytically and the terms that are power suppressed in $R^{2}$ are neglected. 

For the inclusive computation of the photon+2 jet cross-section there are three possible cases that we need to consider:
\begin{itemize}
\item The gluon is inside the cone of either the quark or antiquark,
\item The gluon is outside the cone, 
\item The gluon forms one of the jets, while the other jet is formed by a $q\bar{q}$ pair.
\end{itemize}
We will not consider the third sort of contribution in this list because it does not have a collinear divergence and is hence sub-dominant in the SCA.  A short proof is provided in Appendix~\ref{sec:non-collinear-contributions}.

We will therefore first isolate the singularities from the region of phase space where the real gluon is collinear to the quark or antiquark and shall  use the SCA framework to identify when two partons form a jet. For a jet cone radius $R$ $(R^2 \ll 1)$, two partons $i$ and $k$ with respective momenta $p_{i}$ and $p_{k}$ will form a jet `J'  carrying a momentum equal to the sum of their momenta if both partons satisfy the condition
\begin{equation}
\Delta \phi_{i,k}^{2}+\Delta Y^{2}_{i,k} < R^{2} \, .
\label{eq:small-cone-approximation}
\end{equation}
Here $\Delta \phi_{i,k}$ is the difference of the azimuthal angles between parton $i$ ($k$) and the jet; $\Delta Y_{i,k}$ is the corresponding rapidity difference. In the SCA, the jet constituted of the partons $i$ and $k$ is considered on-shell (upto $O(R)$ corrections) and hence its momentum can be written as
\begin{equation}
p_{J}= (p_{i}^{+}+p_{k}^{+},z_{J}q^{-}, \bm{p}_{i \perp}+\bm{p}_{k \perp}) \, , 
\label{eq:jet-momentum}
\end{equation}
where $z_J = (p_{i}^{-}+p_{k}^{-})/q^-$; $p_{i}^{+}+p_{k}^{+} \sim \bm{p}_{J \perp}^{2} /2z_{J}q^{-}$. The quantities on the l.h.s of the inequality in Eq.~\ref{eq:small-cone-approximation} are given by~\cite{Ivanov:2012ms}
 \begin{equation}
 \Delta \phi_{i,k}= \cos^{-1} \Bigg( \frac{\bm{p}_{J\perp}.\bm{p}_{i,k\perp}}{\vert \bm{p}_{J\perp} \vert \, \vert \bm{p}_{i,k \perp} \vert } \Bigg)  \, , \quad \Delta Y_{i,k}= \frac{1}{2} \, \ln \Bigg( \frac{z_{J}^{2}}{\bm{p}_{J\perp}^{2}}. \frac{\bm{p}_{i,k \perp}^{2}}{z_{i,k}^{2}} \Bigg) \, .
 \label{eq:azimuthal-angle-rapidity-difference}
 \end{equation}
Here $z_{i,k} = p_{i,k}^-/q^-$. We will introduce the ``collinearity'' variable 
\begin{equation}
\bm{\mathcal{C}}_{ik,\perp}= \frac{z_{i}}{z_{i}+z_{k}} \, \bm{p}_{k \perp} - \frac{z_{k}}{z_{i}+z_{k}} \, \bm{p}_{i \perp} \, , 
\label{eq:collinearity-variable}
\end{equation}
which approaches zero when the partons $i$ and $k$ are collinear ($\bm{p}_{k \perp} \rightarrow z_{k}/z_{i} \, \bm{p}_{i \perp}$). It is then possible to rewrite 
\begin{equation}
\bm{p}_{i \perp} = \frac{z_{i}}{z_{i}+z_{k}} \, \bm{p}_{J \perp} - \bm{\mathcal{C}}_{ik, \perp} \, , \quad \bm{p}_{k \perp}= \bm{p}_{J\perp}-\bm{p}_{i \perp} \, , 
\end{equation}
and express the quantities in Eq.~\ref{eq:azimuthal-angle-rapidity-difference} in terms of $\bm{\mathcal{C}}_{ik,\perp}$. The small cone condition in Eq.~\ref{eq:small-cone-approximation} can then be equivalently written in terms of this collinearity variable as
\begin{equation}
\bm{\mathcal{C}}_{ik,\perp}^{2} < R^{2} \bm{p}_{J\perp}^{2} \, \text{min} \Big(\frac{z_{i}^{2}}{z_{J}^{2}}, \frac{z_{k}^{2}}{z_{J}^{2}} \Big) \, .
\label{eq:SCA-in-terms-of-C}
\end{equation}

We will now use the above definitions to extract the collinearly divergent contributions from the processes in which a quark and gluon constitute the first jet `J' and the antiquark constitutes the second jet `K'. The parton level Feynman graphs for these contributions are shown in Fig.~\ref{fig:NLO-real-nogluonscatter}. The corresponding expression for the quark-antiquark interchanged diagrams are obtained simply by $J\leftrightarrow K$ symmetrical replacements. If we look carefully at the diagrams in Fig.~\ref{fig:NLO-real-nogluonscatter}, processes labeled $(R12)-(R15)$ have a gluon emitted by an on-shell quark with the remaining structure identical to those of leading order (LO) processes depicted in Fig.~\ref{fig:LO-diagrams}.

It is physically intuitive that the divergent structure of the amplitude squared for these processes will be proportional to the LO amplitude squared. We will see that this is indeed the case. For the diagram labeled $(R11)$ (see Fig.~\ref{fig:NLO-real-R11}) the collinearity of the gluon with respect to the quark is lost because the photon emitted after the gluon imparts a virtuality  $2k.k_{\gamma}$ to the quark. However in the soft photon limit $k_{\gamma} \rightarrow 0$, the quark goes on-shell and can be collinear to the gluon. In the case of the amplitudes for the graphs $(R12)-(R15)$ in which the gluon is emitted after the photon, we see from  Eqs.~\ref{eq:R-R12}-\ref{eq:R-R15} that there is a possible divergence in the collinear limit $\bm{k}_{g\perp} \rightarrow z_{g}/z_{q} \, \bm{k}_{\perp}$. Because there is a similar term in the numerator of these expressions, the  structure of the singularity becomes transparent only at the level of the squared amplitude. 

It is also easy to check using the expressions for the amplitude that for interference contributions between diagrams in which the gluon is emitted from the quark in the amplitude and from the antiquark in the conjugate amplitude (or vice versa) there are no divergent terms in the collinear limit. Collinear divergences arise only when the gluon is emitted and reabsorbed by the parton to which it is collinear.

For the case where quark and gluon form the jet J, and the antiquark constitutes jet K, we will introduce the jet variables,
\begin{equation}
(z_{J},\bm{p}_{J \perp}, z_{K},\bm{p}_{K\perp}, \bm{\mathcal{C}}_{qg,\perp} )= \Big( z_{q}+z_{g},\bm{k}_{\perp}+\bm{k}_{g\perp},z_{\bar{q}},\bm{p}_{\perp},\frac{z_{q}}{z_{J}} \, (\bm{k}_{g\perp}- z_{g}/z_{q} \, \bm{k}_{\perp} ) \Big) \, ,
\label{eq:jet-variables-T-R2}
\end{equation}
so that the $d$-dimensional phase space differential measure transforms as
\begin{equation}
\frac{\mathrm{d} z_{q}}{(2\pi) \, 2z_{q}}   \, \frac{\mathrm{d}z_{\bar{q}} }{(2\pi) \, 2z_{\bar{q}}} \, \frac{\mathrm{d}z_{g}}{(2\pi) \, 2z_{g}} \, \frac{\mathrm{d}^{d}\bm{k}_{\perp}}{(2\pi)^{d}} \,  \frac{\mathrm{d}^{d}\bm{p}_{\perp}}{(2\pi)^{d}} \, \frac{\mathrm{d}^{d}\bm{k}_{g\perp}  }{(2\pi)^{d}} \rightarrow \frac{\mathrm{d} z_{J}}{(2\pi) \, 2z_{J}} \,  \frac{\mathrm{d}z_{K}}{(2\pi) \, 2z_{K}} \,  \frac{\mathrm{d}z_{g}}{(2\pi) \, 2z_{g}} \,  \frac{\mathrm{d}^{d} \bm{p}_{J\perp}}{(2\pi)^{d}} \, \frac{\mathrm{d}^{d} \bm{p}_{K \perp}}{(2\pi)^{d}} \,\frac{ \mathrm{d}^{d} \bm{\mathcal{C}}_{qg,\perp}}{(2\pi)^{d}}  \, \frac{z_{J}}{z_{J}-z_{g}} \, .
\label{eq:d-dimensional-phase-space-transform}
\end{equation} 
After summing over the spins of the quark and antiquark in the squared amplitude and using
\begin{equation}
\sum_{\rm gluon \enskip  pols.} \epsilon^{i} (\bm{k}_{g}) \, {\epsilon^{*}}^{i} (\bm{k}_{g}) = \delta^{ij} \, ,
\end{equation}
for the sum over gluon polarizations, it is a straightforward exercise to show that the following relation holds in the collinear limit $\bm{\mathcal{C}}_{qg,\perp} \rightarrow 0$,
\begin{align}
\sum_{\rho=12}^{15} \sum_{\sigma=12}^{15} {\mathcal{M}^{(R\rho)}_{\mu \alpha}}^{\dagger} \, {\mathcal{M}^{(R\sigma)}_{\nu}}^{\alpha} \, \Big \vert_{\rm collinear}= 4\pi \alpha_{S}C_{F} \, \frac{\big\{4(z_{J}-z_{g})z_{J}+d \, z_{g}^{2} \big\} \, (z_{J}-z_{g})  }{z_{J}^{3} \, \bm{\mathcal{C}}_{qg,\perp}^{2}} \, {\mathcal{M}^{\rm LO}_{\mu \alpha}}^{\dagger} \, {\mathcal{M}^{\rm LO}_{\nu}}^{\alpha}  \, .
\label{eq:collinear-amp-squared-R12-R15}
\end{align}
%Above we have the quadratic Casimir $C_{F}=(N_{c}^{2}-1)/2N_{c}$, $d=2-\epsilon$ and the leading order amplitude is given by Eq.~\ref{eq:LO-amp-master}. 
When we integrate over the phase space of the gluon, the small cone condition given by Eq.~\ref{eq:SCA-in-terms-of-C} restricts the integration over $\bm{\mathcal{C}}_{qg,\perp}$ to be 
\begin{equation}
\bm{\mathcal{C}}_{qg,\perp}^{2} < \bm{\mathcal{C}}_{qg,\perp}^{2} \big \vert_{\rm max.}= R^{2} \bm{p}_{J\perp}^{2} \, \text{min} \Big( \frac{z_{g}^{2}}{z_{J}^{2}}, \frac{(z_{J}-z_{g})^{2}}{z_{J}^{2}} \Big) \, .
\label{eq:real-coll-limit}
\end{equation}
Here $d=2-\epsilon$ as previously.
We will denote this collinearly divergent (dominant) contribution to the cross-section from the amplitude squared of processes $(R12)-(R15)$ as $\tilde{X}_{\mu \nu; \rm collinear}^{\rm NLO}$. From Eq.~\ref{eq:collinear-amp-squared-R12-R15}, and employing the phase space factors on the r.h.s of Eq.~\ref{eq:d-dimensional-phase-space-transform}, this is given by 
\begin{align}
\tilde{X}_{\mu \nu; \rm collinear}^{\rm NLO}&=\Bigg[ \alpha_{S} C_{F} \int_{z_{0}}^{z_{J}} \mathrm{d}z_{g} \, \Big\{4 \Big( \frac{1}{z_{g}}-\frac{1}{z_{J}} \Big) +(2-\epsilon)\,  \frac{z_{g}}{z^{2}_{J}} \Big\} \times \Big( \mu^{\epsilon} \,  \int^{\bm{\mathcal{C}}_{qg,\perp}^{2} \vert_{\rm max.}} \frac{\mathrm{d}^{2-\epsilon} \bm{\mathcal{C}}_{qg,\perp}}{(2\pi)^{2-\epsilon} \, \bm{\mathcal{C}}_{qg,\perp}^{2} }  \Big) \Bigg]  \, \tilde{X}^{\rm LO; jet}_{\mu \nu} \, ,
\label{eq:collinear-contribution-1}
\end{align}
where $\tilde{X}^{\rm LO; jet}_{\mu \nu}$ is shorthand for the LO hadron tensor for the production of a photon plus the quark-antiquark jets J and K.

This quantity is equivalent to the LO hadron tensor for $\gamma+q\bar{q}$ production if we make the following replacements in Eq.~\ref{eq:H-tensor-LO}:
\begin{equation}
(z_{q},z_{\bar{q}},\bm{k}_{\perp},\bm{p}_{\perp}) \rightarrow (z_{J},z_{K},\bm{p}_{J\perp},\bm{p}_{K\perp}) \, .
\label{eq:parton-to-jets-LO}
\end{equation}
The lower cutoff on the gluon momentum fraction $z_{g}$ is set to $z_{0}=\Lambda_{0}^{-}/q^{-}$
%\begin{equation}
%\approx \Big(Q^{2}_{0}/x_{0}\Big)\times  \Big( x_{Bj}/Q^{2} \Big) \, ,
%\end{equation}
as in the previous sections. We will use the following result in dimensional regularization for the transverse integral,
\begin{align}
\mu^{\epsilon} \,  \int^{\bm{\mathcal{C}}_{qg,\perp}^{2} \vert_{\rm max.}} \frac{\mathrm{d}^{2-\epsilon} \bm{\mathcal{C}}_{qg,\perp}}{(2\pi)^{2-\epsilon} \, \bm{\mathcal{C}}_{qg,\perp}^{2} } &= -\frac{1}{2\pi} \, \Bigg( \frac{1}{\epsilon} +\frac{1}{2} \ln \Big( \frac{\tilde{\mu}^{2}}{Q^{2}} \Big) \Bigg)  +\frac{1}{4\pi } \, \ln \Big( \frac{\bm{\mathcal{C}}^{2}_{qg,\perp} \vert_{\rm max.}}{Q^{2}e^{\gamma_{E}}} \Big)  \, .
\label{eq:collinearity-integral}
\end{align}
The expression for the collinear contribution to the cross-section from the $q\leftrightarrow\bar{q}$ interchanged counterparts of $(R12)-(R15)$ is obtained simply by $J \leftrightarrow K$ symmetry. The net contribution from these processes is
\begin{align}
\tilde{X}_{\mu \nu; \rm collinear}^{\rm NLO} +q\leftrightarrow \bar{q} &= \Bigg\{ - \frac{2\alpha_{S}C_{F}}{\pi} \,  \ln \Big( \frac{1}{z_{0}^{2}} \Big) \, \Bigg( \frac{1}{\epsilon}+\frac{1}{2} \, \ln \Big( \frac{\tilde{\mu}^{2}}{Q^{2}} \Big)  \Bigg)+\frac{2\alpha_{S}C_{F}}{\pi} \, \frac{3}{2} \, \Bigg( \frac{1}{\epsilon}+\frac{1}{2} \, \ln \Big( \frac{\tilde{\mu}^{2}}{Q^{2}} \Big)  \Bigg)  \Bigg\} \,   \tilde{X}^{\rm LO; jet}_{\mu \nu} \nonumber \\
& +\frac{2\alpha_{S}C_{F}}{\pi} \, \Bigg\{ \ln \Big( \frac{1}{z_{0}} \Big) \, \Bigg( \ln \Big( \frac{R^{2} \vert \bm{p}_{J\perp} \vert \, \vert \bm{p}_{K\perp} \vert  }{z_{J} z_{K} Q^{2} e^{\gamma_{E}}} \Big)  -\ln \Big( \frac{1}{z_{0}} \Big) \Bigg) +\Bigg( \frac{1}{2} \, \ln (z_{J}) \ln \Big( \frac{R^{2}\, \bm{p}_{j\perp}^{2}}{z_{J}^{2} Q^{2}e^{\gamma_{E}}} \Big) + J\leftrightarrow K \Bigg) \nonumber \\
&-\frac{3}{4} \ln (z_{J}z_{K})  -\frac{3}{4} \ln \Big(  \frac{R^{2} \vert \bm{p}_{J\perp} \vert \, \vert \bm{p}_{K\perp} \vert  }{4 z_{J} z_{K} Q^{2} e^{\gamma_{E}}} \Big) +\frac{7}{4} -\frac{\pi^{2}}{6} \Bigg\} \, \tilde{X}^{\rm LO; jet}_{\mu \nu} \, .
\label{eq:collinear-divergent-contribution-1}
\end{align}
Now replacing parton momentum definitions with the jet definitions in Eq.~\ref{eq:parton-to-jets-LO} for the virtual contributions, we can combine the above result with the contributions from the interference of virtual and LO graphs in Eq.~\ref{eq:interference-contributions-1} (as well as the rest of the real contributions in Eq.~\ref{eq:real-emission-scattered-gluon-generic-CS}) to obtain 
\begin{align}
&\tilde{X}_{\mu \nu}^{\rm NLO; jet} \Big \vert_{\rm Virtual}+\tilde{X}^{\rm NLO; jet}_{\mu \nu} \Big \vert_{\rm Real;scatter}+ \Big\{ \tilde{X}_{\mu \nu; \rm collinear}^{\rm NLO} +q\leftrightarrow \bar{q}  \Big\} \nonumber \\
 &= \frac{2\alpha_{S}C_{F}}{\pi} \, \Bigg\{  \ln \Big( \frac{1}{z_{0}} \Big) \,  \ln \Big( \frac{R^{2} \vert \bm{p}_{J\perp} \vert \, \vert \bm{p}_{K\perp} \vert  }{z_{J} z_{K} Q^{2} e^{\gamma_{E}}} \Big)  +\Bigg( \frac{1}{2} \, \ln (z_{J}) \ln \Big( \frac{R^{2}\, \bm{p}_{j\perp}^{2}}{z_{J}^{2} Q^{2}e^{\gamma_{E}}} \Big) + J\leftrightarrow K \Bigg) -\frac{3}{4} \ln (z_{J}z_{K})  -\frac{3}{4} \ln \Big(  \frac{R^{2} \vert \bm{p}_{J\perp} \vert \, \vert \bm{p}_{K\perp} \vert  }{4 z_{J} z_{K} Q^{2} e^{\gamma_{E}}} \Big) \nonumber \\
 &+\frac{7}{4} -\frac{\pi^{2}}{6} \Bigg\} \,  \tilde{X}^{\rm LO; jet}_{\mu \nu} + \alpha_{S} \, \ln \Big( \frac{1}{z_{0}} \Big) \, \Big( \mathcal{H}_{V}+\mathcal{H}_{R}^{(1)} \Big) \otimes \tilde{X}^{\rm LO; jet}_{\mu \nu} +\tilde{X}^{\rm NLO; jet}_{\mu \nu;\rm finite} \Big  \vert_{\rm virtual} + \tilde{X}^{\rm NLO; jet}_{\mu \nu; \rm finite} \Big \vert_{\text{real scatter}}  \, .
 \label{eq:virtual-real-collinear-combined}
\end{align}
We observe clearly in the above expression that the collinear divergences cancel and so do the terms proportional to $\ln^{2}(1/z_{0})$. We are however left with a term from Eq.~\ref{eq:collinear-divergent-contribution-1} that contains a soft-collinear divergence proportional to $\ln (1/z_{0}) \, \ln (R)$ (in addition to finite pieces for a fixed cone size $R$). 

Recall however that this result is obtained in the collinear limit $\bm{\mathcal{C}}_{qg,\perp} \rightarrow 0$ of the real graphs as shown in Eq.~\ref{eq:real-coll-limit}. The  dominant contributions to the inclusive cross-section in this limit are when the gluon is inside the cone of the quark or antiquark jet. When one additionally requires {\it the gluon to be soft}, there can be contributions where the gluon is outside the jet cone. As we will now demonstrate, the jet algorithm when applied to the case of a jet formed by a single parton generates soft-collinear terms with opposite sign that exactly cancel those obtained in the collinear limit. 

In the soft gluon limit $k_{g} \rightarrow 0$, the contributions to the amplitude from the five processes in Fig.~\ref{fig:NLO-real-nogluonscatter} and their $q\leftrightarrow \bar{q}$ counterparts (given respectively by Eqs.~\ref{eq:amplitude-R11-R15} and \ref{eq:amplitude-R16-R20}) simplify to 
\begin{align}
&\lim_{k_{g} \rightarrow 0} \Big(  \mathcal{M}^{\rm NLO:Real(2)}_{\mu \alpha;b} +\mathcal{M}^{\rm NLO:Real(3)}_{\mu \alpha;b} \Big)   =(- g) \, 2\pi \, \delta(1-z_{v}^{\rm tot}) \, (eq_{f})^{2} \int \mathrm{d} \Pi_{\perp}^{\rm LO} \overline{u}(\bm{k}) T^{\rm LO}_{\mu \alpha} (\bm{l}_{1\perp}) \nonumber \\
&\times \Bigg\{ \Big( t_{b} \tilde{U}(\bm{x}_{\perp}) \tilde{U}^{\dagger}(\bm{y}_{\perp}) -t_{b} \Big) \, \frac{k^{\beta}}{k.k_{g}}  
 -  \Big( \tilde{U}(\bm{x}_{\perp}) \tilde{U}^{\dagger}(\bm{y}_{\perp})t_{b}  -t_{b} \Big) \, \frac{p^{\beta}}{p.k_{g}} \Bigg\} \epsilon^{*}_{\beta}(\bm{k}_{g}) \, v(\bm{p}) + \text{sub-leading terms in $ k_{g}$} \, .
\label{eq:soft-gluon-limit-NLO-amplitude}
\end{align}
The sub-leading pieces come from the process $(R11)$ (and its $q\leftrightarrow \bar{q}$ counterpart) where the gluon is emitted from an internal fermion line as well as from the next-to-leading soft terms in the amplitudes for $(R12)-(R15)$ (and their $q\leftrightarrow \bar{q}$ counterparts) in the expansion around $k_{g}=0$. 

Squaring this expression, summing over the gluon polarizations, and taking the CGC average over all possible static color charge configurations, the dominant soft gluon contribution to the NLO hadron tensor (and hence the NLO cross-section) is 
\begin{align}
\tilde{X}^{\rm NLO}_{\mu \nu; \rm soft}&=\tilde{X}^{\rm NLO}_{\mu \nu; \rm soft(1)} + \tilde{X}^{\rm NLO}_{\mu \nu; \rm soft(2)}  \, , 
\label{eq:soft-gluon-limit-NLO-cross-section}
\end{align}
where we have broken up the soft gluon contribution into two parts. The first contribution comes from the terms ${T_{R}^{(i)}}^{*} T^{(i)}_{R}$ ($i=2,3$)  in Table~\ref{tab:NLO-assembly} and possess a color structure similar to that of the LO hadron tensor in Eq.~\ref{eq:H-tensor-LO}:
\begin{align}
&\tilde{X}^{\rm NLO}_{\mu \nu; \rm soft(1)} \propto \lim_{k_{g} \rightarrow 0} \Big(  \mathcal{M}^{\rm NLO:Real(2)}_{\mu \alpha;b} \Big)^{\dagger} {\mathcal{M}^{\rm NLO:Real(2)}_{\nu;b}}^{\alpha} +2\rightarrow 3  \, , \nonumber \\
&= g^{2} \, C_{F} \, \Bigg\{ \frac{4}{ \Big(\bm{k}_{g\perp}- \frac{z_{g}}{z_{q}} \, \bm{k}_{\perp} \Big)^{2}  }  +\frac{4}{ \Big( \bm{k}_{g\perp}- \frac{z_{g}}{z_{\bar{q}}} \, \bm{p}_{\perp} \Big)^{2}  } \Bigg\} \, \tilde{X}_{\mu \nu}^{\rm LO;parton} \, .
\label{eq:soft-gluon-contribution-1}
\end{align}

This is however not true for the interference contributions ${T_{R}^{(2)}}^{*} T^{(3)}_{R}+c.c$ in Table~\ref{tab:NLO-assembly}; as noted previously, such interference contributions do not contain collinear divergences. These however have the following dominant contribution to the cross-section in the soft gluon limit\footnote{We should mention here that in decomposing the factors $2k.k_{g}$ and $2p.k_{g}$ in terms of LC coordinates, it is important to keep both terms $\bm{k}_{g\perp}$ and $z_{g}/ z_{q} \, \bm{k}_{\perp}$  (and $ z_{g}/ z_{\bar{q}} \, \bm{p}_{\perp}$) because in the soft gluon limit $z_{g} \ll z_{q}, z_{\bar{q}}$ and $k_{g \perp} \ll k_{\perp},p_{\perp}$.},
\begin{align}
& \tilde{X}^{\rm NLO}_{\mu \nu; \rm soft(2)}  \propto \lim_{k_{g} \rightarrow 0}  \Big(  \mathcal{M}^{\rm NLO:Real(2)}_{\mu \alpha;b} \Big)^{\dagger} {\mathcal{M}^{\rm NLO:Real(3)}_{\nu;b}}^{\alpha} + c.c \, , \nonumber \\
&= -g^{2} C^{\rm LO}_{\mu \nu} \times \Big\{ \frac{N_{c}}{2} (1-D_{xy}-D_{y'x'}+D_{xy}D_{y'x'} ) - \frac{1}{2N_{c}} \Xi (\bm{x}_{\perp},\bm{y}_{\perp};\bm{y'}_{\perp},\bm{x'}_{\perp}) \Big\}  \, \Bigg\{ \frac{8 \Big(\bm{k}_{g\perp}- \frac{z_{g}}{z_{q}} \, \bm{k}_{\perp} \Big). \Big( \bm{k}_{g\perp}- \frac{z_{g}}{z_{\bar{q}}} \, \bm{p}_{\perp} \Big)  }{\Big(\bm{k}_{g\perp}- \frac{z_{g}}{z_{q}} \, \bm{k}_{\perp} \Big)^{2} \,  \Big( \bm{k}_{g\perp}- \frac{z_{g}}{z_{\bar{q}}} \, \bm{p}_{\perp} \Big)^{2}  } \Bigg\} \, ,
\label{eq:soft-gluon-contribution-2}
\end{align}
where $C^{\rm LO}_{\mu \nu}$ is the LO coefficient function,
\begin{align}
C^{\rm LO}_{\mu \nu}= \Big\{ 2\pi (eq_{f})^{2} \delta(1-z_{\rm tot}^{v}) \int \mathrm{d} {\Pi_{\perp}^{\prime}}^{\rm LO} \, \overline{u} (\bm{k}) T^{\rm LO}_{\mu \alpha}(\bm{l}^{\prime}_{1\perp}) v(\bm{p}) \Big\}^{\dagger}  \times \Big\{ 2\pi (eq_{f})^{2} \delta(1-z_{\rm tot}^{v}) \int \mathrm{d} \Pi_{\perp}^{\rm LO} \, \overline{u} (\bm{k}) {T^{\rm LO}(\bm{l}_{1\perp})}_{\nu}^{\alpha} v(\bm{p}) \Big\} \, .
\label{eq:LO-coefficient-function}
\end{align}
There are no soft-collinear divergences in Eq.~\ref{eq:soft-gluon-contribution-2}. The only divergent contributions are those in rapidity that are recovered when taking the slow gluon limit. 

We will therefore focus on extracting the soft gluon contributions to the jet cross-section from Eq.~\ref{eq:soft-gluon-contribution-1}. Since the collinear contribution obtained in Eq.~\ref{eq:collinear-divergent-contribution-1} already includes the soft-collinear limit, adding this contribution to the expression in Eq.~\ref{eq:soft-gluon-contribution-1} will result in a double counting of such divergent pieces. To avoid this, we will restrict the region of integration in the first term on the r.h.s of Eq.~\ref{eq:soft-gluon-contribution-1} to ensure the emitted gluon is outside the cone of the jets formed by the quark/antiquark:
%Because there is no mixing between the two terms the limits of integration can be set independently. 
%With these in mind, we can now write the \textcolor{blue}{first kind of soft gluon contributions to the NLO hadron tensor as
\begin{align}
\tilde{X}^{\rm NLO}_{\mu \nu; \rm soft(1)}&= 4\alpha_{S} C_{F} \, \Bigg\{ \int_{z_{0}}^{z_{J}} \frac{\mathrm{d} z_{g}}{z_{g}} \int_{\bm{\mathcal{C}}_{qg,\perp}^{2} \vert_{\rm min.}} \frac{\mathrm{d}^{2-\epsilon} \bm{\mathcal{C}}_{qg,\perp}}{(2\pi)^{2-\epsilon} \, \bm{\mathcal{C}}_{qg,\perp}^{2} }  +J \rightarrow K \Bigg\} \, \tilde{X}^{\rm LO; jet}_{\mu \nu}  \, , 
\end{align}
where $\bm{\mathcal{C}}_{qg,\perp}^{2} \vert_{\rm min.}=R^{2} \bm{p}_{J \perp}^{2}/z_{J}^{2}$ for the first integral is obtained by imposing the small cone condition in Eq.~\ref{eq:small-cone-approximation} on a jet formed by a single parton~\cite{Ivanov:2012ms}. The upper bound for the integral over the collinearity variable can in principle be infinity. Hence the integration can be performed using dimensional regularization and we will obtain a similar result as in Eq.~\ref{eq:collinearity-integral}. We get finally, 
\begin{align}
\tilde{X}^{\rm NLO}_{\mu \nu; \rm soft(1)} &= -\frac{2\alpha_{S}C_{F}}{\pi} \, \Bigg\{  \ln \Big( \frac{1}{z_{0}} \Big) \,  \ln \Big( \frac{R^{2} \vert \bm{p}_{J\perp} \vert \, \vert \bm{p}_{K\perp} \vert  }{z_{J} z_{K} Q^{2} e^{\gamma_{E}}} \Big)  +\Bigg( \frac{1}{2} \, \ln (z_{J}) \ln \Big( \frac{R^{2}\, \bm{p}_{J\perp}^{2}}{z_{J}^{2} Q^{2}e^{\gamma_{E}}} \Big) + J\leftrightarrow K \Bigg) \Bigg\} \, \tilde{X}^{\rm LO; jet}_{\mu \nu}  \nonumber \\
& +\frac{4\alpha_{S}C_{F}}{\pi} \, \ln \Big( \frac{1}{z_{0}^{2}} \Big) \Big( \frac{1}{\epsilon} + \frac{1}{2} \ln \Big( \frac{\tilde{\mu}^{2}}{Q^{2}} \Big) \Big) \, \tilde{X}^{\rm LO; jet}_{\mu \nu}  \, .
\label{eq:soft-gluon-contribution-1-final}
\end{align}
As promised, we see that the terms in the first line on the r.h.s are identical but have the opposite sign to the corresponding terms in Eq.~\ref{eq:virtual-real-collinear-combined}, therefore canceling the final remaining collinear divergence.

The double log appearing in the second line of the equation above is contained in the ``slow'' gluon limit ($k^{-}_{g} \rightarrow 0 $) of our results.  As we noted previously from our discussion of similar divergent terms in the virtual graphs, this is the double log limit of the BFKL equation. We are therefore double counting here because the soft gluon sector is a subset of the slow limit. In order to obtain finite contributions from the squared real amplitude, we must subtract the pieces from the soft gluon limit that contribute to small $x$ evolution\footnote{The appearance of such BFKL logs in the final state emission of slow gluons outside the jet cone is a concrete illustration of the spacelike-timelike correspondence noted by Mueller~\cite{Mueller:2018llt}. We will discuss this point further in the final section.}. 
%\textcolor{blue}{This contribution can be obtained in $d=2-\epsilon$ dimensions using results that we derive explicitly later in the text in the $k_{g}^{-} \rightarrow 0$ limit as
%\begin{align}
%\mathrm{d}^{3} \sigma^{(1)}_{\rm slow} = \frac{4\alpha_{S}C_{F}}{\pi} \, \ln \Big( \frac{1}{z_{0}^{2}} \Big) \Big( \frac{1}{\epsilon} + \frac{1}{2} \ln \Big( \frac{\tilde{\mu}^{2} \, \vert \bm{r}_{x'x} \vert \, \vert \bm{r}_{y'y} \vert }{4} \Big) \Big) \, \mathrm{d}^{3} \sigma_{\rm LO}^{\rm jet} \, .
%\label{eq:slow-gluon-limit-1}
%\end{align}
%which is precisely the term on the second line of Eq.~\ref{eq:soft-gluon-contribution-1} up to logarithmic accuracy. }
%When this is done,  we get rid of the aforementioned divergent pieces. Coming back to Eq.~\ref{eq:soft-gluon-limit-NLO-cross-section}, we do not expect any soft-collinear pieces from the second terms on the r.h.s. 
%The only kind of divergent contributions can therefore be obtained in the slow gluon limit of these expressions which will be discussed in detail in the next section. We will call this soft contribution to the cross-section $\mathrm{d}^{3} \sigma^{(2)}_{\rm soft} $. We can then write
%Therefore since this contribution is naturally included in the small $x$ evolution of our expressions, we can get rid of it here by adding $\mathrm{d}^{3} \sigma^{(1)}_{\rm slow}$ to $\mathrm{d}^{3} \sigma^{(1)}_{\rm soft}$ \textcolor{blue}{which  will give (up to logarithmic accuracy) the terms on the first line of Eq.~\ref{eq:soft-gluon-contribution-1} that appear after the equality.}

More specifically we will absorb the kernels obtained in the slow gluon limit (of the real unscattered gluon contributions constituting $T_{R}^{(2,3)}$) in a redefinition of $\mathcal{H}_{R}^{(1)}\rightarrow \mathcal{H}_{R}$, whose structure we shall discuss further shortly in the next section. When we combine this sum with the result from Eq.~\ref{eq:virtual-real-collinear-combined} we obtain finally
\begin{align}
\tilde{X}_{\mu \nu}^{\rm NLO; jet}
 &= \frac{2\alpha_{S}C_{F}}{\pi} \, \Bigg\{  -\frac{3}{4} \ln \Big(  \frac{R^{2} \vert \bm{p}_{J\perp} \vert \, \vert \bm{p}_{K\perp} \vert  }{4 z_{J} z_{K} Q^{2} e^{\gamma_{E}}} \Big) +\frac{7}{4} -\frac{\pi^{2}}{6} \Bigg\} \,  \tilde{X}^{\rm LO; jet}_{\mu \nu}+ \alpha_{S} \, \ln \Big( \frac{1}{z_{0}} \Big) \, \Big( \mathcal{H}_{V}+\mathcal{H}_{R} \Big) \otimes  \tilde{X}^{\rm LO; jet}_{\mu \nu}  \nonumber \\
 & +\tilde{X}^{\rm NLO; jet}_{\mu \nu;\rm finite} \Big  \vert_{\rm virtual} + \tilde{X}^{\rm NLO; jet}_{\mu \nu; \rm finite} \Big \vert_{\text{real scatter}}   \, .
 \label{eq:virtual-real-collinear-combined-final}
\end{align}
We see that the soft-collinear pieces cancel leaving terms that have the generic form $A \, \ln (R) + B$ as expected in the SCA. As discussed above the  divergent pieces coming from the slow gluon limit that are contained in the soft gluon limit can be isolated and absorbed in a modification of $\mathcal{H}_{R}^{(1)}$ to form the operator $\mathcal{H}_{R}$. As we will discuss further in Sec.~\ref{sec:JIMWLK-evolution}, $\mathcal{H}_{V}+\mathcal{H}_{R}=\mathcal{H}_{\rm LO}$, where the r.h.s is the LO JIMWLK Hamiltonian that we introduced previously in the introduction. {\it Finally, the finite contributions appearing in the second line together with the first term on the r.h.s of Eq.~\ref{eq:virtual-real-collinear-combined-final} constitute the NLO impact factor for inclusive photon+dijet production. The computation of these terms is the principal result of this work. }

We conclude this section with a few comments. Firstly, we observe that there are no Sudakov double logs in our computation--this also holds for inclusive dijet production case in the soft photon limit of our result. These  logs appear due to the lack of complete real-virtual cancellations arising from a constraint imposed on the process, an example being dijet production in back-to-back correlations\footnote{See~\cite{Mueller:2013wwa} for a comprehensive discussion on Sudakov logs and their resummation for various small $x$ processes.}. Since we do not impose any such kinematic constraints, this explains our observation.

Secondly,  soft gluon factorization is violated. This factorization, in the soft gluon limit of $k_g\rightarrow 0$, corresponds to a convolution of the LO dijet cross-section with the well known eikonal kinematic factor $ \vert k^{\beta}/(k.k_{g}) - p^{\beta} / (p.k_{g}) \vert^{2}$, where the latter is contracted with the sum over gluon polarizations. This violation is a consequence of the differing topologies of the 
color structures that contribute towards soft and collinear divergences. 
%This is expected because our final state partons have color and so the limiting structure of the cross-section depends on the color flow induced by the gluon. 
%The situation is simpler for the case when we project the $q\bar{q}$ onto a singlet such that the color structures get traced over at the amplitude level and we get the LO cross-section factored out in the soft gluon limit. 
However we can rewrite the second term on the r.h.s of Eq.~\ref{eq:soft-gluon-limit-NLO-cross-section} (which violates the soft gluon theorem) as a term that obeys soft gluon factorization and another that violates factorization. With this, we can rewrite Eq.~\ref{eq:soft-gluon-limit-NLO-cross-section} as 
\begin{align}
\tilde{X}^{\rm NLO}_{\mu \nu; \rm soft}&=g^{2} \Bigg[ C_{F} \,  \Bigg\{ \frac{4z_{g}^{2}}{ z^{2}_{q} z^{2}_{\bar{q}} } \frac{\Big( z_{q} \, \bm{p}_{\perp} -z_{\bar{q}} \, \bm{k}_{\perp} \Big)^{2}}{\Big(\bm{k}_{g\perp}- \frac{z_{g}}{z_{q}} \, \bm{k}_{\perp} \Big)^{2} \,  \Big( \bm{k}_{g\perp}- \frac{z_{g}}{z_{\bar{q}}} \, \bm{p}_{\perp} \Big)^{2}}   \Bigg\} \, \tilde{X}^{\rm LO; parton}_{\mu \nu} \nonumber \\
&+ C^{\rm LO}_{\mu \nu}  \Big\{ \frac{N_{c}}{2} (Q_{xy;y'x'}-D_{xy}D_{y'x'}  )  \Big\}  \, \Bigg\{ \frac{8 \Big(\bm{k}_{g\perp}- \frac{z_{g}}{z_{q}} \, \bm{k}_{\perp} \Big). \Big( \bm{k}_{g\perp}- \frac{z_{g}}{z_{\bar{q}}} \, \bm{p}_{\perp} \Big)  }{\Big(\bm{k}_{g\perp}- \frac{z_{g}}{z_{q}} \, \bm{k}_{\perp} \Big)^{2} \,  \Big( \bm{k}_{g\perp}- \frac{z_{g}}{z_{\bar{q}}} \, \bm{p}_{\perp} \Big)^{2}  } \Bigg\}  \Bigg]  \, .
\label{eq:soft-gluon-limit-NLO-cross-section-modified}
\end{align}
%In the above equation the first term on the r.h.s is precisely what one would get if soft gluon factorization were to hold. 
The dipole ($D$) and quadrupole ($Q$) operators defined in Eq.~\ref{eq:dipole-quadrupole-Wilson-line-correlators} correspond to particular boundary conditions at $x^-=\pm \infty$. There have been recent developments that relate soft gluon theorems to the existence of infinite dimensional so-called BMS symmetries~\cite{He:2015zea} and to a color memory effect in Yang-Mills theory~\cite{Pate:2017vwa}. A dictionary between this language in the Regge limit of QCD and that of the CGC was established in \cite{Ball:2018prg}, and involves identifying the spacetime rapidity $\eta = \ln (x^-)$ in the latter with retarded time in the former. An interesting question is whether the soft gluon theorem can be restored by requiring that the factorization violating structure $(Q-DD)$ vanishes by a modification of boundary conditions at $x^-=\pm \infty$. This may be equivalent to defining asymptotic states/propagators that project out this color structure at $x^-=\pm \infty$. We will return to this interesting topic in future.

%A very interesting possibility is whether the soft gluon factorization violating structure $(Q - DD)$ structure vanishes by a modification of boundary conditions at $x^-=\pm \infty$, or equivalently, by defining asymptotic states/propagators that project out this color structure at $x^-=\pm \infty$. 

\section{High energy leading log resummation} \label{sec:JIMWLK-evolution}

In this section, we will consider the ``slow" (relative to the virtual photon LC momentum $q^{-}$) gluon limit of our results for real ($z_{g} \rightarrow 0$) and virtual ($z_{l} \rightarrow 0$) diagrams. This will allow us to isolate the soft divergences in rapidity; we will show explicitly that these terms provide  a nontrivial derivation, in the evolution of the projectile, of the JIMWLK renormalization group equation.  

We begin by examining the amplitudes for virtual processes in the $z_{l} \rightarrow 0$ limit, and subsequently, real emission graphs. 
%The former have the advantage that the virtual gluon loop momentum components are integrated over at the amplitude level in contrast to real gluon emission in which the gluon phase space is integrated over at the amplitude squared level. It is therefore useful to study the $z_{l} \rightarrow 0$ limit for the virtual graphs first. We will use a few representative diagrams and derive generic forms for the limiting expressions for each class of contributing processes.
For the self-energy contributions with the dressed gluon propagator, consider the process $(S1)$ given by Eq.~\ref{eq:M-S1-form-1}:
\begin{align}
\mathcal{M}^{(S1)}_{\mu \alpha}&= 2\pi \delta(1-z_{\text{tot}}^{v}) \, (eq_{f} g)^{2} \,  \int \mathrm{d} \Pi_{\perp}^{v} \, \Bigg[ \Big( t^{b} \tilde{U} (\bm{x}_{\perp}) t^{a} \tilde{U}^{\dagger} (\bm{y}_{\perp}) \Big) U_{ba}(\bm{z}_{\perp}) - C_{F} \mathds{1} \Bigg] \nonumber \\
& \times \int \frac{\mathrm{d}z_{l}}{2\pi} \int_{\bm{l}_{2\perp},\bm{l}_{3\perp}} \!\!\!\!\!\!\!\!\!\!\!\! e^{i\bm{l}_{2\perp}.\bm{r}_{zx}+i\bm{l}_{3\perp}.\bm{r}_{xz}} \,\,  \overline{u} (\bm{k}) \gamma_{\alpha}\frac{\slashed{k}+\slashed{k}_{\gamma}}{2k.k_{\gamma}} \Bigg[ \gamma^{i} \Big( \{\gamma^{+}(1-z_{\bar{q}}-z_{l})q^{-}-\bm{\gamma}_{\perp}.(\bm{k}_{\perp}+\bm{k}_{\gamma \perp}) \} +\gamma^{j}l^{j}_{3} \Big) \nonumber \\
& -\frac{2(1-z_{\bar{q}}-z_{l})}{z_{l}} \, l^{i}_{3} \Bigg] \gamma^{-} \Bigg[ \Big( \{ \gamma^{+}(1-z_{\bar{q}}-z_{l})q^{-} -\bm{\gamma}_{\perp}.\bm{l}_{1\perp} \} +\gamma^{k}l^{k}_{2} \Big) \, \gamma_{i} +\frac{2(1-z_{\bar{q}}-z_{l})}{z_{l}} \, l^{i}_{2} \Bigg]   \nonumber \\
& \times \frac{\gamma^{+}(1-z_{\bar{q}})q^{-}-\gamma^{-}\Big(Q^{2}z_{\bar{q}}+\bm{l}_{1\perp}^{2}\Big)/2z_{\bar{q}}q^{-}-\bm{\gamma}_{\perp}.\bm{l}_{1\perp}}{\Big[(\bm{l}_{2\perp}+\bm{v}_{1\perp}^{(S1)})^{2}+\Delta_{1}^{(S1)} \Big] \, \Big[(\bm{l}_{3\perp}+\bm{v}_{2\perp}^{(S1)})^{2}+\Delta_{2}^{(S1)}\Big]} \gamma_{\mu} \frac{\gamma^{+}z_{\bar{q}}q^{-}+\bm{\gamma}_{\perp}.\bm{l}_{1\perp}}{4(q^{-})^{2} \, (1-z_{\bar{q}})^{2}/z_{l} [\bm{l}_{1\perp}^{2}+Q^{2}z_{\bar{q}}(1-z_{\bar{q}})-i\varepsilon]} \gamma^{-} v(\bm{p}) \, ,
\end{align}
where the parameters $\bm{v}_{1,2\perp}^{(S1)}$ and $\Delta_{1,2}^{(S1)}$ are all proportional to $z_{l}$ as can be seen from Eq.~\ref{eq:S1-integration-parameters}. In the slow gluon limit  $z_{l} \rightarrow 0$, this expression simplifies greatly to give
\begin{align}
\label{eq:S1-softlimit}
\lim_{z_{l}\rightarrow 0} \mathcal{M}^{(S1)}_{\mu \alpha}&=4\alpha_{S} \int \frac{\mathrm{d}z_{l}}{z_{l}} \int_{\bm{z}_{\perp}} \int_{\bm{l}_{2\perp}} \!\!\!\! e^{i\bm{l}_{2\perp}.\bm{r}_{zx}} \frac{l^{i}_{2}}{\bm{l}_{2\perp}^{2}-i\varepsilon} \int_{\bm{l}_{3\perp}} \!\!\!\! e^{i\bm{l}_{3\perp}.\bm{r}_{xz}} \frac{l^{i}_{3}}{\bm{l}_{3\perp}^{2}-i\varepsilon} \Big\{ 2\pi (eq_{f})^{2} \delta(1-z_{\rm tot}^{v}) \int \mathrm{d} \Pi_{\perp}^{\rm LO} \, \overline{u} (\bm{k}) R^{\rm LO:(1)}_{\mu \alpha} (\bm{l}_{1\perp}) v(\bm{p}) \Big\} \nonumber \\
& \times  \Bigg[ \Big( t^{b} \tilde{U} (\bm{x}_{\perp}) t^{a} \tilde{U}^{\dagger} (\bm{y}_{\perp}) \Big) U_{ba}(\bm{z}_{\perp}) - C_{F} \mathds{1} \Bigg] \, ,
\end{align}
where the contribution $R^{\rm LO:(1)}_{\mu \alpha} (\bm{l}_{1\perp})$ to the perturbative amplitude at LO was defined previously in Eq.~\ref{eq:R-LO-1}.

The integrals over $\bm{l}_{2,3 \perp}$ represent the spatial derivative of the two dimensional free gluon propagator. We will use its coordinate space expression\footnote{This result is easily obtained by taking the limits $\bm{v}_{\perp} \rightarrow 0$ and $\Delta \rightarrow 0$ in Eq.~\ref{eq:I-r-11}.} 
\begin{equation}
\int_{\bm{l}_{\perp}} e^{i\bm{l}_{\perp}.\bm{r}_{\perp}} \frac{l^{i}}{\bm{l}_{\perp}^{2}-i\varepsilon} =\frac{i}{2\pi} \frac{r^{i}}{\bm{r}_{\perp}^{2}}  \, ,
\label{eq:derivative-2D-gluon-propagator}
\end{equation}
where the r.h.s represents the well-known Weizs\"{a}cker-Williams field created by the boosted quark-antiquark pair.
Substituting this in Eq.~\ref{eq:S1-softlimit} then gives 
\begin{align}
\lim_{z_{l}\rightarrow 0} \mathcal{M}^{(S1)}_{\mu \alpha}&=\frac{\alpha_{S}}{\pi^{2}} \int \frac{\mathrm{d}z_{l}}{z_{l}} \int_{\bm{z}_{\perp}} \frac{1}{(\bm{x}_{\perp}-\bm{z}_{\perp})^{2}} \Big\{ 2\pi (eq_{f})^{2} \delta(1-z_{\rm tot}^{v}) \int \mathrm{d} \Pi_{\perp}^{\rm LO} \, \overline{u} (\bm{k}) R^{\rm LO:(1)}_{\mu \alpha}(\bm{l}_{1\perp}) v(\bm{p}) \Big\} \nonumber \\
& \times  \Bigg[ \Big( t^{b} \tilde{U} (\bm{x}_{\perp}) t^{a} \tilde{U}^{\dagger} (\bm{y}_{\perp}) \Big) U_{ba}(\bm{z}_{\perp}) - C_{F} \mathds{1} \Bigg] \, .
\label{eq:soft-gluon-limit-S1}
\end{align}
There is a logarithmic divergence in the kernel for $\bm{x}_{\perp} \rightarrow \bm{z}_{\perp}$ under which the color structure also reduces to that for the LO process times the Casimir $C_{F}$. This is precisely the $1/\epsilon$ singularity multiplying  $\ln(1/z_{0})$ that appears in the momentum space result for $(S1)$ given in Eq.~\ref{eq:T-S1}. Note that the color structure there also has the form $C_{F} (\tilde{U}(\bm{x}_{\perp}) \tilde{U}^{\dagger}(\bm{y}_{\perp}) -\mathds{1} )$ for these divergent pieces. These divergences cancel in observables reflecting the fact that there are no  divergences in the LO JIMWLK Hamiltonian in the limit $\bm{x}_{\perp},\bm{y}_{\perp} \rightarrow \bm{z}_{\perp}$.

One can show in a similar fashion that $(S2)$ and $(S3)$ vanish in the $z_{l}\rightarrow 0$ limit. The three remaining processes have an identical structure to Eq.~\ref{eq:soft-gluon-limit-S1} albeit with different $R$-functions corresponding to the structure of their LO counterparts. Combining these results, the six contributions in Fig.~\ref{fig:NLO-self-1} give 
\begin{align}
\lim_{z_{l}\rightarrow 0} \mathcal{M}^{\rm NLO:SE(1)}_{\mu \alpha}&=\frac{\alpha_{S}}{\pi^{2}} \int \frac{\mathrm{d}z_{l}}{z_{l}} \int_{\bm{z}_{\perp}} \frac{1}{(\bm{x}_{\perp}-\bm{z}_{\perp})^{2}} \Big\{ 2\pi (eq_{f})^{2} \delta(1-z_{\rm tot}^{v}) \int \mathrm{d} \Pi_{\perp}^{\rm LO} \, \overline{u} (\bm{k}) T^{\rm LO}_{\mu \alpha}(\bm{l}_{1\perp}) v(\bm{p}) \Big\} \nonumber \\
& \times  \Bigg[ \Big( t^{b} \tilde{U} (\bm{x}_{\perp}) t^{a} \tilde{U}^{\dagger} (\bm{y}_{\perp}) \Big) U_{ba}(\bm{z}_{\perp}) - C_{F} \mathds{1} \Bigg] \, .
\label{eq:soft-gluon-limit-SE1}
\end{align}
The  quark-antiquark interchanged diagrams are obtained by replacing $\bm{x}_{\perp} \leftrightarrow \bm{y}_{\perp}$ in this equation.

We next consider the slow gluon limit of the self-energy corrections containing a free gluon propagator. Half of these 24 processes are depicted in Fig.~\ref{fig:NLO-self-3}, with the other half obtained by interchanging the quark-antiquark lines. In the limit $z_{l} \rightarrow 0$, the loop contribution shown in Fig.~\ref{fig:quark-self-energy}) for the quark self-energy reduces from the expression in Eq.~\ref{eq:loop-contribution-se-generic} to the simpler expression,
\begin{align}
\lim_{z_{l} \rightarrow 0} \Sigma(k_{f})=ig^{2}C_{F} \int \frac{\mathrm{d}z_{l}}{(2\pi) z_{l}} \int_{\bm{l}_{\perp}} \frac{1}{\bm{l}_{\perp}^{2}-i\varepsilon} \, \gamma^{-} \, \frac{k_{f}^{2}}{k_{f}^{-}} \, .
\label{eq:slow-gluon-limit-quark-SE}
\end{align}
Employing the identity
\begin{align}
\int_{\bm{l}_{\perp}} \frac{1}{\bm{l}_{\perp}^{2}-i\varepsilon}= \int_{\bm{z}_{\perp}} \int_{\bm{l}_{\perp}}  e^{i\bm{l}_{\perp}.\bm{r}_{zx}} \frac{l^{i}}{\bm{l}_{\perp}^{2}-i\varepsilon} \int_{\bm{l'}_{\perp}} e^{i\bm{l'}_{\perp}.\bm{r}_{xz}} \frac{{l'}^{i}}{\bm{{l'}}_{\perp}^{2}-i\varepsilon} \, ,
\end{align}
we can deduce an identical form for this limiting expression as in the case of contributions with dressed gluon propagators. Using the above simplification, and the identity in Eq.~\ref{eq:derivative-2D-gluon-propagator}, the amplitude for  $(S13)$ given in Eq.~\ref{eq:amplitude-S13-generic} simplifies to 
\begin{align}
\mathcal{M}^{(S13)}_{\mu \alpha}&= -\frac{\alpha_{S}}{\pi^{2}} \int \frac{\mathrm{d}z_{l}}{z_{l}} \int_{\bm{z}_{\perp}} \frac{1}{(\bm{x}_{\perp}-\bm{z}_{\perp})^{2}} \Big\{ 2\pi (eq_{f})^{2} \delta(1-z_{\rm tot}^{v}) \int \mathrm{d} \Pi_{\perp}^{\rm LO} \, \overline{u} (\bm{k}) R^{\rm LO:(1)}_{\mu \alpha} (\bm{l}_{1\perp}) v(\bm{p}) \Big\} \nonumber \\
& \times C_{F} \Big( \tilde{U}(\bm{x}_{\perp})\tilde{U}^{\dagger}(\bm{y}_{\perp}) -\mathds{1} \Big) \, .
\label{eq:soft-gluon-limit-S13}
\end{align}
The above expression has the same form as Eq.~\ref{eq:soft-gluon-limit-S1} albeit with a negative sign and a different color structure. The corresponding expression for its quark-antiquark interchanged counterpart is obtained by imposing $\bm{x}_{\perp}\leftrightarrow \bm{y}_{\perp}$. A similar simplification occurs for $(S14)$-$(S16)$ with the replacement $R^{\rm LO:(1)}_{\mu \alpha}$ with the corresponding LO $R$-functions for these graphs. 

Recall that as discussed previously in Section~\ref{sec:virtual-self-energy}, the diagrams labeled $(S17)$-$(S20)$ in Fig.~\ref{fig:NLO-self-3} vanish in general. 
More nontrivial is that fact that the limiting forms of the last four processes labeled $(S21)-(S24)$ in Fig.~\ref{fig:NLO-self-3} (and their $q\leftrightarrow\bar{q}$ interchanged counterparts) cancel amongst each another. This can be shown explicitly by using Eq.~\ref{eq:slow-gluon-limit-quark-SE} for $(S21)$ and $(S23)$ in Fig.~\ref{fig:NLO-self-3} and the simplification of 
$\tilde{\Sigma}_{\alpha}(k_{f},k_{\gamma})$ in Eq.~\ref{eq:self-energy-nested-photon-case-A},
\begin{align}
\lim_{z_{l} \rightarrow 0} \tilde{\Sigma}_{\alpha}(k_{f},k_{\gamma}) = -ieq_{f} g^{2} \, C_{F} \int \frac{\mathrm{d}z_{l}}{(2\pi) \, z_{l}} \int_{\bm{l}_{\perp}} \frac{1}{\bm{l}_{\perp}^{2}-i\varepsilon} \, \frac{\gamma^{-} \slashed{k}_{f} \gamma_{\alpha} (\slashed{k}_{f}+\slashed{k}_{\gamma}) \gamma^{-}}{2k_{f}^{-} (k_{f}^{-}+k_{\gamma}^{-})} \, ,
\end{align}
for the  graphs $(S22)$ and $(S24)$ shown in Fig.~\ref{fig:self-energy-nested-photon}.

Thus $(S13)$-$(S16)$ (and their quark-antiquark exchanged counterpart) are the only surviving self-energy contributions with the free gluon propagator; the net result in the slow gluon limit of the 24 self-energy diagrams is simply
\begin{align}
\lim_{z_{l}\rightarrow 0} \mathcal{M}^{\rm NLO:SE(3)}_{\mu \alpha}&=-\frac{\alpha_{S}}{\pi^{2}} \int \frac{\mathrm{d}z_{l}}{z_{l}} \int_{\bm{z}_{\perp}} \Big[\frac{1}{(\bm{x}_{\perp}-\bm{z}_{\perp})^{2}}+\frac{1}{(\bm{y}_{\perp}-\bm{z}_{\perp})^{2}} \Big]  \Big\{ 2\pi (eq_{f})^{2} \delta(1-z_{\rm tot}^{v}) \int \mathrm{d} \Pi_{\perp}^{\rm LO} \, \overline{u} (\bm{k}) T^{\rm LO}_{\mu \alpha} (\bm{l}_{1\perp}) v(\bm{p}) \Big\} \nonumber \\
& \times C_{F} \Big( \tilde{U}(\bm{x}_{\perp})\tilde{U}^{\dagger}(\bm{y}_{\perp}) -\mathds{1} \Big) \, .
\label{eq:soft-gluon-limit-SE3}
\end{align}
We used here the relation  $T^{\text{LO}}_{\mu \alpha} (\bm{l}_{\perp}) =\sum_{i=1}^{4} R^{\text{LO}:(i)}_{\mu \alpha}(\bm{l}_{\perp})$ given in Eq.~\ref{eq:T-LO}.

We now turn to the structure of vertex corrections in the slow gluon limit. As an example, consider the amplitude $(V1)$ we worked out previously for the dressed gluon propagator, the results for which are spelled out in Eqs.~\ref{eq:amplitude-diagV1-modified}, \ref{eq:numerator-V1-modified} and \ref{eq:denominator-V1-modified}. After further redefining $\bm{l}_{1\perp}-\bm{l}_{2\perp} \rightarrow \bm{l}_{1\perp}$, and using Eq.~\ref{eq:derivative-2D-gluon-propagator}, we obtain in the $z_{l} \rightarrow 0$ limit,
\begin{align}
\lim_{z_{l} \rightarrow 0}   \mathcal{M}_{\mu \alpha}^{(V1)}&= -\frac{\alpha_{S}}{\pi^{2}} \int \frac{\mathrm{d}z_{l}}{z_{l}} \int_{\bm{z}_{\perp}} \frac{(\bm{x}_{\perp}-\bm{z}_{\perp}).(\bm{y}_{\perp}-\bm{z}_{\perp})}{(\bm{x}_{\perp}-\bm{z}_{\perp})^{2} \, (\bm{y}_{\perp}-\bm{z}_{\perp})^{2}} \Big\{ 2\pi (eq_{f})^{2} \delta(1-z_{\rm tot}^{v}) \int \mathrm{d} \Pi_{\perp}^{\rm LO} \, \overline{u} (\bm{k}) R^{\rm LO:(1)}_{\mu \alpha} (\bm{l}_{1\perp}) v(\bm{p}) \Big\} \nonumber \\
& \times  \Bigg[ \Big( t^{b} \tilde{U} (\bm{x}_{\perp}) t^{a} \tilde{U}^{\dagger} (\bm{y}_{\perp}) \Big) U_{ba}(\bm{z}_{\perp}) - C_{F} \mathds{1} \Bigg] \, .
\label{eq:soft-gluon-limit-V1}
\end{align}
Similarly, it can be shown that the limiting forms for processes $(V2)$ and $(V5)$ are zero. Combining the contributions for the remaining three processes, we can write the total amplitude for the 6 processes in Fig.~\ref{fig:NLO-vertex-1} under the $z_{l} \rightarrow 0$ limit as
\begin{align}
\lim_{z_{l} \rightarrow 0}   \mathcal{M}_{\mu \alpha}^{\rm NLO:Ver(1)}&= -\frac{\alpha_{S}}{\pi^{2}} \int \frac{\mathrm{d}z_{l}}{z_{l}} \int_{\bm{z}_{\perp}} \frac{(\bm{x}_{\perp}-\bm{z}_{\perp}).(\bm{y}_{\perp}-\bm{z}_{\perp})}{(\bm{x}_{\perp}-\bm{z}_{\perp})^{2} \, (\bm{y}_{\perp}-\bm{z}_{\perp})^{2}} \Big\{ 2\pi (eq_{f})^{2} \delta(1-z_{\rm tot}^{v}) \int \mathrm{d} \Pi_{\perp}^{\rm LO} \, \overline{u} (\bm{k}) T^{\rm LO}_{\mu \alpha} (\bm{l}_{1\perp}) v(\bm{p}) \Big\} \nonumber \\
& \times  \Bigg[ \Big( t^{b} \tilde{U} (\bm{x}_{\perp}) t^{a} \tilde{U}^{\dagger} (\bm{y}_{\perp}) \Big) U_{ba}(\bm{z}_{\perp}) - C_{F} \mathds{1} \Bigg] \, .
\label{eq:soft-gluon-limit-Ver1}
\end{align}
The structure of this kernel is different from that of the self-energy corrections because here the Weizs\"{a}cker-Williams fields are emitted by both the quark and the antiquark. Since the kernel is  symmetric under $\bm{x}_{\perp} \leftrightarrow \bm{y}_{\perp}$, the limiting expression for the amplitude under quark-antiquark interchange has the same form; note however that for the latter,  the color structure is instead $( \tilde{U}(\bm{x}_{\perp})t^{a}\tilde{U}^{\dagger} (\bm{y}_{\perp}) t^{b}   ) \, U_{ba}(\bm{z}_{\perp})-C_{F} \mathds{1} $.

Likewise, one can show that the corresponding virtual amplitudes with free gluon propagators (in Figs.~\ref{fig:NLO-vertex-3} and \ref{fig:NLO-vertex-4}) are 
\begin{align}
\lim_{z_{l} \rightarrow 0}   \mathcal{M}_{\mu \alpha}^{\rm NLO:Ver(3)}&= \frac{\alpha_{S}}{\pi^{2}} \int \frac{\mathrm{d}z_{l}}{z_{l}} \int_{\bm{z}_{\perp}} \frac{(\bm{x}_{\perp}-\bm{z}_{\perp}).(\bm{y}_{\perp}-\bm{z}_{\perp})}{(\bm{x}_{\perp}-\bm{z}_{\perp})^{2} \, (\bm{y}_{\perp}-\bm{z}_{\perp})^{2}} \Big\{ 2\pi (eq_{f})^{2} \delta(1-z_{\rm tot}^{v}) \int \mathrm{d} \Pi_{\perp}^{\rm LO} \, \overline{u} (\bm{k}) T^{\rm LO}_{\mu \alpha}(\bm{l}_{1\perp}) v(\bm{p}) \Big\} \nonumber \\
& \times  C_{F} \Big(\tilde{U}(\bm{x}_{\perp}) \tilde{U}^{\dagger}(\bm{y}_{\perp}) -\mathds{1} \Big)  \, ,
\label{eq:soft-gluon-limit-Ver3}
\end{align}
\begin{align}
\lim_{z_{l} \rightarrow 0}   \mathcal{M}_{\mu \alpha}^{\rm NLO:Ver(4)}&= \frac{\alpha_{S}}{\pi^{2}} \int \frac{\mathrm{d}z_{l}}{z_{l}} \int_{\bm{z}_{\perp}} \frac{(\bm{x}_{\perp}-\bm{z}_{\perp}).(\bm{y}_{\perp}-\bm{z}_{\perp})}{(\bm{x}_{\perp}-\bm{z}_{\perp})^{2} \, (\bm{y}_{\perp}-\bm{z}_{\perp})^{2}} \Big\{ 2\pi (eq_{f})^{2} \delta(1-z_{\rm tot}^{v}) \int \mathrm{d} \Pi_{\perp}^{\rm LO} \, \overline{u} (\bm{k}) T^{\rm LO}_{\mu \alpha}(\bm{l}_{1\perp}) v(\bm{p}) \Big\} \nonumber \\
& \times \Big(t^{a}\tilde{U}(\bm{x}_{\perp}) \tilde{U}^{\dagger}(\bm{y}_{\perp}) t_{a}-C_{F} \mathds{1}  \Big)  \, ,
\label{eq:soft-gluon-limit-Ver4}
\end{align}

At the amplitude squared level, we have to consider the interference contribution of these virtual graphs with the leading order amplitude result in Eq.~\ref{eq:LO-amp-master}. As outlined in the introduction, we then need to perform a CGC averaging over all possible source charge configurations $\rho_{A}$. Using the expressions for the loop contributions derived above in the $z_{l} \rightarrow 0$ limit, and the color structures given Table~\ref{tab:NLO-assembly}, we can obtain the leading logarithmic singular structure (in rapidity) of the squared amplitude for virtual graphs. These are summarized in Table~\ref{tab:amp-squared-LLog-Virtual-LO}. 
\begin{table}[!htbp]
\caption{LL$x$ structure of the amplitude squared from interference contributions of virtual graphs with LO processes. The kernels for the complex conjugates of the squared amplitudes in rows 3 and 4 are obtained by replacing $\bm{x}_{\perp} \rightarrow \bm{x'}_{\perp}$, $\bm{y}_{\perp} \rightarrow \bm{y'}_{\perp}$. The color structure and kernels for the complex conjugates of rows 1 and 2 are obtained by permutations of the coordinates.}
\label{tab:amp-squared-LLog-Virtual-LO}
 \renewcommand{\arraystretch}{2.5}
\begin{center}
    \begin{tabular}{ | l | p{2cm} | p{4.5cm} | p{4.5cm} | }
   \hline
  \multicolumn{4}{|c|}{CGC averaged amplitude squared$=$ (Common factors) $\times$ (Structure of kernel) $\times$ (Color structure)} \\
  \hline
   $\left \langle {\mathcal{M}^{\rm LO}}^{*} \, \Big( \mathcal{M}^{\rm NLO:SE(1)} + \mathcal{M}^{\rm NLO:Ver(1)} \Big) \right \rangle $ & $\frac{\alpha_{S}}{2\pi^{2}} \int \frac{\mathrm{d}z_{l}}{z_{l}} \,  C^{\rm LO} $ & $ \frac{1}{(\bm{x}_{\perp}-\bm{z}_{\perp})^{2}} -\frac{(\bm{x}_{\perp}-\bm{z}_{\perp}).(\bm{y}_{\perp}-\bm{z}_{\perp})}{(\bm{x}_{\perp}-\bm{z}_{\perp})^{2} \, (\bm{y}_{\perp}-\bm{z}_{\perp})^{2}}$& $N_{c}^{2}\Big(1-D_{y'x'}+Q_{zy;y'x'}D_{xz}-D_{xz}D_{zy} \Big) -\Xi(\bm{x}_{\perp},\bm{y}_{\perp};\bm{y'}_{\perp},\bm{x'}_{\perp})$   \\ \hline
    $\left \langle {\mathcal{M}^{\rm LO}}^{*} \, \Big( \mathcal{M}^{\rm NLO:SE(2)} + \mathcal{M}^{\rm NLO:Ver(2)} \Big) \right \rangle $ & $\frac{\alpha_{S}}{2\pi^{2}} \int \frac{\mathrm{d}z_{l}}{z_{l}} \,  C^{\rm LO} $ & $ \frac{1}{(\bm{y}_{\perp}-\bm{z}_{\perp})^{2}} -\frac{(\bm{x}_{\perp}-\bm{z}_{\perp}).(\bm{y}_{\perp}-\bm{z}_{\perp})}{(\bm{x}_{\perp}-\bm{z}_{\perp})^{2} \, (\bm{y}_{\perp}-\bm{z}_{\perp})^{2}}$& $N_{c}^{2}\Big(1-D_{y'x'}+Q_{xz;y'x'}D_{zy}-D_{xz}D_{zy} \Big) -\Xi(\bm{x}_{\perp},\bm{y}_{\perp};\bm{y'}_{\perp},\bm{x'}_{\perp})$   \\ \hline
  $\left \langle {\mathcal{M}^{\rm LO}}^{*} \, \Big( \mathcal{M}^{\rm NLO:SE(3)} + \mathcal{M}^{\rm NLO:Ver(3)} \Big) + \text{c.c}  \right  \rangle $ & $\frac{\alpha_{S}}{2\pi^{2}} \int \frac{\mathrm{d}z_{l}}{z_{l}} \,  C^{\rm LO} $ & $\Big[ - \frac{1}{(\bm{x}_{\perp}-\bm{z}_{\perp})^{2}} - \frac{1}{(\bm{y}_{\perp}-\bm{z}_{\perp})^{2}} + \frac{(\bm{x}_{\perp}-\bm{z}_{\perp}).(\bm{y}_{\perp}-\bm{z}_{\perp})}{(\bm{x}_{\perp}-\bm{z}_{\perp})^{2} \, (\bm{y}_{\perp}-\bm{z}_{\perp})^{2}} \Big] +\Big(\substack{  \bm{x}_{\perp} \rightarrow \bm{x'}_{\perp} \\
\bm{y}_{\perp} \rightarrow \bm{y'}_{\perp}}  \Big)$& $N_{c}^{2}\Big(1-D_{xy}-D_{y'x'}+Q_{xy;y'x'} \Big) -\Xi(\bm{x}_{\perp},\bm{y}_{\perp};\bm{y'}_{\perp},\bm{x'}_{\perp})$   \\ \hline
  $\left \langle {\mathcal{M}^{\rm LO}}^{*} \,  \mathcal{M}^{\rm NLO:Ver(4)} +\text{c.c} \right \rangle $ & $\frac{\alpha_{S}}{2\pi^{2}} \int \frac{\mathrm{d}z_{l}}{z_{l}} \,  C^{\rm LO} $ & $ \frac{(\bm{x}_{\perp}-\bm{z}_{\perp}).(\bm{y}_{\perp}-\bm{z}_{\perp})}{(\bm{x}_{\perp}-\bm{z}_{\perp})^{2} \, (\bm{y}_{\perp}-\bm{z}_{\perp})^{2}}+\Big(\substack{  \bm{x}_{\perp} \rightarrow \bm{x'}_{\perp} \\
\bm{y}_{\perp} \rightarrow \bm{y'}_{\perp}}  \Big)$ & $N_{c}^{2}\Big(1-D_{xy}-D_{y'x'}+D_{xy}D_{y'x'} \Big) -\Xi(\bm{x}_{\perp},\bm{y}_{\perp};\bm{y'}_{\perp},\bm{x'}_{\perp})$   \\ \hline   
    \end{tabular}
\end{center}
\end{table}

The coefficient function $C^{\rm LO}$ in this table was defined previously in Eq.~\ref{eq:LO-coefficient-function}. The ``CGC averaged" leading order squared amplitude constituting the LO hadron tensor in Eq.~\ref{eq:H-tensor} can be expressed as
\begin{equation}
\left \langle {\mathcal{M}^{\rm LO}}^{*} \mathcal{M}^{\rm LO} \right \rangle = N_{c} \, C^{\rm LO} \times \Xi(\bm{x}_{\perp},\bm{y}_{\perp};\bm{y'}_{\perp},\bm{x'}_{\perp}) \, ,
\label{eq:LO-amplitude-squared}
\end{equation}
with $\Xi(\bm{x}_{\perp},\bm{y}_{\perp};\bm{y'}_{\perp},\bm{x'}_{\perp})=1-D_{xy}-D_{y'x'}+Q_{xy;y'x'}$, where the dipole ($D$) and quadrupole ($Q$) traces over the lightlike Wilson lines were 
defined previously in Eq.~\ref{eq:dipole-quadrupole-Wilson-line-correlators}.
%
%where 
%\begin{align}
%D_{xy}& =\frac{1}{N_{c}} \left \langle \text{Tr}\Big( \tilde{U}(\bm{x}_{\perp}) \tilde{U}^{\dagger}(\bm{y}_{\perp}) \Big) \right \rangle \, , \nonumber \\
%Q_{xy;zw} & =\frac{1}{N_{c}} \left \langle \text{Tr} \Big( \tilde{U}(\bm{x}_{\perp}) \tilde{U}^{\dagger}(\bm{y}_{\perp})  \tilde{U}(\bm{z}_{\perp}) \tilde{U}^{\dagger}(\bm{w}_{\perp}) \Big)\right \rangle =Q_{zw;xy} \, ,
%\end{align} 

We will now repeat the above exercise by taking the slow gluon limit of gluon emission diagrams. We will first take the $z_{g} \, (=k_{g}^{-}/q^{-}) \rightarrow 0$ limit in these amplitudes, take their modulus squared, and then integrate over the phase space of the emitted gluon. We begin by considering the contribution from the 10 graphs in which the gluon crosses the nuclear shock wave.  (One half of these are shown in Fig.~\ref{fig:NLO-real-allscatter} and the other half are obtained via quark-antiquark exchange.) Recall that their combined contribution was expressed in Eq.~\ref{eq:amplitude-R1-R10} as
\begin{align}
\mathcal{M}_{\mu \alpha;b}^{\rm NLO:Real(1)}&=\sum_{\beta=1}^{10} \mathcal{M}^{(R\beta)}_{\mu \alpha;b} \nonumber \\
& =2\pi (eq_{f})^{2}g \, \delta(1-z_{\text{tot}}^{r}) \int \mathrm{d}  \Pi_{\perp}^{\rm LO} \,\, \overline{u}(\bm{k})  \Bigg\{  \int_{\bm{z}_{\perp}} \!\! e^{-i \bm{k}_{g\perp}.\bm{z}_{\perp}} \,  T^{(1)}_{R;\mu \alpha}  (\bm{l}_{1\perp})   \Big[ \Big( \tilde{U}(\bm{x}_{\perp}) t^{a} \tilde{U}^{\dagger}(\bm{y}_{\perp}) \Big) U_{ab}(\bm{z}_{\perp})-t_{b} \Big] \Bigg\}  v(\bm{p}) \, ,
\end{align}
where $z_{\text{tot}}^{r}=z_{q}+z_{\bar{q}}+z_{\gamma}+z_{g}$ and  $T^{(1)}_{R}$ is defined to be
\begin{equation}
T_{R,\mu \alpha}^{(1)} (\bm{l}_{1\perp}) =\sum_{\beta=1}^{10} R^{(R\beta)}_{\mu \alpha}(\bm{l}_{1\perp} ) \, .
\end{equation}
As an example, consider the process $(R1)$ for which the $R$-function (in Eq.~\ref{eq:R-R1}) is
\begin{align}
R^{(R1)}_{\mu \alpha} (\bm{l}_{1\perp}) = &  \gamma_{\alpha}\frac{\slashed{k}+\slashed{k}_{\gamma}}{2k.k_{\gamma}} \gamma^{-} \Big[ \Big(  \{ \gamma^{+}(1-z_{\bar{q}}-z_{g})q^{-}-\bm{\gamma}_{\perp}.\bm{l}_{1\perp}\} \, \mathcal{I}^{(1,0)}_{r}(\bm{v}_{\perp}^{(R1)},\Delta^{(R1)};\bm{r}_{zx}) + \gamma^{j} \mathcal{I}^{(1,j)}_{r} (\bm{v}_{\perp}^{(R1)},\Delta^{(R1)}; \bm{r}_{zx}) \Big)    \nonumber \\
& \times \bm{\gamma}_{\perp}.\bm{\epsilon}_{\perp}^{*} (\bm{k}_{g})- 2 \frac{(1-z_{\bar{q}}-z_{g})}{z_{g}} \, \mathcal{I}^{(1,i)}_{r} (\bm{v}_{\perp}^{(R1)},\Delta^{(R1)};\bm{r}_{zx}) \, {\epsilon^{i}}^{*}(\bm{k}_{g}) \Big] \,   \nonumber \\
&  \times \frac{\gamma^{+}(1-z_{\bar{q}})q^{-}-\gamma^{-}\Big( Q^{2}z_{\bar{q}}+\bm{l}_{1\perp}^{2} \Big)/2z_{\bar{q}}q^{-} -\bm{\gamma}_{\perp}.\bm{l}_{1\perp}  }{\bm{l}_{1\perp}^{2}+Q^{2}z_{\bar{q}}(1- z_{\bar{q}}) -i\varepsilon} \gamma_{\mu}   \frac{\gamma^{+}z_{\bar{q}}q^{-}+\bm{\gamma}_{\perp}.\bm{l}_{1\perp}}{2(q^{-})^{2} (1-z_{\bar{q}})/z_{g}} \gamma^{-}  \, ,
\end{align}
where the $\mathcal{I}$ functions are proportional to modified Bessel functions of the second kind and are given in Eq.~\ref{eq:I-factors-R1}. The factors in the arguments of these functions 
multiplying $\bm{r}_{zx}=\bm{z}_{\perp}-\bm{x}_{\perp}$ are
\begin{align}
\bm{v}_{\perp}^{(R1)}& =- \frac{z_{g}}{1-z_{\bar{q}}} \, \bm{l}_{1\perp}  \, , \Delta^{(R1)} = \frac{z_{g}}{(1-z_{\bar{q}})^{2}} \, \frac{1-z_{\bar{q}}-z_{g}}{z_{\bar{q}}} \, \bm{l}_{1\perp}^{2} +\frac{Q^{2}z_{g} (1-z_{\bar{q}}-z_{g})}{1-z_{\bar{q}}}-i\varepsilon \, ,
\end{align}
which vanish in the $z_{g} \rightarrow 0$ limit. Under these conditions, we have 
\begin{equation}
\lim_{z_{g} \rightarrow 0} \mathcal{I}^{(1,i)}_{r} (\bm{v}_{\perp}^{(R1)},\Delta^{(R1)};\bm{r}_{zx}) = \frac{i}{2\pi } \, \frac{r^{i}_{zx}}{\bm{r}_{zx}^{2}} \, ,
\end{equation}
and 
\begin{align}
\lim_{z_{g} \rightarrow 0} R^{(R1)}_{\mu \alpha} (\bm{l}_{1\perp}) &= \frac{ig}{\pi} \, \frac{z^{i}-x^{i}}{(\bm{x}_{\perp}-\bm{z}_{\perp})^{2}} \, {\epsilon^{i}}^{*} (\bm{k}_{g}) \, \Big\{ 2\pi (eq_{f})^{2} \delta(1-z_{\rm tot}^{v}) \int \mathrm{d} \Pi_{\perp}^{\rm LO} \, \overline{u} (\bm{k}) R^{\rm LO:(1)}_{\mu \alpha}(\bm{l}_{1\perp}) v(\bm{p}) \Big\} \, .
\label{eq:soft-gluon-limit-R1}
\end{align}

One can similarly derive limiting expressions for the remaining 4 diagrams in Fig.~\ref{fig:NLO-real-allscatter}. The coordinate space structure of the kernel remains the same for these diagrams but they are proportional to different $R^{\rm LO}$'s. For the quark$\leftrightarrow$ antiquark interchanged diagrams, we can simply replace $\bm{x}_{\perp} \leftrightarrow \bm{y}_{\perp}$ along with an overall change of sign to get the corresponding expressions in the $z_{g} \rightarrow 0$ limit. We obtain the $z_{g} \rightarrow 0$ limit of the amplitude from these 10 contributions as
\begin{align}
\lim_{z_{g} \rightarrow 0} \mathcal{M}_{\mu \alpha;b}^{\rm NLO:Real(1)}&=\frac{ig}{\pi} \int_{\bm{z}_{\perp}} e^{-i\bm{k}_{g\perp}.\bm{z}_{\perp}} \Big( \frac{z^{i}-x^{i}}{(\bm{x}_{\perp}-\bm{z}_{\perp})^{2}} - \frac{z^{i}-y^{i}}{(\bm{y}_{\perp}-\bm{z}_{\perp})^{2}} \Big) \, {\epsilon^{i}}^{*} (\bm{k}_{g}) \Big\{ 2\pi (eq_{f})^{2} \delta(1-z_{\rm tot}^{v})  \nonumber \\
& \times \int \mathrm{d} \Pi_{\perp}^{\rm LO} \, \overline{u} (\bm{k}) T^{\rm LO}_{\mu \alpha}(\bm{l}_{1\perp}) v(\bm{p}) \Big\} \Big[\Big(\tilde{U}(\bm{x}_{\perp}) t^{a} \tilde{U}^{\dagger} (\bm{y}_{\perp}) \Big) U_{ba} (\bm{z}_{\perp}) -t_{b} \Big]  \, .
\label{eq:soft-gluon-limit-R1-R10}
\end{align}

\begin{table}[!htbp]
\caption{LL$x$ structure of the amplitude squared for real gluon emission diagrams. The kernels for the complex conjugates of the squared amplitudes in rows 1 and 2 and their color structures are obtained by permutations of the coordinates.}
\label{tab:amp-squared-LLog-Real-NLO}
 \renewcommand{\arraystretch}{2}
\begin{center}
    \begin{tabular}{ | p{6.5cm} | p{2cm} | p{3.5cm} | p{4.5cm} | }
    \hline
  \multicolumn{4}{| l |}{ \shortstack[l]{CGC averaged amplitude squared\\
 with gluon phase space integrated over $=$ (Common factors) $\times$ (Structure of kernel) $\times$ (Color structure)} }   \\
  \hline
   $\int_{\mathrm{d} \Omega (\bm{k}_{g}) }  \left \langle {\mathcal{M}^{\rm NLO:Real(2)}}^{*} \,  \mathcal{M}^{\rm NLO:Real(1)} \right \rangle $ & $\frac{\alpha_{S}}{2\pi^{2}} \int \frac{\mathrm{d}z_{g}}{z_{g}} \,  C^{\rm LO} $ & $ \Big\{ \frac{(\bm{x'}_{\perp}-\bm{z}_{\perp}).(\bm{y}_{\perp}-\bm{z}_{\perp})}{(\bm{x'}_{\perp}-\bm{z}_{\perp})^{2} \, (\bm{y}_{\perp}-\bm{z}_{\perp})^{2}} - \frac{(\bm{x}_{\perp}-\bm{z}_{\perp}).(\bm{x'}_{\perp}-\bm{z}_{\perp})}{(\bm{x}_{\perp}-\bm{z}_{\perp})^{2} \, (\bm{x'}_{\perp}-\bm{z}_{\perp})^{2}} \Big\} $& $N_{c}^{2}\Big(1-D_{y'x'}+Q_{zy;y'x'}D_{xz}-D_{xz}D_{zy} \Big) -\Xi(\bm{x}_{\perp},\bm{y}_{\perp};\bm{y'}_{\perp},\bm{x'}_{\perp})$   \\ \hline
    $\int_{\mathrm{d} \Omega (\bm{k}_{g}) }  \left \langle {\mathcal{M}^{\rm NLO:Real(3)}}^{*} \,  \mathcal{M}^{\rm NLO:Real(1)} \right \rangle$ & $\frac{\alpha_{S}}{2\pi^{2}} \int \frac{\mathrm{d}z_{g}}{z_{g}} \,  C^{\rm LO} $ & $ \Big\{ \frac{(\bm{x}_{\perp}-\bm{z}_{\perp}).(\bm{y'}_{\perp}-\bm{z}_{\perp})}{(\bm{x}_{\perp}-\bm{z}_{\perp})^{2} \, (\bm{y'}_{\perp}-\bm{z}_{\perp})^{2}} - \frac{(\bm{y}_{\perp}-\bm{z}_{\perp}).(\bm{y'}_{\perp}-\bm{z}_{\perp})}{(\bm{y}_{\perp}-\bm{z}_{\perp})^{2} \, (\bm{y'}_{\perp}-\bm{z}_{\perp})^{2}} \Big\} $& $N_{c}^{2}\Big(1-D_{y'x'}+Q_{xz;y'x'}D_{zy}-D_{xz}D_{zy} \Big) -\Xi(\bm{x}_{\perp},\bm{y}_{\perp};\bm{y'}_{\perp},\bm{x'}_{\perp})$   \\ \hline
  $$\int_{\mathrm{d} \Omega (\bm{k}_{g}) }  \sum_{i=2}^{3} \left \langle {\mathcal{M}^{\rm NLO:Real(i)}}^{*} \,  \mathcal{M}^{\rm NLO:Real(i)} \right \rangle  $$ & $\frac{\alpha_{S}}{2\pi^{2}} \int \frac{\mathrm{d}z_{g}}{z_{g}} \,  C^{\rm LO} $ & $  \Big\{  \frac{(\bm{x}_{\perp}-\bm{z}_{\perp}).(\bm{x'}_{\perp}-\bm{z}_{\perp})}{(\bm{x}_{\perp}-\bm{z}_{\perp})^{2} \, (\bm{x'}_{\perp}-\bm{z}_{\perp})^{2}} + \frac{(\bm{y}_{\perp}-\bm{z}_{\perp}).(\bm{y'}_{\perp}-\bm{z}_{\perp})}{(\bm{y}_{\perp}-\bm{z}_{\perp})^{2} \, (\bm{y'}_{\perp}-\bm{z}_{\perp})^{2}}  \Big\} $& $N_{c}^{2}\Big(1-D_{xy}-D_{y'x'}+Q_{xy;y'x'} \Big) -\Xi(\bm{x}_{\perp},\bm{y}_{\perp};\bm{y'}_{\perp},\bm{x'}_{\perp})$   \\ \hline
   $\int_{\mathrm{d} \Omega (\bm{k}_{g}) }  \left \langle {\mathcal{M}^{\rm NLO:Real(2)}}^{*}   \mathcal{M}^{\rm NLO:Real(3)}+\rm{c.c} \right \rangle$ & $\frac{\alpha_{S}}{2\pi^{2}} \int \frac{\mathrm{d}z_{g}}{z_{g}} \,  C^{\rm LO} $ & $ \Big\{  -\frac{(\bm{x}_{\perp}-\bm{z}_{\perp}).(\bm{y'}_{\perp}-\bm{z}_{\perp})}{(\bm{x}_{\perp}-\bm{z}_{\perp})^{2} \, (\bm{y'}_{\perp}-\bm{z}_{\perp})^{2}} - \frac{(\bm{y}_{\perp}-\bm{z}_{\perp}).(\bm{x'}_{\perp}-\bm{z}_{\perp})}{(\bm{y}_{\perp}-\bm{z}_{\perp})^{2} \, (\bm{x'}_{\perp}-\bm{z}_{\perp})^{2}} \Big\}$ & $N_{c}^{2}\Big(1-D_{xy}-D_{y'x'}+D_{xy}D_{y'x'} \Big) -\Xi(\bm{x}_{\perp},\bm{y}_{\perp};\bm{y'}_{\perp},\bm{x'}_{\perp})$   \\ \hline   
     $\int_{\mathrm{d} \Omega (\bm{k}_{g}) }  \left \langle {\mathcal{M}^{\rm NLO:Real(1)}}^{*} \,  \mathcal{M}^{\rm NLO:Real(1)} \right \rangle $ & $\frac{\alpha_{S}}{2\pi^{2}} \int \frac{\mathrm{d}z_{g}}{z_{g}} \,  C^{\rm LO} $ & $ \Big\{ \frac{(\bm{x}_{\perp}-\bm{z}_{\perp}).(\bm{x'}_{\perp}-\bm{z}_{\perp})}{(\bm{x}_{\perp}-\bm{z}_{\perp})^{2} \, (\bm{x'}_{\perp}-\bm{z}_{\perp})^{2}} - \frac{(\bm{x}_{\perp}-\bm{z}_{\perp}).(\bm{y'}_{\perp}-\bm{z}_{\perp})}{(\bm{x}_{\perp}-\bm{z}_{\perp})^{2} \, (\bm{y'}_{\perp}-\bm{z}_{\perp})^{2}}- \frac{(\bm{x'}_{\perp}-\bm{z}_{\perp}).(\bm{y}_{\perp}-\bm{z}_{\perp})}{(\bm{x'}_{\perp}-\bm{z}_{\perp})^{2} \, (\bm{y}_{\perp}-\bm{z}_{\perp})^{2}} + \frac{(\bm{y}_{\perp}-\bm{z}_{\perp}).(\bm{y'}_{\perp}-\bm{z}_{\perp})}{(\bm{y}_{\perp}-\bm{z}_{\perp})^{2} \, (\bm{y'}_{\perp}-\bm{z}_{\perp})^{2}}\Big\} $& $N_{c}^{2}\Big(1-D_{xz}D_{zy}-D_{y'z}D_{zx'}+D_{xx'}D_{y'y} \Big) -\Xi(\bm{x}_{\perp},\bm{y}_{\perp};\bm{y'}_{\perp},\bm{x'}_{\perp})$   \\ \hline
    \end{tabular}
\end{center}
\end{table}
As a final demonstration, we will consider $(R11)$, a diagram where the gluon is emitted by the quark and does not scatter off the background classical field. Its amplitude 
was given previously in Eq.~\ref{eq:amplitude-R11}. In the $z_{g} \rightarrow 0$ limit, this amplitude reduces to 
\begin{align}
\lim_{z_{g} \rightarrow 0} \mathcal{M}^{(R11)}_{\mu \alpha;b}& = -g \, \frac{2k_{g}^{i} {\epsilon^{i}}^{*}  (\bm{k}_{g}) }{\bm{k}_{g\perp}^{2}-i\varepsilon} \Big\{ 2\pi (eq_{f})^{2} \delta(1-z_{\rm tot}^{v}) \int \mathrm{d} \Pi_{\perp}^{\rm LO} \, \overline{u} (\bm{k}) R^{\rm LO:(1)}_{\mu \alpha}(\bm{l}_{1\perp}) v(\bm{p}) \Big\} e^{-i\bm{k}_{g\perp}.\bm{x}_{\perp}} \nonumber \\
& \times \Big[ \Big( t_{b} \tilde{U}(\bm{x}_{\perp}) \tilde{U}^{\dagger} (\bm{y}_{\perp}) -t_{b} \Big] \, .
\end{align}
The limiting expression for the net contribution from the 5 processes in Fig.~\ref{fig:NLO-real-nogluonscatter} therefore can be written as
\begin{align}
\lim_{z_{l} \rightarrow 0}  \Big( \mathcal{M}_{\mu \alpha;b}^{\rm NLO:Real(2)}=\sum_{\beta=11}^{15} \mathcal{M}^{(R\beta)}_{\mu \alpha;b} \Big) &=  -g \, \frac{2k_{g}^{i} {\epsilon^{i}}^{*}  (\bm{k}_{g}) }{\bm{k}_{g\perp}^{2}-i\varepsilon} \Big\{ 2\pi (eq_{f})^{2} \delta(1-z_{\rm tot}^{v}) \int \mathrm{d} \Pi_{\perp}^{\rm LO} \, \overline{u} (\bm{k}) T^{\rm LO}_{\mu \alpha}(\bm{l}_{1\perp}) v(\bm{p}) \Big\} e^{-i\bm{k}_{g\perp}.\bm{x}_{\perp}} \nonumber \\
& \times \Big[ \Big( t_{b} \tilde{U}(\bm{x}_{\perp}) \tilde{U}^{\dagger} (\bm{y}_{\perp}) \Big)  -t_{b} \Big] \, .
\label{eq:soft-gluon-limit-R11-R15}
\end{align}
Similarly, the contribution from their quark$\leftrightarrow$antiquark interchanged counterparts are,
\begin{align}
\lim_{z_{l} \rightarrow 0}  \Big( \mathcal{M}_{\mu \alpha;b}^{\rm NLO:Real(3)}=\sum_{\beta=16}^{20} \mathcal{M}^{(R\beta)}_{\mu \alpha;b} \Big) &=  g \, \frac{2k_{g}^{i} {\epsilon^{i}}^{*}  (\bm{k}_{g}) }{\bm{k}_{g\perp}^{2}-i\varepsilon} \Big\{ 2\pi (eq_{f})^{2} \delta(1-z_{\rm tot}^{v}) \int \mathrm{d} \Pi_{\perp}^{\rm LO} \, \overline{u} (\bm{k}) T^{\rm LO}_{\mu \alpha}(\bm{l}_{1\perp}) v(\bm{p}) \Big\} e^{-i\bm{k}_{g\perp}.\bm{y}_{\perp}} \nonumber \\
& \times \Big[ \Big( \tilde{U}(\bm{x}_{\perp}) \tilde{U}^{\dagger} (\bm{y}_{\perp}) t_{b} \Big)  -t_{b} \Big] \, .
\label{eq:soft-gluon-limit-R16-R20}
\end{align}
These three amplitude structures therefore give 9 contributions at the level of the cross-section, many of which are Hermitian conjugates of each other. Since the final state process of interest 
is  inclusive photon+dijet production, we will integrate over the phase space of the emitted gluon. To obtain coordinate space kernels, we will liberally use Eq.~\ref{eq:derivative-2D-gluon-propagator} and the  relation between the two dimensional free propagator in momentum and coordinate space:
\begin{equation}
\int_{\bm{k}_{\perp}} \frac{e^{i\bm{k}_{\perp}.\bm{r}_{xy}}}{\bm{k}^{2}_{\perp}-i\varepsilon } = \int_{\bm{z}_{\perp}} \frac{1}{(2\pi)^{2}} \, \frac{(\bm{x}_{\perp}-\bm{z}_{\perp}).(\bm{y}_{\perp}-\bm{z}_{\perp})}{(\bm{x}_{\perp}-\bm{z}_{\perp})^{2} \, (\bm{y}_{\perp}-\bm{z}_{\perp})^{2}} \, .
\label{eq:2D-free-propagator}
\end{equation}
We can now extract the leading logarithm in $x$ (LL$x$) structures of the various real emission contributions to the differential cross-section at NLO. These are summarized in Table~\ref{tab:amp-squared-LLog-Real-NLO}. The quantities in the first column represent the CGC averaged squared amplitudes integrated over the phase space of the emitted gluon,
\begin{equation}
\int_{\mathrm{d} \Omega (\bm{k}_{g}) } \left \langle \mathcal{M}^{*} \mathcal{M} \right \rangle = \int \frac{\mathrm{d}^{2} \bm{k}_{g \perp}}{(2\pi)^{2}} \int \frac{\mathrm{d}z_{g}}{(2\pi) \, 2z_{g}}  \left \langle \mathcal{M}^{*} \mathcal{M} \right \rangle \, .
\end{equation}

We now have all the essential elements we promised in the introduction of the paper to derive the JIMWLK evolution equation. Using the leading logarithmic structures summarized in Tables~\ref{tab:amp-squared-LLog-Virtual-LO} and \ref{tab:amp-squared-LLog-Real-NLO} for virtual and real emissions we can organize the CGC averaged squared amplitudes (and their Hermitian conjugates) in a basis spanned by dipole, $D$ and quadrupole, $Q$ Wilson line correlators and their products $DD$ and $DQ$. (For a general introduction to such basis structures, and how to compute them, we refer the reader to \cite{Dusling:2017aot}.)

After some algebra, and use of the identity
\begin{align}
\frac{(\bm{x}_{\perp}-\bm{z}_{\perp}).(\bm{y}_{\perp}-\bm{z}_{\perp})}{(\bm{x}_{\perp}-\bm{z}_{\perp})^{2} \, (\bm{y}_{\perp}-\bm{z}_{\perp})^{2}} =\frac{1}{2} \Big[ -\frac{(\bm{x}_{\perp}-\bm{y}_{\perp})^{2}}{(\bm{x}_{\perp}-\bm{z}_{\perp})^{2} \, (\bm{y}_{\perp}-\bm{z}_{\perp})^{2}} + \frac{1}{(\bm{x}_{\perp}-\bm{z}_{\perp})^{2}} +\frac{1}{(\bm{y}_{\perp}-\bm{z}_{\perp})^{2}} \Big] \, .
\label{eq:JIMWLK-BFKL-kernel-connection}
\end{align}
we can derive the following leading logarithmic structure for the CGC averaged amplitude squared at NLO
\begin{align}
&  \left \langle {\mathcal{M}^{\rm NLO}}^{*} \, \mathcal{M}^{\rm NLO} \right \rangle \vert_{ LLx}  \nonumber \\
&= \lim_{z_{l} \rightarrow 0}  \Big\{ \int_{\mathrm{d} \Omega (\bm{k}_{g})} \left \langle {\mathcal{M}^{\rm NLO:Real}}^{*} \, \mathcal{M}^{\rm NLO:Real} \right \rangle \Big\}  +\lim_{z_{l} \rightarrow 0}   \Big( \left \langle {\mathcal{M}^{\rm LO}}^{*} \, \mathcal{M}^{\rm NLO:Virtual} +c.c \right \rangle \Big) \nonumber \\
&= \ln \Big( \frac{z_{f}}{z_{0}} \Big) \,   \Bigg\{ \frac{\alpha_{S}N_{c}}{2\pi^{2}} \int_{\bm{z}_{\perp}} \frac{(\bm{x}_{\perp}-\bm{y}_{\perp})^{2}}{(\bm{x}_{\perp}-\bm{z}_{\perp})^{2} \, (\bm{y}_{\perp}-\bm{z}_{\perp})^{2}}  \, D_{xy} + \frac{\alpha_{S}N_{c}}{2\pi^{2}} \int_{\bm{z}_{\perp}} \frac{(\bm{x'}_{\perp}-\bm{y'}_{\perp})^{2}}{(\bm{x'}_{\perp}-\bm{z}_{\perp})^{2} \, (\bm{y'}_{\perp}-\bm{z}_{\perp})^{2}} \, D_{y'x'} \nonumber \\
& -\frac{\alpha_{S} N_{c}}{(2\pi)^{2}} \int_{\bm{z}_{\perp}} \!\! \Big[ \frac{(\bm{x}_{\perp}-\bm{y}_{\perp})^{2}}{(\bm{x}_{\perp}-\bm{z}_{\perp})^{2} \, (\bm{y}_{\perp}-\bm{z}_{\perp})^{2}} +\frac{(\bm{x'}_{\perp}-\bm{y'}_{\perp})^{2}}{(\bm{x'}_{\perp}-\bm{z}_{\perp})^{2} \, (\bm{y'}_{\perp}-\bm{z}_{\perp})^{2}} +\frac{(\bm{x}_{\perp}-\bm{x'}_{\perp})^{2}}{(\bm{x}_{\perp}-\bm{z}_{\perp})^{2} \, (\bm{x'}_{\perp}-\bm{z}_{\perp})^{2}} \nonumber \\
& +\frac{(\bm{y}_{\perp}-\bm{y'}_{\perp})^{2}}{(\bm{y}_{\perp}-\bm{z}_{\perp})^{2} \, (\bm{y'}_{\perp}-\bm{z}_{\perp})^{2}} \Big] Q_{xy;y'x'}   -\frac{\alpha_{S}N_{c}}{2\pi^{2}} \int_{\bm{z}_{\perp}} \frac{(\bm{x}_{\perp}-\bm{y}_{\perp})^{2}}{(\bm{x}_{\perp}-\bm{z}_{\perp})^{2} \, (\bm{y}_{\perp}-\bm{z}_{\perp})^{2}} \, D_{xz} D_{zy} -\frac{\alpha_{S}N_{c}}{2\pi^{2}} \int_{\bm{z}_{\perp}} \frac{(\bm{x'}_{\perp}-\bm{y'}_{\perp})^{2}}{(\bm{x'}_{\perp}-\bm{z}_{\perp})^{2} \, (\bm{y'}_{\perp}-\bm{z}_{\perp})^{2}} \nonumber \\
& \times  D_{y'z}D_{zx'}+\frac{\alpha_{S}N_{c}}{(2\pi)^{2}} \int_{\bm{z}_{\perp}}\Big[  \frac{(\bm{x}_{\perp}-\bm{y'}_{\perp})^{2}}{(\bm{x}_{\perp}-\bm{z}_{\perp})^{2} \, (\bm{y'}_{\perp}-\bm{z}_{\perp})^{2}}+ \frac{(\bm{x'}_{\perp}-\bm{y}_{\perp})^{2}}{(\bm{x'}_{\perp}-\bm{z}_{\perp})^{2} \, (\bm{y}_{\perp}-\bm{z}_{\perp})^{2}}- \frac{(\bm{x}_{\perp}-\bm{x'}_{\perp})^{2}}{(\bm{x}_{\perp}-\bm{z}_{\perp})^{2} \, (\bm{x'}_{\perp}-\bm{z}_{\perp})^{2}} \nonumber \\
&  - \frac{(\bm{y}_{\perp}-\bm{y'}_{\perp})^{2}}{(\bm{y}_{\perp}-\bm{z}_{\perp})^{2} \, (\bm{y'}_{\perp}-\bm{z}_{\perp})^{2}} \Big] \, D_{xx'} D_{y'y} +\frac{\alpha_{S}N_{c}}{(2\pi)^{2}} \int_{\bm{z}_{\perp}}\Big[  \frac{(\bm{x'}_{\perp}-\bm{y'}_{\perp})^{2}}{(\bm{x'}_{\perp}-\bm{z}_{\perp})^{2} \, (\bm{y'}_{\perp}-\bm{z}_{\perp})^{2}}+ \frac{(\bm{x}_{\perp}-\bm{y'}_{\perp})^{2}}{(\bm{x}_{\perp}-\bm{z}_{\perp})^{2} \, (\bm{y'}_{\perp}-\bm{z}_{\perp})^{2}} \nonumber \\
& - \frac{(\bm{x}_{\perp}-\bm{y}_{\perp})^{2}}{(\bm{x}_{\perp}-\bm{z}_{\perp})^{2} \, (\bm{y}_{\perp}-\bm{z}_{\perp})^{2}} - \frac{(\bm{x'}_{\perp}-\bm{y'}_{\perp})^{2}}{(\bm{x'}_{\perp}-\bm{z}_{\perp})^{2} \, (\bm{y'}_{\perp}-\bm{z}_{\perp})^{2}} \Big] \, D_{xy} D_{y'x'} +\frac{\alpha_{S}N_{c}}{(2\pi)^{2}} \int_{\bm{z}_{\perp}} \Big[ \frac{(\bm{x}_{\perp}-\bm{x'}_{\perp})^{2}}{(\bm{x}_{\perp}-\bm{z}_{\perp})^{2} \, (\bm{x'}_{\perp}-\bm{z}_{\perp})^{2}} \nonumber \\
& +\frac{(\bm{x'}_{\perp}-\bm{y'}_{\perp})^{2}}{(\bm{x'}_{\perp}-\bm{z}_{\perp})^{2} \, (\bm{y'}_{\perp}-\bm{z}_{\perp})^{2}} -\frac{(\bm{x}_{\perp}-\bm{y'}_{\perp})^{2}}{(\bm{x}_{\perp}-\bm{z}_{\perp})^{2} \, (\bm{y'}_{\perp}-\bm{z}_{\perp})^{2}} \Big] \, D_{zx'}Q_{xy;y'z}+\frac{\alpha_{S}N_{c}}{(2\pi)^{2}} \int_{\bm{z}_{\perp}} \Big[ \frac{(\bm{y}_{\perp}-\bm{y'}_{\perp})^{2}}{(\bm{y}_{\perp}-\bm{z}_{\perp})^{2} \, (\bm{y'}_{\perp}-\bm{z}_{\perp})^{2}} \nonumber \\
& +\frac{(\bm{x'}_{\perp}-\bm{y'}_{\perp})^{2}}{(\bm{x'}_{\perp}-\bm{z}_{\perp})^{2} \, (\bm{y'}_{\perp}-\bm{z}_{\perp})^{2}} -\frac{(\bm{x'}_{\perp}-\bm{y}_{\perp})^{2}}{(\bm{x'}_{\perp}-\bm{z}_{\perp})^{2} \, (\bm{y}_{\perp}-\bm{z}_{\perp})^{2}} \Big] \, D_{y'z}Q_{xy;zx'}+\frac{\alpha_{S}N_{c}}{(2\pi)^{2}} \int_{\bm{z}_{\perp}} \Big[ \frac{(\bm{x}_{\perp}-\bm{x'}_{\perp})^{2}}{(\bm{x}_{\perp}-\bm{z}_{\perp})^{2} \, (\bm{x'}_{\perp}-\bm{z}_{\perp})^{2}} \nonumber \\
& +\frac{(\bm{x}_{\perp}-\bm{y}_{\perp})^{2}}{(\bm{x}_{\perp}-\bm{z}_{\perp})^{2} \, (\bm{y}_{\perp}-\bm{z}_{\perp})^{2}} -\frac{(\bm{x'}_{\perp}-\bm{y}_{\perp})^{2}}{(\bm{x'}_{\perp}-\bm{z}_{\perp})^{2} \, (\bm{y}_{\perp}-\bm{z}_{\perp})^{2}} \Big] \, D_{xz}Q_{y'x';zy} +\frac{\alpha_{S}N_{c}}{(2\pi)^{2}} \int_{\bm{z}_{\perp}} \Big[ \frac{(\bm{x}_{\perp}-\bm{y}_{\perp})^{2}}{(\bm{x}_{\perp}-\bm{z}_{\perp})^{2} \, (\bm{y}_{\perp}-\bm{z}_{\perp})^{2}} \nonumber \\
& +\frac{(\bm{y}_{\perp}-\bm{y'}_{\perp})^{2}}{(\bm{y}_{\perp}-\bm{z}_{\perp})^{2} \, (\bm{y'}_{\perp}-\bm{z}_{\perp})^{2}} -\frac{(\bm{x}_{\perp}-\bm{y'}_{\perp})^{2}}{(\bm{x}_{\perp}-\bm{z}_{\perp})^{2} \, (\bm{y'}_{\perp}-\bm{z}_{\perp})^{2}} \Big] \, D_{zy}Q_{y'x';xz} \Bigg\} \, \,  N_{c} \, C^{\rm LO} \, .
\label{eq:LLx-structure-NLO-amplitude-squared}
\end{align}
In the above equation, $z_{f}$ is the scale of the longitudinal momentum of the virtual photon probe. 

It is now a straightforward but tedious exercise to show that the set of terms appearing within curly brackets in the above equation can be generated by the action of the LO JIMWLK Hamiltonian on the leading order color structure $\Xi(\bm{x}_{\perp},\bm{y}_{\perp};\bm{y'}_{\perp},\bm{x'}_{\perp})$. The former is defined as~\cite{JalilianMarian:1997gr,JalilianMarian:1997dw,Iancu:2000hn,Ferreiro:2001qy,Weigert:2000gi} 
\begin{align}
H_{\rm LO} = \frac{1}{2} \int_{\bm{u}_{\perp},\bm{v}_{\perp}} \frac{\delta}{\delta A_{\rm cl}^{+,a} (\bm{u}_{\perp})}  \, \eta^{ab} (\bm{u}_{\perp},\bm{v}_{\perp}) \,  \frac{\delta}{\delta A_{\rm cl}^{+,b} (\bm{v}_{\perp})} \, ,
\end{align}
where 
\begin{align}
\eta^{ab} (\bm{u}_{\perp},\bm{v}_{\perp})& = \frac{1}{\pi} \int_{\bm{z}_{\perp}} \frac{1}{(2\pi)^{2}} \frac{(\bm{u}_{\perp}-\bm{z}_{\perp}).(\bm{v}_{\perp}-\bm{z}_{\perp})}{(\bm{u}_{\perp}-\bm{z}_{\perp})^{2} \, (\bm{v}_{\perp}-\bm{z}_{\perp})^{2}} \, [\mathds{1}+U^{\dagger}(\bm{u}_{\perp}) U(\bm{v}_{\perp}) -U^{\dagger} (\bm{u}_{\perp}) U(\bm{z}_{\perp}) -U^{\dagger}(\bm{z}_{\perp}) U(\bm{v}_{\perp}) \, ]^{ab} \, .
\end{align}

The Wilson lines appearing above are in the adjoint representation of $SU(N_{c})$ and written in terms of $A^{+,a}(x^{-},\bm{x}_{\perp})$ as
\begin{equation}
U(\bm{x}_{\perp})= P_{-} \Bigg( \text{exp} \Bigg\{ -ig \int_{-\infty}^{+\infty} \mathrm{d}z^{-} A_{\rm cl}^{+,a}(z^{-},\bm{x}_{\perp}) T^{a} \Bigg\}  \Bigg) \, , 
\end{equation}
where we have defined $A^{+,a}(\bm{x}_{\perp}) = \int_{-\infty}^{+\infty} \mathrm{d}z^{-} A_{\rm cl}^{+,a}(z^{-},\bm{x}_{\perp})$. 
As first outlined in the JIMWLK papers, the proof involves extensive use of the identities,
\begin{align}
\frac{\delta U(\bm{x}_{\perp})}{\delta A_{\rm cl}^{+,a} (\bm{z}_{\perp})}= -ig \delta^{(2)} (\bm{x}_{\perp}-\bm{z}_{\perp}) U(\bm{x}_{\perp}) T^{a} \, ,  \quad 
\frac{\delta U^{\dagger}(\bm{x}_{\perp})}{\delta A_{\rm cl}^{+,a} (\bm{z}_{\perp})}= ig \delta^{(2)} (\bm{x}_{\perp}-\bm{z}_{\perp}) T^{a}U^{\dagger}(\bm{x}_{\perp})  \, ,
\end{align}
and Eq.~\ref{eq:adjoint-to-fundamental-connection} relating the fundamental and adjoint Wilson lines.

Now using Eq.~\ref{eq:LO-amplitude-squared} for the CGC averaged amplitude squared at LO, one can write Eq.~\ref{eq:LLx-structure-NLO-amplitude-squared} as 
\begin{align}
 \left \langle {\mathcal{M}^{\rm NLO}}^{*} \, \mathcal{M}^{\rm NLO} \right \rangle \vert_{ LLx} &= \ln \Big( \frac{z_{f}}{z_{0}} \Big) \, H_{\rm LO} \, \Big( N_{c} \, C^{\rm LO} \, \Xi(\bm{x}_{\perp},\bm{y}_{\perp};\bm{y'}_{\perp},\bm{x'}_{\perp}) \Big)  \,  \nonumber \\
 & = \ln \Big( \frac{z_{f}}{z_{0}} \Big) \, H_{\rm LO} \,  \left \langle {\mathcal{M}^{\rm LO}}^{*} \, \mathcal{M}^{\rm LO} \right \rangle \, . 
\end{align}
While we derived the l.h.s of this identity explicitly here for inclusive photon+dijet production, all the necessary elements to derive the r.h.s were obtained previously in \cite{Dominguez:2011gc}.

In the CGC EFT, the expectation value of an operator $\hat{\mathcal{O}}$ is defined to be 
 \begin{equation}
\langle \hat{\mathcal{O}} \rangle = \int [ \mathcal{D} \rho_{A} ] \, W_{z} [\rho_{A}] \, \hat{\mathcal{O}} [\rho_{A}] \, ,
 \end{equation}
 where $\hat{\mathcal{O}} [\rho_{A}]$ is the expression for the operator for a given charge configuration $\rho_{A}$ and $W_{z} [\rho_{A}]$ is a stochastic weight functional describing the probability density of that charge configuration at a  momentum (fraction) $z$. If we now consider the result for the CGC averaged squared amplitude (or differential cross-section) upto NLO$+$LL$x$ accuracy we can easily see that the leading logarithmic pieces can be absorbed in the weight functional to redefine the EFT at the evolved scale $z_{f}$ as
\begin{align}
 \left \langle {\mathcal{M}^{\rm LO}}^{*} \, \mathcal{M}^{\rm LO} \right \rangle+ \left \langle {\mathcal{M}^{\rm NLO}}^{*} \, \mathcal{M}^{\rm NLO} \right \rangle \vert_{LLx} &= \int [ \mathcal{D} \rho_{A} ] \,  \Big( 1+ \ln \Big( \frac{z_{f}}{z_{0}} \Big) \, H_{\rm LO} \Big) \, W_{z_{0}} [\rho_{A}]  \, \Big(  {\mathcal{M}^{\rm LO}}^{*} \, \mathcal{M}^{\rm LO} \Big) \, [\rho_{A}] \,  , \nonumber \\
 & =  \int [ \mathcal{D} \rho_{A} ] \, W_{z_{f}} [\rho_{A}] \, \Big(  {\mathcal{M}^{\rm LO}}^{*} \, \mathcal{M}^{\rm LO} \Big) \, [\rho_{A}] \,  .
\end{align}
In this equation, the functional dependence on $\rho_{A}$ enters through the dipole and quadrupole Wilson line correlators contained in $\Xi(\bm{x}_{\perp},\bm{y}_{\perp};\bm{y'}_{\perp},\bm{x'}_{\perp})$ appearing in the LO amplitude squared. 

Since the lepton tensor, and other prefactors remain the same, we can extend these arguments to obtain the JIMWLK evolution of the triple differential cross-section for photon+dijet production in small $x$ DIS:
\begin{equation}
\frac{\partial}{\partial \, (\ln \Lambda^{-})} (\mathrm{d}^{3} \sigma_{\text {LL$x$}}^{\rm NLO} ) = \langle \mathcal{H}_{\text{LO}} \, (\mathrm{d}^{3} \sigma^{\rm LO}) \rangle \, .
\end{equation}
%We should mention here that the same structures of the kernels as observed in Eq.~\ref{eq:LLx-structure-NLO-amplitude-squared} were also obtained in \cite{Dominguez:2011gc} albeit by directly acting the JIMWLK Hamiltonian on the color structure obtained for inclusive dijet production in DIS at small $x$. 
%The latter is identical for our case because addition of a photon does not affect the color structure. 
 Our proof is in the spirit of the JIMWLK derivation from the projectile side in \cite{Mueller:2001uk} but is obtained by computing the full cross-section and then taking the slow gluon limit.

\section{Summary and outlook}
We presented in this paper the first computation of the NLO impact factor for the inclusive photon+dijet production in high energy electron-nucleus collisions. The triple differential cross-section for this process 
%following the discussion preceding and including Eq.~\ref{eq:dsigma-NLO-NLLx} in the introduction, 
can be expressed as 
\begin{equation}
\frac{\mathrm{d}^{3} \sigma^{\text{LO+NLO+NLL$x$};\rm jet}}{\mathrm{d}x \mathrm{d}Q^{2} \mathrm{d}^{6} K_{\perp} \mathrm{d}^{3} \eta_{K} }= \frac{\alpha_{em}^{2}q_{f}^{4}y^{2}N_{c}}{512 \pi^{5} Q^{2}} \, \frac{1}{(2\pi)^{4}} \,  \frac{1}{2} \,  L^{\mu \nu}  {\tilde X}_{\mu \nu}^{\text{LO+NLO+NLL$x$}; \rm jet}  \,,
\end{equation}
where $L^{\mu\nu}$ is the lepton tensor defined in Eq.~\ref{eq:L-tensor} and 
\begin{align}
{\tilde X}_{\mu \nu}^{\text{LO+NLO+NLL$x$}; \rm jet} &=   \int [\mathcal{D} \rho_A] \, W_{x_{\rm Bj}}^{ NLLx}[\rho_A] \Bigg[\Bigg(1 + \frac{2\alpha_{S}C_{F}}{\pi} \, \Bigg\{ -\frac{3}{4} \ln \Big(  \frac{R^{2} \vert \bm{p}_{J\perp} \vert \, \vert \bm{p}_{K\perp} \vert  }{4 z_{J} z_{K} Q^{2} e^{\gamma_{E}}} \Big) +\frac{7}{4} -\frac{\pi^{2}}{6} \Bigg\} \Bigg) {\tilde X}_{\mu\nu}^{\rm LO; jet} [\rho_{A}]  \nonumber \\
&+ {\tilde X}_{\mu \nu; \rm finite}^{\rm NLO; \rm jet} [\rho_{A}] \, \Bigg] \,.
\label{eq:Xmunu-final}
\end{align}
Recall that ${\tilde X}_{\mu\nu}^{\rm LO; jet} [\rho_{A}]$ is obtained by implementing the replacements in Eq.~\ref{eq:parton-to-jets-LO} in the LO hadron tensor definition given by Eq.~\ref{eq:H-tensor-LO} of section II. It contains the dipole and quadrupole correlators of lightlike Wilson lines; the latter are functionals of  $\rho_A$ as is clear from Eq.~\ref{eq:Wilson-line} in the introduction.

In this expression, the finite terms ${\tilde X}_{\mu \nu; \rm finite}^{\rm NLO; jet} $ are of order $\alpha_S$ relative to the leading term. The explicit results we derived for these are the principal results of this paper. 
In order to obtain their NLO expressions, we  showed in sections V and VI that one has to first isolate the ultraviolet, collinear and soft divergences respectively in the real gluon emission and virtual self-energy+vertex correction diagrams. For the virtual graphs containing gluon loops, as discussed in section VI, several already cancel between different contributions to the amplitude. Of the rest, the collinear and soft-collinear divergences cancel between the real and virtual graphs at the level of the squared amplitude. 

Our treatment of these in Section VI  necessitated introducing a jet algorithm; the small cone approximation (SCA) framework, corresponding to a small jet cone radius $R$, is particularly convenient for our task. The term proportional to $\alpha_S \ln(R)$ in Eq.~\ref{eq:Xmunu-final} is an $O(1)$ remnant of this procedure. The soft gluon limit ($k_g\rightarrow 0$) is a subset of the slow gluon limit of $k_g^-\rightarrow 0$ but finite $k_\perp$. The large logarithms in $x$ arise in the latter; the resummation of these O($\alpha_S \ln(1/x)$) terms is discussed at length in section VII. Our discussion there provides a nontrivial derivation of the JIMWLK equation. Though this equation has been derived in a variety of ways, it is interesting to see it arise from an explicit Feynman diagram computation of a nontrivial final state.  

An apparently technical point which is however of general interest is our observation in section VI that the JIMWLK kernel already contains pieces of what we isolated as soft-collinear divergences. These have to be subtracted from the jet cross-section to avoid double counting when the NLO impact factor is combined with small $x$ evolution. The fact that slow gluon emission outside the jet cone 
satisfies JIMWLK evolution is consistent with this being a feature of non-global logarithms in jet physics explored by Banfi, Marchesini and Smye~\cite{Banfi:2002hw}, identified with BK/JIMWLK evolution by Marchesini and Mueller~\cite{Marchesini:2003nh} as well as by Weigert~\cite{Weigert:2003mm}, and subsequently significantly developed by Hatta and collaborators in a number of papers~\cite{Hatta:2008st,Avsar:2009yb,Hatta:2013iba,Hatta:2017fwr}. (See also \cite{Neill:2016stq} for a recent discussion of this correspondence.) Our NLO computation of photons+dijets in DIS therefore combines JIMWLK evolution in both the spacelike evolution of the DIS wavefunction and the timelike evolution of dijets in the final state. This spacelike-timelike correspondence in $A^-=0$ gauge was noted previously by Mueller~\cite{Mueller:2018llt} and and is a quantitative implementation of a proposal in his paper.

The JIMWLK evolution equation describes the small $x$ evolution of the gauge invariant weight functional $W_{x_{\rm Bj}}^{\text{LL$x$}}[\rho_A]$ which resums leading logs O(($\alpha_S  \ln(1/x))^n$)
and power corrections\footnote{The saturation scale $Q_s (x)$ is an emergent scale which arises from the evolution of the intrinsic nonperturbative QCD scale $\Lambda_{\rm QCD}$ in the initial 
condition $W_{x_0}[\rho_A]$ at some $x_0$ to small $x$. In the  Balitsky-Kovchegov (BK) equation for the dipole correlator (derived in a large $N_c$ and large $A$ limit of the JIMWLK evolution of the dipole),  $Q_s (x)$ arises from the perturbative unitarization of the dipole cross-section.} O(($Q_s /Q)^n$). An important development is that the NLO JIMWLK evolution equation that resums terms of order O(($\alpha_S^2 \ln(1/x))^n$) is now available (in addition to a significant body of work on the NLO BK equation). If we take advantage of these developments, we can promote $W_{x_{\rm Bj}}^{\text{LL$x$}} [\rho_A] \rightarrow W_{x_{\rm Bj}}^{\text{NLL$x$}} [\rho_A]$, as indicated in Eq.~\ref{eq:dsigma-NLO-NLLx}. 

The finite terms ${\tilde X}_{\mu\nu; \rm finite}^{\rm NLO; jet}$ in Eq.~\ref{eq:Xmunu-final} from the virtual loop contributions are given in Appendix~\ref{sec:finite-pieces-virtual-graphs}. The finite terms from the real gluon emission contributions to the cross-section are obtained by taking the modulus squared of the amplitudes in Eqs.~\ref{eq:amplitude-R1-R10}, \ref{eq:amplitude-R11-R15} and \ref{eq:amplitude-R16-R20}, integrating over the gluon phase space with a cutoff, implementing the SCA in section VI, and subsequently subtracting the pieces that contribute to leading log JIMWLK evolution.

The eventual goal of this project is to provide quantitative predictions for a future Electron-Ion Collider (EIC). As noted in the introduction, the computation of the finite pieces ${\tilde X}_{\mu\nu; \rm finite}^{\rm NLO; jet}$ along with NLO BK/JIMWLK evolution, provide the necessary ingredients to compute photon+dijet production (and the associated measurement channels we identified) in e+A DIS to O($\alpha_S^3  \ln (1/x)$) accuracy. For the $x$ reach of an EIC, this corresponds to an accuracy of O($\alpha_S^2$) or $\sim 10$\%. This level of accuracy in such differential measurements is likely sufficient for the 
unambiguous discovery of gluon saturation. 

The realization of this numerical project while clearly feasible is nevertheless a formidable challenge on several fronts. Firstly, the number and complexity of the finite contributions to the NLO impact factor are far greater than comparable expressions in the collinear factorization framework. This is because, unlike the latter, the large $k_\perp\sim Q_S$ from the target flowing through the diagrams generalizes functional forms in the collinear framework to nontrivial integrals that in many instances have to be performed numerically. 

Further, the results obtained for the NLO JIMWLK Hamiltonian are not yet ripe for numerical evaluation. In addition to the sheer complexity of the NLO evolution kernels, there are subtle conceptual issues, first identified by Salam~\cite{Salam:1998tj}, regarding the regularization of these kernels--recent work in this direction, and references to the extant literature, can be found in \cite{Ducloue:2019ezk,Zheng:2019zul}. A self-consistent treatment of NLO JIMWLK  in our framework can be obtained by computing the leading $\ln(1/x)$ contributions to the next-to-next-to-leading order coefficient function for dijet production in e+A DIS. While challenging, the developments in this paper suggest it can be achieved on the required time scales. 

Both the numerical and analytical developments suggested here are however beyond the scope of this work and will be pursued in parallel in future. As a final note, the methods, computations, and principal results of this paper are summarized in an accompanying letter~\cite{Roy:2019cux}.

\section*{Acknowledgements}
We gratefully acknowledge useful comments from Ian Balitsky, Guillaume Beuf, Renaud Boussarie, Yoshitaka Hatta, Edmond Iancu, Tuomas Lappi, Yacine Mehtar-Tani, Al Mueller, Yair Mulian, Risto Paatelainen, Anna Stasto, Andrey Tarasov, Werner Vogelsang, Bowen Xiao and Feng Yuan that have influenced this work. This material is based on work supported by the U.S. Department of Energy, Office of Science, Office of Nuclear Physics, under Contracts No. de-sc0012704 and within the framework of the TMD Theory Topical Collaboration. K. R is supported by an LDRD grant from Brookhaven Science Associates and by the Joint BNL-Stony Brook Center for Frontiers in Nuclear Science (CFNS).

\newpage
\appendix

\section{Notations and conventions} \label{sec:conventions}
The metric used in this paper is the $-2$ metric, $\hat{g}=diag(+1,-1,-1,-1)$, where the `hat' denotes quantities in usual spacetime coordinates. The light cone coordinates are defined as 
\begin{equation*}
x^{+}=\frac{\hat{x}^{0}+\hat{x}^{3}}{\sqrt{2}}, \quad x^{-}=\frac{\hat{x}^{0}-\hat{x}^{3}}{\sqrt{2}} \enskip ,
\end{equation*}
with the transverse coordinates remaining the same. The same definition holds for the gamma matrices $\gamma^{+}$ and $\gamma^{-}$ with the Dirac algebra given by
\begin{equation}
\{\gamma^{\mu},\gamma^{\nu}\}=2g^{\mu \nu} \enskip ,
\label{eq:dirac-algebra}
\end{equation}
where $g^{+-}=g^{-+}=1$ and $g^{ij}=-\delta^{ij}$ ($i,j=1,2$) are the nonzero entries of the metric tensor. In this convention, $a.b=a^{+}b^{-}+a^{-}b^{+}-\bm{a}_{\perp}.\bm{b}_{\perp}$ and $a_{+}=a^{-}$, $a_{i}=-a^{i}$.

\section{Constituent integrals in computations of gluon emission and virtual gluon loop diagrams} \label{sec:constituent-integrals-real-emission}

In this Appendix, we will derive a generic expression for the tensor integrals that arise in the amplitude (and squared amplitude) computation of real emission diagrams and use it to extract some specific results. By taking a specific limit of this result, we will also get the expression for the tensor integrals that arise in the amplitude computation of virtual gluon exchange diagrams. 

\begin{itemize} 
\item \textit{Generic integral for gluon emissions}: The most general tensor integral in $d$ dimensions appearing in real gluon emission computations has the form 
\begin{align}
\mathcal{I}^{(n,\, i_{1}i_{2}\ldots i_{p})}_{r} (\bm{V}_{\perp},\Delta;\bm{r}_{\perp})&= \int \frac{\mathrm{d}^{d} \, \bm{l}_{\perp}}{(2\pi)^{d}} \, e^{i\bm{l}_{\perp}.\bm{r}_{\perp}} \,  \frac{l^{i_{1}} l^{i_{2}} \ldots l^{i_{p}}}{D_{1} D_{2} \ldots D_{n}}   \, .
\label{eq:generic-tensor-integral-1}
\end{align}
The denominators $D_1,\cdots,D_{n}$ appearing on the r.h.s of the above equation have the form $D_i=(\bm{l}_{\perp}+\bm{v}_{i\perp})^{2}+\Delta_{i}$. By introducing $n$ Feynman parameters $\alpha_1,\cdots,\alpha_n$, we can reexpress these integrals as
\begin{align}
\mathcal{I}^{(n,\, i_{1}i_{2}\ldots i_{p})}_{r} (\bm{V}_{\perp},\Delta;\bm{r}_{\perp})& = \Gamma(n) \int_{0}^{1} \mathrm{d}\alpha_{1} \int_{0}^{1} \mathrm{d} \alpha_{2} \ldots \int_{0}^{1} \mathrm{d} \alpha_{n} \, \delta(1-\alpha_{1}-\alpha_{2}-\ldots \alpha_{n}) \,  \int \frac{\mathrm{d}^{d} \, \bm{l}_{\perp}}{(2\pi)^{d}} \,  e^{i\bm{l}_{\perp}.\bm{r}_{\perp}}  \,  \frac{l^{i_{1}} l^{i_{2}} \ldots l^{i_{p}}}{ \Big( \sum \limits_{k=1}^{n}\alpha_{k} \, D_{k} \Big)^{n}}   \, .
\end{align}
By integrating out one of the Feynman parameters, it is always possible to write the denominator in the above expression as $(\bm{l}_{\perp}+\bm{V}_{\perp})^{2}+\Delta$, where $\bm{V}_{\perp}$ and $\Delta$ can be written in terms of the various $\bm{v}_{i\perp}$'s and $\Delta_{i}$'s with the Feynman parameters acting as coefficients. This will become more clear when we present the expressions for some specific cases below. The integral in Eq.~\ref{eq:generic-tensor-integral-1} can therefore be written as 
\begin{align}
\mathcal{I}^{(n,\, i_{1}i_{2}\ldots i_{p})}_{r} (\bm{V}_{\perp},\Delta;\bm{r}_{\perp})& = \Gamma(n) \int_{0}^{1} \mathrm{d}\alpha_{1} \int  \mathrm{d} \alpha_{2} \ldots \int \mathrm{d} \alpha_{n-1} \,   \int \frac{\mathrm{d}^{d} \, \bm{l}_{\perp}}{(2\pi)^{d}} \,  e^{i\bm{l}_{\perp}.\bm{r}_{\perp}}  \,  \frac{l^{i_{1}} l^{i_{2}} \ldots l^{i_{p}}}{ \Big[( \bm{l}_{\perp}+\bm{V}_{\perp})^{2}+\Delta \Big]^{n}}   \, .
\end{align}
The remaining Feynman parameters satisfy the condition 
\begin{equation}
0\leq \alpha_{1}+\alpha_{2}+\ldots+\alpha_{n-1} \leq 1 \, .
\end{equation}
To reduce this above integral, we first make the observation that
\begin{equation}
\frac{\partial}{\partial V^{i_{p}}} \frac{e^{i\bm{l}_{\perp}.\bm{r}_{\perp}} }{\Big[( \bm{l}_{\perp}+\bm{V}_{\perp})^{2}+\Delta \Big]^{n}} = (-n) \frac{2l^{i_{p}} \, e^{i\bm{l}_{\perp}.\bm{r}_{\perp}}}{\Big[( \bm{l}_{\perp}+\bm{V}_{\perp})^{2}+\Delta \Big]^{n-1}} \, .
\end{equation}
Performing this operation $p$ number of times, we can derive the identity
\begin{equation}
\frac{\partial}{\partial V^{i_{1}}}  \ldots \frac{\partial}{\partial V^{i_{p}}} \frac{e^{i\bm{l}_{\perp}.\bm{r}_{\perp}}}{\Big[( \bm{l}_{\perp}+\bm{V}_{\perp})^{2}+\Delta \Big]^{n-p}} = \frac{(-1)^{p} \Gamma(n)\, 2^{p}} {\Gamma(n-p)} \frac{e^{i\bm{l}_{\perp}.\bm{r}_{\perp}} \, l^{i_{1}} l^{i_{2}} \ldots l^{i_{p}}}{\Big[( \bm{l}_{\perp}+\bm{V}_{\perp})^{2}+\Delta \Big]^{n}} \, .
\end{equation}
We next use the Schwinger parametrization for the term on the l.h.s of the above equation,
\begin{equation}
\frac{1}{\Big[( \bm{l}_{\perp}+\bm{V}_{\perp})^{2}+\Delta \Big]^{n-p}}  = \frac{1}{\Gamma(n-p)} \int_{0}^{\infty} \mathrm{d} t \, t^{n-p-1} e^{-t\Big[( \bm{l}_{\perp}+\bm{V}_{\perp})^{2}+\Delta \Big]} \, ,
\end{equation}
and redefine $\bm{l}_{\perp} +\bm{V}_{\perp} \rightarrow \bm{l}_{\perp} $ to arrive at the following expression for the tensor integral in Eq.~\ref{eq:generic-tensor-integral-1}.
\begin{equation}
\mathcal{I}^{(n, \, i_{1}i_{2}\ldots i_{p})}_{r} (\bm{V}_{\perp},\Delta;\bm{r}_{\perp})=\frac{1}{(4\pi)^{d/2}} \int_{0}^{1} \mathrm{d}\alpha_{1} \int\mathrm{d} \alpha_{2} \ldots \int \mathrm{d} \alpha_{n-1} \int_{0}^{\infty} \frac{\mathrm{d}t}{(-2)^{p}} \, t^{n-p-1-\frac{d}{2}} \frac{\partial}{\partial V^{i_{1}}}  \ldots \frac{\partial}{\partial V^{i_{p}}} \, e^{-t\Delta-\frac{\bm{r}_{\perp}^{2}}{4t}-i \bm{V}_{\perp}.\bm{r}_{\perp}} \, .
\label{eq:generic-tensor-integral-real}
\end{equation}
We now consider some special cases:
\begin{enumerate}
\item $n=1,p=0$: 
\begin{equation}
\mathcal{I}_{r}^{(1,0)}(\bm{v}_{\perp},\Delta;\bm{r}_{\perp})= \int \frac{\mathrm{d}^{d} \bm{l}_{\perp}}{(2\pi)^{d}} \, \frac{e^{i\bm{l}_{\perp}.\bm{r}_{\perp}}}{(\bm{l}_{\perp}+\bm{v}_{\perp})^{2}+\Delta} = \frac{2}{(4\pi)^{d/2}} \, \Big( \frac{\bm{r}_{\perp}^{2}}{4\Delta} \Big)^{\frac{1}{2} -\frac{d}{4}} \, e^{-i\bm{v}_{\perp}.\bm{r}_{\perp}} \, K_{\frac{d}{2}-1} \Big(\sqrt{\bm{r}_{\perp}^{2} \Delta}\Big) \, ,
\label{eq:I-r-10}
\end{equation}
where $K_{\nu}(z)$ represents modified Bessel (or Macdonald) functions of the second kind. For $d=2$, this yields the simple result, $\mathcal{I}_{r}^{(1,0)} (d=2)= \frac{1}{2\pi} \, e^{-i\bm{v}_{\perp}.\bm{r}_{\perp}} \, K_{0}\Big(\sqrt{\bm{r}_{\perp}^{2} \Delta}\Big)$.
\item $n=1,p=1:$
\begin{align}
\mathcal{I}_{r}^{(1,i)} (\bm{v}_{\perp},\Delta;\bm{r}_{\perp})&= \int \frac{\mathrm{d}^{d} \bm{l}_{\perp}}{(2\pi)^{d}} \, \frac{e^{i\bm{l}_{\perp}.\bm{r}_{\perp}} \, l^{i}}{(\bm{l}_{\perp}+\bm{v}_{\perp})^{2}+\Delta} \nonumber \\
& = \frac{2}{(4\pi)^{d/2}} \, \Big( \frac{\bm{r}_{\perp}^{2}}{4\Delta} \Big)^{\frac{1}{2} -\frac{d}{4}} \, e^{-i\bm{v}_{\perp}.\bm{r}_{\perp}} \Bigg\{ \frac{ir^{i}}{2} \, \Big(\frac{\bm{r}_{\perp}^{2}}{4\Delta} \Big)^{-1/2} \, K_{d/2}  \Big(\sqrt{\bm{r}_{\perp}^{2} \Delta}\Big) - v^{i} \, K_{\frac{d}{2}-1}  \Big(\sqrt{\bm{r}_{\perp}^{2} \Delta}\Big) \Bigg\} \, .
\label{eq:I-r-11}
\end{align}
\item $n=2,p=0:$
\begin{align}
\mathcal{I}_{r}^{(2,0)} (\bm{v}_{\perp},\Delta;\bm{r}_{\perp})& = \int \frac{\mathrm{d}^{d} \bm{l}_{\perp}}{(2\pi)^{d}} \, \frac{e^{i\bm{l}_{\perp}.\bm{r}_{\perp}}}{[(\bm{l}_{\perp}+\bm{v}_{1\perp})^{2}+\Delta_{1}]\, [(\bm{l}_{\perp}+\bm{v}_{2\perp})^{2}+\Delta_{2}]} \nonumber \\
&= \frac{2}{(4\pi)^{d/2}} \, \int_{0}^{1} \mathrm{d}\alpha \, \Big( \frac{\bm{r}_{\perp}^{2}}{4\Delta} \Big)^{1 -\frac{d}{4}} \, e^{-i\bm{v}_{\perp}.\bm{r}_{\perp}} \, K_{\frac{d}{2}-2} \Big(\sqrt{\bm{r}_{\perp}^{2} \Delta}\Big) \, ,
\label{eq:I-r-20}
\end{align}
where
\begin{equation}
\bm{v}_{\perp}=\alpha \, \bm{v}_{1\perp}+(1-\alpha) \, \bm{v}_{2\perp} \, , \quad \Delta=\alpha(1-\alpha) \, (\bm{v}_{1\perp}-\bm{v}_{2\perp})^{2}+\alpha \, \Delta_{1}+(1-\alpha) \, \Delta_{2} \, .
\label{eq:v-Delta-two-denominators}
\end{equation}
\item $n=2,p=1:$
\begin{align}
\mathcal{I}_{r}^{(2,i)} (\bm{v}_{\perp},\Delta;\bm{r}_{\perp})&= \int \frac{\mathrm{d}^{d} \bm{l}_{\perp}}{(2\pi)^{d}} \, \frac{e^{i\bm{l}_{\perp}.\bm{r}_{\perp}} \, l^{i}}{[(\bm{l}_{\perp}+\bm{v}_{1\perp})^{2}+\Delta_{1}]\, [(\bm{l}_{\perp}+\bm{v}_{2\perp})^{2}+\Delta_{2}]} \nonumber \\
& = \frac{2}{(4\pi)^{d/2}} \,  \int_{0}^{1} \mathrm{d}\alpha \, \Big( \frac{\bm{r}_{\perp}^{2}}{4\Delta} \Big)^{1 -\frac{d}{4}} \, e^{-i\bm{v}_{\perp}.\bm{r}_{\perp}} \Bigg\{ \frac{ir^{i}}{2} \, \Big(\frac{\bm{r}_{\perp}^{2}}{4\Delta} \Big)^{-1/2} \, K_{\frac{d}{2}-1}  \Big(\sqrt{\bm{r}_{\perp}^{2} \Delta}\Big) - v^{i} \, K_{\frac{d}{2}-2}  \Big(\sqrt{\bm{r}_{\perp}^{2} \Delta}\Big) \Bigg\} \, .
\label{eq:I-r-21}
\end{align}
\item $n=2,p=2:$
\begin{align}
\mathcal{I}_{r}^{(2,ij)} (\bm{v}_{\perp},\Delta;\bm{r}_{\perp})&= \int \frac{\mathrm{d}^{d} \bm{l}_{\perp}}{(2\pi)^{d}} \, \frac{e^{i\bm{l}_{\perp}.\bm{r}_{\perp}} \, l^{i}\, l^{j}}{[(\bm{l}_{\perp}+\bm{v}_{1\perp})^{2}+\Delta_{1}]\, [(\bm{l}_{\perp}+\bm{v}_{2\perp})^{2}+\Delta_{2}]} \nonumber \\
& = \frac{2}{(4\pi)^{d/2}} \,  \int_{0}^{1} \mathrm{d}\alpha \, \Big( \frac{\bm{r}_{\perp}^{2}}{4\Delta} \Big)^{1 -\frac{d}{4}} \, e^{-i\bm{v}_{\perp}.\bm{r}_{\perp}} \Bigg\{ \frac{1}{2} \, [ \, \delta^{ij}-i \, (r^{i}v^{j}+r^{j} v^{i}) \, ] \, \Big(\frac{\bm{r}_{\perp}^{2}}{4\Delta} \Big)^{-1/2} \, K_{\frac{d}{2}-1}  \Big(\sqrt{\bm{r}_{\perp}^{2} \Delta}\Big) \nonumber \\
& -\frac{1}{4} \, r^{i}r^{j} \, \Big(\frac{\bm{r}_{\perp}^{2}}{4\Delta} \Big)^{-1}  \, K_{d/2} \Big(\sqrt{\bm{r}_{\perp}^{2} \Delta}\Big) + v^{i} v^{j} \, K_{\frac{d}{2}-2}  \Big(\sqrt{\bm{r}_{\perp}^{2} \Delta}\Big) \Bigg\} \, .
\label{eq:I-r-22}
\end{align}
Putting $i=j$ above, we can obtain $\mathcal{I}_{r}^{(2,ii)}(\bm{v}_{\perp},\Delta;\bm{r}_{\perp})$. The $\bm{v}_{\perp}$ and $\Delta$ appearing in Eqs.~\ref{eq:I-r-21} and \ref{eq:I-r-22} are same as in Eq.~\ref{eq:v-Delta-two-denominators}.
\end{enumerate}
\item \textit{Generic integral for virtual gluon exchange diagrams}: The most general tensor integral appearing in the computation of virtual graphs has the form 
\begin{equation}
\mathcal{I}^{(n,\, i_{1}i_{2}\ldots i_{p})}_{v}(\bm{V}_{\perp},\Delta) = \int \frac{\mathrm{d}^{d} \, \bm{l}_{\perp}}{(2\pi)^{d}}  \,  \frac{l^{i_{1}} l^{i_{2}} \ldots l^{i_{p}}}{D_{1} D_{2} \ldots D_{n}}  \, .
\label{eq:generic-tensor-integral-2}
\end{equation}
Clearly this can be obtained in the $\bm{r}_{\perp} \rightarrow 0$ limit of the integral in Eq.~\ref{eq:generic-tensor-integral-1}. We have provided explicit expressions for such integrals wherever they appear in the main body of the paper. To derive such expressions from the result given by Eq.~\ref{eq:generic-tensor-integral-real} one needs to carefully take the limit $\bm{r}_{\perp} \rightarrow 0$ in the MacDonald functions $K_{\nu}(\sqrt{\bm{r}_{\perp}^{2} \Delta}) $ that will appear in the computation. The limiting result is given by (see Eq.~10.30.2 of~\cite{NIST:DLMF})
\begin{equation}
\lim_{z \rightarrow 0} K_{\nu}(z) \sim \frac{1}{2} \, \Gamma(\nu) (z/2)^{-\nu} \, , \quad \text{for fixed $\nu$} \, . 
\end{equation}
\end{itemize}  

\section{R-functions appearing in gluon emission amplitudes} \label{sec:R-factors-real-emission}

In this section, we will present expressions for the R-functions (not given in the main text) embedded in the final expression for the NLO amplitude for gluon emissions:
\begin{align}
\mathcal{M}_{\mu \alpha;b}^{\text{NLO;Real}}&=  2\pi (eq_{f})^{2}g \, \delta(1-z_{\text{tot}}^{r}) \, \int \mathrm{d}  \Pi_{\perp}^{\rm LO}  \,\, \overline{u}(\bm{k})  \Bigg\{  \int_{\bm{z}_{\perp}} \!\!\!\! e^{-i \bm{k}_{g\perp}.\bm{z}_{\perp}} \,  T^{(1)}_{R;\mu \alpha}  (\bm{l}_{1\perp})   \Big[ \Big( \tilde{U}(\bm{x}_{\perp}) t^{a} \tilde{U}^{\dagger}(\bm{y}_{\perp}) \Big) U_{ab}(\bm{z}_{\perp})-t_{b} \Big] \nonumber \\
&+ e^{-i\bm{k}_{g\perp}.\bm{x}_{\perp}} \, T^{(2)}_{R;\mu \alpha} (\bm{k}_{g\perp},\bm{l}_{1\perp})\Big[ \Big( t_{b} \tilde{U}(\bm{x}_{\perp}) \tilde{U}^{\dagger}(\bm{y}_{\perp})  \Big) -t_{b} \Big] \nonumber \\
& + e^{-i\bm{k}_{g\perp}.\bm{y}_{\perp}} \, T^{(3)}_{R;\mu \alpha} (\bm{k}_{g\perp},\bm{l}_{1\perp})  \Big[ \Big( \tilde{U}(\bm{x}_{\perp}) \tilde{U}^{\dagger}(\bm{y}_{\perp}) t_{b} \Big) -t_{b} \Big]  \Bigg\}  v(\bm{p}) \, .
\end{align}
Here we defined $T_{R;\mu \alpha}^{(1)} (\bm{l}_{1\perp}) =\sum_{\beta=1}^{10} R^{(R\beta)}_{\mu \alpha}(\bm{l}_{1\perp} )$, $T_{R;\mu \alpha}^{(2)} (\bm{k}_{g\perp},\bm{l}_{1\perp})=\sum_{\beta=11}^{15} R^{(R\beta)}_{\mu \alpha}(\bm{k}_{g\perp},\bm{l}_{1\perp})$ and $T_{R;\mu \alpha}^{(3)} (\bm{k}_{g\perp},\bm{l}_{1\perp})=\sum_{\beta=16}^{20} R^{(R\beta)}_{\mu \alpha}(\bm{k}_{g\perp},\bm{l}_{1\perp}) $.

%The quark-antiquark interchanged counterparts for these can be obtained from the expressions below by employing Eq.~\ref{eq:replacements-qqbar-exchange}.

 The computation of $R^{(R2)}_{\mu \alpha}$ is a bit involved. The expressions for the different transverse momentum integrals are provided in Appendix~\ref{sec:constituent-integrals-real-emission}. We have made extensive use of the fact that any term proportional to $g_{\alpha +}$ is zero after contraction with the polarization vector, ${\epsilon^{\alpha}}^{*}(\bm{k}_{\gamma})$ because ${\epsilon^{*}}^{-}(\bm{k}_{\gamma})=0$ in our choice of gauge. After a fair amount of tedious algebra involving careful manipulation of Dirac matrices, we arrive at the expression,
\begin{align}
R^{(R2)}_{\mu \alpha}(\bm{l}_{1\perp})& = - \frac{1}{2(q^{-})^{2}\, (z_{q}+z_{g})(1-z_{\bar{q}})/z_{g}^{2} \, [ \, \bm{l}_{1\perp}^{2}+\Delta^{\rm LO:(1)}]} \Bigg\{ \frac{z_{q}}{z_{g}} \gamma_{\alpha} \bm{\gamma}_{\perp}.\bm{\epsilon}^{*}_{\perp} (\bm{k}_{g}) \, \mathcal{I}^{(2,ii)}_{r}(\bm{v}^{(R2)}_{\perp},\Delta^{(R2)};\bm{r}_{zx}) \nonumber \\
& -\Big( \frac{2z_{q}}{z_{g}} \gamma^{-} \gamma_{\alpha} \gamma^{j} {\epsilon^{i}}^{*}(\bm{k}_{g}) \,  \mathcal{I}^{(2,ij)}_{r}(\bm{v}^{(R2)}_{\perp},\Delta^{(R2)};\bm{r}_{zx}) +\frac{2(z_{q}+z_{\gamma})}{z_{g}} \gamma^{j} \gamma_{\alpha} \gamma^{-} {\epsilon^{i}}^{*} (\bm{k}_{g}) \, \mathcal{I}^{(2,ij)}_{r}(\bm{v}^{(R2)}_{\perp},\Delta^{(R2)};\bm{r}_{zx}) \Big) \nonumber \\
& +\gamma^{-} \gamma^{j} \gamma_{\alpha}\gamma^{k} \bm{\gamma}_{\perp}.\bm{\epsilon}^{*}_{\perp} (\bm{k}_{g}) \, \mathcal{I}^{(2,jk)}_{r}(\bm{v}^{(R2)}_{\perp},\Delta^{(R2)};\bm{r}_{zx}) + \Big( \gamma^{-}[\gamma^{+}z_{q}q^{-}-\bm{\gamma}_{\perp}.(\bm{l}_{1\perp}-\bm{k}_{\gamma \perp})]\gamma_{\alpha}\gamma^{j} \nonumber \\
& +\gamma^{j}\gamma_{\alpha}\gamma^{-}[\gamma^{+}(z_{q}+z_{\gamma})q^{-} -\bm{\gamma}_{\perp}.\bm{l}_{1\perp}] \Big) \, (\bm{\gamma}_{\perp}.\bm{\epsilon}^{*}_{\perp}(\bm{k}_{g})) \, \mathcal{I}^{(2,j)}_{r}(\bm{v}^{(R2)}_{\perp},\Delta^{(R2)};\bm{r}_{zx}) +\gamma^{-} [\gamma^{+} z_{q} q^{-} \nonumber \\
&-\bm{\gamma}_{\perp}.(\bm{l}_{1\perp}-\bm{k}_{\gamma \perp} ) ] 
  \gamma_{\alpha} [\gamma^{+}(z_{q}+z_{\gamma})q^{-} -\gamma^{-} (Q^{2}z_{\bar{q}} +\bm{l}_{1\perp}^{2})/2z_{\bar{q}}q^{-} -\bm{\gamma}_{\perp}.\bm{l}_{1\perp} ] \Big( \gamma^{i} \, \mathcal{I}^{(2,0)}_{r}(\bm{v}^{(R2)}_{\perp},\Delta^{(R2)};\bm{r}_{zx}) \nonumber \\
& - \frac{\gamma^{-}}{z_{g}q^{-}} \, \mathcal{I}^{(2,i)}_{r}(\bm{v}^{(R2)}_{\perp},\Delta^{(R2)};\bm{r}_{zx}) \Big) \, {\epsilon^{i}}^{*} (\bm{k}_{g}) \Bigg\} [\gamma^{+}(1-z_{\bar{q}})q^{-} -\gamma^{-} (Q^{2}z_{\bar{q}} +\bm{l}_{1\perp}^{2})/2z_{\bar{q}}q^{-} -\bm{\gamma}_{\perp}.\bm{l}_{1\perp} ] \gamma_{\mu} \nonumber \\
& \times [\gamma^{+}z_{\bar{q}}q^{-}+\bm{\gamma}_{\perp}.\bm{l}_{1\perp} ] \gamma^{-} v(\bm{p}) \, ,
\label{eq:R-R2}
\end{align}
where
\begin{align}
\bm{v}_{\perp}^{(R2)}& = \alpha \, \bm{v}_{\perp}^{(R1)}+(1-\alpha) \, \bm{v}_{\perp}^{(R3)} \, , \nonumber \\ 
\Delta^{(R2)}& = \alpha(1-\alpha) \, (\bm{v}_{\perp}^{(R1)}-\bm{v}_{\perp}^{(R3)})^{2}+\alpha \, \Delta^{(R1)}+(1-\alpha) \, \Delta^{(R3)} \, , \nonumber \\
\bm{v}_{\perp}^{(R3)}& = -\frac{z_{g}}{z_{q}+z_{g}} \, (\bm{l}_{1\perp} -\bm{k}_{\gamma \perp}) \, , \nonumber \\
\Delta^{(R3)}& = \frac{z_{q}z_{g}}{(z_{q}+z_{g})^{2}} \, \Big\{ (\bm{l}_{1\perp}-\bm{k}_{\gamma \perp})^{2} +\Big( \frac{z_{q}+z_{g}}{z_{\bar{q}}} \bm{l}_{1\perp}^{2} +\frac{z_{q}+z_{g}}{z_{\gamma}} \bm{k}_{\gamma\perp}^{2} \Big) +Q^{2} (z_{q}+z_{g}) -i\varepsilon \Big\} \, .
\end{align}
The expressions for $\bm{v}_{\perp}^{(R1)}$ and $\Delta^{(R1)}$ are given by Eq.~\ref{eq:v-perp-R1-Delta-R1}. The remaining $R$-functions are similar in structure as $R^{(R1)}_{\mu \alpha}$ and are respectively given by
\begin{align}
R^{(R3)}_{\mu \alpha} (\bm{l}_{1\perp})& = -\gamma^{-} \Bigg[ \Big( \{\gamma^{+}z_{q}q^{-}-\bm{\gamma}_{\perp}.(\bm{l}_{1\perp}-\bm{k}_{\gamma\perp}) \} \, \mathcal{I}_{r}^{(1,0)}(\bm{v}_{\perp}^{(R3)},\Delta^{(R3)};\bm{r}_{zx}) +\gamma^{j} \,  \mathcal{I}_{r}^{(1,j)}(\bm{v}_{\perp}^{(R3)},\Delta^{(R3)};\bm{r}_{zx}) \Big) \nonumber \\
& \times \bm{\gamma}_{\perp}.\bm{\epsilon}^{*}_{\perp}(\bm{k}_{g}) - \frac{2z_{q}}{z_{q}} \, \mathcal{I}_{r}^{(1,i)}(\bm{v}_{\perp}^{(R3)},\Delta^{(R3)};\bm{r}_{zx})\, {\epsilon^{*}}^{i}(\bm{k}_{g}) \Bigg] \, [\gamma^{+}(z_{q}+z_{g})q^{-} -\gamma^{-}(Q^{2}z_{\bar{q}}z_{\gamma}+\bm{l}_{1\perp}^{2}z_{\gamma} \nonumber \\
& +\bm{k}_{\gamma \perp}^{2} z_{\bar{q}})/2z_{\bar{q}}z_{\gamma}q^{-} -\bm{\gamma}_{\perp}.(\bm{l}_{1\perp}-\bm{k}_{\gamma \perp}) ] \gamma_{\alpha} \frac{\gamma^{+}(1-z_{\bar{q}})q^{-}-\gamma^{-}(Q^{2}z_{\bar{q}}+\bm{l}_{1\perp}^{2})/2z_{\bar{q}}q^{-}-\bm{\gamma}_{\perp}.\bm{l}_{1\perp}}{\Big(\bm{l}_{1\perp} + \bm{v}_{\perp}^{\rm LO:(2)}  \Big)^{2} +\frac{z_{\bar{q}}(z_{q}+z_{g}) \, \bm{k}_{\gamma \perp}^{2}}{z_{\gamma}(1-z_{\gamma})^{2}}  +\frac{z_{\bar{q}}(z_{q}+z_{g}) \, Q^{2}}{1-z_{\gamma}} -i\varepsilon} \nonumber \\
& \times \gamma_{\mu} \frac{\gamma^{+}z_{\bar{q}}q^{-}+\bm{\gamma}_{\perp}.\bm{l}_{1\perp}}{2(q^{-})^{2}(1-z_{\gamma})(z_{q}+z_{g})/z_{\bar{q}}z_{g} \, [\bm{l}_{1\perp}^{2}+\Delta^{\rm LO:(1)} ]} \gamma^{-} \, .
\label{eq:R-R3}
\end{align}
For the case of $(R4)$, any choice of contour for the $l_{1}^{+}$ integration encloses two poles and hence the total contribution from this process is written in terms of these individual contributions as
\begin{align}
R^{(R4)}_{\mu \alpha}(\bm{l}_{1\perp})&=(A_{\mu}-B_{\mu}) \, \gamma_{\alpha} \frac{\gamma^{+}z_{\bar{q}}q^{-}+\bm{\gamma}_{\perp}.\bm{l}_{1\perp}}{\frac{2z_{\gamma} (z_{q}+z_{g})}{z_{g}} \, (q^{-})^{2} \, \Big [\Big(\bm{l}_{1\perp}+z_{q}/z_{\gamma} \, \bm{k}_{\gamma \perp}^{2} \Big)^{2}-i\varepsilon\Big]} \gamma^{-} \, ,
\label{eq:R-R4}
\end{align}
with
\begin{align}
A_{\mu}&= (1-z_{q}-z_{g}) \, \gamma^{-} \Bigg[ \Big( \{\gamma^{+}z_{q}q^{-}-\bm{\gamma}_{\perp}.(\bm{l}_{1\perp}-\bm{k}_{\gamma \perp}) \} \, \mathcal{I}_{r}^{(1,0)}(\bm{v}_{\perp}^{(R4)},\Delta^{(R4)};\bm{r}_{zx}) +\gamma^{j} \, \mathcal{I}_{r}^{(1,j)}(\bm{v}_{\perp}^{(R4)},\Delta^{(R4)};\bm{r}_{zx}) \Big) \nonumber \\
& \times  \bm{\gamma}_{\perp}.\bm{\epsilon}^{*}_{\perp}(\bm{k}_{g}) - \frac{2z_{q}}{z_{q}} \, \mathcal{I}_{r}^{(1,i)}(\bm{v}_{\perp}^{(R4)},\Delta^{(R4)};\bm{r}_{zx})\, {\epsilon^{*}}^{i}(\bm{k}_{g}) \Bigg] \, [\gamma^{+}(z_{q}+z_{g})q^{-} -\gamma^{-} \Big( Q^{2}(1-z_{q}-z_{g})\nonumber \\
& +(\bm{l}_{1\perp}-\bm{k}_{\gamma \perp})^{2}\Big)/2(1-z_{q}-z_{g})q^{-}  -\bm{\gamma}_{\perp}.(\bm{l}_{1\perp}-\bm{k}_{\gamma \perp}) ] \gamma_{\mu} \frac{\gamma^{+}(1-z_{q}-z_{g})q^{-}+\gamma^{-}\frac{(\bm{l}_{1\perp}-\bm{k}_{\gamma \perp})^{2}}{2(1-z_{q}-z_{g})q^{-}}+\bm{\gamma}_{\perp}.(\bm{l}_{1\perp}-\bm{k}_{\gamma \perp})}{(\bm{l}_{1\perp}-\bm{k}_{\gamma \perp})^{2}+Q^{2}(z_{q}+z_{g})(1-z_{q}-z_{g}) -i\varepsilon}
\label{eq:R-R4-A-mu}
\end{align}
and 
\begin{align}
B_{\mu}& = \gamma^{-} \Bigg[ \Big( \{\gamma^{+}z_{q}q^{-}-\bm{\gamma}_{\perp}.(\bm{l}_{1\perp}-\bm{k}_{\gamma\perp}) \} \, \mathcal{I}_{r}^{(1,0)}(\bm{v}_{\perp}^{(R3)},\Delta^{(R3)};\bm{r}_{zx}) +\gamma^{j} \,  \mathcal{I}_{r}^{(1,j)}(\bm{v}_{\perp}^{(R3)},\Delta^{(R3)};\bm{r}_{zx}) \Big) \nonumber \\
& \times \bm{\gamma}_{\perp}.\bm{\epsilon}^{*}_{\perp}(\bm{k}_{g}) - \frac{2z_{q}}{z_{q}} \, \mathcal{I}_{r}^{(1,i)}(\bm{v}_{\perp}^{(R3)},\Delta^{(R3)};\bm{r}_{zx})\, {\epsilon^{*}}^{i}(\bm{k}_{g}) \Bigg] \, [\gamma^{+}(z_{q}+z_{g})q^{-} -\gamma^{-}(Q^{2}z_{\bar{q}}z_{\gamma}+\bm{l}_{1\perp}^{2}z_{\gamma} \nonumber \\
& +\bm{k}_{\gamma \perp}^{2} z_{\bar{q}})/2z_{\bar{q}}z_{\gamma}q^{-} -\bm{\gamma}_{\perp}.(\bm{l}_{1\perp}-\bm{k}_{\gamma \perp}) ] \gamma_{\mu} \frac{\gamma^{+}(1-z_{q}-z_{g})q^{-}+\gamma^{-}(\bm{l}_{1\perp}^{2}z_{\gamma}+\bm{k}_{\gamma \perp}^{2}z_{\bar{q}})/2z_{\gamma}z_{\bar{q}}q^{-} +\bm{\gamma}_{\perp}.(\bm{l}_{1\perp}-\bm{k}_{\gamma \perp})   }{\Big(\bm{l}_{1\perp} + \bm{v}_{\perp}^{\rm LO:(2)} \Big)^{2} +\frac{z_{\bar{q}}(z_{q}+z_{g}) \, \bm{k}_{\gamma \perp}^{2}}{z_{\gamma}(1-z_{\gamma})^{2}}  +\frac{z_{\bar{q}}(z_{q}+z_{g}) \, Q^{2}}{1-z_{\gamma}} -i\varepsilon} \nonumber \\
& \times \gamma_{\alpha} \frac{\gamma^{+}z_{\bar{q}}q^{-}+\bm{\gamma}_{\perp}.\bm{l}_{1\perp}}{(1-z_{\gamma})/z_{\bar{q}}}  \gamma^{-} \, ,
\label{eq:R-R4-B-mu}
\end{align}
where
\begin{align}
\bm{v}_{\perp}^{(R4)}& =- \frac{z_{g}}{z_{q}+z_{g}} (\bm{l}_{1\perp}-\bm{k}_{\gamma \perp}) \, , \nonumber \\
\Delta^{(R4)}& = \frac{z_{q}z_{g}}{(z_{q}+z_{g})^{2}(1-z_{q}-z_{g})} [(\bm{l}_{1\perp}-\bm{k}_{\gamma \perp})^{2}+Q^{2}(z_{q}+z_{g})(1-z_{q}-z_{g}) -i\varepsilon] \, .
\end{align}
Finally, we can write 
\begin{align}
R^{(R5)}_{\mu \alpha}&= -A_{\mu} \gamma^{-} \frac{\slashed{p}+\slashed{k}_{\gamma}}{2p.k_{\gamma}} \gamma_{\alpha} \frac{1}{2(q^{-})^{2}(z_{q}+z_{g})(1-z_{q}-z_{g})/z_{g}} \, ,
\label{eq:R-R5}
\end{align}
with $A_{\mu}$ given above in Eq.~\ref{eq:R-R4-A-mu}.

We shall now present the R-functions for the NLO processes wherein the gluon is emitted after the interaction of the dipole off the background classical field; these results were not discussed in the main text.
\begin{align}
R^{(R12)}_{\mu \alpha}(\bm{k}_{g\perp},\bm{l}_{1\perp})&=\Big( k_{g}^{i}-\frac{z_{g}}{z_{q}} \, k^{i} \Big) \, \Big( {\epsilon^{*}}^{i} (\bm{k}_{g}) - \frac{z_{g}}{2z_{q}} \, \gamma^{i} \gamma^{j} \gamma^{-} {\epsilon^{*}}^{j} (\bm{k}_{g}) \, \frac{\slashed{k}_{g}}{2z_{g}q^{-}} \Big) 
\gamma_{\alpha} \, \frac{\slashed{k}+\slashed{k}_{\gamma}+\slashed{k}_{g}}{2k.k_{\gamma}+2k_{g}.(k+k_{\gamma}) }   \gamma^{-} \nonumber \\
& \times \frac{\gamma^{+}(1-z_{\bar{q}})q^{-}-\bm{\gamma}_{\perp}.\bm{l}_{1\perp}}{(q^{-})^{2} \,  \Big[ \Big(\bm{k}_{g\perp}- \frac{z_{g}}{z_{q}} \bm{k}_{\perp} \Big)^{2}-i\varepsilon  \Big]}  \gamma_{\mu} \frac{\gamma^{+}z_{\bar{q}}q^{-}+\bm{\gamma}_{\perp}.\bm{l}_{1\perp}}{\bm{l}_{1\perp}^{2}+ \Delta^{\rm LO:(1)}} \gamma^{-} \, .
\label{eq:R-R12}
\end{align}
In deriving the above expression we have extensively used the anticommutation relations given by Eq.~\ref{eq:dirac-algebra} and the Dirac equation $\overline{u}(\bm{k}) \slashed{k} =0$ for the outgoing quark. Using the same manipulations used to derive Eq.~\ref{eq:R-R12}, we can obtain the expressions for the R-functions for the remaining three processes. These are given by
\begin{align}
R^{(R13)}_{\mu \alpha}(\bm{k}_{g\perp},\bm{l}_{1\perp})&= - \Big( k_{g}^{i}-\frac{z_{g}}{z_{q}} \, k^{i} \Big) \, \Big( {\epsilon^{*}}^{i} (\bm{k}_{g}) - \frac{z_{g}}{2z_{q}} \, \gamma^{i} \gamma^{j}  {\epsilon^{*}}^{j} (\bm{k}_{g})  \Big) \gamma^{-} \, \frac{\gamma^{+}(z_{q}+z_{g})q^{-}-\bm{\gamma}_{\perp}.(\bm{l}_{1\perp}-\bm{k}_{\gamma \perp}) }{(1-z_{\gamma}) /z_{\bar{q}} \,  (q^{-})^{2} \Big[ \Big(\bm{k}_{g\perp}- \frac{z_{g}}{z_{q}} \bm{k}_{\perp} \Big)^{2}-i\varepsilon  \Big]      } \nonumber \\
& \times  \gamma_{\alpha} \frac{\gamma^{+}(1-z_{\bar{q}})q^{-} -\gamma^{-}\Big(Q^{2}z_{\bar{q}}+\bm{l}_{1\perp}^{2}\Big) /2z_{\bar{q}}q^{-}-\bm{\gamma}_{\perp}.\bm{l}_{1\perp}}{\Big(\bm{l}_{1\perp}- \frac{z_{\bar{q}}}{1-z_{\gamma}}  \bm{k}_{\gamma \perp}\Big)^{2} +\frac{z_{\bar{q}}(z_{q}+z_{g}) \, \bm{k}_{\gamma \perp}^{2}}{z_{\gamma}(1-z_{\gamma})^{2}}  +\frac{z_{\bar{q}}(z_{q}+z_{g}) \, Q^{2}}{1-z_{\gamma}} -i\varepsilon} \,  \gamma_{\mu} \frac{\gamma^{+}z_{\bar{q}}q^{-} +\bm{\gamma}_{\perp}.\bm{l}_{1\perp}}{\bm{l}_{1\perp}^{2}+ \Delta^{\rm LO:(1)}} \gamma^{-}  \, ,
\label{eq:R-R13}
\end{align}
\begin{align}
R^{(R14)}_{\mu \alpha} (\bm{k}_{g\perp},\bm{l}_{1\perp}) & =\Big( k_{g}^{i}-\frac{z_{g}}{z_{q}} \, k^{i} \Big) \, \Big( {\epsilon^{*}}^{i} (\bm{k}_{g}) - \frac{z_{g}}{2z_{q}} \, \gamma^{i} \gamma^{j}  {\epsilon^{*}}^{j} (\bm{k}_{g})  \Big) \gamma^{-} \, \frac{\gamma^{+}(z_{q}+z_{g})q^{-}-\bm{\gamma}_{\perp}.(\bm{l}_{1\perp}-\bm{k}_{\gamma \perp}) }{(1-z_{\gamma})/(z_q+z_g) \, (q^{-})^{2} \,  \Big[ \Big(\bm{k}_{g\perp}- \frac{z_{g}}{z_{q}} \bm{k}_{\perp} \Big)^{2}-i\varepsilon  \Big]      }  \gamma_{\mu} \nonumber \\
& \times  \frac{\gamma^{+}(1-z_{q}-z_{g})q^{-}-\gamma^{-}\Big(Q^{2}(z_{q}+z_{g})+(\bm{l}_{1\perp}-\bm{k}_{\gamma \perp})^{2} \Big)/2(z_{q}+z_{g})q^{-} +\bm{\gamma}_{\perp}.(\bm{l}_{1\perp}-\bm{k}_{\gamma \perp}) }{\Big(\bm{l}_{1\perp}- \frac{z_{\bar{q}}}{1-z_{\gamma}}  \bm{k}_{\gamma \perp}\Big)^{2} +\frac{z_{\bar{q}}(z_{q}+z_{g}) \, \bm{k}_{\gamma \perp}^{2}}{z_{\gamma}(1-z_{\gamma})^{2}}  +\frac{z_{\bar{q}}(z_{q}+z_{g}) \, Q^{2}}{1-z_{\gamma}} -i\varepsilon} \nonumber \\
& \times \gamma_{\alpha} \frac{\gamma^{+}z_{\bar{q}}q^{-}+\bm{\gamma}_{\perp}.\bm{l}_{1\perp}}{(\bm{l}_{1\perp}-\bm{k}_{\gamma \perp})^{2}+Q^{2}(z_{q}+z_{g})(1-z_{q}-z_{g})-i\varepsilon} \, ,
\label{eq:R-R14}
\end{align}
\begin{align}
R^{(R15)}_{\mu \alpha}& =  - \Big( k_{g}^{i}-\frac{z_{g}}{z_{q}} \, k^{i} \Big) \, \Big( {\epsilon^{*}}^{i} (\bm{k}_{g}) - \frac{z_{g}}{2z_{q}} \, \gamma^{i} \gamma^{j}  {\epsilon^{*}}^{j} (\bm{k}_{g})  \Big) \gamma^{-} \,   \frac{\gamma^{+}(z_{q}+z_{g})q^{-}-\bm{\gamma}_{\perp}.(\bm{l}_{1\perp}-\bm{k}_{\gamma \perp}) }{ (q^{-})^{2}  \Big[ \Big(\bm{k}_{g\perp}- \frac{z_{g}}{z_{q}} \bm{k}_{\perp} \Big)^{2}-i\varepsilon  \Big]      }  \gamma_{\mu}  \nonumber \\
& \times \frac{\gamma^{+}(1-z_{q}-z_{g})q^{-}+\bm{\gamma}_{\perp}.(\bm{l}_{1\perp}-\bm{k}_{\gamma \perp})}{(\bm{l}_{1\perp}-\bm{k}_{\gamma \perp})^{2}+Q^{2}(z_{q}+z_{g})(1-z_{q}-z_{g})-i\varepsilon} \gamma^{-} \, \frac{\slashed{p}+\slashed{k}_{\gamma}}{2p.k_{\gamma}} \, \gamma_{\alpha} \, .
\label{eq:R-R15}
\end{align}
The usefulness of expressing the amplitude in this form becomes clear in Sec.~\ref{sec:jet-cross-section} where we extract the collinearly divergent structure at the level of the squared amplitude. The factors $k^{i}_{g}-z_{g}/z_{q} \, k^{i}$ in the amplitudes above cancel a collinear structure in the denominator in the squared amplitude. When we then take the collinear limit of the resulting expression, and integrate over the phase space of the emitted gluon, the logarithmic singularity is manifest.

We do not present the R-functions for the diagrams where the quark and antiquark lines are interchanged. These are easily obtained by using Eq.~\ref{eq:replacements-qqbar-exchange}. 

\section{Quark self-energy in $A^{-}=0$ gauge} \label{sec:quark-self-energy}
\begin{figure}[!htbp]
\includegraphics[scale=1]{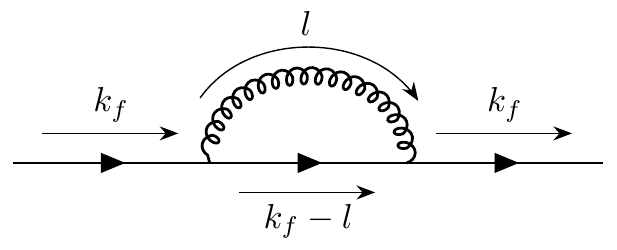}
\caption{Quark self-energy diagram for quark momentum $k_{f}$. This block can appear either in the internal or external legs. \label{fig:quark-self-energy} }
\end{figure}

We will compute here the quark self-energy diagram shown in Fig.~\ref{fig:quark-self-energy}. 
This is useful in the computations of the diagrams $(S13)-(S16)$ and $(S21), (S23)$ (see Fig.~\ref{fig:NLO-self-3}) wherein appears the expression 
\begin{equation*}
S_{0}(k_{f})  \, \Sigma(k_{f}) \, S_{0}(k_{f}) \, ,
\end{equation*}
with $\Sigma(k_{f})$ the contribution from the loop. For diagrams $(S17)-(S20)$, the corresponding generic expression would be $\overline{u}(\bm{k_{f}}) \, \Sigma(k_{f}) \, S_{0}(k_{f})$. In either case, this quark self-energy can be written as 
\begin{align}
\Sigma(k_{f})= -g^{2} C_{F} \int_{\bm{l}_{\perp}} \int_{l^{+},l^{-}} \,  \gamma^{\mu} \frac{\gamma^{+}(k^{-}_{f}-l^{-})+\gamma^{-}(k^{+}_{f}-l^{+})-\bm{\gamma}_{\perp}.(\bm{k}_{f \perp}-\bm{l}_{\perp})}{4l^{-}(k^{-}_{f}-l^{-}) \Big( l^{+}-l^{+}_{a} \Big) \Big( l^{+}-l^{+}_{b} \Big)} \gamma^{\nu} \, \Big( -g_{\mu \nu}+\frac{l_{\mu}n_{\nu}+l_{\nu}n_{\mu}}{l^{-}}\Big) \, ,
\label{eq:quark-selfenergy-loop-contribution}
\end{align}
where
\begin{align}
l^{+}_{a}=\frac{\bm{l}_{\perp}^{2}}{2l^{-}}-\frac{i\varepsilon}{2l^{-}} \,\, , \,\,
l^{+}_{b}=k_{f}^{+}-\frac{(\bm{k}_{f\perp}-\bm{l}_{\perp})^{2}}{2(k_{f}^{-}-l^{-})} +\frac{i\varepsilon}{2(k^{-}_{f}-l^{-})} \, .
\label{eq:pole-locations-quark-selfenergy}
\end{align}
%Similar to Eq.~\ref{eq:gamma-identity}, 
It is a trivial exercise to show that 
\begin{equation}
\gamma^{\mu}(\ldots)  \gamma^{\nu}\, \Big( -g_{\mu \nu}+\frac{l_{\mu}n_{\nu}+l_{\nu}n_{\mu}}{l^{-}}\Big) = \gamma^{i} (\ldots) \gamma^{i}-\gamma^{i} \frac{l^{i}}{l^{-}} (\ldots) \gamma^{-} -\gamma^{-} (\ldots) \gamma^{i} \frac{l^{i}}{l^{-}}+\frac{2l^{+}}{l^{-}} \gamma^{-}(\ldots) \gamma^{-} \, ,
\label{eq:gamma-identity-2}
\end{equation}
where $(\ldots)$ represents the terms in between the gamma matrices in Eq.~\ref{eq:quark-selfenergy-loop-contribution}. Using this identity, we encounter the integrals over $l^{+}$, 
\begin{equation}
\int_{-\infty}^{+\infty} \frac{\mathrm{d}l^{+}}{2\pi} \, \frac{1}{\Big( l^{+}-l^{+}_{a} \Big) \Big( l^{+}-l^{+}_{b} \Big)} \,  \enskip \text{and}  \enskip \int_{-\infty}^{+\infty} \frac{\mathrm{d}l^{+}}{2\pi} \, \frac{l^{+}}{\Big( l^{+}-l^{+}_{a} \Big) \Big( l^{+}-l^{+}_{b} \Big)} \, .
\label{eq:generic-l-plus-integrals-quark-se}
\end{equation}
While the first integral can be done trivially using contour integration, and Cauchy's residue theorem, the second integral should be examined more carefully. This is because unlike the first case where the contribution on the semicircle of the contour vanishes as the radius of the semicircle approaches infinity, it does not for integrals of the second kind. Therefore to obtain the value on the real line (which is the second integral in Eq.~\ref{eq:generic-l-plus-integrals-quark-se}), we must subtract the semicircular contribution from the value obtained using the residue theorem. 

The location of the $l^{+}$ poles given by Eq.~\ref{eq:pole-locations-quark-selfenergy} depend on the sign of $l^{-}$. We need not worry about the $l^{-}=0$ case because that will be regulated by imposing a cutoff at the value $\Lambda_{0}^{-}$. For $0<l^{-}<k_{f}^{-}$ the poles are on opposite sides of the real $l^{+}$ axis whereas for $l^{-}<0$ they are on the same side. Let us discuss the latter case first. When both poles are above the real $l^{+}$ axis, integrals of the first kind in Eq.~\ref{eq:generic-l-plus-integrals-quark-se} give a null result because the contour can be closed in the other direction without enclosing any pole. For integrals of the second kind, if we choose to close the contour above such that both the poles are enclosed, then we have the following result after subtracting the contribution from the semicircle (as the radius approaches infinity).
\begin{equation}
\int_{-\infty}^{+\infty} \frac{\mathrm{d}l^{+}}{2\pi} \, \frac{l^{+}}{\Big( l^{+}-l^{+}_{a} \Big) \Big( l^{+}-l^{+}_{b} \Big)} = i\Bigg( \frac{l^{+}_{a}}{l^{+}_{a}-l^{+}_{b}} + \frac{l^{+}_{b}}{l^{+}_{b}-l^{+}_{a}} \Bigg) - \frac{i}{2 \pi} \int_{0}^{\pi} \mathrm{d}\phi = \frac{i}{2} \, .
\label{eq:second-l-plus-integral-samesidepoles}
\end{equation}
We will obtain the same result if we close the contour clockwise in which case the sole contribution will be the one from the semicircle. We will come back to this result later.

Now for the case $0<l^{-}<k_{f}^{-}$ if we choose to close the contour clockwise enclosing the pole at $l^{+}_{a}$ we get the following result for the second integral,
\begin{equation}
\int_{-\infty}^{+\infty} \frac{\mathrm{d}l^{+}}{2\pi} \, \frac{l^{+}}{\Big( l^{+}-l^{+}_{a} \Big) \Big( l^{+}-l^{+}_{b} \Big)} =-i\frac{l^{+}_{a}}{l^{+}_{a}-l^{+}_{b}} +\frac{i}{2}  = i\frac{l^{+}_{b}}{l^{+}_{b}-l^{+}_{a}} -\frac{i}{2} \, ,
\label{eq:second-l-plus-integral-oppsidepoles}
\end{equation}
where the second equality shows that we will get identical results for either choice of contour.

% as long as we carefully subtract the semicircular contribution (in the limit of infinite radius). 
We will now analytically continue to $d=2-\epsilon$ dimensions for the integration over the transverse momentum and choose the pole at $l^{+}_{a}$ because in this case we get the following compact expression for $\Sigma(k_{f})$,
\begin{align}
\Sigma(k_{f})&= ig^{2} C_{F}  \, \mu^{\epsilon} \,  \int \frac{\mathrm{d}^{2-\epsilon} \bm{l}_{\perp}}{(2\pi)^{2-\epsilon}} \int_{l^{-}} \, \Bigg\{ \Big( \gamma^{i}-\frac{l^{i}}{l^{-}} \gamma^{-} \Big) \frac{\gamma^{+}(k_{f}^{-}-l^{-})+\gamma^{-}\Big( k^{+}_{f} -\bm{l}^{2}_{\perp} /2l^{-} \Big) -\bm{\gamma}_{\perp}.(\bm{k}_{f\perp}-\bm{l}_{\perp})   }{2k^{-}_{f} \Big[ (\bm{l}_{\perp} - \frac{l^{-}}{k^{-}_{f}} \, \bm{k}_{f\perp} )^{2}+\Delta_{s} \Big] } \, \Big( \gamma^{i}-\frac{l^{i}}{l^{-}} \gamma^{-} \Big)  \nonumber \\
& -\frac{1}{2l^{-}} \Bigg(\frac{1}{l^{-}} -\frac{2-\epsilon}{4(k^{-}_{f}-l^{-})} \Bigg) \gamma^{-} \Bigg\}  \, ,
\label{eq:loop-contribution-se-generic}
\end{align}
where 
\begin{equation}
\Delta_{s}= - \frac{l^{-}}{k^{-}_{f}} \Big( 1-\frac{l^{-}}{k^{-}_{f}} \Big) \, k_{f}^{2}  \, .
\label{eq:delta-self-energy}
\end{equation}
The term appearing in the second line of Eq.~\ref{eq:loop-contribution-se-generic} is the contribution on the semicircle which vanishes because of its proportionality to the $d$-dimensional measure; the latter is identically zero within the context of dimensional regularization. In fact, it is a special case of the general identity first conjectured for massive particles by 't Hooft and Veltman~\cite{Leibbrandt:1975dj,Collins:1984xc} and later proven for massless particles~\cite{capper1974conjecture,Capper:1974dc}
\begin{equation}
\int \frac{\mathrm{d}^{d} \bm{l}_{\perp}}{(2\pi)^{d}} \,  (\bm{l}_{\perp}^{2})^{\beta -1} =0 ,  \enskip \beta \geq 1  \enskip \text{for complex $d$} \, . 
\label{eq:veltman-identity}
\end{equation}
%Analyticity in $d$ means that the integral is zero everywhere. 
Using the above arguments, we can redefine $\bm{l}_{\perp}- \frac{l^{-}}{k^{-}_{f}} \, \bm{k}_{f\perp} \rightarrow \bm{l}_{\perp}$ and perform a fair bit of algebra involving gamma matrices to arrive at the following expression for the quark self-energy:
\begin{align}
\Sigma(k_{f})&= ig^{2} C_{F}  \int_{l^{-}} \, \frac{1}{2k^{-}_{f}} \, \Bigg\{  \Bigg(\frac{2+\epsilon}{2l^{-}} +\frac{2(k^{-}_{f}-l^{-})}{(l^{-})^{2}} \Bigg) \gamma^{-} \, \mu^{\epsilon}  \, \mathcal{I}_{v}^{(1,ii)}(\bm{0}_{\perp},\Delta_{s}) \nonumber \\
&+ (2-\epsilon) \, \Bigg[ \slashed{k}_{f} - \Big( \gamma^{+}k^{-}_{f} + \gamma^{-} \frac{\bm{k}_{f\perp}^{2}}{2k^{-}_{f}} - \bm{\gamma}_{\perp}.\bm{k}_{f\perp} \Big) \, \frac{l^{-}}{k^{-}_{f}} \Bigg] \, \mu^{\epsilon}  \, \mathcal{I}_{v}^{(1,0)}(\bm{0}_{\perp},\Delta_{s})  \Bigg\}  \, ,
\label{eq:loop-contribution-final-expression}
\end{align}
 where the constituent integrals are defined as
 \begin{align}
\mu^{\epsilon}  \, \mathcal{I}_{v}^{(1,ii)}(\bm{0}_{\perp},\Delta_{s})&= \mu^{\epsilon}  \, \int \frac{\mathrm{d}^{2-\epsilon} \bm{l}_{\perp}}{(2\pi)^{2-\epsilon}} \, \frac{\bm{l}_{\perp}^{2}}{\bm{l}_{\perp}^{2}+\Delta_{s}}= -\frac{\Delta_{s}}{4\pi} \, \Bigg( \frac{2}{\epsilon} +\ln \Big(\frac{\tilde{\mu}^{2}}{\Delta_{s}} \Big)+O(\epsilon) \Bigg)  \, ,  \enskip \text{for} \, \, \, \epsilon \rightarrow 0 \, , \nonumber \\
\mu^{\epsilon}  \, \mathcal{I}_{v}^{(1,0)}(\bm{0}_{\perp},\Delta_{s})&=\mu^{\epsilon}  \,  \int \frac{\mathrm{d}^{2-\epsilon} \bm{l}_{\perp}}{(2\pi)^{2-\epsilon}} \, \frac{1}{\bm{l}_{\perp}^{2}+\Delta_{s}}= \frac{1}{4\pi} \, \Bigg( \frac{2}{\epsilon} +\ln \Big(\frac{\tilde{\mu}^{2}}{\Delta_{s}} \Big) +O(\epsilon) \Bigg)  \, , \enskip \text{for} \,\, \,  \epsilon \rightarrow 0 \, .
\label{eq:constituent-integrals-self-energy}
 \end{align}
One can now easily check that for diagrams $(S17)-(S20)$ where the loop is on an external line, we will get $\Delta_{s}=0$ (see Eq.~\ref{eq:delta-self-energy}) because the on-shell condition $k^{2}_{f}=2k_{f}^{+}k^{-}_{f}-\bm{k}_{f\perp}^{2}=0$ is satisfied. As a result the constituent integrals in Eq.~\ref{eq:constituent-integrals-self-energy} become scaleless and are zero within the context of dimensional regularization. 

For the first integral in Eq.~\ref{eq:constituent-integrals-self-energy}, this is a special case of the 't Hooft-Veltman identity given by Eq.~\ref{eq:veltman-identity}. Under the condition $\Delta_{s}=0$ the second integral has the form $\mathrm{d}^{2} \bm{l}_{\perp}/\bm{l}_{\perp}^{2}$ which is both UV and IR divergent. In dimensional regularization, we can introduce an arbitrary scale $\Lambda$ to divide the UV and IR regions of loop momentum to write the integral as~\cite{Schwartz:2013pla}
\begin{align}
\int \frac{\mathrm{d}^{d} \bm{l}_{\perp}}{\bm{l}_{\perp}^{2}} &= \Omega_{d} \, \Big( \int_{0}^{\Lambda} \, \mathrm{d} l_{\perp} l_{\perp}^{d-3} + \int_{\Lambda}^{\infty} \, \mathrm{d} l_{\perp} l_{\perp}^{d-3} \Big) \nonumber \\
& = \Omega_{d} \, \Big( \ln \Lambda - \frac{1}{\epsilon_{\rm IR}} \Big) +\Omega_{d} \, \Big( \frac{1}{\epsilon_{\rm UV}} - \ln \Lambda \Big)  \, ,
\end{align}
where  $d=2-\epsilon_{\rm IR}$ for the first integral (assuming $\epsilon_{\rm IR} <0 $) and $d=2-\epsilon_{\rm UV}$ for the second integral (assuming $\epsilon_{\rm UV} > 0 $). Above we have $l_{\perp}= \vert \bm{l}_{\perp} \vert $ and $\Omega_{d}= 2 \pi^{d/2} / \Gamma(d/2)$ is the surface area of a $d$-dimensional sphere of unit radius. Since physical quantities must be independent of $\epsilon_{\rm IR}$ and $\epsilon_{\rm UV}$ we usually set $\epsilon_{\rm IR}=\epsilon_{\rm UV}=\epsilon$ rather than splitting the scaleless integrals each time. The above integral simply vanishes in this case. As noted in 
\cite{Schwartz:2013pla}, a more general argument is that because of the absence of quantities with non-zero mass dimension, scaleless integrals such as the one above must vanish in dimensional regularization in $d$ dimensions. These for diagrams $(S17)-(S20)$ therefore vanish and do not contribute to the amplitude. The same argument holds of course for the quark$\leftrightarrow$antiquark interchanged counterparts of $(S17)-(S20)$.

Returning to the case of $l^{-} < 0$, for which the $l^{+}$ poles are on the same side, Eqs.~\ref{eq:second-l-plus-integral-samesidepoles} and \ref{eq:veltman-identity} tell us that the entire contribution will just vanish. This means that we can only have non-zero contributions to the quark self-energy for $0<l^{-}< k_{f}^{-}$ and $\Delta_{S} \neq 0$. 

We will now use the result obtained in Eq.~\ref{eq:loop-contribution-final-expression} to derive another general expression for the cases in which the virtual gluon is emitted before the scattering and hence appears on the quark line that is off-shell. It is straightforward to show using Dirac algebra that when the loop contribution is sandwiched between two internal quark propagators, we get the expression
\begin{align}
\slashed{k}_{f} \, \Sigma(k_{f}) \, \slashed{k}_{f}  &= ig^{2} C_{F}  \int_{l^{-}} \, \frac{1}{2k^{-}_{f}} \, \Bigg\{  \Bigg(\frac{2+\epsilon}{2l^{-}} +\frac{2(k^{-}_{f}-l^{-})}{(l^{-})^{2}} \Bigg) \, \slashed{k}_{f} \, (2k_{f}^{-}-k^{2}_{f} \gamma^{-}) \, \mu^{\epsilon}  \, \mathcal{I}_{v}^{(1,ii)}(\bm{0}_{\perp},\Delta_{s}) \nonumber \\
&+ (2-\epsilon) \, \Bigg[ k_{f}^{2} \slashed{k}_{f} - k^{2}_{f}  \Big\{ \slashed{k}_{f} - \Big( \gamma^{+}k^{-}_{f} + \gamma^{-} \frac{\bm{k}_{f \perp}^{2}}{2k^{-}_{f}} - \bm{\gamma}_{\perp}.\bm{k}_{f\perp} \Big) \Big\}  \, \frac{l^{-}}{k^{-}_{f}} \Bigg] \,\mu^{\epsilon}  \, \mathcal{I}_{v}^{(1,0)}(\bm{0}_{\perp},\Delta_{s})  \Bigg\}  \, .
\label{eq:self-energy-loop-internal}
\end{align}
This further simplifies when we bring in the $\gamma^{-}$ factor coming from the effective quark vertex. For the processes $(S13)-(S16)$ in which the gluon is absorbed before the shock wave interaction, this gives
\begin{align}
\gamma^{-} \, \slashed{k}_{f} \, \Sigma(k_{f}) \, \slashed{k}_{f}  = ig^{2} C_{F}  \int_{l^{-}}  \, \Bigg\{  \Bigg(\frac{2+\epsilon}{2l^{-}} +\frac{2(k^{-}_{f}-l^{-})}{(l^{-})^{2}} \Bigg)  \, \mu^{\epsilon} \, \mathcal{I}_{v}^{(1,ii)}(\bm{0}_{\perp},\Delta_{s})+ (2-\epsilon) \, \frac{ k_{f}^{2}}{2k^{-}_{f}} \,   \mu^{\epsilon} \, \mathcal{I}_{v}^{(1,0)}(\bm{0}_{\perp},\Delta_{s}) \Bigg\} \,  \gamma^{-} \slashed{k}_{f}  \, .
\label{eq:self-energy-loop-internal-shockwave}
\end{align}
A similar relation can also be obtained for $(S21)$ and $(S23)$ in which the virtual gluon is exchanged after the shock wave interaction. The only difference is that the order in which $\gamma^{-}$ and $\slashed{k}_{f}$ appear on the r.h.s of Eq.~\ref{eq:self-energy-loop-internal-shockwave} is reversed. 

\section{Virtual gluon contributions to the $\gamma qq$ vertex in $A^{-}=0$ gauge}  \label{sec:quark-real photon-quark-vertex-gluon-correction}

\begin{figure}[!htbp]
\includegraphics[scale=1]{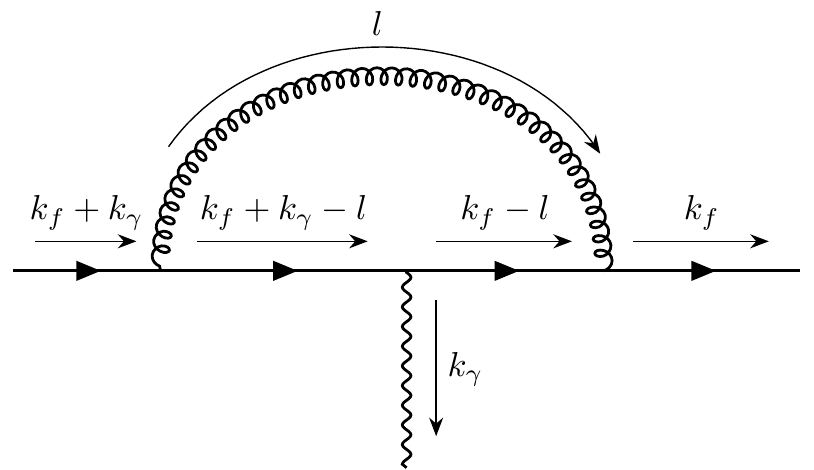}
\caption{Virtual gluon loop contribution to the quark-photon-quark vertex for outgoing quark momentum $k_{f}$ and photon momentum $k_{\gamma}$. This block appears respectively on external and internal quark lines in the diagrams representing processes $(S22)$ and $(S24)$ in Fig.~\ref{fig:NLO-self-3} . \label{fig:self-energy-nested-photon}}
\end{figure}

In a similar fashion to the quark self-energy calculation, we will derive an expression for the gluon loop (with a nested photon) that appears in diagrams $(S22)$ and $(S24)$ and their quark-antiquark interchanged counterparts. The generic diagram is shown in Fig.~\ref{fig:self-energy-nested-photon}.
This contribution to the amplitude can be expressed as 
\begin{equation}
\tilde{\Sigma}_{\alpha} (k_{f},k_{\gamma})= ieq_{f}g^{2} \, \int_{l} \gamma^{\beta} \,S_{0}(k_{f}-l) \gamma_{\alpha} S_{0}(k_{f}+k_{\gamma}-l) \gamma^{\nu}  t^{a} t^{b} \times G^{0}_{\beta \nu;ab} (l) \, .
\label{eq:self-energy-nested-photon-generic}
\end{equation}
If we expand the r.h.s of the above equation using the expressions in Eqs.~\ref{eq:fermion-gluon-propagator} for the free quark and gluon propagators and then apply the identity in  Eq.~\ref{eq:gamma-identity-2}, the numerator in Eq.~\ref{eq:self-energy-nested-photon-generic} is at most proportional to $l^{+}$. This is because terms proportional to $\delta_{\alpha -}$, because of the gauge condition 
$A^-=0$, will yield zero after contraction with the polarization vector for the outgoing photon. 

The contour integration over $l^{+}$ is now well defined and can be computed using the theorem of residues.  The location of the $l^{+}$ poles are given by
\begin{align}
    l^{+} \vert_{a}&= \frac{\bm{l}_{\perp}^{2}}{2l^{-}}-\frac{i \, \varepsilon}{2l^{-}}\, , \nonumber \\
    l^{+} \vert_{b} &= k_{f}^{+}- \frac{(\bm{k}_{f\perp}-\bm{l}_{\perp})^{2}}{2 \, (k_{f}^{-}-l^{-})} +\frac{i \varepsilon}{2 \, (k_{f}^{-}-l^{-})} \, , \nonumber \\
    l^{+} \vert_{c} &= k_{f}^{+}+k_{\gamma}^{+}- \frac{(\bm{k}_{f\perp}+\bm{k}_{\gamma \perp}-\bm{l}_{\perp})^{2}}{2 \, (k_{f}^{-}+k_{\gamma}^{-}-l^{-})} +\frac{i \varepsilon}{2 \, (k_{f}^{-}+k_{\gamma}^{-}-l^{-})}  \, .
    \label{eq:pole-locations-nested-loop}
\end{align}
We can see that for $l^{-}<0$ the three $l^{+}$ poles are located on the same side of the real $l^{+}$ axis and therefore the contour integration yields a null result. For $l^{-} > 0$ we can have two separate contributions depending on the magnitude of $l^{-}$ relative to $k^{-}_{f}$.

\begin{itemize}
 \item Case A: For $l^{-}$ in the range $0<l^{-} < k^{-}_{f}$, we have the poles at $l^{+} \vert_{b}$ and $l^{+} \vert_{c}$ located above the real $l^{+}$ axis whereas the one at $l^{+} \vert_{a}$ is below. We will deform the contour clockwise to enclose this pole.
 
 \item Case B: For $l^{-}$ in the range $k^{-}_{f} < l^{-} < k^{-}_{f}+k^{-}_{\gamma}$, the poles at $l^{+} \vert_{a}$ and $l^{+} \vert_{c}$ are located below the real $l^{+}$ axis. So we will deform the contour anticlockwise to enclose the pole at $l^{+} \vert_{c}$.
 
 \end{itemize}
The total contribution to the amplitude from the loop diagram in Fig.~\ref{fig:self-energy-nested-photon} is therefore the sum of the contributions obtained for these two cases
\begin{equation}
 \tilde{\Sigma}_{\alpha} (k_{f},k_{\gamma})=   \tilde{\Sigma}_{\alpha}^{A} (k_{f},k_{\gamma})+\tilde{\Sigma}_{\alpha}^{B} (k_{f},k_{\gamma}) \, .
\end{equation} 
We will detail below the computation of the divergent and finite pieces for these two cases separately.

Case A: $0<l^{-} < k^{-}_{f}$

The loop contribution in Eq.~\ref{eq:self-energy-nested-photon-generic} becomes
\begin{align}
\tilde{\Sigma}_{\alpha}^{A} (k_{f},k_{\gamma})& = -ieq_{f}g^{2} C_{F} \int_{\bm{l}_{\perp}} \int_{l^{-}} \Big( \gamma^{i}-\frac{l^{i}}{l^{-}} \gamma^{-}  \Big)  \frac{\slashed{k}_{f} - \gamma^{+}l^{-} -\gamma^{-} \frac{\bm{l}_{\perp}^{2}}{2l^{-}} +\bm{\gamma}_{\perp}.\bm{l}_{\perp}    }{2k_{f}^{-}(k^{-}_{f}+k^{-}_{\gamma}) /l^{-} \, \Big[ (\bm{l}_{\perp}+\bm{v}_{1\perp}^{A})^{2}+\Delta_{1}^{A} \Big]} 
\nonumber \\ &\times \gamma_{\alpha}  \frac{\slashed{k}_{f}+\slashed{k}_{\gamma} -\gamma^{+}l^{-} -\gamma^{-} \frac{\bm{l}_{\perp}^{2}}{2l^{-}} +\bm{\gamma}_{\perp}.\bm{l}_{\perp}    }{(\bm{l}_{\perp}+\bm{v}_{2\perp}^{A})^{2}+\Delta_{2}^{A}} \Big( \gamma^{i}-\frac{l^{i}}{l^{-}} \gamma^{-}  \Big)  \, ,
\label{eq:self-energy-nested-photon-case-A}
\end{align}
where the quantities appearing in the denominators are 
\begin{align}
 \bm{v}_{1\perp}^{A}&= -\frac{l^{-}}{k^{-}_{f}+k^{-}_{\gamma}} \, (\bm{k}_{f\perp}+\bm{k}_{\gamma \perp}) \, , \quad \Delta_{1}^{A}= -\frac{l^{-}}{k^{-}_{f}+k^{-}_{\gamma}} \Bigg( 1-\frac{l^{-}}{k^{-}_{f}+k^{-}_{\gamma}}  \Bigg) \, (k_{f}+k_{\gamma})^{2} \, , \nonumber \\
\bm{v}_{2\perp}^{A}& = -\frac{l^{-}}{k^{-}_{f}} \,  \bm{k}_{f\perp} \, , \quad \Delta_{2}^{A}= -\frac{l^{-}}{k^{-}_{f}} \Bigg( 1-\frac{l^{-}}{k^{-}_{f}} \Bigg) \, k^{2}_{f}  \, .
\end{align}
In the following, we will redefine $\bm{l}_{\perp}+\bm{v}_{2\perp}^{A} \rightarrow \bm{l}_{\perp}$ and identify the contributions that will lead to UV divergences in the integration over $\bm{l}_{\perp}$. The convergent pieces will be written later. 

A careful simplification of the numerator in Eq.~\ref{eq:self-energy-nested-photon-case-A} shows that terms proportional to $\bm{l}_{\perp}^{2} l^{i}$ and $l^{i}l^{j}$ potentially have UV divergences. Using dimensional regularization in $d=2-\epsilon$ dimensions, the final expressions for these respective contributions denoted by $\tilde{\Sigma}^{A}_{\alpha;\rm (I)} (k_{f},k_{\gamma})$ and $\tilde{\Sigma}^{A}_{\alpha;\rm (II)} (k_{f},k_{\gamma}) $ are 
\begin{align}
\tilde{\Sigma}^{A}_{\alpha;\rm (I)}  (k_{f},k_{\gamma}) = -ieq_{f} g^{2} C_{F} \int_{l^{-}} \frac{l^{-}}{2k^{-}_{f}(k^{-}_{f}+k^{-}_{\gamma})} \, \Bigg\{ \frac{k^{-}_{f}+k^{-}_{\gamma}-l^{-}}{(l^{-})^{2}} \, \gamma^{i} \gamma_{\alpha} \gamma^{-}  +\frac{k^{-}_{f}-l^{-}   }{(l^{-})^{2}} \, \gamma_{\alpha} \gamma^{i} \gamma^{-} + \frac{\epsilon \gamma^{-}}{l^{-}} \, \delta_{i\alpha} \Bigg\} \, \mu^{\epsilon} \mathcal{I}_{v}^{(2,ijj)}(\bm{V}_{\perp}^{A},\Delta^{A}) \, , 
\label{eq:sigma-tilde-A}
\end{align}
\begin{align}
\tilde{\Sigma}^{A}_{\alpha;\rm (II)} (k_{f},k_{\gamma})& = -ieq_{f}g^{2} C_{F} \int_{l^{-}} \frac{1}{2k_{f}^{-}(k^{-}_{f}+k_{\gamma}^{-})} \, \Bigg\{ \Bigg[ \Big(1-\frac{l^{-}}{k^{-}_{f}} \Big) \, \Big\{\frac{\gamma^{-}\slashed{k}_{f} \gamma_{\alpha} (\slashed{k}_{f}+\slashed{k}_{\gamma}) \gamma^{-}}{l^{-}} -2k_{f\alpha} \gamma^{-}+2k^{-}_{f} \gamma_{\alpha} \Big\} \nonumber \\
& +\gamma_{\alpha} \slashed{k}_{\gamma} \gamma^{-} \Bigg] \, \mu^{\epsilon} \mathcal{I}_{v}^{(2,ii)}( \bm{V}_{\perp}^{A},\Delta^{A}) +  2\Big\{ \gamma_{\alpha}+\delta_{i\alpha} \Big(\gamma^{i}-\frac{k^{i}_{f}}{k^{-}_{f}} \gamma^{-} \Big) \Big\} \, l^{-} \, \mu^{\epsilon} \mathcal{I}_{v}^{(2,ii)}( \bm{V}_{\perp}^{A},\Delta^{A}) -\Big\{ \gamma_{\alpha} \, \mu^{\epsilon} \mathcal{I}_{v}^{(2,ii)}( \bm{V}_{\perp}^{A},\Delta^{A}) \nonumber \\
& + 2\delta_{i\alpha} \Big(\gamma^{j}-\frac{k^{j}_{f}}{k^{-}_{f}} \gamma^{-} \Big) \,  \mu^{\epsilon} \mathcal{I}_{v}^{(2,ij)}( \bm{V}_{\perp}^{A},\Delta^{A})  \Big\} \times  \epsilon\,  l^{-}   +\gamma^{-} \Big\{ \frac{\gamma^{i}}{2} \, \mu^{\epsilon} \mathcal{I}_{v}^{(2,ii)}( \bm{V}_{\perp}^{A},\Delta^{A}) - \gamma^{j} \, \mu^{\epsilon} \mathcal{I}_{v}^{(2,ij)}( \bm{V}_{\perp}^{A},\Delta^{A}) \Big\} \nonumber \\
& \times  \gamma_{\alpha} \Big[ \slashed{k}_{f} + \slashed{k}_{\gamma}  -\frac{l^{-}}{k^{-}_{f}} \Big( \gamma^{+}k^{-}_{f} + \gamma^{-} \frac{\bm{k}_{f\perp}^{2}}{2k^{-}_{f}} -\bm{\gamma}_{\perp}.\bm{k}_{f\perp} \Big) \Big] \, \Big( \gamma^{i} -\frac{k^{i}_{f}}{k^{-}_{f}} \gamma^{-} \Big) + \Big( \gamma^{i} -\frac{k^{i}_{f}}{k^{-}_{f}} \gamma^{-} \Big) \Big[ \slashed{k}_{f} -\frac{l^{-}}{k^{-}_{f}} \Big( \gamma^{+}k^{-}_{f} \nonumber \\
& + \gamma^{-} \frac{\bm{k}_{f\perp}^{2}}{2k^{-}_{f}} -\bm{\gamma}_{\perp}.\bm{k}_{f\perp} \Big) \Big]  \gamma_{\alpha} \, \Big\{ \frac{\gamma^{i}}{2} \, \mu^{\epsilon} \mathcal{I}_{v}^{(2,ii)}( \bm{V}_{\perp}^{A},\Delta^{A}) - \gamma^{j} \, \mu^{\epsilon} \mathcal{I}_{v}^{(2,ij)}( \bm{V}_{\perp}^{A},\Delta^{A}) \Big\} \gamma^{-}  \Bigg\}  \,  .
\label{eq:sigma-tilde-B}
\end{align}
The constituent integrals appearing above are defined in Eqs.~\ref{eq:constituent-integrals-V13-V14} and the arguments are given by
\begin{align}
\bm{V}_{\perp}^{A}&= \alpha \, l^{-} \, \frac{k_{\gamma}^{-} }{k^{-}_{\gamma}+k_{f}^{-}} \, \Big( \frac{\bm{k}_{f\perp}}{k_{f}^{-}} - \frac{\bm{k}_{\gamma \perp}}{k^{-}_{\gamma}} \Big) \, ,  \nonumber \\
\Delta^{A}&= (l^{-})^{2} \, \Bigg\{ \frac{k_{f}^{2}}{(k_{f}^{-})^{2}} - \alpha \Bigg( \frac{k^{2}_{f}}{(k_{f}^{-})^{2}} - \frac{(k_{f}+k_{\gamma})^{2}}{(k_{f}^{-}+k_{\gamma}^{-})^{2}} -\Big( \frac{k_{\gamma}^{-}}{k_{\gamma}^{-}+k_{f}^{-}} \Big)^{2} \, \Big( \frac{\bm{k}_{f\perp}}{k^{-}_{f}} -\frac{\bm{k}_{\gamma \perp}}{k^{-}_{\gamma}} \Big)^{2} \Bigg) \Bigg\} -l^{-} \, \Bigg\{ \frac{k^{2}_{f}}{k^{-}_{f}} -\alpha \, \Big( \frac{k^{2}_{f}}{k^{-}_{f}} - \frac{(k_{f}+k_{\gamma})^{2}}{k^{-}_{f}+k^{-}_{\gamma}} \Big) \Bigg\} \, .
\label{eq:V-perp-Delta-sigma-tilde}
\end{align}
 We will now use the above results, along with the identity
\begin{align}
\gamma_{\alpha} \Bigg\{ \bm{\gamma}_{\perp}. \Big( \frac{\bm{k}_{f\perp}}{k^{-}_{f}} - \frac{\bm{k}_{\gamma \perp}}{k^{-}_{\gamma}} \Big) \Bigg\} \gamma^{-} = -\gamma_{\alpha} \Big( \frac{\slashed{k}_{f}}{k_{f}^{-}} -\frac{\slashed{k}_{\gamma}}{k^{-}_{\gamma}} \Big) \gamma^{-} \, ,
\end{align}
to extract the divergent part of the loop contribution in Eq.~\ref{eq:self-energy-nested-photon-case-A}. This can be written as 
\begin{equation}
\tilde{\Sigma}_{\alpha}^{A; \rm div.} (k_{f},k_{\gamma})=\tilde{\Sigma}_{\alpha; \rm (I) }^{A; \rm div.} (k_{f},k_{\gamma})+\tilde{\Sigma}_{\alpha;\rm (II)}^{A; \rm div.}  (k_{f},k_{\gamma}) \,,
\label{eq:sigma-tilde-divergent-case-A}
\end{equation}
where 
\begin{align}
\tilde{\Sigma}_{\alpha; \rm (I) }^{A; \rm div.} (k_{f},k_{\gamma}) & = -ieq_{f}g^{2}C_{F} \, 
\frac{1}{4\pi^{2}} \Bigg( \frac{1}{\epsilon}+\frac{1}{2} \ln \Big( \frac{\tilde{\mu}^{2}}{Q^{2}} \Big) \Bigg) \, \frac{k_{\gamma}^{-}}{2(k_{\gamma}+k_{f}^{-})^{2}} \, \Big( k_{f\alpha}+\frac{k_{\gamma}^{-}}{k_{f}^{-}} \slashed{k}_{f} \gamma_{\alpha} +\gamma_{\alpha}\slashed{k}_{\gamma} \Big) \, \gamma^{-} \, ,
\label{eq:sigma-tilde-A-divergent}
\end{align}
and 
\begin{align}
\tilde{\Sigma}_{\alpha;\rm (II)}^{A; \rm div.} (k_{f},k_{\gamma})  & = -ieq_{f}g^{2}C_{F} \, \frac{1}{2(k_{\gamma}^{-}+k_{f}^{-})} \, \frac{1}{4\pi^{2}} \, \Bigg[ \Bigg\{ \ln \Big( \frac{z_{f}}{z_{0}} \Big) \, \frac{\gamma^{-}\slashed{k}_{f}\gamma_{\alpha} (\slashed{k}_{f}+\slashed{k}_{\gamma})\gamma^{-}}{k^{-}_{f}} - \frac{\gamma^{-}\slashed{k}_{f}\gamma_{\alpha} (\slashed{k}_{f}+\slashed{k}_{\gamma})\gamma^{-}}{k^{-}_{f}} +k^{-}_{f} \gamma_{\alpha}  \nonumber \\
&+\gamma_{\alpha} \slashed{k}_{\gamma} \gamma^{-} \Bigg\} \, \Bigg( \frac{1}{\epsilon}+\frac{1}{2} \ln \Big( \frac{\tilde{\mu}^{2}}{Q^{2}} \Big) \Bigg) +  \frac{\gamma^{-}\slashed{k}_{f}\gamma_{\alpha} (\slashed{k}_{f}+\slashed{k}_{\gamma})\gamma^{-}}{k^{-}_{f}} \, \frac{1}{2} \ln \Big( \frac{z_{f}}{z_{0}} \Big) \, \Bigg\{ \frac{(z_{f}+z_{\gamma}) \, k^{2}_{f}}{(z_{f}+z_{\gamma}) \, k^{2}_{f}-z_{f} \, (k_{f}+k_{\gamma})^{2}} \, \ln \Bigg(\frac{Q^{2}}{-k_{f}^{2}} \Bigg) \nonumber \\
& -\frac{z_{f} \,(k_{\gamma}+k_{f})^{2} }{(z_{f}+z_{\gamma}) \, k^{2}_{f}-z_{f} \, (k_{f}+k_{\gamma})^{2}}  \, \ln \Bigg(-\frac{Q^{2}\, (z_{f}+z_{\gamma})}{z_{f} \, (k_{f}+k_{\gamma})^{2}} \Bigg)+\frac{1}{2} \ln \Big( \frac{z_{f}}{z_{0}} \Big) \Bigg\} \Bigg] \, ,
\label{eq:sigma-tilde-B-divergent}
\end{align}
and $z_{f}=k_{f}^{-}/q^{-}$.

We will now provide expressions for the finite pieces for the contribution in Eq.~\ref{eq:self-energy-nested-photon-case-A}. These are isolated from Eqs.~\ref{eq:sigma-tilde-A} and \ref{eq:sigma-tilde-B}. In addition, we have to combine these with terms in the integrand proportional to $l^{i}$ and $l_{i}^{0}$ which yield finite results for the integration over $\bm{l}_{\perp}$.  The finite contribution to Eq.~\ref{eq:self-energy-nested-photon-generic} can therefore be written as  
\begin{equation}
\tilde{\Sigma}_{\alpha}^{A;\rm finite} (k_{f},k_{\gamma}) =\hyperref[eq:sigma-tilde-A-finite]{ \tilde{\Sigma}_{\alpha;\rm (I)}^{A; \rm finite}  (k_{f},k_{\gamma}) } + \hyperref[eq:sigma-tilde-B-finite]{\tilde{\Sigma}_{\alpha;\rm (II)}^{A; \rm finite} (k_{f},k_{\gamma})}+ \hyperref[eq:sigma-tilde-C-finite]{\tilde{\Sigma}^{A;\rm finite}_{\alpha;\rm (III)} (k_{f},k_{\gamma})}   \,  , 
\label{eq:self-energy-nested-photon-finite-A}
\end{equation}
where $\tilde{\Sigma}^{A; \rm finite}_{\alpha;\rm (III)}$ is the contribution from terms proportional to $l_{i}^{p}$, $(p=1,0)$. These are respectively obtained as 
\begin{align}
 \tilde{\Sigma}_{\alpha;\rm (I)}^{A; \rm finite} (k_{f},k_{\gamma})&= -ieq_{f}g^{2} C_{F} \int_{0}^{k_{f}^{-}} \frac{\mathrm{d}l^{-}} {4\pi^{2}} \int_{0}^{1} \mathrm{d} \alpha \, \alpha \Big\{ \ln \Big( \frac{Q^{2}}{\Delta^{A}} \Big) +\frac{(\bm{V}^{A}_{\perp})^{2}}{2 \, \Delta^{A}} -\frac{1}{2} \Big\} \, \frac{k^{-}_{\gamma}}{2k^{-}_{f} \, (k_{f}^{-}+k_{\gamma}^{-})^{2}} \nonumber \\
& \times \Big\{ \Big( 1-\frac{l^{-}}{k^{-}_{f}} \Big) 2k_{f\alpha} \gamma^{-} -\Big( \slashed{k}_{\gamma}-\frac{k_{\gamma}^{-}}{k^{-}_{f}} \, \slashed{k}_{f} \Big) \gamma_{\alpha}\gamma^{-} \Big\} 
 -\frac{ieq_{f}g^{2}C_{F}}{16\pi^{2}} \, \frac{k^{-}_{\gamma}}{(k_{\gamma}^{-}+k^{-}_{f})^{2}} \, k_{f\alpha} \gamma^{-} \, ,
\label{eq:sigma-tilde-A-finite}
\end{align}
\begin{align}
\tilde{\Sigma}_{\alpha;\rm (II)}^{A; \rm finite} (k_{f},k_{\gamma}) &=-ieq_{f}g^{2}C_{F} \int_{0}^{k_{f}^{-}} \frac{\mathrm{d}l^{-}} {8 \pi^{2}} \int_{0}^{1} \mathrm{d} \alpha \, \frac{1}{2k_{f}^{-} (k_{f}^{-}+k^{-}_{\gamma}) } \Bigg[ \Big\{ -\frac{\gamma^{-} \slashed{k}_{f} \gamma_{\alpha} \slashed{k}_{\gamma} \gamma^{-}}{k^{-}_{f}} +4 \Big( \frac{l^{-}}{k^{-}_{f}}-2 \Big) k_{f\alpha} \gamma^{-} +\gamma_{\alpha}\slashed{k}_{\gamma} \gamma^{-} \nonumber \\
& + \Big( 1-\frac{l^{-}}{k^{-}_{f}} \Big) \,  (\gamma_{\alpha}\slashed{k}_{f} \gamma^{-}+\gamma^{-}\slashed{k}_{f} \gamma_{\alpha} ) \Big\}  \, \Big\{\ln \Big( \frac{Q^{2}}{\Delta^{A}} \Big) +\frac{(\bm{V}^{A}_{\perp})^{2}}{\Delta^{A}}-1 \Big\}    + \frac{2k_{f\alpha} \, \gamma^{-} \, l^{-} }{k_{f}^{-}} \, \Big\{ \ln \Big( \frac{Q^{2}}{\Delta^{A}} \Big) +\frac{(\bm{V}_{\perp}^{A})^{2}}{\Delta^{A}}-2 \Big\}  \nonumber \\
& +\gamma^{-} \Big\{ \frac{\gamma^{i}}{2} \, \Big(\frac{(\bm{V}^{A}_{\perp})^{2}}{\Delta^{A}}-1 \Big)  - \frac{\gamma^{j} \, (V^{A})^{j} \, (V^{A})^{i}}{\Delta^{A}} \Big\} \gamma_{\alpha} \Big[ \slashed{k}_{f} + \slashed{k}_{\gamma}  -\frac{l^{-}}{k^{-}_{f}} \Big( \gamma^{+}k^{-}_{f} + \gamma^{-} \frac{\bm{k}_{f\perp}^{2}}{2k^{-}_{f}} -\bm{\gamma}_{\perp}.\bm{k}_{f\perp} \Big) \Big] \, \Big( \gamma^{i} -\frac{k^{i}_{f}}{k^{-}_{f}} \gamma^{-} \Big) \nonumber \\
&+ \Big( \gamma^{i} -\frac{k^{i}_{f}}{k^{-}_{f}} \gamma^{-} \Big) \Big[ \slashed{k}_{f} -\frac{l^{-}}{k^{-}_{f}} \Big( \gamma^{+}k^{-}_{f} + \gamma^{-} \frac{\bm{k}_{f\perp}^{2}}{2k^{-}_{f}} -\bm{\gamma}_{\perp}.\bm{k}_{f\perp} \Big) \Big]  \gamma_{\alpha} \, \Big\{ \frac{\gamma^{i}}{2} \, \Big(\frac{(\bm{V}_{\perp}^{A})^{2}}{\Delta^{A}}-1 \Big)  - \frac{\gamma^{j}  \, (V^{A})^{j} \, (V^{A})^{i} }{\Delta^{A}} \Big\} \gamma^{-} \Bigg] \nonumber \\
&+\hyperref[eq:sigma-tilde-B-remainder]{ \tilde{\Sigma}^{A; \rm remainder}_{\alpha; \rm (II)}  (k_{f},k_{\gamma}) } \,  ,
\label{eq:sigma-tilde-B-finite}
\end{align}
where the remainder appearing in the above equation is constituted of finite pieces left from the integration over terms in $\tilde{\Sigma}^{A}_{\alpha; \rm (II)}$ that are proportional to $1/z_{l}$ and produce the rapidity divergent pieces as shown in Eq.~\ref{eq:sigma-tilde-B-divergent}. This can therefore be written as 
\begin{align}
{\tilde{\Sigma}^{A; \rm remainder}_{\alpha; \rm (II)}} (k_{f},k_{\gamma}) &= -\frac{ieq_{f}g^{2}C_{F}}{8\pi^{2}} \, \frac{\gamma^{-}\slashed{k}_{f} \gamma_{\alpha} (\slashed{k}_{f}+\slashed{k}_{\gamma}) \gamma^{-}}{2k^{-}_{f} \, (k^{-}_{\gamma}+k^{-}_{f} ) } \Big\{ \int \frac{\mathrm{d}l^{-}}{l^{-}} \int \mathrm{d} \alpha  \Big( \ln \Big( \frac{Q^{2}}{\Delta^{A}} \Big) +\frac{(\bm{V}^{A}_{\perp})^{2}}{\Delta^{A}}-1 \Big) - \tilde{\Sigma}_{\alpha; \rm log} (k_{f},k_{\gamma}) \Big\}  \, , 
\label{eq:sigma-tilde-B-remainder}
\end{align}
where the rapidity logs that must be subtracted to obtain the finite terms are contained in
\begin{align}
\tilde{\Sigma}_{\alpha; \rm log} (k_{f},k_{\gamma})&=\frac{1}{2} \ln \Big( \frac{z_{f}}{z_{0}} \Big) \, \Bigg\{ \frac{(z_{f}+z_{\gamma}) \, k^{2}_{f}}{(z_{f}+z_{\gamma}) \, k^{2}_{f}-z_{f} \, (k_{f}+k_{\gamma})^{2}} \, \ln \Bigg(\frac{Q^{2}}{-k_{f}^{2}} \Bigg) -\frac{z_{f} \,(k_{\gamma}+k_{f})^{2} }{(z_{f}+z_{\gamma}) \, k^{2}_{f}-z_{f} \, (k_{f}+k_{\gamma})^{2}}  \, \ln \Bigg(-\frac{Q^{2}\, (z_{f}+z_{\gamma})}{z_{f} \, (k_{f}+k_{\gamma})^{2}} \Bigg) \nonumber \\
&+\frac{1}{2} \ln \Big( \frac{z_{f}}{z_{0}} \Big) \Bigg\}  \, .
\end{align}
The final finite contribution is 
\begin{align}
\tilde{\Sigma}_{\alpha;\rm (III)}^{A; \rm finite} (k_{f},k_{\gamma})&= -ieq_{f}g^{2}C_{F} \int_{0}^{k_{f}^{-}} \frac{\mathrm{d}l^{-}} {8 \pi^{2}} \int_{0}^{1} \mathrm{d} \alpha  \, \frac{1}{2k_{f}^{-} (k_{f}^{-}+k^{-}_{\gamma}) } \Bigg[ \Big(1-\frac{l^{-}}{k^{-}_{f}} \Big) \, \gamma^{-} \slashed{k}_{f} \gamma_{\alpha} \Big[ \slashed{k}_{f} + \slashed{k}_{\gamma}  -\frac{l^{-}}{k^{-}_{f}} \Big( \gamma^{+}k^{-}_{f} + \gamma^{-} \frac{\bm{k}_{f\perp}^{2}}{2k^{-}_{f}} \nonumber \\
& -\bm{\gamma}_{\perp}.\bm{k}_{f\perp} \Big) \Big] \,  \Big( \gamma^{i} -\frac{k^{i}_{f}}{k^{-}_{f}} \gamma^{-} \Big) \, V^{i} +  \Big( \gamma^{i} -\frac{k^{i}_{f}}{k^{-}_{f}} \gamma^{-} \Big) \, V^{i}   \Big[ \slashed{k}_{f} -\frac{l^{-}}{k^{-}_{f}} \Big( \gamma^{+}k^{-}_{f} + \gamma^{-} \frac{\bm{k}_{f\perp}^{2}}{2k^{-}_{f}} -\bm{\gamma}_{\perp}.\bm{k}_{f\perp} \Big) \Big]  \gamma_{\alpha} \Big[ \slashed{k}_{\gamma} \nonumber \\
&+ \Big(1-\frac{l^{-}}{k^{-}_{f}} \Big) \slashed{k}_{f} \Big] \gamma^{-}  -   \Big( \gamma^{i} -\frac{k^{i}_{f}}{k^{-}_{f}} \gamma^{-} \Big)  \Bigg\{ l^{-} \,    \Big( \gamma^{j} -\frac{k^{j}_{f}}{k^{-}_{f}} \gamma^{-} \Big)V^{j} \gamma_{\alpha}  \Big[ \slashed{k}_{f} + \slashed{k}_{\gamma}  -\frac{l^{-}}{k^{-}_{f}} \Big( \gamma^{+}k^{-}_{f} + \gamma^{-} \frac{\bm{k}_{f\perp}^{2}}{2k^{-}_{f}} -\bm{\gamma}_{\perp}.\bm{k}_{f\perp} \Big) \Big] \nonumber \\
& + l^{-} \,  \Big[ \slashed{k}_{f} -\frac{l^{-}}{k^{-}_{f}} \Big( \gamma^{+}k^{-}_{f} + \gamma^{-} \frac{\bm{k}_{f\perp}^{2}}{2k^{-}_{f}}  -\bm{\gamma}_{\perp}.\bm{k}_{f\perp} \Big) \Big] \gamma_{\alpha}  \Big( \gamma^{j} -\frac{k^{j}_{f}}{k^{-}_{f}} \gamma^{-} \Big)V^{j} - \Big[ \slashed{k}_{f} -\frac{l^{-}}{k^{-}_{f}} \Big( \gamma^{+}k^{-}_{f} + \gamma^{-} \frac{\bm{k}_{f\perp}^{2}}{2k^{-}_{f}} -\bm{\gamma}_{\perp}.\bm{k}_{f\perp} \Big) \Big] \nonumber \\
& \times  \gamma_{\alpha} \Big[ \slashed{k}_{f} + \slashed{k}_{\gamma}  -\frac{l^{-}}{k^{-}_{f}} \Big( \gamma^{+}k^{-}_{f} + \gamma^{-} \frac{\bm{k}_{f\perp}^{2}}{2k^{-}_{f}}  -\bm{\gamma}_{\perp}.\bm{k}_{f\perp} \Big) \Big] \Bigg\}   \Big( \gamma^{i} -\frac{k^{i}_{f}}{k^{-}_{f}} \gamma^{-} \Big) \Bigg] \, \frac{1}{\Delta^{A}} \, .
\label{eq:sigma-tilde-C-finite}
\end{align}
In the above equations, $\bm{V}_{\perp}^{A}$ and $\Delta^{A}$ can be expressed in terms of the Feynman parameter $\alpha$ and gluon loop momentum $l^{-}$, as shown in Eqs.~\ref{eq:V-perp-Delta-sigma-tilde}. 

Case B: $k^{-}_{f} < l^{-} < k^{-}_{f}+k^{-}_{\gamma}$

The results for this contribution can be obtained following a similar procedure as described above. In this case we can write the loop contribution as
\begin{align}
    \tilde{\Sigma}^{B}_{\alpha}(k_{f},k_{\gamma})&= -i \, eq_{f}g^{2}C_{F} \, \frac{N}{D} \, , 
\end{align}
where the numerator and denominator can be expressed respectively as
\begin{align}
    N&= \int_{l^{-}} \, \int_{\bm{l}_{\perp}} \gamma^{\beta} \, \big[ \gamma^{+}(k^{-}_{f}-l^{-}) +\gamma^{-}  (k^{+}_{f}-l^{+}\vert_{c}) -\bm{\gamma}_{\perp}.(\bm{k}_{f\perp}-\bm{l}_{\perp}) \big] \gamma_{\alpha} \nonumber \\
    & \times  \big[ \gamma^{+}(k^{-}_{f}+k_{\gamma}^{-}-l^{-}) +\gamma^{-}  (k^{+}_{f}+k^{+}_{\gamma}-l^{+}\vert_{c}) -\bm{\gamma}_{\perp}.(\bm{k}_{f\perp}+\bm{k}_{\gamma \perp}-\bm{l}_{\perp}) \big] \nonumber \\
    & \times \gamma^{\rho} \, \Big( -g_{\beta \rho}+ \frac{l_{\beta}n_{\rho}+l_{\rho} n_{\beta}}{l^{-}} \Big \vert_{l^{+} \vert_{c}} \Big) \, , 
\end{align}
and
\begin{align}
    D&= \frac{2 \, k^{-}_{\gamma} \, (k^{-}_{\gamma}+k^{-}_{f})}{k^{-}_{f}+k^{-}_{\gamma}-l^{-}} \, \Bigg[ \Big( \bm{l}_{\perp} +\frac{k^{-}_{f}-l^{-}}{k^{-}_{\gamma}} \, \bm{k}_{\gamma \perp} -\bm{k}_{f\perp} \Big)^{2}-i\varepsilon \Bigg] \nonumber \\
    & \times \Bigg[ \Big( \bm{l}_{\perp} -\frac{l^{-}}{k^{-}_{\gamma}+k^{-}_{f}} \, (\bm{k}_{f\perp}+\bm{k}_{\gamma \perp} )\Big)^{2} - \frac{l^{-}}{k^{-}_{\gamma}+k^{-}_{f}} \Big( 1- \frac{l^{-}}{k^{-}_{\gamma}+k^{-}_{f}} \Big) (k_{f}+k_{\gamma})^{2}-i\varepsilon \Bigg] \, .
\end{align}
The numerator is evaluated at the location of the pole at $l^{+} \vert_{c}$ given by Eq.~\ref{eq:pole-locations-nested-loop}. We can now simplify further by redefining 
\begin{equation}
    \bm{l}_{\perp} +\frac{k^{-}_{f}-l^{-}}{k^{-}_{\gamma}} \, \bm{k}_{\gamma \perp} -\bm{k}_{f\perp} \rightarrow \bm{l}_{\perp} \, , 
\end{equation}
and use the identity in Eq.~\ref{eq:gamma-identity-2} to expand the numerator in terms of four individual contributions. From each of these we collect the pieces proportional to the constituent integrals $\mathcal{I}_{v}^{(2,ijj)}(\bm{V}_{\perp}^{B}, \Delta^{B})$ and $\mathcal{I}_{v}^{(2,ij)}(\bm{V^{B}}_{\perp}, \Delta^{B})$ which will give the divergent contributions as well as finite pieces. The remaining finite contributions are obtained from the terms proportional to the integrals $\mathcal{I}_{v}^{(2,i)}(\bm{V}_{\perp}^{B}, \Delta^{B})$ and $\mathcal{I}_{v}^{(2,0)}(\bm{V}_{\perp}^{B}, \Delta^{B})$. The arguments $\bm{V}_{\perp}^{B}$ and $\Delta^{B}$ for this case can be respectively written in terms of the gluon loop momentum as 
\begin{align}
\bm{V}^{B}_{\perp}&= \alpha \, k^{-}_{f} \, \Big( \frac{\bm{k}_{f\perp}}{k^{-}_{f}} - \frac{\bm{k}_{\gamma \perp}}{k^{-}_{\gamma}} \Big) - l^{-} \Big\{ \alpha \, \frac{k^{-}_{f}}{k_{f}^{-}+k^{-}_{\gamma}} \, \Big( \frac{\bm{k}_{f\perp}}{k^{-}_{f}} - \frac{\bm{k}_{\gamma \perp}}{k^{-}_{\gamma}} \Big) \Big\} \,  , \nonumber \\
\Delta^{B}&= \alpha \, (1-\alpha) \, (k^{-}_{f})^{2} \, \Big( \frac{\bm{k}_{f\perp}}{k^{-}_{f}} - \frac{\bm{k}_{\gamma \perp}}{k^{-}_{\gamma}} \Big)^{2} + l^{-} \, \Big\{ \alpha \, \frac{(k_{f}+k_{\gamma})^{2}}{k^{-}_{f}+k^{-}_{\gamma}} - 2\, \alpha \, (1-\alpha) \, \frac{(k^{-}_{f})^{2}}{k^{-}_{f}+k^{-}_{\gamma}} \, \Big( \frac{\bm{k}_{f\perp}}{k^{-}_{f}} - \frac{\bm{k}_{\gamma \perp}}{k^{-}_{\gamma}} \Big)^{2} \Big\} \nonumber \\
& + (l^{-})^{2} \, \Big\{ \alpha \, (1-\alpha) \, \Big( \frac{k^{-}_{f}}{k^{-}_{f}+k^{-}_{\gamma}} \Big)^{2} \, \Big( \frac{\bm{k}_{f\perp}}{k^{-}_{f}} - \frac{\bm{k}_{\gamma \perp}}{k^{-}_{\gamma}} \Big)^{2} + \alpha \, \frac{(k_{f}+k_{\gamma})^{2}}{(k^{-}_{f}+k^{-}_{\gamma})^{2}} \Big\} \, .
\label{eq:V-perp-Delta-in-terms-of-zl-B}
\end{align}

We can finally write the divergent contribution for this case as
\begin{align}
 \tilde{\Sigma}^{B}_{\alpha; \rm div.}(k_{f},k_{\gamma})& = -ieq_{f} g^{2} C_{F} \, \frac{1}{4\pi^{2}} \, \Bigg( \frac{1}{\epsilon}+\frac{1}{2} \ln \Big( \frac{\tilde{\mu}^{2}}{Q^{2}} \Big) \Bigg) \Bigg\{  \ln \Big( \frac{z_{f}+z_{\gamma}}{z_{f}} \Big)  \, \frac{\gamma^{-} \slashed{k}_{f} \gamma_{\alpha} (\slashed{k}_{f}+\slashed{k}_{\gamma}) \gamma^{-}}{2k^{-}_{f} (k^{-}_{f}+k^{-}_{\gamma}) }  \nonumber \\
 & + \Big( 1- \frac{z_{\gamma}}{2(z_{f}+z_{\gamma})} \Big) \, \Big\{  \frac{k_{f\alpha}}{k^{-}_{\gamma}+k^{-}_{f}} \, \gamma^{-} -\gamma_{\alpha} \Big\} + \frac{k^{-}_{\gamma}}{2 \, (k_{f}^{-}+k^{-}_{\gamma})^{2} } \, \Big( \slashed{k}_{f} \gamma_{\alpha} \gamma^{-} - \gamma_{\alpha} \slashed{k}_{\gamma} \gamma^{-} \Big)  \Bigg\} \, .
 \label{eq:sigma-tilde-divergent-case-B}
\end{align}
Using the expressions in Eqs.~\ref{eq:sigma-tilde-divergent-case-A} and \ref{eq:sigma-tilde-B-divergent}, we can obtain the divergent contributions to the amplitudes for $(S22)$ and $(S24)$ in Fig.~\ref{fig:NLO-self-3}.

The result for $(S22)$ was given by Eq.~\ref{eq:R-S22-div}. Because the computation of the divergent part for $(S24)$ is tedious in comparison, we give it here instead.  We obtain the result
\begingroup
\allowdisplaybreaks
\begin{align}
R^{(S24)}_{\text{div.};\mu \alpha} & (\bm{l}_{1\perp})= \frac{1}{2\pi^{2}} \,  \gamma^{-} \frac{\gamma^{+}z_{q}q^{-}-\bm{\gamma}_{\perp}.(\bm{l}_{1\perp}-\bm{k}_{\gamma \perp})}{2(1-z_{\gamma})/z_{\bar{q}} \, (q^{-})^{2} \, \Big[ \bm{l}_{1\perp}^{2}+\Delta^{\rm LO:(1)} \Big] } \gamma_{\alpha} \frac{\gamma^{+}(1-z_{\bar{q}})q^{-}+\gamma^{-}\, \frac{\bm{l}_{1\perp}^{2}}{2(1-z_{\bar{q}})q^{-}}-\bm{\gamma}_{\perp}.\bm{l}_{1\perp}    }{\Big[ \Big( \bm{l}_{1\perp}+\bm{v}_{\perp}^{\rm LO:(2)} \Big)^{2}+\Delta^{\rm LO:(2)} \Big) \Big] } \, \gamma_{\mu} \nonumber \\
& \times \Big[ \gamma^{+}z_{\bar{q}}q^{-}+\bm{\gamma}_{\perp}.\bm{l}_{1\perp} \Big] \gamma^{-} \times  \, \Bigg\{ \ln \Big( \frac{1}{z_{0}} \Big) \Bigg( \frac{1}{\epsilon}+\frac{1}{2} \ln \Big( \frac{ \tilde{\mu}^{2}   }{Q^{2}} \Big) \Bigg) -\frac{3}{4}  \Bigg( \frac{1}{\epsilon}+\frac{1}{2} \ln \Big( \frac{ \tilde{\mu}^{2}   }{Q^{2}} \Big) \Bigg) + \frac{1}{2} \ln \Big(\frac{1}{z_{0}} \Big)  \Bigg( \ln \frac{Q^{2}z_{\bar{q}}}{\bm{l}_{1\perp}^{2}+\Delta^{\rm LO:(1)}} \nonumber \\
&+\frac{1}{2}  \ln \Big(\frac{1}{z_{0}} \Big)  \Bigg) \Bigg\} +\frac{1}{4\pi^{2}} \,  \Bigg( \frac{1}{\epsilon}+\frac{1}{2} \ln \Big( \frac{ \tilde{\mu}^{2}   }{Q^{2}} \Big) \Bigg) \, \frac{z_{q} \, z_{\bar{q}}}{2 \, (1-z_{\bar{q}}) \, (1-z_{\gamma}) \, q^{-}} \, \Big( 1+\frac{z_{q}}{1-z_{\bar{q}}} \Big)  \, \frac{ \gamma_{\alpha}\gamma_{\mu} \gamma^{-} }{  \Big[(\bm{l}_{1\perp}+\bm{v}_{\perp}^{\rm LO:(2)}  )^{2}+\Delta^{\rm LO:(2)}  \Big]  } 
\, .
\label{eq:R-S24-div}
\end{align}
\endgroup
We will end this section by providing the expressions for the finite contributions from the loop diagram in Fig.~\ref{fig:self-energy-nested-photon} for the case B: $k^{-}<l^{-}<k^{-}+k^{-}_{\gamma}$. We will write this as the sum of finite contributions obtained from terms proportional to the constituent integrals $\mathcal{I}_{v}^{(2,ijj)}(\bm{V}_{\perp}^{B},\Delta^{B})$, $\mathcal{I}_{v}^{(2,ij)}(\bm{V}_{\perp}^{B},\Delta^{B})$ (also for $i=j$)  and $\mathcal{I}_{v}^{(2,i)}(\bm{V}_{\perp}^{B},\Delta^{B})$. The terms proportional to $\mathcal{I}_{v}^{(2,0)}(\bm{V}_{\perp}^{B},\Delta^{B})$ are zero. With this in mind, we can write
\begin{equation}
\tilde{\Sigma}_{\alpha}^{B;\rm finite} (k_{f},k_{\gamma}) =\hyperref[eq:sigma-tilde-finite-B-I]{ \tilde{\Sigma}_{\alpha;\rm (I)}^{B; \rm finite}  (k_{f},k_{\gamma}) } + \hyperref[eq:sigma-tilde-finite-B-II]{\tilde{\Sigma}_{\alpha;\rm (II)}^{B; \rm finite}(k_{f},k_{\gamma})}+ \hyperref[eq:sigma-tilde-finite-B-III]{\tilde{\Sigma}^{B;\rm finite}_{\alpha;\rm (III)} (k_{f},k_{\gamma}) }  \,  , 
\label{eq:self-energy-nested-photon-finite-B}
\end{equation}
where $\tilde{\Sigma}_{\alpha;\rm (I)}^{B; \rm finite} $, $\tilde{\Sigma}_{\alpha;\rm (II)}^{B; \rm finite} $ and $\tilde{\Sigma}_{\alpha;\rm (III)}^{B; \rm finite} $ are respectively constituted of finite pieces from the terms proportional to the constituent integrals $\mathcal{I}_{v}^{(2,ijj)}(\bm{V}_{\perp}^{B},\Delta^{B})$, $\mathcal{I}_{v}^{(2,ij)}(\bm{V}_{\perp}^{B},\Delta^{B})$ (and for $i=j$) and $\mathcal{I}_{v}^{(2,i)}(\bm{V}_{\perp}^{B},\Delta^{B})$. The first contribution is obtained as
\begin{align}
\tilde{\Sigma}_{\alpha;\rm (I)}^{B; \rm finite}  (k_{f},k_{\gamma}) &= \frac{ieq_{f}g^{2}C_{F}}{4\pi^{2}} \Big\{ \frac{k^{-}_{\gamma}}{4\, (k_{\gamma}^{-}+k^{-}_{f})^{2}} \, k_{f \alpha} - \int_{k_{f}^{-}}^{k^{-}_{f}+k^{-}_{\gamma}} \mathrm{d} l^{-} \int_{0}^{1} \mathrm{d} \alpha \, \alpha \, \Big( 1- \frac{l^{-}}{k^{-}_{\gamma}+k^{-}_{f}} \Big) \, \Big[ \, \frac{k_{f}^{-}}{2 \, l^{-} \, (k^{-}_{f}+k^{-}_{\gamma}) } \, \gamma_{\alpha} \, \Big( \frac{\slashed{k}_{f}}{k^{-}_{f}} - \frac{\slashed{k}_{\gamma}}{k^{-}_{\gamma}} \Big) \gamma^{-} \nonumber \\
& +\frac{k_{f\alpha}}{k_{\gamma}^{-} \, (k^{-}_{f}+k^{-}_{\gamma}) } \, \gamma^{-} \Big] \, \Big( \ln \Big( \frac{Q^{2}}{\Delta^{B}} \Big) +\frac{(\bm{V}^{B}_{\perp})^{2}}{2 \, \Delta^{B}} -\frac{1}{2} \Big) \Big\}  \, ,
\label{eq:sigma-tilde-finite-B-I}
\end{align}
where $\bm{V}_{\perp}^{B}$ and $\Delta^{B}$ are expressed in terms of the gluon momentum $l^{-}$ in Eqs.~\ref{eq:V-perp-Delta-in-terms-of-zl-B}. In a similar manner the second term can be obtained as
\begin{align}
\tilde{\Sigma}_{\alpha;\rm (II)}^{B; \rm finite}(k_{f},k_{\gamma})&=\frac{ieq_{f}g^{2}C_{F}}{8\pi^{2}} \, \frac{1}{2k_{\gamma}^{-} \, (k_{\gamma}^{-} +k^{-}_{f})} \int_{k^{-}_{f}}^{k^{-}_{f}+k^{-}_{\gamma}} \!\!\!\! \mathrm{d} l^{-} \int_{0}^{1} \mathrm{d} \alpha \Bigg\{ \Bigg[ \frac{2 \, (k^{-}_{f}-l^{-})}{l^{-}} \, k_{f\alpha} \gamma^{-} + \frac{k^{-}_{\gamma}}{l^{-}} \, ( \gamma^{-} \slashed{k}_{\gamma} \gamma_{\alpha} +\slashed{k}_{f} \gamma_{\alpha} \gamma^{-}  ) \nonumber \\
& - \delta^{i}_{\alpha} \Big(\frac{2 \, (k^{-}_{f}-l^{-})}{k^{-}_{\gamma}}\, k^{i}_{\gamma}+\gamma^{i} \slashed{k}_{\gamma} \Big) \gamma^{-} - 2\Big( k^{-}_{\gamma}+ \frac{(k^{-}_{f}+k^{-}_{\gamma}-l^{-})^{2}}{l^{-}} - \frac{k^{-}_{f} \, k^{-}_{\gamma}}{l^{-}} \Big) \, \gamma_{\alpha} \Bigg]  \times 
\Big( \ln \Big( \frac{Q^{2}}{\Delta^{B}} \Big) +\frac{(\bm{V}^{B}_{\perp})^{2}}{\Delta^{B}}-1 \Big) \nonumber \\
& + \Bigg[ \frac{k^{-}_{f}+k^{-}_{\gamma}-l^{-}}{l^{-}} \, \Big( \gamma^{i} \gamma_{\alpha} \gamma^{j} \bm{\gamma}_{\perp}.(\bm{k}_{f\perp}+\bm{k}_{\gamma \perp}) \gamma^{-} + \bm{\gamma}_{\perp}. \bm{k}_{f\perp} \, \gamma^{i} \gamma_{\alpha} \gamma^{j} \gamma^{-} \Big) + \frac{4 \,(k^{-}_{f}+k^{-}_{\gamma}-l^{-}) }{k_{\gamma}^{-}} \, \delta^{i}_{\alpha} k^{j}_{\gamma} \gamma^{-} \nonumber \\
& - \frac{ (k^{-}_{f}-l^{-}) \,     (k^{-}_{f}+k^{-}_{\gamma}-l^{-})}{l^{-}} \, \gamma^{+} \gamma^{i} \gamma_{\alpha} \gamma^{j} \gamma^{-} -  \frac{(k^{-}_{f}+k^{-}_{\gamma}-l^{-})^{2}}{l^{-}} \, \gamma^{-} \gamma^{i} \gamma_{\alpha} \gamma^{j} \gamma^{+}  \Bigg] \nonumber \\
& \times \Big( \frac{1}{2} \, \ln \Big( \frac{Q^{2}}{\Delta^{B}} \Big) \delta^{ij} + \frac{(V^{B})^{i} (V^{B})^{j} }{\Delta^{B}} \Big) \Bigg\}  \, .
\label{eq:sigma-tilde-finite-B-II}
\end{align}
The third contribution can be written as 
\begin{align}
\tilde{\Sigma}^{B}_{\alpha;\rm (III)} (k_{f},k_{\gamma})&= \frac{ieq_{f}g^{2}C_{F}}{8\pi^{2}} \, \frac{1}{2k_{\gamma}^{-} \, (k_{\gamma}^{-} +k^{-}_{f})} \int_{k^{-}_{f}}^{k^{-}_{f}+k^{-}_{\gamma}} \!\!\!\! \mathrm{d} l^{-} \int_{0}^{1} \mathrm{d} \alpha \,   \alpha \, k^{-}_{f} \,  \Big( 1- \frac{l^{-}}{k^{-}_{\gamma}+k^{-}_{f}} \Big) \frac{1}{\Delta^{B}} \Bigg\{ -\Bigg( \frac{4 \, (k_{f}^{-}+k^{-}_{\gamma}-l^{-})^{2}}{l^{-}} \, \Big( \frac{\slashed{k}_{f}}{k^{-}_{f}} \nonumber \\
& - \frac{\slashed{k}_{\gamma}}{k^{-}_{\gamma}} \Big)  \gamma_{\alpha} \gamma^{-}+ \frac{4 \, (k_{f}^{-}-l^{-}) \, (k^{-}_{f}+k^{-}_{\gamma}-l^{-})}{l^{-}} \gamma_{\alpha} \, \Big( \frac{\slashed{k}_{f}}{k^{-}_{f}}- \frac{\slashed{k}_{\gamma}}{k^{-}_{\gamma}} \Big) \gamma^{-} \Bigg) \, \Big( k_{f}^{+}-\frac{k^{-}_{f}-l^{-}}{k^{-}_{\gamma}} \, k^{+}_{\gamma} \Big) + \frac{(k^{-}_{f}+k^{-}_{\gamma}-l^{-})^{2}}{k^{-}_{\gamma}l^{-}} \gamma^{-} \nonumber \\
& \times  \Big( \frac{\slashed{k}_{f}}{k^{-}_{f}}- \frac{\slashed{k}_{\gamma}}{k^{-}_{\gamma}} \Big) \gamma_{\alpha} \slashed{k}_{\gamma} \, \bm{\gamma}_{\perp}. \Big( \bm{k}_{f\perp}-\frac{k^{-}_{f}-l^{-}}{k^{-}_{\gamma}} \, \bm{k}_{\gamma \perp} \Big)+\frac{(k^{-}_{f}-l^{-}) \, (k^{-}_{f}+k^{-}_{\gamma}-l^{-})}{k^{-}_{\gamma} l^{-}} \slashed{k}_{\gamma} \gamma_{\alpha} \gamma^{-}   \Big( \frac{\slashed{k}_{f}}{k^{-}_{f}}- \frac{\slashed{k}_{\gamma}}{k^{-}_{\gamma}} \Big) \bm{\gamma}_{\perp}. \Big( \bm{k}_{f\perp} \nonumber \\
& - \frac{k^{-}_{f}-l^{-}}{k^{-}_{\gamma}} \, \bm{k}_{\gamma \perp} \Big) - \frac{2 (k^{-}_{f}-l^{-}) \, (k^{-}_{f}+k^{-}_{\gamma}-l^{-})}{l^{-}} \, \gamma_{\alpha} \, \bm{\gamma}_{\perp}. \Big( \frac{\bm{k}_{f\perp}}{k^{-}_{f}} - \frac{\bm{k}_{\gamma \perp}}{k^{-}_{\gamma}} \Big) \,  \bm{\gamma}_{\perp}. \Big( \bm{k}_{f\perp}-\frac{k^{-}_{f}-l^{-}}{k^{-}_{\gamma}} \, \bm{k}_{\gamma \perp} \Big) \nonumber \\
& -\frac{2(k_{f}^{-}+k^{-}_{\gamma}-l^{-})^{2}}{l^{-}} \, \bm{\gamma}_{\perp}. \Big( \bm{k}_{f\perp}-\frac{k^{-}_{f}-l^{-}}{k^{-}_{\gamma}} \, \bm{k}_{\gamma \perp} \Big) \, \bm{\gamma}_{\perp}. \Big( \frac{\bm{k}_{f\perp}}{k^{-}_{f}} - \frac{\bm{k}_{\gamma \perp}}{k^{-}_{\gamma}} \Big) \gamma_{\alpha} -  \frac{2 (k^{-}_{f}-l^{-}) \, (k^{-}_{f}+k^{-}_{\gamma}-l^{-})}{k^{-}_{\gamma} \, l^{-}} \, \gamma^{-} \, \gamma_{\alpha}  \nonumber \\
& \times  \bm{\gamma}_{\perp}. \Big( \bm{k}_{f\perp}-\frac{k^{-}_{f}-l^{-}}{k^{-}_{\gamma}} \, \bm{k}_{\gamma \perp} \Big) \Big( \frac{\bm{k}_{\gamma \perp}.\bm{k}_{f\perp}}{k^{-}_{f}} -\frac{\bm{k}_{\gamma \perp}^{2}}{k^{-}_{\gamma}} \Big) - \frac{2 \, (k_{f}^{-}+k^{-}_{\gamma}-l^{-})^{2}}{k^{-}_{\gamma} \, l^{-}} \,  \bm{\gamma}_{\perp}. \Big( \bm{k}_{f\perp}-\frac{k^{-}_{f}-l^{-}}{k^{-}_{\gamma}} \, \bm{k}_{\gamma \perp} \Big)  \gamma_{\alpha} \gamma^{-} \nonumber \\
& \times  \Big( \frac{\bm{k}_{\gamma \perp}.\bm{k}_{f\perp}}{k^{-}_{f}} -\frac{\bm{k}_{\gamma \perp}^{2}}{k^{-}_{\gamma}} \Big) -\frac{(k^{-}_{f}+k^{-}_{\gamma}-l^{-})^{2}}{k^{-}_{\gamma} \, l^{-}} \bm{\gamma}_{\perp}. \Big( \bm{k}_{f\perp}-\frac{k^{-}_{f}-l^{-}}{k^{-}_{\gamma}} \, \bm{k}_{\gamma \perp} \Big)  \Big( \frac{\slashed{k}_{f}}{k^{-}_{f}}- \frac{\slashed{k}_{\gamma}}{k^{-}_{\gamma}} \Big) \gamma_{\alpha} \gamma^{-} \slashed{k}_{\gamma} \nonumber \\
& +(k^{-}_{f}+k^{-}_{\gamma}-l^{-}) \, \Big[ \, \Bigg(-\frac{4\, (k_{f}^{-}-l^{-})}{k^{-}_{f} \, k^{-}_{\gamma}} \, k_{f\alpha} + 2 \Big( \frac{\bm{\gamma}_{\perp}.\bm{k}_{f\perp}}{k^{-}_{f}}-   \frac{\bm{\gamma}_{\perp}.\bm{k}_{\gamma \perp}}{k^{-}_{\gamma}}- \frac{\bm{k}_{f\perp}.\bm{k}_{\gamma \perp}}{k^{-}_{f} \, k^{-}_{\gamma}} \, \gamma^{-} + \frac{\bm{k}_{\gamma \perp}^{2}}{(k^{-}_{\gamma})^{2}} \, \gamma^{-} \Big) \gamma_{\alpha} \Bigg) \, (\slashed{k}_{\gamma} \nonumber \\
& +\bm{\gamma}_{\perp}.\bm{k}_{\gamma\perp} ) - 2 \Big( \gamma_{\alpha} \slashed{k}_{\gamma} \bm{\gamma}_{\perp}. \Big( \frac{\bm{k}_{f\perp}}{k^{-}_{f}} - \frac{\bm{k}_{\gamma \perp}}{k^{-}_{\gamma}} \Big) + \frac{\bm{\gamma}_{\perp}.\bm{k}_{f\perp}}{k^{-}_{f}}-   \frac{\bm{\gamma}_{\perp}.\bm{k}_{\gamma \perp}}{k^{-}_{\gamma}}- \frac{\bm{k}_{f\perp}.\bm{k}_{\gamma \perp}}{k^{-}_{f} \, k^{-}_{\gamma}} \, \gamma^{-} + \frac{\bm{k}_{\gamma \perp}^{2}}{(k^{-}_{\gamma})^{2}} \, \gamma^{-} \Big) \, \delta^{j}_{\alpha} \slashed{k}_{\gamma} \gamma^{j} \Big] \Bigg\} \, .
\label{eq:sigma-tilde-finite-B-III}
\end{align}
Combining the finite contributions from the two cases we can finally write the net finite contribution from the loop in Fig.~\ref{fig:self-energy-nested-photon} as
\begin{equation}
\tilde{\Sigma}_{\alpha}^{\rm finite}  (k_{f},k_{\gamma})=\hyperref[eq:self-energy-nested-photon-finite-A]{\tilde{\Sigma}_{\alpha}^{A;\rm finite} (k_{f},k_{\gamma}) }+\hyperref[eq:self-energy-nested-photon-finite-B]{\tilde{\Sigma}_{\alpha}^{B;\rm finite} (k_{f},k_{\gamma})} \, .
\label{eq:self-energy-nested-photon-finite}
\end{equation}

\section{Virtual gluon contributions to the $\gamma^{*}q\bar{q}$ vertex in $A^{-}=0$ gauge} \label{sec:vertex-correction-computation}

We will derive here a generic expression, for the building block shown below in Fig.~\ref{fig:vertex-correction}, contributing to the amplitudes $(V13)$ and $(V14)$ in Fig.~\ref{fig:NLO-vertex-3}. 
\begin{figure}[!htbp]
\begin{center}
\includegraphics[scale=1]{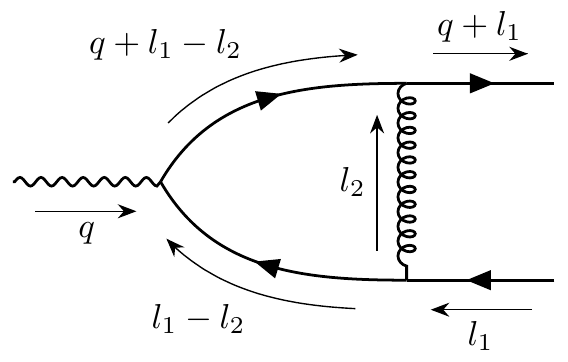}
\caption{Building block for the processes $(V13)$ and $(V14)$ in Fig.~\ref{fig:NLO-vertex-3} (and their quark$\leftrightarrow$antiquark interchanged counterparts) representing O$(\alpha_{S})$  contributions to the $q\bar{q}\gamma$ vertex.\label{fig:vertex-correction}}
\end{center}
\end{figure}
This can in general be written as
\begin{equation*}
S_{0}(q+l_{1}) \, \Gamma_{\mu}(l_{1}^{+},l_{1}^{-}) \, S_{0}(l_{1})\,,
\end{equation*}
where the quantity of interest is 
\begin{align}
\Gamma_{\mu}(l_{1}^{+},l_{1}^{-})= \int_{l_{2}} \, (ig)t^{a} \, \gamma^{\beta} \, S_{0}(q+l_{1}-l_{2}) \, (-ieq_{f}) \gamma_{\mu} \, S_{0}(l_{1}-l_{2}) \, (ig)t^{b} \, \gamma^{\nu} \times G^{0}_{\beta \nu;ab}(l_{2}) \, .
\label{eq:vertex-correction-generic}
\end{align}
The choice of arguments for $\Gamma_{\mu}$ will become clear as we proceed with the calculation. 
%The free quark and gluon propagators are given by Eq.~\ref{eq:fermion-gluon-propagator}. 
We will first perform the integration over $l_{2}^{+}$ using complex contour integration. Using the identity in Eq.~\ref{eq:gamma-identity-2} and the fact that terms proportional to $\gamma^{-} \gamma_{\mu} \gamma^{-}= 2\delta_{\mu -} \gamma^{-}$ yield zero after contraction with the intermediate photon vertex in Eq.~\ref{eq:gamma-contraction-with-virtual-photon-propagator}, we see that the numerator is at most proportional to $l_{2}^{+}$ whereas the denominator is proportional to $(l_{2}^{+})^{3}$. As such, the contour integration can be performed without any additional complications. The three $l_{2}^{+}$ poles are,
\begin{align}
l_{2}^{+} \vert_{a}=\frac{\bm{l}_{2\perp}^{2}}{2l^{-}_{2}} -\frac{i\varepsilon}{2l_{2}^{-}} \, , \,\,  l_{2}^{+} \vert_{b}=l_{1}^{+}-\frac{(\bm{l}_{1\perp}-\bm{l}_{2\perp})^{2}}{2(l^{-}_{1}-l_{2}^{-})} +\frac{i\varepsilon}{2(l_{1}^{-}-l_{2}^{-})} \, , \,\,
l_{2}^{+} \vert_{c}=q^{+}+l_{1}^{+}-\frac{(\bm{l}_{1\perp}-\bm{l}_{2\perp})^{2}}{2(q^{-}+l_{1}^{-}-l_{2}^{-})} + \frac{i\varepsilon}{2(q^{-}+l_{1}^{-}-l_{2}^{-})}  \, ,
\end{align}
where $l_{2}^{+} \vert_{b}$ and $l_{2}^{+} \vert_{c}$ are located below and above the real $l_{2}^{+}$ axis respectively for any $l_{2}^{-}$. The location of $l_{2}^{+} \vert_{a}$ however depends on the sign of $l_{2}^{-}$. In the following, we will obtain the generic expressions for $\Gamma_{\mu}$ for two cases
\begin{itemize}
\item Case A: For $0<l_{2}^{-}<q^{-}+l_{1}^{-}$ we have the poles at $l_{2}^{+} \vert_{a}$ and $l_{2}^{+} \vert_{b}$ located below the real $l_{2}^{+}$ axis whereas $l_{2}^{+} \vert_{c}$ is above. We will therefore deform the contour anticlockwise to enclose the pole at  $l_{2}^{+} \vert_{c}$. Using Cauchy's residue theorem and making the momentum redefinition $\bm{l}_{2\perp}-\bm{l}_{1\perp} \rightarrow \bm{l}_{2\perp}$ with $\bm{l}_{1\perp}$ remaining unchanged we obtain the following expression for $\Gamma_{\mu}$
\begin{align}
\Gamma_{\mu}^{A}(l_{1}^{+},l_{1}^{-})&= -ieq_{f}g^{2} C_{F} \int_{l_{2}^{-}} \int_{\bm{l}_{2\perp}} \gamma^{\beta} \frac{\gamma^{+}(q^{-}+l_{1}^{-}-l_{2}^{-})+\gamma^{-} \frac{\bm{l}_{2\perp}^{2}}{2(q^{-}+l_{1}^{-}-l_{2}^{-})} +\bm{\gamma}_{\perp}.\bm{l}_{2\perp}    }{2q^{-}(q^{-}+l_{1}^{-})/(q^{-}+l_{1}^{-}-l_{2}^{-}) \Big[ (\bm{l}_{2\perp}+\bm{v}_{1\perp}^{V})^{2}+\Delta^{V}_{1}  \Big] } \gamma_{\mu}  \nonumber \\
& \times \frac{\gamma^{+}(l_{1}^{-}-l_{2}^{-}) +\gamma^{-} \Big(\frac{Q^{2}}{2q^{-}}+\frac{\bm{l}_{2\perp}^{2}}{2(q^{-}+l_{1}^{-}-l_{2}^{-})} \Big)+\bm{\gamma}_{\perp}.\bm{l}_{2\perp}   }{\bm{l}_{2\perp}^{2}-\frac{Q^{2}(l_{1}^{-}-l_{2}^{-})(q^{-}+l_{1}^{-}-l_{2}^{-})}{(q^{-})^{2}} -i\varepsilon} \gamma^{\nu}  \Bigg( -g_{\beta \nu}+\frac{(l_{2}+l_{1})_{\beta}n_{\nu}+(l_{2}+l_{1})_{\nu}n_{\beta}}{l^{-}_{2}} \Bigg|_{l_{2}^{+}\vert_{c}} \Bigg) \, ,
\label{eq:vertex-correction-caseA}
\end{align}
where 
\begin{align}
\bm{v}_{1\perp}^{V}&= \frac{q^{-}+l_{1}^{-}-l_{2}^{-}}{q^{-}+l_{1}^{-}} \, \bm{l}_{1\perp}  \, , \quad
\Delta_{1}^{V}= \frac{l_{2}^{-}}{q^{-}+l_{1}^{-}} \Big( 1-\frac{l_{2}^{-}}{q^{-}+l_{1}^{-}} \Big) \Big\{ \bm{l}_{1\perp}^{2} +2(q^{-}+l_{1}^{-}) \Big(\frac{Q^{2}}{2q^{-}} - l_{1}^{+} \Big) \Big\} \, .
\label{eq:denominator-factors-vert-corr-caseA}
\end{align}
The choice of $l_{1}^{+}$ and $l_{1}^{-}$ as the arguments of $\Gamma_{\mu}$ is now transparent from the above expressions. From the identity derived in Eq.~\ref{eq:gamma-identity-2}, it is clear that the numerator in Eq.~\ref{eq:vertex-correction-caseA} above has to be evaluated at $l_{2}^{+} \vert_{c}$ which will have a residual $l_{1}^{+}$ dependence. The value of this will depend on the choice of contour taken for the $l_{1}^{+}$ integration for the full computation. The same is also true for the factor $\Delta_{1}^{V}$ given by Eq.~\ref{eq:denominator-factors-vert-corr-caseA}.

\item Case B: For $0>l_{2}^{-}>l_{1}^{-}$ we have the poles at $l_{2}^{+} \vert_{a}$ and $l_{2}^{+} \vert_{c}$ above the real $l_{2}^{+}$ axis while the pole at $l_{2}^{+} \vert_{b}$ is below. For convenience we therefore choose to deform the contour clockwise. Performing the same momentum redefinition $\bm{l}_{2\perp}-\bm{l}_{1\perp} \rightarrow \bm{l}_{2\perp}$ as above and applying Cauchy's residue theorem we get 
\begin{align}
\Gamma_{\mu}^{B}(l_{1}^{+},l_{1}^{-})&= -ieq_{f}g^{2} C_{F} \int_{l_{2}^{-}} \int_{\bm{l}_{2\perp}} \gamma^{\beta} \frac{\gamma^{+}(q^{-}+l_{1}^{-}-l_{2}^{-})+\gamma^{-} \Big( q^{+}+\frac{\bm{l}_{2\perp}^{2}}{2(l_{1}^{-}-l_{2}^{-})}  \Big) +\bm{\gamma}_{\perp}.\bm{l}_{2\perp}    }{2q^{-}(l_{1}^{-})/(l_{1}^{-}-l_{2}^{-}) \Big[ (\bm{l}_{2\perp}+\bm{v}_{2\perp}^{V})^{2}+\Delta^{V}_{2}  \Big] } \gamma_{\mu}  \nonumber \\
& \times \frac{\gamma^{+}(l_{1}^{-}-l_{2}^{-}) +\gamma^{-} \frac{\bm{l}_{2\perp}^{2}}{2(l_{1}^{-}-l_{2}^{-})} +\bm{\gamma}_{\perp}.\bm{l}_{2\perp}   }{\bm{l}_{2\perp}^{2}-\frac{Q^{2}(l_{1}^{-}-l_{2}^{-})(q^{-}+l_{1}^{-}-l_{2}^{-})}{(q^{-})^{2}} -i\varepsilon} \gamma^{\nu} \times \Bigg( -g_{\beta \nu}+\frac{(l_{2}+l_{1})_{\beta}n_{\nu}+(l_{2}+l_{1})_{\nu}n_{\beta}}{l^{-}_{2}} \Bigg|_{l_{2}^{+}\vert_{b}} \Bigg) \, ,
\label{eq:vertex-correction-caseB}
\end{align}
where 
\begin{align}
\bm{v}_{2\perp}^{V}&= \frac{l_{1}^{-}-l_{2}^{-}}{l_{1}^{-}} \, \bm{l}_{1\perp}  \, ,  \quad
\Delta_{2}^{V}= \frac{l_{2}^{-}}{l_{1}^{-}} \Big( 1-\frac{l_{2}^{-}}{l_{1}^{-}} \Big) \Big\{ \bm{l}_{1\perp}^{2} -2l_{1}^{+}l_{1}^{-} \Big\} \, .
\label{eq:denominator-factors-vert-corr-caseB}
\end{align}
Once again the above expressions are to be evaluated at the $l_{1}^{+}$ pole enclosed by our contour of choice in the full computation.

\end{itemize}

\section{Divergent contributions constituting $T_{V}^{(1)}$} \label{sec:T-V1-div-parts}

In this section, we will discuss the strategy to compute the remaining divergent contributions (those not provided in the main text) that constitute $T_{V}^{(1)}$ in the contributions to the amplitude from the six diagrams in Fig.~\ref{fig:NLO-vertex-1}. Recall that this amplitude can be expressed as 
\begin{align}
\mathcal{M}^{\text{NLO};\text{Vert.}(1)}_{\mu \alpha}& = 2\pi \delta(1-z^{v}_{\text{tot}}) \, (eq_{f} g)^{2} \,  \int \mathrm{d} \Pi_{\perp}^{v} \, \overline{u} (\bm{k}) \Bigg\{ T^{(1)}_{V;\mu \alpha} (\bm{l}_{1\perp}) \, \Big[ \Big( t^{b} \tilde{U} (\bm{x}_{\perp}) t^{a} \tilde{U}^{\dagger} (\bm{y}_{\perp}) \Big) U_{ba}(\bm{z}_{\perp}) - C_{F} \mathds{1} \Big] \Bigg\} v(\bm{p}) \, ,
\end{align}
where 
\begin{equation}
T^{(1)}_{V;\mu \alpha}(\bm{l}_{1\perp}) = \sum_{\beta=1}^{6} R^{(V\beta)}_{\mu \alpha} (\bm{l}_{1\perp})   \, ,
\end{equation}
and each $R^{(V\beta)}$ , $\beta=1,\ldots,6$, can be decomposed as the sum of a divergent and a finite part  as 
\begin{align}
 R^{(V\beta)}_{\mu \alpha} (\bm{l}_{1\perp})   =   R^{(V\beta)}_{\rm div.;\mu \alpha}  (\bm{l}_{1\perp}) +
 R^{(V\beta)}_{\rm finite.;\mu \alpha}(\bm{l}_{1\perp})  
 \equiv  R^{(V\beta)}_{\rm div.;\mu \alpha}  (\bm{l}_{1\perp}) +\big\{\Re^{(V\beta)}_{\mu \alpha}  (\bm{l}_{1\perp})+R^{(V\beta)}_{(\rm II);\mu \alpha}  (\bm{l}_{1\perp}) \big\}  \, .
\end{align}
In Sec.~\ref{sec:virtual-corrections-vertex}, we demonstrated that there are no UV divergent contributions from these diagrams for our choice of gauge. We are only left with singular pieces in rapidity for four out of the six allowed processes in this category. For $(V2)$ and $(V5)$, we have purely finite contributions; we can therefore write,
\begin{align}
  \begin{pmatrix} R^{(V2)}_{\mu \alpha} (\bm{l}_{1\perp})  \\
  R^{(V5)}_{\mu \alpha} (\bm{l}_{1\perp})
  \end{pmatrix}&= \begin{pmatrix}
  R^{(V2)}_{(\rm II);\mu \alpha}  (\bm{l}_{1\perp}) \\
  R^{(V5)}_{(\rm II);\mu \alpha}  (\bm{l}_{1\perp}) 
  \end{pmatrix} \, .
  \label{eq:R-V2-R-V5}
\end{align}
The expressions for these finite pieces will be given in Appendix~\ref{sec:finite-pieces-Ver1}.

We begin with the process $(V6)$ which has a similar structure to the process $(V1)$. We can write
 \begin{equation}
R^{(V6)}_{\mu \alpha} (\bm{l}_{1\perp})   =   \hyperref[eq:R-div-V6]{R^{(V6)}_{\rm div.;\mu \alpha}  (\bm{l}_{1\perp}) }+\big\{\hyperref[eq:remainder-V6]{\Re^{(V6)}_{\mu \alpha}  (\bm{l}_{1\perp})}+\hyperref[eq:R-finite-V6]{R^{(V6)}_{(\rm II);\mu \alpha}  (\bm{l}_{1\perp})} \big\}  \, .
\label{eq:R-V6}
 \end{equation}
 where the divergent contribution is obtained as
\begin{align}
R^{(V6)}_{\rm div.;\mu \alpha}  (\bm{l}_{1\perp}) &=\ln \Big(\frac{1}{z_{0}} \Big) \times \int_{\bm{l}_{2\perp}} \!\!\!\! e^{i\bm{l}_{2\perp}.\bm{r}_{zy}} \, \frac{1}{(q^{-})^{2}}  \, \Bigg[ \big\{4z_{\bar{q}}q^{-} \delta_{i\mu} \gamma^{-}-2q^{-}\gamma_{\mu} \gamma^{i} \gamma^{-} \big\}  \big\{ \mathcal{I}_{v;\rm log}^{(3,ijj)} (\bm{c}_{1\perp}^{(V6)},c_{3}^{(V6)}) \nonumber \\
& + l_{2}^{j} \,  \mathcal{I}_{v;\rm log}^{(3,ij)} (\bm{c}_{1\perp}^{(V6)},c_{3}^{(V6)}) \big\} - \big\{ \gamma^{-} [\gamma^{+}(1-z_{\bar{q}})q^{-}-\bm{\gamma}_{\perp}.\bm{l}_{1\perp}]\gamma_{\mu}  [\gamma^{+}z_{\bar{q}}q^{-}+\bm{\gamma}_{\perp}.\bm{l}_{1\perp}]\gamma^{-} \big\}  \nonumber \\
& \times \big\{ \mathcal{I}_{v;\rm log}^{(3,ii)} (\bm{c}_{1\perp}^{(V6)},c_{3}^{(V6)}) + l_{2}^{i} \,  \mathcal{I}_{v;\rm log}^{(3,i)} (\bm{c}_{1\perp}^{(V6)},c_{3}^{(V6)}) \big\} \Bigg] \, \frac{\slashed{p}+\slashed{k}_{\gamma}}{2p.k_{\gamma}} \, \gamma_{\alpha}   \, .
\label{eq:R-div-V6}
\end{align}
The integrals appearing above are given by Eqs.~\ref{eq:constituent-integrals-V1-log}. For $(V6)$ we obtain the following expressions for the arguments $\bm{c}_{1\perp}$ and $c_{3}$ of these integrals,
\begin{align}
\bm{c}_{1\perp}^{(V6)}&= \alpha_{1} \, \bm{l}_{2\perp} - \alpha_{2} \, (\bm{l}_{1\perp}-\bm{k}_{\gamma \perp}) \, , \nonumber \\
c_{3}^{(V6)}&= \alpha_{1}(1-\alpha_{1}) \, \bm{l}_{2\perp}^{2} +\alpha_{2}(1-\alpha_{2}) \, (\bm{l}_{1\perp}-\bm{k}_{\gamma \perp})^{2}+2\alpha_{1} \alpha_{2} \, (\bm{l}_{1\perp}-\bm{k}_{\gamma \perp}).\bm{l}_{2\perp}+\alpha_{2} \, Q^{2}z_{q}(1-z_{q}) \, .
\label{eq:c1-c3-V6}
\end{align}
We will now present the results for $(V4)$ which has a similar structure to that of $(V3)$. We will not provide expressions for the divergent pieces of $(V3)$ here because they are similar to that of $(V4)$ but more lengthy. The latter is due to the fact that in the amplitude computation of $(V3)$ the contour for the integration over $l_{1}^{+}$ encloses two poles on the same side of the real axis. There are therefore two separate contributions which need to be added in order to obtain the final divergent piece. The divergent terms computed for these processes contribute towards the leading logarithmic evolution of the LO result and can be absorbed in a redefinition of the weight functional $W_{\Lambda_{0}^{-}}[\rho_{A}]$ describing color sources as described in the introduction. This is also explicitly shown in the JIMWLK derivation discussed at length in section~\ref{sec:JIMWLK-evolution}.

For $(V4)$, we can write
\begin{equation}
R^{(V4)}_{\mu \alpha} (\bm{l}_{1\perp})   =   \hyperref[eq:R-div-V4]{R^{(V4)}_{\rm div.;\mu \alpha}  (\bm{l}_{1\perp})} +\big\{\hyperref[eq:remainder-V4]{\Re^{(V4)}_{\mu \alpha}  (\bm{l}_{1\perp})}+\hyperref[eq:R-finite-V4]{R^{(V4)}_{(\rm II);\mu \alpha}  (\bm{l}_{1\perp})} \big\}  \, ,
\label{eq:R-V4}
\end{equation}
where the divergent pieces for $(V4)$ are obtained from terms in the amplitude that are proportional to $1/z_{l}$. These specific terms can be written as
\begingroup
\allowdisplaybreaks
\begin{align}
R^{(V4)}_{\rm (I);\mu \alpha}  (\bm{l}_{1\perp})&=\int \frac{\mathrm{d}z_{l}}{(2\pi) \, z_{l}}\int_{\bm{l}_{2\perp}} \!\!\!\!\! e^{i\bm{l}_{2\perp}.\bm{r}_{zy}} \, \frac{1}{(1-z_{\gamma}) /z_{q} \, (q^{-})^{2}} \,  \Big[ 2z_{\bar{q}}q^{-} \gamma_{\mu} \gamma_{\alpha} \gamma^{-} \,  \Big\{ \mathcal{I}_{v}^{(4,iijj)} (\bm{V}_{\perp}^{(V4)},\Delta^{(V4)}) \nonumber \\
& + l_{2}^{i} \, \mathcal{I}_{v}^{(4,ijj)} (\bm{V}_{\perp}^{(V4)},\Delta^{(V4)}) \Big\} +\big\{ 4z_{q}q^{-} \delta_{i\mu} \gamma_{\alpha} \gamma^{j} \gamma^{-}-2q^{-} \gamma^{i} \gamma_{\mu} \gamma_{\alpha} \gamma^{j} \gamma^{-} -2z_{\bar{q}}q^{-} \gamma^{i} \gamma_{\mu} \gamma^{j} \gamma_{\alpha} \gamma^{-} \big\} \nonumber \\
&\times\Big\{ \mathcal{I}_{v}^{(4,ijkk)} (\bm{V}_{\perp}^{(V4)},\Delta^{(V4)}) +l_{2}^{k} \,  \mathcal{I}_{v}^{(4,ijk)} (\bm{V}_{\perp}^{(V4)},\Delta^{(V4)}) \Big\}  +\big\{ \gamma^{-} \gamma^{i} \gamma_{\mu} [\gamma^{+}(1-z_{q})q^{-}+\bm{\gamma}_{\perp}.(\bm{l}_{1\perp}-\bm{k}_{\gamma \perp}) ] \nonumber \\
& \times \gamma_{\alpha} [\gamma^{+} z_{\bar{q}}q^{-}+\bm{\gamma}_{\perp}.\bm{l}_{1\perp} ]\gamma^{-} -\gamma^{-} [\gamma^{+}z_{q}q^{-}-\bm{\gamma}_{\perp}.(\bm{l}_{1\perp}-\bm{k}_{\gamma \perp}) ]\gamma_{\mu} [\gamma^{+} (1-z_{q})q^{-} +\bm{\gamma}_{\perp}.(\bm{l}_{1\perp}-\bm{k}_{\gamma \perp}) ]\gamma_{\alpha} \gamma^{i} \gamma^{-} \nonumber \\
& -\gamma^{-} [\gamma^{+}z_{q}q^{-} -\bm{\gamma}_{\perp}.(\bm{l}_{1\perp}-\bm{k}_{\gamma \perp}) ] \gamma_{\mu}     \gamma^{i} \gamma_{\alpha} [\gamma^{+} z_{\bar{q}}q^{-}+\bm{\gamma}_{\perp}.\bm{l}_{1\perp} ]\gamma^{-} -4z_{\bar{q}}q^{-} (l_{1}^{i}-k_{\gamma}^{i}) \gamma_{\mu} \gamma_{\alpha} \gamma^{-} \big\} \nonumber \\
&\times  \Big\{ \mathcal{I}_{v}^{(4,ijj)} (\bm{V}_{\perp}^{(V4)},\Delta^{(V4)})+l_{2}^{j} \,  \mathcal{I}_{v}^{(4,ij)} (\bm{V}_{\perp}^{(V4)},\Delta^{(V4)}) \Big\} + \big\{ \gamma^{-} [\gamma^{+}z_{q}q^{-}-\bm{\gamma}_{\perp}.(\bm{l}_{1\perp}-\bm{k}_{\gamma \perp}) ]\nonumber \\
& \times \gamma_{\mu} [\gamma^{+}(1-z_{q})q^{-} -\gamma^{-} \Big( \frac{Q^{2}}{2q^{-}} +\frac{(\bm{l}_{1\perp}-\bm{k}_{\gamma \perp})^{2}}{2z_{q}q^{-} } \Big)  +\bm{\gamma}_{\perp}.(\bm{l}_{1\perp}-\bm{k}_{\gamma \perp} ) ]\gamma_{\alpha} [\gamma^{+}z_{\bar{q}}q^{-}+\bm{\gamma}_{\perp}.\bm{l}_{1\perp} ]\gamma^{-} \big\} \nonumber \\
& \times \Big\{ \mathcal{I}_{v}^{(4,ii)} (\bm{V}_{\perp}^{(V4)},\Delta^{(V4)}) +l_{2}^{i} \,  \mathcal{I}_{v}^{(4,i)} (\bm{V}_{\perp}^{(V4)},\Delta^{(V4)}) \Big\} \,  .
\label{eq:R-V4-propto-zl-inverse}
\end{align}
\endgroup 
The finite expressions for the constituent integrals appearing in the above equation are as follows:
\begin{align}
\mathcal{I}_{v}^{(4,ijkk)}(\bm{V}_{\perp},\Delta)&= \frac{1}{4\pi } \int_{0}^{1} \mathrm{d} \alpha_{1} \int_{0}^{1-\alpha_{1}}  \!\!\!\!\!\!\! \mathrm{d} \alpha_{2} \int_{0}^{1-\alpha_{1}-\alpha_{2}}   \!\!\!\!\!\!\!\!\!\!\!\!\!\! \mathrm{d} \alpha_{3} \,\,\,\,  \Big\{ \Big( \frac{2\bm{V}_{\perp}^{2}}{\Delta^{3}} +\frac{3}{\Delta^{2}} \Big) \, V^{i} V^{j} + \Big( \frac{\bm{V}_{\perp}^{2}}{2\Delta^{2}} +\frac{1}{\Delta} \Big) \, \delta^{ij} \Big\} \, , \nonumber \\
\mathcal{I}_{v}^{(4,ijk)}(\bm{V}_{\perp},\Delta)&=- \frac{1}{4\pi } \int_{0}^{1} \mathrm{d} \alpha_{1} \int_{0}^{1-\alpha_{1}}  \!\!\!\!\!\!\! \mathrm{d} \alpha_{2} \int_{0}^{1-\alpha_{1}-\alpha_{2}}   \!\!\!\!\!\!\!\!\!\!\!\!\!\! \mathrm{d} \alpha_{3} \,\,\,\, \Big\{\frac{2}{\Delta^{3}} \, V^{i} V^{j} V^{k} +\frac{1}{2\Delta^{2}} \, (\delta^{ij} \, V^{k} +\delta^{jk} \,  V^{i} +\delta^{ki} \, V^{j} ) \Big\} \, , \nonumber \\
\mathcal{I}_{v}^{(4,ij)}(\bm{V}_{\perp},\Delta)&= \frac{1}{4\pi } \int_{0}^{1} \mathrm{d} \alpha_{1} \int_{0}^{1-\alpha_{1}}  \!\!\!\!\!\!\! \mathrm{d} \alpha_{2} \int_{0}^{1-\alpha_{1}-\alpha_{2}}   \!\!\!\!\!\!\!\!\!\!\!\!\!\! \mathrm{d} \alpha_{3} \,\,\,\, \Big\{ \frac{2}{\Delta^{3}} \, V^{i} V^{j} +\frac{1}{2\Delta^{2}} \, \delta^{ij} \Big\} \, , \nonumber \\
\mathcal{I}_{v}^{(4,i)}(\bm{V}_{\perp},\Delta)&=- \frac{1}{4\pi } \int_{0}^{1} \mathrm{d} \alpha_{1} \int_{0}^{1-\alpha_{1}}  \!\!\!\!\!\!\! \mathrm{d} \alpha_{2} \int_{0}^{1-\alpha_{1}-\alpha_{2}}   \!\!\!\!\!\!\!\!\!\!\!\!\!\! \mathrm{d} \alpha_{3} \,\,\,\, \frac{2}{\Delta^{3}} \, V^{i} \, , \nonumber \\
\mathcal{I}_{v}^{(4,0)}(\bm{V}_{\perp},\Delta)&= \frac{1}{4\pi } \int_{0}^{1} \mathrm{d} \alpha_{1} \int_{0}^{1-\alpha_{1}}  \!\!\!\!\!\!\! \mathrm{d} \alpha_{2} \int_{0}^{1-\alpha_{1}-\alpha_{2}}   \!\!\!\!\!\!\!\!\!\!\!\!\!\! \mathrm{d} \alpha_{3} \,\,\,\, \frac{2}{\Delta^{3}}  \, .
\label{eq:constituent-integrals-V3-V4}
\end{align}
The remaining integrals can be obtained by equating two of the indices in the expressions provided above. The arguments $\bm{V}_{\perp}$ and $\Delta$ can always be expressed in terms of $z_{l}$ as shown in Eq.~\ref{eq:V-and-Delta-in-terms-of-zl}. The integration over $z_{l}$ can now be performed and we can express the divergent piece proportional to logs in $z_{0}$ in terms of integrals over the Feynman parameters. These integrals only depend on the coefficients $\bm{c}_{1\perp}$ and $c_{3}$ appearing in Eq.~\ref{eq:V-and-Delta-in-terms-of-zl} whose individual expressions in turn depend on the process of interest. 

The divergent term in $(V4)$ can now be finally written as
\begingroup
\allowdisplaybreaks
\begin{align}
R^{(V4)}_{\rm div.;\mu \alpha}  (\bm{l}_{1\perp})&= \ln \Big(\frac{1}{z_{0}} \Big) \times \int_{\bm{l}_{2\perp}} \!\!\!\!\! e^{i\bm{l}_{2\perp}.\bm{r}_{zy}} \, \frac{1}{(1-z_{\gamma}) /z_{q} \, (q^{-})^{2}} \,  \Big[ 2z_{\bar{q}}q^{-} \gamma_{\mu} \gamma_{\alpha} \gamma^{-} \,  \Big\{ \mathcal{I}_{v;\rm log}^{(4,iijj)} (\bm{c}_{1\perp}^{(V4)},c_{3}^{(V4)}) \nonumber \\
& + l_{2}^{i} \, \mathcal{I}_{v;\rm log}^{(4,ijj)} (\bm{c}_{1\perp}^{(V4)},c_{3}^{(V4)}) \Big\} +\big\{ 4z_{q}q^{-} \delta_{i\mu} \gamma_{\alpha} \gamma^{j} \gamma^{-}-2q^{-} \gamma^{i} \gamma_{\mu} \gamma_{\alpha} \gamma^{j} \gamma^{-} -2z_{\bar{q}}q^{-} \gamma^{i} \gamma_{\mu} \gamma^{j} \gamma_{\alpha} \gamma^{-} \big\} \nonumber \\
&\times\Big\{ \mathcal{I}_{v;\rm log}^{(4,ijkk)} (\bm{c}_{1 \perp}^{(V4)},c_{3}^{(V4)}) +l_{2}^{k} \,  \mathcal{I}_{v;\rm log }^{(4,ijk)} (\bm{c}_{1\perp}^{(V4)},c_{3}^{(V4)}) \Big\}  +\big\{ \gamma^{-} \gamma^{i} \gamma_{\mu} [\gamma^{+}(1-z_{q})q^{-}+\bm{\gamma}_{\perp}.(\bm{l}_{1\perp}-\bm{k}_{\gamma \perp}) ] \nonumber \\
& \times \gamma_{\alpha} [\gamma^{+} z_{\bar{q}}q^{-}+\bm{\gamma}_{\perp}.\bm{l}_{1\perp} ]\gamma^{-} -\gamma^{-} [\gamma^{+}z_{q}q^{-}-\bm{\gamma}_{\perp}.(\bm{l}_{1\perp}-\bm{k}_{\gamma \perp}) ]\gamma_{\mu} [\gamma^{+} (1-z_{q})q^{-} +\bm{\gamma}_{\perp}.(\bm{l}_{1\perp}-\bm{k}_{\gamma \perp}) ]\gamma_{\alpha} \gamma^{i} \gamma^{-} \nonumber \\
& -\gamma^{-} [\gamma^{+}z_{q}q^{-} -\bm{\gamma}_{\perp}.(\bm{l}_{1\perp}-\bm{k}_{\gamma \perp}) ] \gamma_{\mu}     \gamma^{i} \gamma_{\alpha} [\gamma^{+} z_{\bar{q}}q^{-}+\bm{\gamma}_{\perp}.\bm{l}_{1\perp} ]\gamma^{-} -4z_{\bar{q}}q^{-} (l_{1}^{i}-k_{\gamma}^{i}) \gamma_{\mu} \gamma_{\alpha} \gamma^{-} \big\} \nonumber \\
&\times  \Big\{ \mathcal{I}_{v;\rm log}^{(4,ijj)} (\bm{c}_{1\perp}^{(V4)},c_{3}^{(V4)})+l_{2}^{j} \,  \mathcal{I}_{v;\rm log}^{(4,ij)} (\bm{c}_{1\perp}^{(V4)},c_{3}^{(V4)}) \Big\} + \big\{ \gamma^{-} [\gamma^{+}z_{q}q^{-}-\bm{\gamma}_{\perp}.(\bm{l}_{1\perp}-\bm{k}_{\gamma \perp}) ]\nonumber \\
& \times \gamma_{\mu} [\gamma^{+}(1-z_{q})q^{-} -\gamma^{-} \Big( \frac{Q^{2}}{2q^{-}} +\frac{(\bm{l}_{1\perp}-\bm{k}_{\gamma \perp})^{2}}{2z_{q}q^{-} } \Big)  +\bm{\gamma}_{\perp}.(\bm{l}_{1\perp}-\bm{k}_{\gamma \perp} ) ]\gamma_{\alpha} [\gamma^{+}z_{\bar{q}}q^{-}+\bm{\gamma}_{\perp}.\bm{l}_{1\perp} ]\gamma^{-} \big\} \nonumber \\
& \times \Big\{ \mathcal{I}_{v;\rm log}^{(4,ii)} (\bm{c}_{1\perp}^{(V4)},c_{3}^{(V4)}) +l_{2}^{i} \,  \mathcal{I}_{v;\rm log }^{(4,i)} (\bm{c}_{1\perp}^{(V4)},c_{3}^{(V4)}) \Big\} \,  .
\label{eq:R-div-V4}
\end{align}
\endgroup
The expressions for some of the integrals appearing in the above equation are provided below. The rest of them can be obtained by putting $i=j$ in the expressions given.
\begingroup
\allowdisplaybreaks
\begin{align}
\mathcal{I}_{v;\rm log}^{(4,ijkk)} \Big(\bm{c}_{1\perp},c_{3}\Big)&=\frac{1}{8\pi^{2}} \int_{0}^{1} \mathrm{d} \alpha_{1} \int_{0}^{1-\alpha_{1}} \!\!\!\!\!\!\!\!\! \mathrm{d} \alpha_{2} \int_{0}^{1-\alpha_{1}-\alpha_{2}} \!\!\!\!\!\!\!\!\!\!\!\! \mathrm{d} \alpha_{3} \, \Big\{\delta^{ij} \Big( \frac{1}{c_{3}} +\frac{\bm{c}_{1\perp}^{2}}{2c_{3}^{2}} \Big) +c_{1}^{i} c_{1}^{j} \Big( \frac{3}{c_{3}^{2}} + \frac{2\bm{c}_{1\perp}^{2}}{c_{3}^{3}} \Big) \Big\} \, , \nonumber \\
\mathcal{I}_{v;\rm log}^{(4,ijk)} \Big(\bm{c}_{1\perp},c_{3}\Big)&= - \frac{1}{8\pi^{2}} \int_{0}^{1} \mathrm{d} \alpha_{1} \int_{0}^{1-\alpha_{1}} \!\!\!\!\!\!\!\!\! \mathrm{d} \alpha_{2} \int_{0}^{1-\alpha_{1}-\alpha_{2}} \!\!\!\!\!\!\!\!\!\!\!\! \mathrm{d} \alpha_{3} \, \Big\{ \big(\delta^{ij} \, c_{1}^{k}+ \delta^{jk} \, c_{1}^{i} +\delta^{ki} \, c_{1}^{j} \big) \, \frac{1}{2c_{3}^{2}} + \frac{2 \, c_{1}^{i} c_{1}^{j} c_{1}^{k}}{c_{3}^{2}} \Big\} \, ,  \nonumber \\
\mathcal{I}_{v;\rm log}^{(4,ij)} \Big(\bm{c}_{1\perp},c_{3}\Big)&= \frac{1}{8\pi^{2}} \int_{0}^{1} \mathrm{d} \alpha_{1} \int_{0}^{1-\alpha_{1}} \!\!\!\!\!\!\!\!\! \mathrm{d} \alpha_{2} \int_{0}^{1-\alpha_{1}-\alpha_{2}} \!\!\!\!\!\!\!\!\!\!\!\! \mathrm{d} \alpha_{3} \, \Big\{ \frac{\delta^{ij}}{2\, c_{3}^{2}} + \frac{2 \, c_{1}^{i} \, c_{1}^{j} }{c_{3}^{3}} \Big\} \, , \nonumber \\
\mathcal{I}_{v;\rm log}^{(4,i)}\Big(\bm{c}_{1\perp},c_{3}\Big)&=- \frac{1}{8\pi^{2}} \int_{0}^{1} \mathrm{d} \alpha_{1} \int_{0}^{1-\alpha_{1}} \!\!\!\!\!\!\!\!\! \mathrm{d} \alpha_{2} \int_{0}^{1-\alpha_{1}-\alpha_{2}} \!\!\!\!\!\!\!\!\!\!\!\! \mathrm{d} \alpha_{3} \, \,  \frac{2 \, c_{1}^{i}}{c_{3}^{3}} \, .
\label{eq:constituent-integrals-V3-V4-log}
\end{align}
\endgroup
Specifically, for $(V4)$, the arguments of these integrals are,
\begin{align}
\bm{c}_{1\perp}^{(V4)}&= \alpha_{1} (\bm{k}_{\gamma \perp} -\bm{l}_{1\perp}) +\alpha_{2} \, \Big( \frac{z_{\bar{q}}}{1-z_{\gamma}} \, \bm{k}_{\gamma \perp} -\bm{l}_{1\perp} \Big) +\alpha_{3} \, \bm{l}_{2\perp}  \, , \nonumber \\
c_{3}^{(V4)}&= \alpha_{1}(1-\alpha_{1}) \, (\bm{l}_{1\perp}-\bm{k}_{\gamma \perp})^{2} +\alpha_{2}(1-\alpha_{2} )  \, \Big( \bm{l}_{1\perp} -\frac{z_{\bar{q}}}{1-z_{\gamma}} \, \bm{k}_{\gamma \perp} \Big)^{2} +\alpha_{3}(1-\alpha_{3} ) \, \bm{l}_{2\perp}^{2} \nonumber \\
&-2\alpha_{1} \alpha_{2} \, (\bm{l}_{1\perp}-\bm{k}_{\gamma \perp} ). \Big( \bm{l}_{1\perp} -\frac{z_{\bar{q}}}{1-z_{\gamma}} \, \bm{k}_{\gamma \perp} \Big)+ 2\alpha_{2} \alpha_{3} \, \Big( \bm{l}_{1\perp} -\frac{z_{\bar{q}}}{1-z_{\gamma}} \, \bm{k}_{\gamma \perp} \Big). \bm{l}_{2\perp} +2 \alpha_{3} \alpha_{1} \,  \bm{l}_{2\perp}. (\bm{l}_{1\perp}-\bm{k}_{\gamma \perp}) \nonumber \\
& +\alpha_{1} \,  z_{q} (1-z_{q} )Q^{2} +\alpha_{2} \, \Big( \frac{z_{q}z_{\bar{q}}}{z_{\gamma}(1-z_{\gamma})^{2}} \, \bm{k}_{\gamma \perp}^{2} +\frac{z_{q}z_{\bar{q}}}{1-z_{\gamma}} \, Q^{2} \Big) \, .
\label{eq:c1-c3-V4}
\end{align}

%%%%%%%%%%%%%%%%%%%%%%%%%%%%%%%%%%%%%%%%%%%%%%%%%%%%%%%%%%%%%%%%%%%%%%%

\section{Divergent contributions constituting $T_{V}^{(4)}$} \label{sec:T-V4-div-parts}

In the main text, we explicitly computed the amplitude for $(V19)$ in Fig.~\ref{fig:NLO-vertex-4}. 
We will present here the divergent pieces in the amplitude for $(V21)$ in this figure (Note that the contribution from $(V20)$ is completely finite). Following the discussion in Sec.~\ref{sec:vertex-corrections-free-gluon}, we can write this amplitude as 
\begin{align}
\mathcal{M}_{\mu \alpha}^{(V21)}& = 2\pi \, (eq_{f}g)^{2} \, \delta(1-z_{\rm tot}^{v}) \int \mathrm{d}\Pi_{\perp}^{\rm LO} \, \overline{u}(\bm{k}) \, R^{(V21)}_{\mu \alpha} (\bm{l}_{1\perp}) \,  \Big[  t^{a} \tilde{U}(\bm{x}_{\perp}) \tilde{U}^{\dagger} (\bm{y}_{\perp})t_{a} -C_{F} \, \mathds{1}  \Big] \, v(\bm{p}) \, , 
\end{align}
where $R^{(V21)}$ is obtained by summing the contributions for the cases when $0<l_{2}^{-}<k^{-} $ and $0>l_{2}^{-}>-p^{-}$ which we will respectively denote by $R^{(V21);A}$ and $R^{(V21);B}$. As we have shown in Sec.~\ref{sec:vertex-corrections-free-gluon}, there are no UV divergences associated with these final state interaction processes owing to our choice of gauge. So to isolate the rapidity singularities, we isolate the terms in these pieces which are proportional to $1/z_{l}$. The contributions from the remaining terms are completely finite. With this in mind, we can write 
\begin{align}
R^{(V21)}_{\mu \alpha}(\bm{l}_{1\perp}) &= \Big\{ R^{(V21);A}_{(\rm I);\mu \alpha} (\bm{l}_{1\perp})+  R^{(V21);B}_{(\rm I);\mu \alpha} (\bm{l}_{1\perp}) \Big\}+\Big\{ R^{(V21);A}_{(\rm II);\mu \alpha} (\bm{l}_{1\perp})+  R^{(V21);B}_{(\rm II);\mu \alpha} (\bm{l}_{1\perp}) \Big\}   \nonumber \\
&= \hyperref[eq:R-div-V21]{R^{(V21)}_{\rm div.;\mu \alpha} (\bm{l}_{1\perp}) }+\hyperref[eq:R-finite-V21]{ R^{(V21)}_{\rm finite;\mu \alpha} (\bm{l}_{1\perp})}  \, .
\label{eq:R-V21}
\end{align}
Above we have the terms proportional to $1/z_{l}$ for the two cases A and B denoted respectively by $R^{(V21);A}_{(\rm I)} $ and $R^{(V21);B}_{(\rm I)} $. From these terms, we will get a logarithmically divergent contribution and a finite remainder. We add this remainder to the finite contributions from terms not proportional to $1/z_{l}$ which are denoted respectively for the two cases by $R^{(V21);A}_{(\rm II)} $ and $R^{(V21);B}_{(\rm II)} $. This constitutes the net finite contribution from these amplitudes. We can write this as 
\begin{align}
R^{(V21)}_{\rm finite;\mu \alpha} (\bm{l}_{1\perp})&= \Big\{ \Re^{(V21);A}_{\mu \alpha} (\bm{l}_{1\perp})  + R^{(V21);A}_{(\rm II);\mu \alpha} (\bm{l}_{1\perp}) \Big\}  +A \rightarrow  B  \, , 
\label{eq:R-finite-V21}
\end{align}
where the remainders are given by
\begin{align}
\Re^{(V21);A,B}_{\mu \alpha} (\bm{l}_{1\perp}) =R^{(V21);A,B}_{(\rm I);\mu \alpha} (\bm{l}_{1\perp}) -  R^{(V21);A,B}_{\rm div.;\mu \alpha} (\bm{l}_{1\perp})  \, .
\label{eq:remainder-V21}
\end{align}

With these definitions in mind, we will now write down the expressions for the contributions proportional to $1/z_{l}$ for the two cases considered above. For $(V21)$, the expression for $R_{(\rm I)}$ for both cases is given in terms of the constituent integrals as
\begin{align}
R^{(V21);A,B}_{(\rm I);\mu \alpha} (\bm{l}_{1\perp}) &=  \frac{1}{(q^{-})^{2}} \!\!  \int \!\! \frac{\mathrm{d}z_{l}}{(2\pi) \, z_{l}}  \,   \Big[\mathcal{R}_{(1);\mu \alpha}^{A,B}  \, \mathcal{I}_{v}^{(4,iijj)}( \bm{V}_{\perp}^{(V21);A,B},\Delta^{(V21);A,B} )  + \mathcal{R}_{(2);\mu \alpha}^{A,B;ij} \, \mathcal{I}_{v}^{(4,ijkk)} ( \bm{V}_{\perp}^{(V21);A,B},\Delta^{(V21);A,B} ) \nonumber \\
& +\mathcal{R}_{(3);\mu \alpha}^{A,B;i} \, \mathcal{I}_{v}^{(4,ijj)} ( \bm{V}_{\perp}^{(V21);A,B},\Delta^{(V21);A,B} ) + \mathcal{R}_{(4);\mu \alpha}^{A,B;ijk} \, \mathcal{I}_{v}^{(4,ijk)} ( \bm{V}_{\perp}^{(V21);A,B},\Delta^{(V21);A,B} ) \nonumber \\
&+ \mathcal{R}_{(5);\mu \alpha}^{A,B} \,  \mathcal{I}_{v}^{(4,ii)} ( \bm{V}_{\perp}^{(V21);A,B},\Delta^{(V21);A,B} )  +\mathcal{R}_{(6);\mu \alpha}^{A,B;ij} \,  \mathcal{I}_{v}^{(4,ij)} ( \bm{V}_{\perp}^{(V21);A,B},\Delta^{(V21);A,B} )  \nonumber \\
& +\mathcal{R}_{(7);\mu \alpha}^{A,B;i} \,  \mathcal{I}_{v}^{(4,i)} ( \bm{V}_{\perp}^{(V21);A,B},\Delta^{(V21);A,B} )\Big]\, \gamma^{-}  \, .
\label{eq:R-V21-propto-zl-inverse-A-B}
\end{align}
The finite expressions for these constituent integrals are given in Eqs.~\ref{eq:constituent-integrals-V3-V4}. It should also be noted that the limits of integration over $z_{l}$ are different for the two cases we are considering. 

 For Case A, we have $z_{l}$ in the limit $[z_{0},z_{q}]$ while for Case B, we have $z_{l}$ in the limit $[-z_{0},-z_{\bar{q}}]$. Any changes in sign occuring due to change of the order of limits of integration have already been accounted for in writing Eq.~\ref{eq:R-V21-propto-zl-inverse-A-B}. 
 The coefficients multiplying the integrals in Eq.~\ref{eq:R-V21-propto-zl-inverse-A-B} can be written as
\begin{align}
\mathcal{R}_{(1);\mu \alpha}^{A,B} &= C^{A,B}_{2} \, C^{A,B}_{3} \, ,   \quad  \mathcal{R}_{(2);\mu \alpha}^{A,B;ij} =C^{A,B}_{2} \, C^{A,B;ij}_{4}  \, ,\quad 
\mathcal{R}_{(3);\mu \alpha}^{A,B;i}=  \Big( C^{A,B}_{2} \, C^{A,B;i}_{5} +C^{A,B;i}_{1} \, C^{A,B}_{3} \Big)  \, , \nonumber \\
\mathcal{R}_{(4);\mu \alpha}^{A,B;ijk} &=  C^{A,B;ij}_{4} \, C^{A,B;k}_{1}  \, ,   \quad  \mathcal{R}_{(5);\mu \alpha}^{A,B} =C^{A,B}_{2} \, C^{A,B}_{6}  \, ,\quad 
\mathcal{R}_{(6);\mu \alpha}^{A,B;ij}= C^{A,B;i}_{1} \, C^{A,B;j}_{5}  \, , \quad \mathcal{R}_{(7);\mu \alpha}^{A,B;i}= C^{A,B;i}_{1} \, C^{A,B}_{6} \, .
\end{align}
The explicit expressions for these are 
\begin{align}
C^{A;i}_{1}&= \frac{z_{\bar{q}}}{(1-z_{\gamma})^{2}} \, \Big(z_{q} \, p^{i} -z_{\bar{q}} \, k^{i} \Big) \, , \quad C^{B;i}_{1}= \frac{z_{q}}{(1-z_{\gamma})^{2}} \, \Big(z_{q} \, p^{i} -z_{\bar{q}} \, k^{i} \Big) \, , \quad
C^{A}_{2}  = \frac{z_{\bar{q}}}{2(1-z_{\gamma})} \, , \quad C^{B}_{2}  = \frac{z_q}{2(1-z_{\gamma})} \, , \nonumber \\
C^{A}_{3}=C^{B}_{3}&= 2z_{q} q^{-} \gamma_{\alpha} \gamma_{\mu} \, , \quad C^{A;ij}_{4}=C^{B;ij}_{4}=2 \Big( z_{\bar{q}} \, \gamma^{i} \gamma_{\alpha} \gamma^{j} \gamma_{\mu} -(1-z_{\bar{q}}) \, \gamma^{i} \gamma_{\alpha} \gamma_{\mu} \gamma^{j} -z_{q} \, \gamma_{\alpha} \gamma^{i} \gamma_{\mu} \gamma^{j} \Big) \, q^{-}  \, , \nonumber \\
 C^{A;i}_{5}= C^{B;i}_{5}&= \gamma^{i} \gamma_{\alpha} \gamma^{-} [\gamma^{+}(1-z_{\bar{q}}) q^{-}-\bm{\gamma}_{\perp}.\bm{l}_{1\perp} ] \gamma_{\mu} [\gamma^{+}z_{\bar{q}}q^{-}+\bm{\gamma}_{\perp}.\bm{l}_{1\perp}]  +\gamma^{-} [\gamma^{+}z_{q}q^{-} -\bm{\gamma}_{\perp}.(\bm{l}_{1\perp}-\bm{k}_{\gamma \perp}) ] \gamma_{\alpha} \gamma^{i} \nonumber \\
 & \times  \gamma_{\mu} [\gamma^{+}z_{\bar{q}}q^{-}+\bm{\gamma}_{\perp}.\bm{l}_{1\perp}]  -4z_{q}q^{-} l_{1}^{i} \gamma_{\alpha} \gamma_{\mu} -\gamma^{-} [\gamma^{+}z_{q}q^{-} -\bm{\gamma}_{\perp}.(\bm{l}_{1\perp}-\bm{k}_{\gamma \perp}) ] \gamma_{\alpha} [\gamma^{+}(1-z_{\bar{q}}) q^{-}-\bm{\gamma}_{\perp}.\bm{l}_{1\perp} ] \gamma_{\mu} \gamma^{i} \, , \nonumber \\
 C^{A}_{6}=C^{B}_{6}&= \gamma^{-}  [\gamma^{+}z_{q}q^{-} -\bm{\gamma}_{\perp}.(\bm{l}_{1\perp}-\bm{k}_{\gamma \perp}) ] \gamma_{\alpha} \Big\{ \gamma^{+}(1-z_{\bar{q}}) q^{-} -\gamma^{-} \, \big( \bm{l}_{1\perp}^{2}+z_{\bar{q}} \, Q^{2} \big) /2z_{\bar{q}}q^{-} -\bm{\gamma}_{\perp}.\bm{l}_{1\perp} \Big\} \gamma_{\mu} [\gamma^{+}z_{\bar{q}}q^{-}+\bm{\gamma}_{\perp}.\bm{l}_{1\perp}]  \, .
\end{align}
We can now express the arguments $\bm{V}_{\perp}$ and $\Delta$ of the constituent integrals appearing in Eq.~\ref{eq:R-V21-propto-zl-inverse-A-B} in an expansion in $z_{l}$, just as in Eq.~\ref{eq:V-and-Delta-in-terms-of-zl}. It is then straightforward to isolate the logarithmically divergent pieces in Eq.~\ref{eq:R-V21-propto-zl-inverse-A-B}. As expected, we will only require the expressions for the coefficients $\bm{c}_{1\perp}$ and $c_{3}$ that appear in Eq.~\ref{eq:V-and-Delta-in-terms-of-zl} for the two cases A and B. The rapidity divergent contribution from $(V21)$ can now be finally written as 
\begin{equation}
R^{(V21)}_{\rm div.;\mu \alpha} (\bm{l}_{1\perp})=\hyperref[eq:R-div-V21-A-B]{R_{\rm div.;\mu \alpha}^{(V21);A} (\bm{l}_{1\perp})}+\hyperref[eq:R-div-V21-A-B]{R_{\rm div.;\mu \alpha}^{(V21);B} (\bm{l}_{1\perp})}  \, , 
\label{eq:R-div-V21}
\end{equation}
where 
\begin{align}
R_{\rm div.;\mu \alpha}^{(V21);A,B} (\bm{l}_{1\perp})&=\frac{1}{(q^{-})^{2}} \, \ln \Big(\frac{1}{z_{0}} \Big) \,   \Big[\mathcal{R}_{(1);\mu \alpha}^{A,B}  \times  \mathcal{I}_{v;\rm log}^{(4,iijj)} \Big(\bm{c}_{1\perp}^{(V21)},c_{3}^{(V21)}\Big)  + \mathcal{R}_{(2);\mu \alpha}^{A,B;ij} \times \mathcal{I}_{v;\rm log}^{(4,ijkk)}  \Big(\bm{c}_{1\perp}^{(V21)},c_{3}^{(V21)}\Big)\nonumber \\
& +\mathcal{R}_{(3);\mu \alpha}^{A,B;i} \times \mathcal{I}_{v;\rm log}^{(4,ijj)}  \Big(\bm{c}_{1\perp}^{(V21)},c_{3}^{(V21)}\Big) + \mathcal{R}_{(4);\mu \alpha}^{A,B;ijk} \times \mathcal{I}_{v;\rm log}^{(4,ijk)}  \Big(\bm{c}_{1\perp}^{(V21)},c_{3}^{(V21)}\Big) \nonumber \\
&+ \mathcal{R}_{(5);\mu \alpha}^{A,B} \,  \mathcal{I}_{v;\rm log}^{(4,ii)}  \Big(\bm{c}_{1\perp}^{(V21)},c_{3}^{(V21)}\Big)+\mathcal{R}_{(6);\mu \alpha}^{A,B;ij} \times \mathcal{I}_{v;\rm log}^{(4,ij)}   \Big(\bm{c}_{1\perp}^{(V21)},c_{3}^{(V21)}\Big) \nonumber \\
& +\mathcal{R}_{(7);\mu \alpha}^{A,B;i} \,  \mathcal{I}_{v;\rm log}^{(4,i)}  \Big(\bm{c}_{1\perp}^{(V21)},c_{3}^{(V21)}\Big) \Big]\, \gamma^{-} \, .
\label{eq:R-div-V21-A-B}
\end{align}
As discussed throughout this paper, the upper limits of integration over $z_{l}$  governed by the large momentum $q^{-}$ have been absorbed in redefining $z_{0}$. The resulting mismatch between the divergent pieces in this redefinition is logarithmic and can be neglected when working to logarithmic accuracy. The expressions for the integrals appearing above are given in Eq.~\ref{eq:constituent-integrals-V3-V4-log}. The arguments of these integrals for the process $(V21)$ are respectively obtained as
\begin{align}
\bm{c}_{1\perp}^{(V21)}& =-\alpha_{1} \, \bm{l}_{1\perp}-\alpha_{2} \, \frac{(z_{\bar{q}} \, \bm{k}_{\perp}-z_{q} \, \bm{p}_{\perp})}{1-z_{\gamma}} + \alpha_{3} \, \Big( \frac{z_{\bar{q}}}{1-z_{\gamma}} \, \bm{k}_{\gamma \perp}-\bm{l}_{1\perp} \Big)  \, , \nonumber \\
c_{3}^{(V21)}&=\alpha_{1} (1-\alpha_{1}) \, \bm{l}_{1\perp}^{2} +\alpha_{2} (1-\alpha_{2}) \, \frac{(z_{\bar{q}} \, \bm{k}_{\perp}-z_{q} \, \bm{p}_{\perp})^{2}}{(1-z_{\gamma})^{2}} +\alpha_{3} (1-\alpha_{3}) \, \Big( \frac{z_{\bar{q}}}{1-z_{\gamma}} \, \bm{k}_{\gamma \perp}-\bm{l}_{1\perp} \Big)^{2}  \nonumber \\
&-2 \, \alpha_{1 } \alpha_{2} \, \frac{\bm{l}_{1\perp}.(z_{\bar{q}} \, \bm{k}_{\perp}-z_{q} \, \bm{p}_{\perp})}{1-z_{\gamma}} +2 \, \alpha_{2} \alpha_{3} \, \frac{(z_{\bar{q}} \, \bm{k}_{\perp}-z_{q} \, \bm{p}_{\perp})}{1-z_{\gamma}} . \Big( \frac{z_{\bar{q}}}{1-z_{\gamma}} \, \bm{k}_{\gamma \perp}-\bm{l}_{1\perp} \Big)+2 \, \alpha_{3} \alpha_{1} \, \bm{l}_{1\perp}. \Big( \frac{z_{\bar{q}}}{1-z_{\gamma}} \, \bm{k}_{\gamma \perp}-\bm{l}_{1\perp} \Big) \nonumber \\
& +\alpha_{1} \, z_{\bar{q}} (1-z_{\bar{q}} ) \, Q^{2} -\alpha_{2} \, \frac{z_{q}z_{\bar{q}}}{(1-z_{\gamma})^{2}} \, (2p.k) +\alpha_{3} \, \frac{z_{q}z_{\bar{q}}}{z_{\gamma} \, (1-z_{\gamma})^{2}} \, \big(\bm{k}_{\gamma \perp}^{2}+z_{\gamma} (1-z_{\gamma}) \, Q^{2} \big) \, .
\label{eq:c1-c3-V21}
\end{align}
Note that these terms are the same for both $A$ and $B$.

\section{Computation of the finite pieces in amplitudes for virtual graphs} \label{sec:finite-pieces-virtual-graphs}

In this section, we will present the generic forms of the finite pieces in the amplitudes for virtual gluon contributions. To facilitate their eventual numerical computation, we will express these explicitly as functions of the gluon momentum fraction $z_{l}$ in the loop and the Feynman parameters $\alpha_{i}$. The resulting expressions can then be easily computed using  \textit{Mathematica}~\cite{Mathematica}. 

The idea here is to present the techniques that we have developed to simplify this tedious computation by presenting a few examples. We will not give the explicit expressions for all the finite pieces in this section since firstly, this will add significantly to the length of the paper, and secondly, because these have to evaluated numerically with some choice of kinematics to gain a sense of the magnitude of such finite contributions. This will therefore be the subject of future quantitative studies  in this framework.

\subsection{Computation of $\mathcal{M}^{\text{SE}(1)}_{\text{finite}}$} \label{sec:finite-pieces-SE-1}

The finite pieces of the amplitude contributed by the six processes in Fig.~\ref{fig:NLO-self-1} are contained in
\begin{align}
\mathcal{M}^{\rm SE(1)}_{\text{finite};\mu \alpha}&= 2\pi \, \delta(1-z_{\text{tot}}^{v}) \, (egq_{f})^{2} \int \mathrm{d} \Pi_{\perp}^{v}  \,  \overline{u}(\bm{k}) \,  R^{\rm SE(1)}_{\rm finite;\mu \alpha} (\bm{l}_{1\perp})  \Bigg[ \Big( t^{b} \tilde{U} (\bm{x}_{\perp}) t^{a} \tilde{U}^{\dagger} (\bm{y}_{\perp}) \Big) U_{ba}(\bm{z}_{\perp}) - C_{F} \mathds{1} \Bigg] \,   v(\bm{p}) \, ,
\end{align}
where
\begin{equation}
R^{\rm SE(1)}_{\rm finite;\mu \alpha} (\bm{l}_{1\perp}) = \sum_{\beta=1}^{6}  \Big( \Re^{(S\beta)}_{\mu \alpha} (\bm{l}_{1\perp})   + R^{(S\beta)}_{(\textrm{II}); \mu \alpha} (\bm{l}_{1\perp}) \, \Big)  \, .
\end{equation}
Recall that what we call the  ``remainder" $\Re^{(S\beta)}$ is comprised of terms in the divergent part $R_{\rm (I); \mu \alpha}^{(S\beta)}$ of the amplitude which do not contain UV and $\ln(1/z_{0})$ singularities. 

Since the divergent pieces are zero for the processes $(S2)$ and $(S3)$ so are the remainder terms for them. We therefore have
\begin{equation}
\Re^{(S\beta)}_{\mu \alpha} (\bm{l}_{1\perp}) = 0, \enskip \text{for} \enskip  \beta= 2,3 \, .
\label{eq:remainder-S2-S3}
\end{equation}
For the processes $(S1)$, $(S4)$ and $(S6)$, the remainder has a generic structure that can be written as
\begin{align}
\begin{pmatrix}
\Re^{(S1)}_{\mu \alpha} (\bm{l}_{1\perp})  \\
\Re^{(S4)}_{\mu \alpha} (\bm{l}_{1\perp}) \\
\Re^{(S6)}_{\mu \alpha} (\bm{l}_{1\perp})
\end{pmatrix} 
& = \frac{1}{4\pi^{2}} \, \int_{\bm{l}_{2\perp}} \!\! e^{i\bm{l}_{2\perp}.\bm{r}_{zx}} \, \begin{pmatrix}
R^{\text{LO}:(1)}_{\mu \alpha}(\bm{l}_{1\perp}) \\ 
R^{\text{LO}:(2)}_{\mu \alpha}(\bm{l}_{1\perp}) \\
R^{\text{LO}:(3)}_{\mu \alpha}(\bm{l}_{1\perp}) 
\end{pmatrix} 
\,   \int \mathrm{d}z_{l} \int \mathrm{d} \alpha \, \Bigg[\frac{1}{z_{l}} \Bigg\{ \frac{(\bm{V}_{\perp}^{(S\beta)})^{2}}{\Delta^{(S\beta)}} +\ln \Big( \frac{\bm{l}_{2\perp}^{2}}{2\Delta^{(S\beta)}} \Big) -\frac{\bm{l}_{2\perp}.\bm{V}_{\perp}^{(S\beta)}}{\Delta^{(S\beta)}} -1 \Bigg\}  \nonumber \\
&+\Big(\frac{z_{l}}{2(a_{1}^{(S\beta)})^{2}} - \frac{1}{a_{1}^{(S\beta)}} \Big) \,  \Bigg( \frac{(\bm{V}_{\perp}^{(S\beta)})^{2}}{\Delta^{(S\beta)}} +\ln \Big( \frac{Q^{2}}{\Delta^{(S\beta)}} \Big)  \Bigg) + \frac{1}{2 a_{1}^{(S\beta)}} \, \Bigg] 
 \, , \quad \quad \beta=1,4,6 \, .
\label{eq:remainder-generic-S1-S4-S6}
\end{align}
%As we discussed earlier, the $\ln(1/z_{0})$ appearing in Eq.~\ref{eq:remainder-generic-S1-S4-S6} exactly cancels a similar term arising from the integrations over $z_{l}$ and $\alpha$. Hence there are no logarithmically divergent terms in the remainder defined by Eq.~\ref{eq:remainder-generic-S1-S4-S6}. 
We can always express $\bm{V}_{\perp}^{(S\beta)}$ and $\Delta^{(S\beta)}$ in terms of $z_{l}$ as
\begin{align}
\bm{V}_{\perp}^{(S\beta)}  =  \bm{c}_{1 \perp}^{(S\beta)}  + z_{l} \, \bm{c}_{2 \perp}^{(S\beta)}  \,\,\,;\,\,\,
\Delta^{(S\beta)}  = c_{3}^{(S\beta)} +z_{l} \, c_{4}^{(S\beta)}+z_{l}^{2} \, c_{5}^{(S\beta)} \, , \quad \beta=1, \ldots,6  \, .
\label{eq:integral-arguments-in-terms-of-zl}
\end{align}
For the processes $(S1)$, $(S4)$ and $(S6)$ these coefficients are respectively given by 
\begin{itemize}
\item For  $(S1)$: 
\begin{align}
a_{1}^{(S1)} &= 1-z_{\bar{q}} \, ,  \quad \bm{c}_{1\perp}^{(S1)}= \alpha \, \bm{l}_{2\perp} \, , \quad \bm{c}_{2\perp}^{(S1)}= -\frac{(\bm{k}_{\perp}+\bm{k}_{\gamma \perp}) - \alpha \, (\bm{k}_{\perp}+\bm{k}_{\gamma \perp} -\bm{l}_{1\perp})}{1-z_{\bar{q}}} \, , \quad c_{3}^{(S1)} = \alpha(1-\alpha) \, \bm{l}_{2\perp}^{2} \, , \nonumber \\
c_{4}^{(S1)}& =  \frac{1}{1-z_{\bar{q}}} \, \Big\{ \alpha (1-\alpha) \, 2 \, \bm{l}_{2\perp}.(\bm{k}_{\perp}+\bm{k}_{\gamma \perp}-\bm{l}_{1\perp}) + \alpha \, \frac{\bm{l}_{1\perp}^{2}+\Delta^{\rm LO:(1)}}{z_{\bar{q}}} -(1-\alpha) \, (2k.k_{\gamma}) \Big\} \, , \nonumber \\
c_{5}^{(S1)}& = \frac{1}{(1-z_{\bar{q}})^{2}} \, \Big\{ \alpha(1-\alpha) \, (\bm{k}_{\perp}+\bm{k}_{\gamma \perp}-\bm{l}_{1\perp})^{2} - \alpha \, \frac{\bm{l}_{1\perp}^{2}+\Delta^{\rm LO:(1)}}{z_{\bar{q}}} +(1-\alpha) \, (2k.k_{\gamma}) \Big\} \, .
\label{eq:coefficients-V-perp-Delta-S1}
\end{align}

\item For $(S4)$: 
\begin{align}
a_{1}^{(S4)} & = z_{q} \, , \quad \bm{c}_{1\perp}^{(S4)}= \alpha \, \bm{l}_{2\perp} \, , \quad \bm{c}^{(S4)}_{2\perp}=  - \frac{\alpha \, (\bm{l}_{1\perp}-\bm{k}_{\perp}-\bm{k}_{\gamma \perp}) }{z_{q}}  \, ,  \quad c_{3}^{(S4)}= \alpha(1-\alpha) \, \bm{l}_{2\perp}^{2} \, , \nonumber \\
 c_{4}^{(S4)}&= \frac{1}{z_{q}} \, \Big\{ -\alpha(1-\alpha) \, 2\, \bm{l}_{2\perp}. (\bm{l}_{1\perp}-\bm{k}_{\perp}-\bm{k}_{\gamma \perp}) +\alpha \, \big[(\bm{l}_{1\perp}-\bm{k}_{\gamma \perp})^{2} +\frac{z_{q}}{z_{\bar{q}}} \, \bm{l}_{1\perp}^{2} +\frac{z_{q}}{z_{\gamma}} \, \bm{k}_{\gamma \perp}^{2} +z_{q} \, Q^{2} \big] \Big\} \, ,  \nonumber \\
 c_{5}^{(S4)}&=\frac{1}{z_{q}^{2}} \, \Big\{ \alpha(1-\alpha) \, (\bm{l}_{1\perp}- \bm{k}_{\perp}-\bm{k}_{\gamma \perp})^{2} -\alpha \, \big[(\bm{l}_{1\perp}-\bm{k}_{\gamma \perp})^{2} +\frac{z_{q}}{z_{\bar{q}}} \, \bm{l}_{1\perp}^{2} +\frac{z_{q}}{z_{\gamma}} \, \bm{k}_{\gamma \perp}^{2} +z_{q} \, Q^{2} \big] \Big\} \, .
\label{eq:coefficients-V-perp-Delta-S4}
\end{align}

\item For $(S6)$:
 \begin{align}
a_{1}^{(S6)} & = a_{1}^{(S4)} \, ,  \quad  \bm{c}_{1\perp}^{(S6)} = \bm{c}_{1\perp}^{(S4)} \, , \quad \bm{c}_{2\perp}^{(S6)} = \bm{c}_{2\perp}^{(S4)} \, , \quad c_{3}^{(S6)}= c_{3}^{(S4)} \, ,   \nonumber \\
c_{4}^{(S6)} & = \frac{1}{z_{q}} \, \Big\{ -\alpha (1-\alpha) \, 2 \, \bm{l}_{2\perp} . (\bm{l}_{1\perp}-\bm{k}_{\perp}-\bm{k}_{\gamma \perp} ) + \alpha \, \frac{(\bm{l}_{1\perp}-\bm{k}_{\gamma \perp})^{2}+\Delta^{\rm LO:(3)}}{1-z_{q}} \Big\} \, , \nonumber \\
 c_{5}^{(S6)}&=\frac{1}{z_{q}^{2}} \, \Big\{ \alpha(1-\alpha) \, (\bm{l}_{1\perp}- \bm{k}_{\perp}-\bm{k}_{\gamma \perp})^{2} -\alpha \, \frac{(\bm{l}_{1\perp}-\bm{k}_{\gamma \perp})^{2}+\Delta^{\rm LO:(3)} }{1-z_{q}} \Big\} \, .
\label{eq:coefficients-V-perp-Delta-S6}
\end{align}\
In the above equations, we have $\Delta^{\rm LO:(1) }= z_{\bar{q}} (1-z_{\bar{q}}) - i \varepsilon $ and its quark-antiquark interchanged counterpart $\Delta^{\rm LO:(3)}= z_{q}(1-z_{q})-i\varepsilon $.
\end{itemize}

Finally for $(S5)$, the contour  integration over $l_{1}^{+}$ encloses two poles; we can write the remainder in terms of these contributions as 
\begin{align}
\Re^{(S5)}_{\mu \alpha} (\bm{l}_{1\perp}) &= \frac{1}{4\pi^{2}} \int_{\bm{l}_{2\perp}} \!\!\!\!e^{i\bm{l}_{2\perp}.\bm{r}_{zx}} \, \gamma^{-} \frac{\gamma^{+}z_{q}q^{-}-\bm{\gamma}_{\perp}.(\bm{l}_{1\perp}-\bm{k}_{\gamma \perp}) }{\Big( \bm{l}_{1\perp}+z_{\bar{q}}/z_{\gamma} \, \bm{k}_{\gamma \perp} \Big)^{2}-i\varepsilon} \gamma_{\mu} \, (C-D) \, \gamma_{\alpha} \frac{\gamma^{+}z_{\bar{q}}q^{-}+\bm{\gamma}_{\perp}.\bm{l}_{1\perp}}{z_{\gamma} (q^{-})^{2}} \gamma^{-} , \enskip \text{where} \, , \nonumber \\[0.3cm]
C& = \frac{\gamma^{+}(1-z_{q})q^{-}+\gamma^{-}(\bm{l}_{1\perp}^{2}z_{\gamma}+\bm{k}_{\gamma \perp}^{2}z_{\bar{q}}) /2z_{\bar{q}}z_{\gamma}q^{-} +\bm{\gamma}_{\perp}.(\bm{l}_{1\perp}-\bm{k}_{\gamma \perp})}{(1-z_{\gamma})/z_{\bar{q}}  \Big[ \Big(\bm{l}_{1 \perp}+\bm{v}_{\perp}^{\rm LO:(2)} \Big)^{2}+\Delta^{\rm LO:(2)} \Big]    } \times \int \mathrm{d}z_{l} \int \mathrm{d} \alpha \, \Bigg[\frac{1}{z_{l}} \Bigg\{ \frac{(\bm{V}_{\perp}^{(S4)})^{2}}{\Delta^{(S4)}} +\ln \Big( \frac{\bm{l}_{2\perp}^{2}}{2\Delta^{(S4)}} \Big) \nonumber \\
& -\frac{\bm{l}_{2\perp}.\bm{V}_{\perp}^{(S4)}}{\Delta^{(S4)}} -1 \Bigg\}   +\Big(\frac{z_{l}}{2(a_{1}^{(S4)})^{2}} - \frac{1}{a_{1}^{(S4)}} \Big) \,  \Bigg( \frac{(\bm{V}_{\perp}^{(S4)})^{2}}{\Delta^{(S4)}} +\ln \Big( \frac{Q^{2}}{\Delta^{(S4)}} \Big)  \Bigg) + \frac{1}{2 a_{1}^{(S4)}} \, \Bigg]  \enskip  , \text{and} \nonumber \\[0.3cm]
D&= \frac{\gamma^{+}(1-z_{q})q^{-}+\gamma^{-} (\bm{l}_{1\perp}-\bm{k}_{\gamma \perp})^{2}/2(1-z_{q})q^{-}+\bm{\gamma}_{\perp}.(\bm{l}_{1\perp}-\bm{k}_{\gamma \perp})}{1/(1-z_{q})    \Big[ (\bm{l}_{1\perp}-\bm{k}_{\gamma \perp})^{2}+\Delta^{\rm LO:(3)} \Big] } \times \int \mathrm{d}z_{l} \int \mathrm{d} \alpha \, \Bigg[\frac{1}{z_{l}} \Bigg\{ \frac{(\bm{V}_{\perp}^{(S6)})^{2}}{\Delta^{(S6)}} +\ln \Big( \frac{\bm{l}_{2\perp}^{2}}{2\Delta^{(S6)}} \Big) \nonumber \\
& -\frac{\bm{l}_{2\perp}.\bm{V}_{\perp}^{(S6)}}{\Delta^{(S6)}} -1 \Bigg\}   +\Big(\frac{z_{l}}{2(a_{1}^{(S6)})^{2}} - \frac{1}{a_{1}^{(S6)}} \Big) \,  \Bigg( \frac{(\bm{V}_{\perp}^{(S6)})^{2}}{\Delta^{(S6)}} +\ln \Big( \frac{Q^{2}}{\Delta^{(S6)}} \Big)  \Bigg) + \frac{1}{2 a_{1}^{(S6)}} \, \Bigg] \, .
\label{eq:remainder-S5}
\end{align}
The coefficients needed for the evaluation of the above terms are given in Eqs.~\ref{eq:coefficients-V-perp-Delta-S4} and \ref{eq:coefficients-V-perp-Delta-S6}.

We will now present the expressions for the pieces $R_{\rm (II)}^{(S\beta)}$ in the finite part of the amplitude given by Eqs.~\ref{eq:amplitude-SE1-finite} and  \ref{eq:R-SE1-finite}.  These are comprised of integrations over the transverse gluon loop momentum, $\bm{l}_{3\perp}$, which are finite in two dimensions. We will first write down these contributions for $(S1)$, $(S4)$, $(S5)$ and $(S6)$ which share a similar structure. This will be followed by the corresponding expressions for processes $(S2)$ and $(S3)$ which are considerably more tedious albeit similar to one another. 

For $(S1)$ we can write the finite piece as
\begin{align}
R^{(S1)}_{\rm (II);\mu \alpha} (\bm{l}_{1\perp}) &=\!\!\! \int_{\bm{l}_{2\perp}} \!\!\!\!\!\! e^{i\bm{l}_{2\perp}.\bm{r}_{zx}} \, \gamma_{\alpha}  \frac{\slashed{k}+\slashed{k}_{\gamma}}{2k.k_{\gamma}} \, \Big(\wp_{a}^{(S1)} +\wp_{b}^{(S1)} \Big) \frac{\gamma^{+}(1-z_{\bar{q}} )q^{-}-\gamma^{-}(\bm{l}_{1\perp}^{2}+Q^{2}z_{\bar{q}})/2z_{\bar{q}}q^{-}-\bm{\gamma}_{\perp}.\bm{l}_{1\perp}}{\bm{l}_{1\perp}^{2}+\Delta^{\rm LO:(1)}} \,  \gamma_{\mu} \frac{\gamma^{+}z_{\bar{q}}q^{-}+\bm{\gamma}_{\perp}.\bm{l}_{1\perp}}{4(q^{-})^{2}(1-z_{\bar{q}})^{2}} \gamma^{-}  \, .
\label{eq:R-II-S1}
\end{align}
The terms appearing above within the parentheses have generic structures for $\beta=1,4,6$ and are given by
\begin{align}
\wp_{a}^{(S\beta)}& =    \int \frac{\mathrm{d}z_{l}}{2\pi} \,   \Big( d_{1}^{(S\beta)} \, z_{l}^{2}+d_{2}^{(S\beta)} \, z_{l} +d_{3}^{(S\beta)} \Big) \, \mathcal{I}_{v}^{(2,i)} (\bm{V}^{(S\beta)}_{\perp}, \Delta^{(S\beta)}), \qquad \beta=1,4,6 \, ,  \nonumber \\
\wp_{b}^{(S\beta)} & = \int \frac{\mathrm{d}z_{l}}{2\pi}  \, \frac{1}{q^{-}} \,  d_{4}^{(S\beta)} \, \Big(d_{5}^{(S\beta)} -d_{6}^{(S\beta)} \, z_{l}\, q^{-} \Big) \, d_{7}^{(S\beta)} \, \Big(d_{8}^{(S\beta)}-d_{9}^{(S\beta)}\, z_{l} \, q^{-} \Big) \, \Big(d_{10}^{(S\beta)} \, z_{l} \, q^{-} -d_{11}^{(S\beta)} \Big)\, \mathcal{I}_{v}^{(2,0)} (\bm{V}^{(S\beta)}_{\perp}, \Delta^{(S\beta)})  \, .
\end{align}
The constituent integrals that appear here have simple (finite) expressions in $d=2$ dimensions:
\begin{equation}
\mathcal{I}_{v}^{(2,i)} (\bm{V}^{(S\beta)}_{\perp}, \Delta^{(S\beta)})= -\frac{1}{4\pi} \int_{0}^{1}  \mathrm{d} \alpha \, \frac{(V^{(S\beta)})^{i}}{ \Delta^{(S\beta)}} , \qquad \mathcal{I}_{v}^{(2,0)} (\bm{V}^{(S\beta)}_{\perp}, \Delta^{(S\beta)})= \frac{1}{4\pi } \int_{0}^{1} \mathrm{d}\alpha \, \frac{1}{\Delta^{(S\beta)}} \, .
\label{eq:I-v-2-i-and-I-v-2-0}
\end{equation}
The decomposition of the  $\bm{V}_{\perp}^{(S\beta)}$ and $\Delta^{(S\beta)}$'s in terms of $z_{l}$ and $\alpha$ and the accompanying coefficients is clearly shown in Eqs.~\ref{eq:coefficients-V-perp-Delta-S1}, \ref{eq:coefficients-V-perp-Delta-S4} and \ref{eq:coefficients-V-perp-Delta-S6}.  The expressions for the coefficients $d_{i}^{(S\beta)}$ ($i=1,\ldots, 11$) for $(S1)$ are as follows:
\begin{align}
d_{1}^{(S1)}& = 4 \gamma^{i} \, (q^{-})^{2}\, , \enskip d_{2}^{(S1)}= -2 \, \bm{\gamma}_{\perp}.\bm{l}_{2\perp} \gamma^{i} \gamma^{-} \, q^{-} - 8(1-z_{\bar{q}}) \, (q^{-})^{2} \, \gamma^{i} - 4\,  l_{2}^{i} \gamma^{-} \, q^{-} \, ,\nonumber \\
d_{3}^{(S1)}& = 4(1-z_{\bar{q}})^{2} \, (q^{-})^{2} \, \gamma^{i} + 8 (1-z_{\bar{q}}) \, q^{-} \, l_{2}^{i} \, \gamma^{-} - 2 (1-z_{\bar{q}}) \, q^{-} \, \big\{ \gamma^{i} \bm{\gamma}_{\perp}. (\bm{k}_{\perp}+\bm{k}_{\gamma \perp}) \gamma^{-} +\bm{\gamma}_{\perp}.(\bm{l}_{1\perp}-2\bm{l}_{2\perp} ) \gamma^{i} \gamma^{-} \big\} \, , \nonumber \\
d_{4}^{(S1)}&= \gamma^{i} \, , \enskip d_{5}^{(S1)}= \slashed{k}+\slashed{k}_{\gamma} \, , \enskip d_{6}^{(S1)}= \gamma^{+} \, ,  \enskip d_{7}^{(S1)}= \gamma^{-} \, , \enskip d_{8}^{(S1)}= \gamma^{+}(1-z_{\bar{q}}) q^{-} -\bm{\gamma}_{\perp}.(\bm{l}_{1\perp}-\bm{l}_{2\perp}) \, , \enskip d_{9}^{(S1)}= \gamma^{+} \, , \nonumber \\
d_{10}^{(S1)}& = \gamma_{i} \, , \enskip d_{11}^{(S1)}= l_{2i} \, \gamma^{-} \, .
\end{align}
Correspondingly, for $(S4)$, $(S5)$ and $(S6)$, we can respectively write
\begin{align}
R^{(S4)}_{\text{II};\mu \alpha} (\bm{l}_{1\perp}) &= \Big(\wp_{a}^{(S4)} +\wp_{b}^{(S4)} \Big) \frac{\gamma^{+}z_{q}q^{-}-\gamma^{-}(Q^{2}z_{\bar{q}}z_{\gamma}+z_{\bar{q}} \, \bm{k}_{\gamma \perp}^{2}+z_{\gamma}\, \bm{l}_{1\perp}^{2})/2z_{\gamma}z_{\bar{q}}q^{-}-\bm{\gamma}_{\perp}.(\bm{l}_{1\perp}-\bm{k}_{\gamma \perp}) }{ \Big(\bm{l}_{1 \perp}+\bm{v}_{\perp}^{\rm LO:(2)} \Big)^{2}+\Delta^{\rm LO:(2)}} \, \gamma^{\alpha} \nonumber \\
& \times \frac{\gamma^{+}(1-z_{\bar{q}})q^{-} -\gamma^{-} (\bm{l}_{1\perp}^{2}+z_{\bar{q}} \, Q^{2}) /2z_{\bar{q}}q^{-} -\bm{\gamma}_{\perp}.\bm{l}_{1\perp}}{\bm{l}_{1\perp}^{2}+\Delta^{\rm LO:(1)}} \gamma_{\mu} \frac{\gamma^{+}z_{\bar{q}}q^{-}+\bm{\gamma}_{\perp}.\bm{l}_{1\perp}}{4(q^{-})^{2} \, z_{q}^{2}(1-z_{\gamma})/z_{\bar{q}} } \, \gamma^{-} \, ,
\label{eq:R-II-S4}
\end{align}
\begin{align}
R^{(S5)}_{\text{II};\mu \alpha} (\bm{l}_{1\perp})  & = (E-F) \, \gamma_{\alpha} \frac{\gamma^{+}z_{\bar{q}}q^{-}+\bm{\gamma}_{\perp}.\bm{l}_{1\perp}}{4\, (q^{-})^{2}\,  z_{\gamma} \, z_{q}^{2} \, \Big[ (\bm{l}_{1\perp}+z_{\bar{q}}/z_{\gamma}\, \bm{k}_{\gamma \perp})^{2}-i\varepsilon\Big] } \, \gamma^{-}, \enskip \text{where}, \nonumber \\[0.4cm]
E&= (\wp^{(S4)}_{a}+\wp^{(S4)}_{b}) \frac{\gamma^{+}z_{q}q^{-}-\gamma^{-}(Q^{2}z_{\bar{q}}z_{\gamma} +z_{\bar{q}} \, \bm{k}_{\gamma \perp}^{2}+z_{\gamma} \, \bm{l}_{1\perp}^{2} )/2z_{\gamma}z_{\bar{q}}q^{-} -\bm{\gamma}_{\perp}.(\bm{l}_{1\perp}-\bm{k}_{\gamma \perp})        }{ (1-z_{\gamma})/z_{\bar{q}} \Big[ \Big(\bm{l}_{1 \perp}+\bm{v}_{\perp}^{\rm LO:(2)} \Big)^{2}+\Delta^{\rm LO:(2)} \Big]} \nonumber\\
& \times \gamma_{\mu} [\gamma^{+}(1-z_{q})q^{-}+\gamma^{-}(z_{\gamma} \, \bm{l}_{1\perp}^{2} +z_{\bar{q}} \, \bm{k}_{\gamma \perp}^{2})/2z_{\bar{q}} z_{\gamma} +\bm{\gamma}_{\perp}.(\bm{l}_{1\perp}-\bm{k}_{\gamma \perp}) ] \, , \enskip \text{and} \nonumber \\[0.4cm]
F&= -(\wp^{(S6)}_{a}+\wp^{(S6)}_{b}) \frac{\gamma^{+}z_{q}q^{-}-\gamma^{-}\Big((\bm{l}_{1\perp}-\bm{k}_{\gamma \perp})^{2}+ (1-z_{q})\, Q^{2} \Big)/2(1-z_{q})q^{-} -\bm{\gamma}_{\perp}.(\bm{l}_{1\perp}-\bm{k}_{\gamma \perp})}{1/(1-z_{q}) \, [(\bm{l}_{1\perp}-\bm{k}_{\gamma \perp})^{2}+\Delta^{\rm LO:(3)}]    } \nonumber \\
& \times \gamma_{\mu} [\gamma^{+}(1-z_{q})q^{-}+\gamma^{-}(\bm{l}_{1\perp}-\bm{k}_{\gamma \perp})^{2}/2(1-z_{q})q^{-}+\bm{\gamma}_{\perp}.(\bm{l}_{1\perp}-\bm{k}_{\gamma \perp})] \, .
\label{eq:R-II-S5}
\end{align}
\begin{align}
R^{(S6)}_{\text{II};\mu \alpha} (\bm{l}_{1\perp})  &= - \Big(\wp_{a}^{(S6)} +\wp_{b}^{(S6)} \Big) \frac{\gamma^{+}z_{q}q^{-}-\gamma^{-}\Big( (\bm{l}_{1\perp}-\bm{k}_{\gamma \perp})^{2}+Q^{2}(1-z_{q})\Big) / 2(1-z_{q})q^{-}-\bm{\gamma}_{\perp}.(\bm{l}_{1\perp}-\bm{k}_{\gamma \perp})    }{(\bm{l}_{1\perp}-\bm{k}_{\gamma \perp})^{2}+\Delta^{\rm LO:(3)}} \gamma_{\mu} \nonumber \\
& \times \frac{\gamma^{+}(1-z_{q})q^{-}+\bm{\gamma}_{\perp}.(\bm{l}_{1\perp}-\bm{k}_{\gamma \perp})}{4(q^{-})^{2} \, z_{q}^{2}} \gamma^{-} \frac{\slashed{p}+\slashed{k}_{\gamma}}{2p.k_{\gamma}} \gamma_{\alpha} \, .
\label{eq:R-II-S6}
\end{align}

The coefficients $d_{i}^{(S\beta)} $ ($i=1, \ldots, 11$) are identical for processes $(S4)$ and $(S6)$ ,
\begin{equation}
d_{i}^{(S4)}= d_{i}^{(S6)} \, , \quad i=1, \ldots, 11 \, ,
\end{equation}
and are given by
\begin{align}
d_{1}^{(S4)}& = 4 \Big( \gamma^{i} -\frac{k^{i}}{z_{q} q^{-}} \, \gamma^{-} \Big) \, (q^{-})^{2} \, , \enskip d_{2}^{(S4)}= -2 \, \bm{\gamma}_{\perp}.\bm{l}_{2\perp} \gamma^{i} \gamma^{-} \, q^{-} - 8z_{q} \, (q^{-})^{2} \, \Big( \gamma^{i}-\frac{k^{i}}{z_{q} q^{-}} \, \gamma^{-} \Big) - 4\,  l_{2}^{i} \gamma^{-} \, q^{-} \, , \nonumber \\
d_{3}^{(S4)}& = 4z_{q}^{2} \, (q^{-})^{2} \, \Big( \gamma^{i} -\frac{k^{i}}{z_{q} q^{-}} \, \gamma^{-} \Big)  + 8 z_{q} \, q^{-} \, l_{2}^{i} \, \gamma^{-} - 2 z_{q} \, q^{-} \, \bm{\gamma}_{\perp}.(\bm{l}_{1\perp}-2\bm{l}_{2\perp}-\bm{k}_{\perp}-\bm{k}_{\gamma \perp}) \gamma^{i} \gamma^{-} \, , \nonumber \\
d_{4}^{(S4)}&= \Big( \gamma^{i} -\frac{k^{i}}{z_{q} q^{-}} \, \gamma^{-} \Big)\, , \enskip d_{5}^{(S4)}= \slashed{k}  \, , \enskip d_{6}^{(S4)}= \gamma^{+} \, ,  \enskip d_{7}^{(S4)}= \gamma^{-} \, , \enskip d_{8}^{(S4)}=\slashed{k}- \bm{\gamma}_{\perp}.(\bm{l}_{1\perp}-\bm{l}_{2\perp}-\bm{k}_{\perp}-\bm{k}_{\gamma \perp})  \, , \enskip d_{9}^{(S4)}= \gamma^{+} \, , \nonumber \\
d_{10}^{(S4)}& = \Big( \gamma_{i} -\frac{k_{i}}{z_{q} q^{-}} \, \gamma^{-} \Big)  \, , \enskip d_{11}^{(S4)}= l_{2i} \, \gamma^{-} \, .
\end{align}

We will now provide expressions for the finite pieces of the amplitudes for the remaining two processes $(S2)$ and $(S3)$. As mentioned in the previous paragraphs, the remainder terms for these processes are zero. We therefore have only the finite terms $R_{\rm (II)}$ for $(S2)$ and $(S3)$. Because there is a real photon nested inside the gluon loop for these graphs, we have two different contributions depending on the magnitude of the momentum of the gluon in the loop, $l_{2}^{-}=l_{3}^{-}$ relative to $k^{-}$. Accordingly, we have to sum the results for the cases $0<l_{3}^{-}<k^{-}$ and $k^{-}<l_{3}^{-}<k^{-}+k^{-}_{\gamma}$ to obtain the net finite contributions for these two diagrams. For each case, these can be written in terms of constituent integrals with coefficients which depend on the process of interest. For $(S2)$, we have
\begin{equation}
R^{(S2)}_{\rm (II);\mu \alpha} (\bm{l}_{1\perp}) = \hyperref[eq:R-II-S2-A-B]{R^{(S2);A}_{\rm (II);\mu \alpha} (\bm{l}_{1\perp})} +\hyperref[eq:R-II-S2-A-B]{R^{(S2);B}_{\rm (II);\mu \alpha} (\bm{l}_{1\perp})} \,,
\label{eq:R-II-S2}
\end{equation}
where $A$ and $B$ denote respectively the contributions to the amplitude for $0<l_{3}^{-}<k^{-}$ and $k^{-}<l_{3}^{-}<k^{-}+k^{-}_{\gamma}$. For each case $p=A,B$, we can write
\begin{align}
R^{(S2);p}_{\rm (II);\mu \alpha} (\bm{l}_{1\perp}) & = \int_{\bm{l}_{2\perp}} e^{i \bm{l}_{2\perp}.\bm{r}_{zx}} \int \frac{\mathrm{d}z_{l}}{2\pi} \Big\{ C_{(1);\alpha}^{(S2);p} \, \mathcal{I}_{v}^{(3,0)}(\bm{V}^{(S2);p}_{\perp},\Delta^{(S2);p}) +( C_{(2);\alpha}^{(S2);p} )^{i} \, \mathcal{I}_{v}^{(3,i)}(\bm{V}^{(S2);p}_{\perp},\Delta^{(S2);p}) \nonumber \\
& +C_{(3);\alpha}^{(S2);p} \, \mathcal{I}_{v}^{(3,ii)}(\bm{V}^{(S2);p}_{\perp},\Delta^{(S2);p}) + (C_{(4);\alpha}^{(S2);p})^{ij} \, \mathcal{I}_{v}^{(3,ij)}(\bm{V}^{(S2);p}_{\perp},\Delta^{(S2);p}) +(C_{(5);\alpha}^{(S2);p})^{i} \, \mathcal{I}_{v}^{(3,ijj)}(\bm{V}^{(S2);p}_{\perp},\Delta^{(S2);p}) \Big\} \nonumber \\
& \times \frac{\gamma^{+}(1-z_{\bar{q}}) q^{-} - \gamma^{-} \frac{\bm{l}_{1\perp}^{2}+z_{\bar{q}} \, Q^{2} }{2z_{\bar{q}} q^{-}} - \bm{\gamma}_{\perp}.\bm{l}_{1\perp}}{4z_{q}(1-z_{\bar{q}})^{2} \, (q^{-})^{2} \, (\bm{l}_{1\perp}^{2}+\Delta^{\rm LO:(1)} ) }  \gamma_{\mu} \, [\gamma^{+}z_{\bar{q}}q^{-}+\bm{\gamma}_{\perp}.\bm{l}_{1\perp} ] \, \gamma^{-} \, , 
\label{eq:R-II-S2-A-B}
\end{align}
where the expressions for the various constituent integrals appearing in the above equation are given in Eq.~\ref{eq:constituent-integrals-V1}. The arguments $\bm{V}_{\perp}$ and $\Delta$ appearing in these integrals can always be expressed in terms of the momentum fraction $z_{l}$ of the gluon, as shown in Eq.~\ref{eq:integral-arguments-in-terms-of-zl}.

For $(S2)$, we obtain the following expressions for the coefficients $c_{i}$ ($i=1,\ldots,5$) that appear in this decomposition for the two cases A and B mentioned above.

Case A: $0<l_{3}^{-}<k^{-}$
\begingroup
\allowdisplaybreaks
\begin{align}
\bm{c}_{1\perp}^{(S2);A}&= \alpha_{2} \, \bm{l}_{2\perp} \, , \enskip \bm{c}_{2\perp}^{(S2);A}= \alpha_{1} \, \Big( \frac{\bm{k}_{\perp}}{z_{q}} -\frac{\bm{k}_{\perp}+\bm{k}_{\gamma \perp}}{1-z_{\bar{q}}} \Big) + \alpha_{2} \, \Big( \frac{\bm{k}_{\perp}}{z_{q}} - \frac{\bm{l}_{1\perp}}{1-z_{\bar{q}}} \Big) \, , \nonumber \\
c_{3}^{(S2);A}& = \alpha_{2}(1-\alpha_{2}) \, \bm{l}_{2\perp}^{2} \, , \nonumber \\
c_{4}^{(S2);A} & = 2 \, \alpha_{2} (1-\alpha_{2}) \, \bm{l}_{2\perp}. \Big( \frac{\bm{k}_{\perp}}{z_{q}} - \frac{\bm{l}_{1\perp}}{1-z_{\bar{q}}} \Big) - 2\, \alpha_{1} \alpha_{2} \, \bm{l}_{2\perp}. \Big( \frac{\bm{k}_{\perp}}{z_{q}} -\frac{\bm{k}_{\perp}+\bm{k}_{\gamma \perp}}{1-z_{\bar{q}}} \Big) -\alpha_{1} \, \frac{2k.k_{\gamma}}{1-z_{\bar{q}}} + \alpha_{2} \, \frac{\bm{l}_{1\perp}^{2}+\Delta^{\rm LO:(1)}}{z_{\bar{q}} \, (1-z_{\bar{q}}) } \, , \nonumber \\
c_{5}^{(S2);A}& = \alpha_{1} (1-\alpha_{1}) \,  \Big( \frac{\bm{k}_{\perp}}{z_{q}} -\frac{\bm{k}_{\perp}+\bm{k}_{\gamma \perp}}{1-z_{\bar{q}}} \Big)^{2} +\alpha_{2} (1-\alpha_{2}) \, \Big( \frac{\bm{k}_{\perp}}{z_{q}} - \frac{\bm{l}_{1\perp}}{1-z_{\bar{q}}} \Big)^{2} -2 \, \alpha_{1} \alpha_{2} \, \Big( \frac{\bm{k}_{\perp}}{z_{q}} - \frac{\bm{l}_{1\perp}}{1-z_{\bar{q}}} \Big).  \Big( \frac{\bm{k}_{\perp}}{z_{q}} -\frac{\bm{k}_{\perp}+\bm{k}_{\gamma \perp}}{1-z_{\bar{q}}} \Big) \nonumber \\
& + \alpha_{1} \, \frac{2k.k_{\gamma}}{(1-z_{\bar{q}})^{2}} -\alpha_{2} \, \frac{\bm{l}_{1\perp}^{2}+\Delta^{\rm LO:(1)}}{z_{\bar{q}} \, (1-z_{\bar{q}})^{2}} \, .
\end{align}
\endgroup

Case B: $k^{-}<l_{3}^{-}< k^{-}+k^{-}_{\gamma}$
\begingroup
\allowdisplaybreaks
\begin{align}
\bm{c}_{1\perp}^{(S2);B}&= (\alpha_{1}+\alpha_{2}) \, \Big( \bm{k}_{\perp}-\frac{z_{q}}{z_{\gamma}} \, \bm{k}_{\gamma\perp} \Big)  +\alpha_{2} \, \bm{l}_{2\perp} \, , \enskip \bm{c}_{2\perp}^{(S2);B}= \alpha_{1} \, \Big(  \frac{z_{q}}{z_{\gamma} \, (1-z_{\bar{q}}) } \, \bm{k}_{\gamma \perp} - \frac{\bm{k}_{\perp}}{z_{q}}\Big) + \alpha_{2} \, \Big( \frac{\bm{k}_{\gamma \perp}}{z_{\gamma}}-\frac{\bm{l}_{1\perp}}{1-z_{\bar{q}}} \Big) \, , \nonumber \\
c_{3}^{(S2);B}& =\alpha_{1}(1-\alpha_{1}) \, \Big( \bm{k}_{\perp}-\frac{z_{q}}{z_{\gamma}} \, \bm{k}_{\gamma\perp} \Big)^{2} +\alpha_{2}(1-\alpha_{2}) \, \Big( \bm{l}_{2\perp}+ \bm{k}_{\perp}-\frac{z_{q}}{z_{\gamma}} \, \bm{k}_{\gamma\perp}  \Big)^{2} - 2\, \alpha_{1} \,  \alpha_{2} \, \Big( \bm{k}_{\perp}-\frac{z_{q}}{z_{\gamma}} \, \bm{k}_{\gamma\perp} \Big) . \Big( \bm{l}_{2\perp}+ \bm{k}_{\perp}-\frac{z_{q}}{z_{\gamma}} \, \bm{k}_{\gamma\perp} \Big)   \, , \nonumber \\
c_{4}^{(S2);B} & = 2 \, \alpha_{1} (1-\alpha_{1}) \, \Big( \bm{k}_{\perp}-\frac{z_{q}}{z_{\gamma}} \, \bm{k}_{\gamma\perp} \Big). \Big(  \frac{z_{q}}{z_{\gamma} \, (1-z_{\bar{q}}) } \, \bm{k}_{\gamma \perp} - \frac{\bm{k}_{\perp}}{z_{q}}\Big) + 2\, \alpha_{2} (1- \alpha_{2} ) \, \Big( \frac{\bm{k}_{\gamma \perp}}{z_{\gamma}}-\frac{\bm{l}_{1\perp}}{1-z_{\bar{q}}} \Big) . \Big( \bm{k}_{\perp}-\frac{z_{q}}{z_{\gamma}} \, \bm{k}_{\gamma\perp} +\bm{l}_{2\perp} \Big) \nonumber \\
&-2 \, \alpha_{1} \alpha_{2} \, \Big\{ \Big( \bm{k}_{\perp}-\frac{z_{q}}{z_{\gamma}} \, \bm{k}_{\gamma\perp} \Big). \Big( \frac{\bm{k}_{\gamma \perp}}{z_{\gamma}}-\frac{\bm{l}_{1\perp}}{1-z_{\bar{q}}} \Big)+ \Big( \bm{k}_{\perp}-\frac{z_{q}}{z_{\gamma}} \, \bm{k}_{\gamma\perp} +\bm{l}_{2\perp} \Big). \Big(  \frac{z_{q}}{z_{\gamma} \, (1-z_{\bar{q}}) } \, \bm{k}_{\gamma \perp} - \frac{\bm{k}_{\perp}}{z_{q}}\Big) \Big\} - \alpha_{1} \, \frac{2k.k_{\gamma}}{1-z_{\bar{q}}} \nonumber \\
&+ \alpha_{2} \, \frac{\bm{l}_{1\perp}^{2}+\Delta^{\rm LO:(1)}}{z_{\bar{q}} \, (1-z_{\bar{q}})}  \, , \nonumber \\
c_{5}^{(S2);B}& = \alpha_{1} (1-\alpha_{1}) \,  \Big(  \frac{z_{q}}{z_{\gamma} \, (1-z_{\bar{q}}) } \, \bm{k}_{\gamma \perp} - \frac{\bm{k}_{\perp}}{z_{q}}\Big)^{2} +\alpha_{2} (1-\alpha_{2}) \, \Big( \frac{\bm{k}_{\gamma \perp}}{z_{\gamma}}-\frac{\bm{l}_{1\perp}}{1-z_{\bar{q}}} \Big)^{2} -2 \, \alpha_{1} \alpha_{2} \, \Big(  \frac{z_{q}}{z_{\gamma} \, (1-z_{\bar{q}}) } \, \bm{k}_{\gamma \perp} \nonumber \\
& - \frac{\bm{k}_{\perp}}{z_{q}}\Big).  \Big( \frac{\bm{k}_{\gamma \perp}}{z_{\gamma}}-\frac{\bm{l}_{1\perp}}{1-z_{\bar{q}}} \Big)+ \alpha_{1} \, \frac{2k.k_{\gamma}}{(1-z_{\bar{q}})^{2}} -\alpha_{2} \, \frac{\bm{l}_{1\perp}^{2}+\Delta^{\rm LO:(1)}}{z_{\bar{q}} \, (1-z_{\bar{q}})^{2}} \, .
\end{align}
\endgroup
The terms proportional to these constituent integrals in Eq.~\ref{eq:R-II-S2-A-B} can be written in terms of coefficients, some of which are also functions of the gluon loop momentum fraction $z_{l}$. The exact expressions of these are lengthy and will be provided in Mathematica scripts upon request.
%%%%%%%%%%%%%%%%%%%%%%%%%%%%%%%%%%%%%%%%%%%%%%%%%%%%%%%%%%%%%%%%%%%%%%%%%%%%%%%%

In a similar manner, the finite piece $R_{\rm (II)}$ for $(S3)$ can be written as
\begin{equation}
R^{(S3)}_{\rm (II);\mu \alpha} (\bm{l}_{1\perp}) = \hyperref[eq:R-II-S3-A-B]{R^{(S3);A}_{\rm (II);\mu \alpha} (\bm{l}_{1\perp})} +\hyperref[eq:R-II-S3-A-B]{R^{(S3);B}_{\rm (II);\mu \alpha} (\bm{l}_{1\perp}) }
\label{eq:R-II-S3}
\end{equation}
where for each case $p=A, B$, we can express
\begin{align}
R^{(S3);p}_{\rm (II);\mu \alpha} (\bm{l}_{1\perp}) & = \int_{\bm{l}_{2\perp}} e^{i \bm{l}_{2\perp}.\bm{r}_{zx}} \int \frac{\mathrm{d}z_{l}}{2\pi} \, \Big\{ \tilde{C}_{(1);\alpha}^{(S3);p} \, \mathcal{I}_{v}^{(3,0)}(\bm{V}^{(S3);p}_{\perp},\Delta^{(S3);p}) + (\tilde{C}_{(2);\alpha}^{(S3);p})^{i} \, \mathcal{I}_{v}^{(3,i)}(\bm{V}^{(S3);p}_{\perp},\Delta^{(S3);p}) \nonumber \\
& +\tilde{C}_{(3);\alpha}^{(S3);p} \, \mathcal{I}_{v}^{(3,ii)}(\bm{V}^{(S3);p}_{\perp},\Delta^{(S3);p}) + (\tilde{C}_{(4);\alpha}^{(S3);p})^{ij} \, \mathcal{I}_{v}^{(3,ij)}(\bm{V}^{(S3);p}_{\perp},\Delta^{(S3);p}) +(\tilde{C}_{(5);\alpha}^{(S3);p})^{i} \, \mathcal{I}_{v}^{(3,ijj)}(\bm{V}^{(S3);p}_{\perp},\Delta^{(S3);p}) \Big\} \nonumber \\
& \frac{\gamma^{+}(1-z_{\bar{q}}) q^{-} - \gamma^{-} \frac{\bm{l}_{1\perp}^{2}+z_{\bar{q}} \, Q^{2}}{ 2z_{\bar{q}} q^{-} }- \bm{\gamma}_{\perp}.\bm{l}_{1\perp}}{4z_{q}^{2} (1-z_{\bar{q}}) \, (q^{-})^{2} \, (\bm{l}_{1\perp}^{2}+\Delta^{\rm LO:(1)} ) } \gamma_{\mu} \, [\gamma^{+}z_{\bar{q}}q^{-}+\bm{\gamma}_{\perp}.\bm{l}_{1\perp} ] \, \gamma^{-} \, , 
\label{eq:R-II-S3-A-B}
\end{align}
with the arguments $\bm{V}_{\perp}^{(S3);p}$and $\Delta^{(S3);p}$ of the constituent integrals for both cases again expressed in terms of the gluon momentum fraction, as in Eq.~\ref{eq:integral-arguments-in-terms-of-zl}. The terms that multiply these constituent integrals in Eq.~\ref{eq:R-II-S3} are polynomials in $z_{l}$ with coefficients constituted of gamma matrices. We will see this behavior in all the finite contributions that we are going to discuss in the upcoming sections.
%%%%%%%%%%%%%%%%%%%%%%%%%%%%%%%%%%%%%%%%%%%%%%%%%%%%%%%%%%%%%%%%%%%%%%%%%%%%%%%%

\subsection{Computation of $\mathcal{M}_{\rm finite}^{\rm SE(2)}$       } \label{sec:finite-pieces-SE-2}  

The finite pieces of the contributions to the amplitudes from the quark-antiquark interchanged counterparts of the six processes in Fig.~\ref{fig:NLO-self-1} are contained in
\begin{align}
\mathcal{M}^{\text{SE}(2)}_{\text{finite};\mu \alpha}&= 2\pi \, \delta(1-z_{\text{tot}}^{v}) \, (egq_{f})^{2} \int \mathrm{d} \Pi_{\perp}^{v} \,\bar{u}(\bm{k}) \, R^{\rm SE(2)}_{ \rm finite;\mu \alpha} (\bm{l}_{1\perp}) \,  \Big[ \Big(  \tilde{U} (\bm{x}_{\perp}) t^{a} \tilde{U}^{\dagger} (\bm{y}_{\perp}) t^{b} \Big) U_{ba}(\bm{z}_{\perp}) - C_{F} \mathds{1} \Big]  \,   v(\bm{p}) \, ,
\end{align}
where 
\begin{equation}
R^{\rm SE(2)}_{ \rm finite;\mu \alpha} (\bm{l}_{1\perp})= \sum_{\beta=7}^{12} \,  \Big( \Re^{(S\beta)}_{\mu \alpha} (\bm{l}_{1\perp})   + R^{(S\beta)}_{(\textrm{II}); \mu \alpha} (\bm{l}_{1\perp}) \, \Big) \, .
\end{equation}
These are obtained by employing the replacements in Eq.~\ref{eq:replacements-qqbar-exchange} in the terms constituting the finite piece in Eq.~\ref{eq:R-SE1-finite}.

\subsection{Computation of $\mathcal{M}_{\rm finite}^{\rm SE(3)}$}  \label{sec:finite-pieces-SE-3}

The finite pieces of the contributions to the amplitude from the 24 processes $(S13)-(S36)$ are contained in 
\begin{align}
\mathcal{M}^{\rm SE(3)}_{\rm finite;\mu \alpha}& = 2\pi \, \delta(1-z_{\rm tot}^{v})  (eq_{f}g)^{2}  \int \mathrm{d} \Pi_{\perp}^{\rm LO}  \, \overline{u}(\bm{k}) \Big\{  \sum_{\beta=13}^{24} R^{(S\beta)}_{\rm finite;\mu \alpha} (\bm{l}_{1\perp}) + q\leftrightarrow \bar{q} \Big\}  \, \Bigg( C_{F} \Big( \tilde{U}(\bm{x}_{\perp}) \tilde{U}^{\dagger}(\bm{y}_{\perp}) -\mathds{1} \Big) \Bigg) v(\bm{p}) \, ,
\end{align}
where $q\leftrightarrow \bar{q}$ in the above equation refers to the expressions for the quark-antiquark interchanged processes $(S25)-(S36)$. The latter are obtained by using Eq.~\ref{eq:replacements-qqbar-exchange} in $R^{(S13)}_{\rm finite}-R^{(S24)}_{\rm finite}$. The expression for the finite contribution from $(S13)$ was given in Eq.~\ref{eq:R-S13-finite}. For $(S14)-(S16)$ we obtain,
\begin{equation}
R^{(S14)}_{\rm finite;\mu \alpha} (\bm{l}_{1\perp})= \frac{1}{2\pi^{2}} \, R^{\rm LO:(2)}_{\mu \alpha} \, \Bigg\{ \frac{7}{8}+\frac{3}{8} \, \ln  \Bigg(\frac{Q^{2}z_{\bar{q}}}{(1-z_{\gamma}) \, \Big[ (\bm{l}_{1\perp} + \bm{v}_{\perp}^{\rm LO:(2)} )^{2}+\Delta^{\rm LO:(2)} \Big]} \Bigg) -\frac{\pi^{2}}{12} \Bigg\} \, ,
\label{eq:R-S14-finite}
\end{equation}
\begin{align}
R^{(S15)}_{\rm finite;\mu \alpha} (\bm{l}_{1\perp})&= \frac{1}{2\pi^{2}} \, R^{\rm LO:(4)}_{\mu \alpha} \,  \Bigg\{ \frac{7}{8} +\frac{3}{8} \, \ln  \Bigg(\frac{Q^{2}z_{\bar{q}}}{(1-z_{\gamma}) \, \Big[ (\bm{l}_{1\perp} + \bm{v}_{\perp}^{\rm LO:(2)} )^{2}+\Delta^{\rm LO:(2)} \Big]} \Bigg) -\frac{\pi^{2}}{12} \Bigg\} +\frac{3}{16 \pi^{2}} \, \overline{A}_{\mu \alpha}(\bm{l}_{1\perp}) \, \ln \Big( \frac{a}{b} \Big) \, ,
\label{eq:R-S15-finite}
\end{align}
where
\begin{align}
\overline{A}_{\mu \alpha}(\bm{l}_{1\perp})&= \gamma^{-} \frac{\gamma^{+}z_{q}q^{-}-\bm{\gamma}_{\perp}.(\bm{l}_{1\perp}-\bm{k}_{\gamma \perp})}{ (\bm{l}_{1\perp} +z_{\bar{q}}/z_{\gamma} \, \bm{k}_{\gamma \perp})^{2}-i\varepsilon } \gamma_{\mu} \frac{\gamma^{+}(1-z_{\gamma})q^{-}+\gamma^{-} \, (\bm{l}_{1\perp}-\bm{k}_{\gamma \perp})^{2} / 2(1-z_{\gamma})q^{-} +\bm{\gamma}_{\perp}.(\bm{l}_{1\perp}-\bm{k}_{\gamma \perp})}{(\bm{l}_{1\perp}-\bm{k}_{\gamma \perp})^{2}+Q^{2}z_{q}(1-z_{q}) -i\varepsilon}  \nonumber \\
& \times \gamma_{\alpha} \, \frac{\gamma^{+}z_{\bar{q}}q^{-}+\bm{\gamma}_{\perp}.\bm{l}_{1\perp}  }{2(q^{-})^{2} z_{\gamma}/(1-z_{q}) } \gamma^{-} \, ,
\end{align}
and 
\begin{align}
a=\frac{(\bm{l}_{1\perp}-\bm{k}_{\gamma \perp})^{2}+\Delta^{\rm LO:(3)}}{1-z_{q}} \, , b=\frac{1-z_{\gamma}}{z_{\bar{q}}} \, \Big[ (\bm{l}_{1\perp}+\bm{v}_{\perp}^{\rm LO:(2)})^{2}+\Delta^{\rm LO:(2)} \Big] \, .
\end{align}
For $(S16)$ we have ,
\begin{equation}
R^{(S16)}_{\rm finite;\mu \alpha} (\bm{l}_{1\perp})= \frac{1}{2\pi^{2}} \, R^{\rm LO:(3)}_{\mu \alpha} \, \Bigg\{ \frac{7}{8}+\frac{3}{8} \, \ln  \Bigg(\frac{Q^{2}\, (1-z_{q})}{ (\bm{l}_{1\perp}- \bm{k}_{\gamma \perp})^{2} +\Delta^{\rm LO:(3)} } \Bigg) -\frac{\pi^{2}}{12} \Bigg\} \, ,
\label{eq:R-S16-finite}
\end{equation}
In the above equations, $\bm{v}_{\perp}^{\rm LO:(2)}$ and $\Delta^{\rm LO:(2)}$ are given by Eq.~\ref{eq:denominator-factors-LO-amplitude} and $\Delta^{\rm LO:(3)}=z_{q}(1-z_{q})$. As shown in detail in Appendix~\ref{sec:quark-self-energy}, the contributions from the diagrams labeled $(S17)-(S20)$ is zero in case of massless quarks. We therefore have
\begin{equation}
R^{(S17)}_{\rm finite;\mu \alpha} (\bm{l}_{1\perp})=R^{(S18)}_{\rm finite;\mu \alpha}(\bm{l}_{1\perp})=R^{(S19)}_{\rm finite;\mu \alpha}(\bm{l}_{1\perp})=R^{(S20)}_{\rm finite;\mu \alpha}(\bm{l}_{1\perp})=0 \, .
\label{eq:R-S17-R-S20-finite}
\end{equation}

The finite pieces for $(S21)$ and $(S23)$ can be obtained in a similar way:
\begin{align}
 R^{(S21)}_{\text{finite};\mu \alpha} (\bm{l}_{1\perp})&= \frac{1}{2\pi^{2}} \, R^{\text{LO}:(1)}_{\mu \alpha}(\bm{l}_{1\perp} ) \, \Bigg\{ \frac{5}{8} -\frac{3}{8} \,  \ln \Bigg(\frac{Q^{2}(1-z_{\bar{q}}) }{-2k.k_{\gamma}} \Bigg) +\frac{\pi^{2}}{12} \Bigg\} \, .
 \label{eq:R-S21-finite}
\end{align}
\begin{align}
R^{(S23)}_{\rm finite;\mu \alpha}  (\bm{l}_{1\perp})&= -\gamma^{-} \, \frac{\gamma^{+}z_{q}q^{-}-\bm{\gamma}_{\perp}.(\bm{l}_{1\perp}-\bm{k}_{\gamma \perp})}{\Big(\bm{l}_{1\perp}  + \bm{v}_{\perp}^{\rm LO:(2)} \Big)^{2}+\Delta^{\rm LO:(2)}    } \gamma_{\alpha} \, \frac{\gamma^{+}(1-z_{\bar{q}})q^{-}+\gamma^{-} \frac{\bm{l}_{1\perp}^{2}}{2(1-z_{\bar{q}})q^{-}}  -\bm{\gamma}_{\perp}.\bm{l}_{1\perp}    }{\bm{l}_{1\perp}^{2}+\Delta^{\rm LO:(1)} }  \gamma_{\mu } \frac{\gamma^{+}z_{\bar{q}}q^{-}+\bm{\gamma}_{\perp}.\bm{l}_{1\perp}}{2(1-z_{\gamma})/z_{\bar{q}} \, (q^{-})^{2}} \, \gamma^{-}  \nonumber \\
& \times \frac{1}{2\pi^{2}}  \Bigg\{ -\frac{7}{8} -\frac{3}{8} \ln \frac{Q^{2}z_{\bar{q}}}{\bm{l}_{1\perp}^{2}+\Delta^{\rm LO:(1)}} +2\ln^{2}(1-z_{\bar{q}}) +\frac{\pi^{2}}{12} \Bigg\} \nonumber \\
& -\frac{\gamma^{-}\gamma_{\alpha}\gamma_{\mu}}{2(1-z_{\gamma})/z_{\bar{q}} \, (q^{-})^{2}  \, \Big[ \Big(\bm{l}_{1\perp}  + \bm{v}_{\perp}^{\rm LO:(2)} \Big)^{2}+\Delta^{\rm LO:(2)} \Big]} \, \frac{z_{q}q^{-}}{2(1-z_{\bar{q}})} \, \frac{1}{2\pi^{2}}  \Bigg\{ \frac{1}{2} +\frac{1}{2}\, \ln \frac{Q^{2}z_{\bar{q}}}{\bm{l}_{1\perp}^{2} +\Delta^{\rm LO:(1)}} \Bigg\} \, ,
\label{eq:R-S23-finite}
\end{align}
Finally, the computation of the finite pieces $(S22)$ and $(S24)$ where the photon is emitted from the quark self-energy loop) is considerably tedious. We will express the finite contributions from these amplitudes in terms of the finite piece of the self-energy contribution that we derived in Eq.~\ref{eq:self-energy-nested-photon-finite} of Appendix~\ref{sec:quark-real photon-quark-vertex-gluon-correction}. These can now be written as 
\begin{equation}
R^{(S22)}_{\rm finite;\mu \alpha}(\bm{l}_{1\perp})= -\frac{i}{eq_{f}g^{2}C_{F}} \tilde{\Sigma}_{\alpha}^{\rm finite}  (k_{f}=k,k_{\gamma}) \, \frac{\slashed{k}+\slashed{k}_{\gamma}}{2k.k_{\gamma}} \gamma^{-} \frac{\gamma^{+}(1-z_{\bar{q}})q^{-}-\bm{\gamma}_{\perp}.\bm{l}_{1\perp}}{2(q^{-})^{2} \, 
\Big( \bm{l}_{1\perp}^{2}+\Delta^{\rm LO:(1)}  \Big) } \gamma_{\mu} [\gamma^{+}z_{\bar{q}}q^{-}+\bm{\gamma}_{\perp}.\bm{l}_{1\perp} ]\gamma^{-} \, ,
\label{eq:R-S22-finite}
\end{equation}
where $\tilde{\Sigma}(k_{f}=k,k_{\gamma})\vert_{\rm finite}$ is obtained by putting $k_{f}=k$ in Eq.~\ref{eq:self-energy-nested-photon-finite}. The expressions can further be simplified by using the Dirac equation $\overline{u}(\bm{k}) \slashed{k}=0$. Similarly we can write
\begin{align}
R^{(S24)}_{\rm finite;\mu \alpha}(\bm{l}_{1\perp})&= \frac{i}{eq_{f}g^{2}C_{F}} \frac{1}{2(q^{-})^{2}} \gamma^{-} \frac{\gamma^{+}z_{q}q^{-} -\bm{\gamma}_{\perp}.(\bm{l}_{1\perp}-\bm{k}_{\gamma \perp})}{(z_{q}+z_{\bar{q}})/z_{\bar{q}} \, \Big(\bm{l}_{1\perp}^{2}+\Delta^{\rm LO:(1)} \Big)} \, \tilde{\Sigma}_{\alpha}^{\rm finite}  (k_{f}=q+l_{1}-k_{\gamma},k_{\gamma}) \,  [ \gamma^{+}(1-z_{\bar{q}})q^{-} \nonumber \\
& -\gamma^{-} \Big( (Q^{2} z_{\bar{q}}+\bm{l}_{\perp}^{2})/2z_{\bar{q}}q^{-} \Big) -\bm{\gamma}_{\perp}.\bm{l}_{1\perp}] \gamma_{\mu} \frac{\gamma^{+}z_{\bar{q}}q^{-}+\bm{\gamma}_{\perp}.\bm{l}_{1\perp}}{\Big(\bm{l}_{1\perp}+\bm{v}_{\perp}^{\text{LO}:(2)} \Big)^{2}+\Delta^{\text{LO}:(2)}    } \gamma^{-}  \, ,
\label{eq:R-S24-finite}
\end{align} 
where $k_{f}= \Big( -\frac{Q^{2}}{2q^{-}} -\frac{\bm{l}_{1\perp}^{2}}{2z_{\bar{q}}q^{-}} -\frac{\bm{k}_{\gamma\perp}^{2}}{2z_{\gamma} q^{-}},k^{-},\bm{l}_{1\perp}-\bm{k}_{\gamma \perp} \Big)$.

\subsection{Computation of $\mathcal{M}_{\rm finite}^{\rm Vert.(1)}$} \label{sec:finite-pieces-Ver1}

The finite pieces for the 6 diagrams in Fig.~\ref{fig:NLO-vertex-allscatter} are contained in 
\begin{align}
\mathcal{M}^{\text{Vert.}(1)}_{\mu \alpha;\rm finite}&=2\pi (eq_{f}g)^{2} \delta(1-z^{v}_{\rm tot}) \, \int \mathrm{d} \Pi_{\perp}^{v} \,  \overline{u}(\bm{k}) \, R^{\rm Vert.(1)}_{\rm finite;\mu \alpha} (\bm{l}_{1\perp})   \, \Big[ \Big( t^{b} \tilde{U} (\bm{x}_{\perp}) t^{a} \tilde{U}^{\dagger} (\bm{y}_{\perp}) \Big) U_{ba}(\bm{z}_{\perp}) -C_{F} \mathds{1} \Big]   \, v(\bm{p}) \, ,
\end{align}
where 
\begin{equation}
R^{\rm Vert.(1)}_{\rm finite;\mu \alpha} (\bm{l}_{1\perp})= \sum_{\beta=1}^{6} \Big( \Re^{(V\beta)}_{\mu \alpha} (\bm{l}_{1\perp}) +R^{(V\beta)}_{(\rm II);\mu \alpha} (\bm{l}_{1\perp})  \Big)  \, .
\label{eq:R-Vertex1-finite}
\end{equation}
In the above equation, for a particular process $\beta$, as previously for the self-energy contributions, $\Re^{(V\beta)}$ is defined as the remainder between terms in the amplitude which are proportional to $1/z_{l}$ and those proportional to logarithms in $z_{0}$. The second term $R_{\rm (II)}^{(V\beta)}$ is comprised of finite terms that are not proportional to $1/z_{l}$. We will represent the latter in terms of constituent integrals described in Appendix~\ref{sec:constituent-integrals-real-emission}. For the computation of this second finite piece, it will be useful to express the amplitude for each process (see below)
\begin{align}
\mathcal{M}^{(V\beta)}_{\mu \alpha}=2\pi (eq_{f}g)^{2} \delta(1-z^{v}_{\rm tot}) \, \int \mathrm{d} \Pi_{\perp}^{v} \,  \overline{u}(\bm{k}) \, R^{(V\beta)}_{\mu \alpha} (\bm{l}_{1\perp})  \,  \Big[ \Big( t^{b} \tilde{U} (\bm{x}_{\perp}) t^{a} \tilde{U}^{\dagger} (\bm{y}_{\perp}) \Big) U_{ba}(\bm{z}_{\perp}) -C_{F} \mathds{1} \Big] \, & v(\bm{p})  \nonumber \\
& \, , \beta=1,\ldots,6 \, ,
\label{eq:amplitude-generic-vertex-shockwave}
\end{align}
in terms of some generic forms which will be identical for processes with similar topology.
We will demonstrate this in detail for a few graphs in the upcoming discussion. The finite pieces for the other graphs can be obtained following similar techniques. These expressions are lengthy and are necessary only in the context of a numerical computation. Mathematica scripts are available that allow for an algorithmic evaluation of these.

We begin our discussion with the processes $(V1)$ and $(V6)$  which have similar structures with respect to the emission vertex of the outgoing photon, which also extends to their amplitudes. This is exhibited in the similar forms for the rapidity divergent structures in Eqs.~\ref{eq:R-div-V1} and \ref{eq:R-div-V6}. 

To compute the first component, $\Re^{(V\beta)}$ ($\beta=1,6$) of the finite pieces we need to subtract the divergent part $R_{\rm div.}^{(V\beta)}$ from the terms in the amplitude proportional to $1/z_{l}$ which is represented by $R_{\rm (I)}^{(V\beta)}$.
\begin{equation}
\Re^{(V1)}_{\mu \alpha} (\bm{l}_{1\perp}) = \hyperref[eq:R-V1-propto-zl-inverse]{R^{(V1)}_{\rm (I); \mu \alpha} (\bm{l}_{1\perp})}  - \hyperref[eq:R-div-V1]{R^{(V1)}_{\mu \alpha;\rm div.} (\bm{l}_{1\perp}) } \, ,
\label{eq:remainder-V1}
\end{equation}
For $(V1)$, these are given respectively by Eqs.~\ref{eq:R-div-V1} and \ref{eq:R-V1-propto-zl-inverse}. While performing the integration over $z_{l}$ in the term $R^{(V1)}_{\rm (I)}$ we will encounter the coefficients $c_{1},\ldots, c_{5}$ in terms of which the arguments $\bm{V}_{\perp}$ and $\Delta$ of the constituent integrals are expressed as polynomials of $z_{l}$ (see Eq.~\ref{eq:V-and-Delta-in-terms-of-zl}).
\begin{align}
\bm{V}_{\perp}^{(V\beta)}& =\bm{c}_{1\perp}^{(V\beta)} +z_{l} \, \bm{c}_{2\perp}^{(V\beta)} \, , \enskip \Delta=c_{3}^{(V\beta)}+c_{4}^{(V\beta)} \, z_{l}+c_{5}^{(V\beta)} \, z_{l}^{2}  \, ,
\end{align} 
 For $(V1)$, $\bm{c}_{1\perp}$ and $c_{3}$ were given in Eqs.~\ref{eq:c1-c3-V1}. Here we provide expressions for the remaining coefficients:
\begin{align}
\bm{c}_{2\perp}^{(V1)} &= \alpha_{1} \, \frac{\bm{k}_{\perp}+\bm{k}_{\gamma \perp}-\bm{l}_{1\perp}-\bm{l}_{2\perp}}{1-z_{\bar{q}}} + \alpha_{2} \, \frac{\bm{k}_{\perp}+\bm{k}_{\gamma \perp}}{1-z_{\bar{q}}} \, , \nonumber \\
c_{4}^{(V1)}&= \frac{1}{1-z_{\bar{q}}} \Big\{ 2 \alpha_{1} (1-\alpha_{1}) \, \bm{l}_{2\perp}.(\bm{k}_{\perp}+\bm{k}_{\gamma \perp}-\bm{l}_{1\perp}-\bm{l}_{2\perp}) -2\alpha_{2} (1-\alpha_{2}) \, \bm{l}_{1\perp}.(\bm{k}_{\perp}+\bm{k}_{\gamma \perp}) \nonumber \\
& +2\alpha_{1} \alpha_{2} \, \bm{l}_{1\perp}. (\bm{k}_{\perp}+\bm{k}_{\gamma \perp}-\bm{l}_{1\perp}-\bm{l}_{2\perp}) -2\alpha_{1} \alpha_{2} \, \bm{l}_{2\perp}. (\bm{k}_{\perp}+\bm{k}_{\gamma \perp}) \nonumber \\
&+\alpha_{1} \, \frac{(\bm{l}_{1\perp}+\bm{l}_{2\perp})^{2}+z_{\bar{q}}(1-z_{\bar{q}}) Q^{2}}{z_{\bar{q}}}  +\alpha_{2} \, (1-z_{\bar{q}})(1-2z_{\bar{q}}) Q^{2} -(1-\alpha_{1}-\alpha_{2}) \, (2k.k_{\gamma})  \Big\} \, , \nonumber \\
c_{5}^{(V1)}&=\frac{1}{(1-z_{\bar{q}})^{2}} \Big\{ \alpha_{1} (1-\alpha_{1}) \, (\bm{l}_{1\perp}+\bm{l}_{2\perp}-\bm{k}_{\perp}-\bm{k}_{\gamma \perp})^{2} +\alpha_{2} (1-\alpha_{2}) \, (\bm{k}_{\perp}+\bm{k}_{\gamma \perp})^{2} \nonumber \\
& + 2\alpha_{1} \alpha_{2} \, (\bm{k}_{\perp}+\bm{k}_{\gamma \perp}).(\bm{l}_{1\perp}+\bm{l}_{2\perp}-\bm{k}_{\perp}-\bm{k}_{\gamma \perp}) -\alpha_{1} \, \frac{(\bm{l}_{1\perp}+\bm{l}_{2\perp})^{2}+z_{\bar{q}}(1-z_{\bar{q}}) Q^{2}}{z_{\bar{q}}} -\alpha_{2} \, (1-z_{\bar{q}})^{2} Q^{2} \nonumber \\
& +(1-\alpha_{1}-\alpha_{2}) \, (2k.k_{\gamma}) \Big\} \, . 
\label{eq:c1-c5-V1}
\end{align}
In case of $(V6)$, the terms in the amplitude proportional to $1/z_{l}$ are contained in 
\begin{align}
R^{(V6)}_{\rm (I);\mu \alpha}&=- \int \frac{\mathrm{d}z_{l}}{(2\pi) z_{l}} \int_{\bm{l}_{2\perp}} \!\!\!\!e^{i\bm{l}_{2\perp}.\bm{r}_{zy}} \, \frac{1}{(q^{-})^{2}} \,  \Bigg[ \big\{2q^{-}\gamma^{i} \gamma_{\mu} \gamma^{-}-4z_{q}q^{-} \delta_{i\mu} \gamma^{-} \big\} \times \Big\{ \mathcal{I}_{v}^{(3,ijj)}(\bm{V}_{\perp}^{(V6)},\Delta^{(V6)})  \nonumber \\
& +l_{2}^{j} \, \mathcal{I}^{(3,ij)}_{v} (\bm{V}_{\perp}^{(V6)},\Delta^{(V6)})  \Big\} +\gamma^{-} [\gamma^{+}z_{q}q^{-}-\bm{\gamma}_{\perp}.(\bm{l}_{1\perp}-\bm{k}_{\gamma \perp})]\gamma_{\mu} [\gamma^{+}(1-z_{q})q^{-}+\bm{\gamma}_{\perp}.(\bm{l}_{1\perp}-\bm{k}_{\gamma \perp}) ]\gamma^{-} \nonumber \\
&  \times  \Big\{ \mathcal{I}^{(3,ii)}_{v} (\bm{V}_{\perp}^{(V6)},\Delta^{(V6)}) +l_{2}^{i} \, \mathcal{I}^{(3,i)}_{v} (\bm{V}_{\perp}^{(V6)},\Delta^{(V6)})   \Big\} \Bigg] \, \frac{\slashed{p}+\slashed{k}_{\gamma}}{2p.k_{\gamma}} \, \gamma_{\alpha} \, ,
\label{eq:R-V6-propto-zl-inverse}
\end{align}
where the finite expressions for the constituent integrals are given in Eqs.~\ref{eq:constituent-integrals-V1}. For $(V6)$ the coefficients $c_{i}$ ($i=2,4,5$) required for the computation of the above term in Eq.~\ref{eq:R-V6-propto-zl-inverse} are given below. The expressions for $\bm{c}_{1\perp}$ and $c_{3}$ are in Eqs.~\ref{eq:c1-c3-V6}.
\begin{align}
\bm{c}_{2\perp}^{(V6)} &= \alpha_{1} \, \frac{\bm{k}_{\perp}+\bm{k}_{\gamma \perp}-\bm{l}_{1\perp}-\bm{l}_{2\perp}}{z_{q}} + \alpha_{2} \, \frac{\bm{k}_{\perp}}{z_{q}} \, , \nonumber \\
c_{4}^{(V6)}&= \frac{1}{z_{q}} \Big\{ -2 \alpha_{1} (1-\alpha_{1}) \, \bm{l}_{2\perp}.(\bm{l}_{1\perp}+\bm{l}_{2\perp}-\bm{k}_{\perp}-\bm{k}_{\gamma \perp}) -2\alpha_{2} (1-\alpha_{2}) \, \bm{k}_{\perp}.(\bm{l}_{1\perp}-\bm{k}_{\gamma \perp}) \nonumber \\
& -2\alpha_{1} \alpha_{2} \, (\bm{l}_{1\perp}+\bm{l}_{2\perp}-\bm{k}_{\perp}-\bm{k}_{\gamma \perp}). (\bm{l}_{1\perp}-\bm{k}_{\gamma \perp}) -2\alpha_{1} \alpha_{2} \, \bm{l}_{2\perp}.\bm{k}_{\perp} \nonumber \\
&+\alpha_{1} \, \frac{(\bm{l}_{1\perp}+\bm{l}_{2\perp}-\bm{k}_{\gamma \perp})^{2}+z_{q}(1-z_{q}) Q^{2}}{1-z_{\bar{q}}}  -\alpha_{2} \, z_{q}(1-2z_{q}) Q^{2} \Big\} \, , \nonumber \\
c_{5}^{(V6)}&=\frac{1}{z_{q}^{2}} \Big\{ \alpha_{1} (1-\alpha_{1}) \, (\bm{l}_{1\perp}+\bm{l}_{2\perp}-\bm{k}_{\perp}-\bm{k}_{\gamma \perp})^{2} +\alpha_{2} (1-\alpha_{2}) \, \bm{k}_{\perp}^{2} +2\alpha_{1} \alpha_{2} \, \bm{k}_{\perp}.(\bm{l}_{1\perp}+\bm{l}_{2\perp}-\bm{k}_{\perp}-\bm{k}_{\gamma \perp}) \nonumber \\
& -\alpha_{1} \, \frac{(\bm{l}_{1\perp}+\bm{l}_{2\perp}-\bm{k}_{\gamma \perp})^{2}+z_{q}(1-z_{q}) Q^{2}}{1-z_{\bar{q}}} -\alpha_{2} \, z_{q}^{2} Q^{2} \Big\} \, . 
\label{eq:c1-c5-V6}
\end{align}
An interesting and important check of these coefficients presented in Eqs.~\ref{eq:c1-c5-V1} and \ref{eq:c1-c5-V6} is that they are identical in the soft photon $k_{\gamma} \rightarrow 0$ limit.

The remainder $\Re^{(V6)}$ is now obtained from
\begin{equation}
\Re^{(V6)}_{\mu \alpha} (\bm{l}_{1\perp})= \hyperref[eq:R-V6-propto-zl-inverse]{R^{(V6)}_{\rm (I); \mu \alpha} (\bm{l}_{1\perp})} - \hyperref[eq:R-div-V6]{R^{(V6)}_{\mu \alpha;\rm div.} (\bm{l}_{1\perp})} \, ,
\label{eq:remainder-V6}
\end{equation}
where the divergent term is given by Eq.~\ref{eq:R-div-V6}. 

We will now compute the second finite piece for these processes. We can express the $R^{(V\beta)}$ ($\beta=1,6$) appearing in Eq.~\ref{eq:amplitude-generic-V1-V6}) in terms of  ${\bar R}_\mu$ which can be   expressed solely in terms of $z_{l}$ and $\bm{l}_{3\perp}$ albeit with coefficients that differ with the diagram of interest.  For $\beta=1,6$, we obtain 
\begin{align}
{\bar R}_\mu^{(V\beta)} (z_{l},\bm{l}_{3\perp}) &= (b_{1}^{(V\beta)} \, z_{l} -b_{2}^{(V\beta)} \,l_{3}^{i} ) \, \{ (b_{3}^{(V\beta)}+b_{4}^{(V\beta)} \, l_{3}^{j} )-b_{5}^{(V\beta)} \, z_{l} \} \, b_{6}^{(V\beta)} \, \{ (b_{7}^{(V\beta)}+b_{8}^{(V\beta)} \, l_{3}^{k} ) -b_{9}^{(V\beta)} \, z_{l} \} \, b_{10}^{(V\beta)} \nonumber \\
& \times \Big\{ (b_{11}^{(V\beta)}-b_{12}^{(V\beta)} \, \bm{l}_{3\perp}^{2} -b_{13}^{(V\beta)} \, l_{3}^{p} ) +(b_{14}^{(V\beta)} +b_{15}^{(V\beta)} \, l_{3}^{w} ) \, z_{l} -b_{16}^{(V\beta)} \, z_{l}^{2} \Big\} \Bigg( b_{17}^{(V\beta)}+ \frac{b_{18}^{(V\beta)}+b^{(V\beta)}_{19} \, l_{3}^{i}}{z_{l}} \Bigg) \, .
\label{eq:R-a-generic}
\end{align}
In terms of this generic structure, the $R$-functions for $(V1)$ and $(V6)$ can be respectively written as
\begin{align}
R^{(V1)}_{\mu \alpha} (\bm{l}_{1\perp}) & = - \int_{\bm{l}_{2\perp}} \!\!\!\!\!\!\! e^{i\bm{l}_{2\perp}.\bm{r}_{zy}} \int \frac{\mathrm{d}z_{l}}{2\pi} \int_{\bm{l}_{3\perp}} \gamma_{\alpha} \frac{\slashed{k}+\slashed{k}_{\gamma}}{2k.k_{\gamma}}  \, {\bar R}_\mu^{(V1)}  (z_{l},\bm{l}_{3\perp}) \, \frac{\gamma^{+}z_{\bar{q}}q^{-}+\bm{\gamma}_{\perp}.(\bm{l}_{1\perp}+\bm{l}_{2\perp})}{ 4z_{\bar{q}}(1-z_{\bar{q}}) \, (q^{-})^{2} } \gamma^{-} \nonumber \\
& \times \frac{1}{ \Big[(\bm{l}_{3\perp}+\bm{v}_{1\perp}^{(V1)})^{2}+\Delta_{1}^{(V1)}\Big] \,  \Big[(\bm{l}_{3\perp}+\bm{v}_{2\perp}^{(V1)})^{2}+\Delta_{2}^{(V1)} \Big] \, \Big[\bm{l}_{3\perp}^{2}+\Delta_{3}^{(V1)}\Big] } \, ,
\end{align}
and 
\begin{align}
R^{(V6)}_{\mu \alpha} (\bm{l}_{1\perp}) & = \int_{\bm{l}_{2\perp}} \!\!\!\!\!\!\! e^{i\bm{l}_{2\perp}.\bm{r}_{zy}} \int \frac{\mathrm{d}z_{l}}{2\pi} \int_{\bm{l}_{3\perp}} {\bar R}_\mu^{(V6)}  (z_{l},\bm{l}_{3\perp}) \, \frac{\gamma^{+} (1-z_{q})q^{-}+\bm{\gamma}_{\perp}.(\bm{l}_{1\perp}+\bm{l}_{2\perp}-\bm{k}_{\gamma \perp}) }{ 4z_{q}(1-z_{q}) \, (q^{-})^{2}}  \gamma^{-} \frac{\slashed{p}+\slashed{k}_{\gamma}}{2p.k_{\gamma}} \gamma_{\alpha}  \nonumber \\
& \times \frac{1}{ \Big[(\bm{l}_{3\perp}+\bm{v}_{1\perp}^{(V6)})^{2}+\Delta_{1}^{(V6)}\Big] \,  \Big[(\bm{l}_{3\perp}+\bm{v}_{2\perp}^{(V6)})^{2}+\Delta_{2}^{(V6)} \Big] \, \Big[\bm{l}_{3\perp}^{2}-i\varepsilon \Big] } \, .
\end{align}
The advantage of this method is that it offers a transparent way to collect terms proportional to a certain power of $\bm{l}_{3\perp
}$ and then organize them in terms of constituent integrals. Moreover, the finite pieces appear as polynomials in $z_{l}$; with the arguments of the integrals expressed in the form given by Eq.~\ref{eq:V-and-Delta-in-terms-of-zl}, one can straightforwardly perform the integration over $z_{l}$.

For the diagrams $(V1)$ and $(V6)$ we finally obtain the finite terms $R_{\rm (II)}$ as
\begin{align}
R^{(V1)}_{\rm (II);\mu \alpha} (\bm{l}_{1\perp}) & = - \int_{\bm{l}_{2\perp}} \!\!\!\!\! e^{i\bm{l}_{2\perp}.\bm{r}_{zy}} \int \frac{\mathrm{d}z_{l}}{2\pi} \,  \gamma_{\alpha} \frac{\slashed{k}+\slashed{k}_{\gamma}}{2k.k_{\gamma}} \, \Bigg\{ F^{(V1)}_{(1);\mu} \, \mathcal{I}_{v}^{(3,0)} (\bm{V}_{\perp}^{(V1)}, \Delta^{(V1)}) + F^{(V1),i}_{(2);\mu} \, \mathcal{I}_{v}^{(3,i)} (\bm{V}_{\perp}^{(V1)}, \Delta^{(V1)}) \nonumber \\
& + F^{(V1)}_{(3);\mu} \, \mathcal{I}_{v}^{(3,ii)} (\bm{V}_{\perp}^{(V1)}, \Delta^{(V1)}) + F^{(V1),ij}_{(4);\mu} \, \mathcal{I}_{v}^{(3,ij)} (\bm{V}_{\perp}^{(V1)}, \Delta^{(V1)}) +F^{(V1),i}_{(5);\mu} \, \mathcal{I}_{v}^{(3,ijj)} (\bm{V}_{\perp}^{(V1)}, \Delta^{(V1)}) \Bigg\}\nonumber \\
& \times  \frac{\gamma^{+}z_{\bar{q}}q^{-}+\bm{\gamma}_{\perp}.(\bm{l}_{1\perp}+\bm{l}_{2\perp})}{ 4z_{\bar{q}}(1-z_{\bar{q}}) \, (q^{-})^{2} } \gamma^{-}  \, ,
\label{eq:R-finite-V1}
\end{align}
and 
\begin{align}
R^{(V6)}_{\rm (II);\mu \alpha} (\bm{l}_{1\perp}) & =  \int_{\bm{l}_{2\perp}} \!\!\!\!\! e^{i\bm{l}_{2\perp}.\bm{r}_{zy}} \int \frac{\mathrm{d}z_{l}}{2\pi}  \Bigg\{ F^{(V6)}_{(1);\mu} \, \mathcal{I}_{v}^{(3,0)} (\bm{V}_{\perp}^{(V6)}, \Delta^{(V6)}) + F^{(V6),i}_{(2);\mu} \, \mathcal{I}_{v}^{(3,i)} (\bm{V}_{\perp}^{(V6)}, \Delta^{(V6)}) \nonumber \\
& + F^{(V6)}_{(3);\mu} \, \mathcal{I}_{v}^{(3,ii)} (\bm{V}_{\perp}^{(V6)}, \Delta^{(V6)}) + F^{(V6),ij}_{(4);\mu} \, \mathcal{I}_{v}^{(3,ij)} (\bm{V}_{\perp}^{(V6)}, \Delta^{(V6)}) +F^{(V6),i}_{(5);\mu} \, \mathcal{I}_{v}^{(3,ijj)} (\bm{V}_{\perp}^{(V6)}, \Delta^{(V6)}) \Bigg\}\nonumber \\
& \times  \frac{\gamma^{+} (1-z_{q})q^{-}+\bm{\gamma}_{\perp}.(\bm{l}_{1\perp}+\bm{l}_{2\perp}-\bm{k}_{\gamma \perp}) }{ 4z_{q}(1-z_{q}) \, (q^{-})^{2}}  \gamma^{-} \frac{\slashed{p}+\slashed{k}_{\gamma}}{2p.k_{\gamma}} \gamma_{\alpha}  \, .
\label{eq:R-finite-V6}
\end{align}
The terms $F_{(j);\mu}$ ($j=1,\ldots,5$) can be expressed in terms of the coefficients $b_{i}$ ($i=1,\ldots,19$) appearing in Eq.~\ref{eq:R-a-generic}. The individual expressions for these process dependent coefficients are provided below for $(V1)$ and $(V6)$.
For $(V1)$ we have:
\begingroup
\allowdisplaybreaks
\begin{align}
b_{1}^{(V1)}&= \gamma^{i} -\frac{k^{i}+k^{i}_{\gamma}}{(1-z_{\bar{q}}) \, q^{-}} \, \gamma^{-} \, , \enskip b_{2}^{(V1)}=\frac{\gamma^{-}}{q^{-}} \, , \enskip b_{3}^{(V1)}= \slashed{k}+\slashed{k}_{\gamma}  \, , \enskip b_{4}^{(V1)}=\gamma^{j} \, , \nonumber \\
b_{5}^{(V1)}& = \frac{\slashed{k}+\slashed{k}_{\gamma}}{1-z_{\bar{q}}} \, , \enskip b_{6}^{(V1)}=\gamma^{-} \, , \enskip b_{7}^{(V1)}=\gamma^{+}(1-z_{\bar{q}}) \, q^{-} -\bm{\gamma}_{\perp}.\bm{l}_{1\perp} \, , \enskip b_{8}^{(V1)}=\gamma^{k} \, ,\enskip  b_{9}^{(V1)}=b_{5}^{(V1)} \, , \nonumber \\
b_{10}^{(V1)}&= \gamma_{\mu} \, , \enskip b_{11}^{(V1)}=\gamma^{+}z_{\bar{q}}q^{-}-\gamma^{-} \Big( \frac{Q^{2}}{2q^{-}}+\frac{\bm{l}_{1\perp}^{2}}{2(1-z_{\bar{q}}) \, q^{-}} \Big) +\bm{\gamma}_{\perp}.\bm{l}_{1\perp} \, , \enskip b_{12}^{(V1)}=\frac{\gamma^{-}}{2(1-z_{\bar{q}}) \, q^{-}} \, , \nonumber \\
b_{13}^{(V1)}&=\gamma^{p} -\frac{l_{1}^{p}}{(1-z_{\bar{q}}) \, q^{-}} \, \gamma^{-} \, , \enskip b_{14}^{(V1)}=\frac{\gamma^{+}(1-2z_{\bar{q}}) \, q^{-}+\gamma^{-} \Big( \frac{Q^{2}}{2q^{-}} + \frac{\bm{l}_{1\perp}.(\bm{k}_{\perp}+\bm{k}_{\gamma \perp})}{(1-z_{\bar{q}}) \, q^{-}} \Big) -\bm{\gamma}_{\perp}. (\bm{l}_{1\perp} +\bm{k}_{\perp}+\bm{k}_{\gamma \perp})  }{1-z_{\bar{q}}} \, ,\nonumber \\
b_{15}^{(V1)}&= \frac{1}{1-z_{\bar{q}}} \Big( \gamma^{w} -\frac{k^{w}+k^{w}_{\gamma}}{(1-z_{\bar{q}}) \, q^{-}} \, \gamma^{-} \Big) \, , b_{16}^{(V1)}= \frac{\gamma^{+}(1-z_{\bar{q}}) \, q^{-} + \gamma^{-} \, \frac{(\bm{k}_{\perp}+\bm{k}_{\gamma \perp})^{2}}{2(1-z_{\bar{q}}) \, q^{-}} -\bm{\gamma}_{\perp}.(\bm{k}_{\perp}+\bm{k}_{\gamma \perp})  }{(1-z_{\bar{q}})^{2}} \, ,\nonumber \\
b_{17}^{(V1)}&=  \gamma_{i} -\frac{k_{i}+k_{\gamma i }}{(1-z_{\bar{q}}) \, q^{-}} \, \gamma^{-} \, , \enskip b_{18}^{(V1)}= \frac{l_{2}^{i}}{q^{-}} \, \gamma^{-} \, , \enskip b_{19}^{(V1)}= \frac{\gamma^{-}}{q^{-}} \, .
\label{eq:coefficients-finite-pieces-V1}
\end{align}
\endgroup
For $(V6)$ we have:
\begingroup
\allowdisplaybreaks
\begin{align}
b_{1}^{(V6)}&= \gamma^{i} -\frac{k^{i}}{z_{q} \, q^{-}} \, \gamma^{-} \, , \enskip b_{2}^{(V6)}=\frac{\gamma^{-}}{q^{-}} \, , \enskip b_{3}^{(V6)}= \slashed{k}  \, , \enskip b_{4}^{(V6)}=\gamma^{j} \, , \nonumber \\
b_{5}^{(V6)}& = \frac{\slashed{k}}{z_{q}} \, , \enskip b_{6}^{(V6)}=\gamma^{-} \, , \enskip b_{7}^{(V6)}=\gamma^{+} z_{q} \, q^{-} -\bm{\gamma}_{\perp}.(\bm{l}_{1\perp}-\bm{k}_{\gamma \perp}) \, , \enskip b_{8}^{(V6)}=\gamma^{k} \, ,\enskip  b_{9}^{(V6)}=b_{5}^{(V1)} \, , \nonumber \\
b_{10}^{(V6)}&= \gamma_{\mu} \, , \enskip b_{11}^{(V6)}=\gamma^{+}(1-z_{q}) \, q^{-}-\gamma^{-} \Big( \frac{Q^{2}}{2q^{-}}+\frac{(\bm{l}_{1\perp}-\bm{k}_{\gamma \perp})^{2}}{2 z_{q} \, q^{-}} \Big) +\bm{\gamma}_{\perp}.(\bm{l}_{1\perp}-\bm{k}_{\gamma \perp})  \, , \enskip b_{12}^{(V6)}=\frac{\gamma^{-}}{2z_{q} \, q^{-}} \, , \nonumber \\
b_{13}^{(V6)}&=\gamma^{p} -\frac{l_{1}^{p}-k_{\gamma}^{p}}{z_{q} \, q^{-}} \, \gamma^{-} \, , \enskip b_{14}^{(V6)}=- \frac{\gamma^{+}(1-2z_{q}) \, q^{-}- \gamma^{-} \Big( \frac{Q^{2}}{2q^{-}} + \frac{(\bm{l}_{1\perp}-\bm{k}_{\gamma \perp}).\bm{k}_{\perp}}{z_{q} \, q^{-}} \Big) +\bm{\gamma}_{\perp}.(\bm{l}_{1\perp}-\bm{k}_{\gamma \perp}+\bm{k}_{\perp})  }{z_{q}} \, ,\nonumber \\
b_{15}^{(V6)}&=\frac{1}{z_{q}} \Big( \gamma^{w} -\frac{k^{w}}{z_{q} \, q^{-}} \, \gamma^{-} \Big) \, , b_{16}^{(V6)}= \frac{\slashed{k}  }{z_{q}^{2}} \, , \enskip b_{17}^{(V6)}=  \gamma_{i} -\frac{k_{i} }{z_{q} \, q^{-}} \, \gamma^{-} \, , \enskip b_{18}^{(V6)}= \frac{l_{2}^{i}}{q^{-}} \, \gamma^{-} \, , \enskip b_{19}^{(V6)}= \frac{\gamma^{-}}{q^{-}} 
 \, .
\label{eq:coefficients-finite-pieces-V6}
\end{align}
\endgroup
We can obtain the expressions for the terms multiplying the constituent integrals in the finite pieces (see Eqs.~\ref{eq:R-finite-V1} and \ref{eq:R-finite-V6}) in terms of these coefficients. Because of the 
particularly lengthy nature of these expressions, we will not provide these here. Mathematica scripts for these are available upon request.

Similarly,  $(V3)$ and $(V4)$ in Fig.~\ref{fig:NLO-vertex-1} have a similar topology with respect to the emission vertex of the final state photon and hence exhibit a similar structure for the divergent pieces. We will present the structures for $(V4)$ here; the corresponding expressions for $(V3)$ are lengthy and will be made available on request as supplementary material. One also needs to include the contribution from the gluon momentum $l_{3}^{-}$ in the range $k^{-}<l_{3}^{-}<k^{-}+k^{-}_{\gamma}$ to obtain the net finite contribution from $(V3)$.

%%%%%%%%%%%%%%%%%%%%%%%%%%%%%%%%%%%%%%%%%%%%%%%%%%%%%%%%%%%%%%
For $(V4)$, the remainder term can be written as
\begin{equation}
\Re^{(V4)}_{\mu \alpha}(\bm{l}_{1\perp}) = \hyperref[eq:R-V4-propto-zl-inverse]{R^{(V4)}_{\rm (I); \mu \alpha}(\bm{l}_{1\perp})  }- \hyperref[eq:R-div-V4]{R^{(V4)}_{\mu \alpha;\rm div.} (\bm{l}_{1\perp}) } \, ,
\label{eq:remainder-V4}
\end{equation}
where the term proportional to $1/z_{l}$ is given by Eq.~\ref{eq:R-V4-propto-zl-inverse} and the divergent piece is contained in Eq.~\ref{eq:R-div-V4}. 
For the computation of the second finite piece $R_{\rm (II)}$ we can express $R^{(V4)}$ in the expression for the amplitude in Eq.~\ref{eq:amplitude-generic-vertex-shockwave} in terms of a generic structure written as a polynomial in $z_{l}$ and $\bm{l}_{3\perp}$ with coefficients which depend on the process of interest. 

%%%%%%%%%%%%%%%%%%%%%%%%%%%%%%%%%%%%%%%%%%%%%%%%%%%%%%%%%%%%%%%%%%%%%%%%%%%
For $(V4)$, we can write 
\begin{align}
R^{(V4)}_{\mu \alpha} (\bm{l}_{1\perp}) &= - \int_{\bm{l}_{2\perp}} \!\!\!\!\! e^{i\bm{l}_{2\perp}.\bm{r}_{zy}} \int \frac{\mathrm{d}z_{l}}{2\pi}  \int_{\bm{l}_{3\perp}}   \, \frac{1}{4z_{\bar{q}}(1-z_{\gamma}) \, (q^{-})^{2}} R^{(V4)}_{(b);\mu \alpha} (z_{l}, \bm{l}_{3\perp})  \, [\gamma^{+}z_{\bar{q}}q^{-}+\bm{\gamma}_{\perp}.(\bm{l}_{1\perp}+\bm{l}_{2\perp})] \gamma^{-} \nonumber \\
& \times \frac{1}{(\bm{l}_{3\perp}^{2}-i\varepsilon) \, \Big[(\bm{l}_{3\perp}+\bm{v}_{1\perp}^{(V4)})^{2}+\Delta_{1}^{(V4)}\Big] \,  \Big[(\bm{l}_{3\perp}+\bm{v}_{2\perp}^{(V4)})^{2}+\Delta_{2}^{(V4)} \Big] \, \Big[(\bm{l}_{3\perp}+\bm{v}_{3\perp}^{(V4)})^{2}+\Delta_{3}^{(V4)} \Big] } \, .
\end{align}
The term $R_{(b)}$ appearing in the above expression is generic for the processes $(V3)$ and $(V4)$ and is given by
\begin{align}
R^{(V\beta)}_{(b);\mu \alpha} (z_{l},\bm{l}_{3\perp}) &= (b_{1}^{(V\beta)} \, z_{l} -b_{2}^{(V\beta)} \, l_{3}^{i} ) \, \big\{ (b_{3}^{(V\beta)}+ b_{4}^{(V\beta)} \, l_{3}^{j} ) -b_{5}^{(V\beta)} \, z_{l} \big\} \, b_{6}^{(V\beta)} \, \big\{ (b_{7}^{(V\beta)}+b_{8}^{(V\beta)} \, l_{3}^{k} ) - b_{9}^{(V\beta)} \, z_{l} \big\} \, b_{10}^{(V\beta)} \nonumber \\
& \times \big\{ (b_{11}^{(V\beta)} -b_{12}^{(V\beta)} \, \bm{l}_{3\perp}^{2}-b_{13}^{(V\beta)} \, l_{3}^{p} ) -z_{l} \, (b_{14}^{(V\beta)}- b_{15}^{(V\beta)} \, l_{3}^{q}) -z_{l}^{2} \, b_{16}^{(V\beta)} \big\} \, b_{17}^{(V\beta)} \, \big\{ (b_{18}^{(V\beta)}-b_{19}^{(V\beta)} \, \bm{l}_{3\perp}^{2} -b_{20}^{(V\beta)} \, l_{3}^{r} ) \nonumber \\
& +z_{l} \, (b_{21}^{(V\beta)}+ b_{22}^{(V\beta)} \, l_{3}^{s} ) -z_{l}^{2} \, b_{23}^{(V\beta)} \big\} \, \Big( b_{24}^{(V\beta)}+\frac{b_{25}^{(V\beta)}+b_{26}^{(V\beta)}  \, l_{3}^{i} }{z_{l}} \Big) \, , \quad \beta=3,4  \, .
\label{eq:R-b-generic}
\end{align}
Expanding the above expression, and organizing the non-zero terms in powers of $\bm{l}_{3\perp}$, we can write the finite pieces $R_{\rm (II)}$ for $(V3)$ and $(V4)$ in terms of constituent integrals with coefficients which can be written explicitly in terms of the various $b_{i}$'s ($i=1,\ldots,26$). 
%%%%%%%%%%%%%%%%%%%%%%%%%%%%%%%%%%%%%%%%

In terms of the constituent integrals given in Eq.~\ref{eq:constituent-integrals-V3-V4} we can now write the finite pieces contained in $R_{\rm (II)}^{(V4)}$ as
\begin{align}
R^{(V4)}_{\rm (II);\mu \alpha} (\bm{l}_{1\perp}) &= - \int_{\bm{l}_{2\perp}} \!\!\!\!\!\! e^{i \bm{l}_{2\perp}.\bm{r}_{zy}} \int \frac{\mathrm{d}z_{l}}{2\pi} \Big\{ \tilde{F}^{(V4)}_{(1);\mu \alpha} \, \mathcal{I}^{(4,0)}_{v} (\bm{V}_{\perp}^{(V4)},\Delta^{(V4)} ) + \tilde{F}^{(V4),i}_{(2);\mu \alpha} \, \mathcal{I}^{(4,i)}_{v} (\bm{V}_{\perp}^{(V4)},\Delta^{(V4)} )  \nonumber \\
&+\tilde{F}^{(V4),ij}_{(3);\mu \alpha} \, \mathcal{I}^{(4,ij)}_{v} (\bm{V}_{\perp}^{(V4)},\Delta^{(V4)} ) +\tilde{F}^{(V4)}_{(4);\mu \alpha} \, \mathcal{I}^{(4,ii)}_{v} (\bm{V}_{\perp}^{(V4)},\Delta^{(V4)} )+ \tilde{F}^{(V4),ijk}_{(5);\mu \alpha} \, \mathcal{I}^{(4,ijk)}_{v} (\bm{V}_{\perp}^{(V4)},\Delta^{(V4)} ) \nonumber \\
&+\tilde{F}^{(V4),i }_{(6);\mu \alpha} \, \mathcal{I}^{(4,ijj)}_{v} (\bm{V}_{\perp}^{(V4)},\Delta^{(V4)} ) +\tilde{F}^{(V4),ij}_{(7);\mu \alpha} \, \mathcal{I}^{(4,ijkk)}_{v} (\bm{V}_{\perp}^{(V4)},\Delta^{(V4)} ) +\tilde{F}^{(V4)}_{(8);\mu \alpha} \, \mathcal{I}^{(4,iijj)}_{v} (\bm{V}_{\perp}^{(V4)},\Delta^{(V4)} ) \Big\} \nonumber \\
& \times \frac{\gamma^{+}z_{\bar{q}}q^{-}+ \bm{\gamma}_{\perp}.(\bm{l}_{1\perp}+\bm{l}_{2\perp}) }{4z_{\bar{q}}(1-z_{\gamma}) \, (q^{-})^{2}} \, \gamma^{-}  \, .
\label{eq:R-finite-V4}
\end{align}
The expressions for the $\tilde{F}$'s can be obtained in terms of the coefficients $b_{i}^{(V\beta)}$ ($i=1,\ldots,26$) for the processes $(V3)$ and $(V4)$. 
In a similar fashion, one can obtain the finite pieces for $(V2)$ and $(V5)$ in Fig.~\ref{fig:NLO-vertex-1}. For these, we need to compute only $R_{\rm (II)}$ which also contain the same constituent integrals as in Eq.~\ref{eq:R-finite-V4}. Similarly to the $(V3)$ case, we need to include the contribution from $l_{3}^{-}$ in the range $k^{-}<l_{3}^{-}<k^{-}+k^{-}_{\gamma}$ in addition to $0<l^{-}_{3}<k^{-}$ to obtain the net finite contribution.

%%%%%%%%%%%%%%%%%%%%%%%%%%%%%%%%%%%%%%%%%%%%%%%%%%%%%%%%%%%%%%%%%%%%%%%%%%%%%%%%%

\subsection{Computation of $\mathcal{M}_{\rm finite}^{\rm Vert.(2)}$} \label{sec:finite-pieces-Ver2}
The finite pieces of the contribution from the quark-antiquark interchanged counterparts of the six processes in Fig.~\ref{fig:NLO-vertex-1} are contained in
\begin{align}
\mathcal{M}^{\text{Vert.}(2)}_{\text{finite};\mu \alpha}&= 2\pi (eq_{f}g)^{2} \delta(1-z^{v}_{\rm tot}) \, \int \mathrm{d} \Pi_{\perp}^{v} \,  \overline{u}(\bm{k}) \, R^{\rm Vert.(2)}_{\rm finite;\mu \alpha} (\bm{l}_{1\perp})  \, \Big[ \Big(  \tilde{U} (\bm{x}_{\perp}) t^{a} \tilde{U}^{\dagger} (\bm{y}_{\perp}) t^{b} \Big) U_{ba}(\bm{z}_{\perp}) - C_{F} \mathds{1} \Big]  \,   v(\bm{p}) \, ,
\label{eq:amplitude-finite-V2-generic}
\end{align}
where 
\begin{equation}
R^{\rm Vert.(2)}_{\rm finite;\mu \alpha} (\bm{l}_{1\perp}) = \sum_{\beta=7}^{12} \Big( \Re^{(V\beta)}_{\mu \alpha} (\bm{l}_{1\perp}) +R^{(V\beta)}_{(\rm II);\mu \alpha} (\bm{l}_{1\perp})  \Big) \, .
\label{eq:R-Vertex2-finite}
\end{equation}
We can obtain the net finite contribution from these diagrams by using Eq.~\ref{eq:replacements-qqbar-exchange} along with a change of sign on the various terms constituting the finite piece given by Eq.~\ref{eq:R-Vertex1-finite}.

\subsection{Computation of $\mathcal{M}_{\rm finite}^{\rm Vert.(3)}$} \label{sec:finite-pieces-Ver3}

The finite contributions to the amplitudes from the processes $(V13)-(V15)$ in Fig.~\ref{fig:NLO-vertex-3} and their $q\leftrightarrow\bar{q}$ counterparts are expressed as
\begin{align}
\mathcal{M}^{\rm Vert.(3)}_{\rm finite;\mu \alpha}& = 2\pi \, \delta(1-z_{\rm tot}^{v})  (eq_{f}g)^{2}  \int \mathrm{d} \Pi_{\perp}^{\rm LO}  \, \overline{u}(\bm{k}) \,  \sum_{\beta=13}^{18} R^{(V\beta)}_{\rm finite;\mu \alpha} (\bm{l}_{1\perp})   \, \Big[ C_{F} \Big( \tilde{U}(\bm{x}_{\perp}) \tilde{U}^{\dagger}(\bm{y}_{\perp}) -\mathds{1} \Big) \Big] \, v(\bm{p}) \, .
\label{eq:amplitude-finite-V3-generic}
\end{align}
In order to obtain these pieces for $(V13)$ and $(V14)$ and their $q\leftrightarrow \bar{q}$ counterparts, we need to compute the finite pieces in the loop contribution described by the graph in Fig.~\ref{fig:vertex-correction} for the cases $l_{2}^{-} >0 $ (denoted by Case A in the discussion of Sec.~\ref{sec:vertex-corrections-free-gluon}) and $l_{2}^{-} <0$ (Case B) and add them up. The computation of the finite contribution from $(V15)$ is considerably tedious and explicit expressions will only be provided for numerical studies in future. Here we sketch the structure of these pieces for the graph $(V13)$ in terms of the constituent integrals that appear in virtual graph computations as we have done in the previous sections. The contributions for $(V14)$ have a similar structure to that of $(V13)$ and can be obtained following the same methods.

For $(V13)$, we discussed the computation of the divergent piece in detail in Sec.~\ref{sec:vertex-corrections-free-gluon}. From the expressions for the gluon loop contribution obtained separately for the two cases in Eqs.~\ref{eq:gamma-A-V13} and \ref{eq:gamma-B-V13}, we can isolate the pieces proportional to the constituent integrals $\mathcal{I}_{v}^{(2,i)} (\bm{V}_{\perp},\Delta)$ and $\mathcal{I}_{v}^{(2,0)} (\bm{V}_{\perp},\Delta)$ which will yield finite results in $d=2$ dimensions. A straightforward application of the identity in Eq.~\ref{eq:gamma-identity-2} in these equations tells us that there are a lot of such contributions proportional to these integrals each with different gamma matrix structures. We can however collect all these pieces and express them as polynomials in $z_{l}$ which will assist in the numerical evaluation of such contributions. With this strategy in mind, we can write the finite contribution to $(V13)$ as
\begin{align}
R^{(V13)}_{\rm finite;\mu \alpha}(\bm{l}_{1\perp})= \hyperref[eq:R-finite-V13-A]{R^{(V13);A}_{\rm finite;\mu \alpha}(\bm{l}_{1\perp})} + \hyperref[eq:R-finite-V13-B]{R^{(V13);B}_{\rm finite;\mu \alpha}(\bm{l}_{1\perp})} \, ,
\label{eq:R-finite-V13}
\end{align}
where 
\begin{align}
R^{(V13);A}_{\rm finite;\mu \alpha}(\bm{l}_{1\perp}) &=  \int_{0}^{1-z_{\bar{q}}} \frac{\mathrm{d}z_{l}}{2\pi} \, \gamma_{\alpha} \frac{\slashed{k}+\slashed{k}_{\gamma}}{2k.k_{\gamma}} \, \gamma^{-} \, \frac{\gamma^{+}(1-z_{\bar{q}}) \, q^{-} - \bm{\gamma}_{\perp}.\bm{l}_{1\perp}}{(q^{-})^{2} \, [\bm{l}_{1\perp}^{2}+\Delta^{\rm LO:(1)} ] } \, \Big\{ \Big( (F1)_{\mu}^{(V13);A} \Big)^{i} \, \mathcal{I}^{(2,i)}_{v} (\bm{V}_{\perp}^{(V13);A},\Delta^{(V13);A}) \nonumber \\
& +(F2)^{(V13);A}_{\mu} \, \mathcal{I}^{(2,0)}_{v} (\bm{V}_{\perp}^{(V13);A},\Delta^{(V13);A})  \Big\} \, \gamma^{-} \, , 
\label{eq:R-finite-V13-A}
\end{align}
and
\begin{align}
R^{(V13);B}_{\rm finite;\mu \alpha}(\bm{l}_{1\perp}) &=  \int_{0}^{-z_{\bar{q}}} \frac{\mathrm{d}z_{l}}{2\pi} \, \gamma_{\alpha} \frac{\slashed{k}+\slashed{k}_{\gamma}}{2k.k_{\gamma}} \, \gamma^{-} \, \frac{\gamma^{+}(1-z_{\bar{q}}) \, q^{-} - \bm{\gamma}_{\perp}.\bm{l}_{1\perp}}{(q^{-})^{2} \, [\bm{l}_{1\perp}^{2}+\Delta^{\rm LO:(1)} ] } \, \Big\{ \Big( (G1)_{\mu}^{(V13);B}\Big)^{i} \, \mathcal{I}^{(2,i)}_{v} (\bm{V}_{\perp}^{(V13);B},\Delta^{(V13);B}) \nonumber \\
& +(G2)^{(V13);B}_{\mu} \, \mathcal{I}^{(2,0)}_{v} (\bm{V}_{\perp}^{(V13);B},\Delta^{(V13);B})  \Big\} \, \gamma^{-} \, .
\label{eq:R-finite-V13-B}
\end{align} 
The finite expressions for these integrals are given in Eqs.~\ref{eq:I-v-2-i-and-I-v-2-0}. The arguments $\bm{V}_{\perp}$ and $\Delta$ of these are in general different for the two cases. As described in Eq.~\ref{eq:V-and-Delta-in-terms-of-zl} we can always express these in terms of $z_{l}$. The coefficients $\bm{c}_{1\perp}$ and $c_{3}$ are identical for the two cases and are obtained as 
\begin{align}
\bm{c}_{1}^{(V13);A}=\bm{c}_{1}^{(V13);B}= \alpha \, \bm{l}_{1\perp} \, , \quad c_{3}^{(V13);A}=c_{3}^{(V13);B} = \alpha (1-\alpha) \, \bm{l}_{1\perp}^{2} +(1-\alpha) \, z_{\bar{q}} (1-z_{\bar{q}} ) \
, Q^{2} \, .
\label{eq:c1-c3-V13}
\end{align}
The expressions for the remaining coefficients are different for the two cases and are provided below separately:
\begin{itemize}
\item Case A: $0<z_{l} <1-z_{\bar{q}}$
\begin{align}
\bm{c}_{2\perp}^{(V13);A}&= -\frac{\alpha}{1-z_{\bar{q}}} \, \bm{l}_{1\perp} \, , \quad  c_{4}^{(V13);A}= \frac{1}{1-z_{\bar{q}}} \, \Big\{ -2 \, \alpha \, (1-\alpha) \, \bm{l}_{1\perp}^{2} + \alpha \, \frac{\bm{l}_{1\perp}^{2}+\Delta^{\rm LO:(1)}}{z_{\bar{q}}} + Q^{2} \, (1-\alpha) \, (1-z_{\bar{q}}) \, (1-2z_{\bar{q}}) \Big\} \, , \nonumber \\
c_{5}^{(V13);A}&= \frac{1}{(1-z_{\bar{q}})^{2}} \Big\{ \alpha \, (1-\alpha) \, \bm{l}_{1\perp}^{2} -\alpha \,\frac{\bm{l}_{1\perp}^{2}+\Delta^{\rm LO:(1)}}{z_{\bar{q}}} -Q^{2} \, (1-\alpha) \, (1-z_{\bar{q}})^{2} \Big\} \, .
\end{align}
\item Case A: $0>z_{l} >-z_{\bar{q}}$
\begin{align}
\bm{c}_{2\perp}^{(V13);B}&= \frac{\alpha}{z_{\bar{q}}} \, \bm{l}_{1\perp} \, , \quad  c_{4}^{(V13);B}= \frac{1}{z_{\bar{q}}} \, \Big\{ 2 \, \alpha \, (1-\alpha) \, \bm{l}_{1\perp}^{2} -  \alpha \, \frac{\bm{l}_{1\perp}^{2}+\Delta^{\rm LO:(1)}}{1-z_{\bar{q}}} + Q^{2} \, (1-\alpha) \, z_{\bar{q}} \, (1-2z_{\bar{q}}) \Big\} \, , \nonumber \\
c_{5}^{(V13);B}&= \frac{1}{z_{\bar{q}}^{2}} \Big\{ \alpha \, (1-\alpha) \, \bm{l}_{1\perp}^{2} -\alpha \,\frac{\bm{l}_{1\perp}^{2}+\Delta^{\rm LO:(1)}}{1-z_{\bar{q}}} -Q^{2} \, (1-\alpha) \, z_{\bar{q}}^{2} \Big\} \, .
\end{align}
\end{itemize}
The terms $(Fi)_{\mu}$ and $(Gi)_{\mu}$ ($i=1,2$) that multiply these constituent integrals can also be expressed in terms of $z_{l}$. This allows for` a straightforward numerical evaluation of these finite pieces.

The structure of the finite pieces in $(V14)$ is exactly similar to that of $(V13)$ with the only difference being the expressions for the terms that multiply the constituent integrals. This is because this process has a different Dirac structure that governs the nature of such terms. Finally, for $(V15)$ we have two contributions for the case $z_{l} >0$ depending on the magnitude of $z_{l}$ relative to $z_{q}$. These two, when added with the contribution for the case $0>z_{l}<-z_{\bar{q}}$, gives the net finite contribution from this graph. For each case, we will encounter a subset of the constituent integrals defined in Eq.~\ref{eq:constituent-integrals-V1} which are finite in $d=2$ dimensions. As in the previous cases considered, Eq.~\ref{eq:replacements-qqbar-exchange} allows us to obtain the corresponding expressions for the finite contributions of the $q\leftrightarrow\bar{q}$ interchanged processes $(V16)-(V18)$.

\subsection{Computation of $\mathcal{M}_{\rm finite}^{\rm Vert.(4)}$} \label{sec:finite-pieces-Ver4}
The finite pieces for the diagrams in Fig.~\ref{fig:NLO-vertex-4} and their $q\leftrightarrow \bar{q}$ counterparts are contained in 
\begin{align}
\mathcal{M}^{\text{Vert.}(4)}_{\mu \alpha;\rm finite}&=2\pi (eq_{f}g)^{2} \delta(1-z^{v}_{\rm tot}) \, \int \mathrm{d} \Pi_{\perp}^{\rm LO} \,  \overline{u}(\bm{k}) \, R^{\rm Vert.(4)}_{\rm finite;\mu \alpha} (\bm{l}_{1\perp})   \,  \Big[  t^{a} \tilde{U}(\bm{x}_{\perp}) \tilde{U}^{\dagger} (\bm{y}_{\perp})t_{a} -C_{F} \, \mathds{1}  \Big]   \, v(\bm{p}) \, ,
\label{eq:amplitude-finite-V4-generic}
\end{align}
where 
\begin{equation}
R^{\rm Vert.(1)}_{\rm finite;\mu \alpha} (\bm{l}_{1\perp})= \sum_{\beta=19}^{24} \Big( \Re^{(V\beta)}_{\mu \alpha} (\bm{l}_{1\perp}) +R^{(V\beta)}_{(\rm II);\mu \alpha} (\bm{l}_{1\perp})  \Big)  \, .
\label{eq:R-Vertex4-finite}
\end{equation}
Again, in the notation used elsewhere, $\Re^{(V\beta)}$ is the remainder between terms in the amplitude proportional to $1/z_{l}$ and those proportional to logarithms in $z_{0}$. The second term $R_{\rm (II)}^{(V\beta)}$ is comprised of finite terms that are not proportional to $1/z_{l}$. In addition for these virtual graphs, we can have two possible cases depending on the sign of $l_{2}^{-}$, the loop momentum carried by the virtual gluon. We therefore have to add the contributions from the two cases in order to obtain the final finite result. 

We will now sketch the computation of the finite pieces by considering the process $(V19)$ in Fig.~\ref{fig:NLO-vertex-4} as an example. For this process, the rapidity divergent pieces were computed for both signs of $l_{2}^{-}$ in Sec.~\ref{sec:vertex-corrections-free-gluon}. The corresponding expressions are given by Eqs.~\ref{eq:R-div-V19-A} and \ref{eq:R-div-V19-B}. These were obtained from the contributions that are proportional to $1/z_{l}$ which are denoted by $R_{(\rm I)}^{(V19);A,B}$ respectively for the cases of $l_{2}^{-} >0$ and $l_{2}^{-} <0$. The remainder term constituting the finite contribution of $(V19)$ can therefore be written as 
\begin{equation}
\Re^{(V19)}_{\mu \alpha} (\bm{l}_{1\perp}) = \Big(  \hyperref[eq:R-V19-propto-zl-inverse-A]{R^{(V19);A}_{(\rm I);\mu \alpha} (\bm{l}_{1\perp})} - \hyperref[eq:R-div-V19-A]{R^{(V19);A}_{\rm div.;\mu \alpha} (\bm{l}_{1\perp})}  \Big) + \Big(  \hyperref[eq:R-V19-propto-zl-inverse-B]{ R^{(V19);B}_{(\rm I);\mu \alpha} (\bm{l}_{1\perp})} - \hyperref[eq:R-div-V19-B]{R^{(V19);B}_{\rm div.;\mu \alpha} (\bm{l}_{1\perp})} \Big) \, , 
\label{eq:remainder-V19}
\end{equation}
where $R_{(\rm I)}^{(V19);A,B}$ are given by Eqs.~\ref{eq:R-V19-propto-zl-inverse-A} and \ref{eq:R-V19-propto-zl-inverse-B} respectively.

The computation of the second term $R_{(\rm II)}$ contributing to the finite piece is very tedious for these processes. For each case, there are several finite pieces in the amplitude that are proportional to the different constituent integrals that appear in virtual graph computations. We can collect all these pieces and express them as polynomials in the gluon loop momentum fraction $z_{l}$ with coefficients that depend on the gamma matrix structure of the various processes. The resulting expressions are lengthy we will not provide them here  but are too available as Mathematica scripts .

Because we express the arguments of our constituent integrals also in terms of $z_{l}$, one only needs to perform the Feynman parameter integrations in these integral definitions which will give the final expressions for these as a function of $z_{l}$. The remaining integration over $z_{l}$ can then be performed easily as there are no singularities left over. With this in mind, we can write down the generic expressions for $R_{\rm (II)}$  for $(V19)-(V21)$ as
\begin{align}
R^{(V\beta)}_{(\rm II);\mu \alpha}( \bm{l}_{1\perp}) = R^{(V\beta);A}_{(\rm II);\mu \alpha}( \bm{l}_{1\perp})+R^{(V\beta);B}_{(\rm II);\mu \alpha}( \bm{l}_{1\perp}) \, , \quad \beta=19,\ldots,21 \, ,
\label{eq:R-II-Vert4}
\end{align}
where $A$ and $B$ denote respectively for each process the contributions from the cases where $l_{2}^{-}>0$ and $l_{2}^{-}<0$. In addition, for the process $(V20)$ there are two contributions for the case  $l_{2}^{-}>0$ namely from $0<l_{2}^{-}<k^{-}$ and $k^{-}<l_{2}^{-}<k^{-}+k^{-}_{\gamma}$ that add to the finite terms.

For $(V19)$ we can write 
\begin{equation}
R^{(V19)}_{(\rm II);\mu \alpha}( \bm{l}_{1\perp})=\hyperref[eq:R-II-V19-A]{R^{(V19);A}_{(\rm II);\mu \alpha}( \bm{l}_{1\perp})}+\hyperref[eq:R-II-V19-B]{R^{(V19);B}_{(\rm II);\mu \alpha}( \bm{l}_{1\perp})} \, , 
\label{eq:R-II-V19}
\end{equation}
where
\begin{align}
R^{(V19);A}_{(\rm II);\mu \alpha}( \bm{l}_{1\perp})& = - \int_{0}^{1-z_{\bar{q}}} \frac{\mathrm{d}z_{l}}{2\pi} \, \gamma_{\alpha} \frac{\slashed{k}+\slashed{k}_{\gamma}}{2k.k_{\gamma}} \, \frac{1}{4 \, (1-z_{\bar{q}} ) \, (q^{-})^{2}} \, \Big\{(F1)_{\mu}^{ijk} \, \mathcal{I}_{v}^{3,ijk} (\bm{V}^{(V19);A}, \Delta^{(V19);A})+ (F2)_{\mu}^{i} \nonumber \\
& \times \mathcal{I}_{v}^{(3,ijj)} (\bm{V}^{(V19);A}, \Delta^{(V19);A}) + (F3)_{\mu}^{ij} \, \mathcal{I}_{v}^{(3,ij)} (\bm{V}^{(V19);A}, \Delta^{(V19);A}) + (F4)_{\mu} \, \mathcal{I}_{v}^{(3,ii)} (\bm{V}^{(V19);A}, \Delta^{(V19);A}) \nonumber \\
& + (F5)_{\mu}^{i} \, \mathcal{I}_{v}^{(3,i)} (\bm{V}^{(V19);A}, \Delta^{(V19);A})  + (F6)_{\mu} \, \mathcal{I}_{v}^{(3,0)} (\bm{V}^{(V19);A}, \Delta^{(V19);A}) \Big\} \,  , 
\label{eq:R-II-V19-A}
\end{align}
and
\begin{align}
R^{(V19);B}_{(\rm II);\mu \alpha}( \bm{l}_{1\perp})& = - \int_{0}^{z_{\bar{q}}}  \frac{\mathrm{d}z_{l}}{2\pi} \, \gamma_{\alpha} \frac{\slashed{k}+\slashed{k}_{\gamma}}{2k.k_{\gamma}} \, \frac{1}{4 \, z_{\bar{q}}  \, (q^{-})^{2}} \, \Big\{(G1)_{\mu}^{ijk} \, \mathcal{I}_{v}^{3,ijk} (\bm{V}^{(V19);B}, \Delta^{(V19);B})+ (G2)_{\mu}^{i} \, \mathcal{I}_{v}^{(3,ijj)} (\bm{V}^{(V19);B}, \Delta^{(V19);B}) \nonumber \\
& + (G3)_{\mu}^{ij} \, \mathcal{I}_{v}^{(3,ij)} (\bm{V}^{(V19);B}, \Delta^{(V19);B}) + (G4)_{\mu} \, \mathcal{I}_{v}^{(3,ii)} (\bm{V}^{(V19);B}, \Delta^{(V19);B}) + (G5)_{\mu}^{i} \, \mathcal{I}_{v}^{(3,i)} (\bm{V}^{(V19);B}, \Delta^{(V19);B}) \nonumber \\
& + (G6)_{\mu} \, \mathcal{I}_{v}^{(3,0)} (\bm{V}^{(V19);B}, \Delta^{(V19);B}) \Big\} \,  .
\label{eq:R-II-V19-B}
\end{align}
In the above expressions, the $(Fi)$ and $(Gi)$'s ($i=1,\ldots,6$) can always be expressed as polynomials in $z_{l}$ with coefficients that are basically products of gamma matrices. The finite expressions for the constituent integrals appearing above are given by Eq.~\ref{eq:constituent-integrals-V1}. As usual we will express the arguments of these integrals in terms of $z_{l}$ as in Eq.~\ref{eq:V-and-Delta-in-terms-of-zl}. The coefficients $\bm{c}_{1\perp}$ and $c_{3}$ for both cases are identical and are given for $(V19)$ by Eqs.~\ref{eq:c1-c3-V19-A}. The remaining coefficients for cases A and B are given below for the process $(V19)$,
\begingroup
\allowdisplaybreaks
\begin{align}
\bm{c}_{2\perp}^{(V19);A}&=  \frac{1}{1-z_{\bar{q}}} \, \Big[ \alpha_{1} \, (\bm{k}_{\perp}+\bm{k}_{\gamma \perp} ) -\alpha_{2} \, \Big\{ (1-z_{\bar{q}}) \, \bm{p}_{\perp} - z_{\bar{q}} \, (\bm{k}_{\perp}+\bm{k}_{\gamma \perp} ) \Big\} \Big] \, , \nonumber \\
c_{4}^{(V19);A} & = - 2 \, \alpha_{1} (1-\alpha_{1}) \, \frac{\bm{l}_{1\perp}.(\bm{k}_{\perp}+\bm{k}_{\gamma \perp})}{1-z_{\bar{q}}} -2 \, \alpha_{2} (1-\alpha_{2}) \,  \frac{ \Big\{ (1-z_{\bar{q}}) \, \bm{p}_{\perp} - z_{\bar{q}} \, (\bm{k}_{\perp}+\bm{k}_{\gamma \perp} ) \Big\}^{2}}{1-z_{\bar{q}}}  - (1-\alpha_{1}-\alpha_{2}) \, \frac{2k.k_{\gamma}}{1-z_{\bar{q}}}   \nonumber \\
& - 2\, \alpha_{1} \alpha_{2}  \, \frac{\bm{l}_{1\perp}+\bm{k}_{\perp}+\bm{k}_{\gamma \perp}}{1-z_{\bar{q}}} .  \Big\{ (1-z_{\bar{q}}) \, \bm{p}_{\perp} - z_{\bar{q}} \, (\bm{k}_{\perp}+\bm{k}_{\gamma \perp} ) \Big\} +\alpha_{1} \, (1-2z_{\bar{q}}) \, Q^{2} - \alpha_{2} \, (1-2 z_{\bar{q}} ) \, (2p.k+2k.k_{\gamma}+2p.k_{\gamma} ) \, , \nonumber \\
c_{5}^{(V19);A} & = \alpha_{1} (1-\alpha_{1}) \, \frac{(\bm{k}_{\perp}+\bm{k}_{\gamma \perp})^{2}}{(1-z_{\bar{q}})^{2}} + \alpha_{2} (1-\alpha_{2} ) \frac{\Big\{ (1-z_{\bar{q}}) \, \bm{p}_{\perp} - z_{\bar{q}} \, (\bm{k}_{\perp}+\bm{k}_{\gamma \perp} ) \Big\}^{2}}{(1-z_{\bar{q}})^{2}} + (1-\alpha_{1}-\alpha_{2}) \, \frac{2k.k_{\gamma}}{(1-z_{\bar{q}})^{2}} \nonumber \\
& + 2 \, \alpha_{1} \alpha_{2} \frac{(\bm{k}_{\perp}+\bm{k}_{\gamma \perp}).\Big\{ (1-z_{\bar{q}}) \, \bm{p}_{\perp} - z_{\bar{q}} \, (\bm{k}_{\perp}+\bm{k}_{\gamma \perp} ) \Big\}   }{(1-z_{\bar{q}})^{2}} - \alpha_{1} \, Q^{2} + \alpha_{2} \, (2p.k+2k.k_{\gamma}+2p.k_{\gamma} )  \, .
\label{eq:c2-c4-c5-V19-A}
\end{align}
\endgroup
\noindent \rule{16cm} {0.5pt}
\begingroup
\allowdisplaybreaks
\begin{align}
\bm{c}_{2\perp}^{(V19);B}&= - \frac{1}{z_{\bar{q}}} \, \Big[ \alpha_{1} \, \bm{p}_{\perp} + \alpha_{2} \, \Big\{ (1-z_{\bar{q}}) \, \bm{p}_{\perp} - z_{\bar{q}} \, (\bm{k}_{\perp}+\bm{k}_{\gamma \perp} ) \Big\} \Big] \, , \nonumber \\
c_{4}^{(V19);B} & =  2 \, \alpha_{1} (1-\alpha_{1}) \, \frac{\bm{l}_{1\perp}.\bm{p}_{\perp}}{z_{\bar{q}}} -2 \, \alpha_{2} (1-\alpha_{2}) \,  \frac{ \Big\{ (1-z_{\bar{q}}) \, \bm{p}_{\perp} - z_{\bar{q}} \, (\bm{k}_{\perp}+\bm{k}_{\gamma \perp} ) \Big\}^{2}}{z_{\bar{q}}} + \alpha_{2} \, (1-2 z_{\bar{q}} ) \, (2p.k+2k.k_{\gamma}+2p.k_{\gamma} )  \nonumber \\
&  2\, \alpha_{1} \alpha_{2}  \, \frac{\bm{p}_{\perp}-\bm{l}_{1\perp}}{z_{\bar{q}}} .  \Big\{ (1-z_{\bar{q}}) \, \bm{p}_{\perp} - z_{\bar{q}} \, (\bm{k}_{\perp}+\bm{k}_{\gamma \perp} ) \Big\} - \alpha_{1} \, (1-2z_{\bar{q}}) \, Q^{2}  \, , \nonumber \\
c_{5}^{(V19);B} & = \alpha_{1} (1-\alpha_{1}) \, \frac{\bm{p}_{\perp}^{2}}{z_{\bar{q}}^{2}} + \alpha_{2} (1-\alpha_{2} ) \frac{\Big\{ (1-z_{\bar{q}}) \, \bm{p}_{\perp} - z_{\bar{q}} \, (\bm{k}_{\perp}+\bm{k}_{\gamma \perp} ) \Big\}^{2}}{z_{\bar{q}}^{2}}  - \alpha_{1} \, Q^{2} + \alpha_{2} \, (2p.k+2k.k_{\gamma}+2p.k_{\gamma} )  \nonumber \\
& -  2 \, \alpha_{1} \alpha_{2} \frac{\bm{p}_{\perp}.\Big\{ (1-z_{\bar{q}}) \, \bm{p}_{\perp} - z_{\bar{q}} \, (\bm{k}_{\perp}+\bm{k}_{\gamma \perp} ) \Big\}   }{z_{\bar{q}}^{2}} \, .
\label{eq:c2-c4-c5-V19-A}
\end{align}
\endgroup
\noindent \rule{16cm} {0.5pt}

The same logic holds for the process $(V21)$ whose divergent pieces for the two signs of $l_{2}^{-}$ are provided in Appendix~\ref{sec:T-V4-div-parts} by Eq.~\ref{eq:R-div-V21-A-B}. The remainder term constituting the finite contribution of $(V21)$ can be written as
\begin{equation}
\Re^{(V21)}_{\mu \alpha} (\bm{l}_{1\perp}) = \Big(  \hyperref[eq:R-V21-propto-zl-inverse-A-B]{R^{(V21);A}_{(\rm I);\mu \alpha} (\bm{l}_{1\perp})} - \hyperref[eq:R-div-V21-A-B]{R^{(V21);A}_{\rm div.;\mu \alpha} (\bm{l}_{1\perp})}  \Big) + \Big(  \hyperref[eq:R-V21-propto-zl-inverse-A-B]{ R^{(V21);B}_{(\rm I);\mu \alpha} (\bm{l}_{1\perp})} - \hyperref[eq:R-div-V21-A-B]{R^{(V21);B}_{\rm div.;\mu \alpha} (\bm{l}_{1\perp})} \Big) \, , 
\label{eq:remainder-V19}
\end{equation}
where the pieces proportional to $1/z_{l}$ denoted by $R_{(\rm I)}^{(V21);A,B}$ are given by Eq.~\ref{eq:R-V21-propto-zl-inverse-A-B} . The constituent integrals that enter the calculation of this process are given in Eq.~\ref{eq:constituent-integrals-V3-V4} and involve three Feynman parameters. Nevertheless the arguments $\bm{V}_{\perp}$ and $\Delta$ can always be expressed as functions of these parameters and the gluon loop momentum fraction $z_{l}$ to facilitate the numerical computation. Similarly to $(V19)$, we can also write down the generic expressions for the second finite piece $R_{\rm (II)}$ for the two cases in terms of some functions which are polynomials in $z_{l}$ in the following manner,
\begin{equation}
R^{(V21)}_{(\rm II);\mu \alpha}( \bm{l}_{1\perp})=\hyperref[eq:R-II-V21-A]{R^{(V21);A}_{(\rm II);\mu \alpha}( \bm{l}_{1\perp})}+\hyperref[eq:R-II-V21-B]{R^{(V21);B}_{(\rm II);\mu \alpha}( \bm{l}_{1\perp})} \, , 
\label{eq:R-II-V21}
\end{equation}
where
\begin{align}
R^{(V21);A}_{(\rm II);\mu \alpha}( \bm{l}_{1\perp})&= - \int_{0}^{z_{q}} \frac{\mathrm{d}z_{l}}{2\pi} \frac{1}{4 \, z_{q} (1-z_{\gamma})^{2} \, (q^{-} )^{2}} \,  \Big[(\tilde{F}1)_{\mu \alpha} \, \mathcal{I}_{v}^{(4,iijj)}( \bm{V}_{\perp}^{(V21);A},\Delta^{(V21);A} )  +(\tilde{F}2)_{\mu \alpha}^{ij} \, \mathcal{I}_{v}^{(4,ijkk)} ( \bm{V}_{\perp}^{(V21);A},\Delta^{(V21);A} ) \nonumber \\
& +(\tilde{F}3)_{\mu \alpha}^{i} \, \mathcal{I}_{v}^{(4,ijj)} ( \bm{V}_{\perp}^{(V21);A},\Delta^{(V21);A} ) +(\tilde{F}4)_{\mu \alpha}^{ijk} \, \mathcal{I}_{v}^{(4,ijk)} ( \bm{V}_{\perp}^{(V21);A},\Delta^{(V21);A} )  \nonumber \\
&+ (\tilde{F}5)_{\mu \alpha} \,  \mathcal{I}_{v}^{(4,ii)} ( \bm{V}_{\perp}^{(V21);A},\Delta^{(V21);A} )  +(\tilde{F}6)_{\mu \alpha}^{ij} \,  \mathcal{I}_{v}^{(4,ij)} ( \bm{V}_{\perp}^{(V21);A},\Delta^{(V21);A} ) +(\tilde{F}7)_{\mu \alpha}^{i} \,  \mathcal{I}_{v}^{(4,i)} ( \bm{V}_{\perp}^{(V21);A},\Delta^{(V21);A} )\Big]  \, , 
\label{eq:R-II-V21-A}
\end{align}
and  
\begin{align}
R^{(V21);B}_{(\rm II);\mu \alpha}( \bm{l}_{1\perp})&=  \int_{0}^{-z_{q}} \frac{\mathrm{d}z_{l}}{2\pi} \frac{1}{4 \, z_{\bar{q}} (1-z_{\gamma})^{2} \, (q^{-} )^{2}} \,  \Big[(\tilde{G}1)_{\mu \alpha} \, \mathcal{I}_{v}^{(4,iijj)}( \bm{V}_{\perp}^{(V21);B},\Delta^{(V21);B} )  +(\tilde{G}2)_{\mu \alpha}^{ij} \, \mathcal{I}_{v}^{(4,ijkk)} ( \bm{V}_{\perp}^{(V21);B},\Delta^{(V21);B} ) \nonumber \\
& +(\tilde{G}3)_{\mu \alpha}^{i} \, \mathcal{I}_{v}^{(4,ijj)} ( \bm{V}_{\perp}^{(V21);B},\Delta^{(V21);B} ) +(\tilde{G}4)_{\mu \alpha}^{ijk} \, \mathcal{I}_{v}^{(4,ijk)} ( \bm{V}_{\perp}^{(V21);B},\Delta^{(V21);B} )  \nonumber \\
&+ (\tilde{G}5)_{\mu \alpha} \,  \mathcal{I}_{v}^{(4,ii)} ( \bm{V}_{\perp}^{(V21);B},\Delta^{(V21);B} )  +(\tilde{G}6)_{\mu \alpha}^{ij} \,  \mathcal{I}_{v}^{(4,ij)} ( \bm{V}_{\perp}^{(V21);B},\Delta^{(V21);B} ) +(\tilde{G}7)_{\mu \alpha}^{i} \,  \mathcal{I}_{v}^{(4,i)} ( \bm{V}_{\perp}^{(V21);B},\Delta^{(V21);B} )\Big]  \, .
\label{eq:R-II-V21-B}
\end{align}
The arguments of the above integrals can always be expressed in the form given by Eq.~\ref{eq:V-and-Delta-in-terms-of-zl}. The finite pieces for $(V20)$ can be expressed in an exactly similar manner as for $(V21)$; we encounter the same constituent integrals in the latter albeit with different expressions for the arguments.

%%%%%%%%%%%%%%%%%%%%%%%%%%%%%%%%%%%%%%%%%%%%%%%%%%%%%%%%%%%%%%%%%%%%%%%%%%%%%%

\section{Proof of the sub-dominance of non-collinearly divergent contributions in the SCA} \label{sec:non-collinear-contributions}

In this section, we will repeat a short proof given in \cite{boussarie_2017} for the power suppression (in powers of jet cone radius $R$) of the non-collinearly divergent contributions. If partons $i$ and $j$ form the first jet `J' and parton $k$ forms the second jet `K' then we introduce new variables such that the differential measure of the final particle phase space transforms as
\begin{equation}
\mathrm{d}z_{i} \, \mathrm{d}z_{j} \, \mathrm{d}z_{k} \, \mathrm{d}^{d} \bm{p}_{i \perp} \, \mathrm{d}^{d} \bm{p}_{j \perp} \, \mathrm{d}^{d} \bm{p}_{k\perp} \rightarrow \mathrm{d}z_{i} \, \mathrm{d}^{d} \bm{\mathcal{C}}_{ij,\perp} \, \mathrm{d}z_{J} \, \mathrm{d} z_{K} \, \mathrm{d}^{d} \bm{p}_{J\perp} \, \mathrm{d}^{d} \bm{p}_{K\perp} \, , 
\end{equation}
where the jet variables are 
\begin{equation}
(z_{J},\bm{p}_{J\perp}, z_{K},\bm{p}_{K\perp}, \bm{\mathcal{C}}_{ij,\perp} ) = \Big( z_{i}+z_{j},\bm{p}_{i \perp}+\bm{p}_{j \perp},z_{k},\bm{p}_{k \perp}, \frac{z_{i}}{z_{J}}\, \bm{p}_{j \perp}-\frac{z_{j}}{z_{J}} \, \bm{p}_{i\perp} \Big) \, ,
\end{equation}
The integration over $\bm{\mathcal{C}}_{ij,\perp}$ is restricted by the small cone condition given by Eq.~\ref{eq:SCA-in-terms-of-C}. We can therefore write the amplitude squared in general as 
\begin{align}
\vert \mathcal{M} \vert^{2} \propto  \int^{\bm{\mathcal{C}}^{2}_{ij,\perp} < R^{2} \bm{p}_{J\perp}^{2}  \text{min} \Big( \frac{z_{i}^{2}}{z_{J}^{2}},\frac{z_{j}^{2}}{z_{J}^{2}} \Big)}  \mathrm{d}^{d} \bm{\mathcal{C}}_{ij,\perp}  \, F(\bm{\mathcal{C}}_{ij,\perp}) \, .
\end{align}
The function $F(\bm{\mathcal{C}}_{ij,\perp})$ can be expanded around the collinear limit $\bm{\mathcal{C}}_{ij,\perp}=0$ by writing 
\begin{equation}
F(\bm{\mathcal{C}}_{ij,\perp})= \frac{[F(\bm{\mathcal{C}}_{ij,\perp})]_{\rm collinear}}{\bm{\mathcal{C}}_{ij,\perp}^{2}} + [F(\bm{\mathcal{C}}_{ij,\perp})]_{\rm collinear=0}+ O(\bm{\mathcal{C}}_{ij,\perp}) \, .
\end{equation}
The contributions without collinear divergences are denoted by $[F(\bm{\mathcal{C}}_{ij,\perp})]_{\rm collinear=0}$. We can now perform the integration over the collinearity variable in $d$ dimensions to obtain the following result
\begin{align}
\vert \mathcal{M} \vert^{2} \propto \frac{2\pi^{d/2}}{\Gamma(d/2)} \, \Bigg\{ \frac{[F(\bm{\mathcal{C}}_{ij,\perp})]_{\rm collinear}}{d-2} + \frac{[F(\bm{\mathcal{C}}_{ij,\perp})]_{\rm collinear=0}}{d} \, R^{2} \bm{p}_{J\perp}^{2} \, \text{min} \Big( \frac{z_{i}^{2}}{z_{J}^{2}} ,  \frac{z_{j}^{2}}{z_{J}^{2}} \Big) \Bigg\} \, \Bigg( R^{2} \bm{p}_{J\perp}^{2} \, \text{min} \Big( \frac{z_{i}^{2}}{z_{J}^{2}} ,  \frac{z_{j}^{2}}{z_{J}^{2}} \Big) \Bigg)^{\frac{d}{2}-1} \, .
\end{align}
It is clear from the above expression that contributions without collinear divergences have a relative suppression of $R^{2}$ and are therefore subdominant in the SCA.

\bibliography{bibliography.bib}

\end{document}